%% file: FLAG_master.tex
\documentclass[11pt,a4paper]{article}
\usepackage{latexsym}
\usepackage{graphicx}
\usepackage{psfrag}
\usepackage{amsmath}
\usepackage{amssymb}
\usepackage{color}
\usepackage{rotating}
\usepackage{multirow}
\usepackage{colortbl}
\usepackage{lscape,epsfig}
\usepackage{a4wide}
\usepackage[]{xcolor}
\usepackage{authblk}
\usepackage{verbatim}
\usepackage{rotating}

\usepackage[dvips,colorlinks=true,a4paper=true,backref=false,linktocpage=true,citecolor=blue,urlcolor=blue,linkcolor=blue,pdfpagemode=UseOutlines]{hyperref}
\usepackage[sort&compress,square,comma,numbers]{natbib}
\usepackage{hypernat}

\input{defs.tex}

\begin{document}

\noindent
\begin{minipage}{0.4\textwidth}
\begin{flushleft}
\vspace{-1cm}
CP3-Origins-2016-023 DNRF90\\ % Michele
DESY 16-111\\ % Rainer
DIAS-2016-23\\ % Michele
Edinburgh 2016/11\\ % Rainer, Roger
FTUAM-16-23\\ % Carlos
HIM-2016-02\\% Hartmut
\end{flushleft}
\end{minipage}
\hfill
\begin{minipage}{0.4\textwidth}
\begin{flushright}
\vspace{-1cm}
IFT-UAM/CSIC-16-057\\ % Carlos
LPT-Orsay-16-47\\ % Damir
MITP/16-059\\ % Hartmut
RM3-TH/16-7\\ % Silvano
ROM2F/2016/05\\ % Petros
YITP-16-77\\ % Sinya
\end{flushright}
\end{minipage}

\vskip 2.75cm
\begin{center}
{\bf \Huge Review of lattice results concerning\\ low-energy particle physics}

\vspace{0.5cm}
%\today
July 1, 2016

\vspace{0.5cm}
{\bf \large FLAG Working Group} 
\renewcommand*{\thefootnote}{\fnsymbol{footnote}}

\vspace{0.5cm}
{\small S.~Aoki,$^1$
Y.~Aoki,$^{2,3}$\footnote{Present address: $^{14}$}
D.~Be\v cirevi\' c,$^{4}$
C.~Bernard,$^5$
T.~Blum,$^{6,3}$ 
G.~Colangelo,$^7$
M.~Della~Morte,$^8$
P.~Dimopoulos,$^{9}$
S.~D\"urr,$^{10}$
H.~Fukaya,$^{11}$
M.~Golterman,$^{12}$
Steven Gottlieb,$^{13}$
S.~Hashimoto,$^{14,15}$
U.~M.~Heller,$^{16}$
R.~Horsley,$^{17}$
A.~J\"uttner,$^{18}$
T.~Kaneko,$^{14,15}$
L.~Lellouch,$^{19}$
H.~Leutwyler,$^7$
C.-J.~D.~Lin,$^{20,19}$
V.~Lubicz,$^{21,22}$
E.~Lunghi,$^{13}$
R.~Mawhinney,$^{23}$		
T.~Onogi,$^{11}$
C.~Pena,$^{24}$
C.~T.~Sachrajda,$^{18}$
S.~R.~Sharpe,$^{25}$
S.~Simula,$^{22}$
R.~Sommer,$^{26}$
A.~Vladikas,$^{27}$
U.~Wenger,$^7$
H.~Wittig$^{28}$}
\end{center}
\renewcommand*{\thefootnote}{\arabic{footnote}}
\setcounter{footnote}{0}

\vskip 0.5cm

{\abstract{ We review lattice results related to pion, kaon, $D$- and
    $B$-meson physics with the aim of making them easily accessible to the
    particle physics community. More specifically, we report on the
    determination of the light-quark masses, the form factor $f_+(0)$,
    arising in the semileptonic $K \to \pi$ transition at zero momentum
    transfer, as well as the decay constant ratio $f_K/f_\pi$ and its 
    consequences for the CKM matrix elements $V_{us}$ and
    $V_{ud}$. Furthermore, we describe the results obtained on the lattice
    for some of the low-energy constants of $SU(2)_L\times SU(2)_R$ and
    $SU(3)_L\times SU(3)_R$ Chiral Perturbation Theory. We review the
    determination of the $B_K$ parameter of neutral kaon mixing
    as well as the additional four $B$ parameters that arise in theories
    of physics beyond the Standard Model. The latter quantities are an
    addition compared to the previous review.
    For the heavy-quark sector, we provide results for $m_c$ and $m_b$
    (also new compared to the previous review),
    as well as those for
    $D$- and $B$-meson decay constants, form factors, and mixing parameters.
    These are the heavy-quark quantities 
    most relevant for the determination of CKM
    matrix elements and the global CKM unitarity-triangle fit.
    Finally, we review the status of lattice determinations of the
    strong coupling constant $\alpha_s$.
}}

\newpage

\begin{flushleft}{\small
$^1$ Center for Gravitational Physics, Yukawa Institute for Theoretical Physics, Kyoto University,
\\ \hskip 0.3cm Kitashirakawa Oiwakecho, Sakyo-ku Kyoto 606-8502, Japan 

$^2$ Kobayashi-Maskawa Institute for the Origin of Particles and the
Universe (KMI), \\ \hskip 0.3cm
Nagoya University, Nagoya 464-8602, Japan

$^3$ RIKEN BNL Research Center, Brookhaven National Laboratory, Upton, NY 11973, USA

$^4$ Laboratoire de Physique Th\'eorique (UMR8627), CNRS, Universit\'e Paris-Sud,\\ \hskip 0.3cm Universit\'e Paris-Saclay, 91405 Orsay, France

$^5$ Department of Physics, Washington University, Saint Louis, MO 63130, USA

$^6$ Physics Department, University of Connecticut, Storrs, CT 06269-3046, USA

$^7$ Albert Einstein Center for Fundamental Physics,
Institut f\"ur Theoretische Physik, \\ \hskip 0.3 cm Universit\"at Bern, Sidlerstr. 5, 3012
Switzerland

$^8$ CP3-Origins \&  Danish IAS, University of Southern
    Denmark, Campusvej 55, \\ \hskip 0.3 cm  DK-5230 Odense M, Denmark
    and \\ \hskip 0.3 cm  IFIC (CSIC),  c/ Catedr\'atico Jos\'e Beltr\'an, 2. E-46980, Paterna, Spain

$^{9}$ Centro Fermi -- Museo Storico della Fisica e Centro Studi e Ricerche Enrico
Fermi Compendio del \\
\hskip 0.3cm Viminale, Piazza del Viminiale 1, I--00184, Rome, Italy,\\
\hskip 0.3cm c/o Dipartimento di Fisica, Universita' di Roma Tor Vergata, Via della Ricerca Scientifica 1, \\
\hskip 0.3cm I--00133 Rome, Italy

\hspace{-0.1cm}$^{10}$ University of Wuppertal, Gau{\ss}stra{\ss}e\,20, 42119 Wuppertal, Germany
   and \\ \hskip 0.3cm J\"ulich Supercomputing Center, Forschungszentrum J\"ulich,
   52425 J\"ulich, Germany 
  
\hspace{-0.1cm}$^{11}$  Department of Physics, Osaka University, Toyonaka, Osaka 560-0043, Japan

\hspace{-0.1cm}$^{12}$ Dept. of Physics and Astronomy, San Francisco State University, San Francisco, CA 94132, USA

\hspace{-0.1cm}$^{13}$  Department of Physics, Indiana University, Bloomington, IN 47405,
  USA
 
\hspace{-0.1cm}$^{14}$ High Energy Accelerator Research Organization (KEK), Tsukuba 305-0801, Japan

\hspace{-0.1cm}$^{15}$ School of High Energy Accelerator Science,
The Graduate University for Advanced Studies \\ \hskip 0.3cm (Sokendai), Tsukuba 305-0801, Japan

\hspace{-0.1cm}$^{16}$ American Physical Society (APS), One Research Road, Ridge, NY 11961, USA

\hspace{-0.1cm}$^{17}$ Higgs Centre for Theoretical Physics, School of Physics and Astronomy, University of Edinburgh, \\ \hskip 0.3cm Edinburgh EH9 3FD, UK

\hspace{-0.1cm}$^{18}$ School of Physics \& Astronomy, University of Southampton, SO17 1BJ, UK

\hspace{-0.1cm}$^{19}$ CNRS, Aix Marseille Universit\'{e}, Universit\'{e} de Toulon, Centre de Physique Th\'{e}orique, \\
\hskip 0.33cm UMR 7332, F-13288 Marseille, France

\hspace{-0.1cm}$^{20}$  Institute of Physics, National Chiao-Tung University, Hsinchu 30010, Taiwan

\hspace{-0.1cm}$^{21}$ Dipartimento di Matematica e Fisica, Universit\`a
                       Roma Tre, Via della Vasca Navale 84, \\ \hskip 0.33 cm 00146 Rome, Italy

\hspace{-0.1cm}$^{22}$ INFN, Sezione di Roma Tre, Via della Vasca Navale 84, 00146 Rome, Italy

\hspace{-0.1cm}$^{23}$ Physics Department, Columbia University, New York, NY 10027, USA
 
\hspace{-0.1cm}$^{24}$ Instituto de F\'{\i}sica Te\'orica UAM/CSIC and
Departamento de F\'{\i}sica Te\'orica, 
\\ \hskip 0.3 cm Universidad Aut\'onoma de Madrid, Cantoblanco 28049 Madrid, Spain 

\hspace{-0.1cm}$^{25}$ Physics Department, University of Washington, Seattle, WA 98195-1560, USA 

\hspace{-0.1cm}$^{26}$ John von Neumann Institute for Computing (NIC), DESY, Platanenallee~6, 15738 Zeuthen, \\ \hskip 0.33 cm Germany

\hspace{-0.1cm}$^{27}$ INFN, Sezione di Tor Vergata, c/o Dipartimento di Fisica,
           Universit\`a di Roma Tor Vergata, \\ \hskip 0.3 cm Via della Ricerca Scientifica 1, 00133 Rome, Italy
  
\hspace{-0.1cm}$^{28}$ PRISMA Cluster of Excellence, Institut f\"ur
Kernphysik and Helmholtz Institute Mainz, \\ \hskip 0.3 cm University of Mainz, 55099 Mainz,
Germany
}
\end{flushleft}

\clearpage
\tableofcontents

\clearpage
\input{introduction.tex}

\input{quality_criteria}

\clearpage
\input{qmass/qmass}

\clearpage
\setcounter{section}{3}
\input{Vudus/VudVus}

\clearpage
\setcounter{section}{4}
\input{LEC/LECs}

\clearpage
\setcounter{section}{5}
\input{BK/BK}

\clearpage
\input{HQ/macros_static.sty}
\setcounter{section}{6}
\clearpage
\input{HQ/HQD.tex}

\clearpage
\setcounter{section}{7}
\input{HQ/HQB.tex}
\clearpage
\setcounter{section}{8}
\input{Alpha_s/alpha}

\input{acknowledgment}

\appendix

\begin{appendix}
\clearpage
\section{Glossary}\label{comm}
\input{Glossary/gloss_app}

\clearpage
\section{Notes}
\input{qmass/qmass_app}
\clearpage
\input{qmass/qmass_mc_app}
\input{qmass/qmass_mb_app}
\setcounter{subsection}{1}

\clearpage
\input{Vudus/VudVus_app}

\clearpage
\input{LEC/LECs_app.tex}

\setcounter{subsection}{3}

\clearpage
\input{BK/BK_app}

\setcounter{subsection}{4}

\clearpage
\input{HQ/HQ_app}

\clearpage
\input{Alpha_s/Alpha_s_app}

\end{appendix}

\clearpage
\bibliography{FLAG} 
\bibliographystyle{JHEP}

\end{document}

%% file: defs.tex
\definecolor{lightgreen}{rgb}{.85,.99,.85}
\definecolor{darkgreen}{rgb}{0,.7,0}
\definecolor{orange}{rgb}{1.0,.6,0}
\newcommand{\tbr}{\hspace{0.25mm}{\color{red}$\blacksquare$}} 
\newcommand{\tbg}{{\color{green}$\bigstar$}}

\newcommand{\rC}{C}
\newcommand{\gA}{A}
\newcommand{\oP}{P}

\newcommand{\bda}{\begin{\displaymath}\begin{array}{rl}}
\newcommand{\eda}{\end{array}\end{displaymath}}
\newcommand{\be}{\begin{equation}}
\newcommand{\ee}{\end{equation}}
\newcommand{\bdm}{\begin{displaymath}}
\newcommand{\edm}{\end{displaymath}}
\newcommand{\bea}{\begin{eqnarray}}
\newcommand{\eea}{\end{eqnarray}}

\newcommand{\fs}{\,.}
\newcommand{\co}{\,,}

\newcommand{\ind}{\scriptscriptstyle}

\newcommand{\qbar}{\overline{\rule[0.42em]{0.4em}{0em}}\hspace{-0.45em}q}
\newcommand{\ubar}{\overline{\rule[0.42em]{0.4em}{0em}}\hspace{-0.5em}u}
\newcommand{\dbar}{\,\overline{\rule[0.65em]{0.4em}{0em}}\hspace{-0.6em}d}
\newcommand{\sbar}{\,\overline{\rule[0.42em]{0.4em}{0em}}\hspace{-0.5em}s}
\newcommand{\lbar}{\bar{\ell}}

\newcommand{\lsim}{\,\raisebox{-0.3em}{$\stackrel{\raisebox{-0.1em}{$<$}}{\sim}$
}\,} 
\newcommand{\gsim}{\,\raisebox{-0.3em}{$\stackrel{\raisebox{-0.1em}{$>$}}{\sim}$
}\,}
\newcommand{\lvac}{\langle 0|\,}

\newcommand{\al}{&\!\!\!}

\newcommand{\Ch}{$\chi$} 

\newcommand{\Mpibar}{\rule{0.05cm}{0cm}\overline{\hspace{-0.08cm}M}_{\hspace{-0.04cm}\pi}}
\newcommand{\MKbar}{\rule{0.05cm}{0cm}\overline{\hspace{-0.08cm}M}_{\hspace{-0.04cm}K}}

\newcommand{\off}[1]{{}}
\newcommand{\flagold}{FLAG 13}

\newcommand{\bi}{\begin{itemize}}
\newcommand{\ei}{\end{itemize}}

\newcommand{\beq}{\begin{equation}}
\newcommand{\eeq}{\end{equation}}

\newcommand{\Mpi}{M_\pi}
\newcommand{\Fpi}{F_\pi}
\newcommand{\Mka}{M_K}
\newcommand{\Fka}{F_K}

\newcommand{\<}{\langle}
\renewcommand{\>}{\rangle}

\newcommand{\lonebar}{\ln\frac{\Lambda_1^2}{M_\pi^2}}
\newcommand{\ltwobar}{\ln\frac{\Lambda_2^2}{M_\pi^2}}
\newcommand{\lthreebar}{\ln\frac{\Lambda_3^2}{M_\pi^2}}
\newcommand{\lfourbar}{\ln\frac{\Lambda_4^2}{M_\pi^2}}
\newcommand{\lsixbar}{\ln\frac{\Lambda_6^2}{M_\pi^2}}
\newcommand{\lMbar}{\ln\frac{\Omega_M^2}{M_\pi^2}}
\newcommand{\lFbar}{\ln\frac{\Omega_F^2}{M_\pi^2}}

\newcommand{\MeV}{\,\mathrm{MeV}}
\newcommand{\Refs}{\,\mathrm{Refs.}}
\newcommand{\Ref}{\,\mathrm{Ref.}}
\newcommand{\GeV}{\,\mathrm{GeV}}
\newcommand{\fm}{\,\mathrm{fm}}

\newcommand{\ep}{\epsilon}

\newcommand{\et}{\eta}

\newcommand{\msbar}{{\overline{{\rm MS}}}}
\newcommand{\lms}{\Lambda_\msbar}

\def\mev{{\rm MeV}}
\def\gev{{\rm GeV}}
\def\tev{{\rm TeV}}
\def\fm{{\rm fm}}
 
\def\qbar{\bar{q}}
\def\psibar{\bar{\psi}}

\def\ubar{\bar{u}} 

\def\csw{c_{\rm sw}}

\def\gbar{\bar{g}}

\newcommand{\bd}{\begin{displaymath}}
\newcommand{\ed}{\end{displaymath}}

\newcommand{\eq}[1]{Eq.~(\ref{#1})}
\newcommand{\fig}[1]{Fig.~\ref{#1}}

\newcommand{\sect}[1]{Sec.~\ref{#1}}

\newcommand{\figurebox}[2]{\fbox{\vbox to#2in{\hbox to #1in{\hfil}\vfil}}}
\newcommand{\bm}[1]{\mbox{\boldmath ${#1}$}}

\newcommand{\gtaeq}{\raisebox{-.6ex}{$\stackrel{\textstyle{>}}{\sim}$}}

\newcommand{\Nf}{N_{\hspace{-0.08 em} f}}
\newcommand{\Tr}{{\rm Tr}\,}

\newcommand{\fKfpicharged}{ \frac{f_{K^\pm}}{f_{\pi^\pm}}}
\newcommand{\fKfpichargedr}{ {f_{K^\pm}}/{f_{\pi^\pm}}}

\newcommand{\mK}{m_{\rm K}}

\newcommand{\fK}{f_{\rm K}}

\newcommand{\half}{\textstyle{1\over2}}

\newcommand{\abar}{\overline{a}}

\newcommand{\rb}[1]{\raisebox{1.5ex}[-1.5ex]{#1}}

\newcommand{\Lo}{\stackrel{\rule[-0.1cm]{0cm}{0cm}\mbox{\tiny LO}}{=}}
\newcommand{\NLo}{\stackrel{\rule[-0.1cm]{0cm}{0cm}\mbox{\tiny NLO}}{=}} 
\newcommand{\epsilonD}{\epsilon}

\definecolor{Gray}{rgb}{0.5,0.5,0.5}
\definecolor{Black}{rgb}{0.0,0.0,0.0}

\def\good{\raisebox{0.35mm}{{\color{green}$\bigstar$}}}
\def\bad{\raisebox{0.35mm}{\hspace{0.65mm}{\color{red}\tiny$\blacksquare$}}} 
\def\soso{\hspace{0.25mm}\raisebox{-0.2mm}{{\color{green}\Large$\circ$}}}
\def\okay{\hspace{0.25mm}\raisebox{-0.2mm}{{\color{green}\large\checkmark}}}

\newcommand{\mr}{\mathrm}

\def\half{{1\over2}}

\def\Tr{\,\mathrm{Tr}}

\def\fm{\mathrm{fm}}
\def\ev{\mathrm{e\kern-0.1em V}}
\def\kev{\mathrm{ke\kern-0.1em V}}
\def\mev{\mathrm{Me\kern-0.1em V}}
\def\gev{\mathrm{Ge\kern-0.1em V}}
\def\tev{\mathrm{Te\kern-0.1em V}}

\let\Re=\re \let\Im=\im

\def\n#1e#2n{{#1}\times 10^{#2}}

\def\bea{\begin{eqnarray}}
\def\eea{\end{eqnarray}}
\def\nn{\nonumber}

\def\oO{\mathcal{Q}}
\def\cO{\mathcal{O}}

\def\ods2{\mathcal{O}_{\Delta S=2}}
\def\zds2{Z_{\Delta S=2}}

\def\msbar{{\overline{\mathrm{MS}}}}

\def\lat{\mathrm{lat}}

\def\spose#1{\hbox to 0pt{#1\hss}}
\def\ltapprox{\mathrel{\spose{\lower 3pt\hbox{$\mathchar"218$}}
 \raise 2.0pt\hbox{$\mathchar"13C$}}}
\def\gtapprox{\mathrel{\spose{\lower 3pt\hbox{$\mathchar"218$}}
 \raise 2.0pt\hbox{$\mathchar"13E$}}}
\def\inapprox{\mathrel{\spose{\lower 3pt\hbox{$\mathchar"218$}}
 \raise 2.0pt\hbox{$\mathchar"232$}}}

\makeatletter
\def\slash#1{{\mathpalette\c@ncel{#1}}} % TeXbook, bottom of p360
\def\big#1{{\hbox{$\left#1\vbox to1.012\ht\strutbox{}\right.\n@space$}}}
\def\Big#1{{\hbox{$\left#1\vbox to1.369\ht\strutbox{}\right.\n@space$}}}
\def\bigg#1{{\hbox{$\left#1\vbox to1.726\ht\strutbox{}\right.\n@space$}}}
\def\Bigg#1{{\hbox{$\left#1\vbox
to2.083\ht\strutbox{}\right.\n@space$}}}
\makeatother

\newcommand{\nl}{\nonumber \\}

\newcommand{\delv}{{\bf \nabla}}
\newcommand{\delvt}{\tilde{{\bf \nabla}}}

\newcommand{\delfour}{{\Delta^{(4)}}}
\newcommand{\delsq}{\Delta^{(2)}}

\newcommand{\Ev}{\tilde{{\bf E}}}
\newcommand{\Bv}{\tilde{{\bf B}}}
\newcommand{\sigmav}{\mbox{\boldmath$\sigma$}}

\def\spose#1{\hbox to 0pt{#1\hss}}
\def\ltapprox{\mathrel{\spose{\lower 3pt\hbox{$\mathchar"218$}}
\raise 2.0pt\hbox{$\mathchar"13C$}}}
\def\gtapprox{\mathrel{\spose{\lower 3pt\hbox{$\mathchar"218$}}
\raise 2.0pt\hbox{$\mathchar"13E$}}}
\def\inapprox{\mathrel{\spose{\lower 3pt\hbox{$\mathchar"218$}}
\raise 2.0pt\hbox{$\mathchar"232$}}}

\newcommand{\alphah}{\alpha_\mathrm{V'}}
\newcommand{\alphav}{\alpha_\mathrm{V}}
\newcommand{\alphap}{\alpha_\mathrm{P}}

\newcommand{\SLfnalmilcBDstar}{FNAL/MILC 14 }   % arXiv:1403.0635
\newcommand{\SLfnalmilcBD}{FNAL/MILC 15C}               % arXiv:1503.07237
\newcommand{\SLfnalmilcBpi}{FNAL/MILC 15 }                              % arXiv:1503.07839
\newcommand{\SLhpqcdBD}{HPQCD 15  }                              % arXiv:1505.03925
\newcommand{\SLhpqcdBsK}{HPQCD 14  }                                              % arXiv:1406.2279
\newcommand{\SLrbcukqcdBpi}{RBC/UKQCD 15}                                       % arXiv:1501.05373
\newcommand{\SLLambdabp}{Detmold 15\\ $\Lambda_b \to p$}                                                % arXiv:1503.01421
\newcommand{\SLLambdabc}{Detmold 15\\ $\Lambda_b \to \Lambda_c$}                                                % arXiv:1503.01421

\newcommand{\FLAGAVBEGIN}{}
\newcommand{\FLAGAVEND}{}

%% file: introduction.tex
\section{Introduction}
\label{sec:introduction}

Flavour physics provides an important opportunity for exploring the
limits of the Standard Model of particle physics and for constraining
possible extensions that go beyond it. As the LHC explores 
a new energy frontier and as experiments continue to extend the precision 
frontier, the importance of flavour physics will grow,
both in terms of searches for signatures of new physics through
precision measurements and in terms of attempts to construct the
theoretical framework behind direct discoveries of new particles. A
major theoretical limitation consists in the precision with which
strong-interaction effects can be quantified. Large-scale numerical
simulations of lattice QCD allow for the computation of these effects
from first principles. The scope of the Flavour Lattice Averaging
Group (FLAG) is to review the current status of lattice results for a
variety of physical quantities in low-energy physics. Set up in
November 2007 it
comprises experts in Lattice Field Theory, Chiral Perturbation
Theory and Standard Model phenomenology. 
Our aim is to provide an answer to the frequently posed
question ``What is currently the best lattice value for a particular
quantity?" in a way that is readily accessible to
nonlattice-experts. This is generally not an easy question to answer;
different collaborations use different lattice actions
(discretizations of QCD) with a variety of lattice spacings and
volumes, and with a range of masses for the $u-$ and $d-$quarks. Not
only are the systematic errors different, but also the methodology
used to estimate these uncertainties varies between collaborations. In
the present work we summarize the main features of each of the
calculations and provide a framework for judging and combining the
different results. Sometimes it is a single result that provides the
``best" value; more often it is a combination of results from
different collaborations. Indeed, the consistency of values obtained
using different formulations adds significantly to our confidence in
the results.

The first two editions of the FLAG review were published in
2011~\cite{Colangelo:2010et} and 2014~\cite{Aoki:2013ldr}.
The second edition reviewed results related to both light 
($u$-, $d$- and $s$-), and heavy ($c$- and $b$-) flavours.
The quantities related to pion and kaon physics  were 
light-quark masses, the form factor $f_+(0)$
arising in semileptonic $K \rightarrow \pi$ transitions 
(evaluated at zero momentum transfer), 
the decay constants $f_K$ and $f_\pi$, 
and the $B_{\rm K}$ parameter from neutral kaon mixing. 
Their implications for
the CKM matrix elements $V_{us}$ and $V_{ud}$ were also discussed.
Furthermore, results
were reported for some of the low-energy constants of $SU(2)_L \times
SU(2)_R$ and $SU(3)_L \times SU(3)_R$ Chiral Perturbation Theory.
The quantities related to $D$- and $B$-meson physics that were
reviewed were the
$B$- and $D$-meson decay constants, form factors, and mixing parameters.
These are the heavy-light quantities most relevant
to the determination of CKM matrix elements and the global
CKM unitarity-triangle fit. Last but not least, the current status of 
lattice results on the QCD coupling  $\alpha_s$ was reviewed.

In the present paper we provide updated results for all the above-mentioned
quantities, but also extend the scope of the review in two ways.
First, we now present results for the charm and bottom quark masses,
in addition to those of the three lightest quarks.
Second, we review results obtained for the kaon mixing matrix
elements of new operators that arise in theories of physics beyond the
Standard Model. Our main results are collected in Tabs.~\ref{tab:summary1} and \ref{tab:summary2}.

Our plan is to continue providing FLAG updates, in the form of a peer
reviewed paper, roughly on a biennial basis. This effort is
supplemented by our more frequently updated
website \href{http://itpwiki.unibe.ch/flag}{{\tt
http://itpwiki.unibe.ch/flag}} \cite{FLAG:webpage}, where figures as well as pdf-files for
the individual sections can be downloaded. The papers reviewed in the
present edition have appeared before the closing date {\bf 30 November 2015}.

\input{summarytable.tex}
\clearpage

This review is organized as follows.  In the remainder of
Sec.~\ref{sec:introduction} we summarize the composition and rules of
FLAG and discuss general issues that arise in modern lattice
calculations.  In Sec.~\ref{sec:qualcrit} we explain our general
methodology for evaluating the robustness of lattice results.  We also
describe the procedures followed for combining results from different
collaborations in a single average or estimate (see
Sec.~\ref{sec:averages} for our definition of these terms). The rest
of the paper consists of sections, each dedicated to a single (or
groups of closely connected) physical quantity(ies). Each of these
sections is accompanied by an Appendix with explicatory notes.

\subsection{FLAG composition, guidelines and rules}

FLAG strives to be representative of the lattice community, both in
terms of the geographical location of its members and the lattice
collaborations to which they belong. We aspire to provide the
particle-physics community with a single source of reliable
information on lattice results.

In order to work reliably and efficiently, we have adopted a formal
structure and a set of rules by which all FLAG members abide.  The
collaboration presently consists of an Advisory Board (AB), an
Editorial Board (EB), and seven Working Groups (WG). The r\^{o}le of
the Advisory Board is that of general supervision and
consultation. Its members may interfere at any point in the process of
drafting the paper, expressing their opinion and offering advice. They
also give their approval of the final version of the preprint before
it is rendered public. The Editorial Board coordinates the activities
of FLAG, sets priorities and intermediate deadlines, and takes care of
the editorial work needed to amalgamate the sections written by the
individual working groups into a uniform and coherent review. The
working groups concentrate on writing up the review of the physical
quantities for which they are responsible, which is subsequently
circulated to the whole collaboration for critical evaluation.

The current list of FLAG members and their Working Group assignments is:
\begin{itemize}
\item
Advisory Board (AB):\hfill
S.~Aoki, C.~Bernard, M.~Golterman, H.~Leutwyler, \\
\hbox{} \hfill and C.~Sachrajda
\item
Editorial Board (EB):\hfill
G.~Colangelo, A.~J\"uttner, S.~Hashimoto, S.~Sharpe, A.~Vladikas, \\
\hbox{} \hfill and U.~Wenger
\item
Working Groups (coordinator listed first):
\begin{itemize}
\item Quark masses \hfill L.~Lellouch, T.~Blum, and V.~Lubicz
\item $V_{us},V_{ud}$ \hfill S.~Simula, P.~Boyle,\footnote{Peter Boyle had participated actively in the early stages of the current FLAG effort. Unfortunately, due to other commitments, it was impossible for him to contribute until the end, and he decided to withdraw from the collaboration.} and T.~Kaneko
\item LEC \hfill S.~D\"urr, H.~Fukaya, and U.M.~Heller
\item $B_K$ \hfill H.~Wittig, P.~Dimopoulos, and R.~Mawhinney
\item $f_{B_{(s)}}$, $f_{D_{(s)}}$, $B_B$ \hfill M.~Della Morte, Y.~Aoki, and D.~Lin
\item $B_{(s)}$, $D$ semileptonic and radiative decays \hfill E.~Lunghi, D.~Becirevic, S.~Gottlieb, \\
\hbox{} \hfill and C.~Pena
\item $\alpha_s$ \hfill R.~Sommer, R.~Horsley, and T.~Onogi
\end{itemize}
\end{itemize}
As some members of the WG on quark masses were faced with unexpected
hindrances, S.~Simula has kindly assisted in the completion of the
relevant section during the final phases of its composition.

The most important FLAG guidelines and rules are the following:
\begin{itemize}
\item
the composition of the AB reflects the main geographical areas in
which lattice collaborations are active, with members from
America, Asia/Oceania and Europe;
\item
the mandate of regular members is not limited in time, but we expect that a
certain turnover will occur naturally;
\item
whenever a replacement becomes necessary this has to keep, and
possibly improve, the balance in FLAG, so that different collaborations, from
different geographical areas are represented;
\item
in all working groups the three members must belong to three different
lattice collaborations;\footnote{The WG on semileptonic $D$ and $B$
decays has currently four members, but only three of them belong to
lattice collaborations.}
\item
a paper is in general not reviewed (nor colour-coded, as described in
the next section) by any of its authors;
\item
lattice collaborations not represented in FLAG will be consulted on the colour coding
of their calculation;
\item
there are also internal rules regulating our work, such as voting procedures.
\end{itemize}

\subsection{Citation policy}
We draw attention to this particularly important point.  As stated
above, our aim is to make lattice QCD results easily accessible to
nonlattice-experts and we are well aware that it is likely that some
readers will only consult the present paper and not the original
lattice literature. It is very important that this paper be not the
only one cited when our results are quoted. We strongly suggest that
readers also cite the original sources. In order to facilitate this,
in Tabs.~\ref{tab:summary1} and \ref{tab:summary2}, besides
summarizing the main results of the present review, we also cite the
original references from which they have been obtained. In addition,
for each figure we make a bibtex-file available on our webpage
\cite{FLAG:webpage} which contains the bibtex-entries of all the
calculations contributing to the FLAG average or estimate. The
bibliography at the end of this paper should also make it easy to cite
additional papers. Indeed we hope that the bibliography will be one of
the most widely used elements of the whole paper.

\subsection{General issues}

Several general issues concerning the present review are thoroughly
discussed in Sec.~1.1 of our initial 2010 paper~\cite{Colangelo:2010et}
and we encourage the reader to consult the relevant pages. In the
remainder of the present subsection, we focus on a few important
points. Though the discussion has been duly updated, it is essentially
that of Sec.~1.2 of the 2013 review~\cite{Aoki:2013ldr}.

The present review aims to achieve two distinct goals:
first, to provide a {\bf description} of the work done on the lattice
concerning low-energy particle physics;
and, second,
to draw {\bf conclusions} on the basis of that work,  summarizing
the results obtained for the various quantities of physical interest.

The core of the information about the work done on the lattice is
presented in the form of tables, which not only list the various
results, but also describe the quality of the data that underlie
them. We consider it important that this part of the review represents
a generally accepted description of the work done. For this reason, we
explicitly specify the quality requirements\footnote{%
We also use terms
like ``quality criteria", ``rating", ``colour coding" etc. when referring to
the classification of results, as described in Sec.~\ref{sec:qualcrit}.}
used and provide sufficient details in appendices so that the reader
can verify the information given in the tables.

On the other hand, the conclusions drawn 
on the basis of the available lattice results
are the responsibility of FLAG alone. Preferring to
err on the side of caution, in several cases we draw
conclusions that are more conservative than those resulting from
a plain weighted average of the available lattice results. This cautious
approach is usually adopted when the average is
dominated by a single lattice result, or when
only one lattice result is available for a given quantity. In such
cases one does not have the same degree of confidence in results and
errors as when there is agreement among several different
calculations using different approaches. The reader should keep
in mind that the degree of confidence cannot be quantified, and
it is not reflected in the quoted errors. 

Each discretization has its merits, but also its shortcomings. For most
topics covered in this review we
have an increasingly broad database, and for most quantities
lattice calculations based on totally different discretizations are
now available. This is illustrated by the dense population of the
tables and figures in most parts of this review. Those
calculations that do satisfy our quality criteria indeed lead to
consistent results, confirming universality within the accuracy
reached. In our opinion, the consistency between independent lattice
results, obtained with different discretizations, methods, and
simulation parameters, is an important test of lattice QCD, and
observing such consistency also provides further evidence that
systematic errors are fully under control.

In the sections dealing with heavy quarks and with $\alpha_s$, the
situation is not the same. Since the $b$-quark mass cannot be resolved
with current lattice spacings, all lattice methods for treating $b$
quarks use effective field theory at some level. This introduces
additional complications not present in the light-quark sector.  An
overview of the issues specific to heavy-quark quantities is given in
the introduction of Sec.~\ref{sec:BDecays}. For $B$ and $D$ meson
leptonic decay constants, there already exists a good number of
different independent calculations that use different heavy-quark
methods, but there are only one or two independent calculations of
semileptonic $B$ and $D$ meson form factors and $B$ meson mixing
parameters. For $\alpha_s$, most lattice methods involve a range of
scales that need to be resolved and controlling the systematic error
over a large range of scales is more demanding. The issues specific to
determinations of the strong coupling are summarized in Sec.~\ref{sec:alpha_s}.
\smallskip
\\{\it Number of sea quarks in lattice simulations:}\\
\noindent
Lattice QCD simulations currently involve two, three or four flavours of
dynamical quarks. Most simulations set
the masses of the two lightest quarks to be equal, while the
strange and charm quarks, if present, are heavier
(and tuned to lie close to their respective physical values). 
Our notation for these simulations indicates which quarks
are nondegenerate, e.g. 
$\Nf=2+1$ if $m_u=m_d < m_s$ and $\Nf =2+1+1$ if $m_u=m_d < m_s < m_c$. 
Calculations with $\Nf =2$, i.e. two degenerate dynamical
flavours, often include strange valence quarks interacting with gluons,
so that bound states with the quantum numbers of the kaons can be
studied, albeit neglecting strange sea-quark fluctuations.  The
quenched approximation ($N_f=0$), in which sea quark contributions 
are omitted, has uncontrolled systematic errors and
is no longer used in modern lattice simulations with relevance to phenomenology.
Accordingly, we will review results obtained with $N_f=2$, $N_f=2+1$,
and $N_f = 2+1+1$, but omit earlier results with $N_f=0$. 
The only exception concerns the QCD coupling constant $\alpha_s$.
Since this observable does not require valence light quarks,
it is theoretically well defined also in the $N_f=0$ theory,
which is simply pure gluodynamics.
The $N_f$-dependence of $\alpha_s$, 
or more precisely of the related quantity $r_0 \Lambda_\msbar$, 
is a theoretical issue of considerable interest; here $r_0$ is a quantity
with the dimension of length, which sets the physical scale, as discussed in
Appendix~\ref{sec_scale}.
We stress, however, that only results with $N_f \ge 3$ 
are used to determine the physical value of $\alpha_s$ at a high scale.
\smallskip
\\{\it Lattice actions, simulation parameters and scale setting:}\\
\noindent
The remarkable progress in the precision of lattice
calculations is due to improved algorithms, better computing resources
and, last but not least, conceptual developments.
Examples of the latter are improved
actions that reduce lattice artifacts and actions that preserve
chiral symmetry to very good approximation.
A concise characterization of
the various discretizations that underlie the results reported in the
present review is given in Appendix~\ref{sec_lattice_actions}.

Physical quantities are computed in lattice simulations in units of the
lattice spacing so that they are dimensionless.
For example, the pion decay constant that is obtained from a simulation
is $f_\pi a$, where $a$ is the spacing between two neighboring lattice sites.
To convert these results to physical units requires knowledge
of the lattice spacing $a$ at the fixed values of the bare QCD parameters
(quark masses and gauge coupling) used in the simulation.
This is achieved by requiring agreement between
the lattice calculation and experimental measurement of a known
quantity, which thus ``sets the scale" of a given simulation. A few details
on this procedure are provided in Appendix~\ref{sec_scale}.
\smallskip
\\{\it Renormalization and scheme dependence:}\\
\noindent
Several of the results covered by this review, such as quark masses,
the gauge coupling, and $B$-parameters, are for quantities defined in a
given renormalization scheme and at a specific renormalization scale. 
The schemes employed (e.g. regularization-independent MOM schemes) are often
chosen because of their specific merits when combined with the lattice
regularization. For a brief discussion of their properties, see
Appendix~\ref{sec_match}. The conversion of the results, obtained in
these so-called intermediate schemes, to more familiar regularization
schemes, such as the $\msbar$-scheme, is done with the aid of
perturbation theory. It must be stressed that the renormalization
scales accessible in simulations are limited, because of the presence
of an ultraviolet (UV) cutoff of $\sim \pi/a$.
To safely match to $\msbar$, a scheme defined in perturbation theory,
Renormalization Group (RG) running to higher scales is performed,
either perturbatively or nonperturbatively (the latter using
finite-size scaling techniques).
\smallskip
\\{\it Extrapolations:}\\
\noindent
Because of limited computing resources, lattice simulations are often
performed at unphysically heavy pion masses, although results at the
physical point have become increasingly common. Further, numerical
simulations must be done at nonzero lattice spacing, and in a finite
(four- dimensional) volume.  In order to obtain physical results,
lattice data are obtained at a sequence of pion masses and a sequence
of lattice spacings, and then extrapolated to the physical pion mass
and to the continuum limit.  In principle, an extrapolation to
infinite volume is also required. However, for most quantities
discussed in this review, finite-volume effects are exponentially
small in the linear extent of the lattice in units of the pion mass
and, in practice, one often verifies volume independence by comparing
results obtained on a few different physical volumes, holding other
parameters equal. To control the associated systematic uncertainties,
these extrapolations are guided by effective theories.  For
light-quark actions, the lattice-spacing dependence is described by
Symanzik's effective theory~\cite{Symanzik:1983dc,Symanzik:1983gh};
for heavy quarks, this can be extended and/or supplemented by other
effective theories such as Heavy-Quark Effective Theory (HQET).  The
pion-mass dependence can be parameterized with Chiral Perturbation
Theory ($\chi$PT), which takes into account the Nambu-Goldstone nature
of the lowest excitations that occur in the presence of light
quarks. Similarly, one can use Heavy-Light Meson Chiral Perturbation
Theory (HM$\chi$PT) to extrapolate quantities involving mesons
composed of one heavy ($b$ or $c$) and one light quark.  One can
combine Symanzik's effective theory with $\chi$PT to simultaneously
extrapolate to the physical pion mass and the continuum; in this case,
the form of the effective theory depends on the discretization.  See
Appendix~\ref{sec_ChiPT} for a brief description of the different
variants in use and some useful references.  Finally, $\chi$PT can
also be used to estimate the size of finite-volume effects measured in
units of the inverse pion mass, thus providing information on the
systematic error due to finite-volume effects in addition to that
obtained by comparing simulations at different volumes.
\smallskip
\\{\it Critical slowing down:}\\
\noindent
The lattice spacings reached in recent simulations go down to 0.05 fm
or even smaller. In this regime, long autocorrelation times slow down
the sampling of the
configurations~\cite{Antonio:2008zz,Bazavov:2010xr,Schaefer:2010hu,Luscher:2010we,Schaefer:2010qh,Chowdhury:2013mea,Brower:2014bqa,Fukaya:2015ara,DelDebbio:2002xa,Bernard:2003gq}.
Many groups check for autocorrelations in a number of observables,
including the topological charge, for which a rapid growth of the
autocorrelation time is observed with decreasing lattice spacing.
This is often referred to as topological freezing. A solution to the
problem consists in using open boundary conditions in time, instead of
the more common antiperiodic ones \cite{Luscher:2011kk}. More recently
two other approaches have been proposed, one based on a multiscale
thermalization algorithm \cite{Endres:2015yca} and another based on
defining QCD on a nonorientable manifold \cite{Mages:2015scv}.  The
problem is also touched upon in Sec.~\ref{s:crit}, where it is
stressed that attention must be paid to this issue. While large scale
simulations with open boundary conditions are already far advanced
\cite{Bruno:2014jqa}, unfortunately so far no results reviewed here
have been obtained with any of the above methods. It is usually {\it
  assumed} that the continuum limit can be reached by extrapolation
from the existing simulations and that potential systematic errors due
to the long autocorrelation times have been adequately controlled.
\smallskip
\\{\it Simulation algorithms and numerical errors:}\\
\noindent
Most of the modern lattice-QCD simulations use exact algorithms such 
as Refs.~\cite{Duane:1987de,Clark:2006wp}, which do not produce any systematic errors when exact 
arithmetic is available. In reality, one uses numerical calculations at 
double (or in some cases even single) precision, and some errors are 
unavoidable. More importantly, the inversion of the Dirac operator is 
carried out iteratively and it is truncated once some accuracy is 
reached, which is another source of potential systematic error. In most 
cases, these errors have been confirmed to be much less than the 
statistical errors. In the following we assume that this source of error 
is negligible. 
Some of the most recent simulations use an inexact algorithm in order to 
speed-up the computation, though it may produce systematic effects. 
Currently available tests indicate that errors from the use of inexact
algorithms are under control.

%% file: summarytable.tex
% \documentclass{article}

% \begin{document}
\begin{sidewaystable}[h]
\vspace{-1cm}
\centering
\begin{tabular}{|l|l||l|l||l|l||l|l|}
\hline
Quantity \rule[-0.2cm]{0cm}{0.6cm}    & \hspace{-1.5mm}Sec.\hspace{-2mm} &$N_f=2+1+1$ & Refs. &  $N_f=2+1$ & Refs. &$N_f=2$ &Refs. \\
\hline \hline
\input{Tables/Summarytable2.tex}
\end{tabular}\\[0.2cm]

\begin{minipage}{\linewidth}
{\footnotesize 
\begin{itemize}
\item[$^\ddagger$] This is a FLAG estimate, based on $\chi$PT and the isospin averaged up- and down-quark mass $m_{ud}$ \cite{Blum:2014tka,Durr:2010vn,Durr:2010aw,McNeile:2010ji,Bazavov:2010yq}.
\end{itemize}
}
\end{minipage}

\caption{\label{tab:summary1} Summary of the main results of this review, grouped in terms of $\Nf$, the number of dynamical quark flavours in lattice simulations. Quark masses and the quark condensate are given in 
the $\msbar$ scheme at running scale
   $\mu=2\,\gev$ or as indicated;  the other quantities listed are specified in the quoted sections.
For each result we list the references that entered the FLAG average or estimate. From the entries in this column one
can also read off the number of results that enter our averages for each quantity. We emphasize that these numbers only give a very rough indication of how thoroughly the quantity in question has been explored on the lattice and recommend to consult the detailed tables and figures in the relevant section for more significant information and for explanations on the source of the quoted errors.}
\end{sidewaystable}
\begin{sidewaystable}[h]
\vspace{-1cm}
\centering
\begin{tabular}{|l|l||l|l||l|l||l|l|}
\hline
Quantity \rule[-0.2cm]{0cm}{0.6cm}    & \hspace{-1.5mm}Sec.\hspace{-2mm} &$N_f=2+1+1$ & Refs. &  $N_f=2+1$ & Refs. &$N_f=2$ &Refs. \\
\hline \hline
\input{Tables/Summarytable1.tex}
\hline
Quantity \rule[-0.2cm]{0cm}{0.6cm}    & \hspace{-1.5mm}Sec.\hspace{-2mm} &\multicolumn{3}{c|}{$N_f=2+1$ and $N_f=2+1+1$} & Refs. & & \\
\hline
\input{Tables/Summarytable1alpha.tex}
\end{tabular}
\caption{\label{tab:summary2}Summary of the main results of this review, grouped in terms of $\Nf$, the number of dynamical quark flavours in lattice simulations.   The  quantities listed are specified in the quoted sections.
For each result we list the references that entered the FLAG average or estimate. From the entries in this column one
can also read off the number of results that enter our averages for each quantity. We emphasize that these numbers only give a very rough indication of how thoroughly the quantity in question has been explored on the lattice and recommend to consult the detailed tables and figures in the relevant section for more significant information and for explanations on the source of the quoted errors. 
}
\end{sidewaystable}

%\end{document}

%% file: Tables/Summarytable2.tex
$ m_s   $[MeV]&\ref{sec:msmud}&$93.9(1.1)$&\cite{Carrasco:2014cwa,Chakraborty:2014aca}&$92.0(2.1)$&\cite{Bazavov:2009fk,Durr:2010vn,Durr:2010aw,McNeile:2010ji,Blum:2014tka}&$ 101(3)$&\cite{Blossier:2010cr,Fritzsch:2012wq}\\[1mm]
$ m_{ud}$[MeV]&\ref{sec:msmud}&$ 3.70 (17)$&\cite{Carrasco:2014cwa}&$ 3.373 (80)$&\cite{Blum:2014tka,Durr:2010vn,Durr:2010aw,McNeile:2010ji,Bazavov:2010yq}&$ 3.6(2 )$&\cite{Blossier:2010cr}\\[1mm]
$ m_s / m_{ud} $&\ref{sec:msovermud}&$ 27.30  (34)$&\cite{Carrasco:2014cwa,Bazavov:2014wgs}&$  27.43  (31 )$&\cite{Blum:2014tka,Burch:2009az,Durr:2010vn,Durr:2010aw}&$ 27.3  (9)$&\cite{Blossier:2010cr}\\[1mm]
$ m_u $[MeV]&\ref{subsec:mumd}&$2.36(24)$&\cite{Carrasco:2014cwa}&$2.16(9)(7)$&${}^\ddagger$&$2.40(23)$&\cite{deDivitiis:2013xla}\\[1mm]
$ m_d $[MeV]&\ref{subsec:mumd}&$ 5.03(26)$&\cite{Carrasco:2014cwa}&$ 4.68(14)(7 )$&${}^\ddagger$&$ 4.80(23)$&\cite{deDivitiis:2013xla}\\[1mm]
$ {m_u}/{m_d} $&\ref{subsec:mumd}&$ 0.470(56)$&\cite{Carrasco:2014cwa}&$ 0.46(2)(2)$&${}^\ddagger$&$ 0.50(4)$&\cite{deDivitiis:2013xla}\\[1mm]
\hline
$\overline{m}_c(\mbox{3 GeV}) $[GeV]&\ref{sec:mcnf2}&$ 0.996  (25)$&\cite{Carrasco:2014cwa,Chakraborty:2014aca}&$ 0.987  (6)$&\cite{McNeile:2010ji,Yang:2014sea}&$ 1.03(4)$&\cite{Blossier:2010cr}\\[1mm]
$ m_c / m_s $&\ref{sec:mcoverms}&$ 11.70  (6)$&\cite{Chakraborty:2014aca,Carrasco:2014cwa,Bazavov:2014wgs}&$ 11.82  (16)$&\cite{Yang:2014sea,Davies:2009ih}&$ 11.74  (35)$&\cite{Blossier:2010cr}\\[1mm]
\hline
$\overline{m}_b(\overline{m}_b) $[GeV]&\ref{s:bmass}&$ 4.190 (21)$&\cite{Chakraborty:2014aca,Colquhoun:2014ica}&$ 4.164 (23 )$&\cite{McNeile:2010ji}&$ 4.256 (81)$&\cite{Carrasco:2013zta,Bernardoni:2013xba}\\[1mm]
\hline
$ f_+(0) $&\ref{sec:Direct}&$ 0.9704(24)(22)$&\cite{Bazavov:2013maa}&$ 0.9677(27 )$&\cite{Bazavov:2012cd,Boyle:2015hfa}&$ 0.9560(57)(62)$&\cite{Lubicz:2009ht}\\[1mm]
$ f_{K^\pm} / f_{\pi^\pm}  $&\ref{sec:Direct}&$  1.193(3)$&\cite{Dowdall:2013rya,Bazavov:2014wgs,Carrasco:2014poa}&$  1.192(5)$&\cite{Follana:2007uv,Bazavov:2010hj,Durr:2010hr,Arthur:2012opa}&$  1.205(6)(17)$&\cite{Blossier:2009bx}\\[1mm]
$ f_{\pi^\pm}$[MeV]&\ref{sec:fKfpi}&&&$ 130.2  (1.4)$&\cite{Follana:2007uv,Bazavov:2010hj,Arthur:2012opa}&&\\[1mm]
$ f_{K^\pm}  $[MeV]&\ref{sec:fKfpi}&$ 155.6  (4)$&\cite{Dowdall:2013rya,Bazavov:2014wgs,Carrasco:2014poa}&$ 155.9  (9)$&\cite{Follana:2007uv,Bazavov:2010hj,Arthur:2012opa}&$ 157.5  (2.4)$&\cite{Blossier:2009bx}\\[1mm]
\hline
$ \Sigma^{1/3}$[MeV]&\ref{sec:SU2_LO}&$ 280(8)(15 )$&\cite{Cichy:2013gja}&$ 274( 3 )$&\cite{Bazavov:2010yq,Borsanyi:2012zv,Durr:2013goa,Blum:2014tka}&$ 266(10 )$&\cite{Baron:2009wt,Cichy:2013gja,Brandt:2013dua,Engel:2014eea}\\[1mm]
$ {\Fpi}/{F}$&\ref{sec:SU2_LO}&$1.076(2)(2  )$&\cite{Baron:2010bv}&$1.064( 7 )$&\cite{Bazavov:2010hj,Beane:2011zm,Borsanyi:2012zv,Durr:2013goa,Blum:2014tka}&$1.073(15 )$&\cite{Frezzotti:2008dr,Baron:2009wt,Brandt:2013dua,Engel:2014eea}\\[1mm]
$ \lbar_3$&\ref{sec:SU2_NLO}&$3.70(7)(26  )$&\cite{Baron:2010bv}&$2.81(64 )$&\cite{Bazavov:2010hj,Beane:2011zm,Borsanyi:2012zv,Durr:2013goa,Blum:2014tka}&$3.41(82 )$&\cite{Frezzotti:2008dr,Baron:2009wt,Brandt:2013dua}\\[1mm]
$ \lbar_4$&\ref{sec:SU2_NLO}&$4.67(3)(10 )$&\cite{Baron:2010bv}&$4.10(45 )$&\cite{Bazavov:2010hj,Beane:2011zm,Borsanyi:2012zv,Durr:2013goa,Blum:2014tka}&$4.51(26 )$&\cite{Frezzotti:2008dr,Baron:2009wt,Brandt:2013dua}\\[1mm]
$ \lbar_6$&\ref{sec:SU2_NLO}&&&&&$15.1(1.2 )$&\cite{Frezzotti:2008dr,Brandt:2013dua}\\[1mm]
\hline
$\hat{B}_{\rm{K}} $&\ref{sec:indCP}&$ 0.717(18)(16)$&\cite{Carrasco:2015pra}&$ 0.7625(97)$&\cite{Durr:2011ap,Laiho:2011np,Blum:2014tka,Jang:2015sla}&$ 0.727(22)(12)$&\cite{Bertone:2012cu}\\[1mm]
\hline

%% file: Tables/Summarytable1.tex
$ f_D$[MeV]&\ref{sec:fD}&$212.15(1.45)$&\cite{Bazavov:2014wgs,Carrasco:2014poa}&$209.2(3.3)$&\cite{Na:2012iu,Bazavov:2011aa}&$208(7)$&\cite{Carrasco:2013zta}\\[1mm]
$ f_{D_s}$[MeV]&\ref{sec:fD}&$ 248.83(1.27)$&\cite{Bazavov:2014wgs,Carrasco:2014poa}&$249.8(2.3)$&\cite{Davies:2010ip,Bazavov:2011aa,Yang:2014sea}&$ 250(7 )$&\cite{Carrasco:2013zta}\\[1mm]
$ {{f_{D_s}}/{f_D}}$&\ref{sec:fD}&$1.1716(32)$&\cite{Bazavov:2014wgs,Carrasco:2014poa}&$1.187(12)$&\cite{Na:2012iu,Bazavov:2011aa}&$1.20(2)$&\cite{Carrasco:2013zta}\\[1mm]
\hline
\hline
$ f_+^{D\pi}(0)$&\ref{sec:DtoPiK}&&&$  0.666(29)$&\cite{Na:2011mc}&&\\[1mm]
$ f_+^{DK}(0)  $&\ref{sec:DtoPiK}&&&$ 0.747(19)$&\cite{Na:2010uf}&&\\[1mm]
\hline
\hline
$ f_B$[MeV]&\ref{sec:fB}&$186(4)$&\cite{Dowdall:2013tga}&$192.0(4.3)$&\cite{Christ:2014uea,Aoki:2014nga,Na:2012sp,McNeile:2011ng,Bazavov:2011aa}&$188(7)$&\cite{Bernardoni:2014fva,Carrasco:2013zta,Carrasco:2013iba}\\[1mm]
$ f_{B_s}$[MeV]&\ref{sec:fB}&$224(5)$&\cite{Dowdall:2013tga}&$228.4(3.7)$&\cite{Christ:2014uea,Aoki:2014nga,Na:2012sp,McNeile:2011ng,Bazavov:2011aa}&$ 227(7)$&\cite{Bernardoni:2014fva,Carrasco:2013zta,Carrasco:2013iba}\\[1mm]
$ {{f_{B_s}}/{f_B}}$&\ref{sec:fB}&$1.205(7 )$&\cite{Dowdall:2013tga}&$1.201(16)$&\cite{Christ:2014uea,Aoki:2014nga,Na:2012sp,McNeile:2011ng,Bazavov:2011aa}&$1.206(23)$&\cite{Bernardoni:2014fva,Carrasco:2013zta,Carrasco:2013iba}\\[1mm]
\hline
\hline
$ f_{B_d}\sqrt{\hat{B}_{B_d}} $[MeV]&\ref{sec:BMix}&&&$  219(14)$&\cite{Gamiz:2009ku,Aoki:2014nga}&$ 216(10)$&\cite{Carrasco:2013zta}\\[1mm]
$ f_{B_s}\sqrt{\hat{B}_{B_s}} $[MeV]&\ref{sec:BMix}&&&$  270(16)$&\cite{Gamiz:2009ku,Aoki:2014nga}&$ 262(10)$&\cite{Carrasco:2013zta}\\[1mm]
$ \hat{B}_{B_d}  $&\ref{sec:BMix}&&&$ 1.26(9)$&\cite{Gamiz:2009ku,Aoki:2014nga}&$ 1.30(6)$&\cite{Carrasco:2013zta}\\[1mm]
$ \hat{B}_{B_s} $&\ref{sec:BMix}&&&$  1.32(6)$&\cite{Gamiz:2009ku,Aoki:2014nga}&$ 1.32(5)$&\cite{Carrasco:2013zta}\\[1mm]
$ \xi  $&\ref{sec:BMix}&&&$  1.239(46)$&\cite{Bazavov:2012zs,Aoki:2014nga}&$  1.225(31)$&\cite{Carrasco:2013zta}\\[1mm]
$ B_{B_s}/B_{B_d}  $&\ref{sec:BMix}&&&$  1.039(63)$&\cite{Bazavov:2012zs,Aoki:2014nga}&$  1.007(21)$&\cite{Carrasco:2013zta}\\[1mm]
\hline
\hline

%% file: Tables/Summarytable1alpha.tex
$ \alpha_{\overline{\rm MS}}^{(5)}(M_Z) $&\ref{s:alpsumm}&\multicolumn{3}{c|}{$ 0.1182(12)$}&\cite{Bazavov:2014soa,Chakraborty:2014aca,McNeile:2010ji,Aoki:2009tf,Maltman:2008bx}&&\\[1mm]
$ \Lambda_{\overline{\rm MS}}^{(5)} $[MeV]&\ref{s:alpsumm}&\multicolumn{3}{c|}{$ 211(14)$}&\cite{Bazavov:2014soa,Chakraborty:2014aca,McNeile:2010ji,Aoki:2009tf,Maltman:2008bx}&&\\[1mm]
\hline

%% file: quality_criteria.tex
\section{Quality criteria, averaging and error estimation}
\label{sec:qualcrit}

The essential characteristics of our approach to the problem of rating
and averaging lattice quantities 
have been outlined in our first publication~\cite{Colangelo:2010et}. 
Our aim is to help the reader
assess the reliability of a particular lattice result without
necessarily studying the original article in depth. This is a delicate
issue, since the ratings may make things appear 
simpler than they are. Nevertheless,
it safeguards against the common practice of using lattice results, and
drawing physics conclusions from them, without a critical assessment
of the quality of the various calculations. We believe that, despite
the risks, it is important to provide some compact information about
the quality of a calculation. We stress, however, the importance of the
accompanying detailed discussion of the results presented in the various
sections of the present review.
 
\subsection{Systematic errors and colour code}
\label{sec:color-code}

The major sources of systematic error are common to most lattice
calculations. These include, as discussed in detail below,
the chiral, continuum and infinite-volume extrapolations.
To each such source of error for which
systematic improvement is possible we
assign one of three coloured symbols: green
star, unfilled green circle
(which replaced in Ref.~\cite{Aoki:2013ldr}
the amber disk used in the original FLAG review~\cite{Colangelo:2010et})
or red square.
These correspond to the following ratings: \\
\hspace{-0.1em}\good \hspace{0.2cm} the parameter values and ranges used 
to generate the datasets allow for a satisfactory control of the systematic uncertainties;\\
\rule{0.1em}{0em}\soso \hspace{0.2cm} the parameter values and ranges used to generate
the datasets allow for a reasonable attempt at estimating systematic uncertainties, which
however could be improved;\\
\rule{0.1em}{0em}\bad \hspace{0.2cm} the parameter values and ranges used to generate
the datasets are unlikely to allow for a reasonable control of systematic uncertainties.\\
The appearance of a red tag, even in a
single source of systematic error of a given lattice result,
disqualifies it from inclusion in the global average.

The attentive reader will notice that these criteria differ from those used in
Refs.~\cite{Colangelo:2010et,Aoki:2013ldr}. 
In the previous FLAG editions we used the three symbols 
in order to rate the reliability of the systematic errors 
attributed to a given result by the paper's authors.
This sometimes proved to be a daunting task,
as the methods used by some collaborations for estimating 
their systematics are not always explained in full detail. 
Moreover, it is sometimes difficult to disentangle and rate 
different uncertainties, since they are interwoven in the error analysis. 
Thus, in the present edition we have opted for a different approach: 
the three symbols rate the quality of a particular simulation, 
based on the values and range of the chosen parameters,
and its aptness to obtain well-controlled systematic uncertainties. 
They do not rate the quality of the analysis performed by the authors 
of the publication. The latter question is
deferred to the relevant sections of the present review, 
which contain detailed discussions of 
the results contributing (or not) to each FLAG average or estimate. As a result 
of this different approach to the rating criteria, as well as changes of the 
criteria themselves, the colour coding of some papers in the current FLAG 
version differs from that of Ref.~\cite{Aoki:2013ldr}.

For most quantities the colour-coding system refers to the following  
sources of systematic errors: (i) chiral extrapolation; 
(ii) continuum extrapolation; (iii) finite volume. 
As we will see below, renormalization is another source of systematic
uncertainties in several quantities. This we also classify using the 
three coloured symbols listed above, but now with
a different rationale:  they express how reliably these quantities are 
renormalized, from a field-theoretic point of view
(namely nonperturbatively, or with 2-loop or 1-loop perturbation theory).

Given the sophisticated status that the field has attained,
several aspects, besides those rated by the coloured symbols,
need to be evaluated before one can conclude
whether a particular analysis leads to results that should be included in an
average or estimate. Some of these aspects are not so easily expressible
in terms of an adjustable parameter such as the lattice spacing, the pion mass
or the volume. As a result of such considerations,
it sometimes occurs, albeit rarely, that a given
result does not contribute to the FLAG average or estimate, 
despite not carrying any red tags.
This happens, for instance, whenever aspects of the analysis appear 
to be incomplete 
(e.g. an incomplete error budget), so that the presence
of inadequately controlled systematic effects cannot be excluded. 
This mostly refers to results with a statistical error only, or results
in which the quoted error budget obviously fails to account 
for an important contribution.

Of course any colour coding has to be treated with caution; we emphasize
that the criteria are subjective and evolving. Sometimes a single
source of systematic error dominates the systematic uncertainty and it
is more important to reduce this uncertainty than to aim for green
stars for other sources of error. In spite of these caveats we hope
that our attempt to introduce quality measures for lattice simulations
will prove to be a useful guide. In addition we would like to
stress that the agreement of lattice results obtained using
different actions and procedures provides further validation.

\subsubsection{Systematic effects and rating criteria}
\label{sec:Criteria}

The precise criteria used in determining the colour coding are
unavoidably time-dependent; as lattice calculations become more
accurate, the standards against which they are measured become
tighter. For this reason, some of the quality criteria related to the 
light-quark sector have been tightened up between the 
first~\cite{Colangelo:2010et}
and second~\cite{Aoki:2013ldr} editions of FLAG. 

In the second edition we have also reviewed quantities related to
heavy quark physics~\cite{Aoki:2013ldr}. 
The criteria used for light- and heavy-flavour
quantities were not always the same. For the continuum limit, the difference
was more a matter of choice: the light-flavour Working Groups 
defined the ratings using conditions involving specific values of the
lattice spacing, whereas
the heavy-flavour Working Groups preferred more data-driven criteria. 
Also, for finite-volume effects,
the heavy-flavour groups slightly relaxed the boundary between \good \, and \soso,
compared to the light-quark case, 
to account for the fact that heavy-quark quantities are less sensitive to 
the finiteness of the volume.

In the present edition we have opted for simplicity and adopted unified
criteria for both light-  and heavy-flavoured quantities.\footnote{
We note, however, 
that the data-driven criteria
can be used by individual working groups in order to rate the
reliability of the analyses for specific quantities.
}
The colour code used in the tables is specified as follows:
\begin{itemize}
\item Chiral extrapolation:\\
\good \hspace{0.2cm}  $M_{\pi,\mathrm{min}}< 200$ MeV  \\
\rule{0.05em}{0em}\soso \hspace{0.2cm}  200 MeV $\le M_{\pi,{\mathrm{min}}} \le$ 400 MeV \\
\rule{0.05em}{0em}\bad \hspace{0.2cm}  400 MeV $ < M_{\pi,\mathrm{min}}$ \\
It is assumed that the chiral extrapolation is performed with at least a
three-point analysis; otherwise this will be explicitly
mentioned. This condition is unchanged from Ref.~\cite{Aoki:2013ldr}.
\item 
Continuum extrapolation:\\
\good \hspace{0.2cm} at least 3 lattice spacings \underline{and} 
at least 2 points below 0.1 fm 
\underline{and} a range of lattice spacings satisfying 
$[a_{\mathrm{max}}/a_{\mathrm{min}}]^2 \geq 2$
\\ 
\rule{0.05em}{0em}\soso \hspace{0.2cm} at least 2 lattice spacings 
\underline{and} at least 1 point below 0.1 fm 
\underline{and} a range of lattice spacings 
satisfying $[a_{\mathrm{max}}/a_{\mathrm{min}}]^2 \geq 1.4$
\\ 
\rule{0.05em}{0em}\bad \hspace{0.2cm}  otherwise\\
It is assumed that the lattice action is $\cO(a)$-improved (i.e. the
discretization errors vanish quadratically with the lattice spacing);
otherwise this will be explicitly mentioned. For
unimproved actions an additional lattice spacing is required.
This condition has been tightened compared to that of Ref.~\cite{Aoki:2013ldr}
by the requirements concerning the range of lattice spacings.
\item 
Finite-volume effects:\\ 
\good \hspace{0.2cm}  $[M_{\pi,\mathrm{min}} / M_{\pi,\mathrm{fid}}]^2 \exp\{4-M_{\pi,\mathrm{min}}[L(M_{\pi,\mathrm{min}})]_{\mathrm{max}}\} < 1$,
\underline{or} at least 3 volumes\\
\rule{0.05em}{0em}\soso \hspace{0.2cm} $[M_{\pi,\mathrm{min}} / M_{\pi,\mathrm{fid}}]^2 \exp\{3-M_{\pi,\mathrm{min}}[L(M_{\pi,\mathrm{min}})]_{\mathrm{max}}\} < 1$,
\underline{or} at least 2 volumes\\
\rule{0.05em}{0em}\bad \hspace{0.2cm}  otherwise 
\\
It is assumed here that calculations are in the $p$-regime\footnote{We refer to Sec.~\ref{sec:chPT} and Appendix \ref{sec_ChiPT} in the Glossary for an explanation of the various regimes of chiral perturbation theory.} of chiral perturbation
theory, and that all volumes used exceed 2 fm. 
Here we are using a more sophisticated condition than that of
Ref.~\cite{Aoki:2013ldr}.
The new condition involves the quantity
$[L(M_{\pi,\mathrm{min}})]_{\mathrm{max}}$,
which is the maximum box size used in the simulations performed at 
smallest pion mass $M_{\pi,\mathrm{min}}$,
as well as a fiducial pion mass $M_{\pi,\mathrm{fid}}$, which we set to
200 MeV (the cutoff value for a green star in the chiral extrapolation).

The rationale for this condition is as follows.
Finite volume effects contain the universal factor $\exp\{- L~M_\pi\}$,
and if this were the only contribution a criterion based on
the values of $M_{\pi,\textrm{min}} L$ would be appropriate. 
This is what we used in Ref.~\cite{Aoki:2013ldr} 
(with $M_{\pi,\textrm{min}} L>4$ for \good \,
and $M_{\pi,\textrm{min}} L>3$ for \soso).
However, as pion masses decrease, one must also account for
the weakening of the pion couplings. In particular,
1-loop chiral perturbation theory~\cite{Colangelo:2005gd} 
reveals a behaviour proportional to
$M_\pi^2 \exp\{- L~M_\pi\}$. 
Our new condition includes this weakening of the coupling, 
and ensures for example, that simulations with
$M_{\pi,\mathrm{min}} = 135~{\rm MeV}$ and $L~M_{\pi,\mathrm{min}} =
3.2$ are rated equivalently to those with $M_{\pi,\mathrm{min}} = 200~{\rm MeV}$
and $L~M_{\pi,\mathrm{min}} = 4$.

\item Renormalization (where applicable):\\
\good \hspace{0.2cm}  nonperturbative\\
\rule{0.05em}{0em}\soso \hspace{0.2cm}  1-loop perturbation theory or higher  
with a reasonable estimate of truncation errors\\
\rule{0.05em}{0em}\bad \hspace{0.2cm}  otherwise \\
In Ref.~\cite{Colangelo:2010et}, we assigned a red square to all
results which were renormalized at 1-loop in perturbation theory. In 
Ref.~\cite{Aoki:2013ldr} we decided that this was too restrictive, since 
the error arising from renormalization constants, calculated in perturbation theory at
1-loop, is often estimated conservatively and reliably.

\item Renormalization Group (RG) running (where applicable): \\ 
For scale-dependent quantities, such as quark masses or $B_K$, it is
essential that contact with continuum perturbation theory can be established.
Various different methods are used for this purpose
(cf.~Appendix \ref{sec_match}): Regularization-independent Momentum
Subtraction (RI/MOM), the Schr\"odinger functional, and direct comparison with
(resummed) perturbation theory. Irrespective of the particular method used,
the uncertainty associated with the choice of intermediate
renormalization scales in the construction of physical observables
must be brought under control. This is best achieved by performing
comparisons between nonperturbative and perturbative running over a
reasonably broad range of scales. These comparisons were initially
only made in the Schr\"odinger functional approach, but are now
also being performed in RI/MOM schemes.  We mark the data for which
information about nonperturbative running checks is available and
give some details, but do not attempt to translate this into a
colour code.
\end{itemize}

The pion mass plays an important role in the criteria relevant for
chiral extrapolation and finite volume.  For some of the
regularizations used, however, it is not a trivial matter to identify
this mass. 

In the case of twisted-mass fermions, discretization
effects give rise to a mass difference between charged and neutral
pions even when the up- and down-quark masses are equal: the charged pion
is found to be the heavier of the two for twisted-mass Wilson fermions
(cf.~Ref.~\cite{Boucaud:2007uk}).
In early works, typically
referring to $N_f=2$ simulations (e.g.~Refs.~\cite{Boucaud:2007uk}
and~\cite{Baron:2009wt}), chiral extrapolations are based on chiral
perturbation theory formulae which do not take these regularization
effects into account. After the importance of keeping the isospin
breaking when doing chiral fits was shown in Ref.~\cite{Bar:2010jk},
later works, typically referring to $N_f=2+1+1$ simulations, have taken
these effects into account~\cite{Carrasco:2014cwa}.
We use $M_{\pi^\pm}$ for $M_{\pi,\mathrm{min}}$
in the chiral-extrapolation rating criterion. On the
other hand, sea quarks (corresponding to both charged and neutral ``sea pions``
in an effective-chiral-theory logic) 
as well as valence quarks are intertwined with
finite-volume effects. Therefore, we identify $M_{\pi,\mathrm{min}}$ with
the root mean square (RMS) of $M_{\pi^+}$,
$M_{\pi^-}$ and $M_{\pi^0}$ in the finite-volume rating criterion.\footnote{
This is a change from Ref.~\cite{Aoki:2013ldr}, where we used
the charged pion mass when evaluating both chiral-extrapolation
and finite-volume effects.}

In the case of staggered fermions,
discretization effects give rise to several light states with the
quantum numbers of the pion.\footnote{
We refer the interested reader to a number of good reviews on the
subject~\cite{Durr:2005ax,Sharpe:2006re,Kronfeld:2007ek,Golterman:2008gt,Bazavov:2009bb}.}
The mass splitting among these ``taste'' partners represents a
discretization effect of $\cO(a^2)$, which can be significant at large
lattice spacings but shrinks as the spacing is reduced. In the
discussion of the results obtained with staggered quarks given in the
following sections, we assume that these artefacts are under
control. We conservatively identify $M_{\pi,\mathrm{min}}$ with the root mean
square (RMS) average of the masses of all the taste partners, 
both for chiral-extrapolation and finite-volume criteria.\footnote{
In Ref.~\cite{Aoki:2013ldr}, the RMS value
was used in the chiral-extrapolation criteria throughout the paper. 
For the finite-volume rating, however,
$M_{\pi,\mathrm{min}}$ was identified with the RMS value only in
Secs.~\ref{sec:vusvud} and \ref{sec:BK}, while in Secs.~\ref{sec:qmass},
\ref{sec:LECs}, \ref{sec:DDecays} and \ref{sec:BDecays} it
was identified with the mass of the lightest pseudoscalar state.}

The strong coupling $\alpha_s$ is computed in lattice QCD with methods
differing substantially
from those used in the calculations of the other quantities 
discussed in this review. Therefore we have established separate criteria for
$\alpha_s$ results, which will be discussed in Sec.~\ref{s:crit}.

\subsubsection{Heavy-quark actions}
\label{sec:HQCriteria}

In most cases, and in particular for the $b$ quark,
the discretization of the
heavy-quark action follows a very different approach to that used for light
flavours. There are several different methods for
treating heavy quarks on the lattice, each with their own issues and
considerations.  All of these methods use
Effective Field Theory (EFT) at some point in the computation, either
via direct simulation of the EFT, or by using EFT
as a tool to estimate the size of cutoff errors, 
or by using EFT to extrapolate from the simulated
lattice quark masses up to the physical $b$-quark mass. 
Because of the use of an EFT, truncation errors must be
considered together with discretization errors. 

The charm quark lies at an intermediate point between the heavy
and light quarks. In our previous review, the bulk of the calculations
involving charm quarks treated it using one of the approaches adopted
for the $b$ quark. Many recent calculations, however, simulate
the charm quark using light-quark actions, 
in particular the $N_f=2+1+1$ calculations.
This has become possible thanks to the increasing availability of
dynamical gauge field ensembles with fine lattice spacings.
But clearly, when charm quarks are treated relativistically, discretization
errors are more severe than those of
the corresponding light-quark quantities.

In order to address these complications, we add a new heavy-quark
treatment category to the rating system. The purpose of this
criterion is to provide a guideline for the level of action and
operator improvement needed in each approach to make reliable
calculations possible, in principle. 

A description of the different approaches to treating heavy quarks on
the lattice is given in Appendix~\ref{app:HQactions}, including a
discussion of the associated discretization, truncation, and matching
errors.  For truncation errors we use HQET power counting throughout,
since this review is focused on heavy quark quantities involving $B$
and $D$ mesons rather than bottomonium or charmonium quantities.  
Here we describe the criteria for how each approach
must be implemented in order to receive an acceptable (\okay) rating
for both the heavy quark actions and the weak operators.  Heavy-quark
implementations without the level of improvement described below are
rated not acceptable (\bad). The matching is evaluated together with
renormalization, using the renormalization criteria described in
Sec.~\ref{sec:Criteria}.  We emphasize that the heavy-quark
implementations rated as acceptable and described below have been
validated in a variety of ways, such as via phenomenological agreement
with experimental measurements, consistency between independent
lattice calculations, and numerical studies of truncation errors.
These tests are summarized in Sec.~\ref{sec:BDecays}.  \smallskip
\\ {\it Relativistic heavy quark actions:} \\
\noindent 
\okay \hspace{0.2cm}   at least tree-level $\cO(a)$ improved action and 
weak operators  \\
This is similar to the requirements for light quark actions. All
current implementations of relativistic heavy quark actions satisfy
this criterion. \smallskip \\
{\it NRQCD:} \\
\noindent 
\okay \hspace{0.2cm}   tree-level matched through $\cO(1/m_h)$ 
and improved through $\cO(a^2)$ \\
The current implementations of NRQCD satisfy this criterion, and also
include tree-level corrections of $\cO(1/m_h^2)$ in the action. 
\smallskip \\
{\it HQET: }\\
\noindent 
\okay \hspace{0.2cm}  tree-level  matched through $\cO(1/m_h)$ 
with discretization errors starting at $\cO(a^2)$ \\
The current implementation of HQET by the ALPHA collaboration
satisfies this criterion, since both action and weak operators are
matched nonperturbatively through $\cO(1/m_h)$.  Calculations that
exclusively use a static-limit action do not satisfy this criterion,
since the static-limit action, by definition, does not include $1/m_h$
terms.  However for $SU(3)$-breaking ratios,
such as $\xi$ and $f_{B_s}/f_B$, truncation
errors start at $\cO((m_s - m_d)/m_h)$. We therefore consider lattice
calculations of such ratios that use a static-limit action to still
have controllable truncation errors. \smallskip \\
{\it Light-quark actions for heavy quarks:}  \\
\noindent 
\okay \hspace{0.2cm}  discretization errors starting at $\cO(a^2)$ or higher \\
This applies to calculations that use the tmWilson action, a
nonperturbatively improved Wilson action, or the HISQ action for charm
quark quantities. It also applies to calculations that use these light
quark actions in the charm region and above together with either the
static limit or with an HQET inspired extrapolation to obtain results
at the physical $b$ quark mass. In these cases, the continuum
extrapolation criteria described earlier 
must be applied to the entire range of heavy-quark masses used in 
the calculation.

\subsubsection{Conventions for the figures}
\label{sec:figurecolours}

For a coherent assessment of the present situation, the quality of the
data plays a key role, but the colour coding cannot be carried over to
the figures. On the other hand, simply showing all data on equal
footing would give the misleading impression that the overall
consistency of the information available on the lattice is
questionable. Therefore, in the figures we indicate the quality of the data
in a rudimentary way, using the following symbols:
\\ \raisebox{0.35mm}{\hspace{0.65mm}{\color{darkgreen}$\blacksquare$}} \hspace{0.2cm}
corresponds to results included in the average or estimate (i.e. results that contribute to the black square below);
\\ \raisebox{0.35mm}{\hspace{0.65mm}{\color{lightgreen}$\blacksquare$\hspace{-0.3cm}\color{darkgreen}$\square$}} \hspace{0.2cm}
corresponds to results that are not included in the average but pass all quality
criteria;
\\ \raisebox{0.35mm}{\hspace{0.65mm}{\color{red}$\square$}} \hspace{0.2cm}
corresponds to all other results;
\\ \raisebox{0.35mm}{\hspace{0.65mm}{\color{black}$\blacksquare$}} \hspace{0.2cm}
corresponds to FLAG averages or estimates; they are also highlighted by a gray vertical band.
\\ The reason for not including a given result in
the average is not always the same: the result may fail one of the
quality criteria; the paper may be unpublished; 
it may be superseded by newer results;
or it may not offer a complete error budget. 

Symbols other than squares are
used to distinguish results with specific properties and are always
explained in the caption.\footnote{%
For example, for quark mass results we
distinguish between perturbative and nonperturbative renormalization, 
for low-energy constants we distinguish between the $p$- and $\epsilon$-regimes, 
and for heavy flavour results we distinguish between
those from leptonic and semi-leptonic decays.}

Often nonlattice data are also shown in the figures for comparison. 
For these we use the following symbols:
\\ \raisebox{0.15mm}{\hspace{0.65mm}\color{blue}\Large\textbullet} \hspace{0.2cm}
corresponds to nonlattice results;
\\ \raisebox{0.35mm}{\hspace{0.65mm}{\color{black}$\blacktriangle$}} \hspace{0.2cm}
corresponds to Particle Data Group (PDG) results.

\subsection{Averages and estimates}\label{sec:averages}

FLAG results of a given quantity are denoted either as {\it averages} or as {\it estimates}. Here we clarify this distinction. To start with, both {\it averages} and {\it estimates} are based on results without any red tags in their colour coding. For many observables there are enough independent lattice calculations of good quality, with all sources of error (not merely those related to the colour-coded criteria), as analyzed in the original papers, appearing to be under control. In such cases it makes sense to average these results and propose such an {\it average} as the best current lattice number. The averaging procedure applied to this data and the way the error is obtained is explained in detail in Sec.~\ref{sec:error_analysis}. In those cases where only a sole result passes our rating criteria (colour coding), we refer to it as our FLAG {\it average}, provided it also displays adequate control of all other sources of systematic uncertainty.

On the other hand, there are some cases in which this procedure leads to a result that, in our opinion, does not cover all uncertainties. Systematic error estimates are by their nature often subjective and difficult to estimate, and may thus end up being underestimated in one or more results that receive green symbols for all explicitly tabulated criteria.   
Adopting a conservative policy, in these cases we opt for an {\it estimate} (or a range), which we consider as a fair assessment of the knowledge acquired on the lattice at present. This {\it estimate} is not obtained with a prescribed mathematical procedure, but reflects what we consider the best possible analysis of the available information. The hope is that this will encourage more detailed investigations by the lattice community.

There are two other important criteria that also play a role in this
respect, but that cannot be colour coded, because a systematic
improvement is not possible. These are: {\em i)} the publication
status, and {\em ii)} the number of sea-quark flavours $\Nf$. As far as the
former criterion is concerned, we adopt the following policy: we
average only results that have been published in peer-reviewed
journals, i.e.~they have been endorsed by referee(s). The only
exception to this rule consists in straightforward updates of previously
published results, typically presented in conference proceedings. Such
updates, which supersede the corresponding results in the published
papers, are included in the averages. 
Note that updates of earlier results rely, at least partially, on the
same gauge-field-configuration ensembles. For this reason, we do not
average updates with earlier results. 
Nevertheless, all results are
listed in the tables,\footnote{%
Whenever figures turn out to be overcrowded,
older, superseded results are omitted. However, all the most recent results
from each collaboration are displayed.}
and their publication status is identified by the following
symbols:
\begin{itemize}
\item Publication status:\\
\gA  \hspace{0.2cm}published or plain update of published results\\
\oP  \hspace{0.2cm}preprint\\ 
\rC  \hspace{0.2cm}conference contribution
\end{itemize}
In the present edition, the
publication status on the {\bf 30th of November 2015} is relevant. If the paper
appeared in print after that date, this is accounted for in the
bibliography, but does not affect the averages.

As noted above,
in this review we present results from simulations with $N_f=2$,
$N_f=2+1$ and $N_f=2+1+1$ (except for $ r_0 \Lambda_\msbar$ where we
also give the $N_f=0$ result). We are not aware of an {\em a priori} way
to quantitatively estimate the difference between results produced in
simulations with a different number of dynamical quarks. We therefore
average results at fixed $\Nf$ separately; averages of calculations
with different $\Nf$ will not be provided.

To date, no significant differences between results with different
values of $N_f$ have been observed in the quantities 
listed in Tabs.~\ref{tab:summary1} and \ref{tab:summary2}.
In the future, as the accuracy
and the control over systematic effects in lattice calculations 
increases, it will hopefully be possible to see a difference between results
from simulations with $\Nf = 2$ and $\Nf = 2 + 1$, 
and thus determine the size of the
Zweig-rule violations related to strange-quark loops. This is a very
interesting issue {\em per se}, and one which can be quantitatively 
addressed only with lattice calculations.

The question of differences between results with $\Nf=2+1$ and
$\Nf=2+1+1$ is more subtle.
The dominant effect of including the charm sea quark is to
shift the lattice scale, an effect that is accounted for by
fixing this scale nonperturbatively using physical quantities.
For most of the quantities discussed in this review, it is 
expected that residual effects are small in the continuum limit,
suppressed by $\alpha_s(m_c)$ and powers of $\Lambda^2/m_c^2$.
Here $\Lambda$ is a hadronic scale that can only be
roughly estimated and depends on the process under consideration.
Note that the $\Lambda^2/m_c^2$ effects have been addressed in~Ref.~\cite{Bruno:2014ufa}.
Assuming that such effects are small, it might be reasonable to
average the results from $\Nf=2+1$ and $\Nf=2+1+1$ simulations.
This is not yet a pressing issue in this review, since there are
relatively few results with $\Nf=2+1+1$, but it will become a
more important question in the future.

\subsection{Averaging procedure and error analysis}
\label{sec:error_analysis}

In the present report we repeatedly average results
obtained by different collaborations and estimate the error on the resulting
averages. We follow the procedure of the previous  edition~\cite{Aoki:2013ldr},
which we describe here in full detail.

One of the problems arising when forming averages is that not all
of the datasets are independent.
In particular, the same gauge-field configurations,
produced with a given fermion descretization, are often used by
different research teams with different valence-quark lattice actions,
obtaining results that are not really independent.  
Our averaging procedure takes such correlations into account. 

Consider a given measurable quantity $Q$, measured by $M$ distinct,
not necessarily uncorrelated, numerical experiments (simulations). The result
of each of these measurement is expressed as
\begin{equation}
Q_i \,\, = \,\, x_i \, \pm \, \sigma^{(1)}_i \pm \, \sigma^{(2)}_i \pm \cdots
\pm \, \sigma^{(E)}_i  \,\,\, ,
\label{eq:resultQi}
\end{equation}
where $x_i$ is the value obtained by the $i^{\rm th}$ experiment
($i = 1, \cdots , M$) and $\sigma^{(k)}_i$ (for $k = 1, \cdots , E$) 
are the various errors.
Typically $\sigma^{(1)}_i$ stands for the statistical error 
and $\sigma^{(k)}_i$ ($k \ge 2$) are the different
systematic errors from various sources. 
For each individual result, we estimate the total
error $\sigma_i $ by adding statistical and systematic errors in quadrature:
\begin{eqnarray}
Q_i \,\, &=& \,\, x_i \, \pm \, \sigma_i \,\,\, ,
\nonumber \\
\sigma_i \,\, &\equiv& \,\, \sqrt{\sum_{k=1}^E \Big [\sigma^{(k)}_i \Big ]^2} \,\,\, .
\label{eq:av-err-Qi}
\end{eqnarray}
With the weight factor of each total error estimated in standard fashion:
\begin{equation}
\omega_i \,\, = \,\, \dfrac{\sigma_i^{-2}}{\sum_{i=1}^M \sigma_i^{-2}} \,\,\, ,
\label{eq:weighti}
\end{equation}
the central value of the average over all simulations is given by
\begin{eqnarray}
x_{\rm av} \,\, &=& \,\, \sum_{i=1}^M x_i\, \omega_i \,\, . 
\end{eqnarray}
The above central value corresponds to a $\chi_{\rm min}^2$ weighted
average, evaluated by adding statistical and systematic errors in quadrature.
If the fit is not of good quality ($\chi_{\rm min}^2/dof > 1$),
the statistical and systematic error bars are stretched by a factor
$S = \sqrt{\chi^2/dof}$.

Next we examine error budgets for
individual calculations and look for potentially correlated
uncertainties. Specific problems encountered in connection with
correlations between different data sets are described in the text
that accompanies the averaging.
If there is reason to believe that a source of error is correlated
between two calculations, a $100\%$ correlation is assumed. 
The correlation matrix $C_{ij}$ for the set of correlated lattice results is
estimated by a prescription due to Schmelling~\cite{Schmelling:1994pz}.
This consists in defining
\begin{equation}
\sigma_{i;j} \,\, = \,\, \sqrt{\sum^\prime_{(k)} \Big[ \sigma_i^{(k)} \Big]^2 } \,\,\, ,
\end{equation}
with $\sum_{(k)}^\prime$ running only over those errors of $x_i$ that
are correlated with the corresponding errors of measurement $x_j$. 
This expresses the part of the uncertainty in $x_i$
that is correlated with the uncertainty in $x_j$. 
If no such correlations are known to exist, then
we take $\sigma_{i;j} =0$. 
The diagonal and off-diagonal elements of the correlation
matrix are then taken to be
\begin{eqnarray}
C_{ii} \,\,&=& \,\, \sigma_i^2 \qquad \qquad (i = 1, \cdots , M) \,\,\, ,
\nonumber \\
C_{ij} \,\,&=& \,\, \sigma_{i;j} \, \sigma_{j;i} \qquad \qquad (i \neq j) \,\,\, .
\end{eqnarray}
Finally the error of the average is estimated by
\begin{equation}
\sigma^2_{\rm av} \,\, = \,\, \sum_{i=1}^M \sum_{j=1}^M \omega_i \,\omega_j \,C_{ij}\,\,,
\end{equation}
and the FLAG average is
\begin{equation}
Q_{\rm av} \,\, = \,\, x_{\rm av} \, \pm \, \sigma_{\rm av} \,\,\, .
\end{equation}

%% file: qmass/qmass.tex
% !TEX root = /Users/tblum/Dropbox/FLAG3_Review/working_copy_v1.2/FLAG_master.tex
\section{Quark masses}
\label{sec:qmass}

Quark masses are fundamental parameters of the Standard Model. An
accurate determination of these parameters is important for both
phenomenological and theoretical applications. The charm and bottom
masses, for instance, enter the theoretical expressions of several
cross sections and decay rates in heavy-quark expansions. The up-,
down- and strange-quark masses govern the amount of explicit chiral
symmetry breaking in QCD. From a theoretical point of view, the values
of quark masses provide information about the flavour structure of
physics beyond the Standard Model. The Review of Particle Physics of
the Particle Data Group contains a review of quark masses
\cite{Manohar_and_Sachrajda}, which covers light as well as heavy
flavours. Here we also consider light- and heavy- quark masses, but
focus on lattice results and discuss them in more detail. We do not
discuss the top quark, however, because it decays weakly before it can
hadronize, and the nonperturbative QCD dynamics described by present
day lattice simulations is not relevant. The lattice determination of
light- (up, down, strange), charm- and bottom-quark masses is
considered below in Secs.~\ref{sec:lqm}, \ref{s:cmass},
and \ref{s:bmass}, respectively.

Quark masses cannot be measured directly in experiment because
quarks cannot be isolated, as they are confined inside hadrons. On the
other hand, quark masses are free parameters of the theory and, as
such, cannot be obtained on the basis of purely theoretical
considerations. Their values can only be determined by comparing the
theoretical prediction for an observable, which depends on the quark
mass of interest, with the corresponding experimental value.

In the last edition of this review \cite{Aoki:2013ldr}, quark-mass
determinations came from two- and three-flavour QCD
calculations. Moreover, these calculations were most often performed
in the isospin limit, where the up- and down-quark masses (especially
those in the sea) are set equal. In addition, some of the results
retained in our light-quark mass averages were based on simulations
performed at values of $m_{ud}$ which were still substantially larger
than its physical value imposing a significant extrapolation to reach
the physical up- and down-quark mass point. Among the calculations
performed near physical $m_{ud}$ by
PACS-CS~\cite{Aoki:2008sm,Aoki:2009ix,Aoki:2010wm},
BMW~\cite{Durr:2010vn,Durr:2010aw} and
RBC/UKQCD~\cite{Arthur:2012opa}, only the ones in Refs.~\cite{Durr:2010vn,Durr:2010aw} did so while controlling all other
sources of systematic error.

Today, however, the effects of the charm quark in the sea are more and
more systematically considered and most of the new quark-mass results
discussed below have been obtained in $N_f=2+1+1$ simulations by ETM
\cite{Carrasco:2014cwa}, HPQCD \cite{Bazavov:2014wgs} and FNAL/MILC
\cite{Chakraborty:2014aca}. In addition, RBC/UKQCD
\cite{Blum:2014tka}, HPQCD \cite{Bazavov:2014wgs} and FNAL/MILC
\cite{Chakraborty:2014aca} are extending their calculations down to
up-down-quark masses at or very close to their physical values while
still controlling other sources of systematic error. Another aspect
that is being increasingly addressed are electromagnetic and
$(m_d-m_u)$, strong isospin-breaking effects. As we will see below
these are particularly important for determining the individual up- and
down-quark masses. But with the level of precision being reached in
calculations, these effects are also becoming important for other
quark masses.

\medskip

Three-flavour QCD has four free parameters: the strong coupling,
$\alpha_s$ (alternatively $\Lambda_\mathrm{QCD}$) and the up-, down- and
strange-quark masses, $m_u$, $m_d$ and $m_s$. Four-flavour calculations
have an additional parameter, the charm-quark mass $m_c$. When the
calculations are performed in the isospin limit, up- and down-quark
masses are replaced by a single parameter: the isospin-averaged up-
and down-quark mass, $m_{ud}=\frac12(m_u+m_d)$.  A lattice
determination of these parameters, and in particular of the quark
masses, proceeds in two steps:
\begin{enumerate}
\item
  One computes as many experimentally measurable quantities as there
  are quark masses. These observables should obviously be sensitive to
  the masses of interest, preferably straightforward to compute and
  obtainable with high precision. They are usually computed for a
  variety of input values of the quark masses which are then adjusted
  to reproduce experiment. Another observable, such as the pion decay constant 
  or the mass of a member of the baryon octet, must be used to fix the overall
  scale. Note that the mass of a quark, such as the $b$, which is not
  accounted for in the generation of gauge configurations, can still
  be determined. For that an additional valence-quark observable
  containing this quark must be computed and the mass of that quark
  must be tuned to reproduce experiment.

\item The input quark masses are bare parameters which depend on the
  lattice spacing and particulars of the lattice regularization used
  in the calculation. To compare their values at different lattice
  spacings and to allow a continuum extrapolation they must be
  renormalized.  This renormalization is a short-distance calculation,
  which may be performed perturbatively.  Experience shows that
  1-loop calculations are unreliable for the renormalization of
  quark masses: usually at least two loops are required to have
  trustworthy results. Therefore, it is best to perform the
  renormalizations nonperturbatively to avoid potentially large
  perturbative uncertainties due to neglected higher-order
  terms. Nevertheless we will include in our averages 1-loop results
  if they carry a solid estimate of the systematic uncertainty due to
  the truncation of the series.
\end{enumerate}
In the absence of electromagnetic corrections, the renormalization
factors for all quark masses are the same at a given lattice
spacing. Thus, uncertainties due to renormalization are absent in
ratios of quark masses if the tuning of the masses to their physical
values can be done lattice spacing by lattice spacing and
significantly reduced otherwise.

We mention that lattice QCD calculations of the $b$-quark mass have an additional complication which is not present in the case of the charm- and light-quarks.
At the lattice spacings currently used in numerical simulations the direct treatment of the $b$ quark with the fermionic actions commonly used for light quarks will result in large cutoff effects, because the $b$-quark mass is of order one in lattice units.
There are a few widely used approaches to treat the $b$ quark on the lattice, which have been already discussed in the FLAG 13 review (see Section 8 of Ref.~\cite{Aoki:2013ldr}).
Those relevant for the determination of the $b$-quark mass will be briefly described in Sec.~\ref{s:bmass}.

%\newpage

\input{qmass/qmass_mudms}

%\newpage

\input{qmass/qmass_mc}

\newpage

\input{qmass/qmass_mb}

%% file: qmass/qmass_mudms.tex
% !TEX root = /Users/tblum/Dropbox/FLAG3_Review/working_copy_v1.2/qmass/qmass.tex
\subsection{Masses of the light quarks}
\label{sec:lqm}

Light-quark masses are particularly difficult to determine because
they are very small (for the up and down quarks) or small (for
the strange quark) compared to typical hadronic scales. Thus, their impact
on typical hadronic observables is minute, and it is difficult to
isolate their contribution accurately.

Fortunately, the spontaneous breaking of $SU(3)_L\times SU(3)_R$
chiral symmetry provides observables which are particularly sensitive
to the light-quark masses: the masses of the resulting Nambu-Goldstone
bosons (NGB), i.e.~pions, kaons and etas. Indeed, the
Gell-Mann-Oakes-Renner relation~\cite{GellMann:1968rz} predicts that
the squared mass of a NGB is directly proportional to the sum of the
masses of the quark and antiquark which compose it, up to higher-order
mass corrections. Moreover, because these NGBs are light and are
composed of only two valence particles, their masses have a
particularly clean statistical signal in lattice-QCD calculations. In
addition, the experimental uncertainties on these meson masses are
negligible. Thus, in lattice calculations, light-quark masses are
typically obtained by renormalizing the input quark mass and tuning
them to reproduce NGB masses, as described above.

\subsubsection{Contributions from the electromagnetic interaction}
\label{subsec:electromagnetic interaction}

As mentioned in Sec.~\ref{sec:color-code}, the present review
relies on the hypothesis that, at low energies, the Lagrangian ${\cal
  L}_{\mbox{\tiny QCD}}+{\cal L}_{\mbox{\tiny QED}}$ describes nature
to a high degree of precision. However, most of the results presented
below are obtained in pure QCD calculations, which do not include
QED. Quite generally, when comparing QCD calculations with experiment,
radiative corrections need to be applied. In pure QCD simulations,
where the parameters are fixed in terms of the masses of some of the
hadrons, the electromagnetic contributions to these masses must be
accounted for. Of course, once QED is included in lattice calculations, the subtraction
of e.m.~contributions is no longer necessary. 

The electromagnetic interaction plays a particularly important role in
determinations of the ratio $m_u/m_d$, because the isospin-breaking
effects generated by this interaction are comparable to those from
$m_u\neq m_d$ (see Subsection \ref{subsec:mumd}). In determinations of
the ratio $m_s/m_{ud}$, the electromagnetic interaction is less
important, but at the accuracy reached, it cannot be neglected. The
reason is that, in the determination of this ratio, the pion mass
enters as an input parameter. Because $M_\pi$ represents a small
symmetry-breaking effect, it is rather sensitive to the perturbations
generated by QED.

The decomposition of the sum ${\cal L}_{\mbox{\tiny QCD}}+{\cal
  L}_{\mbox{\tiny QED}}$ into two parts is not unique and specifying the
QCD part requires a convention. In order to give results for the quark
masses in the Standard Model at scale $\mu=2\,\mbox{GeV}$, on the
basis of a calculation done within QCD, it is convenient to match the
parameters of the two theories at that scale. We use this convention throughout the
present review~\footnote{Note that a different convention is used in the
analysis of the precision measurements carried out in low-energy pion
physics (e.g. Ref.~\cite{BlochDevaux:2008zz}). When comparing lattice
results with experiment, it is important to fix the QCD parameters in
accordance with the convention used in the analysis of the
experimental data (for a more detailed discussion, see Refs.~\cite{Gasser:2003hk,Rusetsky:2009ic,Gasser:2007de,Leutwyler:2009jg}).}.

Such a convention allows us to distinguish the physical mass $M_P$,
$P\in\{\pi^+,$ $\pi^0$, $K^+$, $K^0\}$, from the mass $\hat{M}_P$
within QCD. The e.m.~self-energy is the difference between the
two, $M_P^\gamma\equiv M_P-\hat{M}_P$. Because the self-energy of the
Nambu-Goldstone bosons diverges in the chiral limit, it is convenient
to replace it by the contribution of the e.m.~interaction to the {\it
  square} of the mass,
\be \label{eq:DeltaP}
\Delta_{P}^\gamma\equiv M_P^2-\hat{M}_P^2= 2\,M_P M_P^\gamma+\cO(e^4)\,.\ee 
The main
effect of the e.m.\ interaction is an increase in the mass of the
charged particles, generated by the photon cloud that surrounds
them. The self-energies of the neutral ones are comparatively small,
particularly for the Nambu-Goldstone bosons, which do not have a
magnetic moment. Dashen's theorem~\cite{Dashen:1969eg} confirms this
picture, as it states that, to leading order (LO) of the chiral expansion,
the self-energies of the neutral NGBs vanish, while the charged ones
obey $\Delta_{K^+}^\gamma = \Delta_{\pi^+}^\gamma $. It is convenient
to express the self-energies of the neutral particles as well as the
mass difference between the charged and neutral pions within QCD in
units of the observed mass difference, $\Delta_\pi\equiv
M_{\pi^+}^2-M_{\pi^0}^2$:
\be\label{eq:epsilon1}
\Delta_{\pi^0}^\gamma \equiv
\epsilon_{\pi^0}\,\Delta_\pi\co\hspace{0.2cm}\Delta_{K^0}^\gamma \equiv
\epsilon_{K^0}\,\Delta_\pi\co\hspace{0.2cm}\hat{M}_{\pi^+}^2-
\hat{M}_{\pi^0}^2\equiv
\epsilon_m\,\Delta_\pi\fs\ee
In this notation, the self-energies of the charged particles are given
by 
\be\label{eq:epsilon2}
\Delta_{\pi^+}^\gamma=(1+\epsilon_{\pi^0}-\epsilon_m)\,\Delta_\pi\co\hspace{0.5cm}
\Delta_{K^+}^\gamma=(1+\epsilonD+\epsilon_{K^0}-\epsilon_m)\,\Delta_\pi\co\ee
where the dimensionless coefficient $\epsilonD$ parameterizes the
violation of Dashen's theorem, \footnote{\label{fn1}Sometimes,
  e.g.~in~Ref.~\cite{Blum:2010ym}, the violation of Dashen's theorem is given in
  terms of a different quantity, $\bar\epsilon\equiv
  (\Delta_{K^+}^\gamma-\Delta_{K^0}^\gamma)/(\Delta_{\pi^+}^\gamma-\Delta_{\pi^0}^\gamma)-1$. This
  parameter is related to $\epsilonD$ used here through
  $\epsilonD=(1-\epsilon_m)\bar\epsilon$. Given the value of
  $\epsilon_m$ (see Eq.~(\ref{eq:epsilon num})), these two quantities differ
  by 4\% only.}
\be\label{eq:epsilon3}
\Delta_{K^+}^\gamma-\Delta_{K^0}^\gamma-
\Delta_{\pi^+}^\gamma+\Delta_{\pi^0}^\gamma\equiv\epsilonD\,\Delta_\pi\fs\ee
Any determination of the light-quark masses based on a calculation of
the masses of $\pi^+,K^+$ and $K^0$ within QCD requires an estimate
for the coefficients $\epsilonD$, $\epsilon_{\pi^0}$, $\epsilon_{K^0}$
and $\epsilon_m$.

The first determination of the self-energies on the lattice was
carried out by Duncan, Eichten and Thacker~\cite{Duncan:1996xy}. Using
the quenched approximation, they arrived at
$M_{K^+}^\gamma-M_{K^0}^\gamma= 1.9\,\mbox{MeV}$. Actually, the
parameterization of the masses given in that paper yields an estimate
for all but one of the coefficients introduced above (since the mass
splitting between the charged and neutral pions in QCD is neglected,
the parameterization amounts to setting $\epsilon_m=0$ ab
initio). Evaluating the differences between the masses obtained at the
physical value of the electromagnetic coupling constant and at $e=0$,
we obtain $\epsilonD = 0.50(8)$, $\epsilon_{\pi^0} = 0.034(5)$ and
$\epsilon_{K^0} = 0.23(3)$. The errors quoted are statistical only: an
estimate of lattice systematic errors is not possible from the limited
results of~Ref.~\cite{Duncan:1996xy}. The result for $\epsilonD$ indicates
that the violation of Dashen's theorem is sizeable: according to this
calculation, the nonleading contributions to the self-energy
difference of the kaons amount to 50\% of the leading term. The result
for the self-energy of the neutral pion cannot be taken at face value,
because it is small, comparable to the neglected mass difference
$\hat{M}_{\pi^+}-\hat{M}_{\pi^0}$. To illustrate this, we note that
the numbers quoted above are obtained by matching the parameterization
with the physical masses for $\pi^0$, $K^+$ and $K^0$. This gives a
mass for the charged pion that is too high by 0.32 MeV. Tuning the
parameters instead such that $M_{\pi^+}$ comes out correctly, the
result for the self-energy of the neutral pion becomes larger:
$\epsilon_{\pi^0}=0.10(7)$ where, again, the error is statistical
only.

In an update of this calculation by the RBC collaboration~\cite{Blum:2007cy} (RBC 07), the electromagnetic interaction is still
treated in the quenched approximation, but the strong interaction is
simulated with $\Nf=2$ dynamical quark flavours. The quark masses are fixed
with the physical masses of $\pi^0$, $K^+$ and $K^0$. The outcome for the
difference in the electromagnetic self-energy of the kaons reads
$M_{K^+}^\gamma-M_{K^0}^\gamma= 1.443(55)\,\mbox{MeV}$. This corresponds to
a remarkably small violation of Dashen's theorem. Indeed, a recent
extension of this work to $\Nf=2+1$ dynamical flavours~\cite{Blum:2010ym}
leads to a significantly larger self-energy difference:
$M_{K^+}^\gamma-M_{K^0}^\gamma= 1.87(10)\,\mbox{MeV}$, in good agreement
with the estimate of Eichten et al. Expressed in terms of the coefficient
$\epsilonD$ that measures the size of the violation of Dashen's theorem, it
corresponds to $\epsilonD=0.5(1)$.

The input for the electromagnetic corrections used by MILC is
specified in~Ref.~\cite{Aubin:2004he}. In their analysis of the lattice data, 
$\epsilon_{\pi^0}$, $\epsilon_{K^0}$ and $\epsilon_m$ are set
equal to zero. For the remaining coefficient, which plays a crucial
role in determinations of the ratio $m_u/m_d$, the very conservative
range $\epsilonD=1(1)$ was used in MILC 04~\cite{Aubin:2004fs}, while 
in MILC 09~\cite{Bazavov:2009bb}
and MILC 09A~\cite{Bazavov:2009fk} this input has been replaced by
$\epsilonD=1.2(5)$, as suggested by phenomenological estimates for
the corrections to Dashen's theorem~\cite{Bijnens:1996kk,Donoghue:1996zn}. Results of an evaluation of the
electromagnetic self-energies based on $\Nf=2+1$ dynamical quarks in
the QCD sector and on the quenched approximation in the QED sector have been
also reported by MILC in Refs.~\cite{Basak:2008na,Basak:2012zx,Basak:2013iw} and updated recently in~Refs.~\cite{Basak:2014vca,Basak:2015lla}. 
Their latest (preliminary) result is $\bar\epsilon = 0.84(5)(19)$, where the first error is statistical and the second systematic, coming from discretization and finite-volume uncertainties added in quadrature. With the estimate for $\epsilon_m$ given in Eq.~(\ref{eq:epsilon num}), this result corresponds to $\epsilon = 0.81(5)(18)$.
  
Preliminary results have been also reported by the BMW collaboration in conference proceedings~\cite{Portelli:2010yn,Portelli:2012pn,Portelli:2015wna}, with the updated result  being $\epsilon = 0.57(6)(6)$, where the first error is statistical and the second systematic.

The RM123 collaboration employs a new technique to compute e.m.~shifts
in hadron masses in 2-flavour QCD: the effects are included at
leading order in the electromagnetic coupling $\alpha$ through simple
insertions of the fundamental electromagnetic interaction in quark
lines of relevant Feynman graphs~\cite{deDivitiis:2013xla}. They find
$\epsilon=0.79(18)(18)$, where the first error is statistical and the
second is the total systematic error resulting from chiral, finite-volume, 
discretization, quenching and fitting errors all added in
quadrature.

Recently~\cite{Horsley:2015vla} the QCDSF/UKQCD collaboration has presented results for several pseudoscalar meson masses obtained from $N_f = 2+1$ dynamical simulations of QCD + QED (at a single lattice spacing $ a \simeq 0.07$ fm).
Using the experimental values of the $\pi^0$, $K^0$ and $K^+$ mesons masses to fix the three light-quark masses, they find $\epsilon = 0.50 (6)$, where the error is statistical only.

The effective Lagrangian that governs the self-energies to next-to-leading
order (NLO) of the chiral expansion was set up in~Ref.~\cite{Urech:1994hd}. The
estimates made in Refs.~\cite{Bijnens:1996kk,Donoghue:1996zn} are obtained by
replacing QCD with a model, matching this model with the effective theory
and assuming that the effective coupling constants obtained in this way
represent a decent approximation to those of QCD. For alternative model
estimates and a detailed discussion of the problems encountered in models
based on saturation by resonances, see~Refs.~\cite{Baur:1995ig,Baur:1996ya,Moussallam:1997xx}.  In the present review of
the information obtained on the lattice, we avoid the use of models
altogether.

There is an indirect phenomenological determination of $\epsilonD$,
which is based on the decay $\eta\rightarrow 3\pi$ and does not rely
on models.  The result for the quark-mass ratio $Q$, defined in
Eq.~(\ref{eq:Qm}) and obtained from a dispersive analysis of this decay,
implies $\epsilonD = 0.70(28)$ (see Sec.~\ref{subsec:mumd}).  {
  While the values found in older lattice calculations~\cite{Duncan:1996xy,Blum:2007cy,Blum:2010ym} are a little less than
  one standard deviation lower, the most recent determinations~\cite{Basak:2008na,Portelli:2010yn,Portelli:2012pn,llconfx12,Basak:2014vca,Basak:2015lla,Basak:2012zx,Basak:2013iw,deDivitiis:2013xla},
  though still preliminary, are in excellent agreement with this
  result and have significantly smaller error bars. However, even in
  the more recent calculations, e.m.\ effects are treated in the
  quenched approximation. Thus, we choose to quote $\epsilonD =
  0.7(3)$, which is essentially the $\eta\rightarrow 3\pi$ result and
  covers the range of post-2010 lattice results. Note that
  this value has an uncertainty which is reduced by about 40\%
  compared to the result quoted in the first edition of the FLAG review~\cite{Colangelo:2010et}.}

We add a few comments concerning the physics of the self-energies and then
specify the estimates used as an input in our analysis of the data. The
Cottingham formula~\cite{Cottingham} represents the self-energy of a
particle as an integral over electron scattering cross sections; elastic as
well as inelastic reactions contribute. For the charged pion, the term due
to elastic scattering, which involves the square of the e.m.~form factor,
makes a substantial contribution. In the case of the $\pi^0$, this term is
absent, because the form factor vanishes on account of charge conjugation
invariance. Indeed, the contribution from the form factor to the
self-energy of the $\pi^+$ roughly reproduces the observed mass difference
between the two particles. Furthermore, the numbers given in~Refs.~\cite{Socolow:1965zz,Gross:1979ur,Gasser:1982ap} indicate that the
inelastic contributions are significantly smaller than the elastic
contributions to the self-energy of the $\pi^+$. The low-energy theorem of
Das, Guralnik, Mathur, Low and Young~\cite{Das:1967it} ensures that, in the
limit $m_u,m_d\rightarrow 0$, the e.m.~self-energy of the $\pi^0$ vanishes,
while the one of the $\pi^+$ is given by an integral over the difference
between the vector and axial-vector spectral functions. The estimates for
$\epsilon_{\pi^0}$ obtained in~Ref.~\cite{Duncan:1996xy} and more recently in~Ref.~\cite{Horsley:2015vla}
are consistent with the suppression of the self-energy of the $\pi^0$ implied by chiral
$SU(2)\times SU(2)$.
In our opinion, as already done in the FLAG 13 review~\cite{Aoki:2013ldr}, the value $\epsilon_{\pi^0}=0.07(7)$ still represents a
quite conservative estimate for this coefficient. The self-energy of the $K^0$ is
suppressed less strongly, because it remains different from zero if $m_u$
and $m_d$ are taken massless and only disappears if $m_s$ is turned off as
well. Note also that, since the e.m.~form factor of the $K^0$ is different
from zero, the self-energy of the $K^0$ does pick up an elastic
contribution. The recent lattice result $\epsilon_{K^0} = 0.2(1)$ obtained 
in~Ref.~\cite{Horsley:2015vla} indicates that the violation of Dashen's theorem 
is smaller than in the case of $\epsilonD$.
Following the FLAG 13 review~\cite{Aoki:2013ldr} we confirm the choice of 
the conservative value $\epsilon_{K^0} = 0.3(3)$.

Finally, we consider the mass splitting between the charged and neutral
pions in QCD. This effect is known to be very small, because it is of
second order in $m_u-m_d$. There is a parameter-free prediction, which
expresses the difference $\hat{M}_{\pi^+}^2-\hat{M}_{\pi^0}^2$ in terms of
the physical masses of the pseudoscalar octet and is valid to NLO of the
chiral perturbation series. Numerically, the relation yields
$\epsilon_m=0.04$~\cite{Gasser:1984gg}, indicating that this contribution
does not play a significant role at the present level of accuracy. We
attach a conservative error also to this coefficient: $\epsilon_m=0.04(2)$.
The lattice result for the self-energy difference of the pions, reported in~Ref.~\cite{Blum:2010ym}, $M_{\pi^+}^\gamma-M_{\pi^0}^\gamma=
4.50(23)\,\mbox{MeV}$, agrees with this estimate: expressed in terms of the
coefficient $\epsilon_m$ that measures the pion-mass splitting in QCD, the
result corresponds to $\epsilon_m=0.04(5)$. The corrections of
next-to-next-to-leading order (NNLO) have been worked out~in Ref.~\cite{Amoros:2001cp}, but the numerical evaluation of the 
formulae again meets with the problem that the relevant effective coupling
constants are not reliably known. 

In summary, we use the following estimates for the
e.m.~corrections:
\be\label{eq:epsilon num}\epsilonD={ 0.7(3)}
\co\hspace{0.5cm}\epsilon_{\pi^0}=0.07(7)\co\hspace{0.5cm}
\epsilon_{K^0}=0.3(3)\co\hspace{0.5cm}\epsilon_m=0.04(2)\fs\ee
While the range used for the coefficient $\epsilonD$ affects our
analysis in a significant way, the numerical values of the other
coefficients only serve to set the scale of these contributions. The
range given for $\epsilon_{\pi^0}$ and $\epsilon_{K^0}$ may be overly
generous, but because of the exploratory nature of the 
lattice determinations, we consider it advisable to use a conservative
estimate.

Treating the uncertainties in the four coefficients as statistically
independent and adding errors in quadrature, the numbers in Eq.~(\ref{eq:epsilon num}) yield the following estimates for the
e.m.~self-energies,
{ 
\bea\label{eq:Mem}&&\hspace{-1cm} M_{\pi^+}^\gamma= 4.7(3)\,
\mbox{MeV}\co\hspace{0.45cm} M_{\pi^0}^\gamma = 0.3(3)\,\mbox{MeV}
\co\hspace{0.5cm} M_{\pi^+}^\gamma-M_{\pi^0}^\gamma=4.4(1)\, \mbox{MeV}\co\\
&&\hspace{-1cm} M_{K^+}^\gamma= 2.5(5)\,\mbox{MeV}\co\hspace{0.35cm}
M_{K^0}^\gamma
=0.4(4)\,\mbox{MeV}\co\hspace{0.3cm}M_{K^+}^\gamma-M_{K^0}^\gamma= 2.1(4)\,
\mbox{MeV}\,,\nonumber \eea
and for the pion and kaon masses occurring in the QCD sector of the
Standard Model, 
\bea\label{eq:MQCD}&&\hspace{-1cm} \hat{M}_{\pi^+}= 134.8(3)\,
\mbox{MeV}\co\hspace{0.2cm} \hat{M}_{\pi^0} = 134.6(3)\,\mbox{MeV}
\co\hspace{0.5cm} \hat{M}_{\pi^+}-\hat{M}_{\pi^0}=\hspace{0.25cm}0.2(1)\,
\mbox{MeV}\co\\ 
&&\hspace{-1.1cm} \hat{M}_{K^+}= 491.2(5)\,\mbox{MeV}\co\hspace{0.12cm}
\hat{M}_{K^0}
=497.2(4)\,\mbox{MeV}\co\hspace{0.3cm}\hat{M}_{K^+}-\hat{M}_{K^0}=-6.1(4)\,
\mbox{MeV}\fs\nonumber \eea
}
The self-energy difference between the charged and neutral pion involves
the same coefficient $\epsilon_m$ that describes the mass difference in QCD
-- this is why the estimate for $ M_{\pi^+}^\gamma-M_{\pi^0}^\gamma$ is so
precise.

\subsubsection{Pion and kaon masses in the isospin limit}
\label{subsec:mu not equal md}

As mentioned above, most of the lattice calculations concerning the
properties of the light mesons are performed in the isospin limit of QCD 
($m_u-m_d\rightarrow0$ at fixed $m_u+m_d$). We
denote the pion and kaon masses in that limit by $\Mpibar$ and
$\MKbar$, respectively. Their numerical values can be estimated as
follows. Since the operation $u\leftrightarrow d$ interchanges $\pi^+$
with $\pi^-$ and $K^+$ with $K^0$, the expansion of the quantities
$\hat{M}_{\pi^+}^2$ and $\frac{1}{2}(\hat{M}_{K^+}^2+\hat{M}_{K^0}^2)$
in powers of $m_u-m_d$ only contains even powers. As shown in~Ref.~\cite{Gasser:1983yg}, the effects generated by $m_u-m_d$ in the mass
of the charged pion are strongly suppressed: the difference
$\hat{M}_{\pi^+}^2-\Mpibar^{\,2}$ represents a quantity of
$\cO[(m_u-m_d)^2(m_u+m_d)]$ and is therefore small compared to the
difference $\hat{M}_{\pi^+}^2-\hat{M}_{\pi^0}^2$, for which an
estimate was given above. In the case of 
$\frac{1}{2}(\hat{M}_{K^+}^2+\hat{M}_{K^0}^2)-\MKbar^{\,2}$, the
expansion does contain a contribution at NLO, determined by the
combination $2L_8-L_5$ of low-energy constants, but the lattice results 
for that combination show that this contribution is very
small, too. Numerically, the effects generated by $m_u-m_d$ in
$\hat{M}_{\pi^+}^2$ and in
$\frac{1}{2}(\hat{M}_{K^+}^2+\hat{M}_{K^0}^2)$ are negligible compared
to the uncertainties in the electromagnetic self-energies. The
estimates for these given in Eq.~(\ref{eq:MQCD}) thus imply
{
\be
\Mpibar = \hat{M}_{\pi^+}=134.8(3)\,\mev\ ,\hspace{1cm} \MKbar=
\sqrt{\frac{1}{2}(\hat{M}_{K^+}^2+\hat{M}_{K^0}^2)}= 494.2(3)\,\mev\ . 
\label{eq:MpiMKiso}
\ee
}
This shows that, for the convention used above to specify the QCD sector of
the Standard Model, and within the accuracy to which this convention can
currently be implemented, the mass of the pion in the isospin limit agrees
with the physical mass of the neutral pion: $\Mpibar-M_{\pi^0}=-0.2(3)$
MeV.

\subsubsection{Lattice determination of $m_s$ and $m_{ud}$}
\label{sec:msmud}

We now turn to a review of the lattice calculations of the light-quark
masses and begin with $m_s$, the isospin-averaged up- and down-quark
mass, $m_{ud}$, and their ratio. Most groups quote only $m_{ud}$, not
the individual up- and down-quark masses. We then discuss the ratio
$m_u/m_d$ and the individual determination of $m_u$ and $m_d$.

Quark masses have been calculated on the lattice since the
mid-nineties. However early calculations were performed in the quenched
approximation, leading to unquantifiable systematics. Thus in the following,
we only review modern, unquenched calculations, which include the effects of
light sea quarks.

Tabs.~\ref{tab:masses2}, \ref{tab:masses3} and \ref{tab:masses4} list
the results of $\Nf=2$, $\Nf=2+1$ and $\Nf=2+1+1$ lattice calculations
of $m_s$ and $m_{ud}$. These results are given in the $\msbar$ scheme
at $2\,\gev$, which is standard nowadays, though some groups are
starting to quote results at higher scales
(e.g.~Ref.~\cite{Arthur:2012opa}). The tables also show the colour coding
of the calculations leading to these results. As indicated earlier in
this review, we treat calculations with different numbers, $N_f$, of
dynamical quarks separately.

\bigskip
\noindent
{\em $\Nf=2$ lattice calculations}
\medskip

For $\Nf=2$, no new calculations have been performed since the
previous edition of the FLAG review~\cite{Aoki:2013ldr}. A quick inspection of
Tab.~\ref{tab:masses2} indicates that only the more recent
calculations, ALPHA~12~\cite{Fritzsch:2012wq} and ETM~10B~\cite{Blossier:2010cr}, control all systematic effects -- the
special case of D\"urr 11~\cite{Durr:2011ed} is discussed below. Only
ALPHA 12~\cite{Fritzsch:2012wq}, ETM 10B~\cite{Blossier:2010cr} and
ETM 07~\cite{Blossier:2007vv} really enter the chiral regime, with
pion masses down to about 270~MeV for ALPHA and ETM. Because this pion
mass is still quite far from the physical pion mass, ALPHA 12 refrain
from determining $m_{ud}$ and give only $m_s$. All the other
calculations have significantly more massive pions, the lightest being
about 430~MeV, in the calculation by CP-PACS 01~\cite{AliKhan:2001tx}.
Moreover, the latter calculation is performed on very coarse lattices,
with lattice spacings $a\ge 0.11\,\fm$ and only 1-loop perturbation
theory is used to renormalize the results.

ETM 10B's~\cite{Blossier:2010cr} calculation of $m_{ud}$ and $m_s$ is
an update of the earlier twisted mass determination of ETM
07~\cite{Blossier:2007vv}. In particular, they have added ensembles
with a larger volume and three new lattice spacings, $a = 0.054,
0.067$ and $0.098\,\fm$, allowing for a continuum extrapolation. In
addition, it features analyses performed in $SU(2)$ and $SU(3)$
$\chi$PT.

The ALPHA 12~\cite{Fritzsch:2012wq} calculation of $m_s$ is an
update of ALPHA 05~\cite{DellaMorte:2005kg}, which pushes computations
to finer lattices and much lighter pion masses. It also importantly
includes a determination of the lattice spacing with the
decay constant $F_K$, whereas ALPHA 05 converted results to physical
units using the scale parameter $r_0$~\cite{Sommer:1993ce}, defined
via the force between static quarks. In particular, the conversion
relied on measurements of $r_0/a$ by QCDSF/UKQCD 04~\cite{Gockeler:2004rp} which differ significantly from the new
determination by ALPHA 12. As in ALPHA 05, in ALPHA 12 both
nonperturbative running and nonperturbative renormalization are
performed in a controlled fashion, using Schr\"odinger functional
methods.

\begin{table}[!t]
{\footnotesize{
\begin{tabular*}{\textwidth}[l]{l@{\extracolsep{\fill}}rllllllll}
Collaboration & Ref. & \hspace{0.15cm}\begin{rotate}{60}{publication status}\end{rotate}\hspace{-0.15cm} &
 \hspace{0.15cm}\begin{rotate}{60}{chiral extrapolation}\end{rotate}\hspace{-0.15cm} &
 \hspace{0.15cm}\begin{rotate}{60}{continuum  extrapolation}\end{rotate}\hspace{-0.15cm}  &
 \hspace{0.15cm}\begin{rotate}{60}{finite volume}\end{rotate}\hspace{-0.15cm}  &  
 \hspace{0.15cm}\begin{rotate}{60}{renormalization}\end{rotate}\hspace{-0.15cm} &  
 \hspace{0.15cm}\begin{rotate}{60}{running}\end{rotate}\hspace{-0.15cm}  & 
\rule{0.6cm}{0cm}$m_{ud} $ & \rule{0.6cm}{0cm}$m_s $ \\
&&&&&&&&& \\[-0.1cm]
\hline
\hline
&&&&&&&&& \\[-0.1cm]
{ALPHA 12}& \cite{Fritzsch:2012wq} & \gA & \soso & \good & \good & \good & $\,a,b$
 &  & 102(3)(1) \\ 

{D\"urr 11$^\ddagger$}& \cite{Durr:2011ed} & \gA & \soso & \good & \soso & $-$ & $-$
 & 3.52(10)(9) & 97.0(2.6)(2.5) \\ 

{ETM 10B}& \cite{Blossier:2010cr} & \gA & \soso & \good & \soso & \good & $\,c$
 & 3.6(1)(2) & 95(2)(6) \\ 

{JLQCD/TWQCD 08A}& \cite{Noaki:2008iy} & \gA& \soso&\bad&\bad&\good& $-$& 4.452(81)(38)$\binom{+0}{-227}$ &\rule{0.6cm}{0cm}--\\                  

{RBC 07$^\dagger$} & \cite{Blum:2007cy} & \gA & \bad & \bad & \good  & \good &
$-$       & $4.25(23)(26)$        & 119.5(5.6)(7.4)              \\

{ETM 07} & \cite{Blossier:2007vv} & \gA &  \soso & \bad & \soso & \good &$-$
& $3.85(12)(40)$        & $105(3)(9)$                  \\

\hspace{-0.2cm}{\begin{tabular}{l}QCDSF/\\
UKQCD 06\end{tabular}} & \cite{Gockeler:2006jt} & \gA &  \bad  & \good & \bad &
\good &$-$      & $4.08(23)(19)(23)$ &  $111(6)(4)(6)$ \\

{SPQcdR 05} & \cite{Becirevic:2005ta} & \gA & \bad & \soso & \soso & \good &
$-$& $4.3(4)(^{+1.1}_{-0.0})$ & $101(8)(^{+25}_{-0})$        \\

{ALPHA 05} & \cite{DellaMorte:2005kg} & \gA &  \bad & \soso & \good  & \good &
$\,a$  &                      & 97(4)(18)$^\S$           \\

\hspace{-0.2cm}{\begin{tabular}{l}QCDSF/\\
UKQCD 04\end{tabular}} & \cite{Gockeler:2004rp} & \gA &  \bad  & \good & \bad &
\good & $-$       & $4.7(2)(3)$ & $119(5)(8)$    \\

{JLQCD 02} & \cite{Aoki:2002uc} & \gA &  \bad  & \bad & \soso & \bad & $-$   
& $3.223(^{+46}_{-69})$ & $84.5(^{+12.0}_{-1.7})$        \\

{CP-PACS 01} & \cite{AliKhan:2001tx} & \gA & \bad & \bad & \good & \bad &$-$ &
$3.45(10)(^{+11}_{-18})$ & $89(2)(^{+2}_{-6})^\star$    \\
&&&&&&&&& \\[-0.1cm] 
\hline
\hline\\
\end{tabular*}\\[-2mm]
}}
\begin{minipage}{\linewidth}
{\footnotesize 
\begin{itemize}
\item[$^\ddagger$] What is calculated is
  $m_c/m_s=11.27(30)(26)$. $m_s$ is then obtained using lattice and
  phenomenological determinations of $m_c$ which rely on perturbation
  theory. Finally, $m_{ud}$ is determined from $m_s$ using BMW 10A,
  10B's $N_f=2+1$ result for $m_s/m_{ud}$~\cite{Durr:2010vn,Durr:2010aw}. Since $m_c/m_s$ is renormalization
  group invariant in QCD, the renormalization and running of the quark
  masses enter indirectly through that of $m_c$, a mass that we do not
  review here.\\[-5mm]
\item[$^\dagger$] The calculation includes quenched e.m. effects.\\[-5mm]
\item[$^\S$] The data used to obtain the bare value of $m_s$ are from UKQCD/QCDSF 04~\cite{Gockeler:2004rp}.\\[-5mm]
\item[$^\star$] This value of $m_s$ was obtained using the kaon mass
  as input. If the $\phi$-meson mass is used instead,
  the authors find $m_s =90(^{+5}_{-11}).$\\[-5mm]
\item[$a$] The masses are renormalized and run nonperturbatively up to
  a scale of $100\,\gev$ in the $N_f=2$ SF scheme. In this scheme,
  nonperturbative and NLO running for the quark masses are shown to
  agree well from 100 GeV all the way down to 2~GeV~\cite{DellaMorte:2005kg}.\\[-5mm]
\item[$b$] The running and renormalization results of~Ref.~\cite{DellaMorte:2005kg} are improved in~Ref.~\cite{Fritzsch:2012wq} with
  higher statistical and systematic accuracy.\\[-5mm]
\item[$c$] The masses are renormalized nonperturbatively at scales
  $1/a\sim 2\div3\,\gev$ in the $N_f=2$ RI/MOM scheme.  In this
  scheme, nonperturbative and N$^3$LO running for the quark masses are
  shown to agree from 4~GeV down to 2~GeV to better than 3\%~\cite{Constantinou:2010gr}.
\end{itemize}
}
\end{minipage}

\caption{\label{tab:masses2} $\Nf=2$ lattice results for the masses $m_{ud}$ and $m_s$ (MeV, running masses in the $\msbar$ scheme  at scale 2 GeV). The significance of the colours is explained in Sec.~\ref{sec:qualcrit}. If information about nonperturbative running is available, this is indicated in
  the column ``running'', with details given at the bottom of the table.}
\end{table}

The conclusion of our analysis of $\Nf=2$ calculations is that the
results of ALPHA 12~\cite{Fritzsch:2012wq} and ETM 10B~\cite{Blossier:2010cr} (which update and extend ALPHA 05~\cite{DellaMorte:2005kg} and ETM 07~\cite{Blossier:2007vv},
respectively), are the only ones to date which satisfy our selection
criteria. Thus we average those two results for $m_s$, obtaining
101(3)~MeV. Regarding $m_{ud}$, for which only ETM
10B~\cite{Blossier:2010cr} gives a value, we do not offer an average
but simply quote ETM's number. Thus, we quote as our estimates:
%
%FLAGRESULT BEGIN
% TAG      &ms & mud   &END
% REFS     &\cite{Blossier:2010cr,Fritzsch:2012wq}& \cite{Blossier:2010cr}  &END
% UNITS    & '(MeV)' & '(MeV)'  &END
% NUMRESULTS & 2 & 1 &END
% FLAVOURs & 2 & 2 &END
%FLAGRESULT END
%FLAGRESULTFORMULA BEGIN
\begin{align}
\label{eq:quark masses Nf=2} 
&&\FLAGAVBEGIN m_s   &= 101(3)\FLAGAVEND ~\mbox{MeV} &&\Refs~\mbox{\cite{Blossier:2010cr,Fritzsch:2012wq}},\nonumber \\[-3mm]
&\Nf=2 :&\\[-3mm]
&&\FLAGAVBEGIN m_{ud}&= 3.6(2) \FLAGAVEND ~\mbox{MeV}&&\Ref~\mbox{\cite{Blossier:2010cr}}.\nonumber
\end{align}
%FLAGRESULTFORMULA END
%
The errors on these results are 3\% and 6\%, respectively. However,
these errors do not account for the fact that sea strange-quark mass
effects are absent from the calculation, a truncation of the theory
whose systematic effects cannot be estimated {\em a priori}. Thus,
these results carry an additional unknown systematic arror. 
It is worth remarking that the difference between ALPHA 12's~\cite{Fritzsch:2012wq} central value for
$m_s$ and that of ETM 10B~\cite{Blossier:2010cr} is 7(7) MeV.

We have not included the results of D\"urr 11~\cite{Durr:2011ed} in
the averages of Eq.~(\ref{eq:quark masses Nf=2}), despite the fact that
they satisfy our selection criteria. The reason for this is that the
observable which they actually compute on the lattice is
$m_c/m_s=11.27(30)(26)$, reviewed in Sec.~\ref{sec:mcoverms}.  They
obtain $m_s$ by combining that value of $m_c/m_s$ with already
existing phenomenological calculations of $m_c$. Subsequently they
obtain $m_{ud}$ by combining this result for $m_s$ with the $N_f=2+1$
calculation of $m_s/m_{ud}$ of BMW 10A, 10B~\cite{Durr:2010vn,Durr:2010aw} discussed below. Thus, their results
for $m_s$ and $m_{ud}$ are not {\em per se} lattice results, nor do
they correspond to $N_f=2$. The value of the charm-quark mass which
they use is an average of phenomenological determinations, which they
estimate to be $m_c(2\,\gev)=1.093(13)\,\gev$, with a 1.2\% total
uncertainty. This value for $m_c$ leads to the results for $m_s$ and
$m_{ud}$ in Tab.~\ref{tab:masses2} which are compatible with the
averages given in Eq.~(\ref{eq:quark masses Nf=2}) and have similar
uncertainties.  Note, however, that their determination of $m_c/m_s$
is about 1.5 combined standard deviations below the only other $\Nf=2$
result which satisfies our selection criteria, ETM 10B's~\cite{Blossier:2010cr} result, as discussed in
Sec.~\ref{sec:mcoverms}.
\clearpage

\bigskip
\noindent
{\em $\Nf=2+1$ lattice calculations}
\medskip

We turn now to $\Nf=2+1$ calculations. These and the corresponding
results for $m_{ud}$ and $m_s$ are summarized in
Tab.~\ref{tab:masses3}. Given the very high precision of a number of
the results, with total errors on the order of 1\%, it is important to
consider the effects neglected in these calculations.  Since isospin
breaking and e.m.\ effects are small on $m_{ud}$ and $m_s$, and have
been approximately accounted for in the calculations that will be
retained for our averages, the largest potential source of
uncontrolled systematic error is that due to the omission of the charm
quark in the sea. Beyond the small perturbative corrections that come
from matching the $\Nf=3$ to the $\Nf=4$ $\msbar$ scheme at $m_c$
($\sim -0.2\%$), the charm sea-quarks affect the determination of the
light-quark masses through contributions of order $1/m_c^2$. As these
are further suppressed by the Okubo-Zweig-Iizuka rule, they are also
expected to be small, but are difficult to quantify {\em a priori}.
Fortunately, as we will see below, $m_s$ has been directly computed
with $N_f=2+1+1$ simulations. In particular,
HPQCD~14~\cite{Chakraborty:2014aca} has computed $m_s$ in QCD$_4$ with
very much the same approach as it had used to obtain the QCD$_3$
result of HPQCD~10~\cite{McNeile:2010ji}. Their results for
$m_s(N_f=3, 2~\gev)$ are $93.8(8)\,\mev$~\cite{Chakraborty:2014aca}
and $92.2(1.3)\,\mev$~\cite{McNeile:2010ji}, where the $N_f=4$ result
has been converted perturbatively to $N_f=3$
in~Ref.~\cite{Chakraborty:2014aca}. This leads to a relative
difference of $1.7(1.6)\%$. While the two results are compatible
within one combined standard deviation, a $\sim 2\%$ effect cannot be
excluded. Thus, we will retain this 2\% uncertainty and add it to the
averages for $m_s$ and $m_{ud}$ given below.

\begin{table}[!ht]
\vspace{2mm}
{\footnotesize{
\begin{tabular*}{\textwidth}[l]{l@{\extracolsep{\fill}}rllllllll}
Collaboration & Ref. & \hspace{0.15cm}\begin{rotate}{60}{publication status}\end{rotate}\hspace{-0.15cm} &
 \hspace{0.15cm}\begin{rotate}{60}{chiral extrapolation}\end{rotate}\hspace{-0.15cm} &
 \hspace{0.15cm}\begin{rotate}{60}{continuum  extrapolation}\end{rotate}\hspace{-0.15cm}  &
 \hspace{0.15cm}\begin{rotate}{60}{finite volume}\end{rotate}\hspace{-0.15cm}  &  
 \hspace{0.15cm}\begin{rotate}{60}{renormalization}\end{rotate}\hspace{-0.15cm} &  
 \hspace{0.15cm}\begin{rotate}{60}{running}\end{rotate}\hspace{-0.15cm}  & 
\rule{0.6cm}{0cm}$m_{ud} $ & \rule{0.6cm}{0cm}$m_s $ \\
&&&&&&&&& \\[-0.1cm]
\hline
\hline
&&&&&&&&& \\[-0.1cm]

{RBC/UKQCD 14B$^\ominus$}& \cite{Blum:2014tka} & \oP & \good & \good & \good &
\good & $d$  & 3.31(4)(4)  & 90.3(0.9)(1.0)\\

{RBC/UKQCD 12$^\ominus$}& \cite{Arthur:2012opa} & \gA & \good & \soso & \good &
\good & $d$  &  3.37(9)(7)(1)(2) & 92.3(1.9)(0.9)(0.4)(0.8)\\

{PACS-CS 12$^\star$}& \protect{\cite{Aoki:2012st}} & \gA & \good & \bad & \bad & \good & $\,b$
&  3.12(24)(8) &  83.60(0.58)(2.23) \\

{Laiho 11} & \cite{Laiho:2011np} & \rC & \soso & \good & \good & \soso
& $-$ & 3.31(7)(20)(17)
& 94.2(1.4)(3.2)(4.7)\\

{BMW 10A, 10B$^+$} & \cite{Durr:2010vn,Durr:2010aw} & \gA & \good & \good & \good & \good &
$\,c$ & 3.469(47)(48)& 95.5(1.1)(1.5)\\

{PACS-CS 10}& \cite{Aoki:2010wm} & \gA & \good & \bad & \bad & \good & $\,b$
&  2.78(27) &  86.7(2.3) \\

{MILC 10A}& \cite{Bazavov:2010yq} & \rC & \soso  & \good & \good &
\soso  &$-$& 3.19(4)(5)(16)&\rule{0.6cm}{0cm}-- \\

{HPQCD~10$^\ast$}&  \cite{McNeile:2010ji} &\gA & \soso & \good & \good & $-$
&$-$& 3.39(6)$ $ & 92.2(1.3) \\

{RBC/UKQCD 10A}& \cite{Aoki:2010dy} & \gA & \soso & \soso & \good &
\good & $\,a$  &  3.59(13)(14)(8) & 96.2(1.6)(0.2)(2.1)\\

{Blum~10$^\dagger$}&\cite{Blum:2010ym}& \gA & \soso & \bad & \soso & \good &
$-$ &3.44(12)(22)&97.6(2.9)(5.5)\\

{PACS-CS 09}& \cite{Aoki:2009ix}& \gA &\good   &\bad   & \bad & \good  &  $\,b$
 & 2.97(28)(3) &92.75(58)(95)\\

{HPQCD 09A$^\oplus$}&  \cite{Davies:2009ih}&\gA & \soso & \good & \good & $-$
& $-$& 3.40(7) & 92.4(1.5) \\

{MILC 09A} & \cite{Bazavov:2009fk} & \rC &  \soso & \good & \good & \soso &
$-$ & 3.25 (1)(7)(16)(0) & 89.0(0.2)(1.6)(4.5)(0.1)\\

{MILC 09} & \cite{Bazavov:2009bb} & \gA & \soso & \good & \good & \soso & $-$
& 3.2(0)(1)(2)(0) & 88(0)(3)(4)(0)\\

{PACS-CS 08} & \cite{Aoki:2008sm} &  \gA & \good & \bad & \bad  & \bad & $-$ &
2.527(47) & 72.72(78)\\

{RBC/UKQCD 08} & \cite{Allton:2008pn} & \gA & \soso & \bad & \good & \good &
$-$ &$3.72(16)(33)(18)$ & $107.3(4.4)(9.7)(4.9)$\\

\hspace{-0.2cm}{\begin{tabular}{l}CP-PACS/\\JLQCD 07\end{tabular}} 
& \cite{Ishikawa:2007nn}& \gA & \bad & \good & \good  & \bad & $-$ &
$3.55(19)(^{+56}_{-20})$ & $90.1(4.3)(^{+16.7}_{-4.3})$ \\

{HPQCD 05}
& 
\cite{Mason:2005bj}& \gA & \soso & \soso & \soso & \soso &$-$&
$3.2(0)(2)(2)(0)^\ddagger$ & $87(0)(4)(4)(0)^\ddagger$\\

\hspace{-0.2cm}{\begin{tabular}{l}MILC 04, HPQCD/\\MILC/UKQCD 04\end{tabular}} 
& \cite{Aubin:2004fs,Aubin:2004ck} & \gA & \soso & \soso & \soso & \bad & $-$ &
$2.8(0)(1)(3)(0)$ & $76(0)(3)(7)(0)$\\
&&&&&&&&& \\[-0.1cm]
\hline
\hline\\
\end{tabular*}\\[-0.2cm]
}}
\begin{minipage}{\linewidth}
{\footnotesize 
\begin{itemize}
\item[$^\ominus$] The results are given in the $\msbar$ scheme at 3
  instead of 2~GeV. We run them down to 2~GeV using numerically
  integrated 4-loop running~\cite{vanRitbergen:1997va,Chetyrkin:1999pq} with $N_f=3$ and with
  the values of $\alpha_s(M_Z)$, $m_b$ and $m_c$ taken from~Ref.~\cite{Agashe:2014kda}. The running factor is 1.106. At three loops
  it is only 0.2\% smaller, indicating that running uncertainties are
  small. We neglect them here.\\[-5mm]
\item[$^\star$] The calculation includes e.m. and $m_u\ne m_d$ effects
  through reweighting.\\ [-5mm]
\item[$^+$] The fermion action used is tree-level improved.\\[-5mm]
\item[$^\ast$] What is calculated 
       is then obtained by combining
      this result with HPQCD 09A's $m_c/m_s=11.85(16)$~\cite{Davies:2009ih}.
      Finally, $m_{ud}$
      is determined from $m_s$ with the MILC 09 result for
      $m_s/m_{ud}$. Since $m_c/m_s$ is renormalization group invariant
      in QCD, the renormalization and running of the quark masses
      enter indirectly through that of $m_c$ (see below).\\[-5mm]
\item[$^\dagger$] The calculation includes quenched e.m. effects.\\[-5mm]
\item[$^\oplus$] What is calculated is $m_c/m_s=11.85(16)$. $m_s$ is then obtained by combing
      this result with the determination $m_c(m_c) = 1.268(9)$~GeV
      from~Ref.~\cite{Allison:2008xk}. Finally, $m_{ud}$
      is determined from $m_s$ with the MILC 09 result for
      $m_s/m_{ud}$.\\[-5mm]
\item[$^\ddagger$] The bare numbers are those of MILC 04. The masses are simply rescaled, using the
ratio of the 2-loop to 1-loop renormalization factors.\\[-5mm]
\item[$a$] The masses are renormalized nonperturbatively at a scale of
  2~GeV in a couple of $N_f=3$ RI/SMOM schemes. A careful study of
  perturbative matching uncertainties has been performed by comparing
  results in the two schemes in the region of 2~GeV to 3~GeV~\cite{Aoki:2010dy}.\\[-5mm]
\item[$b$] The masses are renormalized and run nonperturbatively up to
  a scale of $40\,\gev$ in the $N_f=3$ SF scheme. In this scheme,
  nonperturbative and NLO running for the quark masses are shown to
  agree well from 40 GeV all the way down to 3 GeV~\cite{Aoki:2010wm}.\\[-5mm]
\item[$c$] The masses are renormalized and run nonperturbatively up to
  a scale of 4 GeV in the $N_f=3$ RI/MOM scheme.  In this scheme,
  nonperturbative and N$^3$LO running for the quark masses are shown
  to agree from 6~GeV down to 3~GeV to better than 1\%~\cite{Durr:2010aw}.  \\[-5mm]
\item[$d$] All required running is performed nonperturbatively.
\end{itemize}
}
\end{minipage}
\caption{$\Nf=2+1$ lattice results for the masses $m_{ud}$ and $m_s$ (see Tab.~\ref{tab:masses2} for notation).}
\label{tab:masses3}
\end{table}

The only new calculation since the last FLAG report~\cite{Aoki:2013ldr} is that of
RBC/UKQCD~14~\cite{Blum:2014tka}.  It significantly improves on
their RBC/UKQCD~12~\cite{Arthur:2012opa} work by adding three new
domain wall fermion simulations to three used previously. Two of the
new simulations are performed at essentially physical pion masses
($M_\pi\simeq 139\,\mev$) on lattices of about $5.4\,\fm$ in size and
with lattice spacings of $0.114\,\fm$ and $0.084\,\fm$. It is
complemented by a third simulation with $M_\pi\simeq 371\,\mev$,
$a\simeq 0.063$ and a rather small $L\simeq 2.0\,\fm$. Altogether,
this gives them six simulations with six unitary $M_\pi$'s in the
range of $139$ to $371\,\mev$ and effectively three lattice spacings
from $0.063$ to $0.114\,\fm$. They perform a combined global continuum
and chiral fit to all of their results for the $\pi$ and $K$ masses
and decay constants, the $\Omega$ baryon mass and two Wilson-flow
parameters.  Quark masses in these fits are renormalized and run
nonperturbatively in the RI/SMOM scheme. This is done by computing the
relevant renormalization constant for a reference ensemble and
determining those for other simulations relative to it by adding
appropriate parameters in the global fit. This new calculation passes
all of our selection criteria. Its results will replace the older
RBC/UKQCD~12 results in our averages.

$\Nf=2+1$ MILC results for light-quark masses go back to
2004~\cite{Aubin:2004fs,Aubin:2004ck}. They use rooted staggered
fermions.  By 2009 their simulations covered an impressive range of
parameter space, with lattice spacings which go down to 0.045~fm and
valence-pion masses down to approximately
180~MeV~\cite{Bazavov:2009fk}.  The most recent MILC $\Nf=2+1$
results, i.e. MILC 10A~\cite{Bazavov:2010yq} and MILC
09A~\cite{Bazavov:2009fk}, feature large statistics and 2-loop
renormalization.  Since these data sets subsume those of their
previous calculations, these latest results are the only ones that
must be kept in any world average.

The PACS-CS 12~\cite{Aoki:2012st} calculation represents an
important extension of the collaboration's earlier 2010 computation~\cite{Aoki:2010wm}, which already probed pion masses down to
$M_\pi\simeq 135\,\mev$, i.e.\ down to the physical-mass point. This
was achieved by reweighting the simulations performed in PACS-CS
08~\cite{Aoki:2008sm} at $M_\pi\simeq 160\,\mev$. If adequately
controlled, this procedure eliminates the need to extrapolate to the
physical-mass point and, hence, the corresponding systematic
error. The new calculation now applies similar reweighting techniques
to include electromagnetic and $m_u\ne m_d$ isospin-breaking effects
directly at the physical pion mass. Further, as in PACS-CS 10~\cite{Aoki:2010wm}, renormalization of quark masses is implemented
nonperturbatively, through the Schr\"odinger functional
method~\cite{Luscher:1992an}. As it stands, the main drawback of the
calculation, which makes the inclusion of its results in a world
average of lattice results inappropriate at this stage, is that for
the lightest quark mass the volume is very small, corresponding to
$LM_\pi\simeq 2.0$, a value for which finite-volume effects will be
difficult to control. Another problem is that the calculation was
performed at a single lattice spacing, forbidding a continuum
extrapolation. Further, it is unclear at this point what might be
the systematic errors associated with the reweighting procedure.

The BMW 10A, 10B~\cite{Durr:2010vn,Durr:2010aw}
calculation still satisfies our stricter selection criteria. They
reach the physical up- and down-quark mass
by {\it interpolation} instead of by extrapolation. Moreover, their
calculation was performed at five lattice spacings ranging from 0.054
to 0.116~fm, with full nonperturbative renormalization and running
and in volumes of up to (6~fm)$^3$ guaranteeing that the continuum
limit, renormalization and infinite-volume extrapolation are
controlled. It does neglect, however, isospin-breaking effects, which
are small on the scale of their error bars.

Finally we come to another calculation which satisfies our selection
criteria, HPQCD~10~\cite{McNeile:2010ji}. It updates the staggered
fermions calculation of HPQCD~09A~\cite{Davies:2009ih}. In these
papers the renormalized mass of the strange quark is obtained by
combining the result of a precise calculation of the renormalized
charm-quark mass, $m_c$, with the result of a calculation of the
quark-mass ratio, $m_c/m_s$. As described in Ref.~\cite{Allison:2008xk} and
in Sec.~\ref{s:cmass}, HPQCD determines $m_c$ by fitting
Euclidean-time moments of the $\bar cc$ pseudoscalar density two-point
functions, obtained numerically in lattice QCD, to fourth-order,
continuum perturbative expressions. These moments are normalized and
chosen so as to require no renormalization with staggered
fermions. Since $m_c/m_s$ requires no renormalization either, HPQCD's
approach displaces the problem of lattice renormalization in the
computation of $m_s$ to one of computing continuum perturbative
expressions for the moments. To calculate $m_{ud}$
HPQCD~10~\cite{McNeile:2010ji} use the MILC 09 determination of the
quark-mass ratio $m_s/m_{ud}$~\cite{Bazavov:2009bb}.

HPQCD~09A~\cite{Davies:2009ih} obtains
$m_c/m_s=11.85(16)$~\cite{Davies:2009ih} fully nonperturbatively,
with a precision slightly larger than 1\%. HPQCD~10's determination of the
charm-quark mass, $m_c(m_c)=1.268(6)$,~\footnote{To obtain this number, 
we have used the conversion from $\mu=3\,$ GeV to $m_c$ given in Ref.~\cite{Allison:2008xk}.} is even more precise, achieving an accuracy
better than 0.5\%. While these errors are, perhaps, surprisingly small, we take
them at face value as we do those of RBC/UKQCD~14, since we will add a 2\% error
due to the quenching of the charm on the final result.

This discussion leaves us with four results for our final average for
$m_s$: MILC~09A~\cite{Bazavov:2009fk}, BMW~10A,
10B~\cite{Durr:2010vn,Durr:2010aw}, HPQCD~10~\cite{McNeile:2010ji} and
RBC/UKQCD~14~\cite{Blum:2014tka}. Assuming that the result from HPQCD~10 is
100\% correlated with that of MILC~09A, as it is based on a subset of the MILC~09A
configurations, we find $m_s=92.0(1.1)\,\mev$ with a $\chi^2/$dof = 1.8.

For the light quark mass $m_{ud}$, the results satisfying our criteria are RBC/UKQCD 14B, BMW 10A, 10B, HPQCD 10, and MILC 10A. For the error, we include the same 100\% correlation between statistical errors for the latter two as for the strange case, resulting in $m_{ud}=3.373(43)$ at 2 GeV in the $\overline{\rm MS}$ scheme ($\chi^2/$d.of.=1.5). Adding the 2\% estimate for the missing charm contribution, our final estimates for the light-quark masses are
% 
%FLAGRESULT BEGIN
% TAG      &mud & ms   &END
% REFS     &\cite{Blum:2014tka,Durr:2010vn,Durr:2010aw,McNeile:2010ji,Bazavov:2010yq}& \cite{Bazavov:2009fk,Durr:2010vn,Durr:2010aw,McNeile:2010ji,Blum:2014tka}  &END
% UNITS    & '(MeV)' & '(MeV)'  &END
% NUMRESULTS & 5 & 5 &END
% FLAVOURs & 2+1 & 2+1 &END
%FLAGRESULT END
%FLAGRESULTFORMULA BEGIN
\begin{align}\label{eq:nf3msmud}
&& \FLAGAVBEGIN m_{ud}&= 3.373 (80)\FLAGAVEND\;\mev&&\Refs~\mbox{\cite{Blum:2014tka,Durr:2010vn,Durr:2010aw,McNeile:2010ji,Bazavov:2010yq}},\,\nonumber \\[-3mm]
&\Nf=2+1 :&\\[-3mm]
&&\FLAGAVBEGIN m_s    &=92.0(2.1)\FLAGAVEND\;\;\mev&&\Refs~\mbox{\cite{Bazavov:2009fk,Durr:2010vn,Durr:2010aw,McNeile:2010ji,Blum:2014tka}}. \nonumber
\end{align}
%FLAGRESULTFORMULA END
%

\bigskip
\noindent
{\em $\Nf=2+1+1$ lattice calculations}
\medskip

One of the novelties since the last edition of this review~\cite{Aoki:2013ldr} is the fact that $N_f=2+1+1$ results for the light-quark masses have been published. These and the features of the
corresponding calculations are summarized in
Tab.~\ref{tab:masses4}. Note that the results of Ref.~\cite{Chakraborty:2014aca} are reported as $m_s(2\,\gev;N_f=3)$ and those of Ref.~\cite{Carrasco:2014cwa} as $m_{ud(s)}(2\,\gev;N_f=4)$. We convert the former to $N_f=4$ and obtain $m_s(2\,\gev;N_f=4)=93.7(8)\mev$. The average of ETM 14 and HPQCD 14A  is 93.9(1.1)$\mev$ with $\chi^2/$d.o.f.=1.8. 
For the light0quark average we use the sole available value from ETM 14A. Our averages are
 % 
%FLAGRESULT BEGIN
% TAG      &mud & ms   &END
% REFS     &\cite{Carrasco:2014cwa}& \cite{Carrasco:2014cwa,Chakraborty:2014aca}  &END
% UNITS    & '[MeV]' & '[MeV]'  &END
% NUMRESULTS & 1 & 2 &END
% FLAVOURs & 2+1+1 & 2+1+1 &END
%FLAGRESULT END
%FLAGRESULTFORMULA BEGIN
\begin{align}\label{eq:nf4msmud}
&&\FLAGAVBEGIN m_{ud}&= 3.70 (17)\FLAGAVEND\;\mev&& \Ref~\mbox{\cite{Carrasco:2014cwa}},\nonumber\\[-3mm]% \\
&\Nf=2+1+1 :& \\[-3mm]
&&\FLAGAVBEGIN m_s   &=93.9(1.1)\FLAGAVEND\; \mev&& \Refs~\mbox{\cite{Carrasco:2014cwa,Chakraborty:2014aca}}.\nonumber%\,. 
\end{align}
%FLAGRESULTFORMULA END
%

\begin{table}[!htb]
\vspace{2.5cm}
{\footnotesize{
\begin{tabular*}{\textwidth}[l]{l@{\extracolsep{\fill}}rllllllll}
Collaboration & Ref. & \hspace{0.15cm}\begin{rotate}{60}{publication status}\end{rotate}\hspace{-0.15cm} &
 \hspace{0.15cm}\begin{rotate}{60}{chiral extrapolation}\end{rotate}\hspace{-0.15cm} &
 \hspace{0.15cm}\begin{rotate}{60}{continuum  extrapolation}\end{rotate}\hspace{-0.15cm}  &
 \hspace{0.15cm}\begin{rotate}{60}{finite volume}\end{rotate}\hspace{-0.15cm}  &  
 \hspace{0.15cm}\begin{rotate}{60}{renormalization}\end{rotate}\hspace{-0.15cm} &  
 \hspace{0.15cm}\begin{rotate}{60}{running}\end{rotate}\hspace{-0.15cm}  & 
\rule{0.6cm}{0cm}$m_{ud} $ & \rule{0.6cm}{0cm}$m_s $ \\
&&&&&&&&& \\[-0.1cm]
\hline
\hline
&&&&&&&&& \\[-0.1cm]

{HPQCD 14A $^\oplus$} & \cite{Chakraborty:2014aca} & \gA & \good & \good & \good &
$-$ & $-$  &  & 93.7(8) \\

{ETM 14$^\oplus$}& \cite{Carrasco:2014cwa} & \gA & \soso & \good & \good &
\good & $-$  & 3.70(13)(11)  & 99.6(3.6)(2.3)\\
&&&&&&&&& \\[-0.1cm] 
\hline
\hline\\[-2mm]
\end{tabular*}
}}
\begin{minipage}{\linewidth}
{\footnotesize 
\begin{itemize}
\item[$^\oplus$] As explained in the text, $m_s$ is obtained by combining the
        results $m_c(5\,\gev;N_f=4)=0.8905(56)$~GeV and
        $(m_c/m_s)(N_f=4)=11.652(65)$, determined on the same data
        set. A subsequent scale and scheme conversion, performed by
        the authors leads, to the value 93.6(8). In the table we have converted this
        to $m_s(2\,\gev;N_f=4)$, which makes a very small change. 
\end{itemize}
}
\end{minipage}

\caption{$\Nf=2+1+1$ lattice results for the masses $m_{ud}$ and $m_s$ (see Tab.~\ref{tab:masses2} for notation).}
\label{tab:masses4}
\end{table}

\begin{figure}[!htb]
\begin{center}
\psfig{file=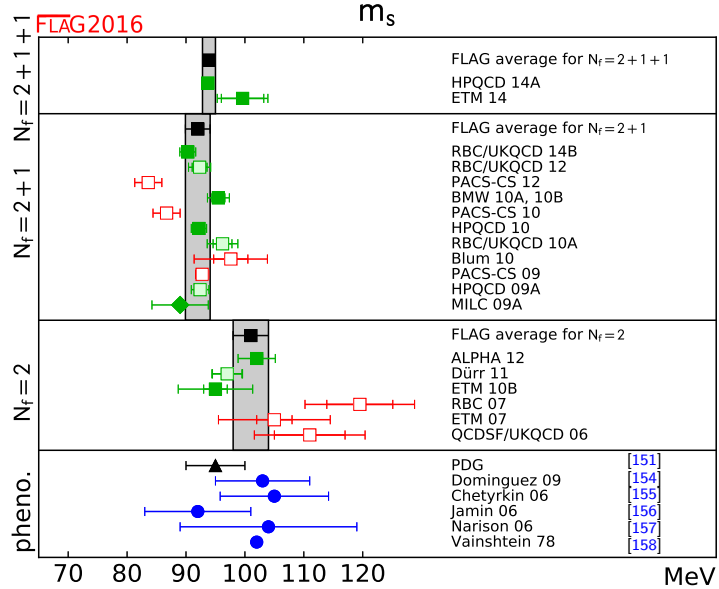,width=11.5cm}
\end{center}
\vspace{-2.47cm}\hspace{11.2cm}\parbox{6cm}{\sffamily\tiny

\vspace{-1.0em} \cite{Agashe:2014kda}\\

\vspace{-1.2em}\cite{Dominguez:2008jz}\\

\vspace{-1.2em}\cite{Chetyrkin:2005kn}\\

\vspace{-1.2em}\cite{Jamin:2006tj}\\

\vspace{-1.2em}\cite{Narison:2005ny}\\

\vspace{-1.2em}\cite{Vainshtein:1978nn}}

\vspace{0.2cm}
\begin{center}
\caption{ \label{fig:ms} $\msbar$ mass of the strange quark (at 2 GeV scale) in MeV. 
 The upper three panels show the lattice results 
  listed in Tabs.~\ref{tab:masses2}, \ref{tab:masses3} and \ref{tab:masses4}, while 
  the bottom panel collects a few sum rule results and also indicates the current PDG estimate. 
  Diamonds and squares represent results based on perturbative and nonperturbative
  renormalization, respectively. 
 The black squares and the grey bands represent our estimates (\ref{eq:quark masses Nf=2}) ,
  (\ref{eq:nf3msmud}) and (\ref{eq:nf4msmud}). The significance of the colours is explained in Sec.~\ref{sec:qualcrit}.
}\end{center}

\end{figure}

\begin{figure}[!htb]

\begin{center}
\psfig{file=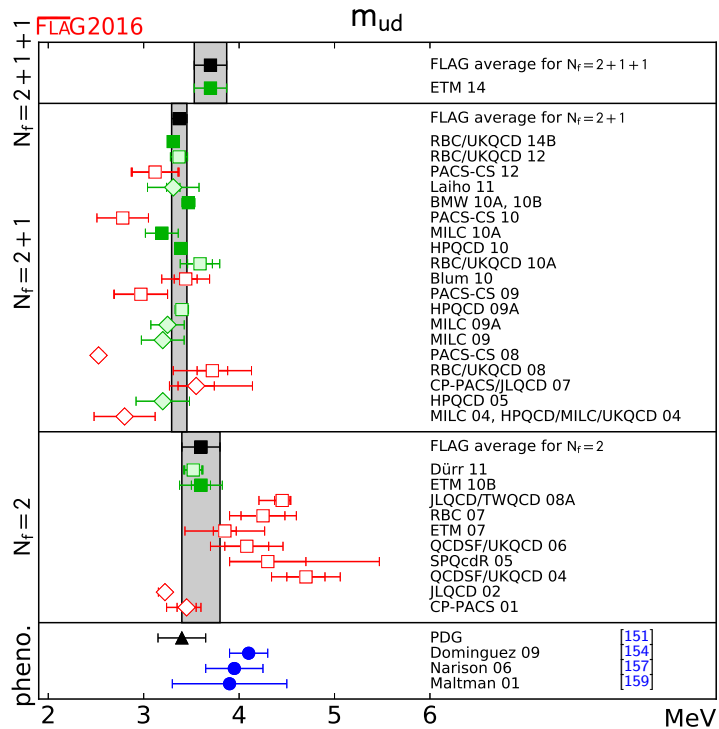,width=11.5cm}
\end{center}
\vspace{-2.48cm}\hspace{11.1cm}\parbox{6cm}{\sffamily\tiny  \cite{Agashe:2014kda}\\

\vspace{-1.3em}\cite{Dominguez:2008jz}\\

\vspace{-1.3em}\cite{Narison:2005ny}\\

\vspace{-1.3em}\cite{Maltman:2001nx}}

\vspace{0.2cm}
\begin{center}
\caption{ \label{fig:mud} Mean mass of the two lightest quarks,
 $m_{ud}=\frac{1}{2}(m_u+m_d)$ (for details see Fig.~\ref{fig:ms}).}\end{center}

\end{figure}

In Figs.~\ref{fig:ms} and \ref{fig:mud} the lattice results listed in Tabs.~\ref{tab:masses2}, \ref{tab:masses3} and \ref{tab:masses4} and the FLAG averages obtained at each value of $N_f$ are presented and compared with various phenomenological results. 

\subsubsection{Lattice determinations of $m_s/m_{ud}$}
\label{sec:msovermud}

\begin{table}[!htb]
\vspace{3cm}
{\footnotesize{
\begin{tabular*}{\textwidth}[l]{l@{\extracolsep{\fill}}rllllll}
Collaboration & Ref. & $\Nf$ & \hspace{0.15cm}\begin{rotate}{60}{publication status}\end{rotate}\hspace{-0.15cm}  &
 \hspace{0.15cm}\begin{rotate}{60}{chiral extrapolation}\end{rotate}\hspace{-0.15cm} &
 \hspace{0.15cm}\begin{rotate}{60}{continuum  extrapolation}\end{rotate}\hspace{-0.15cm}  &
 \hspace{0.15cm}\begin{rotate}{60}{finite volume}\end{rotate}\hspace{-0.15cm}  & \rule{0.1cm}{0cm} 
$m_s/m_{ud}$ \\
&&&&&& \\[-0.1cm]
\hline
\hline
&&&&&& \\[-0.1cm]

{FNAL/MILC 14A} & \cite{Bazavov:2014wgs} & 2+1+1 & \gA & \good & \good & \good & $27.35(5)^{+10}_{-7}$\\

{ETM 14}& \cite{Carrasco:2014cwa} & 2+1+1 & \gA & \soso & \good & \soso & 26.66(32)(2)\\

&&&&&& \\[-0.1cm]
\hline
&&&&&& \\[-0.1cm]

{RBC/UKQCD 14B}& \cite{Blum:2014tka} &2+1  & \oP & \good & \good & \good & 27.34(21)\\

{RBC/UKQCD 12$^\ominus$}& \cite{Arthur:2012opa} &2+1  & \gA & \good & \soso & \good & 27.36(39)(31)(22)\\

{PACS-CS 12$^\star$}& \cite{Aoki:2012st}       &2+1  & \gA & \good & \bad & \bad & 26.8(2.0)\\

{Laiho 11} & \cite{Laiho:2011np}              &2+1  & \rC & \soso & \good & \good & 28.4(0.5)(1.3)\\

{BMW 10A, 10B$^+$}& \cite{Durr:2010vn,Durr:2010aw} &2+1  & \gA & \good & \good & \good & 27.53(20)(8) \\

{RBC/UKQCD 10A}& \cite{Aoki:2010dy}           &2+1  & \gA & \soso & \soso & \good & 26.8(0.8)(1.1) \\

{Blum 10$^\dagger$}&\cite{Blum:2010ym}         &2+1  & \gA & \soso & \bad & \soso & 28.31(0.29)(1.77)\\

{PACS-CS 09}  & \cite{Aoki:2009ix}            &2+1  &  \gA &\good   &\bad   & \bad & 31.2(2.7)  \\

{MILC 09A}    & \cite{Bazavov:2009fk}       &2+1  & \rC & \soso & \good & \good & 27.41(5)(22)(0)(4)  \\
{MILC 09}      & \cite{Bazavov:2009bb}      &2+1  & \gA & \soso & \good & \good & 27.2(1)(3)(0)(0)  \\

{PACS-CS 08}   & \cite{Aoki:2008sm}           &2+1  & \gA & \good & \bad  & \bad & 28.8(4)\\

{RBC/UKQCD 08} & \cite{Allton:2008pn}         &2+1  & \gA & \soso & \bad  & \good & 28.8(0.4)(1.6) \\

\hspace{-0.2cm}{\begin{tabular}{l}MILC 04, HPQCD/\\MILC/UKQCD 04\end{tabular}} 
& \cite{Aubin:2004fs,Aubin:2004ck}            &2+1  & \gA & \soso & \soso & \soso & 27.4(1)(4)(0)(1)  \\
&&&&&& \\[-0.1cm]
\hline
&&&&&& \\[-0.1cm]

{ETM 14D}& \cite{Abdel-Rehim:2014nka} &2 & \rC & \good & \bad & \bad & 27.63(13) \\

{ETM 10B}& \cite{Blossier:2010cr}            &2  & \gA & \soso & \good  & \soso & 27.3(5)(7) \\

{RBC 07}$^\dagger$ & \cite{Blum:2007cy}       &2  & \gA & \bad  & \bad  & \good & 28.10(38) \\

{ETM 07}       & \cite{Blossier:2007vv}      &2  & \gA & \soso & \bad  & \soso & 27.3(0.3)(1.2) \\

{QCDSF/UKQCD 06} & \cite{Gockeler:2006jt}    &2  & \gA &  \bad & \good & \bad & 27.2(3.2)\\
&&&&&& \\[-0.1cm]
\hline
\hline\\
\end{tabular*}\\[-0.2cm]
}}
\begin{minipage}{\linewidth}
{\footnotesize 
\begin{itemize}
\item[$^\ominus$] The errors are statistical, chiral and finite volume.\\[-5mm]
\item[$^\star$] The calculation includes e.m. and $m_u\ne m_d$ effects through reweighting.\\[-5mm]
\item[$^+$] The fermion action used is tree-level improved.\\[-5mm]
\item[$^\dagger$] The calculation includes quenched e.m. effects.
\end{itemize}
}
\end{minipage}
\caption{Lattice results for the ratio $m_s/m_{ud}$.}
\label{tab:ratio_msmud}
\end{table}

The lattice results for $m_s/m_{ud}$ are summarized in Tab.~\ref{tab:ratio_msmud}.
In the ratio $m_s/m_{ud}$, one of the sources of systematic error -- the
uncertainties in the renormalization factors -- drops out. Also, we can
compare the lattice results with the leading-order formula of {\Ch}PT,
\be\label{eq:LO1}\frac{m_s}{m_{ud}}\Lo\frac{\hat{M}_{K^+}^2+
\hat{M}_{K^0}^2-\hat{M}_{\pi^+}^2}{\hat{M}_{\pi^+}^2}\co\ee
which relates the quantity $m_s/m_{ud}$ to a ratio of meson masses in QCD.
Expressing these in terms of the physical masses and the four coefficients
introduced in Eqs.~(\ref{eq:epsilon1})-(\ref{eq:epsilon3}), linearizing the
result with respect to the corrections and inserting the observed mass
values, we obtain 
\be\label{eq:LO1 num} \frac{m_s}{m_{ud}} \Lo 25.9 - 0.1\,
\epsilonD + 1.9\, \epsilon_{\pi^0} - 0.1\, \epsilon_{K^0} -1.8
\,\epsilon_m\fs\ee 
If the coefficients $\epsilonD$, $\epsilon_{\pi^0}$, $\epsilon_{K^0}$
and $\epsilon_m$ are set equal to zero, the right hand side reduces to
the value $m_s/m_{ud}=25.9$ that follows from Weinberg's leading-order
formulae for $m_u/m_d$ and $m_s/m_d$~\cite{Weinberg:1977hb}, in
accordance with the fact that these do account for the
e.m.\ interaction at leading chiral order, and neglect the mass
  difference between the charged and neutral pions in QCD.  Inserting
  the estimates (\ref{eq:epsilon num}) gives the effect of chiral
  corrections to the e.m.\ self-energies and of the mass difference
  between the charged and neutral pions in QCD. With these, the LO
  prediction in QCD becomes 
\be\label{eq:LO
  ms/mud}\frac{m_s}{m_{ud}}\Lo 25.9(1)\ ,\ee
leaving the central value unchanged at 25.9. The corrections
parameterized by the coefficients of Eq.~(\ref{eq:epsilon num}) are small
for this quantity.  Note that the quoted uncertainty does not include
an estimate of higher-order chiral contributions to this LO QCD
formula, but only accounts for the error bars in the
coefficients. However, even this small uncertainty is no longer irrelevant
given the the high precision reached in lattice determinations of the
ratio $m_s/m_{ud}$.

The lattice results in Tab.~\ref{tab:ratio_msmud}, which satisfy our selection 
criteria, indicate that the corrections generated by the nonleading terms of the 
chiral perturbation series are remarkably small, in the range 3--10\%. 
Despite the fact that the $SU(3)$-flavour-symmetry breaking effects in the 
Nambu-Goldstone boson masses are very large ($M_K^2\simeq 13\, M_\pi^2$), 
the mass spectrum of the pseudoscalar octet obeys the $SU(3)\times SU(3)$
formula (\ref{eq:LO1}) very well.

\bigskip
\noindent
{\em $\Nf=2$ lattice calculations}
\medskip

With respect to the FLAG 13 review~\cite{Aoki:2013ldr} there is only one new result, ETM 14D~\cite{Abdel-Rehim:2014nka}, based on recent ETM gauge ensembles generated close to the physical point with the addition of a clover term to the tmQCD action.
The new simulations are performed at a single lattice spacing of $\simeq 0.09$ fm and at a single box size $L \simeq 4$ fm and therefore their calculations do not pass our criteria for the continuum extrapolation and finite-volume effects.

Therefore the FLAG average at $N_f = 2$ is still obtained by considering only the ETM 10B result (described already in the previous Section), namely
%FLAGRESULT BEGIN
% TAG      &msomud &END
% REFS     &\cite{Blossier:2010cr}&END
% UNITS    & 1   &END
% NUMRESULTS & 1  &END
% FLAVOURs & 2  &END
%FLAGRESULT END
%FLAGRESULTFORMULA BEGIN
 \be
     \label{eq:msovmud2} 
      \mbox{$N_f = 2$ :}\qquad \FLAGAVBEGIN m_s / m_{ud} = 27.3 ~ (9)\FLAGAVEND\qquad\Ref~\mbox{\cite{Blossier:2010cr}},
 \ee
%FLAGRESULTFORMULA END
with an overall uncertainty equal to 3.3\%.

\newpage
\bigskip
\noindent
{\em $\Nf=2+1$ lattice calculations}
\medskip

For $N_f = 2+1$ our average of $m_s/m_{ud}$ is based on the new result RBC/UKQCD 14B, which replaces 
RBC/UKQCD 12 (see Sec.~\ref{sec:msmud}), and on the results MILC 09A and BMW 10A, 10B. 
The value quoted by HPQCD 10 does not represent independent information as it relies 
on the result for $m_s/m_{ud}$ obtained by the MILC collaboration. Averaging these
results according to the prescriptions of Sec.~\ref{sec:error_analysis} gives
$m_s / m_{ud} = 27.43(13)$ with $\chi^2/\mbox{dof} \simeq 0.2$. Since the errors associated 
with renormalization drop out in the ratio, the uncertainties are even
smaller than in the case of the quark masses themselves: the above
number for $m_s/m_{ud}$ amounts to an accuracy of 0.5\%.

At this level of precision, the uncertainties in the electromagnetic
and strong isospin-breaking corrections are not completely
negligible. The error estimate in the LO result (\ref{eq:LO ms/mud})
indicates the expected order of magnitude. In view of this, we ascribe conservatively 
a 1.0\% uncertainty to this source of error. Thus, our final conservative 
estimate is
%FLAGRESULT BEGIN
% TAG      &msomud &END
% REFS     &\cite{Blum:2014tka,Burch:2009az,Durr:2010vn,Durr:2010aw}&END
% UNITS    & 1  &END
% NUMRESULTS & 3  &END
% FLAVOURs & 2+1  &END
%FLAGRESULT END
%FLAGRESULTFORMULA BEGIN
 \be
     \label{eq:msovmud3} 
     %\mbox{$N_f = 2+1$ :} \qquad\FLAGAVBEGIN \frac{m_s}{m_{ud}} =  27.43 ~ (31) \FLAGAVEND 
     \mbox{$N_f = 2+1$ :} \qquad {m_s}/{m_{ud}} = 27.43 ~ (13) ~ (27) = 27.43 ~ (31) \qquad\Ref~\mbox{\cite{Blum:2014tka,Bazavov:2009fk,Durr:2010vn,Durr:2010aw}}, 
 \ee 
%FLAGRESULTFORMULA END
with a total 1.1\% uncertainty. It is also fully consistent with the ratio computed 
from our individual quark masses in Eq.~(\ref{eq:nf3msmud}), $m_s / m_{ud} = 27.6(6)$, 
which has a larger 2.2\% uncertainty. In Eq.~(\ref{eq:msovmud3}) the first error comes 
from the averaging of the lattice results, and the second is the one that we add to 
account for the neglect of isospin-breaking effects.

\bigskip
\noindent
{\em $\Nf=2+1+1$ lattice calculations}
\medskip

For $N_f = 2+1+1$ there are two results, ETM 14~\cite{Carrasco:2014cwa} and FNAL/MILC 14A~\cite{Bazavov:2014wgs}, both of which satisfy our selection criteria.

ETM 14 uses 15 twisted mass gauge ensembles at 3 lattice spacings ranging from 0.062 to 0.089 fm (using $f_\pi$ as input), in boxes of size ranging from 2.0 to 3.0 fm and pion masses from 210 to 440 MeV (explaining the tag \soso\ in the chiral extrapolation and the tag \good\ for the continuum extrapolation).
The value of $M_\pi L$ at their smallest pion mass is 3.2 with more than two volumes (explaining the tag \soso\ in the finite-volume effects).
They fix the strange mass with the kaon mass.

FNAL/MILC 14A employs HISQ staggered fermions.
Their result is based on 21 ensembles at 4 values of the coupling $\beta$ corresponding to lattice spacings in the range from 0.057 to 0.153 fm, in boxes of sizes up to 5.8 fm and with taste-Goldstone pion masses down to 130 MeV and RMS pion masses down to 143 MeV.
They fix the strange mass with $M_{\bar ss}$, corrected for e.m.~effects with $\bar\epsilon = 0.84(20)$~\cite{Basak:2014vca}. 
All of our selection criteria are satisfied with the tag \good\ .
Thus our average is given by $m_s / m_{ud} = 27.30 ~ (20)$, where the error includes a large stretching factor equal to $\sqrt{\chi^2/\mbox{dof}} \simeq 2.1$, coming from our rules for the averages discussed in Sec.~\ref{sec:averages}.
Nevertheless the above number amounts still to an accuracy of 0.7\%.
As in the case of our average for $N_f = 2+1$, we add a 1.0\% uncertainty related to the neglect of isospin-breaking effects, leading to
%FLAGRESULT BEGIN
% TAG      &msomud &END
% REFS     &\cite{Carrasco:2014cwa,Bazavov:2014wgs}&END
% UNITS    & 1  &END
% NUMRESULTS & 2  &END
% FLAVOURs & 2+1+1  &END
%FLAGRESULT END
%FLAGRESULTFORMULA BEGIN
 \be
      \label{eq:msovmud4} 
      %\mbox{$N_f = 2+1+1$ :}\qquad \FLAGAVBEGIN m_s / m_{ud} = 27.30 ~ (34)\FLAGAVEND ~ ,
      \mbox{$N_f = 2+1+1$ :}\qquad  m_s / m_{ud} = 27.30 ~ (20) ~ (27) = 27.30 ~ (34) \qquad\Refs~\mbox{\cite{Carrasco:2014cwa,Bazavov:2014wgs}},
 \ee
%FLAGRESULTFORMULA END
which corresponds to an overall uncertainty equal to 1.3\%.

\begin{figure}[!htb]
\begin{center}
\psfig{file=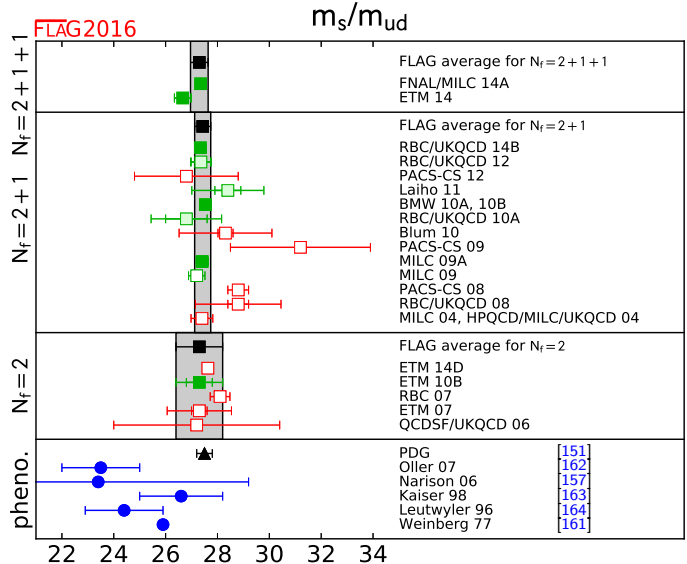,width=11cm}
\end{center}
\vspace{-2.54cm}\hspace{10.5cm}\parbox{6cm}{\sffamily\tiny  \cite{Agashe:2014kda}\\[-0.04cm]

\vspace{-1.15em}\cite{Oller:2006xb}\\[-0.04cm]

\vspace{-1.15em}\cite{Narison:2005ny}\\[-0.04cm]

\vspace{-1.15em}\cite{Kaiser}\\[-0.04cm]

\vspace{-1.15em}\cite{Leutwyler:1996qg}\\[-0.04cm]

\vspace{-1.15em}\cite{Weinberg:1977hb}}

\vspace{0.5cm}
\begin{center}
\caption{ \label{fig:msovmud}Results for the ratio $m_s/m_{ud}$. The upper part indicates the lattice results listed in Tab.~\ref{tab:ratio_msmud} together with the FLAG averages for each value of $N_f$. The lower part shows results obtained from $\chi$PT and sum rules, together with the current PDG estimate.}
\end{center}
\end{figure}

All the lattice results listed in Tab.~\ref{tab:ratio_msmud} as well as the FLAG averages for each value of $N_f$ are reported in Fig.~\ref{fig:msovmud} and compared with $\chi$PT, sum rules and the updated PDG estimate $m_s / m_{ud} = 27.5(3)$~\cite{Agashe:2014kda}.

Note that our averages (\ref{eq:msovmud2}), (\ref{eq:msovmud3}) and (\ref{eq:msovmud4}), obtained for $N_f = 2$, $2+1$ and $2+1+1$, respectively, agree very well within the quoted errors.
They also show that the LO prediction of {\Ch}PT in Eq.~(\ref{eq:LO ms/mud}) receives only small corrections from higher
orders of the chiral expansion: according to Eqs.~(\ref{eq:msovmud3})  
and (\ref{eq:msovmud4}), these generate shifts of $5.9(1.1) \%$ 
and $5.4(1.2) \%$ relative to Eq.~(\ref{eq:LO ms/mud}), respectively.

The ratio $m_s/m_{ud}$ can also be extracted from the masses of the neutral
Nambu-Goldstone bosons: neglecting effects of order $(m_u-m_d)^2$ also
here, the leading-order formula reads 
$m_s / m_{ud} \Lo \frac{3}{2} \hat{M}_\eta^2 / \hat{M}_\pi^2 - \frac{1}{2}$.
Numerically, this gives $m_s / m_{ud} \Lo 24.2$. The relation has the advantage
that the e.m.~corrections are expected to be much smaller here, but it is
more difficult to calculate the $\eta$-mass on the lattice. The comparison with
Eqs.~(\ref{eq:msovmud3}) and (\ref{eq:msovmud4}) shows that, in this case, the NLO 
contributions are somewhat larger: $11.9(9) \%$ and $11.4( 1.1) \%$.

\subsubsection{Lattice determination of $m_u$ and $m_d$}
\label{subsec:mumd}

Since FLAG 13, two new results have been reported for nondegenerate light-quark masses, ETM 14~\cite{Carrasco:2014cwa}, and QCDSF/UKQCD~15~\cite{Horsley:2015eaa}, for $N_f=2+1+1$, and $3$ flavours respectively. The former uses simulations in pure QCD, but determines $m_u-m_d$ from the slope of the square of the kaon mass and the neutral-charged mass-squares difference, evaluated at the isospin-symmetric point.  The latter uses QCD+QED dynamical simulations performed at the $SU(3)$-flavour-symmetric point, but at a single lattice spacing, so they do not enter our average. While QCDSF/UKQCD~15 use three volumes, the smallest has linear size roughly 1.7 fm, and the smallest partially quenched pion mass is greater than 200 MeV, so our finite-volume and chiral-extrapolation criteria require $\soso$ ratings. In Ref.~\cite{Horsley:2015eaa} results for $\epsilon$ and $m_{u}/m_{d}$ are computed in the so-called Dashen scheme. A subsequent paper~\cite{Horsley:2015vla} gives formulae to convert the $\epsilon$ parameters to the $\overline{\rm MS}$ scheme.

As the above implies, the determination of $m_u$ and $m_d$ separately requires additional
input.  MILC 09A~\cite{Bazavov:2009fk} uses the mass difference
between $K^0$ and $K^+$, from which they subtract electromagnetic
effects using Dashen's theorem with corrections, as discussed in
Sec.~\ref{subsec:electromagnetic interaction}.  The up  and down  
sea quarks remain degenerate in their calculation, fixed to the value of
$m_{ud}$ obtained from $M_{\pi^0}$.

To determine $m_u/m_d$, BMW 10A, 10B~\cite{Durr:2010vn,Durr:2010aw}
follow a slightly different strategy. They obtain this ratio from
their result for $m_s/m_{ud}$ combined with a phenomenological
determination of the isospin-breaking quark-mass ratio $Q=22.3(8)$,
defined below in Eq.~(\ref{eq:Qm}), from $\eta\to3\pi$
decays~\cite{Leutwyler:2009jg} (the decay $\eta\to3\pi$ is very
sensitive to QCD isospin breaking but fairly insensitive to QED
isospin breaking).  As discussed in Sec.~\ref{sec:RandQ}, the
central value of the e.m.~parameter $\epsilon$ in Eq.~(\ref{eq:epsilon
  num}) is taken from the same source.

RM123 11~\cite{deDivitiis:2011eh} actually uses the e.m.~parameter
$\epsilon=0.7(5)$ from the first edition of  the FLAG review~\cite{Colangelo:2010et}. However they estimate the effects of
strong isospin breaking at first nontrivial order, by inserting the
operator $\frac12(m_u-m_d)\int(\bar uu-\bar dd)$ into correlation
functions, while performing the gauge averages in the isospin
limit. Applying these techniques, they obtain $(\hat M_{K^0}^2-\hat
M_{K^+}^2)/(m_d-m_u)=2.57(8)\,\mev$. Combining this result with the
phenomenological $(\hat M_{K^0}^2-\hat M_{K^+}^2)=6.05(63)\times10^3$
determined with the above value of $\epsilon$, they get
$(m_d-m_u)=2.35(8)(24)\,\mev$, where the first error corresponds to the
lattice statistical and systematic uncertainties combined in
quadrature, while the second arises from the uncertainty on
$\epsilon$. Note that below we quote results from RM123 11 for $m_u$,
$m_d$ and $m_u/m_d$. As described in Tab.~\ref{tab:mu_md_grading}, we
obtain them by combining RM123 11's result for $(m_d-m_u)$ with ETM 10B's
result for $m_{ud}$.

\begin{table}[!htb]
\vspace{1.8cm}
{\footnotesize{
\begin{tabular*}{\textwidth}[l]{l@{\extracolsep{\fill}}r@{\hspace{1mm}}l@{\hspace{1mm}}l@{\hspace{1mm}}l@{\hspace{1mm}}l@{\hspace{1mm}}l@{\hspace{1mm}}l@{\hspace{1mm}}l@{\hspace{1mm}}l@{\hspace{1mm}}l}
Collaboration \al  Ref. \al \hspace{0.15cm}\begin{rotate}{60}{publication status}\end{rotate}\hspace{-0.15cm}  \al 
\hspace{0.15cm}\begin{rotate}{60}{chiral extrapolation}\end{rotate}\hspace{-0.15cm} \al 
\hspace{0.15cm}\begin{rotate}{60}{continuum  extrapolation}\end{rotate}\hspace{-0.15cm}  \al 
\hspace{0.15cm}\begin{rotate}{60}{finite volume}\end{rotate}\hspace{-0.15cm}  \al   
\hspace{0.15cm}\begin{rotate}{60}{renormalization}\end{rotate}\hspace{-0.15cm} \al   
\hspace{0.15cm}\begin{rotate}{60}{running}\end{rotate}\hspace{-0.15cm}  \al  
\rule{0.6cm}{0cm}$m_u$\al 
\rule{0.6cm}{0cm}$m_d$ \al \rule{0.3cm}{0cm} $m_u/m_d$\\
\al \al \al \al \al \al \al \al \al \al  \\[-0.1cm]
\hline
\hline
\al \al \al \al \al \al \al \al \al \al  \\[-0.1cm]
{MILC 14} \al \cite{Basak:2014vca} \al \rC \al \good \al \good \al \good \al
$-$ \al $-$ \al \al \al $0.4482(48)({}^{+\phantom{0}21}_{-115})(1)(165)$\\

{ETM 14}& \cite{Carrasco:2014cwa}  \al \gA \al \good \al \good \al \good \al \good \al
$\,b$ \al 2.36(24) \al 5.03(26) \al 0.470(56) \\[0.5ex]
\hline
\al \al \al \al \al \al \al \al \al \al  \\[-0.1cm]

{QCDSF/UKQCD 15$^{\ominus}$} \al \protect{\cite{Horsley:2015eaa}} \al \oP \al \soso \al \bad \al \soso \al $-$\al $-$
\al  \al   \al 0.52(5)\\

{PACS-CS 12$^\star$} \al \protect{\cite{Aoki:2012st}} \al \gA \al \good \al \bad \al \bad \al \good \al $\,a$
\al  2.57(26)(7) \al  3.68(29)(10) \al 0.698(51)\\

{Laiho 11} \al \cite{Laiho:2011np} \al \rC \al \soso \al \good \al
\good \al \soso \al $-$ \al 1.90(8)(21)(10) \al
4.73(9)(27)(24) \al 0.401(13)(45)\\

{HPQCD~10$^\ddagger$}\al \cite{McNeile:2010ji} \al \gA \al \soso \al \good \al \good \al \good \al
$-$ \al 2.01(14) \al 4.77(15) \al  \\

{BMW 10A, 10B$^+$}\al \cite{Durr:2010vn,Durr:2010aw} \al \gA \al \good \al \good \al \good \al \good \al
$\,b$ \al 2.15(03)(10) \al 4.79(07)(12) \al 0.448(06)(29) \\

{Blum~10$^\dagger$}\al\cite{Blum:2010ym}\al \gA \al \soso \al \bad \al \soso \al \good \al $-$ \al 2.24(10)(34)\al 4.65(15)(32)\al 0.4818(96)(860)\\

{MILC 09A} \al  \cite{Bazavov:2009fk} \al  \rC \al  \soso \al \good \al \good \al \soso \al $-$
\al 1.96(0)(6)(10)(12)
\al  4.53(1)(8)(23)(12)  \al   0.432(1)(9)(0)(39) \\

{MILC 09} \al  \cite{Bazavov:2009bb} \al  \gA \al  \soso \al  \good \al  \good \al  \soso \al 
$-$\al  1.9(0)(1)(1)(1)
\al  4.6(0)(2)(2)(1) \al  0.42(0)(1)(0)(4) \\

\hspace{-0.2cm}{\begin{tabular}{l}MILC 04, HPQCD/\rule{0.1cm}{0cm}\\MILC/UKQCD 04\end{tabular}} \al  {\begin{tabular}{r}\cite{Aubin:2004fs}\\\cite{Aubin:2004ck}\end{tabular}} \al  \gA \al  \soso \al  \soso \al  \soso \al 
\bad \al$-$\al  1.7(0)(1)(2)(2)
\al  3.9(0)(1)(4)(2)  \al  0.43(0)(1)(0)(8) \\

\al \al \al \al \al \al \al \al \al \al  \\[-0.1cm]
\hline
\al \al \al \al \al \al \al \al \al \al  \\[-0.1cm]

{RM123 13} \al \cite{deDivitiis:2013xla}  \al \gA \al \soso \al \good \al \soso \al \good \al $\,c$
\al 2.40(15)(17) \al  4.80 (15)(17) \al 0.50(2)(3)\\

{RM123 11$^\oplus$} & \cite{deDivitiis:2011eh} \al \gA \al \soso \al \good \al \soso \al \good \al $\,c$
\al { {\em 2.43(11)(23)}} \al { {\em 4.78(11)(23)}} \al { {\em 0.51(2)(4)}}\\

{D\"urr 11$^\ast$}\al \cite{Durr:2011ed} \al \gA \al \soso \al \good \al \soso \al $-$ \al $-$
 \al 2.18(6)(11) \al 4.87(14)(16) \al  \\ 

{RBC 07$^\dagger$} \al  \cite{Blum:2007cy} \al  \gA \al  \bad \al  \bad \al  \good  \al  \good \al  $-$
\al  3.02(27)(19) \al  5.49(20)(34)  \al  0.550(31)\\

\al \al \al \al \al \al \al \al \al \al  \\[-0.1cm]
\hline
\hline\\
\end{tabular*}\\[-0.2cm]
}}
\begin{minipage}{\linewidth}
{\footnotesize 
\begin{itemize}
\item[$^{\ominus}$] Results are computed in QCD+QED and quoted in an
  unconventional ``Dashen scheme".\\[-5mm]
\item[$^\star$] The calculation includes e.m. and $m_u\ne m_d$ effects
  through reweighting.\\[-5mm]
\item[$^\ddagger$]Values obtained by combining the HPQCD 10 result
  for $m_s$ with the MILC 09 results for $m_s/m_{ud}$ and
  $m_u/m_d$.\\[-5mm]
\item[$^+$] The fermion action used is tree-level improved.\\[-5mm]
\item[$^\ast$] Values obtained by combining the D\"urr 11 result for
  $m_s$ with the BMW 10A, 10B results for $m_s/m_{ud}$ and
  $m_u/m_d$.\\[-5mm]
\item[$^\oplus$] $m_u$, $m_d$ and $m_u/m_d$ are obtained by combining
  the result of RM123 11 for $(m_d-m_u)$~\cite{deDivitiis:2011eh} with
  $m_{ud}=3.6(2)\,\mev$ from ETM 10B. $(m_d-m_u)=2.35(8)(24)\,\mev$ in Ref.~\cite{deDivitiis:2011eh} was obtained assuming $\epsilonD =
  0.7(5)$~\cite{Colangelo:2010et} and
  $\epsilon_m=\epsilon_{\pi^0}=\epsilon_{K^0}=0$. In the quoted
  results, the first error corresponds to the lattice statistical and
  systematic errors combined in quadrature, while the second arises
  from the uncertainties associated with $\epsilonD$.\\[-5mm]
\item[$^\dagger$] The calculation includes quenched e.m. effects.\\[-5mm]
\item[$a$] The masses are renormalized and run nonperturbatively up to
  a scale of $100\,\gev$ in the $N_f=2$ SF scheme. In this scheme,
  nonperturbative and NLO running for the quark masses are shown to
  agree well from 100 GeV all the way down to 2 GeV~\cite{DellaMorte:2005kg}.\\[-5mm]
\item[$b$] The masses are renormalized and run nonperturbatively up to
  a scale of 4 GeV in the $N_f=3$ RI/MOM scheme.  In
  this scheme, nonperturbative and N$^3$LO running for the quark
  masses are shown to agree from 6~GeV down to 3~GeV to
  better than 1\%~\cite{Durr:2010aw}. \\[-5mm]
\item[$c$] The masses are renormalized nonperturbatively at scales $1/a\sim 2\div3\,\gev$ in the $N_f=2$ RI/MOM scheme.  In this
scheme, nonperturbative and N$^3$LO running for the quark masses
are shown to agree from 4~GeV down 2 GeV to better than 3\%~\cite{Constantinou:2010gr}.
\end{itemize}
}
\end{minipage}
\caption{Lattice results for $m_u$, $m_d$ (MeV) and for the ratio $m_u/m_d$. The values refer to the 
$\msbar$ scheme  at scale 2 GeV.  The top part of the table lists the result obtained with $\Nf=2+1+1$,  
while the middle  and lower part presents calculations with $N_f = 2+1 $ and $N_f = 2$, respectively.}
\label{tab:mu_md_grading}
\end{table}

Instead of subtracting electromagnetic effects using phenomenology,
RBC~07~\cite{Blum:2007cy} and Blum~10~\cite{Blum:2010ym} actually
include a quenched electromagnetic field in their calculation. This
means that their results include corrections to Dashen's theorem,
albeit only in the presence of quenched electromagnetism. Since the up 
and down quarks in the sea are treated as degenerate, very small
isospin corrections are neglected, as in MILC's calculation.

PACS-CS 12~\cite{Aoki:2012st} takes the inclusion of isospin-breaking
effects one step further. Using reweighting techniques, it also
includes electromagnetic and $m_u-m_d$ effects in the sea.

Lattice results for $m_u$, $m_d$ and $m_u/m_d$ are summarized in
Tab.~\ref{tab:mu_md_grading}. In order to discuss them, we consider
the LO formula
\be\label{eq:LO2}\frac{m_u}{m_d}\Lo\frac{\hat{M}_{K^+}^2-\hat{M}_{K^0}^2+\hat{M}_{\pi^+}^2}
{\hat{M}_{K^0}^2-\hat{M}_{K^+}^2+\hat{M}_{\pi^+}^2} \fs\ee
Using Eqs.~(\ref{eq:epsilon1})--(\ref{eq:epsilon3}) to express
the meson masses in QCD in terms of the physical ones and linearizing
in the corrections, this relation takes the form
\be\label{eq:LO2 num}\frac{m_u}{m_d}\Lo 0.558 - 0.084\, \epsilonD - 0.02\,
\epsilon_{\pi^0} + 0.11\, \epsilon_m \fs\ee
Inserting the estimates
(\ref{eq:epsilon num}) and adding errors in quadrature, the LO
prediction becomes
{ 
\be\label{eq:mu/md LO}\frac{m_u}{m_d}\Lo
0.50(3)\fs\ee 
}
Again, the quoted error exclusively accounts for the errors attached
to the estimates (\ref{eq:epsilon num}) for the epsilons --
contributions of nonleading order are ignored. The uncertainty in the
leading-order prediction is dominated by the one in the coefficient
$\epsilonD$, which specifies the difference between the meson
squared-mass splittings generated by the e.m. interaction in the kaon
and pion multiplets. { The reduction in the error on this coefficient
since the previous review~\cite{Colangelo:2010et} results in a
reduction of a factor of a little less than 2 in the uncertainty on
the LO value of $m_u/m_d$ given in Eq.~(\ref{eq:mu/md LO}).}

It is interesting to compare the assumptions made or results obtained
by the different collaborations for the violation of Dashen's
theorem. The input used in MILC 09A is $\epsilonD=1.2(5)$~\cite{Bazavov:2009fk}, 
while the $N_f=2$ computation of RM123 13 finds
$\epsilonD=0.79(18)(18)$~\cite{deDivitiis:2013xla}.  As discussed in
Sec.~\ref{sec:RandQ}, the value of $Q$ used by BMW 10A,
10B~\cite{Durr:2010vn,Durr:2010aw} gives $\epsilonD=0.70(28)$ at NLO
(see Eq.~(\ref{eq:epsilon eta})). On the other hand, RBC 07~\cite{Blum:2007cy} and Blum~10~\cite{Blum:2010ym} obtain the results
$\epsilonD=0.13(4)$ and $\epsilonD=0.5(1)$. The new results from QCDSF/UKQCD 15 give $\epsilon=0.50(6)$~\cite{Horsley:2015vla}. Note that PACS-CS 12
\cite{Aoki:2012st} do not provide results which allow us to determine
$\epsilon$ directly. However, using their result for $m_u/m_d$,
together with Eq.~(\protect{\ref{eq:LO2 num}}), and neglecting NLO terms,
one finds $\epsilonD=-1.6(6)$, which is difficult to reconcile with
what is known from phenomenology (see Secs.~\ref{subsec:electromagnetic interaction} and \ref{sec:RandQ}).  Since
the values assumed or obtained for $\epsilonD$ differ, it does not
come as a surprise that the determinations of $m_u/m_d$ are different.

These values of $\epsilonD$ are also interesting because they allow us
to estimate the chiral corrections to the LO prediction (\ref{eq:mu/md
  LO}) for $m_u/m_d$. Indeed, evaluating the relation (\ref{eq:LO2
  num}) for the values of $\epsilonD$ given above, and neglecting all
other corrections in this equation, yields the LO values
$(m_u/m_d)^\mathrm{LO}=0.46(4)$, 0.547(3), 0.52(1), 0.50(2), 0.49(2)
and 0.51(1)
for MILC 09A, RBC 07, Blum 10, BMW 10A, 10B, RM123 13, and QCDSF/UKQCD 15,
respectively. However, in comparing these numbers to the
nonperturbative results of Tab.~\ref{tab:mu_md_grading} one must be
careful not to double count the uncertainty arising from
$\epsilonD$. One way to obtain a sharp comparison is to consider the
ratio of the results of Tab.~\ref{tab:mu_md_grading} to the LO values $(m_u/m_d)^{\rm LO}$,
in which the uncertainty from $\epsilon$ cancels to good
accuracy. Here we will assume for simplicity that they cancel
completely and will drop all uncertainties related to $\epsilon$. For
$N_f = 2$ we consider RM123 13~\cite{deDivitiis:2013xla}, which
updates RM123 11 and has no red dots.  Since the uncertainties common
to $\epsilon$ and $m_u/m_d$ are not explicitly given in Ref.~\cite{deDivitiis:2013xla}, we have to estimate them. For that we use
the leading-order result for $m_u/m_d$, computed with RM123 13's value
for $\epsilon$. Its error bar is the contribution of the uncertainty
on $\epsilon$ to $(m_u/m_d)^{\rm LO}$. To good approximation this
contribution will be the same for the value of $m_u/m_d$ computed in Ref.~\cite{deDivitiis:2013xla}. Thus, we subtract it in quadrature from
RM123 13's result in Tab.~\ref{tab:mu_md_grading} and compute
$(m_u/m_d)/(m_u/m_d)^{\rm LO}$, dropping uncertainties related to
$\epsilon$. We find $(m_u/m_d)/(m_u/m_d)^{\rm LO} = 1.02(6)$. This
result suggests that chiral corrections in the case of $\Nf=2$ are
negligible. For the two most accurate $\Nf=2+1$ calculations, those of
MILC 09A and BMW 10A, 10B, this ratio of ratios is 0.94(2) and
0.90(1), respectively. { Though these two numbers are not fully
  consistent within our rough estimate of the errors, they indicate
  that higher-order corrections to Eq.~(\ref{eq:mu/md LO}) are negative
  and about 8\% when $\Nf=2+1$. In the following, we will take them to
  be -8(4)\%. The fact that these corrections are seemingly larger and
  of opposite sign than in the $\Nf=2$ case is not understood at this
  point. It could be an effect associated with the quenching of the
  strange quark. It could also be due to the fact that the RM123 13
  calculation does not probe deeply enough into the chiral regime -- it
  has $M_\pi\gsim 270\,\mev$ -- to pick up on important chiral
  corrections. Of course, being less than a two-standard-deviation
  effect, it may be that there is no problem at all and that
  differences from the LO result are actually small.

Given the exploratory nature of the RBC 07 calculation, its results do
not allow us to draw solid conclusions about the e.m.\,contributions
to $m_u/m_d$ for $\Nf=2$.  As discussed 
in Sec.~\ref{sec:msmud} and here,
the $\Nf=2+1$ results of Blum 10, PACS-CS 12, and QCDSF/UKQCD~15 do not pass our
selection criteria either. We therefore resort to the phenomenological
estimates of the electromagnetic self-energies discussed in Sec.~\ref{subsec:electromagnetic interaction}, which are validated by
recent, preliminary lattice results. 

Since RM123 13~\cite{deDivitiis:2013xla} includes a lattice estimate of
  e.m.\ corrections, for the $\Nf=2$ final results we simply quote the values
  of $m_u$, $m_d$, and $m_{u}/m_{d}$ from RM123 13 given in
  Tab.~\ref{tab:mu_md_grading}:
%FLAGRESULT BEGIN
% TAG      &mu&md&muomd &END
% REFS     &\cite{deDivitiis:2013xla}&\cite{deDivitiis:2013xla}&\cite{deDivitiis:2013xla}&END
% UNITS    & '[MeV]' & '[MeV]' &1  &END
% NUMRESULTS & 1& 1& 1  &END
% FLAVOURs & 2& 2& 2  &END
%FLAGRESULT END
%FLAGRESULTFORMULA BEGIN
\begin{align}
&&\FLAGAVBEGIN m_u &=2.40(23)\FLAGAVEND   \,\mev&\Ref~\mbox{\cite{deDivitiis:2013xla}},\nonumber\\
\label{eq:mumdNf2} \hspace{0cm}\Nf = 2:\hspace{0.2cm}
&&\FLAGAVBEGIN m_d &= 4.80(23)\FLAGAVEND  \,\mev&\Ref~\mbox{\cite{deDivitiis:2013xla}},\\
&&\FLAGAVBEGIN {m_u}/{m_d} &= 0.50(4)\FLAGAVEND &\Ref~\mbox{\cite{deDivitiis:2013xla}},\nonumber
\end{align}
%FLAGRESULTFORMULA END
with errors of roughly 10\%, 5\% and 8\%, respectively. In these results, the
errors are obtained by combining the lattice statistical and
systematic errors in quadrature.

For $\Nf=2+1$ there is to date no final, published
  computation of e.m.\ corrections. Thus, we take the LO estimate
  for $m_u/m_d$ of Eq.~(\ref{eq:mu/md LO}) and use the -8(4)\% obtained
  above as an estimate of the size of the corrections from higher
  orders in the chiral expansion. This gives $m_u/m_d=0.46(3)$. The
  two individual masses can then be worked out from the estimate
  (\ref{eq:nf3msmud}) for their mean. Therefore, for $\Nf=2+1$ we
  obtain:
%
%FLAGRESULT BEGIN
% TAG      &mu&md&muomd &END
% REFS     &${}^\ddagger$&${}^\ddagger$&${}^\ddagger$&END
% UNITS    & '[MeV]' & '[MeV]' &1  &END
% NUMRESULTS & 1& 1& 1  &END
% FLAVOURs & 2+1& 2+1& 2+1  &END
%FLAGRESULT END
%FLAGRESULTFORMULA BEGIN
\begin{align}
&&\FLAGAVBEGIN m_u &=2.16(9)(7)\FLAGAVEND\,\mev\,, \nonumber\\
\label{eq:mumd} \hspace{0cm}\Nf = 2+1:\hspace{0.2cm}
&&\FLAGAVBEGIN m_d &= 4.68(14)(7) \FLAGAVEND\,\mev\,,\\
&&\FLAGAVBEGIN {m_u}/{m_d} &= 0.46(2)(2)\FLAGAVEND\,.\nonumber
\end{align}
%FLAGRESULTFORMULA END
%
In these results, the
first error represents the lattice statistical and systematic errors, combined
in quadrature, while the second arises from the uncertainties
associated with e.m.\ corrections of Eq.~(\ref{eq:epsilon num}).
The estimates in Eq.~(\ref{eq:mumd}) have uncertainties of order 5\%, 3\% and 7\%,
respectively.

Finally, for four flavours we simply adopt the results of ETM~14A which meet all of our criteria.
%
%FLAGRESULT BEGIN
% TAG      &mu&md&muomd &END
% REFS     &\cite{Carrasco:2014cwa}&\cite{Carrasco:2014cwa}&\cite{Carrasco:2014cwa}&END
% UNITS    & '[MeV]' & '[MeV]' &1  &END
% NUMRESULTS & 1& 1& 1  &END
% FLAVOURs & 2+1+1& 2+1+1& 2+1+1  &END
%FLAGRESULT END
%FLAGRESULTFORMULA BEGIN
\begin{align}
	&&\FLAGAVBEGIN m_u &=2.36(24)\FLAGAVEND \,\mev&\Ref~\mbox{\cite{Carrasco:2014cwa}}\,,\nonumber\\
\label{eq:mumd 4 flavour} \Nf = 2+1+1:\hspace{0.15cm}
        &&\FLAGAVBEGIN m_d &= 5.03(26)\FLAGAVEND \,\mev&\Ref~\mbox{\cite{Carrasco:2014cwa}}\,,\\
	&&\FLAGAVBEGIN {m_u}/{m_d} &= 0.470(56)\FLAGAVEND&\Ref~\mbox{\cite{Carrasco:2014cwa}}\,.\nonumber
\end{align}
%FLAGRESULTFORMULA END
%

Naively propagating errors to the end, we obtain $(m_u/m_d)_{N_f=2}/(m_u/m_d)_{N_f=2+1}=1.09(10)$. If instead of Eq.~(\ref{eq:mumdNf2}) we use the results from RM123 11, modified by the e.m. corrections in Eq.~(\ref{eq:epsilon num}), as was done in our previous review, we obtain $(m_u/m_d)_{N_f=2}/(m_u/m_d)_{N_f=2+1}=1.11(7)(1)$,  confirming again the strong cancellation of e.m.\ uncertainties in the ratio. 
The $N_f=2$ and $2+1$ results are compatible at the 1 to 1.5~$\sigma$ level. Clearly the difference between three and four flavours is even smaller, and completely covered by the quoted uncertainties.

It is interesting to note that in the results above, the errors
  are no longer dominated by the uncertainties in the input used for
  the electromagnetic corrections, though these are still significant
  at the level of precision reached in the $N_f=2+1$ results. This is
  due to the reduction in the error on $\epsilonD$ discussed in
  Sec.~\ref{subsec:electromagnetic interaction}. Nevertheless, the
  comparison of Eqs.~(\ref{eq:mu/md LO}) and (\ref{eq:mumd})
  indicates that more than half of the difference between the
  prediction $m_u/m_d=0.558$ obtained from Weinberg's mass formulae~\cite{Weinberg:1977hb} and the result for $m_u/m_d$ obtained on the
  lattice stems from electromagnetism, the higher orders in the chiral
  perturbation generating a comparable correction.

In view of the fact that a {\it massless up-quark} would solve the
strong CP-problem, many authors have considered this an attractive
possibility, but the results presented above exclude this possibility:
the value of $m_u$ in Eq.~(\ref{eq:mumd}) differs from zero by 20
standard deviations. We conclude that nature solves the strong
CP-problem differently. This conclusion relies on lattice calculations
of kaon masses and on the phenomenological estimates of the
e.m.~self-energies discussed in Sec.~\ref{subsec:electromagnetic
  interaction}. The uncertainties therein currently represent the
limiting factor in determinations of $m_u$ and $m_d$. As demonstrated
in Refs.~\cite{Duncan:1996xy,Blum:2007cy,Blum:2010ym,Basak:2008na,Portelli:2010yn,Portelli:2012pn,llconfx12,Basak:2014vca,Basak:2015lla,Basak:2012zx,deDivitiis:2013xla,Basak:2013iw},
lattice methods can be used to calculate the
e.m.~self-energies. Further progress on the determination of the light-quark 
masses hinges on an improved understanding of the e.m.~effects.

\subsubsection{Estimates for $R$ and $Q$}\label{sec:RandQ}

The quark-mass ratios
\be\label{eq:Qm}
R\equiv \frac{m_s-m_{ud}}{m_d-m_u}\hspace{0.5cm} \mbox{and}\hspace{0.5cm}Q^2\equiv\frac{m_s^2-m_{ud}^2}{m_d^2-m_u^2}
\ee
compare $SU(3)$ breaking  with isospin breaking. The quantity $Q$ is of
particular interest because of a low-energy theorem~\cite{Gasser:1984pr},
which relates it to a ratio of meson masses,  
\begin{equation}\label{eq:QM}
 Q^2_M\equiv \frac{\hat{M}_K^2}{\hat{M}_\pi^2}\cdot\frac{\hat{M}_K^2-\hat{M}_\pi^2}{\hat{M}_{K^0}^2-
   \hat{M}_{K^+}^2}\co\hspace{1cm}\hat{M}^2_\pi\equiv\mbox{$\frac{1}{2}$}( \hat{M}^2_{\pi^+}+ \hat{M}^2_{\pi^0})
 \co\hspace{0.5cm}\hat{M}^2_K\equiv\mbox{$\frac{1}{2}$}( \hat{M}^2_{K^+}+ \hat{M}^2_{K^0})\fs\end{equation}
Chiral symmetry implies that the expansion of $Q_M^2$ in powers of the
quark masses (i) starts with $Q^2$ and (ii) does not receive any
contributions at NLO:
\be\label{eq:LET Q}Q_M\NLo Q \fs\ee

{
Inserting the estimates for the mass ratios $m_s/m_{ud}$, and
$m_u/m_d$ given for $\Nf=2$ in Eqs.~(\ref{eq:quark masses Nf=2})
and (\ref{eq:mumdNf2}) respectively, we obtain
\be\label{eq:RQresNf2} R=40.7(3.7)(2.2)\co\hspace{2cm}Q=24.3(1.4)(0.6)\ ,\ee 
where the errors have been propagated naively and the e.m.\ uncertainty
has been separated out, as discussed in the third paragraph after
Eq.~(\ref{eq:mu/md LO}). Thus, the meaning of the errors is the same as in
Eq.~(\ref{eq:mumd}). These numbers agree within errors with those reported
in Ref.~\cite{deDivitiis:2013xla} where values for $m_s$ and $m_{ud}$ are taken
from ETM 10B~\cite{Blossier:2010cr}.

For $\Nf=2+1$, we use Eqs.~(\ref{eq:msovmud3}) and (\ref{eq:mumd}) and obtain
\be\label{eq:RQres} R=35.7(1.9)(1.8)\co\hspace{2cm}Q=22.5(6)(6)\ ,\ee 
where the meaning of the errors is the same as above. The
$\Nf=2$ and $\Nf=2+1$ results are compatible within 
2$\sigma$, even taking the correlations between e.m. effects into account.
}

Again, for $\Nf=2+1+1$, we simply take values from ETM 14A,
\be\label{eq:RQresNf4} R=35.6(5.1)\co\hspace{2cm}Q=22.2(1.6)\ ,\ee 
which are quite compatible with two and three flavour results.

It is interesting to use these results to study the size of
chiral corrections in the relations of $R$ and $Q$ to their
expressions in terms of meson masses. To investigate this issue, we
use {\Ch}PT to express the quark-mass ratios in terms of the pion and
kaon masses in QCD and then again use Eqs.~(\ref{eq:epsilon1})--(\ref{eq:epsilon3}) to relate the QCD masses to
the physical ones. Linearizing in the corrections, this leads to {
\bea\label{eq:R epsilon}R\al \Lo\al R_M = 43.9 - 10.8\, \epsilonD +
0.2\, \epsilon_{\pi^0} - 0.2\, \epsilon_{K^0}- 10.7\, \epsilon_m\co\\
Q\al\NLo\al Q_M = 24.3 - 3.0\, \epsilonD + 0.9\,
\epsilon_{\pi^0} - 0.1\, \epsilon_{K^0} + 2.6 \,\epsilon_m
\fs
\label{eq:Q epsilon} 
\eea
}
While
the first relation only holds to LO of the chiral perturbation series, the
second remains valid at NLO, on account of the low-energy theorem mentioned
above. The first terms on the right hand side represent the values of $R$
and $Q$ obtained with the Weinberg leading-order formulae for the quark-mass 
ratios~\cite{Weinberg:1977hb}. Inserting the estimates
(\ref{eq:epsilon num}), we find that the e.m.~corrections lower the
Weinberg values to { $R_M= 36.7(3.3)$ and $Q_M= 22.3(9)$}, respectively.

{ Comparison of $R_M$ and $Q_M$ with the full results quoted above
  gives a handle on higher-order terms in the chiral
  expansion. Indeed, the ratios $R_M/R$ and $Q_M/Q$ give NLO and NNLO
  (and higher)-corrections to the relations $R \Lo R_M$ and $Q\NLo
  Q_M$, respectively. The uncertainties due to the use of the
  e.m.\ corrections of Eq.~(\ref{eq:epsilon num}) are highly correlated in
  the numerators and denominators of these ratios, and we make the
  simplifying assumption that they cancel in the ratio.  { Thus, for
    $N_f=2$ we evaluate Eqs.~(\ref{eq:R epsilon}) and (\ref{eq:Q epsilon})
    using $\epsilon=0.79(18)(18)$ from RM123 13~\cite{deDivitiis:2013xla} and the other corrections from
    Eq.~(\ref{eq:epsilon num}), dropping all uncertainties. We divide them
    by the results for $R$ and $Q$ in Eq.~(\ref{eq:RQresNf2}), omitting
    the uncertainties due to e.m. We obtain $R_M/R\simeq 0.88(8)$ and
    $Q_M/Q\simeq 0.91(5)$. We proceed analogously for $N_f=2+1$ and 2+1+1, using
    $\epsilon=0.70(3)$ from Eq.~(\ref{eq:epsilon num}) and $R$ and $Q$
    from Eqs.~(\ref{eq:RQres}) and (\ref{eq:RQresNf4}), and find $R_M/R\simeq 1.02(5)$ and 1.03(17), and
    $Q_M/Q\simeq 0.99(3)$ and 1.00(8).}
    The chiral corrections appear to be small
  for three and four flavours, especially those in the relation of $Q$ to
  $Q_M$. This is less true for $N_f=2$, where the NNLO and higher
  corrections to $Q=Q_M$ could be significant. However, as for other
  quantities which depend on $m_u/m_d$, this difference is not
  significant.}

As mentioned in Sec.~\ref{subsec:electromagnetic interaction}, there is
a phenomenological determination of $Q$ based on the decay $\eta\rightarrow
3\pi$~\cite{Kambor:1995yc,Anisovich:1996tx}. The key point is that the
transition $\eta\rightarrow 3\pi$ violates isospin conservation. The
dominating contribution to the transition amplitude stems from the mass
difference $m_u-m_d$. At NLO of {\Ch}PT, the QCD part of the amplitude can
be expressed in a parameter-free manner in terms of $Q$.  It is well-known
that the electromagnetic contributions to the transition amplitude are
suppressed (a thorough recent analysis is given in Ref.~\cite{Ditsche:2008cq}).
This implies that the result for $Q$ is less sensitive to the
electromagnetic uncertainties than the value obtained from the masses of
the Nambu-Goldstone bosons.  For a recent update of this determination and
for further references to the literature, we refer to Ref.~\cite{Colangelo:2009db}. Using dispersion theory to pin down the 
momentum dependence of the amplitude, the observed decay rate implies $Q=22.3(8)$
(since the uncertainty quoted in Ref.~\cite{Colangelo:2009db} does not include
an estimate for all sources of error, we have retained the error estimate
given in Ref.~\cite{Leutwyler:1996qg}, which is twice as large). The formulae
for the corrections of NNLO are available also in this case~\cite{Bijnens:2007pr} -- the poor knowledge of the effective coupling
constants, particularly of those that are relevant for the dependence on
the quark masses, is currently the limiting factor encountered in the
application of these formulae.

As was to be expected, the central value of $Q$ obtained from
$\eta$-decay agrees exactly with the central value obtained from the
low-energy theorem: we have used that theorem to estimate the
coefficient $\epsilonD$, which dominates the e.m.~corrections. Using
the numbers for $\epsilon_m$, $\epsilon_{\pi^0}$ and $\epsilon_{K^0}$
in Eq.~(\ref{eq:epsilon num}) and adding the corresponding uncertainties
in quadrature to those in the phenomenological result for $Q$, we
obtain
\be\label{eq:epsilon eta}
\epsilonD\NLo 0.70(28)\fs\ee 
The estimate (\ref{eq:epsilon num}) for the size of the coefficient
$\epsilonD$ is taken from here, { as it is confirmed by the most
  recent, preliminary lattice determinations~\cite{Basak:2008na,Portelli:2010yn,Portelli:2012pn,Basak:2012zx,Basak:2013iw,deDivitiis:2013xla}.}

Our final results for the masses $m_u$, $m_d$, $m_{ud}$, $m_s$ and the mass ratios
$m_u/m_d$, $m_s/m_{ud}$, $R$, $Q$ are collected in Tabs.~\ref{tab:mudms} and
\ref{tab:mumdRQ}. We separate $m_u$, $m_d$, $m_u/m_d$, $R$ and $Q$
from $m_{ud}$, $m_s$ and $m_s/m_{ud}$, because the latter are
completely dominated by lattice results while the former still include
some phenomenological input.

\begin{table}[!thb]\vspace{0.5cm}
{
\begin{tabular*}{\textwidth}[l]{@{\extracolsep{\fill}}cccc}
\hline\hline
$\Nf$ & $m_{ud}$ & $ m_s $ & $m_s/m_{ud}$ \\ 
&&& \\[-2ex]
\hline\rule[-0.1cm]{0cm}{0.5cm}
&&& \\[-2ex]
2+1+1 & 3.70(17) & 93.9(1.1) & 27.30(34)\\ 
&&& \\[-2ex]
\hline\rule[-0.1cm]{0cm}{0.5cm}
&&& \\[-2ex]
2+1 & 3.373(80) & 92.0(2.1) & 27.43(31)\\ 
&&& \\[-2ex]
\hline\rule[-0.1cm]{0cm}{0.5cm}
&&& \\[-2ex]
2 & 3.6(2) & 101(3) & 27.3(9)\\ 
&&& \\[-2ex]
\hline
\hline
\end{tabular*}
\caption{\label{tab:mudms} Our estimates for the strange-quark and the average
  up-down-quark masses in the $\msbar$ scheme  at running scale
  $\mu=2\,\gev$. Numerical values are given in MeV. In the
  results presented here, the error is the one which we obtain
  by applying the averaging procedure of Sec.~\ref{sec:error_analysis} to the
  relevant lattice results. We have added an uncertainty to the
  $N_f=2+1$ results, associated with the neglect of the charm sea-quark 
  and isospin-breaking effects, as discussed around
  Eqs.~(\ref{eq:nf3msmud}) and (\ref{eq:msovmud3}). This uncertainty is not
  included in the $N_f=2$ results, as it should be smaller than the
  uncontrolled systematic associated with the neglect of strange
  sea-quark effects.}  }
\end{table}

\begin{table}[!thb]%\hspace{-0.5cm}
{
\begin{tabular*}{\textwidth}[l]{@{\extracolsep{\fill}}cccccc}
\hline\hline
$\Nf$ & $m_u  $ & $m_d $ & $m_u/m_d$ & $R$ & $Q$\\ 
&&&&& \\[-2ex]
\hline\rule[-0.1cm]{0cm}{0.5cm}
&&&&& \\[-2ex]
2+1+1 & 2.36(24) & 5.03(26)& 0.470(56) & 35.6(5.1) & 22.2 (1.6) \\ 
&&&&& \\[-2ex]
\hline\rule[-0.1cm]{0cm}{0.5cm}
&&&&& \\[-2ex]
2+1 & 2.16(9)(7) & 4.68(14)(7) & 0.46(2)(2) & 35.0(1.9)(1.8) & 22.5(6)(6) \\ 
&&&&& \\[-2ex]
\hline\rule[-0.1cm]{0cm}{0.5cm}
&&&&& \\[-2ex]
2 & 2.40(23) & 4.80(23) & 0.50(4) & 40.7(3.7)(2.2) & 24.3(1.4)(0.6)\\ 
&&&&& \\[-2ex]
\hline
\hline
\end{tabular*}
\caption{\label{tab:mumdRQ} Our estimates for the masses of the
  two lightest quarks and related, strong isospin-breaking
  ratios. Again, the masses refer to the $\msbar$ scheme  at running
  scale $\mu=2\,\gev$. Numerical values are given
  in MeV. In the results presented here, the first error is the one
  that comes from lattice computations while the second for $N_f=2+1$
  is associated with the phenomenological estimate of e.m.\
  contributions, as discussed after Eq.~(\ref{eq:mumd}). The
  second error on the $N_f=2$ results for $R$ and $Q$ is also an
  estimate of the e.m.\ uncertainty, this time associated with the
  lattice computation of Ref.~\cite{deDivitiis:2013xla}, as
  explained after Eq.~(\ref{eq:RQresNf2}). We present these
  results in a separate table, because they are less firmly
  established than those in Tab.~\protect\ref{tab:mudms}. For
  $N_f=2+1$ and 2+1+1 they still include information coming from phenomenology,
  in particular on e.m.\ corrections, and for $N_f=2$ the e.m.\
  contributions are computed neglecting the feedback of sea quarks on
  the photon field.}  }
\end{table}

%% file: qmass/qmass_mc.tex
% !TEX root = /Users/tblum/Dropbox/FLAG3_Review/working_copy_v1.2/qmass/qmass.tex
\subsection{Charm-quark mass}
\label{s:cmass}

In the present review we collect and discuss for the first time the lattice determinations of the $\overline{\rm MS}$ charm-quark mass $\overline{m}_c$.
Most of the results have been obtained by analyzing the lattice-QCD simulations of 2-point heavy-light- or 
heavy-heavy-meson correlation functions, using as input the experimental values of the $D$, $D_s$ and charmonium mesons.
The exceptions are represented by the HPQCD 14A~\cite{Chakraborty:2014aca} result at $N_f = 2+1+1$, the HPQCD 08B~\cite{Allison:2008xk}, HPQCD 10~\cite{McNeile:2010ji} and JLQCD 15B~\cite{Nakayama:2015hrn} results at $N_f = 2 +1$, and the ETM 11F~\cite{Jansen:2011vr} result at $N_f = 2$, where the moments method has been employed.
The latter is based on the lattice calculation of the Euclidean time moments of pseudoscalar-pseudoscalar correlators for heavy-quark currents followed by an OPE expansion dominated by perturbative QCD effects, which provides the determination of both the heavy-quark mass and the strong coupling constant $\alpha_s$.

The heavy-quark actions adopted by the various lattice collaborations have been reviewed already in the FLAG 13 review~\cite{Aoki:2013ldr}, and their descriptions can be found in Sec.~\ref{app:HQactions}.
While the charm mass determined with the moments method does not need any lattice evaluation of the mass renormalization constant $Z_m$, the extraction of $\overline{m}_c$  from 2-point heavy-meson correlators does require the nonperturbative calculation of $Z_m$.
The lattice scale at which $Z_m$ is obtained, is usually at least of the order $2 - 3$ GeV, and therefore it is natural in this review to provide the values of $\overline{m}_c(\mu)$ at the renormalization scale $\mu = 3~\gev$.
Since the choice of a renormalization scale equal to $\overline{m}_c$ is still commonly adopted (as by PDG~\cite{Agashe:2014kda}), we have collected in Tab.~\ref{tab:mc} the lattice results for both $\overline{m}_c(\overline{m}_c)$ and $\overline{m}_c(\mbox{3 GeV})$, obtained at $N_f = 2$, $2+1$ and $2+1+1$.
When not directly available in the publications, we apply a conversion factor equal either to $0.900$ between the scales $\mu = 2$ GeV and $\mu = 3$ GeV or to $0.766$ between the scales $\mu = \overline{m}_c$ and $\mu = 3$ GeV, obtained using perturbative QCD evolution at four loops assuming $\Lambda_{QCD} = 300$ MeV for $N_f = 4$.

\input{qmass/qmass_tab_mc}

In the next subsections we review separately the results of $\overline{m}_c(\overline{m}_c)$ for the various values of $N_f$.

\subsubsection{$N_f = 2+1+1$ results}
\label{sec:mcnf4}

There are three recent results employing four dynamical quarks in the sea.
ETM 14~\cite{Carrasco:2014cwa} uses 15 twisted mass gauge ensembles at 3 lattice spacings ranging from 0.062 to 0.089 fm (using $f_\pi$ as input), in boxes of size ranging from 2.0 to 3.0 fm and pion masses from 210 to 440 MeV (explaining the tag \soso\ in the chiral extrapolation and the tag \good\ for the continuum extrapolation).
The value of $M_\pi L$ at their smallest pion mass is 3.2 with more than two volumes (explaining the tag \soso\ in the finite-volume effects).
They fix the strange mass with the kaon mass and the charm one with that of the $D_s$ and $D$ mesons.

ETM 14A~\cite{Alexandrou:2014sha} uses 10 out of the 15 gauge ensembles adopted in ETM 14 spanning the same range of values for the pion mass and the lattice spacing, but the latter is fixed using the nucleon mass. 
Two lattice volumes with size larger than 2.0 fm are employed.
The physical strange and the charm mass are obtained using the masses of the $\Omega^-$ and $\Lambda_c^+$ baryons, respectively.

HPQCD 14A~\cite{Chakraborty:2014aca} works with the moments method adopting HISQ staggered fermions. 
Their results are based on 9 out of the 21 ensembles carried out by the MILC collaboration~\cite{Bazavov:2014wgs} at 4 values of the coupling $\beta$ corresponding to lattice spacings in the range from 0.057 to 0.153 fm, in boxes of sizes up to 5.8 fm and with taste-Goldstone-pion masses down to 130 MeV and RMS-pion masses down to 173 MeV. 
The strange- and charm-quark masses are fixed using as input the lattice result $M_{\bar ss} = 688.5 (2.2)~\mev$, calculated without including $\bar ss$ annihilation effects, and $M_{\eta_c} = 2.9863(27)~\gev$, obtained from the experimental $\eta_c$ mass after correcting for $\bar cc$ annihilation and e.m.~effects.
All of the selection criteria of Sec.~\ref{sec:Criteria} are satisfied with the tag \good\ \footnote{Note that in Section 9.7.2 different coding criteria are adopted and the HPQCD 14A  paper is tagged differently for the continuum extrapolation.}.

According to our rules on the publication status all the three results can enter the FLAG average at $N_f = 2+1+1$.
The determinations of $\overline{m}_c$ obtained by ETM 14 and 14A agree quite well with each other, but they are not compatible with the HPQCD 14A  result.
Therefore we first combine the two ETM results with a 100$\%$ correlation in the statistical error, yielding $\overline{m}_c(\overline{m}_c) = 1.348 (20) \gev$.
Then  we perform the average with the HPQCD 14A  result, obtaining the final FLAG averages
%FLAGRESULT BEGIN
% TAG      &mc&END
% REFS     &\cite{Carrasco:2014cwa,Chakraborty:2014aca}&END
% UNITS    & '[GeV]'  &END
% NUMRESULTS & 2  &END
% FLAVOURs & 2+1+1 &END
%FLAGRESULT END
%FLAGRESULTFORMULA BEGIN
 \begin{align}
      \label{eq:mcmcnf4} 
&& \overline{m}_c(\overline{m}_c)           & = 1.286 ~ (30) ~ \gev          &&\Refs~\mbox{\cite{Carrasco:2014cwa,Chakraborty:2014aca}}, \\[-3mm]
&\mbox{$N_f = 2+1+1$:}& \nonumber\\[-3mm]
&&  \FLAGAVBEGIN\overline{m}_c(\mbox{3 GeV})& = 0.996 ~ (25)\FLAGAVEND ~ \gev&&\Refs~\mbox{\cite{Carrasco:2014cwa,Chakraborty:2014aca}},
 \end{align}
%FLAGRESULTFORMULA END
where the errors include a quite large value ($3.5$ and $4.4$, respectively) for the stretching factor $\sqrt{\chi^2/\mbox{dof}}$ coming from our rules for the averages discussed in Sec.~\ref{sec:averages}.

\subsubsection{$N_f = 2+1$ results}
\label{sec:mcnf3}

The HPQCD 10~\cite{McNeile:2010ji} result is based on the moments method adopting a subset of $N_f = 2+1$ Asqtad-staggered-fermion ensembles from MILC~\cite{Bazavov:2009bb}, on which HISQ valence fermions are studied. 
The charm mass is fixed from that of the $\eta_c$ meson, $M_{\eta_c} = 2.9852 (34) ~ \gev$ corrected for $\bar cc$ annihilation and e.m.~effects. 
HPQCD 10 replaces the result HPQCD 08B~\cite{Allison:2008xk}, in which Asqtad staggered fermions have been used also for the valence quarks.

$\chi$QCD 14~\cite{Yang:2014sea} uses a mixed-action approach based on overlap fermions for the valence quarks and on domain-wall fermions for the sea quarks.
They adopt six of the gauge ensembles generated by the RBC/UKQCD collaboration~\cite{Aoki:2010dy} at two values of the lattice spacing (0.087 and 0.11 fm) with unitary pion masses in the range from 290 to 420 MeV.
For the valence quarks no light-quark masses are simulated.
At the lightest pion mass $M_\pi \simeq$ 290 MeV, the value of $M_\pi L$ is 4.1, which satisfies the tag \soso\ for the finite-volume effects.
The strange- and charm-quark masses are fixed together with the lattice scale by using the experimental values of the $D_s$, $D_s^*$ and $J/\psi$ meson masses. 

JLQCD 15B~\cite{Nakayama:2015hrn} determines the charm mass through the moments method using M\"obius domain-wall fermions at three values of the lattice spacing, ranging from 0.044 to 0.083 fm.
The lightest pion mass is $\simeq 230$ MeV and the corresponding value of $M_\pi L$ is $\simeq 4.4$.

Thus, according to our rules on the publication status, the FLAG average for the charm-quark mass at $N_f = 2+1$ is obtained by combining the two results HPQCD 10 and $\chi$QCD 14, leading to
%FLAGRESULT BEGIN
% TAG      &mc&END
% REFS     &\cite{McNeile:2010ji,Yang:2014sea}&END
% UNITS    & '[GeV]'  &END
% NUMRESULTS & 2  &END
% FLAVOURs & 2+1 &END
%FLAGRESULT END
%FLAGRESULTFORMULA BEGIN
\begin{align}
      \label{eq:mcmcnf3} 
&& \overline{m}_c(\overline{m}_c)         & = 1.275 ~ (8) ~ \gev          &&\Refs~\mbox{\cite{McNeile:2010ji,Yang:2014sea}}, \\[-3mm]
&\mbox{$N_f = 2+1$:}&\nonumber\\[-3mm]
&&\FLAGAVBEGIN\overline{m}_c(\mbox{3 GeV})& = 0.987 ~ (6)\FLAGAVEND ~ \gev&&\Refs~\mbox{\cite{McNeile:2010ji,Yang:2014sea}},
\end{align}
%FLAGRESULTFORMULA END
where the error on $ \overline{m}_c(\overline{m}_c)$ includes a stretching factor $\sqrt{\chi^2/\mbox{dof}} \simeq 1.4$ as discussed in Sec.~\ref{sec:averages}.

\subsubsection{$N_f = 2$ results}
\label{sec:mcnf2}

We turn now to the three results at $N_f = 2$.

ETM 10B~\cite{Blossier:2010cr} is based on tmQCD simulations at four values of the lattice spacing in the range from 0.05 fm to 0.1 fm, with pion masses as low as 270 MeV at two lattice volumes. 
They fix the strange-quark mass with either $M_K$ or $M_{\bar{s}s}$ and the charm mass using alternatively the $D$, $D_s$ and $\eta_c$ masses.

ETM 11F~\cite{Jansen:2011vr} is based on the same gauge ensemble as ETM 10B, but the moments method is adopted. 

ALPHA 13B uses a subset of the CLS gauge ensembles with $\cO(a)$-improved Wilson fermions generated at two values of the lattice spacing (0.048 fm and 0.065 fm), using the kaon decay constant to fix the scale. 
The pion masses are as low as 190 MeV with the value of $M_\pi L$ equal to $\simeq 4$ at the lightest pion mass (explaining the tag \good\ for finite-volume effects).

According to our rules on the publication status ETM 10B becomes the FLAG average at $N_f = 2$, n-mely
%FLAGRESULT BEGIN
% TAG      &mc&END
% REFS     &\cite{Blossier:2010cr}&END
% UNITS    & '[GeV]'  &END
% NUMRESULTS & 2  &END
% FLAVOURs & 2 &END
%FLAGRESULT END
%FLAGRESULTFORMULA BEGIN
\begin{align}
      \label{eq:mcmcnf2} 
&& \overline{m}_c(\overline{m}_c)           &= 1.28~(4) ~ \gev          &&\Ref~\mbox{\cite{Blossier:2010cr}}, \\[-3mm]
&\mbox{$N_f = 2$:}&\nonumber\\[-3mm]
&& \FLAGAVBEGIN\overline{m}_c(\mbox{3 GeV}) &= 1.03~(4)\FLAGAVEND ~ \gev&&\Ref~\mbox{\cite{Blossier:2010cr}}.
\end{align}
%FLAGRESULTFORMULA END

\begin{figure}[!htb]
\begin{center}
\includegraphics[width=11.5cm]{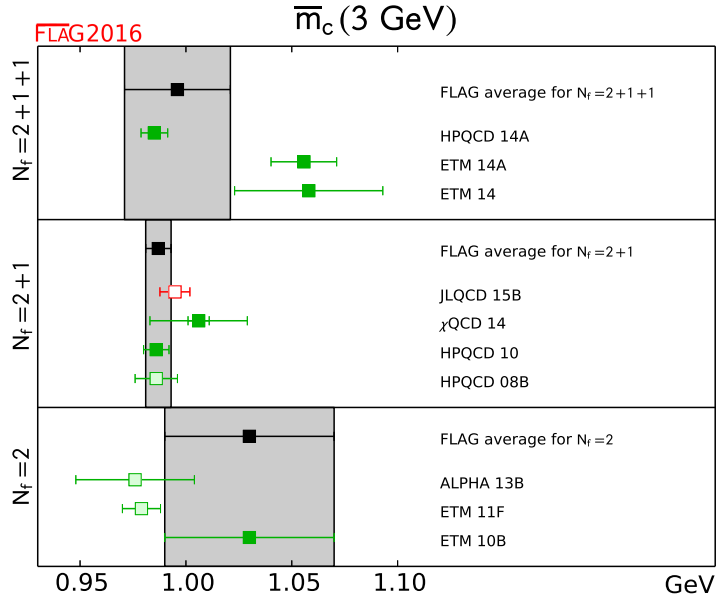}
\end{center}
\vspace{-1cm}
\caption{\label{fig:mc} Lattice results and FLAG averages at $N_f = 2$, $2+1$, and $2+1+1$ for the charm-quark mass $\overline{m}_c(3~\gev)$.}
\end{figure}

In Fig.~\ref{fig:mc} the lattice results of Tab.~\ref{tab:mc} and the FLAG averages obtained at $N_f = 2$, $2+1$ and $2+1+1$ are presented.

\subsubsection{Lattice determinations of the ratio $m_c/m_s$}
\label{sec:mcoverms}

Because some of the results for the light-quark masses given in this review are obtained via the quark-mass ratio $m_c/m_s$, we now review also these lattice calculations, which are listed in Tab.~\ref{tab:mcoverms}.

\input{qmass/qmass_tab_mcoverms}

We begin with the $N_f = 2$ results.
Besides the result ETM 10B, already discussed in Sec.~\ref{sec:mcnf2}, there are two more results, D\"urr 11~\cite{Durr:2011ed} and ETM 14D~\cite{Abdel-Rehim:2014nka}.
D\"urr 11~\cite{Durr:2011ed} is based on QCDSF $N_f = 2$ $\cO(a)$-improved Wilson-fermion simulations~\cite{Gockeler:2006jt,Bietenholz:2010az} on which valence, Brillouin-improved Wilson quarks~\cite{Durr:2010ch} are considered.
It features only 2 ensembles with $M_\pi < 400~\mev$. 
The bare axial-Ward-identity (AWI) masses for $m_s$ and $m_c$ are tuned to simultaneously reproduce the physical values of $M_{\bar ss}^2/(M_{D_s^*}^2-M_{D_s}^2)$ and $(2M_{D_s^*}^2-M_{\bar ss}^2)/(M_{D_s^*}^2-M_{D_s}^2)$, where $M_{\bar ss}^2 = 685.8 (8)$ MeV is the quark-connected-$\bar ss$ pseudoscalar mass.

The ETM 14D result~\cite{Abdel-Rehim:2014nka} is based on recent ETM gauge ensembles generated close to the physical point with the addition of a clover term to the tmQCD action.
The new simulations are performed at a single lattice spacing of $\simeq 0.09$ fm and at a single box size $L \simeq 4$ fm and therefore their calculations do not pass our criteria for the continuum extrapolation and finite-volume effects.
The FLAG average at $N_f = 2$ can be therefore obtained by averaging ETM 10B and D\"urr 11, obtaining
%FLAGRESULT BEGIN
% TAG      &mcoms&END
% REFS     &\cite{Blossier:2010cr}&END
% UNITS    & 1  &END
% NUMRESULTS & 2  &END
% FLAVOURs & 2 &END
%FLAGRESULT END
%FLAGRESULTFORMULA BEGIN
 \be
      \label{eq:mcmsnf2} 
      \mbox{$N_f = 2$:} \qquad \FLAGAVBEGIN m_c / m_s = 11.74 ~ (35)\FLAGAVEND\qquad\Ref~\mbox{\cite{Blossier:2010cr,Durr:2011ed}}, 
 \ee
%FLAGRESULTFORMULA END
where the error includes the stretching factor $\sqrt{\chi^2/\mbox{dof}} \simeq 1.5$.

The situation is similar also for the $N_f = 2+1$ results, as besides $\chi$QCD 14 there is only the result HPQCD 09A~\cite{Davies:2009ih}. 
The latter is based on a subset of $N_f  = 2+1$ Asqtad-staggered-fermion simulations from MILC, on which HISQ-valence fermions are studied.
The strange mass is fixed with $M_{\bar ss} = 685.8(4.0),\mev$ and the charm's from that of the $\eta_c$, $M_{\eta_c} = 2.9852(34)~\gev$ corrected for $\bar cc$
annihilation and e.m.~effects.
By combing the results $\chi$QCD 14 and HPQCD 09A  we obtain
%FLAGRESULT BEGIN
% TAG      &mcoms&END
% REFS     &\cite{Yang:2014sea,Davies:2009ih}&END
% UNITS    & 1  &END
% NUMRESULTS & 2  &END
% FLAVOURs & 2+1 &END
%FLAGRESULT END
%FLAGRESULTFORMULA BEGIN
 \be
      \label{eq:mcmsnf3} 
      \mbox{$N_f = 2+1$:} \qquad\FLAGAVBEGIN m_c / m_s = 11.82 ~ (16)\FLAGAVEND\qquad\Refs~\mbox{\cite{Yang:2014sea,Davies:2009ih}},
 \ee
%FLAGRESULTFORMULA END
with a $\chi^2/\mbox{dof} \simeq 0.85$.

Turning now to the $N_f = 2+1+1$ results, in addition to the HPQCD 14A  and ETM 14 calculations, already described in Sec.~\ref{sec:mcnf4}, we consider the recent FNAL/MILC 14 result~\cite{Bazavov:2014wgs}, where HISQ staggered fermions are employed.
Their result is based on the use of 21 gauge ensembles at 4 values of the coupling $\beta$ corresponding to lattice spacings in the range from 0.057 to 0.153 fm, in boxes of sizes up to 5.8 fm and with taste-Goldstone-pion masses down to 130 MeV and RMS-pion masses down to 143 MeV.
They fix the strange mass with $M_{\bar ss}$, corrected for e.m.~effects with $\bar\epsilon = 0.84(20)$~\cite{Basak:2014vca}. 
The charm mass is fixed with the mass of the $D_s$ meson.
As for the HPQCD 14A  result, all of our selection criteria are satisfied with the tag \good\ .
However a slight tension exists between the two results. 
Indeed by combining HPQCD 14A  and FNAL/MILC 14 results, assuming a 100 $\%$ correlation between the statistical errors (since the two results share the same gauge configurations), we obtain $m_c / m_s = 11.71 (6)$, where the error includes the stretching factor $\sqrt{\chi^2/\mbox{dof}} \simeq 1.35$.
A further average with the ETM 14A result leads to our final average
%FLAGRESULT BEGIN
% TAG      &mcoms&END
% REFS     &\cite{Chakraborty:2014aca,Carrasco:2014cwa,Bazavov:2014wgs}&END
% UNITS    & 1  &END
% NUMRESULTS & 3  &END
% FLAVOURs & 2+1+1 &END
%FLAGRESULT END
%FLAGRESULTFORMULA BEGIN
 \be
      \label{eq:mcmsnf4} 
      \mbox{$N_f = 2+1+1$:} \qquad \FLAGAVBEGIN m_c / m_s = 11.70 ~ (6)\FLAGAVEND\qquad\Refs~\mbox{\cite{Chakraborty:2014aca,Carrasco:2014cwa,Bazavov:2014wgs}},
 \ee
%FLAGRESULTFORMULA END
which has a remarkable overall precision of 0.5 $\%$.

\begin{figure}[!htb]
\begin{center}
\psfig{file=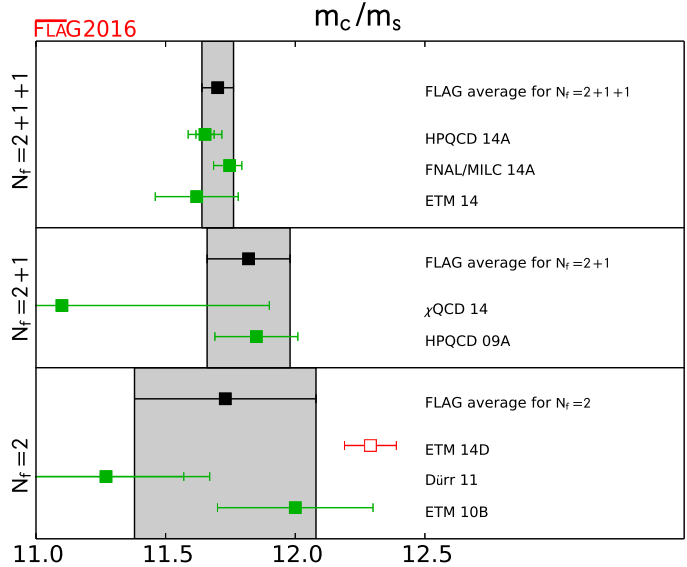,width=11cm}
\end{center}
\begin{center}
\caption{ \label{fig:mcoverms} Lattice results for the ratio $m_c / m_s$ listed in Tab.~\ref{tab:mcoverms} and the FLAG averages corresponding to $N_f =2$, $2+1$ and $2+1+1$.}
\end{center}
\end{figure}

All of the results for $m_c/m_s$ discussed above are shown in Fig.~\ref{fig:mcoverms} together with the FLAG averages corresponding to $N_f =2$, $2+1$ and $2+1+1$. 

%% file: qmass/qmass_tab_mc.tex
%%%%%%%%%%%%%%%%%%%%%%%%%%%%%%%%%%%%%%%%%%%%%%%%%%%%%%%%%%%%%%%%%%%%%%%%%
\begin{table}[!htb]
\vspace{3cm}
{\footnotesize{
\begin{tabular*}{\textwidth}[l]{l@{\extracolsep{\fill}}rllllllll}
Collaboration & Ref. & $N_f$ & \hspace{0.15cm}\begin{rotate}{60}{publication status}\end{rotate}\hspace{-0.15cm} &
 \hspace{0.15cm}\begin{rotate}{60}{chiral extrapolation}\end{rotate}\hspace{-0.15cm} &
 \hspace{0.15cm}\begin{rotate}{60}{continuum  extrapolation}\end{rotate}\hspace{-0.15cm} &
 \hspace{0.15cm}\begin{rotate}{60}{finite volume}\end{rotate}\hspace{-0.15cm} &  
 \hspace{0.15cm}\begin{rotate}{60}{renormalization}\end{rotate}\hspace{-0.15cm} & 
  \rule{0.5cm}{0cm}$\overline{m}_c(\overline{m}_c)$ & 
  \rule{0.3cm}{0cm}$\overline{m}_c(\mbox{3 GeV})$ \\
&&&&&&&&& \\[-0.1cm]
\hline
\hline
%%%%%%%%%%%%%%%%%%%
&&&&&&&&& \\[-0.1cm]
HPQCD 14A  & \cite{Chakraborty:2014aca} & 2+1+1 & \gA & \good & \good & \good & $-$ & 1.2715(95) & 0.9851(63) \\ 
ETM 14A & \cite{Alexandrou:2014sha} & 2+1+1 & \gA & \soso & \good & \soso & \good & 1.3478(27)(195) & 1.0557(22)(153) \\ 
ETM 14 & \cite{Carrasco:2014cwa} & 2+1+1 & \gA & \soso & \good & \soso & \good & 1.348(46) &1.058(35) \\ 
&&&&&&&&& \\[-0.1cm]
%%%%%%%%%%%%%%%%%%%
\hline
&&&&&&&&& \\[-0.1cm]
JLQCD 15B  & \cite{Nakayama:2015hrn} & 2+1 & \rC & \soso & \good & \good & $-$ & 1.2769(21)(89) & 0.9948(16)(69) \\
$\chi$QCD 14 & \cite{Yang:2014sea} & 2+1 & \gA& \soso & \soso & \soso & \good & 1.304(5)(20) & 1.006(5)(22) \\                  
HPQCD 10  & \cite{McNeile:2010ji} & 2+1 & \gA & \soso & \good & \soso  & $-$ & 1.273(6) & 0.986(6) \\
HPQCD 08B & \cite{Allison:2008xk} & 2+1 & \gA &  \soso & \good & \soso & $-$ & 1.268(9) & 0.986(10) \\
&&&&&&&&& \\[-0.1cm]
%%%%%%%%%%%%%%%%%%%
\hline
&&&&&&&&& \\[-0.1cm]
ALPHA 13B & \cite{Heitger:2013oaa} & 2 & \rC & \good & $\soso$ & \good & \good & 1.274(36) & 0.976(28) \\
ETM 11F & \cite{Jansen:2011vr} & 2 & \rC &  \soso & \good & \soso  & $-$ & 1.279(12)/1.296(18)$^\star$ & 0.979(09)/0.998(14)$^\star$ \\
ETM 10B & \cite{Blossier:2010cr} & 2 & \gA &   \soso  & \good &  \soso & \good & 1.28(4) & 1.03(4) \\
&&&&&&&&& \\[-0.1cm] 
\hline \hline
&&&&&&&&& \\[-0.1cm]
PDG & \cite{Agashe:2014kda} & & & & & & & 1.275(25)& \\[1.0ex]
\hline \hline
&&&&&&&&& \\
\end{tabular*}\\[-0.2cm]
}}
%%%%%%%%%%%%%%%%%%%%%%%%%%%%%%%%%%%%%%%%%%%%%%%%%%%%%%%%%%%%%%%%%%%%%%%%%
\begin{minipage}{\linewidth}
{\footnotesize 
\begin{itemize}
\item[$^\star$] Two results are quoted.
\end{itemize}
}
\end{minipage}
%%%%%%%%%%%%%%%%%%%%%%%%%%%%%%%%%%%%%%%%%%%%%%%%%%%%%%%%%%%%%%%%%%%%%%%%%
\caption{\label{tab:mc} Lattice results for the $\msbar$-charm-quark mass $\overline{m}_c(\overline{m}_c)$ and $\overline{m}_c(\mbox{3 GeV})$ in GeV, together with the colour coding of the calculations used to obtain these. When not directly available in the publications, a conversion factor equal to $0.900$ between the scales $\mu = 2$ GeV and $\mu = 3$ GeV (or equal to $0.766$ between the scales $\mu = \overline{m}_c$ and $\mu = 3$ GeV) has been considered.}
\end{table}

%% file: qmass/qmass_tab_mcoverms.tex
%%%%%%%%%%%%%%%%%%%%%%%%%%%%%%%%%%%%%%%%%%%%%%%%%%%%%%%%%%%%%%%%%%%%%%%%%
\begin{table}[!htb]
\vspace{3cm}
{\footnotesize{
\begin{tabular*}{\textwidth}[l]{l@{\extracolsep{\fill}}rllllll}
Collaboration & Ref. & $\Nf$ & \hspace{0.15cm}\begin{rotate}{60}{publication status}\end{rotate}\hspace{-0.15cm}  &
 \hspace{0.15cm}\begin{rotate}{60}{chiral extrapolation}\end{rotate}\hspace{-0.15cm} &
 \hspace{0.15cm}\begin{rotate}{60}{continuum  extrapolation}\end{rotate}\hspace{-0.15cm}  &
 \hspace{0.15cm}\begin{rotate}{60}{finite volume}\end{rotate}\hspace{-0.15cm}  & \rule{0.1cm}{0cm} 
$m_c/m_s$ \\
&&&&&& \\[-0.1cm]
\hline
\hline
%%%%%%%%%%%%%%%%%%%
&&&&&& \\[-0.1cm]
HPQCD 14A  & \cite{Chakraborty:2014aca} & 2+1+1  & \gA & \good & \good & \good  & 11.652(35)(55) \\
FNAL/MILC 14A & \cite{Bazavov:2014wgs}  & 2+1+1  & \gA & \good & \good & \good  & 11.747(19)($^{+59}_{-43}$) \\
ETM 14 & \cite{Carrasco:2014cwa}  & 2+1+1  & \gA & \soso & \good & \soso & 11.62(16) \\
&&&&&& \\[-0.1cm]  
%%%%%%%%%%%%%%%%%%%
\hline 
&&&&&& \\[-0.1cm]
$\chi$QCD 14 & \cite{Yang:2014sea} & 2+1  & \gA & \soso & \soso & \soso & 11.1(8) \\
HPQCD 09A & \cite{Davies:2009ih}  & 2+1  & \gA & \soso & \good & \good & 11.85(16) \\
&&&&&& \\[-0.1cm]  
%%%%%%%%%%%%%%%%%%%
\hline 
&&&&&& \\[-0.1cm]
ETM 14D &  \cite{Abdel-Rehim:2014nka}  & 2  & \rC & \good & \bad & \bad & 12.29(10)   \\
D\"urr 11 & \cite{Durr:2011ed}  & 2  &  \gA & \soso   & \good & \soso & 11.27(30)(26) \\
ETM 10B  & \cite{Blossier:2010cr}  & 2 & \gA & \soso & \good & \soso & 12.0(3) \\
&&&&&& \\[-0.1cm]
\hline
\hline
\end{tabular*}
}}
\caption{Lattice results for the quark-mass ratio $m_c/m_s$, together with the colour coding of the calculations used to obtain these.}
\label{tab:mcoverms}
\end{table}

%% file: qmass/qmass_mb.tex
\subsection{Bottom-quark mass}
\label{s:bmass}

We now give the lattice results for the $\overline{\rm MS}$-bottom-quark mass $\overline{m}_b$ for the first time as part of this review.
Related heavy-quark actions and observables have been discussed in the FLAG 13 review \cite{Aoki:2013ldr}, and descriptions can be found in Sec.~\ref{app:HQactions}.
In Tab.~\ref{tab:mb} we have collected the lattice results for $\overline{m}_b(\overline{m}_b)$ obtained at $N_f = 2$, $2+1$ and $2+1+1$, which in the following we review separately.
Available results for the quark-mass ratio $m_b / m_c$ are also reported.
Afterwards we evaluate the corresponding FLAG averages.

\input{qmass/qmass_tab_mb}

\subsubsection{$N_f=2+1+1$}
Results have been published by HPQCD using NRQCD and HISQ-quark actions  (HPQCD 14B ~\cite{Colquhoun:2014ica} and HPQCD 14A~\cite{Chakraborty:2014aca}, respectively).  In both works the $b$-quark mass is computed with the moments method, that is, from Euclidean-time moments of two-point, heavy-heavy meson correlation functions (see Sec.~\ref{s:curr} for a description of the method).

In HPQCD 14B  the $b$-quark mass is computed from ratios of the moments $R_n$ of heavy current-current correlation functions, namely
 \be
      \left[\frac{R_n r_{n-2}}{R_{n-2}r_n}\right]^{1/2} \frac{\bar{M}_{\rm kin}}{2 m_b} = \frac{\bar{M}_{\Upsilon,\eta_b}}{2 \bar m_b(\mu)} ~ ,
      \label{eq:moments}
 \ee
 where $r_n$ are the perturbative moments calculated at N$^3$LO, $\bar{M}_{\rm kin}$ is the spin-averaged kinetic mass of the heavy-heavy vector and pseudoscalar mesons and $\bar{M}_{\Upsilon,\eta_b}$ is the experimental spin average of the $\Upsilon$ and $\eta_b$ masses. 
The kinetic mass $\bar{M}_{\rm kin}$ is chosen since in the lattice calculation the splitting of the $\Upsilon$ and $\eta_b$ states is inverted. 
In Eq.~(\ref{eq:moments}) the bare mass $m_b$ appearing on the left hand side is tuned so that the spin-averaged mass agrees with experiment, while the mass $\overline{m}_b$ at the fixed scale $\mu = 4.18$ GeV is extrapolated to the continuum limit using three HISC (MILC) ensembles with $a \approx$ 0.15, 0.12 and 0.09 fm and two pion masses, one of which is the physical one. Therefore according to our rules on the chiral extrapolation a warning must be given. Their final result is $\overline{m}_b(\mu = 4.18 \gev) = 4.207(26)$ GeV, where the error is from adding systematic uncertainties in quadrature only (statistical errors are smaller than $0.1 \%$ and ignored). The errors arise from renormalization, perturbation theory, lattice spacing, and NRQCD systematics. The finite-volume uncertainty is not estimated, but at the lowest pion mass they have $ m_\pi L \simeq 4$, which leads to the tag \good\ .

In HPQCD 14A  the quark mass is computed using a similar strategy as above but with HISQ heavy quarks instead of NRQCD. The gauge-field ensembles are the same as in HPQCD 14B  above plus the one with $a = 0.06$ fm (four lattice spacings in all).  Bare heavy-quark masses are tuned to their physical values using the $\eta_h$ mesons, and ratios of ratios yield $m_h/m_c$. The $\overline{\rm MS}$-charm-quark mass determined as described in Sec.~\ref{s:cmass} then gives $m_b$. The moment ratios are expanded using the OPE, and the quark masses and $\alpha_S$ are determined from fits of the lattice ratios to this expansion. The fits are complicated: HPQCD uses cubic splines for valence- and sea-mass dependence, with several knots, and many priors for 21 ratios to fit 29 data points. Taking this fit at face value results in a $\good$ rating for the continuum limit since they use four lattice spacings down to 0.06 fm. See however the detailed discussion of the continuum limit given in Sec.~\ref{s:curr} on $\alpha_S$.

The third four-flavour result is from the ETM Collaboration and appears in a conference proceedings, so it is not included in our final average. The calculation is performed on a set of configurations generated with twisted Wilson fermions with three lattice spacings in the range 0.06 to 0.09 fm and with pion masses in the range 210 to 440 MeV. The $b$-quark mass is determined from a ratio of heavy-light pseudoscalar meson masses designed to yield the quark pole mass in the static limit. The pole mass is related to the $\overline{\rm MS}$ mass through perturbation theory at N$^3$LO. The key idea is that by taking ratios of ratios, the $b$-quark mass is accessible through fits to 
heavy-light(strange)-meson correlation functions computed on the lattice in the range $\sim 1-2\times m_c$ and the static limit, the latter being exactly 1. By simulating below $\overline{m}_b$, taking the continuum limit is easier. They find $\overline{m}_b(\overline{m}_b) = 4.26(7)(14)$ GeV, where the first error is statistical and the second systematic. The dominant errors come from setting the lattice scale and fit systematics.

\subsubsection{$N_f=2+1$}

HPQCD 13B ~\cite{Lee:2013mla} extracts $\overline{m}_b$ from a lattice determination of the $\Upsilon$ energy in NRQCD and the experimental value of the meson mass. The latter quantities yield the pole mass which is related to the $\overline{\rm MS}$ mass in 3-loop perturbation theory. The MILC coarse (0.12 fm) and fine (0.09 fm) Asqtad-2+1-flavour ensembles are employed in the calculation. The bare light-(sea)-quark masses correspond to a single, relatively heavy, pion mass of about 300 MeV. No estimate of the finite-volume error is given.

The value of $\overline{m}_b(\overline{m}_b)$ reported in HPQCD 10 \cite{McNeile:2010ji} is computed in a very similar fashion to the one in HPQCD 14A described in the last section, except that MILC 2+1-flavour-Asqtad ensembles are used under HISQ-heavy-valence quarks. The lattice spacings of the ensembles range from 0.18 to 0.045 fm and pion masses down to about 165 MeV. In all, 22 ensembles were fit simultaneously. An estimate of the finite-volume error based on leading-order perturbation theory for the moment ratio is also provided. Details of perturbation theory and renormalization systematics are given in Sec.~\ref{s:curr}.

\subsubsection{$N_f=2$}

The ETM Collaboration computes $\overline{m}_b(\overline{m}_b)$ using the ratio method described above on two-flavour twisted-mass gauge ensembles with four values of the lattice spacing in the range 0.10 to 0.05 fm and pion masses between 280 and 500 MeV (ETM 13B updates ETM 11). The heavy-quark masses cover a range from charm to a little more than three GeV, plus the exact static-limit point. They find $\overline{m}_b(\overline{m}_b) = 4.31(9)(8)$ GeV for two-flavour running, while $\overline{m}_b(\overline{m}_b) = 4.27(9)(8)$ using four-flavour running, from the 3 GeV scale used in the N$^3$LO perturbative matching calculation from the pole mass to the $\overline{\rm MS}$ mass. The latter are computed nonperturbatively in the RI-MOM scheme at 3 GeV and matched to $\overline{\rm MS}$. The dominant errors are combined statistical+fit(continuum+chiral limits) and the uncertainty in setting the lattice scale. ETM quotes the average of two- and five-flavour results, $\overline{m}_b(\overline{m}_b) = 4.29(9)(8)(2)$ where the last error is one-half the difference between the two. In our average (see below), we use the two-flavour result.

The Alpha Collaboration uses HQET for heavy-light mesons to obtain $m_b$~\cite{Bernardoni:2013xba} (ALPHA 13C). They employ CLS, nonperturbatively improved, Wilson gauge field ensembles with three lattice spacings (0.075-0.048 fm), pion masses from 190 to 440 MeV, and three or four volumes at each lattice spacing, with $m_\pi L > 4.0$. The bare-quark mass is related to the RGI-scheme mass using the Schr\"odinger Functional technique with conversion to $\overline{\rm MS}$ through four-loop anomalous dimensions for the mass. The final result, extrapolated to the continuum and chiral limits, is $\overline{m}_b(\overline{m}_b) = 4.21(11)$ with two-flavour running, where the error combines statistical and systematic uncertainties. The value includes all corrections in HQET through $\Lambda^2/m_b$, but repeating the calculation in the static limit yields the identical result, indicating the HQET expansion is under very good control.

\subsubsection{Averages for $\overline{m}_b(\overline{m}_b)$}

Taking the results that meet our rating criteria, $\soso$, or better, we compute the averages from HPQCD 14A  and 14B  for $N_f = 2+1+1$, ETM 13B and ALPHA 13C for $N_f = 2$, and we take HPQCD 10 as estimate for $N_f = 2+1$, obtaining
%FLAGRESULT BEGIN
% TAG      &mb&mb&mb &END
% REFS     &\cite{Chakraborty:2014aca,Colquhoun:2014ica}&\cite{McNeile:2010ji}&\cite{Carrasco:2013zta,Bernardoni:2013xba}&END
% UNITS    & '[MeV]' & '[MeV]' & '[GeV]'  &END
% NUMRESULTS & 2& 1& 2  &END
% FLAVOURs & 2+1+1& 2+1& 2  &END
%FLAGRESULT END
%FLAGRESULTFORMULA BEGIN
\begin{align}
&N_f= 2+1+1 :&\FLAGAVBEGIN\overline{m}_b(\overline{m}_b)& = 4.190 (21)\FLAGAVEND&&\Refs~\mbox{\cite{Chakraborty:2014aca,Colquhoun:2014ica}},\\ 
&N_f= 2+1 :  &\FLAGAVBEGIN\overline{m}_b(\overline{m}_b)& = 4.164 (23) \FLAGAVEND&&\Ref ~\mbox{\cite{McNeile:2010ji}}, \\  
&N_f= 2 :    &\FLAGAVBEGIN\overline{m}_b(\overline{m}_b)& = 4.256 (81)\FLAGAVEND &&\Refs~\mbox{\cite{Carrasco:2013zta,Bernardoni:2013xba}} .
\end{align}
%FLAGRESULTFORMULA END
Since HPQCD quotes $\overline{m}_b(\overline{m}_b)$ values using $N_f = 5$ running, we used those values directly in these $N_f=2+1+1$ and 2+1 averages.
The results ETM 13B and ALPHA 13C, entering the average at $N_f = 2$, correspond to the $N_f =2 $ running.

\begin{figure}[!htb]
\begin{center}
\psfig{file=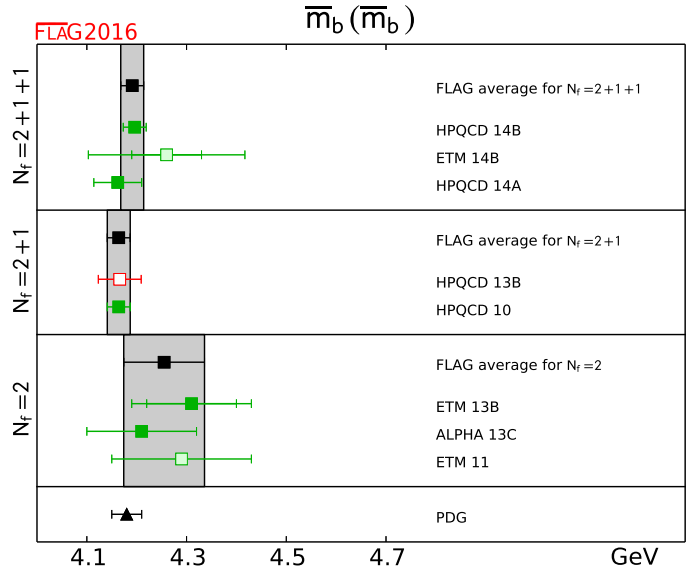,width=11cm}
\end{center}
\vspace{-1cm}
\caption{ \label{fig:mb}
Lattice results and FLAG averages at $N_f = 2$, $2+1$, and $2+1+1$ for the $b$-quark mass $\overline{m}_b(\overline{m}_b)$. The updated PDG value from Ref.~\cite{Agashe:2014kda} is reported for comparison.}
\end{figure}

All the results for $\overline{m}_b(\overline{m}_b)$ discussed above are shown in Fig.~\ref{fig:mb} together with the FLAG averages corresponding to $N_f = 2$, $2+1$ and $2+1+1$. 

%% file: qmass/qmass_tab_mb.tex
%%%%%%%%%%%%%%%%%%%%%%%%%%%%%%%%%%%%%%%%%%%%%%%%%%%%%%%%%%%%%%%%%%%%%%%%%
\begin{table}[!htb]
\vspace{3cm}
{\footnotesize{
\begin{tabular*}{\textwidth}[l]{l@{\extracolsep{\fill}}rlllllllll}
Collaboration & Ref. & $N_f$ & \hspace{0.15cm}\begin{rotate}{60}{publication status}\end{rotate}\hspace{-0.15cm} &
 \hspace{0.15cm}\begin{rotate}{60}{chiral extrapolation}\end{rotate}\hspace{-0.15cm} &
 \hspace{0.15cm}\begin{rotate}{60}{continuum extrapolation}\end{rotate}\hspace{-0.15cm} &
 \hspace{0.15cm}\begin{rotate}{60}{finite volume}\end{rotate}\hspace{-0.15cm} &  
 \hspace{0.15cm}\begin{rotate}{60}{renormalization}\end{rotate}\hspace{-0.15cm} &  
 \hspace{0.15cm}\begin{rotate}{60}{heavy quark treatment}\end{rotate}\hspace{-0.15cm} & 
 \rule{0.2cm}{0cm}$\overline{m}_b(\overline{m}_b)$ & 
 \rule{0.2cm}{0cm}$m_b / m_c$ \\
&&&&&&&&&& \\[-0.1cm]
\hline
\hline
&&&&&&&&&& \\[-0.1cm]
HPQCD 14B  & \cite{Colquhoun:2014ica} & 2+1+1 & \gA &\good & \good & \good & \good & \okay & 4.196(23)$^\dagger$ & \\
ETM 14B & \cite{Bussone:2014cha} & 2+1+1 & \rC &\soso &\good & \soso &\good & \okay & 4.26(7)(14) & 4.40(6)(5) \\ 
HPQCD 14A  & \cite{Chakraborty:2014aca} & 2+1+1 & \gA & \good &\good & \good & $-$ & \okay & 4.162(48) & 4.528(14)(52) \\ 
&&&&&&&&&& \\[-0.1cm]
\hline
&&&&&&&&&& \\[-0.1cm]
HPQCD 13B  & \cite{Lee:2013mla} & 2+1 & \gA &\bad &\soso & $-$ & $-$ & \okay & 4.166(43) & \\ 
HPQCD 10 & \cite{McNeile:2010ji} & 2+1 & \gA &\good & \good & \good& $-$ & \okay & 4.164(23)$^\star$ & 4.51(4) \\ 
&&&&&&&&&& \\[-0.1cm]
\hline
&&&&&&&&&& \\[-0.1cm]
ETM 13B & \cite{Carrasco:2013zta} & 2 & \gA & \soso & \good & \soso & \good &\okay & 4.31(9)(8) & \\
ALPHA 13C & \cite{Bernardoni:2013xba} & 2 & \gA & \good & \good & \good & \good & \okay & 4.21(11) & \\
ETM 11A & \cite{Dimopoulos:2011gx} & 2 & \gA & \soso & \good & \soso & \good & \okay & 4.29(14) & \\[1.0ex]
\hline \hline
&&&&&&&&&& \\[-0.1cm]
PDG & \cite{Agashe:2014kda} & & & & & & & & 4.18(3) & \\[1.0ex]
\hline \hline
&&&&&&&&&& \\
\end{tabular*}\\[-0.2cm]
}}
\begin{minipage}{\linewidth}
{\footnotesize 
\begin{itemize}
\item[$^\dagger$] Warning: only two pion points are used for chiral extrapolation. \\[-5mm]
\item[$^{\star}$] The number that is given is $m_b(10~\gev, N_f = 5) = 3.617(25)~\gev$.
\end{itemize}
}
\end{minipage}
\caption{\label{tab:mb} Lattice results for the $\msbar$-bottom-quark mass $\overline{m}_b(\overline{m}_b)$ in GeV,
  together with the systematic error ratings for each. Available results for the quark mass ratio $m_b / m_c$ are also reported.}
\end{table}

%% file: Vudus/VudVus.tex
\section{Leptonic and semileptonic kaon and pion decay and $|V_{ud}|$ and $|V_{us}|$}\label{sec:vusvud}
This section summarizes state-of-the-art lattice calculations of the leptonic kaon and pion decay constants and the kaon semileptonic-decay form factor and provides an analysis in view of the Standard Model.
With respect to the previous edition of the FLAG review \cite{Aoki:2013ldr} the data in this section has been updated.
As in Ref.~\cite{Aoki:2013ldr}, when combining lattice data with experimental results, we take into account the strong $SU(2)$ isospin correction, either obtained in lattice calculations or estimated by using chiral perturbation theory, both for the kaon leptonic decay constant $f_{K^\pm}$ and for the ratio $f_{K^\pm} / f_{\pi^\pm}$.

%%%%%%%%%%%%%%%%%%%%%%%%%%%%%%%%%%%%%%%%%%%%%%%%%%%%%%%%%%%%%%%%%%%%%%%%%%%%%%%%
\subsection{Experimental information concerning $|V_{ud}|$, $|V_{us}|$, $f_+(0)$ and $\fKfpichargedr$}\label{sec:Exp} 
%%%%%%%%%%%%%%%%%%%%%%%%%%%%%%%%%%%%%%%%%%%%%%%%%%%%%%%%%%%%%%%%%%%%%%%%%%%%%%%%

The following review relies on the fact that precision 
experimental data on kaon decays
very accurately determine the product $|V_{us}|f_+(0)$ \cite{Moulson:2014cra} and the ratio
$|V_{us}/V_{ud}|f_{K^\pm}/f_{\pi^\pm}$ \cite{Moulson:2014cra,Rosner:2015wva}: 
\be\label{eq:products}
|V_{us}| f_+(0) = 0.2165(4)\co \hspace{1cm} \;
\left|\frac{V_{us}}{V_{ud}}\right|\frac{ f_{K^\pm}}{ f_{\pi^\pm}} \;
=0.2760(4)\fs\ee 
Here and in the following $f_{K^\pm}$ and $f_{\pi^\pm}$ are the isospin-broken 
decay constants, respectively, in QCD
(the
electromagnetic effects have already been subtracted in the experimental
analysis using chiral perturbation theory). We will refer to the decay 
constants in the $SU(2)$ isospin-symmetric limit as $f_K$ and $f_\pi$ 
(the latter at leading order in the mass difference ($m_u - m_d$) coincides with $f_{\pi^\pm}$).
$|V_{ud}|$ and $|V_{us}|$ are
elements of the Cabibbo-Kobayashi-Maskawa matrix and $f_+(t)$ represents
one of the form factors relevant for the semileptonic decay
$K^0\rightarrow\pi^-\ell\,\nu$, which depends on the momentum transfer $t$
between the two mesons.  What matters here is the value at $t=0$:
$f_+(0)\equiv
f_+^{K^0\pi^-}\hspace{-0.1cm}(t)\,\rule[-0.15cm]{0.02cm}{0.5cm}_{\;t\rightarrow
  0}$. The pion and kaon decay constants are defined by\footnote{The pion
  decay constant represents a QCD matrix element -- in the full Standard
  Model, the one-pion state is not a meaningful notion: the correlation
  function of the charged axial current does not have a pole at
  $p^2=M_{\pi^+}^2$, but a branch cut extending from $M_{\pi^+}^2$ to
  $\infty$. The analytic properties of the correlation function and the
  problems encountered in the determination of $f_\pi$ are thoroughly
  discussed in Ref.~\cite{Gasser:2010wz}. The ``experimental'' value of $f_\pi$
  depends on the convention used when splitting the sum ${\cal
    L}_{\mbox{\tiny QCD}}+{\cal L}_{\mbox{\tiny QED}}$ into two parts
  (compare Sec.~\ref{subsec:electromagnetic interaction}).  The lattice
  determinations of $f_\pi$ do not yet reach the accuracy where this is of
  significance, but at the precision claimed by the Particle Data Group
  \cite{Agashe:2014kda,Rosner:2015wva}, the numerical value does depend on the convention used
  \cite{Gasser:2003hk,Rusetsky:2009ic,Gasser:2007de,Gasser:2010wz}. }  \bdm
\lvac \dbar\gamma_\mu\gamma_5 \hspace{0.05cm}u|\pi^+(p)\rangle=i
\hspace{0.05cm}p_\mu f_{\pi^+}\co\hspace{1cm} \lvac \sbar\gamma_\mu\gamma_5
\hspace{0.05cm} u|K^+(p)\rangle=i \hspace{0.05cm}p_\mu f_{K^+}\fs\edm In this
normalization, $f_{\pi^\pm} \simeq 130$~MeV, $f_{K^\pm}\simeq 155$~MeV.
 
The measurement of $|V_{ud}|$ based on superallowed nuclear $\beta$
transitions has now become remarkably precise. The result of the 
update of Hardy and Towner \cite{Hardy:2014qxa}, which is based on 20
different superallowed transitions, reads\footnote{It is not a trivial
  matter to perform the data analysis at this precision. In particular,
  isospin-breaking effects need to be properly accounted for
  \cite{Towner:2007np,Miller:2008my,Auerbach:2008ut,Liang:2009pf,Miller:2009cg,Towner:2010bx}.
  For a review of recent work on this issue, we refer to
  Ref.~\cite{Hardy:2014qxa}.}
\be\label{eq:Vud beta}
|V_{ud}| = 0.97417(21)\fs\ee 

The matrix element $|V_{us}|$ can be determined from semiinclusive 
$\tau$ decays
\cite{Gamiz:2002nu,Gamiz:2004ar,Maltman:2008na,Pich_Kass}. Separating the
inclusive decay $\tau\rightarrow \mbox{hadrons}+\nu$ into nonstrange and
strange final states, e.g. HFAG 14~\cite{Amhis:2014hma} obtain
\be\label{eq:Vus tau}|V_{us}|=0.2176(21) \fs \ee Maltman et
al.~\cite{Maltman:2008ib,Maltman:2008na,Maltman:2009bh} and Gamiz et al.~\cite{Gamiz:2007qs,Gamiz:2013wn}
arrive at very similar values.

Inclusive hadronic $\tau$ decay offers an interesting way to measure
$|V_{us}|$, but a number of open issues yet remain to be clarified. In
particular, the value of $|V_{us}|$ as determined from $\tau$ decays
differs from the result one obtains from assuming three-flavour
SM-unitarity by more than three standard deviations~\cite{Amhis:2014hma}. 
It is important to understand this apparent tension better. A possibility 
is that at the current level of precision the treatment of higher orders 
in the operator product expansion and violations of quark-hadron duality 
may play a role. Very recently \cite{Hudspith:2015kfx} a new implementation 
of the relevant sum rules has been elaborated suggesting a much larger 
value of $|V_{us}|$ with respect to the result~(\ref{eq:Vus tau}), namely 
$|V_{us}| = 0.2228 (23)$, which is in much better agreement with CKM 
unitarity. Another possibility is that $\tau$ decay involves new physics, 
but more work both on the theoretical and experimental side is required.

The experimental results in Eq.~(\ref{eq:products}) are for the 
semileptonic decay of a neutral kaon into a negatively charged pion and the
charged pion and kaon leptonic decays, respectively, in QCD. In the case of
the semileptonic decays the corrections for strong
and electromagnetic isospin breaking in chiral perturbation
theory at NLO have allowed for averaging the different experimentally
measured isospin channels~\cite{Antonelli:2010yf}. 
This is quite a convenient procedure as long as lattice QCD does not include
strong or QED isospin-breaking effects. 
Lattice results for $f_K/f_\pi$ are typically quoted for QCD with (squared)
pion and kaon masses of $M_\pi^2=M_{\pi^0}^2$ and $M_K^2=\frac 12
	\left(M_{K^\pm}^2+M_{K^0}^2-M_{\pi^\pm}^2+M_{\pi^0}^2\right)$
for which the leading strong and electromagnetic isospin violations cancel.
While progress
is being made for including strong and electromagnetic isospin breaking in 
the simulations
(e.g. Ref.~\cite{Aoki:2008sm,deDivitiis:2011eh,Ishikawa:2012ix,TakuLat12,deDivitiis:2013xla,Tantalo:2013maa,Portelli:2015gda}),
for now contact to experimental results is made
by correcting leading $SU(2)$ isospin breaking 
guided either by chiral perturbation theory or by lattice calculations.

%%%%%%%%%%%%%%%%%%%%%%%%%%%%%%%%%%%%%%%%%%%%%%%%%%%%%%%%%%%%%%%%%%%%%%%%%%%%%%%%
\subsection{Lattice results for $f_+(0)$ and $f_{K^\pm}/f_{\pi^\pm}$}
%%%%%%%%%%%%%%%%%%%%%%%%%%%%%%%%%%%%%%%%%%%%%%%%%%%%%%%%%%%%%%%%%%%%%%%%%%%%%%%%

\begin{table}[t]
\centering 
\vspace{2.8cm}
{\footnotesize\noindent
\begin{tabular*}{\textwidth}[l]{@{\extracolsep{\fill}}llllllll}
Collaboration & Ref. & $\Nf$ & 
\hspace{0.15cm}\begin{rotate}{60}{publication status}\end{rotate}\hspace{-0.15cm}&
\hspace{0.15cm}\begin{rotate}{60}{chiral extrapolation}\end{rotate}\hspace{-0.15cm}&
\hspace{0.15cm}\begin{rotate}{60}{continuum extrapolation}\end{rotate}\hspace{-0.15cm}&
\hspace{0.15cm}\begin{rotate}{60}{finite-volume errors}\end{rotate}\hspace{-0.15cm}&\rule{0.3cm}{0cm}
$f_+(0)$ \\
&&&&&&& \\[-0.1cm]
\hline
\hline&&&&&&& \\[-0.1cm]
ETM 15C                       &\cite{Carrasco:2015wzu} &2+1+1  &\rC&\soso&\good&\soso& 0.9709(45)(9)\\[-0.5mm]
FNAL/MILC 13E               &\cite{Bazavov:2013maa} &2+1+1  &\gA&\good&\good&\good& {0.9704(24)(22)}\\[-0.5mm]
FNAL/MILC 13C             &\cite{Gamiz:2013xxa} &2+1+1  &\rC&\good&\good&\good& 0.9704(24)(32)\\[-0.5mm]
&&&&&&& \\[-0.1cm]
\hline
&&&&&&& \\[-0.1cm]
RBC/UKQCD 15A              &\cite{Boyle:2015hfa}  &2+1  &\gA&\good&\soso&\soso& {0.9685(34)(14)}\\[-0.5mm]
RBC/UKQCD 13              & \cite{Boyle:2013gsa}  &2+1  &\gA&\good&\soso&\soso& 0.9670(20)($^{+18}_{-46}$)\\[-0.5mm]
FNAL/MILC 12I                 & \cite{Bazavov:2012cd} &2+1  &\gA&\soso&\soso&\tbg& {0.9667(23)(33)}\\[-0.5mm]
JLQCD 12                        & \cite{Kaneko:2012cta} &2+1  &\rC&\soso&\tbr&\tbg& 0.959(6)(5)\\[-0.5mm]
JLQCD 11                        & \cite{Kaneko:2011rp}  &2+1  &\rC&\soso&\tbr&\tbg& 0.964(6)\\[-1.5mm]
RBC/UKQCD 10              & \cite{Boyle:2010bh}   &2+1  &\gA&\soso&\tbr&\tbg& 0.9599(34)($^{+31}_{-47}$)(14)\rule{0cm}{0.4cm}\\ 
RBC/UKQCD 07              & \cite{Boyle:2007qe}   &2+1  &\gA&\soso&\tbr&\tbg& 0.9644(33)(34)(14)\\
&&&&&&& \\[-0.1cm]
\hline
&&&&&&& \\[-0.1cm]
ETM 10D                   & \cite{Lubicz:2010bv}  &2 &\rC&\soso&\tbg&\soso& 0.9544(68)$_{stat}$\\
ETM 09A 	                 & \cite{Lubicz:2009ht}  &2 &\gA&\soso&\soso&\soso& {0.9560(57)(62)}\\	
QCDSF 07	         & \cite{Brommel:2007wn} &2 &\rC&\tbr&\tbr&\tbg& 0.9647(15)$_{stat}$ \\
RBC 06  	                 & \cite{Dawson:2006qc}  &2 &\gA&\tbr&\tbr&\tbg& 0.968(9)(6)\\	
JLQCD 05 	         & \cite{Tsutsui:2005cj} &2 &\rC&\tbr&\tbr&\tbg& 0.967(6), 0.952(6)\\ 
 &&&&&&& \\[-0.1cm]
\hline
\hline
\end{tabular*}}
\caption{Colour code for the data on $f_+(0)$.\hfill}\label{tab:f+(0)}
\end{table}

The traditional way of determining $|V_{us}|$ relies on using estimates for
the value of $f_+(0)$, invoking the Ademollo-Gatto theorem
\cite{Ademollo_Gatto}.  Since this theorem only holds to leading order of
the expansion in powers of $m_u$, $m_d$ and $m_s$, theoretical models are
used to estimate the corrections. Lattice methods have now reached the
stage where quantities like $f_+(0)$ or $f_K/f_\pi$ can be determined to
good accuracy. As a consequence, the uncertainties inherent in the
theoretical estimates for the higher order effects in the value of $f_+(0)$
do not represent a limiting factor any more and we shall therefore not
invoke those estimates. Also, we will use the experimental results based on
nuclear $\beta$ decay and $\tau$ decay exclusively for comparison -- the
main aim of the present review is to assess the information gathered with
lattice methods and to use it for testing the consistency of the SM and its
potential to provide constraints for its extensions.

The database underlying the present review of the semileptonic form factor 
and the ratio of decay constants is
listed in Tabs.~\ref{tab:f+(0)} and \ref{tab:FKFpi}. The properties of the
lattice data play a crucial role for the conclusions to be drawn from these
results: range of $M_\pi$, size of $L M_\pi$, continuum extrapolation,
extrapolation in the quark masses, finite-size effects, etc. The key
features of the various data sets are characterized by means of the 
colour code specified in Sec.~\ref{sec:color-code}. More detailed information
on individual computations are compiled in appendix~\ref{app:VusVud}.

The quantity $f_+(0)$ represents a matrix element of a strangeness-changing
null-plane charge, $f_+(0)=\langle K|Q^{us}|\pi \rangle$. The vector charges obey the
commutation relations of the Lie algebra of $SU(3)$, in particular
$[Q^{us},Q^{su}]=Q^{uu-ss}$. This relation implies the sum rule $\sum_n
|\langle K|Q^{us}|n \rangle|^2-\sum_n |\langle K|Q^{su}|n \rangle|^2=1$. Since the contribution from
the one-pion intermediate state to the first sum is given by $f_+(0)^2$,
the relation amounts to an exact representation for this quantity
\cite{Furlan}: \be \label{eq:Ademollo-Gatto} f_+(0)^2=1-\sum_{n\neq \pi}
|\langle K|Q^{us}|n \rangle|^2+\sum_n |\langle K |Q^{su}|n \rangle|^2\fs\ee While the first sum on the
right extends over nonstrange intermediate states, the second runs over
exotic states with strangeness $\pm 2$ and is expected to be small compared
to the first.

The expansion of $f_+(0)$ in $SU(3)$ chiral perturbation theory
in powers of $m_u$, $m_d$ and $m_s$ starts with
$f_+(0)=1+f_2+f_4+\ldots\,$ \cite{Gasser:1984gg}.  Since all of the low-energy constants occurring in $f_2$ can be expressed in terms of $M_\pi$,
$M_K$, $M_\eta$ and $f_\pi$ \cite{Gasser:1984ux}, the NLO correction is
known. In the language of the sum rule (\ref{eq:Ademollo-Gatto}), $f_2$
stems from nonstrange intermediate states with three mesons. Like all
other nonexotic intermediate states, it lowers the value of $f_+(0)$:
$f_2=-0.023$ when using the experimental value of $f_\pi$ as input.  
The corresponding expressions have also been derived in
quenched or partially quenched (staggered) chiral perturbation theory
\cite{Bernard:2013eya,Bazavov:2012cd}.  At the same order in the $SU(2)$ expansion
\cite{Flynn:2008tg}, $f_+(0)$ is parameterized in terms of $M_\pi$ and two
\textit{a priori} unknown parameters. The latter can be determined from the
dependence of the lattice results on the masses of the quarks.  Note that
any calculation that relies on the {\Ch}PT formula for $f_2$ is subject to
the uncertainties inherent in NLO results: instead of using the physical
value of the pion decay constant $f_\pi$, one may, for instance, work with
the constant $f_0$ that occurs in the effective Lagrangian and represents
the value of $f_\pi$ in the chiral limit. Although trading $f_\pi$ for
$f_0$ in the expression for the NLO term affects the result only at NNLO,
it may make a significant numerical difference in calculations where the
latter are not explicitly accounted for (the lattice results concerning the
value of the ratio $f_\pi/f_0$ are reviewed in Sec.~\ref{sec:SU3results}).

\begin{figure}[ht]
\psfrag{y}{\tiny $\star$}
\hspace{-9mm}\includegraphics[height=6.8cm]{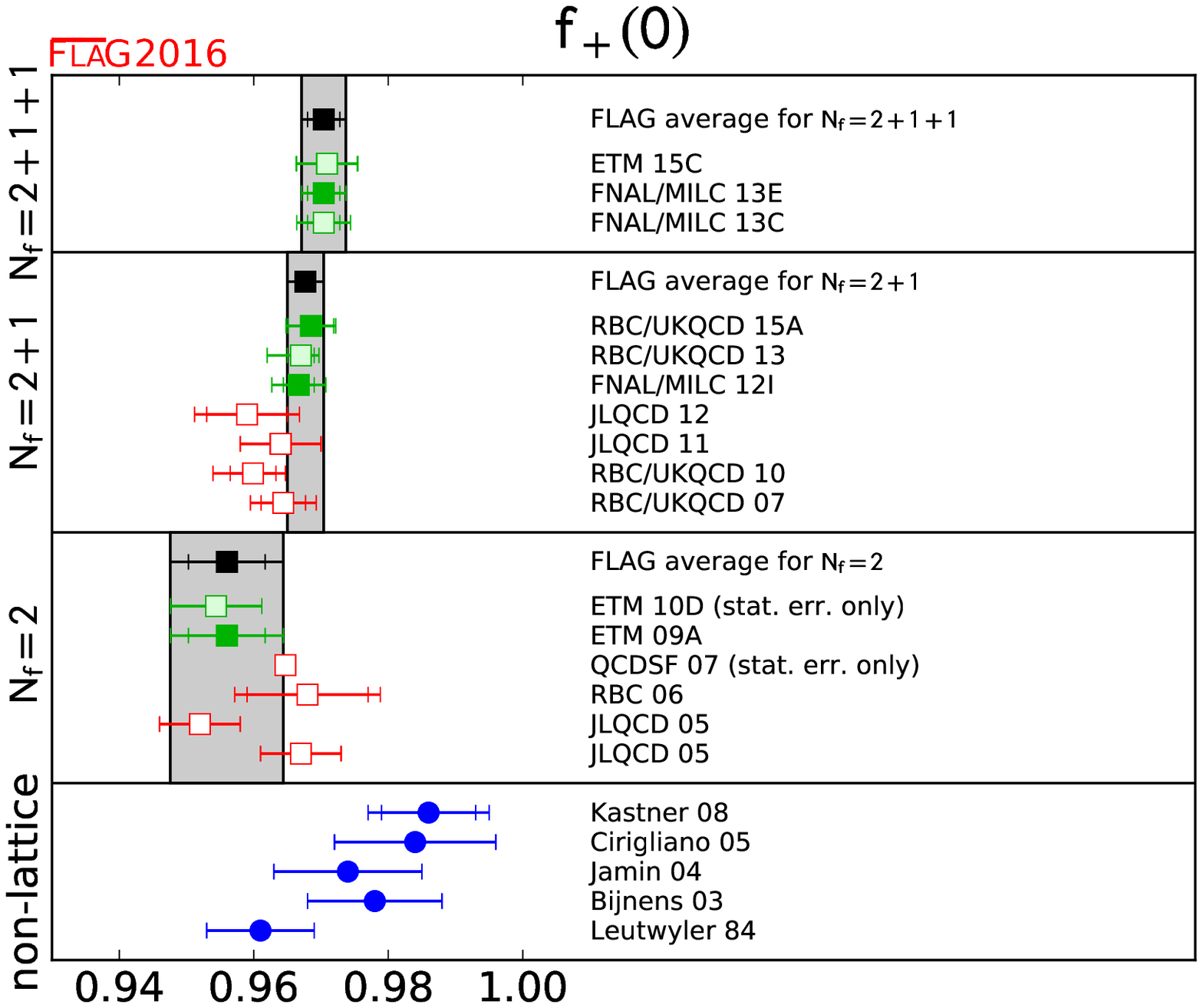}
\hspace{-1cm}
\includegraphics[height=6.8cm]{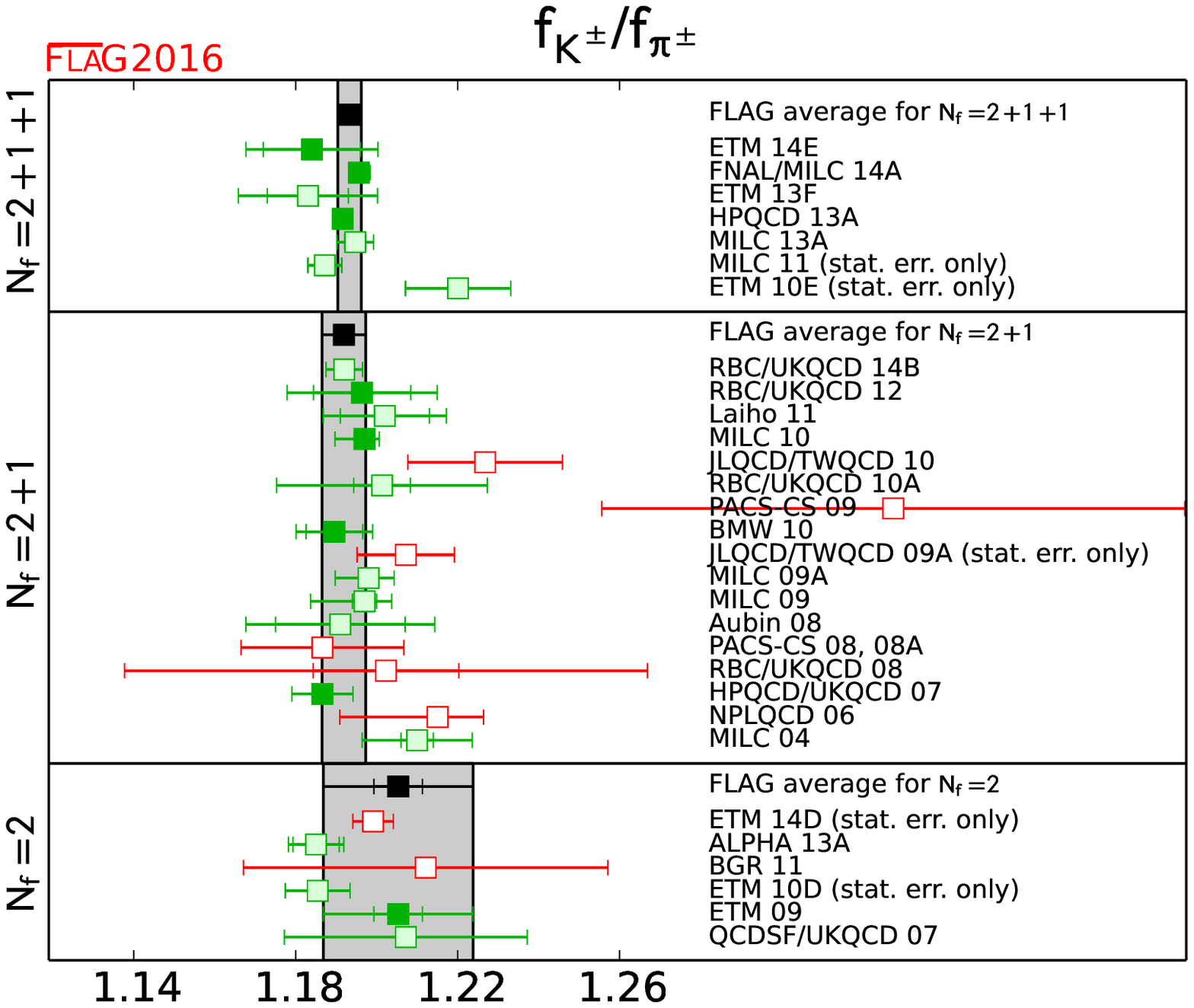}
  
\vspace{-1.79cm}\hspace{5.5cm}\parbox{6cm}{\sffamily\tiny  \cite{Kastner:2008ch}\\

\vspace{-1.29em}\cite{Cirigliano:2005xn}\\

\vspace{-1.29em}\cite{Jamin:2004re}\\

\vspace{-1.29em}\cite{Bijnens:2003uy}\\

\vspace{-1.29em}\cite{Leutwyler:1984je}}
\vspace{0.5cm}

\caption{\label{fig:lattice data}Comparison of lattice results (squares) for $f_+(0)$ and $f_{K^\pm}/ f_{\pi^\pm}$ with various model estimates based on {\Ch}PT (blue circles). The ratio $f_{K^\pm}/f_{\pi^\pm}$ is obtained in pure QCD including the $SU(2)$ isospin-breaking correction (see Sec.~\ref{sec:Direct}). The black squares and grey bands indicate our estimates. The significance of the colours is explained in Sec.~\ref{sec:qualcrit}.}

\end{figure}

The lattice results shown in the left panel of Fig.~\ref{fig:lattice 
data} indicate that the higher order contributions $\Delta f\equiv
f_+(0)-1-f_2$ are negative and thus amplify the effect generated by $f_2$.
This confirms the expectation that the exotic contributions are small. The
entries in the lower part of the left panel represent various model
estimates for $f_4$. In Ref.~\cite{Leutwyler:1984je} the symmetry-breaking
effects are estimated in the framework of the quark model. The more recent
calculations are more sophisticated, as they make use of the known explicit
expression for the $K_{\ell3}$ form factors to NNLO in {\Ch}PT
\cite{Post:2001si,Bijnens:2003uy}. The corresponding formula for $f_4$
accounts for the chiral logarithms occurring at NNLO and is not subject to
the ambiguity mentioned above.\footnote{Fortran programs for the
  numerical evaluation of the form factor representation in
  Ref.~\cite{Bijnens:2003uy} are available on request from Johan Bijnens.} 
 The numerical result, however, depends on
the model used to estimate the low-energy constants occurring in $f_4$
\cite{Bijnens:2003uy,Jamin:2004re,Cirigliano:2005xn,Kastner:2008ch}. The
figure indicates that the most recent numbers obtained in this way
correspond to a positive or an almost vanishing rather than a negative value for $\Delta f$.
We note that FNAL/MILC 12I \cite{Bazavov:2012cd} have made an attempt 
at determining a combination of some of the low-energy constants appearing 
in $f_4$ from lattice data.

%%%%%%%%%%%%%%%%%%%%%%%%%%%%%%%%%%%%%%%%%%%%%%%%%%%%%%%%%%%%%%%%%%%%%%%%%%%%%%%%
\subsection{Direct determination of $f_+(0)$ and $f_{K^\pm}/f_{\pi^\pm}$}\label{sec:Direct} 
%%%%%%%%%%%%%%%%%%%%%%%%%%%%%%%%%%%%%%%%%%%%%%%%%%%%%%%%%%%%%%%%%%%%%%%%%%%%%%%%

All lattice results for the form factor $f_+(0)$ and many available results for the ratio of decay constants, that we summarize here in Tabs.~\ref{tab:f+(0)} and~\ref{tab:FKFpi}, respectively, have been computed in isospin-symmetric QCD. 
The reason for this unphysical parameter choice is that there are only few simulations of $SU(2)$ isospin-breaking effects in lattice QCD, which is ultimately the cleanest way for predicting these effects
~\cite{Duncan:1996xy,Basak:2008na,Blum:2010ym,Portelli:2010yn,deDivitiis:2011eh,deDivitiis:2013xla,Tantalo:2013maa,Portelli:2015gda}. 
In the meantime one relies either on chiral perturbation theory~\cite{Gasser:1984gg,Aubin:2004fs} to estimate the correction to the isospin limit or one calculates the breaking at leading order in $(m_u-m_d)$ in the valence quark sector by extrapolating the lattice data for the charged kaons to the physical value of the $up$($down$)-quark mass (the result for the pion decay constant is always extrapolated to the value of the average light-quark mass $\hat m$).
This defines the prediction for $f_{K^\pm}/f_{\pi^\pm}$.

\begin{table}[!htb]
\centering
\vspace{3.0cm}{\footnotesize\noindent
\begin{tabular*}{\textwidth}[l]{@{\extracolsep{\fill}}lrlllllll}
Collaboration & Ref. & $\Nf$ &
\hspace{0.15cm}\begin{rotate}{60}{publication status}\end{rotate}\hspace{-0.15cm}&
\hspace{0.15cm}\begin{rotate}{60}{chiral extrapolation}\end{rotate}\hspace{-0.15cm}&
\hspace{0.15cm}\begin{rotate}{60}{continuum extrapolation}\end{rotate}\hspace{-0.15cm}&
\hspace{0.15cm}\begin{rotate}{60}{finite-volume errors}\end{rotate}\hspace{-0.15cm}&
\rule{0.2cm}{0cm} $f_K/f_\pi$ &
\rule{0.2cm}{0cm} $f_{K^\pm}/f_{\pi^\pm}$ \\  
&&&&&&& \\[-0.1cm]
\hline
\hline
&&&&&&& \\[-0.1cm]
ETM 14E       &\cite{Carrasco:2014poa}          &2+1+1&\gA&\soso &\good&\soso    		&	1.188(11)(11)   &{1.184(12)(11)} \\
FNAL/MILC 14A &\cite{Bazavov:2014wgs}	     &2+1+1&\gA&\good &\good&\good    		&					      &{1.1956(10)($_{-18}^{+26}$)} \\
ETM 13F       &\cite{Dimopoulos:2013qfa}      &2+1+1&\rC&\soso &\good&\soso    		&	 1.193(13)(10)    	      &1.183(14)(10)	\\
HPQCD 13A       &\cite{Dowdall:2013rya}	     &2+1+1&\gA&\good &\soso&\good    		&	 1.1948(15)(18)&{1.1916(15)(16)} \\
MILC 13A        &\cite{Bazavov:2013cp}	     &2+1+1&\gA&\good &\good&\good    		&					      &1.1947(26)(37) \\
MILC 11        &\cite{Bazavov:2011fh}	             &2+1+1&\rC&\soso &\soso&\soso    		&					      &1.1872(42)$^\dagger_{\rm stat.}$ \\
ETM 10E       &\cite{Farchioni:2010tb}            &2+1+1&\rC&\soso&\soso&\soso		        &       1.224(13)$_{\rm stat}$   &						\\
&&&&&&& \\[-0.1cm]                                                                                                              
\hline                                                                                                                          
&&&&&&& \\[-0.1cm]                                                                                                              
RBC/UKQCD 14B   &\cite{Blum:2014tka}    &2+1&\gA&\good    & \good	 &  \good  	&1.1945(45)					&					\\
RBC/UKQCD 12   &\cite{Arthur:2012opa}    &2+1&\gA&\good    & \soso	 &  \good  	&{1.199(12)(14)}					&					\\
Laiho 11       &\cite{Laiho:2011np}       &2+1&\rC&\soso    & \good   &  \soso  	&                                       	&$1.202(11)(9)(2)(5)$$^{\dagger\dagger}$	\\
MILC 10        &\cite{Bazavov:2010hj}&2+1&\rC&\soso&\good&\good			&                             			&{1.197(2)($^{+3}_{-7}$)}			\\
JLQCD/TWQCD 10 &\cite{Noaki:2010zz}&2+1&\rC&\soso&\tbr&\tbg			&1.230(19)					&                               		\\
RBC/UKQCD 10A  &\cite{Aoki:2010dy}   &2+1&\gA&\soso&\soso&\good			&1.204(7)(25)					&                               		\\
PACS-CS 09     &\cite{Aoki:2009ix}   &2+1&\gA&\good&\tbr&\tbr			&1.333(72)					&                               		\\
BMW 10         &\cite{Durr:2010hr}         &2+1&\gA&\good &\tbg&\tbg			&{1.192(7)(6)}					&                               		\\
JLQCD/TWQCD 09A&\cite{JLQCD:2009sk}  &2+1&\rC&\soso&\tbr&\tbr			&$1.210(12)_{\rm stat}$				&                      				\\
MILC 09A       &\cite{Bazavov:2009fk}&2+1&\rC&\soso&\tbg&\tbg			&                                               &1.198(2)($^{\hspace{0.01cm}+6}_{-8}$)	\\
MILC 09        &\cite{Bazavov:2009bb}&2+1&\gA&\soso&\tbg&\tbg			&                                               &1.197(3)($^{\;+6}_{-13}$)		\\
Aubin 08       &\cite{Aubin:2008ie}  &2+1&\rC&\soso&\soso&\soso			&                                               &1.191(16)(17)					\\
PACS-CS 08, 08A&\cite{Aoki:2008sm, Kuramashi:2008tb} &2+1&\gA&\tbg&\tbr&\tbr	&1.189(20)					&                                               \\
RBC/UKQCD 08   &\cite{Allton:2008pn} &2+1&\gA&\soso&\tbr&\tbg			&1.205(18)(62)					&                                               \\
HPQCD/UKQCD 07 &\cite{Follana:2007uv}&2+1&\gA&\soso&\soso&\soso	&{1.189(2)(7)}		&                                               \\
NPLQCD 06      &\cite{Beane:2006kx}  &2+1&\gA&\soso&\tbr&\tbr			&1.218(2)($^{+11}_{-24}$)			&                                               \\
MILC 04 &\cite{Aubin:2004fs}&2+1&\gA&\soso&\soso&\soso				&						&1.210(4)(13)				\\
&&&&&&& \\[-0.1cm]                                                                                                              
\hline                                                                                                                          
&&&&&&& \\[-0.1cm]                                                                                                              
ETM 14D         &\cite{Abdel-Rehim:2014nka} &2  &\rC&\good&\tbr&\soso			&1.203(5)$_{\rm stat}$		&                                       	\\
ALPHA 13A       &\cite{Lottini:2013rfa}&2  &\rC&\tbg    &\tbg   &\tbg    	&1.1874(57)(30)					&                                       	\\
BGR 11	       &\cite{Engel:2011aa}  &2  &\gA&\soso    &\tbr   &\tbr    		&1.215(41)					&                                       	\\
ETM 10D        &\cite{Lubicz:2010bv} &2  &\rC&\soso&\tbg&\soso			&1.190(8)$_{\rm stat}$ 				&                                       	\\
ETM 09         &\cite{Blossier:2009bx}         &2  &\gA&\soso&\tbg&\soso			&{1.210(6)(15)(9)}				&                                       	\\
QCDSF/UKQCD 07 &\cite{QCDSFUKQCD}    &2  &\rC&\soso&\soso&\tbg			&1.21(3)					&                                       	\\
&&&&&&& \\[-0.1cm]
\hline
\hline
&&&&&&& \\[-0.1cm]
\end{tabular*}}\\[-2mm]
\begin{minipage}{\linewidth}
{\footnotesize 
\begin{itemize}
   \item[$^\dagger$] Result with statistical error only from polynomial interpolation to the physical point.\\[-5mm]
\item[$^{\dagger\dagger}$] This work is the continuation of Aubin 08.
\end{itemize}
}
\end{minipage}
\vspace{-0.3cm}
\caption{Colour code for the data on the ratio of decay constants: $f_K/f_\pi$ is the pure QCD $SU(2)$-symmetric ratio, while $f_{K^\pm}/f_{\pi^\pm}$ is in pure QCD including
the $SU(2)$ isospin-breaking correction.\hfill}
\label{tab:FKFpi}
\end{table}

%%%%%
Since the majority of the collaborations present their newest results including the strong $SU(2)$ isospin-breaking correction (as we will see this comprises the majority of results which qualify for inclusion into the FLAG average), we prefer to provide in Fig.~\ref{fig:lattice data} the overview of the world data of $f_{K^\pm}/f_{\pi^\pm}$, at variance with the choice made in the previous edition of the FLAG review \cite{Aoki:2013ldr}.
For all the results of Tab.~\ref{tab:FKFpi} provided only in the isospin-symmetric limit we apply individually an isospin correction which will be described later on (see equations Eqs.~(\ref{eq:convert}-\ref{eq:iso})).
%%%%%

The plots in Fig.~\ref{fig:lattice data} illustrate our compilation of data for $f_+(0)$ and $f_{K^\pm}/f_{\pi^\pm}$.
The lattice data for the latter quantity are largely consistent even when comparing simulations with different $N_f$, while in the case of $f_+(0)$ a slight tendency to get higher values for increasing $N_f$ seems to be visible, even if it does not exceed one standard deviation.
We now proceed to form the corresponding averages, separately for the data with $\Nf=2+1+1$, $\Nf=2+1$ and $\Nf=2$ dynamical flavours and in the following we will refer to these averages as the ``direct'' determinations. 
 
For $f_+(0)$ there are currently two computational strategies: 
FNAL/MILC uses the Ward identity to relate the $K\to\pi$ form factor at zero momentum transfer to the matrix element $\langle \pi|S|K\rangle$ of the flavour-changing scalar current. 
Peculiarities of the staggered fermion discretization used by FNAL/MILC (see Ref.~\cite{Bazavov:2012cd}) makes this the favoured choice. 
The other collaborations are instead computing the vector current matrix element $\langle \pi |V_\mu|K\rangle$. 
Apart from FNAL/MILC 13C and the recent FNAL/MILC 13E all simulations in Tab.~\ref{tab:f+(0)} involve unphysically heavy quarks and therefore the lattice data needs to be extrapolated to the physical pion and kaon masses corresponding to the $K^0\to\pi^-$ channel. 
We note also that the recent computations of $f_+(0)$ obtained by the FNAL/MILC and RBC/UKQCD collaborations make use of the partially-twisted boundary conditions to determine the form-factor results directly at the relevant kinematical point $q^2=0$ \cite{Guadagnoli:2005be,Boyle:2007wg}, avoiding in this way any uncertainty due to the momentum dependence of the vector and/or scalar form factors. 
The ETM collaboration uses partially-twisted boundary conditions to compare the momentum dependence of the scalar and vector form factors with the one of the experimental data \cite{Lubicz:2010bv}, while keeping at the same time the advantage of the high-precision determination of the scalar form factor at the kinematical end-point $q_{max}^2 = (M_K - M_\pi)^2$ \cite{Becirevic:2004ya,Lubicz:2009ht} for the interpolation at $q^2 = 0$.

According to the colour codes reported in Tab.~\ref{tab:f+(0)} and to the FLAG rules of Sec.~\ref{sec:averages}, only the result ETM 09A with $\Nf =2$, the results FNAL/MILC 12I and  RBC/UKQCD 15A with $\Nf=2+1$ and the result FNAL/MILC 13E with $\Nf=2+1+1$ dynamical flavours of fermions, respectively, can enter the FLAG averages.

At $\Nf=2+1+1$ the new result from the FNAL/MILC collaboration, $f_+(0) = 0.9704 (24) (22)$ (FNAL/MILC 13E), is based on the use of the Highly Improved Staggered Quark (HISQ) action (for both valence and sea quarks), which has been taylored to reduce staggered taste-breaking effects, and includes simulations with three lattice spacings and physical light-quark masses.
These features allow to keep the uncertainties due to the chiral extrapolation and to the discretization artifacts well below the statistical error.
The remaining largest systematic uncertainty comes from finite-size effects.

At $\Nf=2+1$ there is a new result from the RBC/UKQCD collaboration, $f_+(0) = 0.9685 (34) (14)$ \cite{Boyle:2015hfa} (RBC/UKQCD 15A), which satisfies all FLAG criteria for entering the average.
RBC/UKQCD 15A superseeds RBC/UKQCD 13 thanks to two new simulations at the physical point.
The other result eligible to enter the FLAG average at $\Nf=2+1$ is the one from FNAL/MILC 12I, $f_+(0)=0.9667(23)(33)$. 
The two results, based on different  fermion discretizations (staggered fermions in the case of FNAL/MILC and domain wall fermions in the case of RBC/UKQCD) are in nice agreement. 
Moreover, in the case of FNAL/MILC the form factor has been determined from the scalar current matrix element, while in the case of RBC/UKQCD it has been determined including also the matrix element of the vector current. 
To a certain extent both simulations are expected to be affected by different systematic effects.

RBC/UKQCD 15A has analyzed results on ensembles with pion masses down to 140~MeV, mapping out the complete range from the $SU(3)$-symmetric limit to the physical point. 
No significant cut-off effects (results for two lattice spacings) were observed in the simulation results.
Ensembles with unphysical light-quark masses are weighted to work as a guide for small corrections toward the physical point, reducing in this way the model dependence in the fitting ansatz.
The systematic uncertainty turns out to be dominated by finite-volume effects, for which an estimate based on effective theory arguments is provided. 

The result FNAL/MILC 12I is from simulations reaching down to a lightest RMS pion mass of about 380~MeV (the lightest valence pion mass for one of their ensembles is about 260~MeV).
Their combined chiral and continuum extrapolation (results for two lattice spacings) is based on NLO staggered chiral perturbation theory supplemented by the continuum NNLO expression~\cite{Bijnens:2003uy} and a phenomenological parameterization of the breaking of the Ademollo-Gatto theorem at finite lattice spacing inherent in their approach.
The $p^4$ low-energy constants entering the NNLO expression have been fixed in terms of external input~\cite{Amoros:2001cp}. 

The ETM collaboration uses the twisted-mass discretization and provides at $\Nf=2$ a comprehensive study of the systematics \cite{Lubicz:2009ht,Lubicz:2010bv}, by presenting results for four lattice spacings and by simulating at light pion masses (down to $M_\pi = 260$~MeV).  
This makes it possible to constrain the chiral extrapolation, using both $SU(3)$ \cite{Gasser:1984ux} and $SU(2)$ \cite{Flynn:2008tg} chiral perturbation theory. 
Moreover, a rough estimate for the size of the effects due to quenching the strange quark is given, based on the comparison of the result for $\Nf=2$ dynamical quark flavours \cite{Blossier:2009bx} with the one in the quenched approximation, obtained earlier by the SPQcdR collaboration \cite{Becirevic:2004ya}. 

We now compute the $N_f =2+1$ FLAG-average for $f_+(0)$ based on FNAL/MILC 12I and RBC/UKQCD 15A, which we consider uncorrelated, while for $N_f = 2+1+1$ and $N_f = 2$ we consider directly the FNAL/MILC 13E and ETM 09A results, respectively:
%FLAGRESULT BEGIN
% TAG      & fKpi    & fKpi	&fKpi &END
% REFS     & \cite{Bazavov:2013maa} &\cite{Bazavov:2012cd,Boyle:2015hfa} &\cite{Lubicz:2009ht} &END
% UNITS    & 1 & 1 & 1 &END
% NUMRESULTS & 1 & 2 & 1 &END
% FLAVOURs & 2+1+1 & 2+1 & 2 &END
%FLAGRESULT END
%FLAGRESULTFORMULA BEGIN
\begin{align}
&\label{eq:fplus_direct_2p1p1}
\mbox{direct},\,\Nf=2+1+1:&\FLAGAVBEGIN f_+(0) &= 0.9704(24)(22)\FLAGAVEND  &&\Ref~\mbox{\cite{Bazavov:2013maa}},\\
&\label{eq:fplus_direct_2p1}                                                               
\mbox{direct},\,\Nf=2+1:  &\FLAGAVBEGIN f_+(0) &= 0.9677(27) \FLAGAVEND     &&\Refs~\mbox{\cite{Bazavov:2012cd,Boyle:2015hfa}},   \\
&\label{eq:fplus_direct_2}                                                                  
\mbox{direct},\,\Nf=2:    &\FLAGAVBEGIN f_+(0) &= 0.9560(57)(62)\FLAGAVEND  &&\Ref~\mbox{\cite{Lubicz:2009ht}},
\end{align}
%FLAGRESULTFORMULA END
where the brackets in the first and third lines indicate the statistical and systematic errors, respectively.
We stress that the results (\ref{eq:fplus_direct_2p1p1}) and (\ref{eq:fplus_direct_2p1}), corresponding to $N_f = 2+1+1$ and $N_f = 2+1$ respectively, include already simulations with physical light-quark masses.

In the case of the ratio of decay constants the data sets that meet the criteria formulated in the introduction are HPQCD 13A~\cite{Dowdall:2013rya}, FNAL/MILC 14A \cite{Bazavov:2014wgs} (which updates MILC 13A~\cite{Bazavov:2013cp}) and ETM 14E \cite{Carrasco:2014poa} with $N_f=2+1+1$, MILC 10~\cite{Bazavov:2010hj}, BMW 10~\cite{Durr:2010hr},  HPQCD/UKQCD 07~\cite{Follana:2007uv} and RBC/UKQCD 12~\cite{Arthur:2012opa} (which is an update of RBC/UKQCD 10A~\cite{Aoki:2010dy}) with $\Nf=2+1$ and ETM 09 \cite{Blossier:2009bx} with $\Nf=2$ dynamical flavours.

ETM 14E uses the twisted-mass discretization and provides a comprehensive study of the systematics by presenting results for three lattice spacings in the range $0.06 - 0.09$ fm and and for pion masses in the range $210 - 450$ MeV.  
This makes it possible to constrain the chiral extrapolation, using both $SU(2)$ \cite{Flynn:2008tg} chiral perturbation theory and polynomial fits.
The ETM collaboration always includes the spread in the central values obtained from different ans\"atze into the systematic errors.
The final result of their analysis is $\fKfpichargedr = 1.184(12)_{\rm stat+fit}(3)_{\rm Chiral}(9)_{\rm a^2}(1)_{Z_P}(3)_{FV}(3)_{IB}$ where the errors are (statistical + the error due to the fitting procedure), due to the chiral extrapolation, the continuum extrapolation, the mass-renormalization constant, the finite-volume and (strong) isospin-breaking effects.

FNAL/MILC 14A has determined the ratio of the decay constants from a comprehensive set of HISQ ensembles with $N_f = 2+1+1$ dynamical flavours. 
They have generated ensembles for four values of the lattice spacing ($0.06 - 0.15$ fm, scale set with $f_{\pi^+}$) and with both physical and unphysical values of the light sea-quark masses, controlling in this way the systematic uncertainties due to chiral and continuum extrapolations.
With respect to MILC 13A they have increased the statistics and added an important ensemble at the finest lattice spacing and for physical values of the light-quark mass.
The final result of their analysis is $\fKfpichargedr=1.1956(10)_{\rm stat}($$_{-14}^{+23}$$)_{\rm a^2} (10)_{FV} (5)_{EM}$ where the errors are statistical, due to the continuum extrapolation, finite-volume and electromagnetic effects.
With respect to MILC 13A a factor of $\simeq 2.6,~ 1.8$ and $\simeq 1.7$ has been gained for the statistical, the discretization and the finite-volume errors. 

%%%%
HPQCD 13A analyzes ensembles generated by MILC and therefore its study of $\fKfpichargedr$ is based on the same set of ensembles bar the one for the finest lattice spacing ($a = 0.09 - 0.15$ fm, scale set with $f_{\pi^+}$ and relative scale set with the Wilson flow~\cite{Luscher:2010iy,Borsanyi:2012zs}) supplemented by some simulation points with heavier quark masses.
HPQCD employs a global fit based on continuum NLO $SU(3)$ chiral perturbation theory for the decay constants supplemented by a model for higher-order terms including discretization and finite-volume effects (61 parameters for 39 data points supplemented by Bayesian priors). 
Their final result is $f_{K^\pm}/f_{\pi^\pm}=1.1916(15)_{\rm stat}(12)_{\rm a^2}(1)_{FV}(10)$, where the errors are statistical, due to the continuum extrapolation, due to finite-volume effects and the last error contains the combined uncertainties from the chiral extrapolation, the scale-setting uncertainty, the experimental input in terms of $f_{\pi^+}$ and from the uncertainty in $m_u/m_d$.
%%%%

In the previous edition of the FLAG review \cite{Aoki:2013ldr} the error budget of HPQCD 13A was compared with the one of MILC 13A and discussed in details.
It was pointed out that, despite the large overlap in primary lattice data, both collaborations arrive at surprisingly different error budgets. 
The same still holds when the comparison is made between HPQCD 13A and FNAL/MILC 14A.

Concerning the cutoff dependence, the finest lattice included into MILC's analysis is $a = 0.06$ fm while the finest lattice in HPQCD's case is $a = 0.09$ fm and both collaborations allow for taste-breaking terms in their analyses.
MILC estimates the residual systematic after extrapolating to the continuum limit by taking the split between the result of an extrapolation with up to quartic and only up to quadratic terms in $a$ as their systematic. 
HPQCD on the other hand models cutoff effects within their global fit ansatz up to including terms in $a^8$, using priors for the unknown coefficients and without including the spread in the central values obtained from different ans\"atze into the systematic errors.
In this way HPQCD arrives at a systematic error due to the continuum limit which is smaller than MILC's estimate by about a factor $\simeq 1.8$.

Turning to finite-volume effects, NLO staggered chiral perturbation theory (MILC) or continuum chiral perturbation theory (HPQCD) was used for correcting the lattice data towards the infinite-volume limit. 
MILC then compared the finite-volume correction to the one obtained by the NNLO expression and took the difference as their estimate for the residual finite-volume error.
In addition they checked the compatibility of the effective-theory predictions (NLO continuum, staggered and NNLO continuum chiral perturbation theory) against lattice data of different spacial extent.
The final verdict is that the related residual systematic uncertainty on $\fKfpichargedr$ made by MILC is larger by an order of magnitude than the one made by HPQCD.

Adding in quadrature all the uncertainties one gets: $f_{K^\pm}/f_{\pi^\pm} = 1.1916(22)$ (HPQCD 13A) and $\fKfpichargedr=1.1960(24)$\footnote{Here we have
symmetrized the asymmetric systematic error and shifted the central value by half the difference as will be done throughout this section.} (FNAL/MILC 14A).
It can be seen that the total errors turn out to be very similar, but the central values seem to show a slight tension of about two standard deviations.
While FLAG is looking forward to independent confirmations of the result for $\fKfpichargedr$ at the same level of precision, we evaluate the FLAG average using a two-step procedure.
First, the HPQCD 13A and FNAL/MILC 14A are averaged assuming a $100 \%$ statistical correlation, obtaining $\fKfpichargedr=1.1936(29)$, where, following the prescription of Sec.~\ref{sec:error_analysis}, the error has been inflated by the factor $\sqrt{(\chi^2/{\rm dof})} \simeq \sqrt{2.5}$ as a result of the tension between the two central values. 
Then, the above finding is averaged with the (uncorrelated) ETM 14E result, obtaining 
\begin{align}
\mbox{direct},\,\Nf=2+1+1:  &\quad \quad \FLAGAVBEGIN \fKfpichargedr=1.1933(29)\FLAGAVEND    &&\Refs~\mbox{\cite{Dowdall:2013rya,Bazavov:2014wgs,Carrasco:2014poa}} \, .
\end{align}

For both $N_f=2+1$ and $N_f=2$ no new result enters the corresponding FLAG averages with respect to the previous edition of the FLAG review \cite{Aoki:2013ldr} and before the closing date specified in Sec.~\ref{sec:introduction}.
Here we limit ourselves to note that for $N_f=2+1$ MILC 10 and HPQCD/UKQCD 07 are based on staggered fermions, BMW 10 has used improved Wilson fermions and RBC/UKQCD 12's result is based on the domain-wall formulation. 
Concerning simulations with $N_f=2$ the FLAG average remains the ETM 09 result, which has simulated twisted-mass fermions. 
In contrast to FNAL/MILC 14A all these simulations are for unphysical values of the light-quark masses (corresponding to smallest pion masses in the range $240 - 260$ MeV in the case of MILC 10, HPQCD/UKQCD 07 and  ETM 09 and around $170$ MeV for RBC/UKQCD 12) and therefore slightly more sophisticated extrapolations needed to be controlled.
Various ans\"atze for the mass and cutoff dependence comprising $SU(2)$ and $SU(3)$ chiral perturbation theory or simply polynomials were used and compared in order to estimate the model dependence.
While BMW 10 and RBC/UKQCD 12 are entirely independent computations, subsets of the MILC gauge ensembles used by MILC 10 and HPQCD/UKQCD 07 are the same.
MILC 10 is certainly based on a larger and more advanced set of gauge configurations than HPQCD/UKQCD 07. 
This allows them for a more reliable estimation of systematic effects. 
In this situation we consider only their statistical but not their systematic uncertainties to be correlated.  

Before determining the average for $f_{K^\pm}/f_{\pi^\pm}$, which should be used for applications to Standard-Model phenomenology, we apply the isospin correction individually to all those results which have been published in the isospin-symmetric limit, i.e.~BMW 10, HPQCD/UKQCD 07 and RBC/UKQCD 12 at $N_f = 2+1$ and ETM 09 at $N_f = 2$. 
To this end, as in the previous edition of the FLAG review \cite{Aoki:2013ldr}, we make use of NLO $SU(3)$ chiral perturbation theory~\cite{Gasser:1984gg,Cirigliano:2011tm}, which predicts
\begin{equation}\label{eq:convert}
	\fKfpicharged = \frac{f_K}{f_\pi} ~ \sqrt{1 + \delta_{SU(2)}} ~ ,
\end{equation}
where~\cite{Cirigliano:2011tm}
\begin{equation}\label{eq:iso}
 \begin{array}{rcl}
	 \delta_{SU(2)}& \approx&
	\sqrt{3}\,\epsilon_{SU(2)}
	\left[-\frac{4}{3} \left(f_K/f_\pi-1\right)+\frac 2{3 (4\pi)^2 f_0^2}
        \left(M_K^2-M_\pi^2-M_\pi^2\ln\frac{M_K^2}{M_\pi^2}\right)
        \right]\,.
  \end{array}
 \end{equation}
We use as input $\epsilon_{SU(2)}=\sqrt{3}/4/R$ with the FLAG result for $R$ of Eq.~(\ref{eq:RQres}), $F_0=f_0/\sqrt{2}=80(20)$~MeV,
$M_\pi=135$ MeV and $M_K=495$ MeV (we decided to choose a conservative uncertainty on $f_0$ in order to reflect the magnitude of potential higher-order 
corrections).
The results are reported in Tab.~\ref{tab:correctedfKfPi}, where in the last column the first error is statistical and the second error is due to the isospin correction (the remaining errors are quoted in the same order as in the original data). 

\begin{table}[!htb]
\begin{center}
\begin{tabular}{llll}
\hline\hline\\[-4mm]
		&$f_K/f_\pi$	&$\delta_{SU(2)}$&$f_{K^\pm}/f_{\pi^\pm}$\\
\hline\\[-4mm]
HPQCD/UKQCD 07	&1.189(2)(7)	&-0.0040(7)&1.187(2)(2)(7)\\
BMW 10		        &1.192(7)(6)	&-0.0041(7)&1.190(7)(2)(6)\\
RBC/UKQCD 12	&1.199(12)(14)	&-0.0043(9)&1.196(12)(2)(14)\\
\hline\hline
\end{tabular}
\caption{Values of the $SU(2)$ isospin-breaking correction $\delta_{SU(2)}$ applied to the lattice data for $f_K/f_\pi$ , entering the FLAG average at $N_f=2+1$, for obtaining the corrected charged ratio $f_{K^\pm}/f_{\pi^\pm}$.}
\label{tab:correctedfKfPi}
\end{center}
\end{table}

For $N_f=2$ a dedicated study of the strong-isospin correction in lattice QCD does exist. 
The (updated) result of the RM123 collaboration~\cite{deDivitiis:2013xla} amounts to $\delta_{SU(2)}=-0.0080(4)$ and we use this result for the isospin correction of the ETM 09 result at $N_f=2$.

Note that the RM123 value for the strong-isospin correction is almost incompatible with the results based on $SU(3)$ chiral perturbation theory, $\delta_{SU(2)}=-0.004(1)$ (see Tab.~\ref{tab:correctedfKfPi}).
Moreover, for $N_f=2+1+1$ HPQCD 13A \cite{Dowdall:2013rya} and ETM 14E \cite{Carrasco:2014poa} estimate a value for $\delta_{SU(2)}$ equal to $-0.0054(14)$ and $-0.0080(38)$, respectively.
One would not expect the strange and heavier sea-quark contributions to be responsible for such a large effect. 
Whether higher-order effects in chiral perturbation theory or other sources are responsible still needs to be understood. 
More lattice QCD simulations of $SU(2)$ isospin-breaking effects are therefore required.
To remain on the conservative side we add a $100 \%$ error to the correction based on $SU(3)$ chiral perturbation theory. 
For further analyses we add (in quadrature) such an uncertainty to the systematic error.

Using the results of Tab.~\ref{tab:correctedfKfPi} for $N_f = 2+1$ we obtain
%FLAGRESULT BEGIN
% TAG      & fKopi    & fKopi	&fKopi &END
% REFS     & \cite{Dowdall:2013rya,Bazavov:2014wgs,Carrasco:2014poa} &\cite{Follana:2007uv,Bazavov:2010hj,Durr:2010hr,Arthur:2012opa} &\cite{Blossier:2009bx} &END
% UNITS    & 1 & 1 & 1 &END
% NUMRESULTS & 3 & 4 & 1 &END
% FLAVOURs & 2+1+1 & 2+1 & 2 &END
%FLAGRESULT END
%FLAGRESULTFORMULA BEGIN
\begin{align}
    \label{eq:fKfpi_direct_broken_2p1p1} 
&\mbox{direct},\,\Nf=2+1+1:&\FLAGAVBEGIN f_{K^\pm} / f_{\pi^\pm} & =  1.193(3)\FLAGAVEND     &&\Refs~\mbox{\cite{Dowdall:2013rya,Bazavov:2014wgs,Carrasco:2014poa}},        \\
    \label{eq:fKfpi_direct_broken_2p1}                                                                             
&\mbox{direct},\,\Nf=2+1:  &\FLAGAVBEGIN f_{K^\pm} / f_{\pi^\pm} & =  1.192(5)\FLAGAVEND     &&\Refs~\mbox{\cite{Follana:2007uv,Bazavov:2010hj,Durr:2010hr,Arthur:2012opa}},\\
    \label{eq:fKfpi_direct_broken_2}                                                                               
&\mbox{direct},\,\Nf=2:    &\FLAGAVBEGIN f_{K^\pm} / f_{\pi^\pm} & =  1.205(6)(17)\FLAGAVEND &&\Ref~\mbox{\cite{Blossier:2009bx}},
\end{align}
%FLAGRESULTFORMULA END
for QCD with broken isospin.

\begin{figure}[!htb]
\vspace{0.2cm}
\centering
\includegraphics[width=12cm]{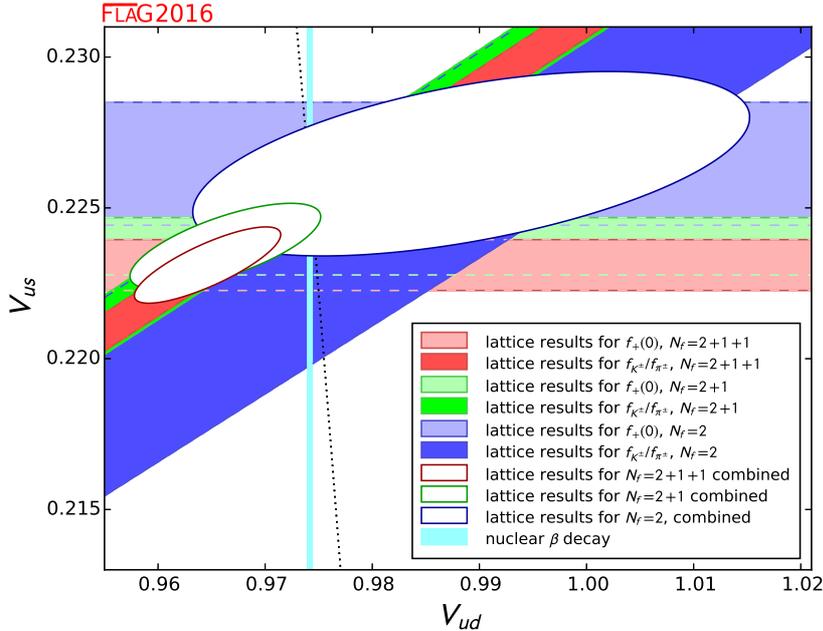}  
\caption{\label{fig:VusVersusVud} The plot compares the information for $|V_{ud}|$, $|V_{us}|$ obtained on the lattice with the experimental result extracted from nuclear $\beta$ transitions. The dotted line indicates the correlation between $|V_{ud}|$ and $|V_{us}|$ that follows if the CKM-matrix is unitary.}
\end{figure}

It is instructive to convert the above results for $f_+(0)$ and $\fKfpichargedr$ into a corresponding range for the CKM matrix elements $|V_{ud}|$ and $|V_{us}|$, using the relations (\ref{eq:products}). 
Consider first the results for $\Nf=2+1+1$. 
The range for $f_+(0)$ in Eq.~(\ref{eq:fplus_direct_2p1p1}) is mapped into the interval $|V_{us}|=0.2231(9)$, depicted as a horizontal red band in Fig.~\ref{fig:VusVersusVud}, while the one for $\fKfpichargedr$ in Eq.~(\ref{eq:fKfpi_direct_broken_2p1p1}) is converted into $|V_{us}|/|V_{ud}|= 0.2313(7)$, shown as a tilted red band. 
The red ellipse is the intersection of these two bands and represents the 68\% likelihood contour,\footnote{Note that the ellipses shown in Fig.~5 of both Ref.~\cite{Colangelo:2010et} and Ref.~\cite{Aoki:2013ldr} correspond instead to the 39\% likelihood contours. Note also that in Ref.~\cite{Aoki:2013ldr} the likelihood was erroneously stated to be $68 \%$ rather than $39 \%$.} obtained by treating the above two results as independent measurements. 
Repeating the exercise for $\Nf=2+1$ and $\Nf=2$ leads to the green and blue ellipses, respectively.
The plot indicates a slight tension between the $N_f=2+1+1$ and the nuclear $\beta$ decay results.

%%%%%%%%%%%%%%%%%%%%%%%%%%%%%%%%%%%%%%%%%%%%%%%%%%%%%%%%%%%%%%%%%%%%%%%%%%%%%%%%
\subsection{Tests of the Standard Model}\label{sec:testing}
%%%%%%%%%%%%%%%%%%%%%%%%%%%%%%%%%%%%%%%%%%%%%%%%%%%%%%%%%%%%%%%%%%%%%%%%%%%%%%%%
  
In the Standard Model, the CKM matrix is unitary. In particular, the elements of the first row obey
\be
   \label{eq:CKM unitarity}
   |V_u|^2\equiv |V_{ud}|^2 + |V_{us}|^2 + |V_{ub}|^2 = 1\fs
\ee 
The tiny contribution from $|V_{ub}|$ is known much better than needed in the present context: $|V_{ub}|= 4.13 (49) \cdot 10^{-3}$ \cite{Agashe:2014kda}. 
In the following, we first discuss the evidence for the validity of the relation (\ref{eq:CKM unitarity}) and only then use it to analyse the lattice data within the 
Standard Model.

In Fig.~\ref{fig:VusVersusVud}, the correlation between $|V_{ud}|$ and $|V_{us}|$ imposed by the unitarity of the CKM matrix is indicated by a dotted line (more precisely, in view of the uncertainty in $|V_{ub}|$, the correlation corresponds to a band of finite width, but the effect is too small to be seen here).
The plot shows that there is a slight tension with unitarity in the data for $N_f = 2 + 1 + 1$: Numerically, the outcome for the sum of the squares of the first row of the CKM matrix reads $|V_u|^2 = 0.980(9)$, which deviates from unity at the level of two standard deviations. 
Still, it is fair to say that at this level the Standard Model passes a nontrivial test that exclusively involves lattice data and well-established kaon decay branching ratios. 
Combining the lattice results for $f_+(0)$ and $\fKfpichargedr$ in Eqs.~(\ref{eq:fplus_direct_2p1p1}) and (\ref{eq:fKfpi_direct_broken_2p1p1}) with the $\beta$ decay value of $|V_{ud}|$ quoted in Eq.~(\ref{eq:Vud beta}), the test sharpens considerably: the lattice result for $f_+(0)$ leads to $|V_u|^2 = 0.9988(6)$, which highlights again a $2\sigma$-tension with unitarity, while the one for $\fKfpichargedr$ implies $|V_u|^2 = 0.9998(5)$, confirming the first-row CKM unitarity below the permille level. 

The situation is similar for $\Nf=2+1$: $|V_u|^2 = 0.984(11)$ with the lattice data alone. 
Combining the lattice results for $f_+(0)$ and $\fKfpichargedr$ in Eqs.~(\ref{eq:fplus_direct_2p1}) and (\ref{eq:fKfpi_direct_broken_2p1}) with the $\beta$ decay value of $|V_{ud}|$, the test sharpens again considerably: the lattice result for $f_+(0)$ leads to $|V_u|^2 = 0.9991(6)$, while the one for $\fKfpichargedr$ implies $|V_u|^2 = 0.9999(6)$, thus confirming 
again CKM unitarity below the permille level. 

Repeating the analysis for $N_f = 2$, we find $|V_u|^2 = 1.029(34)$ with the lattice data alone. 
This number is fully compatible with unity and perfectly consistent with the value of $|V_{ud}|$ found in nuclear $\beta$ decay: combining 
this value with the result (\ref{eq:fplus_direct_2}) for $f_+(0)$ yields $|V_u|^2=1.0003(10)$, combining it with the data (\ref{eq:fKfpi_direct_broken_2}) on $\fKfpichargedr$ gives $|V_u|^2= 0.9988(15)$.

Note that the above tests also offer a check of the basic hypothesis that underlies our analysis: we are assuming that the weak interaction between the quarks and the leptons is governed by the same Fermi constant as the one that determines the strength of the weak interaction among the leptons and determines the lifetime of the muon. 
In certain modifications of the Standard Model, this is not the case. 
In those models it need not be true that the rates of the decays $\pi\rightarrow \ell\nu$, $K\rightarrow\ell\nu$ and $K\rightarrow \pi\ell \nu$ can be used to determine the matrix elements $|V_{ud}f_\pi|$, $|V_{us}f_K|$ and $|V_{us}f_+(0)|$, respectively and that $|V_{ud}|$ can be measured in nuclear $\beta$ decay. 
The fact that the lattice data are consistent with unitarity and with the value of $|V_{ud}|$ found in nuclear $\beta$ decay indirectly also checks the equality of the Fermi constants.

%%%%%%%%%%%%%%%%%%%%%%%%%%%%%%%%%%%%%%%%%%%%%%%%%%%%%%%%%%%%%%%%%%%%%%%%%%%%%%%%
\subsection{Analysis within the Standard Model} \label{sec:SM} 
 %%%%%%%%%%%%%%%%%%%%%%%%%%%%%%%%%%%%%%%%%%%%%%%%%%%%%%%%%%%%%%%%%%%%%%%%%%%%%%%%
 
The Standard Model implies that the CKM matrix is unitary. 
The precise experimental constraints quoted in (\ref{eq:products}) and the unitarity condition (\ref{eq:CKM unitarity}) then reduce the four quantities $|V_{ud}|,|V_{us}|,f_+(0),\fKfpichargedr$ to a single unknown: any one of these determines the other three within narrow uncertainties.
 
\begin{figure}[!htb]
\psfrag{y}{\tiny $\star$}
\begin{center}
\vspace{0.5cm} 
\includegraphics[width=14.5cm]{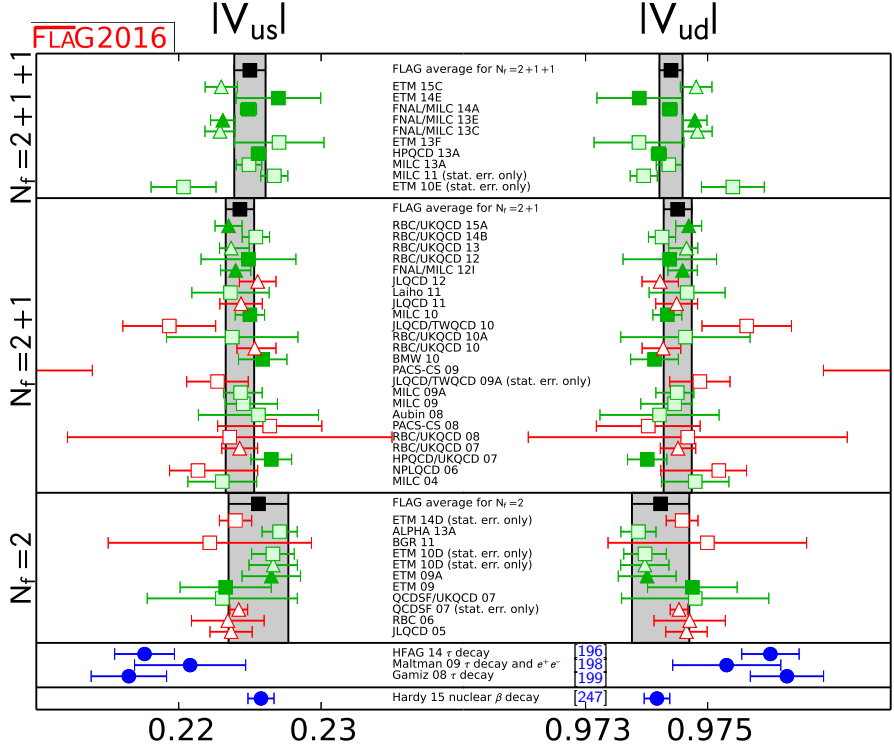}
\end{center}
\vspace{-2.42cm}\hspace{9.4cm}\parbox{6cm}{\sffamily\tiny  \cite{Amhis:2014hma}\\

\vspace{-1.39em}\cite{Maltman:2009bh}\\

\vspace{-1.39em}\cite{Gamiz:2007qs}\\

\vspace{-1.07em}\hspace{0em}\cite{Hardy:2008gy}}
\vspace{0.8cm}
\caption{\label{fig:Vus Vud} Results for $|V_{us}|$ and $|V_{ud}|$ that follow from the lattice data for $f_+(0)$ (triangles) and $\fKfpichargedr$ (squares), on the basis of the assumption that the CKM matrix is unitary. 
The black squares and the grey bands represent our estimates, obtained by combining these two different ways of measuring $|V_{us}|$ and $|V_{ud}|$ on a lattice.
For comparison, the figure also indicates the results obtained if the data on nuclear $\beta$ decay and $\tau$ decay are analysed within the Standard Model.}
\end{figure}

Fig.~\ref{fig:Vus Vud} shows that the results obtained for $|V_{us}|$ and $|V_{ud}|$ from the data on $\fKfpichargedr$ (squares) are quite consistent with the determinations via $f_+(0)$ (triangles). 
In order to calculate the corresponding average values, we restrict ourselves to those determinations that we have considered best in Sec.~\ref{sec:Direct}.
The corresponding results for $|V_{us}|$ are listed in Tab.~\ref{tab:Vus} (the error in the experimental numbers used to convert the values of $f_+(0)$ and $\fKfpichargedr$ into values for $|V_{us}|$ is included in the statistical error).
   
\begin{table}[!htb]
\centering
\noindent
\begin{tabular*}{\textwidth}[l]{@{\extracolsep{\fill}}lclcll}
Collaboration & Ref. &\rule{0.5cm}{0cm}$\Nf$&from&\rule{0.6cm}{0cm}$|V_{us}|$&\rule{0.6cm}{0cm}$|V_{ud}|$\\
&&&&& \\[-2ex]
\hline \hline &&&&&\\[-2ex]
FNAL/MILC 13E &\cite{Bazavov:2013maa}&$2+1+1$&$f_+(0)$  \rule{0cm}{0.45cm} &0.2231(7)(5)&0.97479(16)(12)\\
ETM 14E &\cite{Carrasco:2014poa}&$2+1+1$&$\fKfpichargedr$ \rule{0cm}{0.45cm} &0.2270(22)(20)&0.97388(51)(47)\\
FNAL/MILC 14A &\cite{Bazavov:2014wgs}&$2+1+1$&$\fKfpichargedr$ \rule{0cm}{0.45cm} &0.2249(4)(4)&0.97438(8)(9)\\
HPQCD 13A &\cite{Dowdall:2013rya}&$2+1+1$&$\fKfpichargedr$ \rule{0cm}{0.45cm} &0.2256(4)(3)&0.97420(10)(7)\\
&&&&& \\[-2ex]
\hline
&&&&& \\[-2ex]
RBC/UKQCD 15A &\cite{Boyle:2015hfa}&$2+1$&$f_+(0)$          \rule{0cm}{0.45cm} &0.2235(9)(3)&0.97469(20)(7)\\
FNAL/MILC 12I    &\cite{Bazavov:2012cd}&$2+1$&$f_+(0)$       \rule{0cm}{0.45cm} &0.2240(7)(8)&0.97459(16)(18)\\
MILC 10 &\cite{Bazavov:2010hj}&$2+1$&$\fKfpichargedr$ \rule{0cm}{0.45cm} &0.2250(5)(9)&0.97434(11)(21)\\
RBC/UKQCD 12  &\cite{Aoki:2010dy}   &$2+1$&$\fKfpichargedr$ \rule{0cm}{0.45cm} &0.2249(22)(25)&0.97438(50)(58)\\ 
BMW 10 &\cite{Durr:2010hr}  & $2+1$ \rule{0cm}{0.45cm}& $\fKfpichargedr$ & $0.2259(13)(11)$&0.97413(30)(25)\\
HPQCD/UKQCD 07 &\cite{Follana:2007uv}\rule{0cm}{0.4cm}& $2+1$ & $\fKfpichargedr$&  $  0.2265(6)(13)$&0.97401(14)(29)\\
&&&&& \\[-2ex]
\hline
&&&&& \\[-2ex]
ETM 09A & \cite{Lubicz:2009ht}\rule{0cm}{0.4cm}&2&$f_+(0)$&   $ 0.2265 (14) (15)$&0.97401(33)(34)\\
ETM 09  &\cite{Blossier:2009bx}\rule{0cm}{0.4cm}&2&$\fKfpichargedr$& $ 0.2233 (11) (30)$&0.97475(25)(69)\\
&&&&& \\[-2ex]
\hline \hline 
\end{tabular*}
\caption{\label{tab:Vus} Values of $|V_{us}|$ and $|V_{ud}|$ obtained from the lattice determinations of either $f_+(0)$ or $\fKfpichargedr$ assuming CKM unitarity. 
The first (second) number in brackets represents the statistical (systematic) error.} 
\end{table} 

For $\Nf=2+1+1$ we consider the data both for $f_+(0)$ and $\fKfpichargedr$, treating FNAL/MILC 13E, FNAL/MILC 14A and HPQCD 13A as statistically correlated (according to the prescription of Sec.~\ref{sec:error_analysis}). 
We obtain $|V_{us}|=0.2250(11)$, where the error includes the inflation factor due the value of $\chi^2/{\rm dof} \simeq 2.3$.
This result is indicated on the left hand side of Fig.~\ref{fig:Vus Vud} by the narrow vertical band. 
In the case $N_f = 2+1$ we consider MILC 10, FNAL/MILC 12I and HPQCD/UKQCD 07 on the one hand and RBC/UKQCD 12 and RBC/UKQCD 15A on the other hand, as mutually statistically correlated, since the analysis in the two cases starts from partly the same set of gauge ensembles.
In this way we arrive at $|V_{us}| = 0.2243(10)$ with $\chi^2/{\rm dof} \simeq 1.0$. 
For $\Nf=2$ we consider ETM 09A and ETM 09 as statistically correlated, obtaining $|V_{us}|=0.2256(21)$ with $\chi^2/{\rm dof} \simeq 0.7$.
The figure shows that the result obtained for the data with $\Nf=2$, $\Nf=2+1$ and $\Nf=2+1+1$ are consistent with each other.
 
Alternatively, we can solve the relations for $|V_{ud}|$ instead of $|V_{us}|$. 
Again, the result $|V_{ud}|=0.97440(19)$ which follows from the lattice data with $\Nf=2+1+1$ is perfectly consistent with the values $|V_{ud}|=0.97451(23)$ and $|V_{ud}|=0.97423(47)$ obtained from the data with $\Nf=2+1$ and $\Nf=2$, respectively.  
The reduction of the uncertainties in the result for $|V_{ud}|$ due to CKM unitarity is to be expected from Fig.~\ref{fig:VusVersusVud}: the unitarity condition reduces the region allowed by the lattice results to a nearly vertical interval.

Next, we determine the values of $f_+(0)$ and  $\fKfpichargedr$ that follow from our determinations of $|V_{us}|$ and $|V_{ud}|$ obtained from the lattice data within the Standard Model.
We find $f_+(0) = 0.9622(50)$ for $\Nf=2+1+1$, $f_+(0) = 0.9652(47)$ for $\Nf=2+1$, $f_+(0) = 0.9597(91)$ for $\Nf=2$ and $\fKfpichargedr = 1.195(5)$ for $\Nf=2+1+1$, $\fKfpichargedr = 1.199(5)$ for $\Nf=2+1$, $\fKfpichargedr = 1.192(9) $ for $\Nf=2$, respectively.
These results are collected in the upper half of Tab.~\ref{tab:Final results}. 
In the lower half of the table, we list the analogous results found by working out the consequences of the CKM unitarity using the values of $|V_{ud}|$ and $|V_{us}|$ obtained from nuclear $\beta$ decay and $\tau$ decay, respectively. 
The comparison shows that the lattice result for $|V_{ud}|$ not only agrees very well with the totally independent determination based on nuclear $\beta$ transitions, but is also
remarkably precise. 
On the other hand, the values of $|V_{ud}|$, $f_+(0)$ and $\fKfpichargedr$ which follow from the $\tau$-decay data if the Standard Model is assumed to be valid, are not in good agreement with the lattice results for these quantities. 
The disagreement is reduced considerably if the analysis of the $\tau$ data is supplemented with experimental results on electroproduction \cite{Maltman:2009bh}: the discrepancy then amounts to little more than one standard deviation.  

\begin{table}[!htb]
\centering
\begin{tabular*}{\textwidth}[l]{@{\extracolsep{\fill}}llllll}
\rule[-0.2cm]{0cm}{0.5cm}& Ref. & \rule{0.3cm}{0cm} $|V_{us}|$&\rule{0.3cm}{0cm} $|V_{ud}|$&\rule{0.25cm}{0cm} $f_+(0)$&$\fKfpichargedr$\\
&&&& \\[-2ex]
\hline \hline
&&&& \\[-2ex]
$\Nf= 2+1+1$& &\rule{0cm}{0.4cm}0.2250(11)& 0.97440(19)  & 0.9622(50)   & 1.195(5)\\
&&&& \\[-2ex]
\hline
$\Nf= 2+1$&   &\rule{0cm}{0.4cm}0.2243(10)& 0.97451(23)  & 0.9652(47)   & 1.199(5)\\
&&&& \\[-2ex]
\hline
&&&& \\[-2ex]
$\Nf=2$ & &\rule{0cm}{0.4cm}0.2256(21) &0.97423(47)  &0.9597(91) &1.192(9)\\
&&&& \\[-2ex]
\hline\hline
&&&& \\[-2ex]
$\beta$ decay &\cite{Hardy:2014qxa}&0.2258(9)& 0.97417(21) & 0.9588(42)&
1.191(4) \\ 
&&&& \\[-2ex]
$\tau$ decay &\cite{Gamiz:2007qs}&0.2165(26)&0.9763(6)& 1.0000(122)&
1.245(12)\\ 
&&&& \\[-2ex]
$\tau$ decay  &\cite{Maltman:2009bh}&0.2208(39)&0.9753(9)& 0.9805(174)&
1.219(18)\\ 
&&&& \\[-2ex]
\hline\hline
\end{tabular*}
\caption{\label{tab:Final results}The upper half of the table shows our final results for $|V_{us}|$, $|V_{ud}|$,  $f_+(0)$ and $\fKfpichargedr$, which are obtained by analysing the lattice data within the Standard Model. 
For comparison, the lower half lists the values that follow if the lattice results are replaced by the experimental results on nuclear $\beta$ decay and $\tau$ decay, respectively.}
\end{table}

%%%%%%%%%%%%%%%%%%%%%%%%%%%%%%%%%%%%%%%%%%%%%%%%%%%%%%%%%%%%%%%%%%%%%%%%%%%%%%%%
\subsection{Direct determination of $f_{K^\pm}$ and $f_{\pi^\pm}$}\label{sec:fKfpi}
%%%%%%%%%%%%%%%%%%%%%%%%%%%%%%%%%%%%%%%%%%%%%%%%%%%%%%%%%%%%%%%%%%%%%%%%%%%%%%%%

It is useful for flavour physics studies to provide not only the lattice average of $f_{K^\pm} / f_{\pi^\pm}$, but also the average of the decay constant $f_{K^\pm}$. 
The case of the decay constant $f_{\pi^\pm}$ is different, since the experimental value of this quantity is often used for setting the scale in lattice QCD (see Appendix A.2).
However, the physical  scale can be set in different ways, namely by using as input the mass of the $\Omega$-baryon ($m_\Omega$) or the $\Upsilon$-meson spectrum ($\Delta M_\Upsilon$), which are less sensitive to the uncertainties of the chiral extrapolation in the light-quark mass with respect to $f_{\pi^\pm}$. 
In such cases the value of the decay constant $f_{\pi^\pm}$ becomes a direct prediction of the lattice-QCD simulations.
It is therefore interesting to provide also the average of the decay constant $f_{\pi^\pm}$, obtained when the physical scale is set through another hadron observable, in order to check the consistency of different scale setting procedures.

Our compilation of the values of $f_{\pi^\pm}$ and $f_{K^\pm}$ with the corresponding colour code is presented in Tab.~\ref{tab:FK Fpi}.
With respect to the case of $f_{K^\pm} / f_{\pi^\pm}$ we have added two columns indicating which quantity is used to set the physical scale and the possible use of a renormalization constant for the axial current.
Indeed, for several lattice formulations the use of the nonsinglet axial-vector Ward identity allows to avoid the use of any renormalization constant.

One can see that the determinations of $f_{\pi^\pm}$ and $f_{K^\pm}$ suffer from larger uncertainties with respect to the ones of the ratio $f_{K^\pm} / f_{\pi^\pm}$, which is less sensitive to various systematic effects (including the uncertainty of a possible renormalization constant) and, moreover, is not exposed to the uncertainties of the procedure used to set the physical scale.

According to the FLAG rules, for $N_f = 2 + 1 + 1$ three data sets can form the average of $f_{K^\pm}$ only: ETM 14E \cite{Carrasco:2014poa}, FNAL/MILC 14A \cite{Bazavov:2014wgs} and HPQCD 13A \cite{Dowdall:2013rya}.
Following the same procedure already adopted in Sec.~\ref{sec:Direct} in the case of the ratio of the decay constant we treat FNAL/MILC 14A and HPQCD 13A as statistically correlated.
For $N_f = 2 + 1$ three data sets can form the average of $f_{\pi^\pm}$ and $f_{K^\pm}$ : RBC/UKQCD 12 \cite{Arthur:2012opa} (update of RBC/UKQCD 10A), HPQCD/UKQCD 07 \cite{Follana:2007uv} and MILC 10 \cite{Bazavov:2010hj}, which is the latest update of the MILC program.
We consider HPQCD/UKQCD 07 and MILC 10 as statistically correlated and use the prescription of Sec.~\ref{sec:error_analysis} to form an average.
For $N_f = 2$ the average cannot be formed for $f_{\pi^\pm}$, and only one data set (ETM 09) satisfies the FLAG rules in the case of $f_{K^\pm}$.

\begin{figure}[!htb]
\begin{center}
\includegraphics[height=10cm]{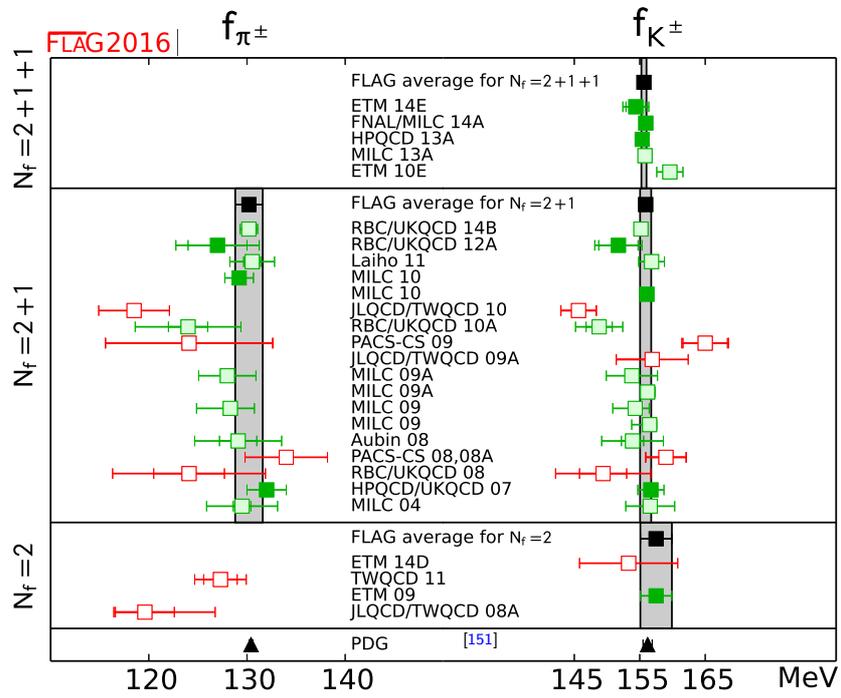}
\end{center}
  \vspace{-2.08cm}\hspace{8.2cm}\parbox{6cm}{\sffamily\tiny  \cite{Agashe:2014kda}}
\vspace{1cm}
\caption{\label{fig:latticedata_decayconstants}
Values of $f_\pi$ and $f_K$.
The black squares and grey bands indicate our estimates (\ref{eq:fPi}) and (\ref{eq:fK}).
The black triangles represent the experimental values quoted by the PDG, see Eq.~(\ref{eq:fps_PDG}).}
\end{figure}

Thus, our estimates read
%FLAGRESULT BEGIN
% TAG      &fpi & fK    & fK	&fK &END
% REFS     &\cite{Follana:2007uv,Bazavov:2010hj,Arthur:2012opa}& \cite{Dowdall:2013rya,Bazavov:2014wgs,Carrasco:2014poa}  &\cite{Follana:2007uv,Bazavov:2010hj,Arthur:2012opa}&\cite{Blossier:2009bx} &END
% UNITS    & '[MeV]' & '[MeV]' & '[MeV]'& '[MeV]' &END
% NUMRESULTS & 3 & 3 & 3&1 &END
% FLAVOURs & 2+1 & 2+1+1 & 2+1 & 2 &END
%FLAGRESULT END
%FLAGRESULTFORMULA BEGIN
\begin{align}
  \label{eq:fPi}
&N_f = 2 + 1:     &\FLAGAVBEGIN f_{\pi^\pm}&= 130.2 ~ (1.4)\FLAGAVEND  ~ \mbox{MeV} &&\Refs~\mbox{\cite{Follana:2007uv,Bazavov:2010hj,Arthur:2012opa}},\\ \nonumber 
                \\                                                       
&N_f = 2 + 1 + 1: &\FLAGAVBEGIN f_{K^\pm} & = 155.6 ~ (0.4)\FLAGAVEND  ~ \mbox{MeV} &&\Refs~\mbox{\cite{Dowdall:2013rya,Bazavov:2014wgs,Carrasco:2014poa}}         ,\nonumber\\ 
&N_f = 2 + 1:     &\FLAGAVBEGIN f_{K^\pm} & = 155.9 ~ (0.9)\FLAGAVEND  ~ \mbox{MeV} &&\Refs~\mbox{\cite{Follana:2007uv,Bazavov:2010hj,Arthur:2012opa}},\label{eq:fK}\\ 
\nonumber
&N_f = 2:         &\FLAGAVBEGIN f_{K^\pm} & = 157.5 ~ (2.4)\FLAGAVEND  ~ \mbox{MeV} &&\Ref~\mbox{\cite{Blossier:2009bx}}.\\\nonumber
 \end{align}
%FLAGRESULTFORMULA END
The lattice results of Tab.~\ref{tab:FK Fpi} and our estimates (\ref{eq:fPi}-\ref{eq:fK}) are reported in Fig.~\ref{fig:latticedata_decayconstants}. 
The latter ones agree within the errors with the latest experimental determinations of $f_\pi$ and $f_K$ from the PDG \cite{Agashe:2014kda}:
 \begin{eqnarray}
     \label{eq:fps_PDG}
     f_{\pi^\pm}^{(PDG)} = 130.41 ~ (0.20) ~ \mbox{MeV} \qquad , \qquad f_{K^\pm}^{(PDG)} = 156.2 ~ (0.7) ~ \mbox{MeV} ~ .
 \end{eqnarray}
Moreover the values of $f_{\pi^\pm}$ and $f_{K^\pm}$ quoted by the PDG are obtained assuming Eq.~(\ref{eq:Vud beta}) for the value of $|V_{ud}|$ and adopting the average of FNAL/MILC 12I and  RBC-UKQCD 10 results for $f_+(0)$.

\begin{table}[!htb]
       {\centering
\vspace{1.5cm}{\footnotesize\noindent
\begin{tabular*}{\textwidth}[l]{@{\extracolsep{\fill}}l@{\hspace{1mm}}r@{\hspace{1mm}}l@{\hspace{1mm}}l@{\hspace{1mm}}l@{\hspace{1mm}}l@{\hspace{1mm}}l@{\hspace{3mm}}l@{\hspace{1mm}}l@{\hspace{1mm}}l@{\hspace{5mm}}l@{\hspace{1mm}}l}
Collaboration & Ref. & $\Nf$ &
\hspace{0.15cm}\begin{rotate}{40}{publication status}\end{rotate}\hspace{-0.15cm}&
\hspace{0.15cm}\begin{rotate}{40}{chiral extrapolation}\end{rotate}\hspace{-0.15cm}&
\hspace{0.15cm}\begin{rotate}{40}{continuum extrapolation}\end{rotate}\hspace{-0.15cm}&
\hspace{0.15cm}\begin{rotate}{40}{finite-volume errors}\end{rotate}\hspace{-0.15cm}& 
\hspace{0.15cm}\begin{rotate}{40}{renormalization}\end{rotate}\hspace{-0.15cm}&
\hspace{0.05cm}\begin{rotate}{40}{physical scale}\end{rotate}\hspace{-0.15cm}&\rule{0cm}{0cm}
\hspace{0.0cm}\begin{rotate}{40}{$SU(2)$ breaking}\end{rotate}\hspace{-0.15cm}&\rule{0.5cm}{0cm}
$f_{\pi^\pm}$&\rule{0.5cm}{0cm}$f_{K^\pm}$ \\
&&&&&&& \\[-0.1cm]
\hline
\hline
&&&&&&& \\[-0.1cm]
ETM 14E &\cite{Carrasco:2014poa}&2+1+1&\gA&\soso&\good&\soso&na&$f_\pi$&&--&{154.4(1.5)(1.3)}\\
FNAL/MILC 14A&\cite{Bazavov:2014wgs}&2+1+1&\gA&\good&\good&\good&na&$f_\pi$&&--&{155.92(13)($_{-23}^{+34}$)}\\
HPQCD 13A&\cite{Dowdall:2013rya}&2+1+1&\gA&\good&\soso&\good&na&$f_\pi$&&--&{155.37(20)(27)}\\
MILC 13A&\cite{Bazavov:2013cp}&2+1+1&\gA&\good&\soso&\good&na&$f_\pi$&&--&155.80(34)(54)\\
ETM 10E &\cite{Farchioni:2010tb}&2+1+1&\rC&\soso&\soso&\soso&na&$f_\pi$&\checkmark&--&159.6(2.0)\\
&&&&&&& \\[-0.1cm]
\hline
&&&&&&& \\[-0.1cm]
RBC/UKQCD 14B   &\cite{Blum:2014tka}&2+1&\gA&\good&\good&\good&NPR&$m_\Omega$ &\checkmark& 130.19(89) & 155.18(89) \\
RBC/UKQCD 12   &\cite{Arthur:2012opa}&2+1&\gA&\tbg&\soso&\good&NPR&$m_\Omega$ &\checkmark& 127.1(2.7)(2.7)& 152.1(3.0)(1.7) \\
Laiho 11       &\cite{Laiho:2011np}   &2+1&\rC&\soso&\good&\soso&na&${}^\dagger$ && $130.53(87)(210)$&$156.8(1.0)(1.7)$\\
MILC 10 &\cite{Bazavov:2010hj}&2+1&\rC&\soso&\good&\good&na&${}^\dagger$ & &{129.2(4)(14)}&--\\
MILC 10 &\cite{Bazavov:2010hj}&2+1&\rC&\soso&\good&\good&na&$f_\pi$ &&--          &{156.1(4)($_{-9}^{+6}$)}\\
JLQCD/TWQCD 10 &\cite{Noaki:2010zz}&2+1&\rC&\soso&\tbr&\tbg&na&$m_\Omega$&\checkmark&118.5(3.6)$_{\rm stat}$&145.7(2.7)$_{\rm stat}$\\
RBC/UKQCD 10A  &\cite{Aoki:2010dy} &2+1&\gA&\soso&\soso&\good&NPR&$m_\Omega$&\checkmark&124(2)(5)&148.8(2.0)(3.0)\\
PACS-CS 09     &\cite{Aoki:2009ix} &2+1&\gA&\good&\tbr&\tbr  &NPR &$m_\Omega$&\checkmark&124.1(8.5)(0.8) & 165.0(3.4)(1.1)\\
JLQCD/TWQCD 09A&\cite{JLQCD:2009sk} &2+1&\rC&\soso&\tbr&\tbr &na&$f_\pi$&\checkmark&--&156.9(5.5)$_{\rm stat}$\\
MILC 09A &\cite{Bazavov:2009fk}&2+1&\rC&\soso&\tbg&\tbg &na&$\Delta M_\Upsilon$ &&128.0(0.3)(2.9)&          153.8(0.3)(3.9)\\
MILC 09A &\cite{Bazavov:2009fk}&2+1&\rC&\soso&\tbg&\tbg &na&$f_\pi$&&--&156.2(0.3)(1.1)\\
MILC 09 &\cite{Bazavov:2009bb}&2+1&\gA&\soso&\tbg&\tbg &na&$\Delta M_\Upsilon$&&128.3(0.5)($^{+2.4}_{-3.5}$)&154.3(0.4)($^{+2.1}_{-3.4}$) \\
MILC 09 &\cite{Bazavov:2009bb}&2+1&\gA&\soso&\tbg&\tbg &na&$f_\pi$&&&156.5(0.4)($^{+1.0}_{-2.7}$)\\
Aubin 08       &\cite{Aubin:2008ie} &2+1&\rC&\soso&\soso&\soso&na&$\Delta M_\Upsilon$     && 129.1(1.9)(4.0)   & 153.9(1.7)(4.4)  \\
PACS-CS 08, 08A&\cite{Aoki:2008sm, Kuramashi:2008tb} &2+1&\gA&\tbg&\tbr&\tbr&1lp&$m_\Omega$&\checkmark&134.0(4.2)$_{\rm stat}$& 159.0(3.1)$_{\rm stat}$\\
RBC/UKQCD 08   &\cite{Allton:2008pn} &2+1&\gA&\soso&\tbr&\tbg&NPR&$m_\Omega$&\checkmark&124.1(3.6)(6.9) &        149.4(3.6)(6.3)\\
HPQCD/UKQCD 07 &\cite{Follana:2007uv}&2+1&\gA&\soso&\soso&\soso&na&$\Delta M_\Upsilon$&\checkmark& {132(2)}                & {156.7(0.7)(1.9)}\\
MILC 04 &\cite{Aubin:2004fs}&2+1&\gA&\soso&\soso&\soso&na&$\Delta M_\Upsilon$&&129.5(0.9)(3.5)     &     156.6(1.0)(3.6)\\[-1mm]
&&&&&&& \\[-0.1cm]
\hline
&&&&&&& \\[-0.1cm]
ETM 14D &\cite{Abdel-Rehim:2014nka}&2&\rC&\good&\tbr&\soso&na&$f_\pi$&\checkmark&--&153.3(7.5)$_{\rm stat}$\\
TWQCD 11       &\cite{Chiu:2011bm}   &2 &\oP&\tbg&\tbr&\tbr&na&${r_0}^\ast$&&127.3(1.7)(2.0)$^{\ast\ast}$&--\\
ETM 09         &\cite{Blossier:2009bx}         &2 &\gA&\soso&\tbg&\soso&na&$f_\pi$&\checkmark& --& {157.5(0.8)(2.0)(1.1)}$^{\dagger\dagger}$\\
JLQCD/TWQCD 08A&\cite{Noaki:2008iy} &2  &\gA  &\soso &\tbr   &\tbr &na&$r_0$&&119.6(3.0)($^{+6.5}_{-1.0}$)$^{\ast\ast}$&--\\[-1mm]
&&&&&&& \\[-0.1cm]
\hline
\hline
&&&&&&& \\[-0.1cm]
\end{tabular*}}\\[-2mm]
}

\begin{minipage}{\linewidth}
\footnotesize The label 'na' indicates the lattice calculations which do not require the use of any renormalization constant for the axial current, while the label 'NPR' ('1lp') signals the use of a renormalization constant calculated nonperturbatively (at 1-loop order in perturbation theory).  
\begin{itemize}
{\footnotesize 
\item[$^{\dagger}$] The ratios of lattice spacings within the ensembles were determined using the quantity $r_1$. 
The conversion to physical units was made on the basis of Ref.~\cite{Davies:2009tsa} and we note that such a determination depends on the experimental value of the pion decay constant\\[-5mm]
\item[$^{\dagger\dagger}$] Errors are (stat+chiral)($a\neq 0$)(finite size).
\\[-5mm]
\item[$^\ast$] The ratio $f_\pi/M_\pi$ was used as experimental input to fix the light-quark mass.
\\[-5mm]
\item[$^{\ast\ast}$] $L_{\rm min}<2$fm in these simulations.
\\[-5mm]
}
\end{itemize}
\end{minipage}
\caption{Colour code for the lattice data on $f_{\pi^\pm}$ and $f_{K^\pm}$ together with information on the way the lattice spacing was converted to physical units and on whether or not an isospin-breaking correction has been applied to the quoted result (see Sec.~\ref{sec:Direct}). The numerical values are listed in MeV units. \hfill}
\label{tab:FK Fpi}
\end{table}

\clearpage

%% file: LEC/LECs.tex
%%%%%%%%%%%%%%%%%%%%%%%%%%%%%%%%%%%%%%%%%%%%%%%%%%%%%%%%%%%%%%%%%%%%%%%%%%%%%%%

\section{Low-energy constants\label{sec:LECs}}

%%%%%%%%%%%%%%%%%%%%%%%%%%%%%%%%%%%%%%%%%%%%%%%%%%%%%%%%%%%%%%%%%%%%%%%%%%%%%%%

In the study of the quark-mass dependence of QCD observables calculated on
the lattice, it is common practice to invoke chiral perturbation theory
({\Ch}PT).  For a given quantity this framework predicts the nonanalytic
quark-mass dependence  and it provides symmetry relations among different
observables.  These relations are best expressed with the help of a set of
linearly independent and universal (i.e.\ process-independent) low-energy
constants (LECs), which appear as coefficients of the polynomial terms (in
$m_q$ or $\Mpi^2$) in different observables. When numerical simulations are
done at heavier than physical (light) quark masses, {\Ch}PT is usually invoked
in the extrapolation to physical quark masses.

%%%%%%%%%%%%%%%%%%%%%%%%%%%%%%%%%%%%%%%%%%%%%%%%%%%%%%%%%%%%%%%%%%%%%%%%%%%%%%%

\subsection{Chiral perturbation theory \label{sec:chPT}}

%%%%%%%%%%%%%%%%%%%%%%%%%%%%%%%%%%%%%%%%%%%%%%%%%%%%%%%%%%%%%%%%%%%%%%%%%%%%%%%

{\Ch}PT is an effective field theory approach to the low-energy properties
of QCD based on the spontaneous breaking of chiral symmetry,
$SU(\Nf)_L \times SU(\Nf)_R \to SU(\Nf)_{L+R}$, and its soft explicit
breaking by quark-mass terms.
In its original implementation, in infinite volume, it is an expansion in
$m_q$ and $p^2$ with power counting $\Mpi^2 \sim m_q \sim p^2$.

If one expands around the $SU(2)$ chiral limit, there appear two LECs at
order $p^2$ in the chiral effective Lagrangian,
\be
F\equiv F_\pi\,\rule[-0.3cm]{0.01cm}{0.7cm}_{\;m_u,m_d\rightarrow 0}
\quad \mbox{and} \qquad
B\equiv \frac{\Sigma}{F^2} \; , \quad\mbox{where} \quad
\Sigma\equiv-\<\ubar u\>\,\Big|_{\;m_u,m_d\rightarrow 0} \; ,
\ee
and seven at order $p^4$, indicated by $\bar \ell_i$ with $i=1,\ldots,7$.
In the analysis of the $SU(3)$ chiral limit there are also just two LECs at
order $p^2$,
\be
F_0\equiv F_\pi\,\rule[-0.3cm]{0.01cm}{0.7cm}_{\;m_u,m_d,m_s\rightarrow 0}
\quad \mbox{and} \qquad
B_0\equiv \frac{\Sigma_0}{F_0^2} \; , \quad\mbox{where} \quad
\Sigma_0\equiv-\<\ubar u\>\,\Big|_{\;m_u,m_d,m_s\rightarrow 0} \; ,
\ee
but ten at order $p^4$, indicated by the capital letter $L_i(\mu)$ with
$i=1,\ldots,10$. These constants are independent of the quark masses,%
\footnote{More precisely, they are independent of the 2 or 3 light quark
masses which are explicitly considered in the respective framework.
However, all low-energy constants depend on the masses of the remaining
quarks $s,c,b,t$ or $c,b,t$ in the $SU(2)$ and $SU(3)$ framework, respectively,
although the dependence on the masses of the $c,b,t$ quarks is expected to
be small.}
but they become scale dependent after renormalization (sometimes a
superscript $r$ is added).  The $SU(2)$ constants $\lbar_i$ are scale
independent, since they are defined at scale $\mu=\Mpi$ (as indicated by
the bar).  For the precise definition of these constants and their scale
dependence we refer the reader to Refs.~\cite{Gasser:1983yg,Gasser:1984gg}.

If the box volume is finite but large compared to the Compton wavelength
of the pion, $L \gg 1/\Mpi$, the power counting generalizes to
$m_q \sim p^2 \sim 1/L^2$, as one would assume based on the fact that
$p_\mr{min}=2\pi/L$ is the minimum momentum in a finite box.
This is the so-called $p$-regime of {\Ch}PT.
It coincides with the setting that is used for standard phenomenologically
oriented lattice-QCD computations, and we shall consider the $p$-regime
the default in the following.
However, if the pion mass is so small that the box-length $L$ is no longer
large compared to the Compton wavelength that the pion would have, at the
given $m_q$, in infinite volume, then the chiral series must be reordered.
Such finite-volume versions of {\Ch}PT with correspondingly adjusted power
counting schemes, referred to as $\epsilon$- and $\delta$-regime, are described
in Secs.~\ref{sec_eps} and \ref{sec_su2_delta}, respectively.

Lattice calculations can be used to test if chiral symmetry is indeed
spontaneously broken along the path $SU(\Nf)_L \times SU(\Nf)_R \to SU(\Nf)_{L+R}$
by measuring nonzero chiral condensates and by verifying the validity of
the GMOR relation $\Mpi^2\propto m_q$ close to the chiral limit.  If the
chiral extrapolation of quantities calculated on the lattice is made with
the help of fits to their {\Ch}PT forms, apart from determining the
observable at the physical value of the quark masses, one also obtains the
relevant LECs.  This is a very important by-product for two reasons:
\begin{enumerate}
\itemsep-2pt
\item
All LECs up to order $p^4$ (with the exception of $B$ and $B_0$, since only
the product of these times the quark masses can be estimated from
phenomenology) have either been determined by comparison to experiment or
estimated theoretically, e.g.\ in large-$N_c$ QCD.  A lattice
determination of the better known LECs thus provides a test of the {\Ch}PT
approach.
\item
The less well-known LECs are those which describe the quark-mass dependence
of observables -- these cannot be determined from experiment, and therefore
the lattice, where quark masses can be varied, provides unique quantitative
information.  This information is essential for improving phenomenological
{\Ch}PT predictions in which these LECs play a role.
\end{enumerate}
We stress that this program is based on the nonobvious assumption that
{\Ch}PT is valid in the region of masses and momenta used in the lattice
simulations under consideration, something that can and should be checked.
In the end one wants to compare lattice and phenomenological determinations
of LECs, much in the spirit of Ref.~\cite{Bijnens:2014lea}.
An overview of many of the conceptual issues involved in matching
lattice data to an effective field theory framework like {\Ch}PT is
given in Refs.~\cite{Sharpe:2006pu,Golterman:2009kw,Bernard:2015wda}.

The fact that, at large volume, the finite-size effects, which occur if a
system undergoes spontaneous symmetry breakdown, are controlled by the
Nambu-Goldstone modes, was first noted in solid state physics, in
connection with magnetic systems \cite{Fisher:1985zz,Brezin:1985xx}.  As
pointed out in Ref.~\cite{Gasser:1986vb} in the context of QCD, the thermal
properties of such systems can be studied in a systematic and
model-independent manner by means of the corresponding effective field
theory, provided the temperature is low enough.  While finite volumes are
not of physical interest in particle physics, lattice simulations are
necessarily carried out in a finite box.  As shown in
Refs.~\cite{Gasser:1987ah,Gasser:1987zq,Hasenfratz:1989pk}, the ensuing
finite-size effects can be studied on the basis of the effective theory --
{\Ch}PT in the case of QCD -- provided the simulation is close enough to
the continuum limit, the volume is sufficiently large and the explicit
breaking of chiral symmetry generated by the quark masses is sufficiently
small.  Indeed, {\Ch}PT represents a useful tool for the analysis of the
finite-size effects in lattice simulations.

In the remainder of this subsection we collect the relevant {\Ch}PT formulae
that will be used in the two following subsections to extract $SU(2)$ and $SU(3)$
LECs from lattice data.

%%%%%%%%%%%%%%%%%%%%%%%%%%%%%%%%%%%%%%%%%%%%%%%%%%%%%%%%%%%%%%%%%%%%%%%%%%%%%%%

\subsubsection{Quark-mass dependence of pseudoscalar masses and decay
constants\label{sec_MF}}

\noindent A. $SU(2)$ formulae
\vskip 0.3cm

\noindent The expansions%
\footnote{Here and in the following, we stick to the notation used in the
papers where the {\Ch}PT formulae were established, i.e.\ we work with
$F_\pi\equiv f_\pi/\sqrt{2}=92.2(1)\MeV$ and $F_K\equiv f_K/\sqrt{2}$. The
occurrence of different normalization conventions is not convenient, but
avoiding it by reformulating the formulae in terms of $f_\pi$, $f_K$ is not
a good way out. Since we are using different symbols, confusion cannot
arise.  \label{foot:fpi}}
of $\Mpi^2$ and $F_\pi$ in powers of the quark mass are known to
next-to-next-to-leading order (NNLO) in the $SU(2)$ chiral effective theory.
In the isospin limit, $m_u=m_d=m$, the explicit expressions may be written
in the form \cite{Colangelo:2001df}
\begin{eqnarray}
\Mpi^2 & = & M^2\left\{1-\frac{1}{2}x\ln\frac{\Lambda_3^2}{M^2}
  +\frac{17}{8}x^2 \left(\ln\frac{\Lambda_M^2}{M^2}  \right)^2 +x^2 k_M
  +\cO(x^3)             \right\},
\label{eq:MF}
\\
F_\pi & = & F\left\{1+x\ln\frac{\Lambda_4^2}{M^2} -\frac{5}{4}x^2
  \left(\ln\frac{\Lambda_F^2}{M^2}  \right)^2 +x^2k_F   +\cO(x^3)
\right\}.
\nonumber
\end{eqnarray}
Here the expansion parameter is given by
\begin{equation}
x=\frac{M^2}{(4\pi F)^2},\;\;\;\;\;\;\;\;\;\;M^2=2Bm=\frac{2\Sigma m}{F^2},
\label{eq:xM2}
\end{equation}
but there is another option as discussed below.
The scales $\Lambda_3,\Lambda_4$ are related to the effective coupling
constants $\lbar_3,\lbar_4$ of the chiral Lagrangian at scale
$\Mpi \equiv \Mpi^\mr{phys}$ by
\begin{equation}
\lbar_n=\ln\frac{\Lambda_n^2}{\Mpi^2},\;\;\;\;\;\;\;\;\;\;\;n=1,...,7.
\end{equation}
Note that in Eq.\,(\ref{eq:MF}) the logarithms are evaluated at $M^2$, not at
$\Mpi^2$.
The coupling constants $k_M,k_F$ in Eq.\,(\ref{eq:MF}) are mass-independent.
The scales of the squared logarithms can be expressed in terms of the
$\cO(p^4)$ coupling constants as
\begin{eqnarray}
  \ln\frac{\Lambda_M^2}{M^2} & = &
  \frac{1}{51}\left(28\ln\frac{\Lambda_1^2}{M^2}
    +32\ln\frac{\Lambda_2^2}{M^2}    -9 \ln\frac{\Lambda_3^2}{M^2}+49
  \right),
\\
  \ln\frac{\Lambda_F^2}{M^2} & = &
  \frac{1}{30}\left(14\ln\frac{\Lambda_1^2}{M^2}
    +16\ln\frac{\Lambda_2^2}{M^2}    +6 \ln\frac{\Lambda_3^2}{M^2}
    - 6 \ln\frac{\Lambda_4^2}{M^2}      +23  \right).
\nonumber
\end{eqnarray}
Hence by analysing the quark-mass dependence of $\Mpi^2$ and $F_\pi$ with
Eq.\,(\ref{eq:MF}), possibly truncated at NLO, one can determine%
\footnote{Notice that one could analyse the quark-mass dependence entirely in
terms of the parameter $M^2$ defined in Eq.\,(\ref{eq:xM2}) and determine
equally well all other LECs. Using the determination of the quark masses
described in Sec.~\ref{sec:qmass} one can then extract $B$ or $\Sigma$.
No matter the strategy of extraction, determination of $B$ or $\Sigma$
requires knowledge of the scale and scheme dependent quark mass
renormalization factor $Z_m(\mu)$.}
the $\cO(p^2)$ LECs $B$ and $F$, as well as the $\cO(p^4)$ LECs $\bar \ell_3$
and $\bar \ell_4$.  The quark condensate in the chiral limit is given by
$\Sigma=F^2B$.  With precise enough data at several small enough pion
masses, one could in principle also determine $\Lambda_M$, $\Lambda_F$ and
$k_M$, $k_F$.  To date this is not yet possible.  The results for the LO
and NLO constants will be presented in Sec.~\ref{sec:SU2results}.

Alternatively, one can invert Eq.\,(\ref{eq:MF}) and express $M^2$ and $F$ as
an expansion in
\be
\xi \equiv \frac{\Mpi^2}{16 \pi^2 F_\pi^2} \; \; ,
\label{eq:xi}
\ee
and the corresponding expressions then take the form
\bea
\label{eq:MpiFpi}
M^2&=& \Mpi^2\,\left\{
1+\frac{1}{2}\,\xi\,\lthreebar-
\frac{5}{8}\,\xi^2 \left(\!\lMbar\!\right)^2+
\xi^2 c_{\ind M}+\cO(\xi^3)\right\} \co
\\
F&=& F_\pi\,\left\{1-\xi\,\lfourbar-\frac{1}{4}\,\xi^2
\left(\!\lFbar\!\right)^2
+\xi^2 c_{\ind F}+\cO(\xi^3)\right\} \fs \nn
\eea
The scales of the quadratic logarithms are determined by
$\Lambda_1,\ldots,\Lambda_4$ through
\bea
\lMbar&=&\frac{1}{15}\left(28\,\lonebar+32\,\ltwobar-
33\,\lthreebar-12\,\lfourbar +52\right) \co \\
\lFbar&=&\frac{1}{3}\,\left(-7\,\lonebar-8\,\ltwobar+
18\,\lfourbar- \frac{29}{2}\right)\nonumber \fs
\eea

\noindent B. $SU(3)$ formulae
\vskip 0.3cm

\noindent While the formulae for the pseudoscalar masses and decay
constants are known to NNLO for $SU(3)$ as well
\cite{Amoros:1999dp}, they are rather complicated and we restrict ourselves
here to next-to-leading order (NLO).  In the isospin limit, the relevant $SU(3)$
formulae take the form \cite{Gasser:1984gg}
\bea
\Mpi^2\!\!&\!\!\NLo\!\!&\!\! 2B_0m_{ud}
\Big\{
1+\mu_\pi-\frac{1}{3}\mu_\et+\frac{B_0}{F_0^2}
\Big[16m_{ud}(2L_8\!-\!L_5)+16(m_s\!+\!2m_{ud})(2L_6\!-\!L_4)\Big]
\Big\}\;,\nn
\\
M_{\!K}^2\!\!&\!\!\NLo\!\!&\!\! B_0(m_s\!\!+\!m_{ud})
\Big\{
1\!+\!\frac{2}{3}\mu_\et\!+\!\frac{B_0}{F_0^2}
\Big[8(m_s\!\!+\!m_{ud})(2L_8\!-\!L_5)\!+\!16(m_s\!\!+\!2m_{ud})(2L_6\!-\!L_4)\Big]
\Big\}\;,\quad\nonumber
\\
\Fpi\!\!&\!\!\NLo\!\!&\!\!F_0
\Big\{
1-2\mu_\pi-\mu_K+\frac{B_0}{F_0^2}
\Big[8m_{ud}L_5+8(m_s\!+\!2m_{ud})L_4\Big]
\Big\}\;,
\\
\Fka\!\!&\!\!\NLo\!\!&\!\!F_0
\Big\{
1-\frac{3}{4}\mu_\pi-\frac{3}{2}\mu_K-\frac{3}{4}\mu_\et+\frac{B_0}{F_0^2}
\Big[4(m_s\!+\!m_{ud})L_5+8(m_s\!+\!2m_{ud})L_4\Big]
\Big\}\;,\nonumber
\eea
where $m_{ud}$ is the common up and down quark mass (which may be different
from the one in the real world), and $B_0=\Sigma_0/F_0^2$, $F_0$ denote the
condensate parameter and the pseudoscalar decay constant in the $SU(3)$ chiral
limit, respectively.
In addition, we use the notation
\beq
\mu_P=\frac{M_P^2}{32\pi^2F_0^2}
\ln\!\Big(\frac{M_P^2}{\mu^2}\Big)\;.
\label{def_muP}
\eeq
At the order of the chiral expansion used in these formulae, the quantities
$\mu_\pi$, $\mu_K$, $\mu_\eta$ can equally well be evaluated with the
leading-order expressions for the masses,
\beq
\Mpi^2\Lo 2B_0\,m_{ud}\;,\quad
M_K^2\Lo B_0(m_s\!+\!m_{ud})\;,\quad
M_\et^2\Lo \mbox{$\frac{2}{3}$}B_0(2m_s\!+\!m_{ud})
\;.
\eeq
Throughout, $L_i$ denotes the renormalized low-energy constant/coupling
(LEC) at scale $\mu$, and we adopt the convention which is standard in
phenomenology, $\mu=M_\rho=770\MeV$.  The normalization used for the decay
constants is specified in footnote \ref{foot:fpi}.

%%%%%%%%%%%%%%%%%%%%%%%%%%%%%%%%%%%%%%%%%%%%%%%%%%%%%%%%%%%%%%%%%%%%%%%%%%%%%%%

\subsubsection{Pion form factors and charge radii\label{sec:pion_form}}

The scalar and vector form factors of the pion are defined by the matrix
elements
\bea
\langle \pi^i(p_2) |\, \qbar\, q  \, | \pi^k(p_1) \rangle \al = \al
\delta^{ik} F_S^\pi(t) \co
\\
\langle \pi^i(p_2) | \,\qbar\, \mbox{$\frac{1}{2}$}\tau^j \gamma^\mu q\,| \pi^k(p_1)
\rangle \al = \al
\mr{i} \,\epsilon^{ijk} (p_1^\mu + p_2^\mu) F_V^\pi(t) \co\nonumber
\eea
where the operators contain only the lightest two quark flavours, i.e.\
$\tau^1$, $\tau^2$, $\tau^3$ are the Pauli matrices, and $t\equiv (p_1-p_2)^2$
denotes the momentum transfer.

The vector form factor has been measured by several experiments for
time-like as well as for space-like values of $t$.  The scalar form factor is
not directly measurable, but it can be evaluated theoretically from data on
the $\pi \pi$ and $\pi K$ phase shifts \cite{Donoghue:1990xh} by means of
analyticity and unitarity, i.e.\ in a model-independent way.  Lattice
calculations can be compared with data or model-independent theoretical
evaluations at any given value of $t$.  At present, however, most lattice
studies concentrate on the region close to $t=0$ and on the evaluation of
the slope and curvature which are defined as
\begin{eqnarray}
  F^\pi_V(t) & = & 1+\mbox{$\frac{1}{6}$}\langle r^2 \rangle^\pi_V t +
c_V\hspace{0.025cm} t^2+\ldots \;\co
\\
  F^\pi_S(t) & = & F^\pi_S(0) \left[1+\mbox{$\frac{1}{6}$}\langle r^2
    \rangle^\pi_S t + c_S\, t^2+ \ldots \right] \; \; . \nn
\end{eqnarray}
The slopes are related to the mean-square vector and scalar radii which are
the quantities on which most experiments and lattice calculations
concentrate.

In {\Ch}PT, the form factors are known at NNLO for $SU(2)$
\cite{Bijnens:1998fm}.
The corresponding formulae are available in fully analytical form and are
compact enough that they can be used for the chiral extrapolation of the
data (as done, for example in Refs.~\cite{Frezzotti:2008dr,Kaneko:2008kx}).  The
expressions for the scalar and vector radii and for the $c_{S,V}$
coefficients at two-loop level read
\bea
\langle r^2 \rangle^\pi_S &=& \frac{1}{(4\pi\Fpi)^2} \left\{6 \lfourbar-\frac{13}{2}
-\frac{29}{3}\,\xi \left(\!\ln\frac{\Omega_{r_S}^2}{\Mpi^2} \!\right)^2+ 6
\xi \, k_{r_S}+\cO(\xi^2)\right\} \co\nn
\\
\langle r^2 \rangle^\pi_V &=& \frac{1}{(4\pi\Fpi)^2} \left\{ \lsixbar-1
+2\,\xi \left(\!\ln\frac{\Omega_{r_V}^2}{\Mpi^2} \!\right)^2+6 \xi \,k_{r_V}+\cO(\xi^2)\right\}\co
\label{formula_rsqu}
\\
c_S &=&\frac{1}{(4\pi\Fpi\Mpi)^2} \left\{\frac{19}{120}  + \xi \left[ \frac{43}{36} \left(\!
      \ln\frac{\Omega_{c_S}^2}{\Mpi^2} \!\right)^2 + k_{c_S} \right]
\right\} \co \nn
\\
c_V &=&\frac{1}{(4\pi\Fpi\Mpi)^2} \left\{\frac{1}{60}+\xi \left[\frac{1}{72} \left(\!
      \ln\frac{\Omega_{c_V}^2}{\Mpi^2} \!\right)^2 + k_{c_V} \right]
\right\} \co \nn
\eea
where
\bea
\ln\frac{\Omega_{r_S}^2}{\Mpi^2}&=&\frac{1}{29}\,\left(31\,\lonebar+34\,\ltwobar -36\,\lfourbar
  +\frac{145}{24}\right)  \co \nn\\
\ln\frac{\Omega_{r_V}^2}{\Mpi^2}&=&\frac{1}{2}\,\left(\lonebar-\ltwobar+\lfourbar+\lsixbar
-\frac{31}{12}\right) \co  \\
\ln\frac{\Omega_{c_S}^2}{\Mpi^2}&=&\frac{43}{63}\,\left(11\,\lonebar+14\,\ltwobar+18\,\lfourbar
  -\frac{6041}{120}\right)
\co \nn \\
\ln\frac{\Omega_{c_V}^2}{\Mpi^2}&=&\frac{1}{72}\,\left(2\lonebar-2\ltwobar-\lsixbar
-\frac{26}{30}\right) \co \nn
\eea
and $k_{r_S},k_{r_V}$ and $k_{c_S},k_{c_V}$ are independent of the quark
masses.  Their expression in terms of the $\ell_i$ and of the $\cO(p^6)$
constants $c_M,c_F$ is known but will not be reproduced here.

The $SU(3)$ formula for the slope of the pion vector form factor reads, to
NLO \cite{Gasser:1984ux},
\beq
\<r^2\>_V^\pi\;\NLo\;-\frac{1}{32\pi^2F_0^2}
\Big\{
3+2\ln\frac{\Mpi^2}{\mu^2}+\ln\frac{\Mka^2}{\mu^2}
\Big\}
+\frac{12L_9}{F_0^2}
\;,
\eeq
while the expression $\<r^2\>_S^\mathrm{oct}$ for the octet part of the
scalar radius does not contain any NLO low-energy constant at one-loop
order \cite{Gasser:1984ux} -- contrary to the situation in $SU(2)$, see
Eq.\,(\ref{formula_rsqu}).

The difference between the quark-line connected and the full (i.e.\
containing the connected and the disconnected pieces) scalar pion form
factor has been investigated by means of {\Ch}PT in
Ref.~\cite{Juttner:2011ur}.  It is expected that the technique used can be
applied to a large class of observables relevant in QCD phenomenology.

As a point of practical interest let us remark that there are no
finite-volume correction formulae for the mean-square radii $\<r^2\>_{V,S}$
and the curvatures $c_{V,S}$.  The lattice data for $F_{V,S}(t)$ need to be
corrected, point by point in $t$, for finite-volume effects.  In fact, if a
given $t$ is realized through several inequivalent $p_1\!-\!p_2$
combinations, the level of agreement after the correction has been applied
is indicative of how well higher-order effects are under control.

%%%%%%%%%%%%%%%%%%%%%%%%%%%%%%%%%%%%%%%%%%%%%%%%%%%%%%%%%%%%%%%%%%%%%%%%%%%%%%%

\subsubsection{Partially quenched and mixed action formulations}

The term ``partially quenched QCD'' is used in two ways.  For heavy quarks
($c,b$ and sometimes $s$) it usually means that these flavours are included
in the valence sector, but not into the functional determinant, i.e.\ the
sea sector.  For the light quarks ($u,d$ and sometimes $s$) it means that
they are present in both the valence and the sea sector of the theory, but
with different masses (e.g.\ a series of valence quark masses is evaluated
on an ensemble with fixed sea-quark masses).

The program of extending the standard (unitary) $SU(3)$ theory to the (second
version of) ``partially quenched QCD'' has been completed at the two-loop
(NNLO) level for masses and decay constants~\cite{Bijnens:2006jv}.  These
formulae tend to be complicated, with the consequence that a
state-of-the-art analysis with $\cO(2000)$ bootstrap samples on $\cO(20)$
ensembles with $\cO(5)$ masses each [and hence $\cO(200\,000)$ different fits]
will require significant computational resources.  For
a summary of recent developments in {\Ch}PT
relevant to lattice QCD we refer to Ref.~\cite{Bijnens:2011tb}.
The $SU(2)$ partially quenched formulae can be obtained from the $SU(3)$ ones
by ``integrating out the strange quark.'' At NLO, they can be found in
Ref.~\cite{Du:2009ih} by setting the lattice artifact terms from the staggered
{\Ch}PT form to zero.

The theoretical underpinning of how ``partial quenching'' is to be understood
in the (properly extended) chiral framework is given in
Ref.~\cite{Bernard:2013kwa}. Specifically, for partially quenched QCD with
staggered quarks it is shown that a transfer matrix can be constructed
which is not Hermitian but bounded, and can thus be used to construct
correlation functions in the usual way.
The program of calculating all observables in the $p$-regime in finite-volume to two
loops, first completed in the unitary theory \cite{Bijnens:2013doa,Bijnens:2014dea},
has been carried out for the partially quenched case, too \cite{Bijnens:2015dra}.

A further extension of the {\Ch}PT framework concerns the lattice effects
that arise in partially quenched simulations where sea and valence quarks
are implemented with different lattice fermion actions
\cite{Bar:2002nr,Bar:2003mh,Bar:2005tu,Chen:2009su,Bae:2010ki,Bailey:2012wb,Bernard:2013eya,Bailey:2015zga}.

%%%%%%%%%%%%%%%%%%%%%%%%%%%%%%%%%%%%%%%%%%%%%%%%%%%%%%%%%%%%%%%%%%%%%%%%%%%%%%%

\subsubsection{Correlation functions in the $\epsilon$-regime\label{sec_eps}}

\def\ltap{\raisebox{-.4ex}{\rlap{$\sim$}} \raisebox{.4ex}{$<$}}   % < or ~
\def\gtap{\raisebox{-.4ex}{\rlap{$\sim$}} \raisebox{.4ex}{$>$}}   % > or ~

The finite-size effects encountered in lattice calculations can be used
to determine some of the LECs of QCD. In order to illustrate this point,
we focus on the two lightest quarks, take the isospin limit $m_u=m_d=m$
and consider a box of size $L_s$ in the three space directions and size
$L_t$ in the time direction. If $m$ is sent to zero at fixed box size,
chiral symmetry is restored, and the zero-momentum mode of the pion field
becomes nonperturbative. An intuitive way to understand the regime with
$ML<1$ ($L=L_s\,\ltap\,L_t$) starts from considering the pion propagator
$G(p)=1/(p^2+M^2)$ in finite volume. For $ML\,\gtap\,1$ and $p\sim 1/L$,
$G(p)\sim L^2$ for small momenta, including $p=0$. But when $M$ becomes
of order $1/L^2$, $G(0)\propto L^4\gg G(p\ne 0)\sim L^2$. The $p=0$ mode
of the pion field becomes nonperturbative, and the integration over this
mode restores chiral symmetry in the limit $m\to 0$.

The pion effective action for the zero-momentum field depends only on the
combination $\mu=m\Sigma V$, the symmetry-restoration parameter, where
$V=L_s^3 L_t$. In the $\epsilon$-regime, in which $m\sim 1/V$, all other
terms in the effective action are sub-dominant in powers of $\epsilon\sim 1/L$,
leading to a reordering of the usual chiral expansion, which assumes that
$m\sim\epsilon^2$ instead of $m\sim\epsilon^4$. In the $p$-regime, with
$m\sim\epsilon^2$ or equivalently $ML\,\gtap\, 1$, finite-volume corrections
are of order $\int d^4p\,e^{ipx}\,G(p)|_{x\sim L}\sim e^{-ML}$, while in the
$\epsilon$-regime, the chiral expansion is an expansion in powers of
$1/(\Lambda_\mathrm{QCD}L)\sim 1/(FL)$.

As an example, we consider the correlator of the axial charge carried by
the two lightest quarks, $q(x)=\{u(x),d(x)\}$.  The axial current and the
pseudoscalar density are given by
\be
A_\mu^i(x)=
\qbar(x)\mbox{$\frac{1}{2}$} \tau^i\,\gamma_\mu\gamma_5\,q(x)\,,
\hspace{1cm}P^i(x) = \qbar(x)\mbox{$\frac{1}{2}$} \tau^i\,\mr{i} \gamma_5\,q(x)\,,
\ee
where $\tau^1, \tau^2,\tau^3$ are the Pauli matrices in flavour space.  In
Euclidean space, the correlators of the axial charge and of the space
integral over the pseudoscalar density are given by
\begin{eqnarray}\label{eq:correlators}
\delta^{ik}C_{AA}(t)\al = \al L_s^3\int \hspace{-0.12cm}d^3\hspace{-0.04cm}\vec{x}\;\langle A_4^i(\vec{x},t)
A_4^k(0)\rangle\,,
\\
\delta^{ik}C_{PP}(t)\al  =\al L_s^3\int \hspace{-0.12cm}d^3\hspace{-0.04cm}\vec{x}\;\langle P^i(\vec{x},t)
P^k(0)\rangle\,.\nonumber
\end{eqnarray}
{\Ch}PT yields explicit finite-size scaling formulae for these quantities
\cite{Hasenfratz:1989pk,Hansen:1990un,Hansen:1990yg}.  In the
$\epsilon$-regime, the expansion starts with
\begin{eqnarray}  \label{aa-eps}
C_{AA}(t) \al = \al \frac{F^2L_s^3}{L_t}\left[a_A+
  \frac{L_t}{F^2L_s^3}\,b_A\,h_1\hspace{-0.1cm}\left(\frac{t}{L_t}  \right)
+\cO(\epsilon^4)\right],
\\
C_{PP}(t) \al = \al
\Sigma^2L_s^6\left[a_P+\frac{L_t}{F^2L_s^3}\,b_P\,h_1\hspace{-0.1cm}\left(\frac{t}{L_t}  \right)
+\cO(\epsilon^4)\right],\nonumber
\end{eqnarray}
where the coefficients $a_A$, $b_A$, $a_P$, $b_P$ stand for quantities of
$\cO(\ep^0)$.
They can be expressed in terms of the variables $L_s$, $L_t$ and $m$ and
involve only the two leading low-energy constants $F$ and $\Sigma$.  In
fact, at leading order only the combination $\mu=m\,\Sigma\,L_s^3 L_t$
matters, the correlators are $t$-independent and the dependence on $\mu$ is
fully determined by the structure of the groups involved in the 
pattern of spontaneous symmetry breaking.  In the case of $SU(2)\times SU(2)$ $\rightarrow$ $SU(2)$, relevant
for QCD in the symmetry restoration region with two light quarks, the
coefficients can be expressed in terms of Bessel functions.  The
$t$-dependence of the correlators starts showing up at $\cO(\ep^2)$, in the
form of a parabola, viz.\
$h_1(\tau)=\frac{1}{2}\left[\left(\tau-\frac{1}{2} \right)^2-\frac{1}{12}
\right]$.  Explicit expressions for $a_A$, $b_A$, $a_P$, $b_P$ can be found
in Refs.~\cite{Hasenfratz:1989pk,Hansen:1990un,Hansen:1990yg}, where some of the
correlation functions are worked out to NNLO.  By matching the finite-size
scaling of correlators computed on the lattice with these predictions one
can extract $F$ and $\Sigma$.  A way to deal with the numerical challenges
germane to the $\ep$-regime has been described \cite{Giusti:2004yp}.

The fact that the representation of the correlators to NLO is not
``contaminated'' by higher-order unknown LECs, makes the $\ep$-regime
potentially convenient for a clean extraction of the LO couplings.  The
determination of these LECs is then affected by different systematic
uncertainties with respect to the standard case; simulations in this regime
yield complementary information which can serve as a valuable cross-check
to get a comprehensive picture of the low-energy properties of QCD.

The effective theory can also be used to study the distribution of the
topological charge in QCD \cite{Leutwyler:1992yt} and the various
quantities of interest may be defined for a fixed value of this charge. The
expectation values and correlation functions then not only depend on the
symmetry restoration parameter $\mu$, but also on the topological charge
$\nu$. The dependence on these two variables can explicitly be calculated.
It turns out that the two-point correlation functions considered above
retain the form (\ref{aa-eps}), but the coefficients $a_A$, $b_A$, $a_P$,
$b_P$ now depend on the topological charge as well as on the symmetry
restoration parameter (see
Refs.~\cite{Damgaard:2001js,Damgaard:2002qe,Aoki:2009mx} for explicit
expressions).

A specific issue with $\ep$-regime calculations is the scale setting.
Ideally one would perform a $p$-regime study with the same bare parameters
to measure a hadronic scale (e.g.\ the proton mass).  In the literature,
sometimes a gluonic scale (e.g.\ $r_0$) is used to avoid such expenses.
Obviously the issues inherent in scale setting are aggravated if the
$\ep$-regime simulation is restricted to a fixed sector of topological
charge.

It is important to stress that in the $\epsilon$-expansion higher-order
finite-volume corrections might be significant, and the physical box size
(in fm) should still be large in order to keep these distortions under
control.  The criteria for the chiral extrapolation and finite-volume
effects are obviously different with respect to the $p$-regime.  For these
reasons we have to adjust the colour coding defined in
Sec.\,\ref{sec:color-code} (see Sec.\,\ref{sec:SU2results} for more
details).

Recently, the effective theory has been extended to the ``mixed regime''
where some quarks are in the $p$-regime and some in the $\ep$-regime
\cite{Bernardoni:2008ei,Hernandez:2012tw}.  In Ref.~\cite{Damgaard:2008zs} a
technique is proposed to smoothly connect the $p$- and $\ep$-regimes.  In
Ref.~\cite{Aoki:2011pza} the issue is reconsidered with a counting rule which is
essentially the same as in the $p$-regime. In this new scheme, one can
treat the IR fluctuations of the zero-mode nonperturbatively,
while keeping the logarithmic quark mass dependence of the $p$-regime.

Also first steps towards calculating higher $n$-point functions in the
$\ep$-regime have been taken. For instance the electromagnetic pion form
factor in QCD has been calculated to NLO in the $\ep$-expansion, and a way
to get rid of the pion zero-momentum part has been proposed \cite{Fukaya:2014bna}.

%%%%%%%%%%%%%%%%%%%%%%%%%%%%%%%%%%%%%%%%%%%%%%%%%%%%%%%%%%%%%%%%%%%%%%%%%%%%%%%

\subsubsection{Energy levels of the QCD Hamiltonian in a box and $\delta$-regime\label{sec_su2_delta}}

At low temperature, the properties of the partition function are governed
by the lowest eigenvalues of the Hamiltonian.  In the case of QCD, the
lowest levels are due to the Nambu-Goldstone bosons and can be worked out
with {\Ch}PT \cite{Leutwyler:1987ak}.  In the chiral limit the level
pattern follows the one of a quantum-mechanical rotator, i.e.\
$E_\ell=\ell(\ell+1)/(2\,\Theta)$ with $\ell = 0, 1,2,\ldots$.  For a cubic
spatial box and to leading order in the expansion in inverse powers of the
box size $L_s$, the moment of inertia is fixed by the value of the pion
decay constant in the chiral limit, i.e.\ $\Theta=F^2L_s^3$.

In order to analyse the dependence of the levels on the quark masses and on
the parameters that specify the size of the box, a reordering of the chiral
series is required, the so-called $\delta$-expansion; the region where the
properties of the system are controlled by this expansion is referred to as
the $\delta$-regime. Evaluating the chiral series in this
regime, one finds that the expansion of the partition function goes in even
inverse powers of $FL_s$, that the rotator formula for the energy levels
holds up to NNLO and the expression for the moment of inertia is now also
known up to and including terms of order $(FL_s)^{-4}$
\cite{Hasenfratz:2009mp,Niedermayer:2010mx,Weingart:2010yv}. Since the
level spectrum is governed by the value of the pion decay constant in the
chiral limit, an evaluation of this spectrum on the lattice can be used to
measure $F$. More generally, the evaluation of various observables in the
$\delta$-regime offers an alternative method for a determination of some of
the low-energy constants occurring in the effective Lagrangian. At present,
however, the numerical results obtained in this way
\cite{Hasenfratz:2006xi,Bietenholz:2010az} are not yet competitive with
those found in the $p$- or $\epsilon$-regime.

%%%%%%%%%%%%%%%%%%%%%%%%%%%%%%%%%%%%%%%%%%%%%%%%%%%%%%%%%%%%%%%%%%%%%%%%%%%%%%%

\subsubsection{Other methods for the extraction of the low-energy constants\label{sec_su2_extra}}

An observable that can be used to extract LECs is the topological
susceptibility
\begin{equation}
\chi_t=\int d^4\!x\; \langle \omega(x) \omega(0)\rangle,
\end{equation}
where $\omega(x)$ is the topological charge density,
\begin{equation}
\omega(x)=\frac{1}{32\pi^2}
\epsilon^{\mu\nu\rho\sigma}{\rm Tr}\left[F_{\mu\nu}(x)F_{\rho\sigma}(x)\right].
\end{equation}
At infinite volume, the expansion of $\chi_t$ in powers of the quark masses
starts with \cite{DiVecchia:1980ve}
\begin{equation}\label{chi_t}
\chi_t=\overline{m}\,\Sigma \,\{1+\cO(m)\}\,,\hspace{2cm}
\overline{m}\equiv\left(
\frac{1}{m_u}+\frac{1}{m_d}+\frac{1}{m_s}+\ldots
\right)^{-1}.
\end{equation}
The condensate $\Sigma$ can thus be extracted from the properties of the
topological susceptibility close to the chiral limit.  The behaviour at
finite volume, in particular in the region where the symmetry is restored,
is discussed in Ref.~\cite{Hansen:1990yg}.  The dependence on the vacuum angle
$\theta$ and the projection on sectors of fixed $\nu$ have been studied in
Ref.~\cite{Leutwyler:1992yt}.  For a discussion of the finite-size effects at
NLO, including the dependence on $\theta$, we refer to
Refs.~\cite{Mao:2009sy,Aoki:2009mx}.

The role that the topological susceptibility plays in attempts to determine
whether there is a large paramagnetic suppression when going from the
$\Nf=2$ to the $\Nf=2+1$ theory has been highlighted in
Ref.\,\cite{Bernard:2012fw}.  And the potential usefulness of higher
moments of the topological charge distribution to determine LECs has been
investigated in Ref.~\cite{Bernard:2012ci}.

Another method for computing the quark condensate has been proposed in
Ref.~\cite{Giusti:2008vb}, where it is shown that starting from the Banks-Casher
relation \cite{Banks:1979yr} one may extract the condensate from suitable
(renormalizable) spectral observables, for instance the number of Dirac
operator modes in a given interval.  For those spectral observables
higher-order corrections can be systematically computed in terms of the
chiral effective theory.  For recent implementations of this strategy, see
Refs.~\cite{Cichy:2013gja,Engel:2014cka,Engel:2014eea}.  As an
aside let us remark that corrections to the Banks-Casher relation that come
from a finite quark mass, a finite four-dimensional volume and (with
Wilson-type fermions) a finite lattice spacing can be parameterized in a
properly extended version of the chiral framework \cite{Sharpe:2006ia,Necco:2013sxa}.

An alternative strategy is based on the fact that at LO in the
$\ep$-expansion the partition function in a given topological sector $\nu$
is equivalent to the one of a chiral Random Matrix Theory (RMT)
\cite{Shuryak:1992pi,Verbaarschot:1993pm,Verbaarschot:1994qf,Verbaarschot:2000dy}.
In RMT it is possible to extract the probability distributions of
individual eigenvalues
\cite{Nishigaki:1998is,Damgaard:2000ah,Basile:2007ki} in terms of two
dimensionless variables $\zeta=\lambda\Sigma V$ and $\mu=m\Sigma V$, where
$\lambda$ represents the eigenvalue of the massless Dirac operator and $m$
is the sea quark mass.  More recently this approach has been extended to
the Hermitian (Wilson) Dirac operator \cite{Kieburg:2013tca} which is
easier to study in numerical simulations.  Hence, if it is possible to
match the QCD low-lying spectrum of the Dirac operator to the RMT
predictions, then one may extract%
\footnote{By introducing an imaginary isospin chemical potential, the
framework can be extended such that the low-lying spectrum of the Dirac
operator is also sensitive to the pseudoscalar decay constant $F$ at LO
\cite{Akemann:2006ru}.}
the chiral condensate $\Sigma$.  One issue with this method is that for the
distributions of individual eigenvalues higher-order corrections are still
not known in the effective theory, and this may introduce systematic
effects which are hard%
\footnote{Higher-order systematic effects in the matching with RMT have
been investigated in Refs.~\cite{Lehner:2010mv,Lehner:2011km}.}
to control.  Another open question is that, while it is clear how the
spectral density is renormalized \cite{DelDebbio:2005qa}, this is not the
case for the individual eigenvalues, and one relies on assumptions.  There
have been many lattice studies \cite{Fukaya:2007yv,Lang:2006ab,
DeGrand:2006nv,Hasenfratz:2007yj,DeGrand:2007tm} which investigate the
matching of the low-lying Dirac spectrum with RMT.  In this review the
results of the LECs obtained in this way%
\footnote{The results for
$\Sigma$ and $F$ lie in the same range as the determinations reported in
Tables~\ref{tab:sigma} and \ref{tab:f}.}
are not included.

%%%%%%%%%%%%%%%%%%%%%%%%%%%%%%%%%%%%%%%%%%%%%%%%%%%%%%%%%%%%%%%%%%%%%%%%%%%%%%%

%\subsubsection{Gradient flow observables}

%Recently purely gluonic observables defined at fixed flow time have drawn a lot
%of attention \cite{Luscher:2010iy,Luscher:2011bx}. The gradient flow
%suppresses UV fluctuations, irrelevant for low-energy physics.
%The chiral Lagrangian for scalar and pseudoscalar densities defined with the
%gradient flow has been constructed at NLO, and it has been established that the
%non-analytic terms in the quark mass dependence of the scales
%$t_0$ \cite{Luscher:2010iy} and $w_0$ \cite{Borsanyi:2012zs},
%which are defined from the scalar gluon density, enter at NNLO
%order \cite{Bar:2013ora}.

%%%%%%%%%%%%%%%%%%%%%%%%%%%%%%%%%%%%%%%%%%%%%%%%%%%%%%%%%%%%%%%%%%%%%%%%%%%%%%%

\subsection{Extraction of $SU(2)$ low-energy constants \label{sec:SU2results}}

%%%%%%%%%%%%%%%%%%%%%%%%%%%%%%%%%%%%%%%%%%%%%%%%%%%%%%%%%%%%%%%%%%%%%%%%%%%%%%%

In this and the following subsections we summarize the lattice results for
the $SU(2)$ and $SU(3)$ LECs, respectively.  In either case we first discuss
the $\cO(p^2)$ constants and then proceed to their $\cO(p^4)$ counterparts.
The $\cO(p^2)$ LECs are determined from the chiral extrapolation of masses
and decay constants or, alternatively, from a finite-size study of
correlators in the $\ep$-regime.  At order $p^4$ some LECs affect two-point
functions while others appear only in three- or four-point functions; the
latter need to be determined from form factors or scattering amplitudes.
The {\Ch}PT analysis of the (nonlattice) phenomenological quantities is
nowadays%
\footnote{Some of the $\cO(p^6)$ formulae presented below have been derived
in an unpublished note by three of us (GC, SD and HL) and J\"urg Gasser. We
thank him for allowing us to publish them here.}
based on $\cO(p^6)$ formulae.  At this level the number of LECs explodes and
we will not discuss any of these.  We will, however, discuss how comparing
different orders and different expansions (in particular the $x$ versus
$\xi$-expansion) can help to assess the theoretical uncertainties of the
LECs determined on the lattice.

The lattice results for the $SU(2)$ LECs are summarized in Tabs.~(\ref{tab:sigma}--\ref{tab:radii}) and 
Figs.~(\ref{fig:sigma}--\ref{fig:l3l4l6}).  The tables present our usual colour
coding which summarizes the main aspects related to the treatment of the
systematic errors of the various calculations.

A delicate issue in the lattice determination of chiral LECs (in particular
at NLO) which cannot be reflected by our colour coding is a reliable
assessment of the theoretical error that comes from the chiral expansion.
We add a few remarks on this point:
\begin{enumerate}
\item
Using \emph{both} the $x$ and the $\xi$ expansion is a good way to test how
the ambiguity of the chiral expansion (at a given order) affects the
numerical values of the LECs that are determined from a particular set of
data \cite{Noaki:2008iy,Durr:2013goa}.
For instance, to determine $\bar{\ell}_4$ (or $\Lambda_4$) from
lattice data for $\Fpi$ as a function of the quark mass, one may compare
the fits based on the parameterisation $\Fpi=F\{1+x\ln(\Lambda_4^2/M^2)\}$
[see Eq.\,(\ref{eq:MF})] with those obtained from
$\Fpi=F/\{1-\xi\ln(\Lambda_4^2/\Mpi^2)\}$ [see Eq.\,(\ref{eq:MpiFpi})].
The difference between the two results provides an estimate of the
uncertainty due to the truncation of the chiral series.  Which central
value one chooses is in principle arbitrary, but we find it advisable to
use the one obtained with the $\xi$ expansion,%
\footnote{There are theoretical arguments suggesting that the $\xi$
expansion is preferable to the $x$ expansion, based on the observation that
the coefficients in front of the squared logs in Eq.\,(\ref{eq:MF}) are somewhat
larger than in Eq.\,(\ref{eq:MpiFpi}). This can be traced to the fact that a
part of every formula in the $x$ expansion is concerned with locating the
position of the pion pole (at the previous order) while in the $\xi$
expansion the knowledge of this position is built in exactly. Numerical
evidence supporting this view is presented in Ref.~\cite{Noaki:2008iy}.}
in particular because it makes the comparison with phenomenological
determinations (where it is standard practice to use the $\xi$ expansion)
more meaningful.
\item
Alternatively one could try to estimate the influence of higher chiral
orders by reshuffling irrelevant higher-order terms.  For instance, in the
example mentioned above one might use $\Fpi=F/\{1-x\ln(\Lambda_4^2/M^2)\}$
as a different functional form at NLO.  Another way to establish such an
estimate is through introducing by hand ``analytical'' higher-order terms
(e.g.\ ``analytical NNLO'' as done, in the past, by MILC
\cite{Bazavov:2009bb}).  In principle it would be preferable to include all
NNLO terms or none, such that the structure of the chiral expansion is
preserved at any order (this is what ETM \cite{Baron:2009wt} and
JLQCD/TWQCD \cite{Noaki:2008iy} have done for $SU(2)$ {\Ch}PT and MILC for
both $SU(2)$ and $SU(3)$ {\Ch}PT
\cite{Bazavov:2009fk,Bazavov:2010yq,Bazavov:2010hj}).  There are different
opinions in the field as to whether it is advisable to include terms to
which the data are not sensitive.  In case one is willing to include
external (typically: nonlattice) information, the use of priors is a
theoretically well founded option (e.g.\ priors for NNLO LECs if one is
interested exclusively in LECs at LO/NLO).
\item
Another issue concerns the $s$-quark mass dependence of the LECs
$\bar{\ell}_i$ or $\Lambda_i$ of the $SU(2)$ framework.  As far as variations
of $m_s$ around $m_s^\mr{phys}$ are concerned (say for
$0<m_s<1.5m_s^\mr{phys}$ at best) the issue can be studied in $SU(3)$
{\Ch}PT, and this has been done in a series of papers
\cite{Gasser:1984gg,Gasser:2007sg,Gasser:2009hr}.  However, the effect of
sending $m_s$ to infinity, as is the case in $\Nf=2$ lattice studies of
$SU(2)$ LECs, cannot be addressed in this way.  A way to analyse this
difference is to compare the numerical values of LECs determined in $\Nf=2$
lattice simulations to those determined in $\Nf=2+1$ lattice simulations
(see e.g. Ref.~\cite{Durr:2013koa} for a discussion).
\item
Last but not least let us recall that the determination of the LECs is
affected by discretisation effects, and it is important that these are
removed by means of a continuum extrapolation.  In this step invoking an
extended version of the chiral Lagrangian
\cite{Rupak:2002sm,Aoki:2003yv,Aubin:2003mg,Aubin:2003uc,Bar:2003mh,Bar:2014bda} may be
useful%
\footnote{This means that for any given lattice formulation one needs to
determine additional lattice-artifact low-energy constants.  For certain
formulations, e.g.\ the twisted-mass approach, first steps in this
direction have already been taken \cite{Herdoiza:2013sla}, while with
staggered fermions MILC routinely does so, see e.g.
Refs.~\cite{Aubin:2004fs,Bazavov:2009bb}.}
in case one aims for a global fit of lattice data involving several $\Mpi$
and $a$ values and several chiral observables.
\end{enumerate}

In the tables and figures we summarize the results of various lattice
collaborations for the $SU(2)$ LECs at LO ($F$ or $F/F_\pi$, $B$ or $\Sigma$)
and at NLO ($\lbar_1-\lbar_2$, $\lbar_3$, $\lbar_4$, $\lbar_6$).
Throughout we group the results into those
which stem from $\Nf=2+1+1$ calculations, those which come from $\Nf=2+1$
calculations and those which stem from $\Nf=2$ calculations (since, as
mentioned above, the LECs are logically distinct even if the current
precision of the data is not sufficient to resolve the differences).
Furthermore, we make a distinction whether the results are obtained from
simulations in the $p$-regime or whether alternative methods ($\ep$-regime,
spectral densities, topological susceptibility, etc.) have been used (this
should not affect the result).  For comparison we add, in each case, a few
representative phenomenological determinations.

A generic comment applies to the issue of the scale setting.  In the past
none of the lattice studies with $\Nf\geq2$ involved simulations in the
$p$-regime at the physical value of $m_{ud}$.  Accordingly, the setting of
the scale $a^{-1}$ via an experimentally measurable quantity did
necessarily involve a chiral extrapolation, and as a result of this
dimensionful quantities used to be particularly sensitive to this
extrapolation uncertainty, while in dimensionless ratios such as $F_\pi/F$,
$F/F_0$, $B/B_0$, $\Sigma/\Sigma_0$ this particular problem is much reduced
(and often finite lattice-to-continuum renormalization factors drop out).
Now, there is a new generation of lattice studies with
$\Nf=2$ \cite{Abdel-Rehim:2015pwa},
$\Nf=2+1$ \cite{Aoki:2009ix,Durr:2010vn,Durr:2010aw,Borsanyi:2012zv,Bazavov:2012xda,Bazavov:2012cd,Arthur:2012opa,Durr:2013goa,Blum:2014tka,Boyle:2015exm}, and
$\Nf=2+1+1$ \cite{Dowdall:2013rya,Koponen:2015tkr},
which does involve simulations at physical pion masses.  In such studies
the uncertainty that the scale setting has on dimensionful quantities
is much mitigated.

It is worth repeating here that the standard colour-coding scheme of our
tables is necessarily schematic and cannot do justice to every calculation.
In particular there is some difficulty in coming up with a fair adjustment
of the rating criteria to finite-volume regimes of QCD.  For instance, in
the $\epsilon$-regime%
\footnote{Also in case of Refs.~\cite{Fukaya:2009fh,Fukaya:2010na} the
colour-coding criteria for the $\epsilon$-regime have been applied.}
we re-express the ``chiral extrapolation'' criterion in terms of
$\sqrt{2m_\mr{min}\Sigma}/F$, with the same threshold values (in MeV)
between the three categories as in the $p$-regime.  Also the ``infinite
volume'' assessment is adapted to the $\ep$-regime, since the $\Mpi L$
criterion does not make sense here; we assign a green star if at least 2
volumes with $L>2.5\,\fm$ are included, an open symbol if at least 1 volume
with $L>2\,\fm$ is invoked and a red square if all boxes are smaller than
$2\,\fm$.  Similarly, in the calculation of form factors and charge radii the
tables do not reflect whether an interpolation to the desired $q^2$ has
been performed or whether the relevant $q^2$ has been engineered by means
of ``twisted boundary conditions'' \cite{Boyle:2008yd}.
In spite of these limitations we feel that these tables give an adequate
overview of the qualities of the various calculations.

%%%%%%%%%%%%%%%%%%%%%%%%%%%%%%%%%%%%%%%%%%%%%%%%%%%%%%%%%%%%%%%%%%%%%%%%%%%%%%%

\subsubsection{Results for the LO $SU(2)$ LECs \label{sec:SU2_LO}}

\begin{table}[!tbp] %%% \Sigma
\vspace*{3cm}
\centering
\footnotesize
\begin{tabular*}{\textwidth}[l]{l@{\extracolsep{\fill}}rlllllll}
Collaboration & Ref. & $\Nf$ &
\hspace{0.15cm}\begin{rotate}{60}{publication status}\end{rotate}\hspace{-0.15cm} &
\hspace{0.15cm}\begin{rotate}{60}{chiral extrapolation}\end{rotate}\hspace{-0.15cm}&
\hspace{0.15cm}\begin{rotate}{60}{continuum  extrapolation}\end{rotate}\hspace{-0.15cm} &
\hspace{0.15cm}\begin{rotate}{60}{finite volume}\end{rotate}\hspace{-0.15cm} &
\hspace{0.15cm}\begin{rotate}{60}{renormalization}\end{rotate}\hspace{-0.15cm} & \rule{0.4cm}{0cm}$\Sigma^{1/3}$ \\[2mm]
\hline
\hline
\\[-2mm]
ETM 13                  & \cite{Cichy:2013gja}       &2+1+1& \gA & \soso & \good & \good & \good & 280(8)(15)                           \\[2mm]
\hline
\\[-2mm]
RBC/UKQCD 15E           & \cite{Boyle:2015exm}       & 2+1 & \oP & \good & \good & \good & \good & 274.2(2.8)(4.0)                      \\
RBC/UKQCD 14B           & \cite{Blum:2014tka}        & 2+1 & \gA & \good & \good & \good & \good & 275.9(1.9)(1.0)                      \\
BMW 13                  & \cite{Durr:2013goa}        & 2+1 & \gA & \good & \good & \good & \good & 271(4)(1)                            \\
Borsanyi 12             & \cite{Borsanyi:2012zv}     & 2+1 & \gA & \good & \good & \good & \good & 272.3(1.2)(1.4)                      \\
MILC 10A                & \cite{Bazavov:2010yq}      & 2+1 & \rC & \soso & \good & \good & \good & 281.5(3.4)$\binom{+2.0}{-5.9}$(4.0)  \\
JLQCD/TWQCD 10A         & \cite{Fukaya:2010na}       & 2+1 & \gA & \good & \bad  & \soso & \good & 234(4)(17)                           \\
RBC/UKQCD 10A           & \cite{Aoki:2010dy}         & 2+1 & \gA & \soso & \soso & \soso & \good & 256(5)(2)(2)                         \\
JLQCD 09                & \cite{Fukaya:2009fh}       & 2+1 & \gA & \good & \bad  & \soso & \good & 242(4)$\binom{+19}{-18}$             \\
MILC 09A, $SU(3)$-fit   & \cite{Bazavov:2009fk}      & 2+1 & \rC & \soso & \good & \good & \good & 279(1)(2)(4)                         \\
MILC 09A, $SU(2)$-fit   & \cite{Bazavov:2009fk}      & 2+1 & \rC & \soso & \good & \good & \good & 280(2)$\binom{+4}{-8}$(4)            \\
MILC 09                 & \cite{Bazavov:2009bb}      & 2+1 & \gA & \soso & \good & \good & \good & 278(1)$\binom{+2}{-3}$(5)            \\
TWQCD 08                & \cite{Chiu:2008jq}         & 2+1 & \gA & \bad  & \bad  & \bad  & \good & 259(6)(9)                            \\
JLQCD/TWQCD 08B         & \cite{Chiu:2008kt}         & 2+1 & \rC & \soso & \bad  & \bad  & \good & 249(4)(2)                            \\
PACS-CS 08, $SU(3)$-fit & \cite{Aoki:2008sm}         & 2+1 & \gA & \good & \bad  & \bad  & \bad  & 312(10)                              \\
PACS-CS 08, $SU(2)$-fit & \cite{Aoki:2008sm}         & 2+1 & \gA & \good & \bad  & \bad  & \bad  & 309(7)                               \\
RBC/UKQCD 08            & \cite{Allton:2008pn}       & 2+1 & \gA & \soso & \bad  & \soso & \good & 255(8)(8)(13)                        \\[2mm]
\hline
\\[-2mm]
Engel 14                & \cite{Engel:2014eea}       &  2  & \gA & \good & \good & \good & \good & 263(3)(4)                            \\
Brandt 13               & \cite{Brandt:2013dua}      &  2  & \gA & \soso & \good & \soso & \good & 261(13)(1)                           \\
ETM 13                  & \cite{Cichy:2013gja}       &  2  & \gA & \soso & \good & \soso & \good & 283(7)(17)                           \\
ETM 12                  & \cite{Burger:2012ti}       &  2  & \gA & \soso & \good & \soso & \good & 299(26)(29)                          \\
Bernardoni 11           & \cite{Bernardoni:2011kd}   &  2  & \rC & \soso & \bad  & \bad  & \good & 306(11)                              \\
TWQCD 11                & \cite{Chiu:2011bm}         &  2  & \gA & \soso & \bad  & \bad  & \good & 230(4)(6)                            \\
TWQCD 11A               & \cite{Chiu:2011dz}         &  2  & \gA & \soso & \bad  & \bad  & \good & 259(6)(7)                            \\
JLQCD/TWQCD 10A         & \cite{Fukaya:2010na}       &  2  & \gA & \good & \bad  & \bad  & \good & 242(5)(20)                           \\
Bernardoni 10           & \cite{Bernardoni:2010nf}   &  2  & \gA & \soso & \bad  & \bad  & \good & 262$\binom{+33}{-34}\binom{+4}{-5}$  \\
ETM 09C                 & \cite{Baron:2009wt}        &  2  & \gA & \soso & \good & \soso & \good & 270(5)$\binom{+3}{-4}$               \\
ETM 09B                 & \cite{Jansen:2009tt}       &  2  & \rC & \good & \bad  & \soso & \good & 245(5)                               \\
ETM 08                  & \cite{Frezzotti:2008dr}    &  2  & \gA & \soso & \soso & \soso & \good & 264(3)(5)                            \\
CERN 08                 & \cite{Giusti:2008vb}       &  2  & \gA & \soso & \bad  & \soso & \good & 276(3)(4)(5)                         \\
Hasenfratz 08           & \cite{Hasenfratz:2008ce}   &  2  & \gA & \soso & \bad  & \good & \good & 248(6)                               \\
JLQCD/TWQCD 08A         & \cite{Noaki:2008iy}        &  2  & \gA & \soso & \bad  & \bad  & \good & 235.7(5.0)(2.0)$\binom{+12.7}{-0.0}$ \\
JLQCD/TWQCD 07          & \cite{Fukaya:2007pn}       &  2  & \gA & \good & \bad  & \bad  & \good & 239.8(4.0)                           \\
JLQCD/TWQCD 07A         & \cite{Aoki:2007pw}         &  2  & \gA & \good & \bad  & \bad  & \good & 252(5)(10)                           \\[2mm]
\hline
\hline
\end{tabular*}
\normalsize
\vspace*{-2mm}
\caption{\label{tab:sigma}
Cubic root of the $SU(2)$ quark condensate $\Sigma\equiv-\langle\ubar u\rangle|_{m_u,m_d\to0}$ in $\MeV$ units, in
the $\overline{\rm MS}$-scheme, at the renormalization scale $\mu=2$ GeV. Horizontal lines separate different $\Nf$.
All ETM values which were available only in $r_0$ units were converted on the basis of $r_0=0.48(2)\,\fm$
\cite{Aoki:2009sc,Bazavov:2014pvz,Abdel-Rehim:2015pwa}, with this error being added in quadrature to any existing systematic error.}
\end{table}

\begin{figure}[!tb]
\centering
\includegraphics[width=12.0cm]{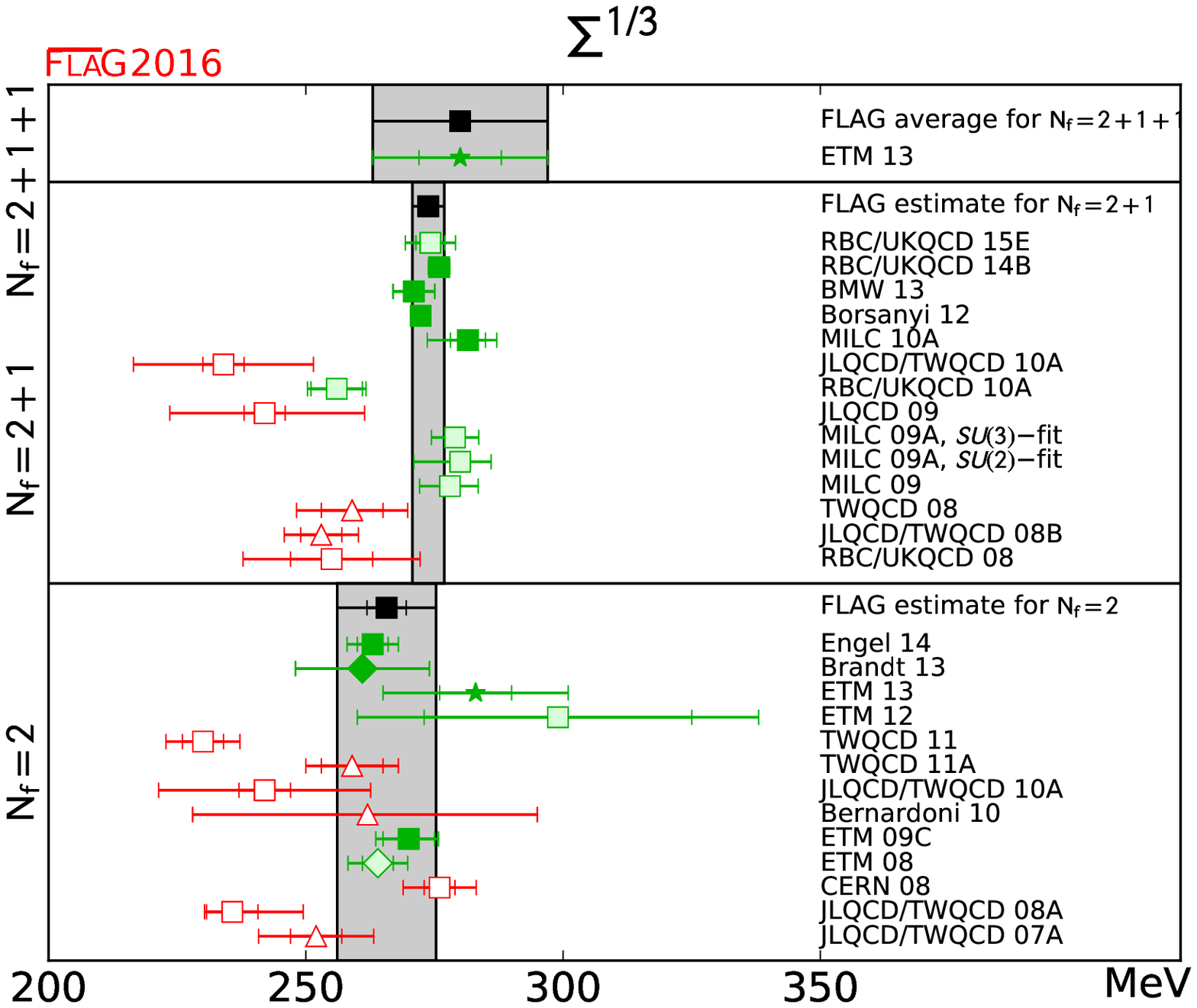}%
\vspace*{-2mm}
\caption{\label{fig:sigma}
Cubic root of the $SU(2)$ quark condensate $\Sigma\equiv-\langle\ubar u\rangle|_{m_u,m_d\to0}$
in the $\overline{\rm MS}$-scheme, at the renormalization scale $\mu=2$ GeV.
Squares indicate determinations from correlators in the
$p$-regime. Up triangles refer to extractions
from the topological susceptibility, diamonds to determinations from the pion
form factor, and star symbols refer to the spectral density method.
}
\end{figure}

We begin with a discussion of the lattice results for the $SU(2)$ LEC $\Sigma$.
We present the results in Tab.~\ref{tab:sigma} and Fig.~\ref{fig:sigma}.
We add that results which include only a statistical error are listed in the table but omitted from the plot.
Regarding the $\Nf=2$ computations there are six entries without a red tag.
We form the average based on ETM~09C, ETM~13 (here we deviate from our ``superseded'' rule,
since the two works use different methods), Brandt~13, and Engel~14.
We add that the last one (with numbers identical to those given in
Ref.\,\cite{Engel:2014cka}) is new compared to FLAG 13.
Here and in the following we take into account that ETM~09C, ETM~13
share configurations, and the same statement holds true for Brandt~13 and Engel~14.
Regarding the $\Nf=2+1$ computations there are four published or updated papers
(MILC~10A, Borsanyi~12, BMW~13, and RBC/UKQCD~14B) which qualify for the $\Nf=2+1$ average.
The last one is new compared to FLAG 13, and the last but one was not included in
the FLAG 13 average, since at the time it was only a preprint.

In slight deviation from the general recipe outlined in Sec.\,\ref{sec:averages}
we use these values as a basis for our \emph{estimates}
(as opposed to \emph{averages}) of the $\Nf=2$ and $\Nf=2+1$ condensates.
In each case the central value is obtained from our standard averaging procedure, but the
(symmetrical) error is just the median of the overall uncertainties of all
contributing results (see the comment below for details).
This leads to the values
%FLAGRESULT BEGIN
% TAG      &Sigma13 & Sigma13 & Sigma13   &END
% REFS     &\cite{Baron:2009wt,Cichy:2013gja,Brandt:2013dua,Engel:2014eea}& \cite{Bazavov:2010yq,Borsanyi:2012zv,Durr:2013goa,Blum:2014tka} & \cite{Cichy:2013gja} &END
% UNITS    & '[MeV]' & '[MeV]'  & '[MeV]'  &END
% NUMRESULTS & 4 & 4 & 1 &END
% FLAVOURs & 2 & 2+1 & 2+1+1 &END
%FLAGRESULT END
%FLAGRESULTFORMULA BEGIN
\begin{align}
\label{eq:condensates}
&N_f=2  :&\FLAGAVBEGIN \Sigma^{1/3}&= 266(10) \FLAGAVEND \MeV &&\Refs~\mbox{\cite{Baron:2009wt,Cichy:2013gja,Brandt:2013dua,Engel:2014eea}},\nonumber\\[-3mm]  
\\[-3mm]
&N_f=2+1:&\FLAGAVBEGIN \Sigma^{1/3}&= 274( 3) \FLAGAVEND \MeV &&\Refs~\mbox{\cite{Bazavov:2010yq,Borsanyi:2012zv,Durr:2013goa,Blum:2014tka}},\nonumber 
%&N_f=2+1+1:&\FLAGAVBEGIN \Sigma^{1/3}&= 280(8)(15) \FLAGAVEND \MeV &&\Refs~\mbox{\cite{Cichy:2013gja}},\nonumber 
\end{align}
%FLAGRESULTFORMULA END
in the $\msbar$ scheme at the renormalization scale $2\GeV$, where the errors
include both statistical and systematic uncertainties.
In accordance with our guidelines we ask the reader to cite
the appropriate set of references as indicated in Eq.\,(\ref{eq:condensates})
when using these numbers.
Finally, for $\Nf=2+1+1$ there is only one calculation available, the result
of Ref.~\cite{Cichy:2013gja} as given in Tab.~\ref{tab:sigma}.
According to the conventions of Sec.~\ref{sec:averages} this will be denoted as the
``FLAG average'' for $N_f=2+1+1$ in Fig.~\ref{fig:sigma}.

As a rationale for using \emph{estimates} (as opposed to \emph{averages}) for $\Nf=2$
and $\Nf=2+1$, we add that for $\Sigma^{1/3}|_{\Nf=2}$ and $\Sigma^{1/3}|_{\Nf=2+1}$
the standard averaging method would yield central values as quoted in
Eq.\,(\ref{eq:condensates}), but with (overall) uncertainties of $4\MeV$
and $1\MeV$, respectively.
It is not entirely clear to us that the scale is sufficiently well known
in all contributing works to warrant a precision of up to 0.36\% on our $\Sigma^{1/3}$,
and a similar statement can be made about the level of control over the convergence
of the chiral expansion.
The aforementioned uncertainties would suggest an $\Nf$-dependence of the $SU(2)$
chiral condensate which (especially in view of similar issues with other LECs,
see below) seems premature to us.
Therefore we choose to form the central value of our estimate with
the standard averaging procedure, but its uncertainty is taken as the median
of the uncertainties of the participating results.
We hope that future high-quality determinations with both $N_f=2$, $N_f=2+1$,
and in particular with $N_f=2+1+1$, will help determine whether there is a
noticeable $N_f$-dependence of the $SU(2)$ chiral condensate or not.

%%%%%%%%%%%%%%%%%%%%%%%%%%%%%%%%%%%%%%%%%%%%%%%%%%%%%%%%%%%%%%%%%%%%%%%%%%%%%%%

\begin{table}[!tbp] %%% F and Fpi/F
\vspace*{3cm}
\centering
\footnotesize
\begin{tabular*}{\textwidth}[l]{l@{\extracolsep{\fill}}rlllllll}
Collaboration & Ref. & $\Nf$ &
\hspace{0.15cm}\begin{rotate}{60}{publication status}\end{rotate}\hspace{-0.15cm}&
\hspace{0.15cm}\begin{rotate}{60}{chiral extrapolation}\end{rotate}\hspace{-0.15cm}&
\hspace{0.15cm}\begin{rotate}{60}{continuum  extrapolation}\end{rotate}\hspace{-0.15cm} &
\hspace{0.15cm}\begin{rotate}{60}{finite volume}\end{rotate}\hspace{-0.15cm} &
\rule{0.2cm}{0cm} $F$  &\rule{0.2cm}{0cm} $F_\pi/F$ \\[2mm]
\hline
\hline
\\[-2mm]
ETM 11                  & \cite{Baron:2011sf}        &2+1+1& \rC & \soso & \good & \soso & 85.60(4)                            & {\sl 1.077(1)}                        \\
ETM 10                  & \cite{Baron:2010bv}        &2+1+1& \gA & \soso & \soso & \good & 85.66(6)(13)                        & 1.076(2)(2)                           \\[2mm]
\hline
\\[-2mm]
RBC/UKQCD 15E           & \cite{Boyle:2015exm}       & 2+1 & \oP & \good & \good & \good & 85.8(1.1)(1.5)                      & 1.0641(21)(49)                        \\
RBC/UKQCD 14B           & \cite{Blum:2014tka}        & 2+1 & \gA & \good & \good & \good & 86.63(12)(13)                       & 1.0645(15)(0)                         \\
BMW 13                  & \cite{Durr:2013goa}        & 2+1 & \gA & \good & \good & \good & 88.0(1.3)(0.3)                      & 1.055(7)(2)                           \\
Borsanyi 12             & \cite{Borsanyi:2012zv}     & 2+1 & \gA & \good & \good & \good & 86.78(05)(25)                       & 1.0627(06)(27)                        \\
NPLQCD 11               & \cite{Beane:2011zm}        & 2+1 & \gA & \soso & \soso & \soso & {\sl 86.8(2.1)$\binom{+3.3}{-3.4}$} & 1.062(26)$\binom{+42}{-40}$           \\
MILC 10                 & \cite{Bazavov:2010hj}      & 2+1 & \rC & \soso & \good & \good & 87.0(4)(5)                          & {\sl 1.060(5)(6)}                     \\
MILC 10A                & \cite{Bazavov:2010yq}      & 2+1 & \rC & \soso & \good & \good & 87.5(1.0)$\binom{+0.7}{-2.6}$       & {\sl 1.054(12)$\binom{+31}{-09}$}     \\
MILC 09A, $SU(3)$-fit   & \cite{Bazavov:2009fk}      & 2+1 & \rC & \soso & \good & \good & 86.8(2)(4)                          & 1.062(1)(3)                           \\
MILC 09A, $SU(2)$-fit   & \cite{Bazavov:2009fk}      & 2+1 & \rC & \soso & \good & \good & 87.4(0.6)$\binom{+0.9}{-1.0}$       & {\sl 1.054(7)$\binom{+12}{-11}$}      \\
MILC 09                 & \cite{Bazavov:2009bb}      & 2+1 & \gA & \soso & \good & \good & {\sl 87.66(17)$\binom{+28}{-52}$}  & 1.052(2)$\binom{+6}{-3}$               \\
PACS-CS 08, $SU(3)$-fit & \cite{Aoki:2008sm}         & 2+1 & \gA & \good & \bad  & \bad  & 90.3(3.6)                           & 1.062(8)                              \\
PACS-CS 08, $SU(2)$-fit & \cite{Aoki:2008sm}         & 2+1 & \gA & \good & \bad  & \bad  & 89.4(3.3)                           & 1.060(7)                              \\
RBC/UKQCD 08            & \cite{Allton:2008pn}       & 2+1 & \gA & \soso & \bad  & \soso & 81.2(2.9)(5.7)                      & 1.080(8)                              \\[2mm]
\hline
\\[-2mm]
ETM 15A                 & \cite{Abdel-Rehim:2015pwa} &  2  & \oP & \good & \bad  & \soso & 86.3(2.8)                           & {\sl 1.069(35)}                       \\
Engel 14                & \cite{Engel:2014eea}       &  2  & \gA & \good & \good & \good & 85.8(0.7)(2.0)                      & {\sl 1.075(09)(25)}                   \\
Brandt 13               & \cite{Brandt:2013dua}      &  2  & \gA & \soso & \good & \soso & 84(8)(2)                            & 1.080(16)(6)                          \\
QCDSF 13                & \cite{Horsley:2013ayv}     &  2  & \gA & \good & \soso & \soso & 86(1)                               & {\sl 1.07(1)}                         \\
TWQCD 11                & \cite{Chiu:2011bm}         &  2  & \gA & \soso & \bad  & \bad  & 83.39(35)(38)                       & {\sl 1.106(5)(5)}                     \\
ETM 09C                 & \cite{Baron:2009wt}        &  2  & \gA & \soso & \good & \soso & 85.91(07)$\binom{+78}{-07}$         & 1.0755(6)$\binom{+08}{-94}$           \\
ETM 09B                 & \cite{Jansen:2009tt}       &  2  & \rC & \good & \bad  & \soso & 92.1(4.9)                           & {\sl 1.00(5)}                         \\ % was wrong
ETM 08                  & \cite{Frezzotti:2008dr}    &  2  & \gA & \soso & \soso & \soso & 86.6(7)(7)                          & 1.067(9)(9)                           \\
Hasenfratz 08           & \cite{Hasenfratz:2008ce}   &  2  & \gA & \soso & \bad  & \good & 90(4)                               & {\sl 1.02(5)}                         \\
JLQCD/TWQCD 08A         & \cite{Noaki:2008iy}        &  2  & \gA & \soso & \bad  & \bad  & 79.0(2.5)(0.7)$\binom{+4.2}{-0.0}$  & {\sl 1.167(37)(10)$\binom{+02}{-62}$} \\
JLQCD/TWQCD 07          & \cite{Fukaya:2007pn}       &  2  & \gA & \good & \bad  & \bad  & 87.3(5.6)                           & {\sl 1.06(7)}                         \\[2mm]
\hline
\\[-2mm]
Colangelo 03            & \cite{Colangelo:2003hf}    &     &     &       &       &       & 86.2(5)                             & 1.0719(52)                            \\[2mm]
\hline
\hline
\end{tabular*}
\normalsize
\vspace*{-2mm}
\caption{\label{tab:f}
Results for the $SU(2)$ low-energy constant $F$ (in MeV) and for the ratio $F_\pi/F$.
Horizontal lines separate different $\Nf$.
All ETM values which were available only in $r_0$ units were converted on the basis of $r_0=0.48(2)\,\fm$
\cite{Aoki:2009sc,Bazavov:2014pvz,Abdel-Rehim:2015pwa}, with
this error being added in quadrature to any existing systematic error.
Numbers in slanted fonts have been calculated by us, based on $\sqrt{2}\Fpi^\mr{phys}=130.41(20)\MeV$ \cite{Agashe:2014kda}, with
this error being added in quadrature to any existing systematic error.}
\end{table}

\begin{figure}[!tb]
\centering
\includegraphics[width=12.0cm]{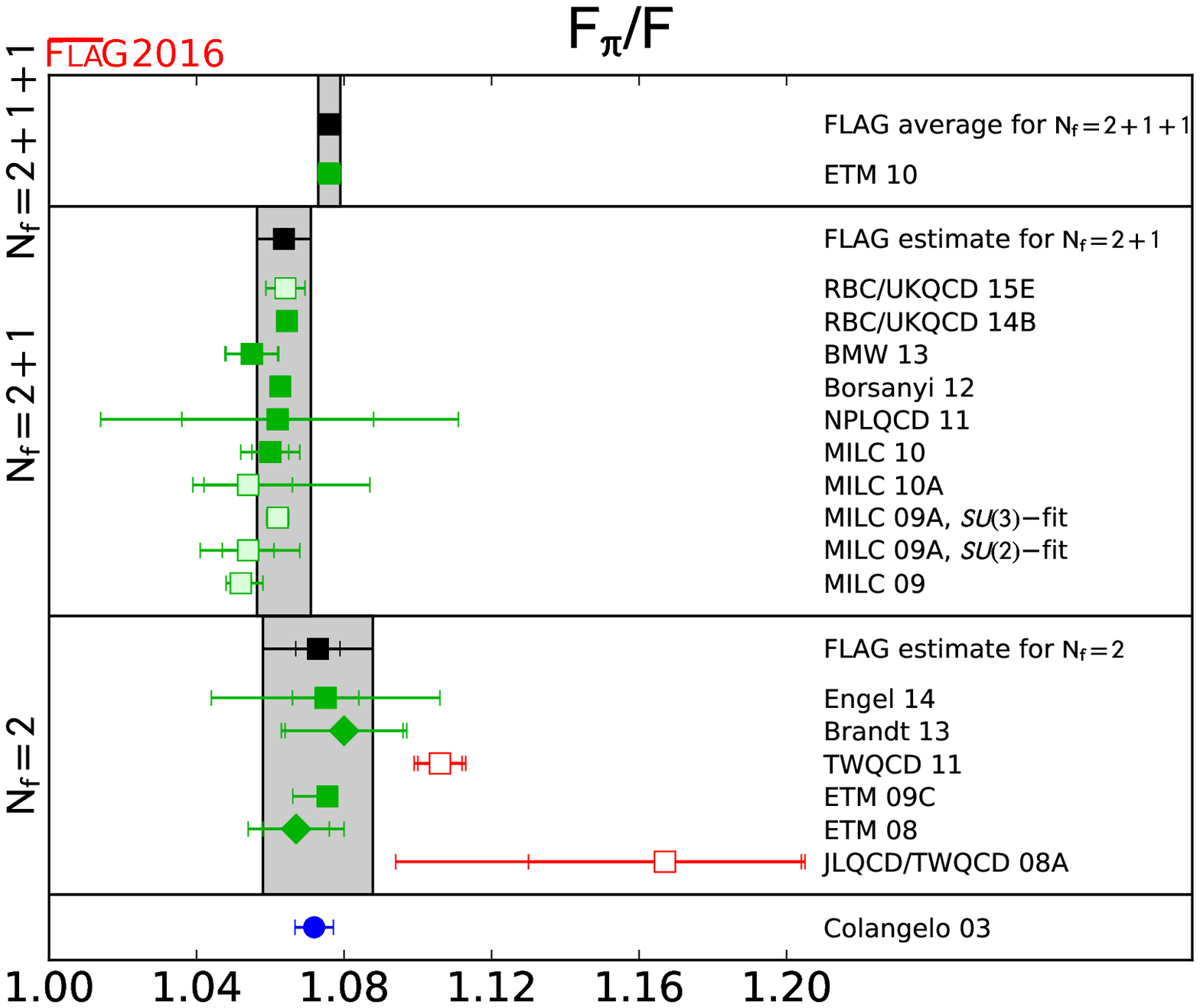}%
\vspace*{-2mm}
\caption{\label{fig:f}
Comparison of the results for the ratio of the physical pion decay constant
$F_\pi$ and  the leading-order $SU(2)$ low-energy constant $F$.
The meaning of the symbols is the same as in Fig.~\ref{fig:sigma}. }
\end{figure}

The next quantity considered is $F$, i.e.\ the pion decay constant in the
$SU(2)$ chiral limit ($m_{ud}\to0$, at fixed physical $m_s$ for $N_f > 2$
simulations).  As argued on previous occasions
we tend to give preference to $\Fpi/F$ (here the numerator is meant to
refer to the physical-pion-mass point) wherever it is available, since
often some of the systematic uncertainties are mitigated.  We collect the
results in Tab.~\ref{tab:f} and Fig.~\ref{fig:f}.  In those cases where
the collaboration provides only $F$, the ratio is computed on the basis of
the phenomenological value of $\Fpi$, and the respective entries in
Tab.~\ref{tab:f} are in slanted fonts. We encourage authors to provide both
$F$ and $F_\pi/F$ from their analysis, since the ratio is less dependent on
the scale setting, and errors tend to partially cancel. Among the $\Nf=2$ determinations
five (ETM 08, ETM 09C, QCDSF 13, Brandt 13 and Engel 14) are without red tags.
Since the third one is without systematic error, only four of them
enter the average.
Compared to FLAG 13 the last work is the only one which is new.
Among the  $\Nf=2+1$ determinations five values (MILC 10 as an update of MILC 09,
NPLQCD 11, Borsanyi 12, BMW 13, and RBC/UKQCD 14B) contribute to the average.
Compared to FLAG 13 the last work is a new addition, and the last but one is
included in the average for the first time.
Here and in the following we take into account that MILC 10 and NPLQCD 11
share configurations.
Finally, there is a single $\Nf=2+1+1$ determination (ETM 10) which forms
the current best estimate in this category.

In analogy to the condensates discussed above, we use these values as a basis
for our \emph{estimates} (as opposed to \emph{averages}) of the decay constant ratios
%FLAGRESULT BEGIN
% TAG      &FpioF & FpioF   & FpioF   &END
% REFS     &\cite{Frezzotti:2008dr,Baron:2009wt,Brandt:2013dua,Engel:2014eea}& \cite{Bazavov:2010hj,Beane:2011zm,Borsanyi:2012zv,Durr:2013goa,Blum:2014tka}  & \cite{Baron:2010bv}  &END
% UNITS    & 1 & 1  & 1  &END
% NUMRESULTS & 4 & 4 & 1 &END
% FLAVOURs & 2 & 2+1 & 2+1+1 &END
%FLAGRESULT END
%FLAGRESULTFORMULA BEGIN
\begin{align}
\label{eq:decayratios}
&N_f=2  :&\FLAGAVBEGIN {\Fpi}/{F}&=1.073(15) \FLAGAVEND&&\Refs~\mbox{\cite{Frezzotti:2008dr,Baron:2009wt,Brandt:2013dua,Engel:2014eea}}            ,\nonumber\\[-3mm] 
\\[-3mm]
&N_f=2+1:&\FLAGAVBEGIN {\Fpi}/{F}&=1.064( 7) \FLAGAVEND&&\Refs~\mbox{\cite{Bazavov:2010hj,Beane:2011zm,Borsanyi:2012zv,Durr:2013goa,Blum:2014tka}},\nonumber 
%&N_f=2+1+1:&\FLAGAVBEGIN {\Fpi}/{F}&=1.076(2)(2)  \FLAGAVEND&&\Refs~\mbox{\cite{Baron:2010bv}},\nonumber 
\end{align}
%FLAGRESULTFORMULA END
where the errors include both statistical and systematic uncertainties.
These numbers are obtained through the well-defined procedure described next
to Eq.\,(\ref{eq:condensates}).
We ask the reader to cite
the appropriate set of references as indicated in Eq.\,(\ref{eq:decayratios})
when using these numbers. Finally, for $\Nf=2+1+1$ the result of Ref.~\cite{Baron:2010bv} is the
only one available; see Tab.~\ref{tab:f} for the numerical value.

For this observable the standard averaging method would yield the central
values as quoted in Eq.\,(\ref{eq:decayratios}), but with (overall) uncertainties
of $6$ and $1$, respectively, on the last digit quoted.
In this particular case the single $\Nf=2+1+1$ determination lies significantly
higher than the $\Nf=2+1$ \emph{average} (with the small error-bar), basically on
par with the $\Nf=2$ \emph{average} (with the small error-bar), and this makes
such a standard \emph{average} look even more suspicious to us.
At the least, one should wait for one more qualifying $N_f=2+1+1$ determination
before attempting any conclusions about the $N_f$ dependence of $F_\pi/F$.
While we are not aware of any theorem which excludes a nonmonotonic behavior in
$N_f$ of a LEC, standard physics reasoning would suggest that quark-loop effects
become smaller with increasing quark mass, hence a dynamical charm quark
will influence LECs less significantly than a dynamical strange quark,
and even the latter one seems to bring rather small shifts.
As a result, we feel that a nonmonotonic behavior of $\Fpi/F$
with $\Nf$, once established, would represent a noteworthy finding.
We hope this reasoning explains why we prefer to stay in
Eq.\,(\ref{eq:decayratios}) with \emph{estimates} which
obviously are on the conservative side.

%%%%%%%%%%%%%%%%%%%%%%%%%%%%%%%%%%%%%%%%%%%%%%%%%%%%%%%%%%%%%%%%%%%%%%%%%%%%%%%

\subsubsection{Results for the NLO $SU(2)$ LECs \label{sec:SU2_NLO}}

\begin{table}[!tbp] %%% \lbar_3 and \lbar_4
\vspace*{3cm}
\centering
\footnotesize
\begin{tabular*}{\textwidth}[l]{l@{\extracolsep{\fill}}rlllllll}
Collaboration & Ref. & $\Nf$ &
\hspace{0.15cm}\begin{rotate}{60}{publication status}\end{rotate}\hspace{-0.15cm} &
\hspace{0.15cm}\begin{rotate}{60}{chiral extrapolation}\end{rotate}\hspace{-0.15cm}&
\hspace{0.15cm}\begin{rotate}{60}{continuum  extrapolation}\end{rotate}\hspace{-0.15cm} &
\hspace{0.15cm}\begin{rotate}{60}{finite volume}\end{rotate}\hspace{-0.15cm} &\rule{0.3cm}{0cm} $\lbar_3$ & \rule{0.3cm}{0cm}$\lbar_4$     \\[2mm]
\hline
\hline
\\[-2mm]
ETM 11                  & \cite{Baron:2011sf}        &2+1+1& \rC & \soso & \good & \soso & 3.53(5)                        & 4.73(2)                        \\
ETM 10                  & \cite{Baron:2010bv}        &2+1+1& \gA & \soso & \soso & \good & 3.70(7)(26)                    & 4.67(3)(10)                    \\[2mm]
\hline
\\[-2mm]
RBC/UKQCD 15E           & \cite{Boyle:2015exm}       & 2+1 & \oP & \good & \good & \good & 2.81(19)(45)                   & 4.02(8)(24)                    \\
RBC/UKQCD 14B           & \cite{Blum:2014tka}        & 2+1 & \gA & \good & \good & \good & 2.73(13)(0)                    & 4.113(59)(0)                   \\
BMW 13                  & \cite{Durr:2013goa}        & 2+1 & \gA & \good & \good & \good & 2.5(5)(4)                      & 3.8(4)(2)                      \\
RBC/UKQCD 12            & \cite{Arthur:2012opa}      & 2+1 & \gA & \good & \good & \good & 2.91(23)(07)                   & 3.99(16)(09)                   \\
Borsanyi 12             & \cite{Borsanyi:2012zv}     & 2+1 & \gA & \good & \good & \good & 3.16(10)(29)                   & 4.03(03)(16)                   \\
NPLQCD 11               & \cite{Beane:2011zm}        & 2+1 & \gA & \soso & \soso & \soso & 4.04(40)$\binom{+73}{-55}$     & 4.30(51)$\binom{+84}{-60}$     \\
MILC 10                 & \cite{Bazavov:2010hj}      & 2+1 & \rC & \soso & \good & \good & 3.18(50)(89)                   & 4.29(21)(82)                   \\
MILC 10A                & \cite{Bazavov:2010yq}      & 2+1 & \rC & \soso & \good & \good & 2.85(81)$\binom{+37}{-92}$     & 3.98(32)$\binom{+51}{-28}$     \\
RBC/UKQCD 10A           & \cite{Aoki:2010dy}         & 2+1 & \gA & \soso & \soso & \soso & 2.57(18)                       & 3.83(9)                        \\
MILC 09A, $SU(3)$-fit   & \cite{Bazavov:2009fk}      & 2+1 & \rC & \soso & \good & \good & 3.32(64)(45)                   & 4.03(16)(17)                   \\
MILC 09A, $SU(2)$-fit   & \cite{Bazavov:2009fk}      & 2+1 & \rC & \soso & \good & \good & 3.0(6)$\binom{+9}{-6}$         & 3.9(2)(3)                      \\
PACS-CS 08, $SU(3)$-fit & \cite{Aoki:2008sm}         & 2+1 & \gA & \good & \bad  & \bad  & 3.47(11)                       & 4.21(11)                       \\
PACS-CS 08, $SU(2)$-fit & \cite{Aoki:2008sm}         & 2+1 & \gA & \good & \bad  & \bad  & 3.14(23)                       & 4.04(19)                       \\
RBC/UKQCD 08            & \cite{Allton:2008pn}       & 2+1 & \gA & \soso & \bad  & \soso & 3.13(33)(24)                   & 4.43(14)(77)                   \\[2mm]
\hline
\\[-2mm]
ETM 15A                 & \cite{Abdel-Rehim:2015pwa} &  2  & \oP & \good & \bad  & \soso &                                & 3.3(4)                         \\
G\"ulpers 15            & \cite{Gulpers:2015bba}     &  2  & \oP & \good & \good & \good &                                & 4.54(30)(0)                    \\
G\"ulpers 13            & \cite{Gulpers:2013uca}     &  2  & \gA & \soso & \bad  & \soso &                                & 4.76(13)                       \\
Brandt 13               & \cite{Brandt:2013dua}      &  2  & \gA & \soso & \good & \soso & 3.0(7)(5)                      & 4.7(4)(1)                      \\
QCDSF 13                & \cite{Horsley:2013ayv}     &  2  & \gA & \good & \soso & \soso &                                & 4.2(1)                         \\
Bernardoni 11           & \cite{Bernardoni:2011kd}   &  2  & \rC & \soso & \bad  & \bad  & 4.46(30)(14)                   & 4.56(10)(4)                    \\
TWQCD 11                & \cite{Chiu:2011bm}         &  2  & \gA & \soso & \bad  & \bad  & 4.149(35)(14)                  & 4.582(17)(20)                  \\
ETM 09C                 & \cite{Baron:2009wt}        &  2  & \gA & \soso & \good & \soso & 3.50(9)$\binom{+09}{-30}$      & 4.66(4)$\binom{+04}{-33}$      \\
JLQCD/TWQCD 09          & \cite{JLQCD:2009qn}        &  2  & \gA & \soso & \bad  & \bad  &                                & 4.09(50)(52)                   \\
ETM 08                  & \cite{Frezzotti:2008dr}    &  2  & \gA & \soso & \soso & \soso & 3.2(8)(2)                      & 4.4(2)(1)                      \\
JLQCD/TWQCD 08A         & \cite{Noaki:2008iy}        &  2  & \gA & \soso & \bad  & \bad  & 3.38(40)(24)$\binom{+31}{-00}$ & 4.12(35)(30)$\binom{+31}{-00}$ \\
CERN-TOV 06             & \cite{DelDebbio:2006cn}    &  2  & \gA & \soso & \bad  & \bad  & 3.0(5)(1)                      &                                \\[2mm]
\hline
\\[-2mm]
Colangelo 01            & \cite{Colangelo:2001df}    &     &     &       &       &       &                                & 4.4(2)                         \\
Gasser 84               & \cite{Gasser:1983yg}       &     &     &       &       &       & 2.9(2.4)                       & 4.3(9)                         \\[2mm]
\hline
\hline
\end{tabular*}
\normalsize
\vspace*{-2mm}
\caption{\label{tab:l3and4}
Results for the $SU(2)$ NLO low-energy constants $\lbar_3$ and $\lbar_4$.
For comparison, the last two lines show results from phenomenological analyses.}
\end{table}

\begin{figure}[!tbp]
\centering
\includegraphics[width=9.5cm]{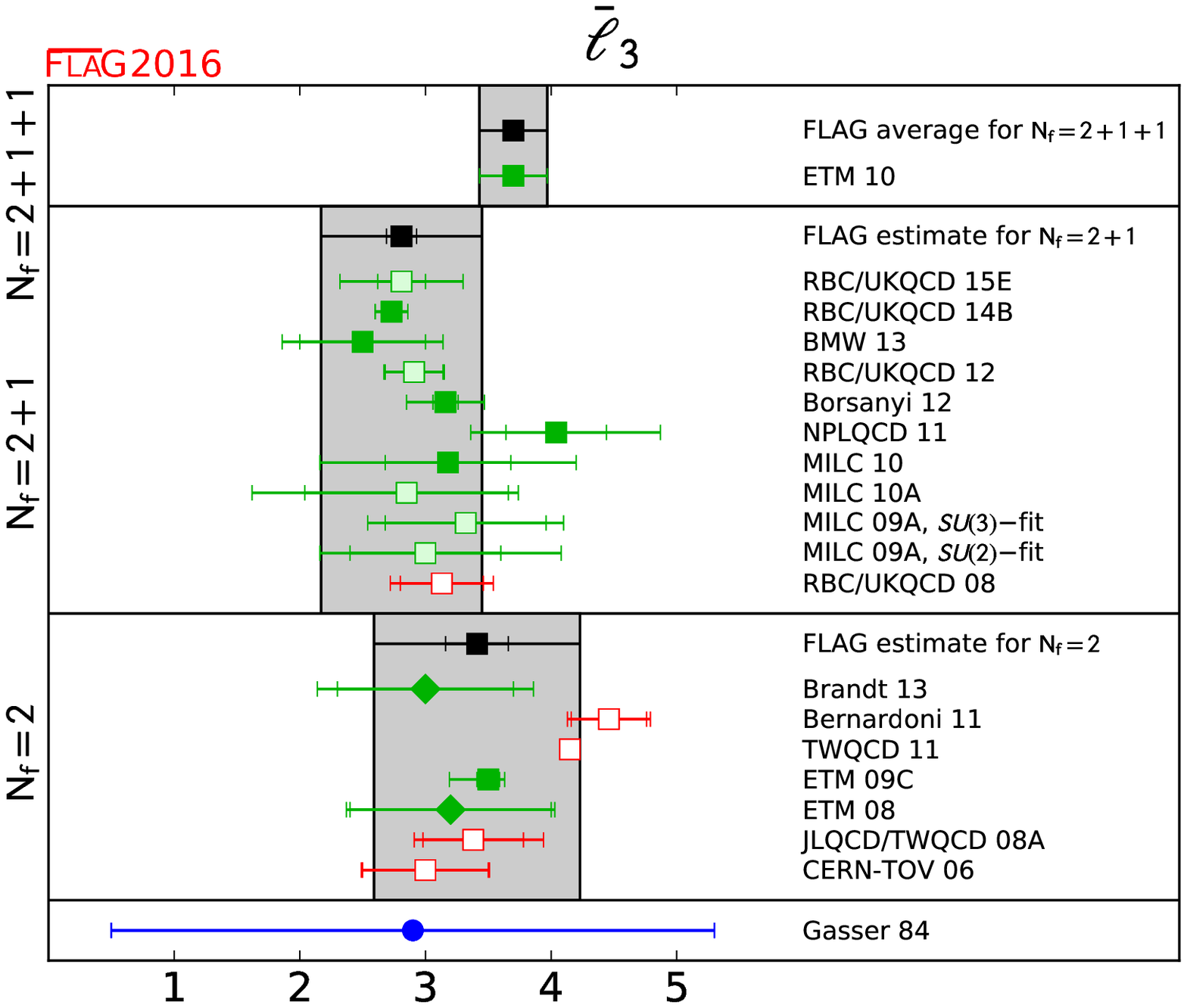}\\[-2mm]
\includegraphics[width=9.5cm]{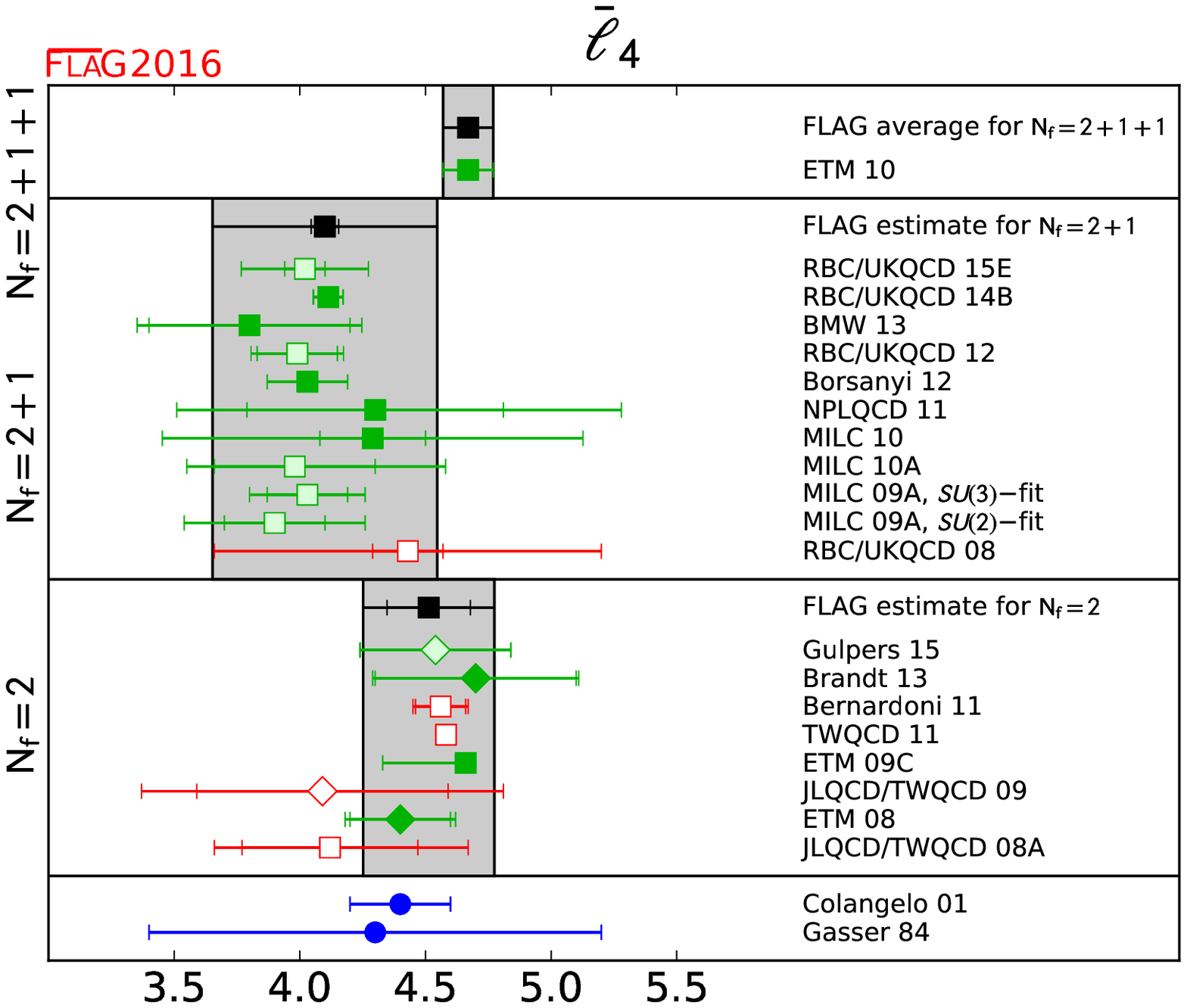}\\[-2mm]
\includegraphics[width=9.5cm]{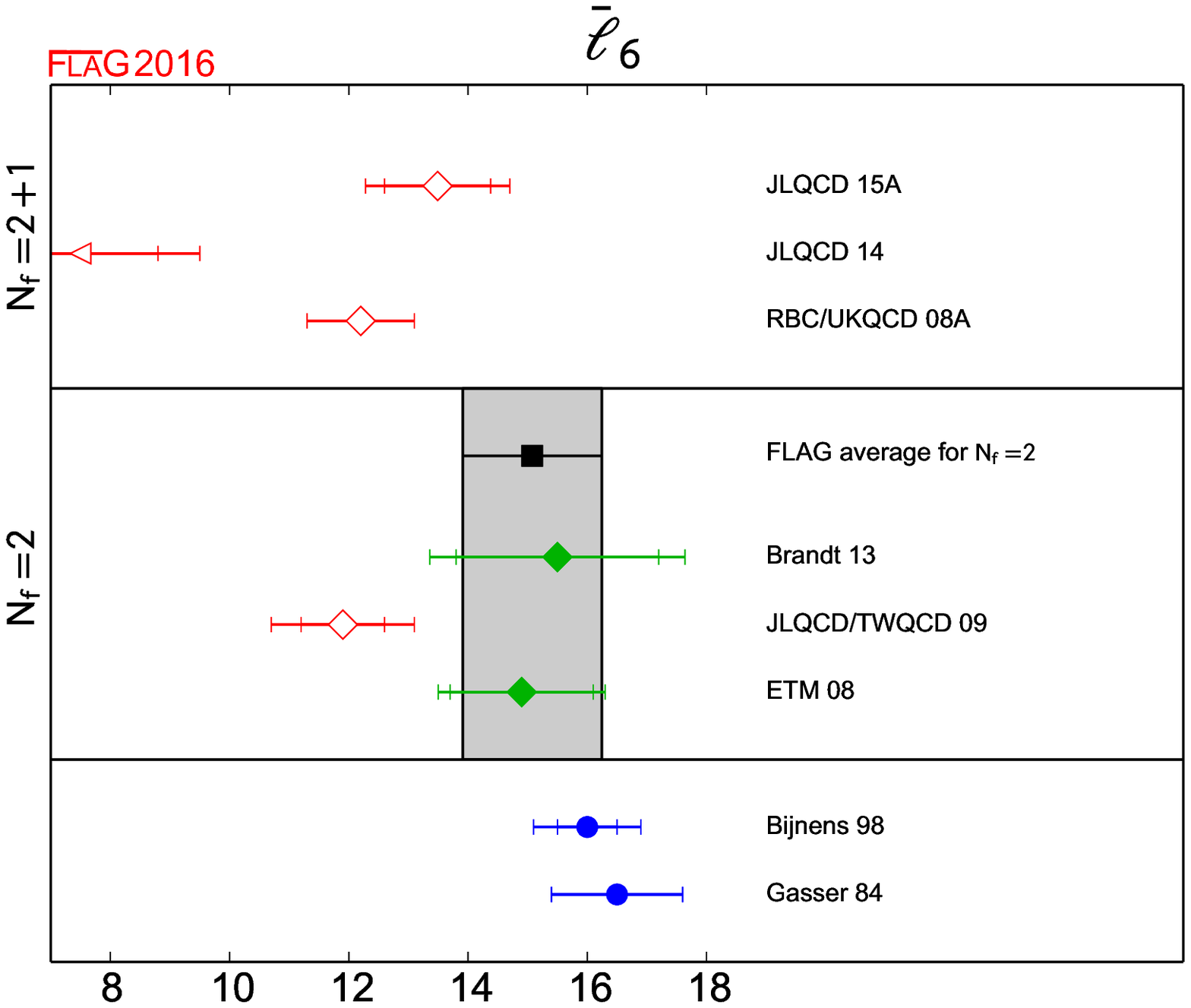}%
\vspace*{-2mm}
\caption{\label{fig:l3l4l6}
Effective coupling constants $\lbar_3$, $\lbar_4$ and $\lbar_6$.
Squares indicate determinations from correlators in the $p$-regime, diamonds
refer to determinations from the pion form factor.}
\end{figure}

We move on to a discussion of the lattice results for the NLO LECs
$\lbar_3$ and $\lbar_4$.  We remind the reader that on the lattice
the former LEC is obtained as a result of the tiny deviation from linearity
seen in $\Mpi^2$ versus $Bm_{ud}$, whereas the latter LEC is extracted from
the curvature in $\Fpi$ versus $Bm_{ud}$.  The available determinations are
presented in Tab.~\ref{tab:l3and4} and Fig.~\ref{fig:l3l4l6}.
Among the $\Nf=2$ determinations ETM 08, ETM 09C and Brandt 13 are published
prior to the deadline, with a systematic uncertainty, and without red tags.
Given that the former two use different approaches, all three determinations
enter our average.
The colour coding of the $\Nf=2+1$ results looks very promising; there is a
significant number of lattice determinations without any red tag.
Applying our superseding rule, MILC 10, NPLQCD 11, Borsanyi 12, BMW 13,
and RBC/UKQCD 14B contribute to the average.  Compared to the previous
edition of our review, the last one is a new addition, and the last but
one is included for the first time in the average.
For $\Nf=2+1+1$ there is only the single work ETM 10.

In analogy to our processing of the LECs at LO, we use these determinations as the
basis of our \emph{estimate} (as opposed to \emph{average}) of the NLO quantities
%FLAGRESULT BEGIN
% TAG      &lbar3 & lbar3   & lbar3   &lbar4 & lbar4 & lbar4 &END
% REFS     &\cite{Frezzotti:2008dr,Baron:2009wt,Brandt:2013dua}& \cite{Bazavov:2010hj,Beane:2011zm,Borsanyi:2012zv,Durr:2013goa,Blum:2014tka}& \cite{Baron:2010bv}& \cite{Frezzotti:2008dr,Baron:2009wt,Brandt:2013dua}& \cite{Bazavov:2010hj,Beane:2011zm,Borsanyi:2012zv,Durr:2013goa,Blum:2014tka}  & \cite{Baron:2010bv} &END
% UNITS    & 1 & 1& 1 & 1  & 1 & 1  &END
% NUMRESULTS & 3 & 5 & 1 & 3 & 5 & 1 &END
% FLAVOURs & 2 &2+1&2+1+1& 2 & 2+1 & 2+1+1 &END
%FLAGRESULT END
%FLAGRESULTFORMULA BEGIN
\begin{align}
&N_f=2  :&\FLAGAVBEGIN \lbar_3&=3.41(82) \FLAGAVEND&&\Refs~\mbox{\cite{Frezzotti:2008dr,Baron:2009wt,Brandt:2013dua}},\nonumber\\[-3mm]
\label{results_l3}\\[-3mm]
&N_f=2+1:&\FLAGAVBEGIN \lbar_3&=2.81(64) \FLAGAVEND&&\Refs~\mbox{\cite{Bazavov:2010hj,Beane:2011zm,Borsanyi:2012zv,Durr:2013goa,Blum:2014tka}},\nonumber
%&N_f=2+1+1:&\FLAGAVBEGIN \lbar_3&=3.70(7)(26)  \FLAGAVEND&&\Refs~\mbox{\cite{Baron:2010bv}},\nonumber
\end{align}
\begin{align}
&N_f=2  :&\FLAGAVBEGIN \lbar_4&=4.51(26) \FLAGAVEND&&\Refs~\mbox{\cite{Frezzotti:2008dr,Baron:2009wt,Brandt:2013dua}},\nonumber\\[-3mm]
\label{results_l4}\\[-3mm]
&N_f=2+1:&\FLAGAVBEGIN \lbar_4&=4.10(45) \FLAGAVEND&&\Refs~\mbox{\cite{Bazavov:2010hj,Beane:2011zm,Borsanyi:2012zv,Durr:2013goa,Blum:2014tka}},\nonumber
%&N_f=2+1+1:&\FLAGAVBEGIN \lbar_4&=4.67(3)(10) \FLAGAVEND&&\Refs~\mbox{\cite{Baron:2010bv}},\nonumber
\end{align}
%FLAGRESULTFORMULA END
where the errors include both statistical and systematic uncertainties.
These numbers are obtained through the well-defined procedure described next
to Eq.\,(\ref{eq:condensates}).
Again we ask the reader to cite
the appropriate set of references as indicated in Eq.~(\ref{results_l3}) or Eq.~(\ref{results_l4})
when using these numbers.  For $\Nf=2+1+1$ once again
Ref.~\cite{Baron:2010bv} is the single reference available, see
Tab.~\ref{tab:l3and4} for the numerical values.

We remark that our preprocessing procedure%
\footnote{There are two naive procedures to symmetrize an asymmetric
systematic error: ($i$) keep the central value untouched and enlarge the
smaller error, ($ii$) shift the central value by half of the difference
between the two original errors and enlarge/shrink both errors by the same
amount. Our procedure ($iii$) is to average the results of ($i$) and
($ii$). In other words a result $c(s)\binom{+u}{-\ell}$ with $\ell>u$ is
changed into $c+(u-\ell)/4$ with statistical error $s$ and a symmetric
systematic error $(u+3\ell)/4$.  The case $\ell<u$ is handled accordingly.}
symmetrizes the asymmetric error of ETM 09C with a slight adjustment of the
central value.
Regarding the difference between the \emph{estimates} as given in
Eqs.~(\ref{results_l3}, \ref{results_l4}) and the result of the
standard \emph{averaging} procedure we add that the latter would
yield the overall uncertainties $25$ and $12$ for $\bar\ell_3$,
and the overall uncertainties $17$ and $5$ for $\bar\ell_4$.
In all cases the central value would be unchanged.
Especially for $\bar\ell_4$ such numbers would suggest a clear
difference between the value with $\Nf=2$ dynamical flavours
and the one at $\Nf=2+1$.
Similarly to what happened with $\Fpi/F$, the single determination
with $\Nf=2+1+1$ is more on the $\Nf=2$ side which, if confirmed,
would suggest a nonmonotonicity of a {\Ch}PT LEC with $\Nf$.
Again we think that currently such a conclusion would be premature,
and this is why we give preference to the \emph{estimates}
quoted in Eqs.~(\ref{results_l3}, \ref{results_l4}).

From a more phenomenological point of view there is a notable difference
between $\lbar_3$ and $\lbar_4$ in Fig.~\ref{fig:l3l4l6}.  For
$\lbar_4$ the precision of the phenomenological determination achieved
in Colangelo 01 \cite{Colangelo:2001df} represents a significant
improvement compared to Gasser 84 \cite{Gasser:1983yg}.  Picking any $\Nf$,
the lattice estimate of $\lbar_4$ is consistent with both of the
phenomenological values and comes with an error-bar which is roughly comparable
to or somewhat larger than the one in Colangelo 01 \cite{Colangelo:2001df}.
By contrast, for $\lbar_3$ the error of an individual lattice computation is
usually much smaller than the error of the estimate given in Gasser 84
\cite{Gasser:1983yg}, and even our conservative estimates (\ref{results_l3}) have
uncertainties which represent a significant improvement on the error-bar
of Gasser 84 \cite{Gasser:1983yg}.
Evidently, our hope is that future determinations of $\bar\ell_3,\bar\ell_4$,
with $\Nf=2$, $\Nf=2+1$ and $\Nf=2+1+1$, will allow us to further shrink our
error-bars in a future edition of FLAG.

%%%%%%%%%%%%%%%%%%%%%%%%%%%%%%%%%%%%%%%%%%%%%%%%%%%%%%%%%%%%%%%%%%%%%%%%%%%%%%%

\begin{table}[!tbp] %%% <r^2> of pion
\vspace*{3cm}
\centering
\footnotesize
\begin{tabular*}{\textwidth}[l]{l@{\extracolsep{\fill}}rllllllll}
Collaboration & Ref. & $\Nf$ &
\hspace{0.15cm}\begin{rotate}{60}{publication status}\end{rotate}\hspace{-0.15cm} &
\hspace{0.15cm}\begin{rotate}{60}{chiral extrapolation}\end{rotate}\hspace{-0.15cm}&
\hspace{0.15cm}\begin{rotate}{60}{continuum  extrapolation}\end{rotate}\hspace{-0.15cm} &
\hspace{0.15cm}\begin{rotate}{60}{finite volume}\end{rotate}\hspace{-0.15cm} &
\rule{0.3cm}{0cm}$\<r^2\>_V^\pi$ & \rule{0.3cm}{0cm}$c_V$ & \rule{0.3cm}{0cm}$\lbar_6$ \\[2mm]
\hline
\hline
\\[-2mm]
HPQCD 15B               & \cite{Koponen:2015tkr}     &2+1+1& \oP & \good & \good & \good & 0.403(18)(6)      &              &                \\[2mm]
\hline
\\[-2mm]
JLQCD 15A , $SU(2)$-fit & \cite{Aoki:2015pba}        & 2+1 & \oP & \soso & \bad  & \soso & 0.395(26)(32)     &              & 13.49(89)(82)  \\
JLQCD 14                & \cite{Fukaya:2014jka}      & 2+1 & \gA & \good & \bad  & \bad  & 0.49(4)(4)        &              &  7.5(1.3)(1.5) \\
PACS-CS 11A             & \cite{Nguyen:2011ek}       & 2+1 & \gA & \soso & \bad  & \soso & 0.441(46)         &              &                \\
RBC/UKQCD 08A           & \cite{Boyle:2008yd}        & 2+1 & \gA & \soso & \bad  & \soso & 0.418(31)         &              & 12.2(9)        \\
LHP 04                  & \cite{Bonnet:2004fr}       & 2+1 & \gA & \soso & \bad  & \bad  & 0.310(46)         &              &                \\[2mm]
\hline
\\[-2mm]
Brandt 13               & \cite{Brandt:2013dua}      &  2  & \gA & \soso & \good & \soso & 0.481(33)(13)     &              & 15.5(1.7)(1.3) \\
JLQCD/TWQCD 09          & \cite{JLQCD:2009qn}        &  2  & \gA & \soso & \bad  & \bad  & 0.409(23)(37)     & 3.22(17)(36) & 11.9(0.7)(1.0) \\
ETM 08                  & \cite{Frezzotti:2008dr}    &  2  & \gA & \soso & \soso & \soso & 0.456(30)(24)     & 3.37(31)(27) & 14.9(1.2)(0.7) \\
QCDSF/UKQCD 06A         & \cite{Brommel:2006ww}      &  2  & \gA & \soso & \good & \soso & 0.441(19)(63)     &              &                \\[2mm]
\hline
\\[-2mm]
Bijnens 98              & \cite{Bijnens:1998fm}      &     &     &       &       &       & 0.437(16)         & 3.85(60)     & 16.0(0.5)(0.7) \\
NA7 86                  & \cite{Amendolia:1986wj}    &     &     &       &       &       & 0.439(8)          &              &                \\
Gasser 84               & \cite{Gasser:1983yg}       &     &     &       &       &       &                   &              & 16.5(1.1)      \\[2mm]
\hline
\hline
\end{tabular*}
\\[3.5cm]
\begin{tabular*}{\textwidth}[l]{l@{\extracolsep{\fill}}rlllllll}
Collaboration & Ref. & $\Nf$ &
\hspace{0.15cm}\begin{rotate}{60}{publication status}\end{rotate}\hspace{-0.15cm} &
\hspace{0.15cm}\begin{rotate}{60}{chiral extrapolation}\end{rotate}\hspace{-0.15cm}&
\hspace{0.15cm}\begin{rotate}{60}{continuum  extrapolation}\end{rotate}\hspace{-0.15cm} &
\hspace{0.15cm}\begin{rotate}{60}{finite volume}\end{rotate}\hspace{-0.15cm} &
\rule{0.3cm}{0cm}$\<r^2\>_S^\pi$ & \rule{0.3cm}{0cm}$\lbar_1-\lbar_2$ \\[2mm]
\hline
\hline
\\[-2mm]
HPQCD 15B               & \cite{Koponen:2015tkr}     &2+1+1& \oP & \good & \good & \good &  0.481(37)(50)    &                \\[2mm]
\hline
\\[-2mm]
RBC/UKQCD 15E           & \cite{Boyle:2015exm}       & 2+1 & \oP & \good & \good & \good &                   & -9.2(4.9)(6.5) \\[2mm]
\hline
\\[-2mm]
G\"ulpers 15            & \cite{Gulpers:2015bba}     &  2  & \oP & \good & \good & \good & 0.600(52)(0)      &                \\
G\"ulpers 13            & \cite{Gulpers:2013uca}     &  2  & \gA & \soso & \bad  & \soso & 0.637(23)         &                \\
JLQCD/TWQCD 09          & \cite{JLQCD:2009qn}        &  2  & \gA & \soso & \bad  & \bad  & 0.617(79)(66)     & -2.9(0.9)(1.3) \\[2mm]
\hline
\\[-2mm]
Colangelo 01            & \cite{Colangelo:2001df}    &     &     &       &       &       & 0.61(4)           & -4.7(6)        \\[2mm]
\hline
\hline
\end{tabular*}
\normalsize
\vspace*{-2mm}
\caption{\label{tab:radii}
Top (vector form factor of the pion): Lattice results for the charge
radius $\<r^2\>_V^\pi$ (in $\mathrm{fm}^2$), the curvature $c_V$ (in $\mathrm{GeV}^{-4}$)
and the effective coupling constant $\lbar_6$ are compared with the experimental value,
as obtained by NA7, and some phenomenological estimates.
Bottom (scalar form factor of the pion): Lattice results for the scalar
radius $\< r^2 \>_S^\pi$ (in $\mathrm{fm}^2$) and the combination $\lbar_1-\lbar_2$
are compared with a dispersive calculation of these quantities.}
\end{table}

We finish with a discussion of the lattice results for $\lbar_6$ and
$\lbar_1-\lbar_2$. The LEC $\lbar_6$ determines the leading
contribution in the chiral expansion of the pion vector charge radius,
see Eq.\,(\ref{formula_rsqu}).  Hence from a lattice study of the vector
form factor of the pion with several $\Mpi$ one may extract the radius
$\<r^2\>_V^\pi$, the curvature $c_V$ (both at the physical pion-mass point)
and the LEC $\lbar_6$ in one go.  Similarly, the leading contribution in
the chiral expansion of the scalar radius of the pion determines
$\lbar_4$, see Eq.\,(\ref{formula_rsqu}).  This LEC is also present in
the pion-mass dependence of $\Fpi$, as we have seen.  The difference
$\lbar_1-\lbar_2$, finally, may be obtained  from the momentum dependence
of the vector and scalar pion form factors, based on the two-loop formulae
of Ref.~\cite{Bijnens:1998fm}.  The top part of Tab.~\ref{tab:radii} collects the
results obtained from the vector form factor of the pion (charge radius,
curvature and $\lbar_6$).  Regarding this low-energy constant two $\Nf=2$
calculations are published works without a red tag; we thus arrive at the
\emph{average} (actually the first one in the LEC section)
%FLAGRESULT BEGIN
% TAG      &lbar6 &END
% REFS     &\cite{Frezzotti:2008dr,Brandt:2013dua}  &END
% UNITS    & 1 &END
% NUMRESULTS & 2 &END
% FLAVOURs & 2 &END
%FLAGRESULT END
%FLAGRESULTFORMULA BEGIN
\beq
N_f=2:\hspace{1cm}\FLAGAVBEGIN \lbar_6=15.1(1.2) \FLAGAVEND \qquad \Refs~\mbox{\cite{Frezzotti:2008dr,Brandt:2013dua}},
\eeq
%FLAGRESULTFORMULA END
which is represented as a grey band in the last panel of
Fig.~\ref{fig:l3l4l6}.  Here we ask the reader to cite
Refs.~\cite{Frezzotti:2008dr,Brandt:2013dua} when using this number.

The experimental information concerning the charge radius is excellent and
the curvature is also known very accurately, based on $e^+e^-$ data and
dispersion theory.  The vector form factor calculations thus present an
excellent testing ground for the lattice methodology.  The first data column of
Tab.~\ref{tab:radii} shows that most of the available lattice results pass the test.
There is, however, one worrisome point.
For $\lbar_6$ the agreement seems less convincing than for the charge radius,
even though the two quantities are closely related.
In particular the $\lbar_6$ value of JLQCD 14   \cite{Fukaya:2014jka}
seems inconsistent with the phenomenological determinations of Refs.~\cite{Gasser:1983yg,Bijnens:1998fm},
even though its value for $\<r^2\>_V^\pi$ is consistent.
So far we have no explanation (other than observing that lattice computations
which disagree with the phenomenological determination of $\lbar_6$ tend
to have red tags), but we urge the groups to pay special attention to this point.
Similarly, the bottom part of Tab.~\ref{tab:radii} collects the results obtained
for the scalar form factor of the pion and the combination $\lbar_1-\lbar_2$
that is extracted from it.
A new feature is that the (yet unpublished) paper \cite{Koponen:2015tkr}
gives both the (flavour) octet and singlet part in $SU(3)$, finding
$\<r^2\>_{S,\mr{octet}}^\pi=0.431(38)(46)$ and
$\<r^2\>_{S,\mr{singlet}}^\pi=0.506(38)(53)$.
For reasons of backward compatibility they also give $\<r^2\>_{S,ud}^\pi$
defined with a $\bar{u}u+\bar{d}d$ density, and this number is shown in
Tab.~\ref{tab:radii}.
Last but not least they find the ordering
$\<r^2\>_{S,\mr{conn}}^\pi < \<r^2\>_{S,\mr{octet}}^\pi < \<r^2\>_{S,ud}^\pi < \<r^2\>_{S,\mr{singlet}}^\pi$
\cite{Koponen:2015tkr}.

%%%%%%%%%%%%%%%%%%%%%%%%%%%%%%%%%%%%%%%%%%%%%%%%%%%%%%%%%%%%%%%%%%%%%%%%%%%%%%%

\subsubsection{Epilogue}

In this subsection there are several quantities for which only one qualifying
(``all-green'') determination is available for a given $SU(2)$ LEC.
Obviously the phenomenologically oriented reader is encouraged to use
such a value (as provided in our tables) and to cite the original work.
We hope that the lattice community will
come up with further computations, in particular for $\Nf=2+1+1$, such
that a fair comparison of different works is possible at any $\Nf$,
and eventually a statement can be made about the presence or absence
of an $\Nf$-dependence of $SU(2)$ LECs.

What can be learned about the convergence pattern of $SU(2)$ {\Ch}PT from
varying the fit ranges (in $m_{ud}$) of the pion mass and decay
constant (i.e.\ the quantities from which $\lbar_3,\lbar_4$ are derived) is
discussed in Ref.\,\cite{Durr:2014oba}, where also the usefulness of comparing
results from the $x$ and the $\xi$ expansion (with material taken from
Ref.\,\cite{Durr:2013goa}) is emphasized.

Perhaps the most important physics result of this subsection is that the
lattice simulations confirm the approximate validity of the
Gell-Mann-Oakes-Renner formula and show that the square of the pion mass
indeed grows in proportion to $m_{ud}$.  The formula represents the leading
term of the chiral series and necessarily receives corrections
from higher orders.  At first nonleading order, the correction is
determined by the effective coupling constant $\lbar_3$.  The results
collected in Tab.~\ref{tab:l3and4} and in the top panel of
Fig.~\ref{fig:l3l4l6} show that $\lbar_3$ is now known quite well.  They
corroborate the conclusion drawn already in Ref.\,\cite{Durr:2002zx}: the
lattice confirms the estimate of $\lbar_3$ derived in
Ref.~\cite{Gasser:1983yg}.  In the graph of $\Mpi^2$ versus $m_{ud}$, the
values found on the lattice for $\lbar_3$ correspond to remarkably little
curvature: the Gell-Mann-Oakes-Renner formula represents a reasonable first
approximation out to values of $m_{ud}$ that exceed the physical value by
an order of magnitude.

As emphasized by Stern and collaborators \cite{Fuchs:1991cq,Stern:1993rg,
DescotesGenon:1999uh}, the analysis in the framework of {\Ch}PT is
coherent only if ($i$) the leading term in the chiral expansion of
$\Mpi^2$ dominates over the remainder and ($ii$) the ratio $m_s/m_{ud}$ is
close to the value $25.6$ that follows from Weinberg's leading-order
formulae.  In order to investigate the possibility that one or both of
these conditions might fail, the authors proposed a more general framework,
referred to as ``generalized {\Ch}PT'', which includes {\Ch}PT as a special
case.  The results found on the lattice demonstrate that QCD does satisfy
both of the above conditions -- in the context of QCD, the proposed
generalization of the effective theory does not appear to be needed.  There
is a modified version, however, referred to as ``re-summed {\Ch}PT''
\cite{Bernard:2010ex}, which is motivated by the possibility that the
Zweig-rule violating couplings $L_4$ and $L_6$ might be larger than
expected.  The available lattice data do not support this possibility, but
they do not rule it out either (see Sec.~\ref{sec:SU3results} for
details).

%%%%%%%%%%%%%%%%%%%%%%%%%%%%%%%%%%%%%%%%%%%%%%%%%%%%%%%%%%%%%%%%%%%%%%%%%%%%%%%

\subsection{Extraction of $SU(3)$ low-energy constants \label{sec:SU3results}}

%%%%%%%%%%%%%%%%%%%%%%%%%%%%%%%%%%%%%%%%%%%%%%%%%%%%%%%%%%%%%%%%%%%%%%%%%%%%%%%

To date, there are three comprehensive $SU(3)$ papers with results based on
lattice QCD with $\Nf\!=\!2+1$ dynamical flavours \cite{Allton:2008pn,
Aoki:2008sm,Bazavov:2009bb}, and one more with results based on
$\Nf\!=\!2+1+1$ dynamical flavours \cite{Dowdall:2013rya}.  It is
an open issue whether the data collected at
$m_s \simeq m_s^\mathrm{phys}$ allow for an unambiguous determination of
$SU(3)$ low-energy constants (cf.\ the discussion in Ref.~\cite{Allton:2008pn}).
To make definite statements one needs data at considerably smaller $m_s$,
and so far only MILC has some \cite{Bazavov:2009bb}.  We are aware of a few
papers with a result on one $SU(3)$ low-energy constant each
which we list for completeness.  Some particulars of the computations are listed in
Tab.~\ref{tab:SU3_overview}.

\begin{table}[!tbp] %%% F0 and B0
\vspace*{3cm}
\centering
\footnotesize
\begin{tabular*}{\textwidth}[l]{l@{\extracolsep{\fill}}rcllllllllll}
 Collaboration & Ref. & $\Nf$ &
\hspace{0.15cm}\begin{rotate}{60}{publication status}\end{rotate}\hspace{-0.15cm} &
\hspace{0.15cm}\begin{rotate}{60}{chiral extrapolation}\end{rotate}\hspace{-0.15cm} &
\hspace{0.15cm}\begin{rotate}{60}{continuum  extrapolation}\end{rotate}\hspace{-0.15cm} &
\hspace{0.15cm}\begin{rotate}{60}{finite volume}\end{rotate}\hspace{-0.15cm} &
\rule{0.3cm}{0cm}$F_0$ & \rule{0.1cm}{0cm} $F/F_0$ & \rule{0.2cm}{0cm}$B/B_0$ & \hspace{2.5cm} \\[2mm]
\hline
\hline
\\[-2mm]
JLQCD/TWQCD 10A         & \cite{Fukaya:2010na}       &  3  & \gA & \bad  & \bad  & \bad  & 71(3)(8)       &                           &                                  \\[2mm]
\hline
\\[-2mm]
MILC 10                 & \cite{Bazavov:2010hj}      & 2+1 & \rC & \soso & \good & \good & 80.3(2.5)(5.4) &                           &                                  \\
MILC 09A                & \cite{Bazavov:2009fk}      & 2+1 & \rC & \soso & \good & \good & 78.3(1.4)(2.9) & {\sl 1.104(3)(41)}        & {\sl 1.21(4)$\binom{+5}{-6}$}    \\
MILC 09                 & \cite{Bazavov:2009bb}      & 2+1 & \gA & \soso & \good & \good &                & 1.15(5)$\binom{+13}{-03}$ & {\sl 1.15(16)$\binom{+39}{-13}$} \\
PACS-CS 08              & \cite{Aoki:2008sm}         & 2+1 & \gA & \good & \bad  & \bad  & 83.8(6.4)      & 1.078(44)                 & 1.089(15)                        \\
RBC/UKQCD 08            & \cite{Allton:2008pn}       & 2+1 & \gA & \soso & \bad  & \soso & 66.1(5.2)      & 1.229(59)                 & 1.03(05)                         \\[2mm]
\hline
\hline
\end{tabular*}
\newline
\vspace*{3.5cm}
\begin{tabular*}{\textwidth}[l]{l@{\extracolsep{\fill}}rclllllll}
 Collaboration & Ref. & $\Nf$ &
\hspace{0.15cm}\begin{rotate}{60}{publication status}\end{rotate}\hspace{-0.15cm} &
\hspace{0.15cm}\begin{rotate}{60}{chiral extrapolation}\end{rotate}\hspace{-0.15cm} &
\hspace{0.15cm}\begin{rotate}{60}{continuum  extrapolation}\end{rotate}\hspace{-0.15cm} &
\hspace{0.15cm}\begin{rotate}{60}{finite volume}\end{rotate}\hspace{-0.15cm} &
\hspace{0.15cm}\begin{rotate}{60}{renormalization}\end{rotate}\hspace{-0.15cm} &
\rule{0.3cm}{0cm}$\Sigma_0^{1/3}$ & \rule{0.1cm}{0cm} $\Sigma/\Sigma_0$ \\[2mm]
\hline
\hline
\\[-2mm]
JLQCD/TWQCD 10A         & \cite{Fukaya:2010na}       &  3  & \gA & \bad  & \bad  & \bad  & \good & 214(6)(24)                  & {\sl 1.31(13)(52)}         \\[2mm]
\hline
\\[-2mm]
MILC 09A                & \cite{Bazavov:2009fk}      & 2+1 & \rC & \soso & \good & \good & \good & 245(5)(4)(4)                & {\sl 1.48(9)(8)(10)}       \\
MILC 09                 & \cite{Bazavov:2009bb}      & 2+1 & \gA & \soso & \good & \good & \good & 242(9)$\binom{+05}{-17}$(4) & 1.52(17)$\binom{+38}{-15}$ \\
PACS-CS 08              & \cite{Aoki:2008sm}         & 2+1 & \gA & \good & \bad  & \bad  & \bad  & 290(15)                     & 1.245(10)                  \\
RBC/UKQCD 08            & \cite{Allton:2008pn}       & 2+1 & \gA & \soso & \bad  & \soso & \good &                             & 1.55(21)                   \\[2mm]
\hline
\hline
\end{tabular*}
\normalsize
\vspace*{-2mm}
\caption{\label{tab:SU3_overview}
Lattice results for the low-energy constants $F_0$, $B_0$ (in MeV) and
$\Sigma_0\!\equiv\!F_0^2B_0$, which specify the effective $SU(3)$ Lagrangian at
leading order. The ratios $F/F_0$, $B/B_0$, $\Sigma/\Sigma_0$, which
compare these with their $SU(2)$ counterparts, indicate the strength of the
Zweig-rule violations in these quantities (in the large-$N_c$ limit, they
tend to unity). Numbers in slanted fonts are calculated by us, from the
information given in the references.}
\end{table}

\begin{table}[!tbp] %%% NLO in SU(3)
\vspace*{5mm}
\centering
\footnotesize
%\begin{tabular*}{\textwidth}[l]{l@{\extracolsep{\fill}}rcllllllll}
%{} & Ref. & $\Nf$ &
\begin{tabular*}{\textwidth}[l]{l@{\extracolsep{\fill}}r@{\hspace{1mm}}c@{\hspace{1mm}}l@{\hspace{1mm}}l@{\hspace{1mm}}l@{\hspace{1mm}}l@{\hspace{1mm}}l@{\hspace{1mm}}l@{\hspace{1mm}}l@{\hspace{1mm}}l}
Collaboration & Ref. & $\Nf$ &
\hspace{0.15cm}\begin{rotate}{60}{publication status}\end{rotate}\hspace{-0.15cm} &
\hspace{0.15cm}\begin{rotate}{60}{chiral extrapolation}\end{rotate}\hspace{-0.15cm} &
\hspace{0.15cm}\begin{rotate}{60}{continuum  extrapolation}\end{rotate}\hspace{-0.15cm} &
\hspace{0.15cm}\begin{rotate}{60}{finite volume}\end{rotate}\hspace{-0.15cm} &
\rule{0.1cm}{0cm}$10^3L_4$ &$\rule{0.1cm}{0cm}10^3L_6$ & \hspace{-0.3cm} $10^3(2L_6\!-\!L_4)$ \\[2mm]
\hline
\hline
\\[-2mm]
HPQCD 13A               & \cite{Dowdall:2013rya}     &2+1+1& \gA & \good & \good & \good & 0.09(34)                    & 0.16(20)                         & 0.22(17)                    \\[2mm]
\hline
\\[-2mm]
JLQCD/TWQCD 10A         & \cite{Fukaya:2010na}       &  3  & \gA & \bad  & \bad  & \bad  &                             & 0.03(7)(17)                      &                             \\[2mm]
\hline
\\[-2mm]
MILC 10                 & \cite{Bazavov:2010hj}      & 2+1 & \rC & \soso & \good & \good & -0.08(22)$\binom{+57}{-33}$ & {\sl-0.02(16)$\binom{+33}{-21}$} & 0.03(24)$\binom{+32}{-27}$  \\
MILC 09A                & \cite{Bazavov:2009fk}      & 2+1 & \rC & \soso & \good & \good & 0.04(13)(4)                 & 0.07(10)(3)                      & 0.10(12)(2)                 \\
MILC 09                 & \cite{Bazavov:2009bb}      & 2+1 & \gA & \soso & \good & \good & 0.1(3)$\binom{+3}{-1}$      & 0.2(2)$\binom{+2}{-1}$           & 0.3(1)$\binom{+2}{-3}$      \\
PACS-CS 08              & \cite{Aoki:2008sm}         & 2+1 & \gA & \good & \bad  & \bad  & -0.06(10)(--)               & {\sl0.02(5)(--)}                 & 0.10(2)(--)                 \\
RBC/UKQCD 08            & \cite{Allton:2008pn}       & 2+1 & \gA & \soso & \bad  & \soso & 0.14(8)(--)                 & 0.07(6)(--)                      & 0.00(4)(--)                 \\[2mm]
\hline
\\[-2mm]
Bijnens 11              & \cite{Bijnens:2011tb}      &     &     &       &       &       & 0.75(75)                    & 0.29(85)                         & {\sl-0.17(1.86)}            \\
Gasser 85               & \cite{Gasser:1984gg}       &     &     &       &       &       & -0.3(5)                     & -0.2(3)                          & {\sl-0.1(8)}                \\[2mm]
\hline
\hline
\\
Collaboration  & Ref. & $\Nf$ & & & & &
\rule{0.1cm}{0cm} $10^3L_5$ &\rule{0.05cm}{0cm}  $10^3L_8$ &\hspace{-0.3cm} $10^3(2L_8\!-\!L_5)$ \\[2mm]
\hline
\hline
\\[-2mm]
HPQCD 13A               & \cite{Dowdall:2013rya}     &2+1+1& \gA & \good & \good & \good & 1.19(25)                    & 0.55(15)                         & -0.10(20)                   \\[2mm]
\hline
\\[-2mm]
MILC 10                 & \cite{Bazavov:2010hj}      & 2+1 & \rC & \soso & \good & \good & 0.98(16)$\binom{+28}{-41}$  & {\sl0.42(10)$\binom{+27}{-23}$}  & -0.15(11)$\binom{+45}{-19}$ \\
MILC 09A                & \cite{Bazavov:2009fk}      & 2+1 & \rC & \soso & \good & \good & 0.84(12)(36)                & 0.36(5)(7)                       & -0.12(8)(21)                \\
MILC 09                 & \cite{Bazavov:2009bb}      & 2+1 & \gA & \soso & \good & \good & 1.4(2)$\binom{+2}{-1}$      & 0.8(1)(1)                        & 0.3(1)(1)                   \\
PACS-CS 08              & \cite{Aoki:2008sm}         & 2+1 & \gA & \good & \bad  & \bad  & 1.45(7)(--)                 & {\sl0.62(4)(--)}                 & -0.21(3)(--)                \\
RBC/UKQCD 08            & \cite{Allton:2008pn}       & 2+1 & \gA & \soso & \bad  & \soso & 0.87(10)(--)                & 0.56(4)(--)                      & 0.24(4)(--)                 \\
NPLQCD 06               & \cite{Beane:2006kx}        & 2+1 & \gA & \soso & \bad  & \bad  & 1.42(2)$\binom{+18}{-54}$   &                                  &                             \\[2mm]
\hline
\\[-2mm]
Bijnens 11              & \cite{Bijnens:2011tb}      &     &     &       &       &       & 0.58(13)                    & 0.18(18)                         & {\sl-0.22(38)}              \\
Gasser 85               & \cite{Gasser:1984gg}       &     &     &       &       &       & 1.4(5)                      & 0.9(3)                           & {\sl0.4(8)}                 \\[2mm]
\hline
\hline
\\
Collaboration  & Ref. & $\Nf$ & & & & &
\rule{0.1cm}{0cm} $10^3L_9$ &\rule{0.1cm}{0cm} $10^3L_{10}$ & \\[2mm]
\hline
\hline
\\[-2mm]
Boito 15                & \cite{Boito:2015fra}       & 2+1 & \oP & \good & \soso & \good &                             & -3.50(17)                        &                             \\
JLQCD 15A               & \cite{Aoki:2015pba}        & 2+1 & \oP & \soso & \bad  & \soso & 4.6(1.1)$\binom{+0.1}{-0.5}$(0.4) &                            &                             \\
Boyle 14                & \cite{Boyle:2014pja}       & 2+1 & \gA & \good & \soso & \good &                             & -3.46(32)                        &                             \\
JLQCD 14                & \cite{Fukaya:2014jka}      & 2+1 & \gA & \good & \bad  & \bad  & 2.4(0.8)(1.0)               &                                  &                             \\
RBC/UKQCD 09            & \cite{Boyle:2009xi}        & 2+1 & \gA & \soso & \bad  & \soso &                             & -5.7(11)(07)                     &                             \\
RBC/UKQCD 08A           & \cite{Boyle:2008yd}        & 2+1 & \gA & \soso & \bad  & \soso & 3.08(23)(51)                &                                  &                             \\[2mm]
\hline
\\[-2mm]
JLQCD 08A               & \cite{Shintani:2008qe}     &  2  & \gA & \soso & \bad  & \bad  &                             & -5.2(2)$\binom{+5}{-3}$          &                             \\[2mm]
\hline
\\[-2mm]
Bijnens 02              & \cite{Bijnens:2002hp}      &     &     &       &       &       & 5.93(43)                    &                                  &                             \\
Davier 98               & \cite{Davier:1998dz}       &     &     &       &       &       &                             & -5.13(19)                        &                             \\
Gasser 85               & \cite{Gasser:1984gg}       &     &     &       &       &       & 6.9(7)                      & -5.5(7)                          &                             \\[2mm]
\hline
\hline
\end{tabular*}
\normalsize
\vspace*{-2mm}
\caption{\label{tab:SU3_NLO}
Low-energy constants of the $SU(3)$ Lagrangian at NLO with running scale $\mu\!=\!770\MeV$
(the values in Refs.~\cite{Dowdall:2013rya,Bazavov:2010hj,Bazavov:2009bb,Bazavov:2009fk,Gasser:1984gg}
are evolved accordingly). The MILC 10 entry for $L_6$ is obtained from their results for
$2L_6\!-\!L_4$ and $L_4$ (similarly for other entries in slanted fonts).
The JLQCD 08A result for $\ell_5(770\MeV)$ [despite the paper saying $L_{10}(770\MeV)$]
was converted to $L_{10}$ with the GL one-loop formula, assuming that the difference
between $\lbar_5(m_s\!=\!m_s^\mr{phys})$ [needed in the formula]
and $\lbar_5(m_s\!=\!\infty)$ [computed by JLQCD] is small.}
\end{table}

%%%  results from \bibitem{Bijnens:2011tb}
%%%    10^3 L_1 (mu=0.77 GeV) = 0.88(9)
%%%    10^3 L_2 (mu=0.77 GeV) = 0.61(20)
%%%    10^3 L_3 (mu=0.77 GeV) = -3.04(43)
%%%    10^3 L_4 (mu=0.77 GeV) = 0.75(75)
%%%    10^3 L_5 (mu=0.77 GeV) = 0.58(13)
%%%    10^3 L_6 (mu=0.77 GeV) = 0.29(85)
%%%    10^3 L_7 (mu=0.77 GeV) = -0.11(15)
%%%    10^3 L_8 (mu=0.77 GeV) = 0.18(18)

%%% oneL4_cen:=  0.75; oneL4_err:=  0.75;
%%% oneL5_cen:=  0.58; oneL5_err:=  0.13;
%%% twoL6_cen:=2*0.29; twoL6_err:=2*0.85;
%%% twoL8_cen:=2*0.18; twoL8_err:=2*0.18;
%%% twoL6_cen-oneL4_cen; sqrt(twoL6_err^2+oneL4_err^2); % --> -0.17(1.86)
%%% twoL8_cen-oneL5_cen; sqrt(twoL8_err^2+oneL5_err^2); % --> -0.22(0.38)

Results for the $SU(3)$ low-energy constants of leading order are found in
Tab.~\ref{tab:SU3_overview} and analogous results for some of the
effective coupling constants that enter the chiral $SU(3)$ Lagrangian at NLO
are collected in Tab.~\ref{tab:SU3_NLO}.  From PACS-CS \cite{Aoki:2008sm}
only those results are quoted which have been \emph{corrected} for
finite-size effects (misleadingly labelled ``w/FSE'' in their tables).  For
staggered data our colour-coding rule states that $\Mpi$ is to be
understood as $\Mpi^\mr{RMS}$.  The rating of
Refs.~\cite{Bazavov:2009bb,Bazavov:2010hj} is based on the information regarding
the RMS masses given in Ref.~\cite{Bazavov:2009fk}.
Finally, Refs.\,\cite{Boyle:2014pja,Boito:2015fra} are ``hybrids'' in the
sense that they combine lattice data and experimental information.

\begin{figure}[!tbp]
\centering
\includegraphics[width=8.1cm]{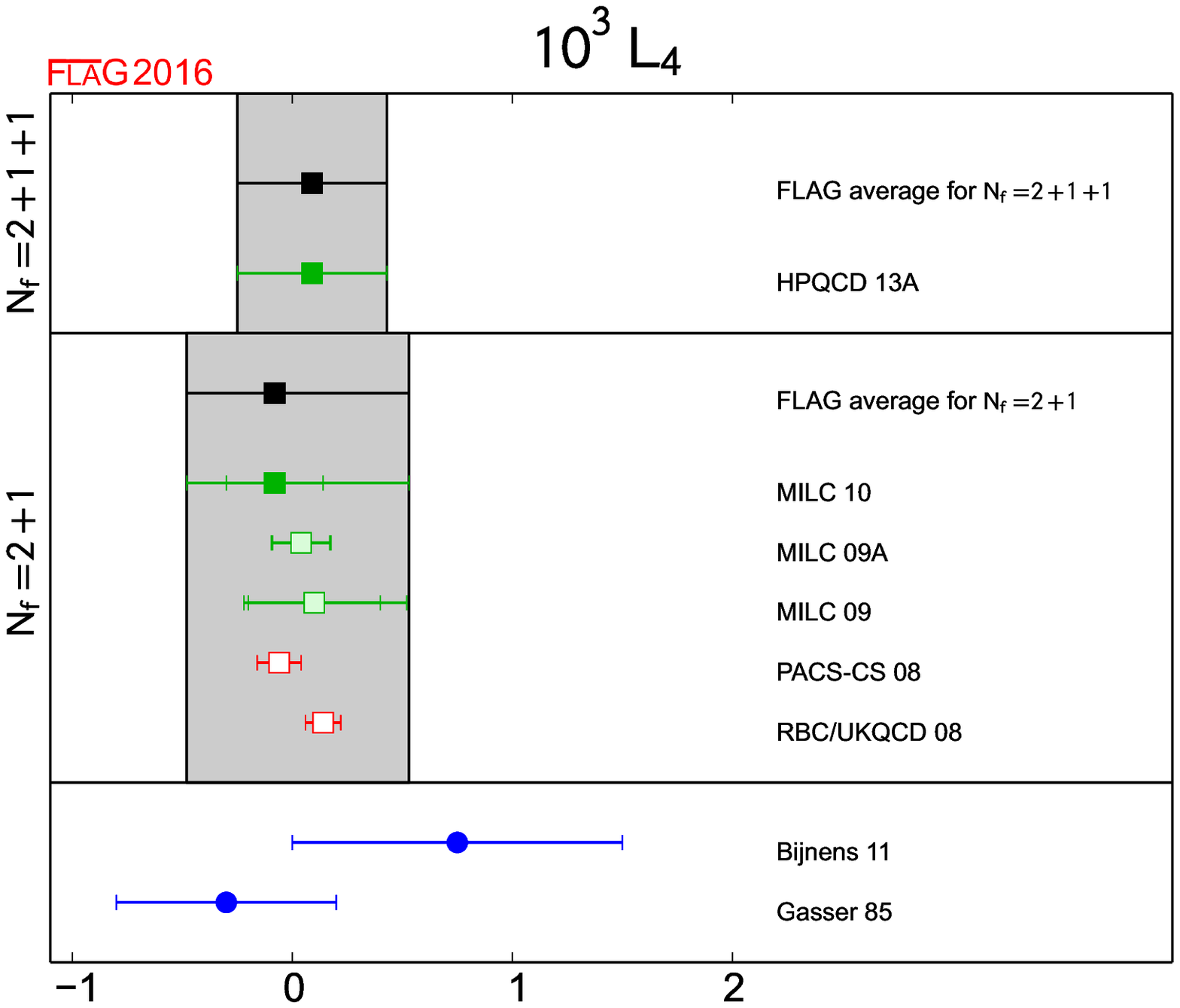}%
\includegraphics[width=8.1cm]{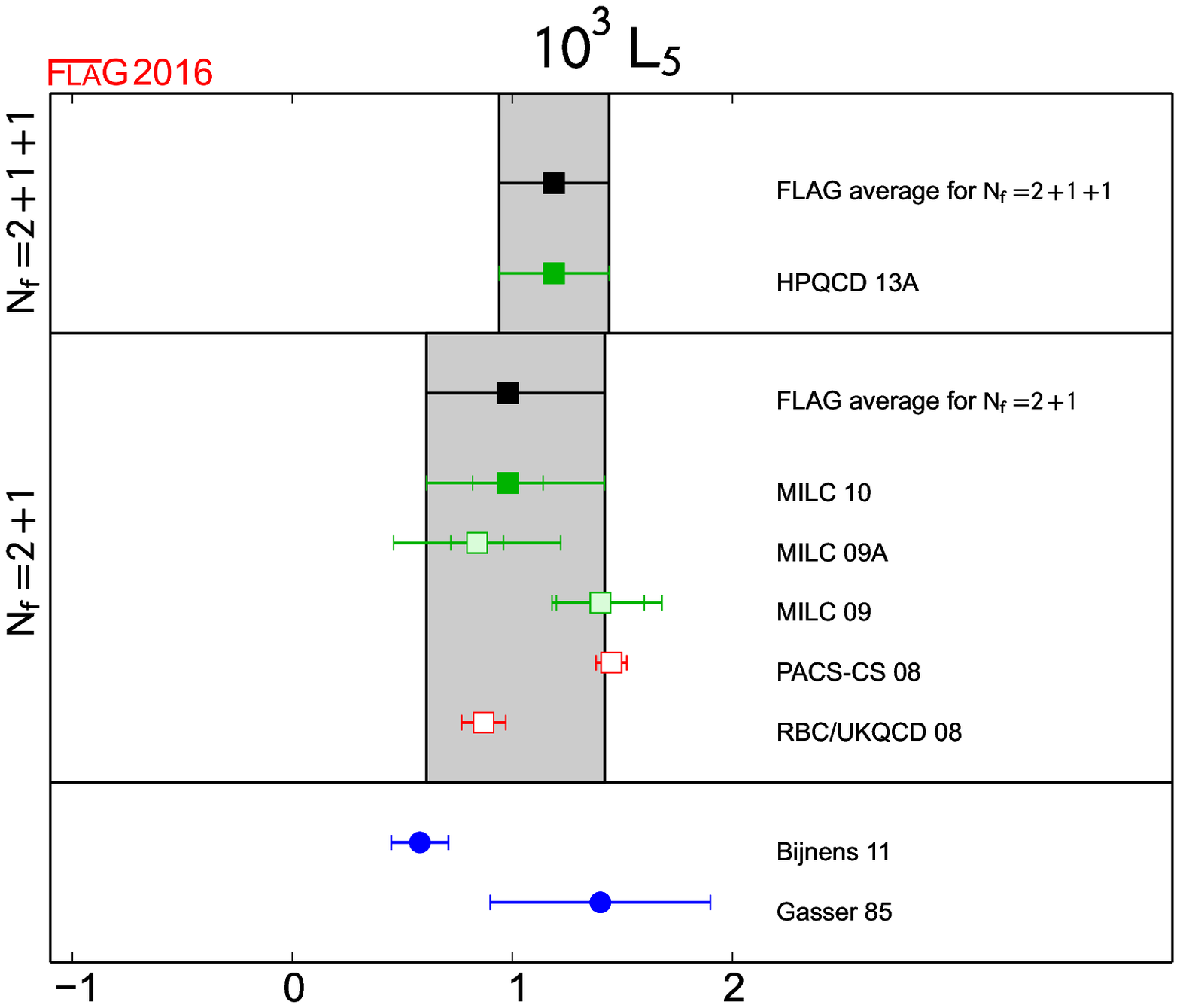}\\
\includegraphics[width=8.1cm]{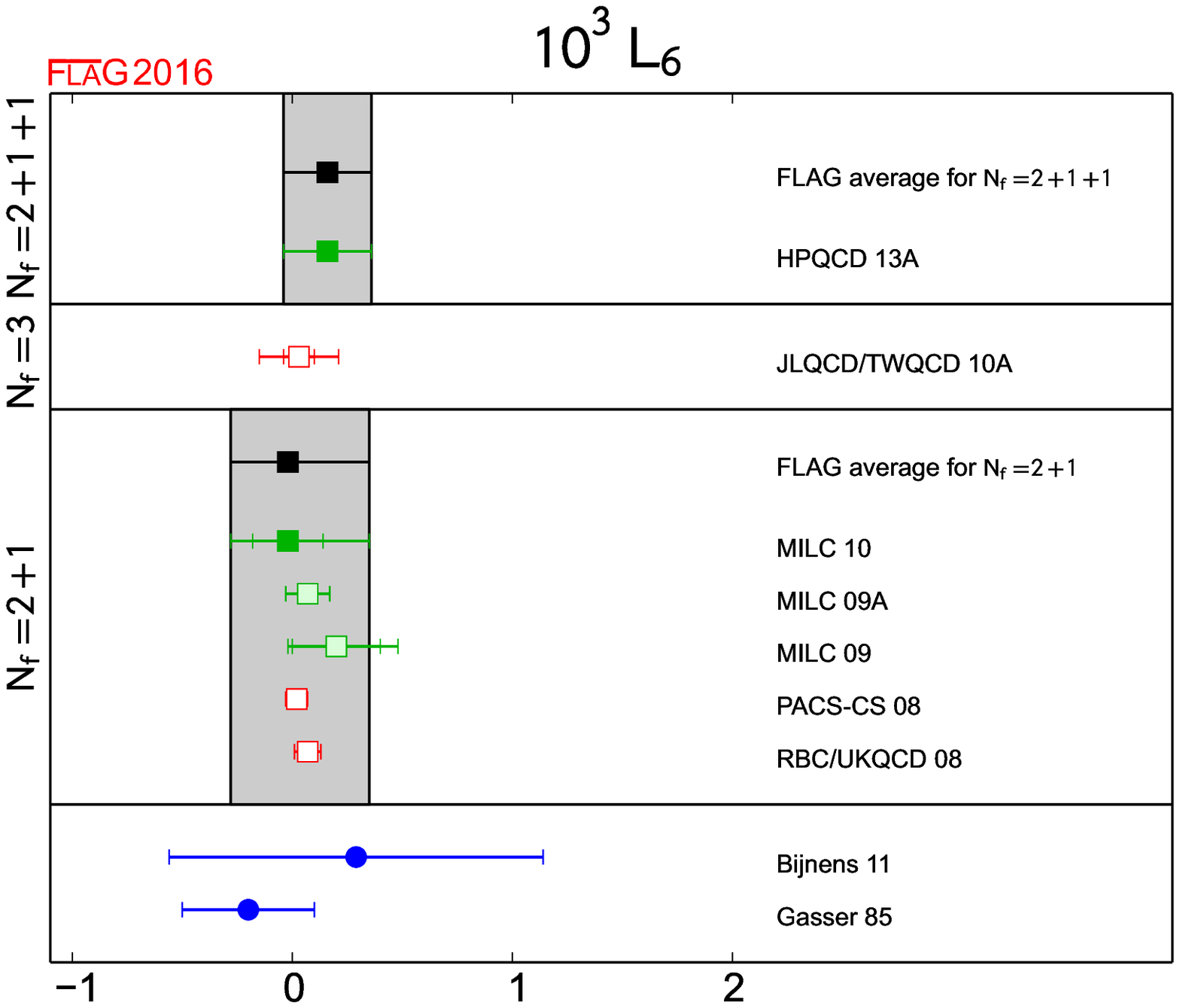}%
\includegraphics[width=8.1cm]{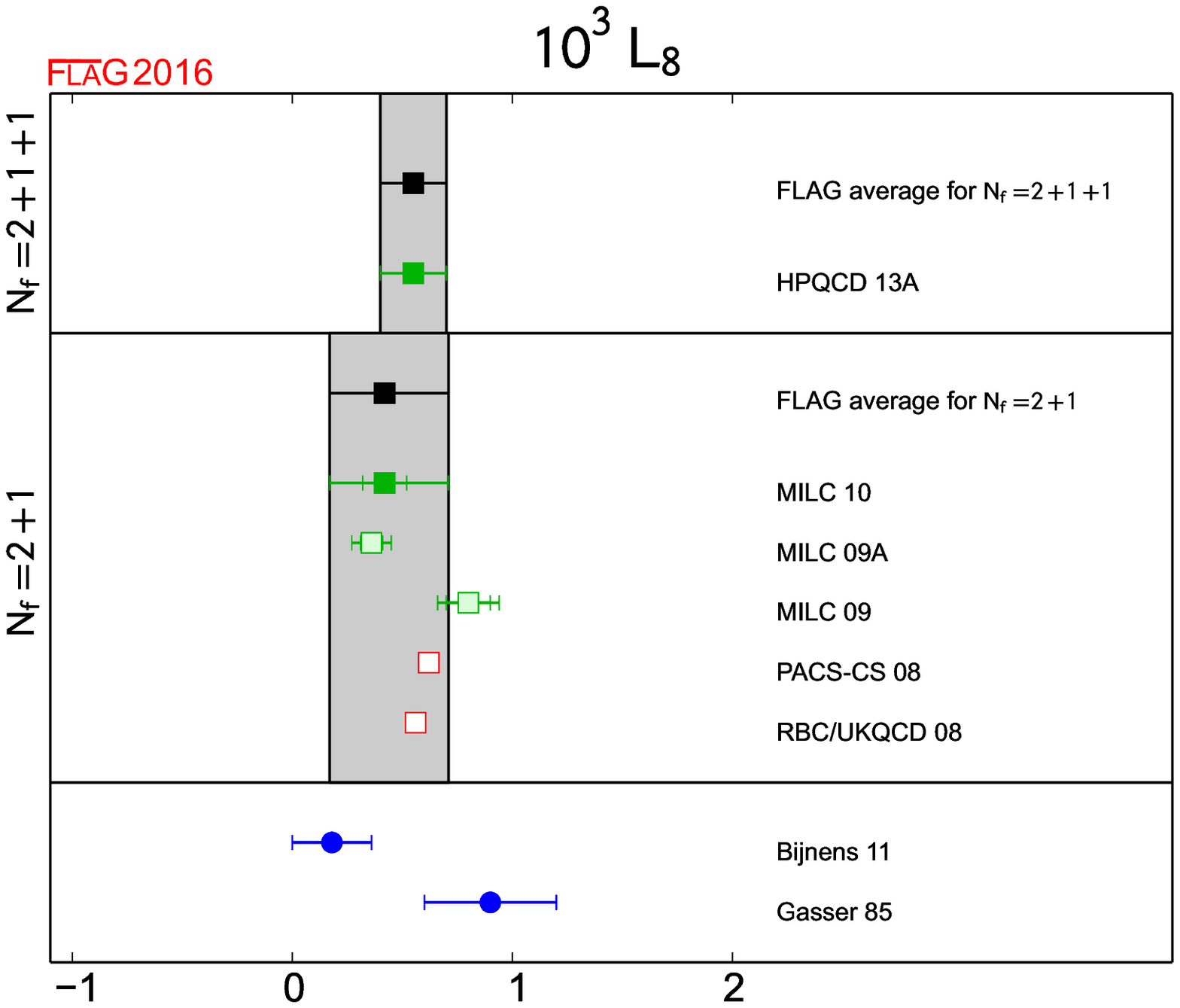}\\
\includegraphics[width=8.1cm]{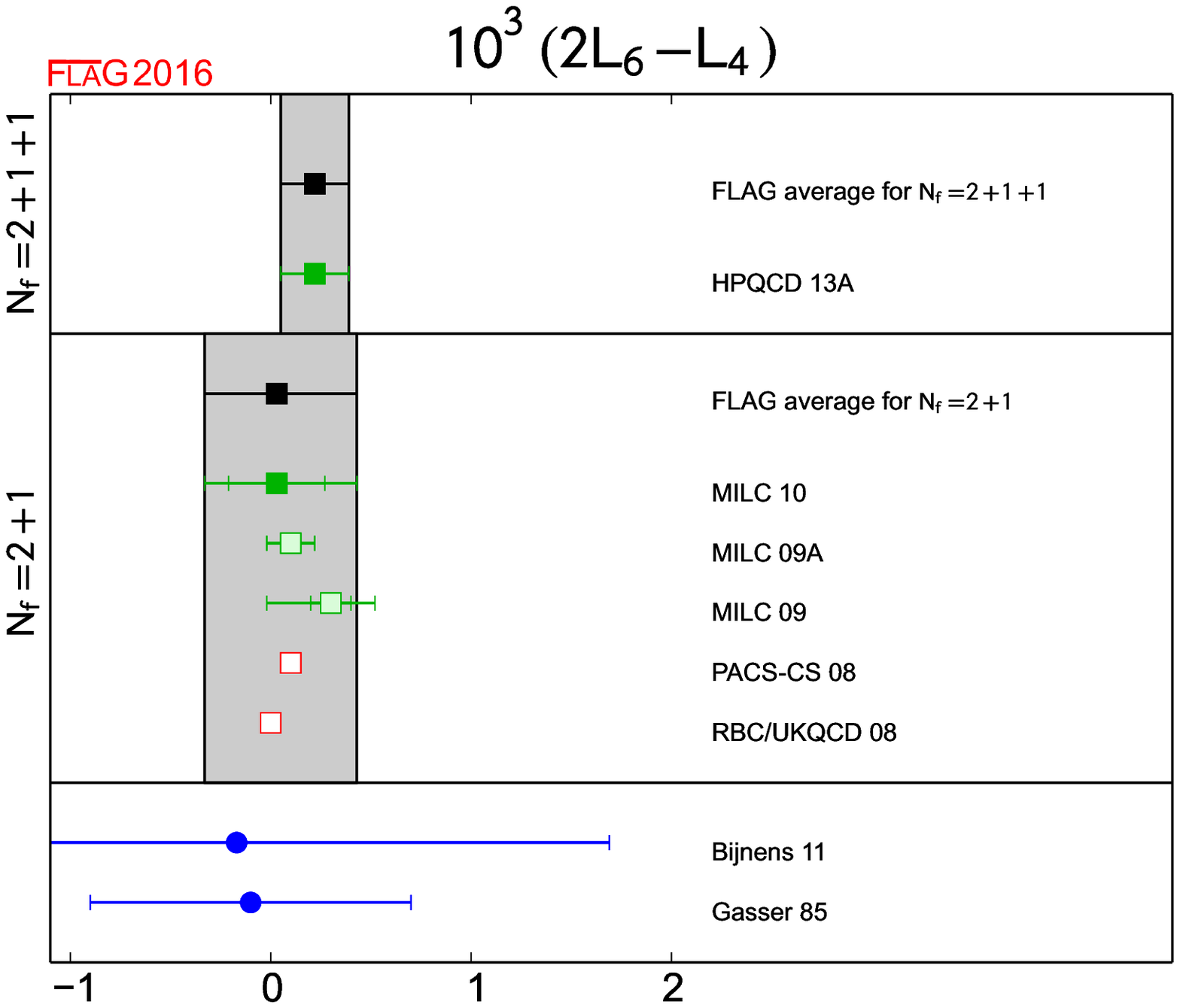}%
\includegraphics[width=8.1cm]{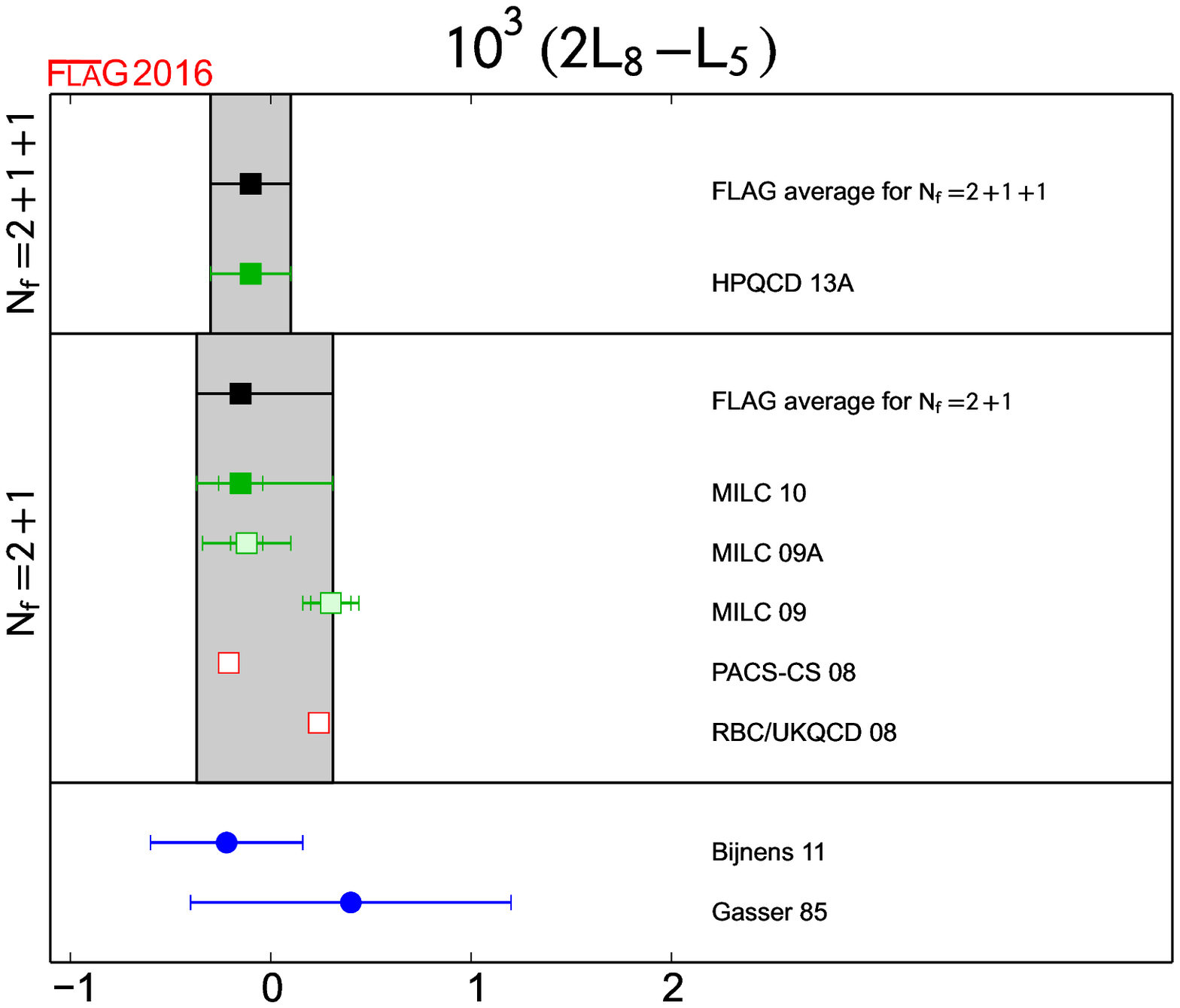}%
\vspace*{-2mm}
\caption{\label{fig:SU3}
Low-energy constants that enter the effective $SU(3)$ Lagrangian at NLO,
with scale $\mu=770\MeV$.
The grey bands labelled as ``FLAG average'' coincide with the results of
MILC 10 \cite{Bazavov:2010hj} for $\Nf=2+1$ and with HPQCD 13A
\cite{Dowdall:2013rya} for $\Nf=2+1+1$, respectively.}
\end{figure}

A graphical summary of the lattice results for the coupling constants
$L_4$, $L_5$, $L_6$ and $L_8$, which determine the masses and the decay
constants of the pions and kaons at NLO of the chiral $SU(3)$ expansion, is
displayed in Fig.~\ref{fig:SU3}, along with the two
phenomenological determinations quoted in the above tables.  The overall
consistency seems fairly convincing.  In spite of this apparent
consistency, there is a point which needs to be clarified as soon as
possible. Some collaborations (RBC/UKQCD and PACS-CS) find that they are
having difficulties in fitting their partially quenched data to the
respective formulas for pion masses above $\simeq$ 400 MeV.  Evidently,
this indicates that the data are stretching the regime of validity of these
formulas. To date it is, however, not clear which subset of the data causes
the troubles, whether it is the unitary part extending to too large values
of the quark masses or whether it is due to $m^\mathrm{val}/m^\mathrm{sea}$
differing too much from one.  In fact, little is known, in the framework of
partially quenched {\Ch}PT, about the \emph{shape} of the region of
applicability in the $m^\mathrm{val}$ versus $m^\mathrm{sea}$ plane for
fixed $\Nf$.  This point has also been emphasized in Ref.~\cite{Durr:2013koa}.

To date only the computations MILC 10 \cite{Bazavov:2010hj} (as an obvious
update of MILC 09 and MILC 09A) and HPQCD 13A \cite{Dowdall:2013rya} are
free of red tags.  Since they use different $\Nf$ (in the former case
$\Nf=2+1$, in the latter case $\Nf=2+1+1$) we stay away from averaging
them.  Hence the situation remains unsatisfactory in the sense that for
each $\Nf$ only a single determination of high standing is available.
Accordingly, for the phenomenologically oriented reader there is no
alternative to using the results of MILC 10 \cite{Bazavov:2010hj} for $\Nf=2+1$ and
HPQCD 13A \cite{Dowdall:2013rya} for $\Nf=2+1+1$, as given in Tab.~\ref{tab:SU3_NLO}.

\subsubsection{Epilogue}

In this subsection we find ourselves again in the unpleasant situation
that only one qualifying (``all-green'') determination is available (at a
given $\Nf$) for several LECs in the $SU(3)$ framework, both at LO and at NLO.
Obviously the phenomenologically oriented reader is encouraged to use
such a value (as provided in our tables) and to cite the original work.
Again our hope is that further computations would become available in
forthcoming years, such that a fair comparison of different works will
become possible both at $\Nf=2+1$ and $\Nf=2+1+1$.

In the large-$N_c$ limit, the Zweig rule becomes exact, but the quarks have
$N_c=3$.  The work done on the lattice is ideally suited to confirm or
disprove the approximate validity of this rule for QCD.  Two of the coupling
constants entering the effective $SU(3)$ Lagrangian at NLO disappear when
$N_c$ is sent to infinity: $L_4$ and $L_6$.  The upper part of
Tab.~\ref{tab:SU3_NLO} and the left panels of Fig.~\ref{fig:SU3} show
that the lattice results for these quantities are in good agreement.
At the scale $\mu=M_\rho$, $L_4$ and $L_6$ are consistent
with zero, indicating that these constants do approximately obey the
Zweig rule.  As mentioned above, the ratios $F/F_0$, $B/B_0$ and
$\Sigma/\Sigma_0$ also test the validity of this rule.  Their expansion in
powers of $m_s$ starts with unity and the contributions of first order in
$m_s$ are determined by the constants $L_4$ and $L_6$, but they also
contain terms of higher order.  Apart from measuring the Zweig-rule
violations, an accurate determination of these ratios will thus also allow
us to determine the range of $m_s$ where the first few terms of the
expansion represent an adequate approximation.  Unfortunately, at present,
the uncertainties in the lattice data on these ratios are too large to draw
conclusions, both concerning the relative size of the subsequent terms in
the chiral series and concerning the magnitude of the
Zweig-rule violations.  The data seem to confirm the {\it paramagnetic
inequalities} \cite{DescotesGenon:1999uh}, which require $F/F_0>1$,
$\Sigma/\Sigma_0>1$, and it appears that the ratio $B/B_0$ is also larger
than unity, but the numerical results need to be improved before further
conclusions can be drawn.

The matching formulae in Ref.~\cite{Gasser:1984gg} can be used to calculate the
$SU(2)$ couplings $\bar{l}_i$ from the $SU(3)$ couplings $L_j$. Results
obtained in this way are included in Tab.~\ref{tab:l3and4}, namely the
entries explicitly labelled ``$SU(3)$-fit'' as well as MILC 10. Within the
still rather large errors, the converted LECs from the $SU(3)$ fits agree
with those directly determined within $SU(2)$ {\Ch}PT.  We plead with every
collaboration performing $\Nf=2+1$ simulations to also \emph{directly}
analyse their data in the $SU(2)$ framework.  In practice, lattice
simulations are performed at values of $m_s$ close to the physical value
and the results are then corrected for the difference of $m_s$ from its
physical value.  If simulations with more than one value of $m_s$ have been
performed, this can be done by interpolation.  Alternatively one can use
the technique of \emph{re-weighting} (for a review see e.g.
Ref.~\cite{Jung:2010jt}) to shift $m_s$ to its physical value.

%% file: BK/BK.tex
\section{Kaon mixing}
\label{sec:BK}

The mixing of neutral pseudoscalar mesons plays an important role in
the understanding of the physics of CP violation. In this section we
discuss $K^0 - \bar K^0$ oscillations, which probe the physics of
indirect CP violation. Extensive reviews on the subject can be found
in Refs.~\cite{Branco:1999fs,Buchalla:1995vs,Buras:1998raa}. For the
most part we shall focus on kaon mixing in the SM. The case of
Beyond-the-Standard-Model (BSM) contributions is discussed in
section~\ref{sec:Bi}.

\subsection{Indirect CP violation and $\epsilon_{K}$ in the
  SM \label{sec:indCP}} 

Indirect CP violation arises in $K_L \rightarrow \pi \pi$ transitions
through the decay of the $\rm CP=+1$ component of $K_L$ into two pions
(which are also in a $\rm CP=+1$ state). Its measure is defined as
\be 
\epsilon_{K} \,\, = \,\, \dfrac{{\cal A} [ K_L \rightarrow
(\pi\pi)_{I=0}]}{{\cal A} [ K_S \rightarrow (\pi\pi)_{I=0}]} \,\, ,
\ee
with the final state having total isospin zero. The parameter
$\epsilon_{K}$ may also be expressed in terms of $K^0 - \bar K^0$
oscillations. In particular, to lowest order in the electroweak
theory, the contribution to these oscillations arises from so-called
box diagrams, in which two $W$ bosons and two ``up-type" quarks
(i.e. up, charm, top) are exchanged between the constituent down and
strange quarks of the $K$ mesons. The loop integration of the box
diagrams can be performed exactly. In the limit of vanishing external
momenta and external quark masses, the result can be identified with
an effective four-fermion interaction, expressed in terms of the
``effective Hamiltonian"
\be
  {\cal H}_{\rm eff}^{\Delta S = 2} \,\, = \,\,
  \frac{G_F^2 M_{\rm{W}}^2}{16\pi^2} {\cal F}^0 Q^{\Delta S=2} \,\, +
   \,\, {\rm h.c.} \,\,.
\ee
In this expression, $G_F$ is the Fermi coupling, $M_{\rm{W}}$ the
$W$-boson mass, and
\be
   Q^{\Delta S=2} =
   \left[\bar{s}\gamma_\mu(1-\gamma_5)d\right]
   \left[\bar{s}\gamma_\mu(1-\gamma_5)d\right]
   \equiv O_{\rm VV+AA}-O_{\rm VA+AV} \,\, ,
\label{eq:Q1def}
\ee
is a dimension-six, four-fermion operator. The function ${\cal F}^0$
is given by
\be
{\cal F}^0 \,\, = \,\, \lambda_c^2 S_0(x_c) \, + \, \lambda_t^2
S_0(x_t) \, + \, 2 \lambda_c  \lambda_t S_0(x_c,x_t)  \,\, , 
\ee
where $\lambda_a = V^\ast_{as} V_{ad}$, and $a=c\,,t$ denotes a
flavour index. The quantities $S_0(x_c),\,S_0(x_t)$ and $S_0(x_c,x_t)$
with $x_c=m_c^2/M_{\rm{W}}^2$, $x_t=m_t^2/M_{\rm{W}}^2$ are the
Inami-Lim functions \cite{Inami:1980fz}, which express the basic
electroweak loop contributions without QCD corrections. The
contribution of the up quark, which is taken to be massless in this
approach, has been taken into account by imposing the unitarity
constraint $\lambda_u + \lambda_c + \lambda_t = 0$.

When strong interactions are included, $\Delta{S}=2$ transitions can
no longer be discussed at the quark level. Instead, the effective
Hamiltonian must be considered between mesonic initial and final
states. Since the strong coupling is large at typical hadronic scales,
the resulting weak matrix element cannot be calculated in perturbation
theory. The operator product expansion (OPE) does, however, factorize
long- and short- distance effects. For energy scales below the charm
threshold, the $K^0-\bar K^0$ transition amplitude of the effective
Hamiltonian can be expressed as
\begin{eqnarray}
\label{eq:Heff}
\langle \bar K^0 \vert {\cal H}_{\rm eff}^{\Delta S = 2} \vert K^0
\rangle  \,\, = \,\, \frac{G_F^2 M_{\rm{W}}^2}{16 \pi^2}  
\Big [ \lambda_c^2 S_0(x_c) \eta_1  \, + \, \lambda_t^2 S_0(x_t)
  \eta_2 \, + \, 2 \lambda_c  \lambda_t S_0(x_c,x_t) \eta_3
  \Big ]  \nn \\ 
\times 
  \left(\frac{\gbar(\mu)^2}{4\pi}\right)^{-\gamma_0/(2\beta_0)}
  \exp\bigg\{ \int_0^{\gbar(\mu)} \, dg \, \bigg(
  \frac{\gamma(g)}{\beta(g)} \, + \, \frac{\gamma_0}{\beta_0g} \bigg)
  \bigg\} 
   \langle \bar K^0 \vert  Q^{\Delta S=2}_{\rm R} (\mu) \vert K^0
   \rangle \,\, + \,\, {\rm h.c.} \,\, ,
\end{eqnarray}
where $\gbar(\mu)$ and $Q^{\Delta S=2}_{\rm R}(\mu)$ are the
renormalized gauge coupling and four-fermion operator in some
renormalization scheme. The factors $\eta_1, \eta_2$ and $\eta_3$
depend on the renormalized coupling $\gbar$, evaluated at the various
flavour thresholds $m_t, m_b, m_c$ and $ M_{\rm{W}}$, as required by
the OPE and RG-running procedure that separate high- and low-energy
contributions. Explicit expressions can be found
in Refs.~\cite{Buchalla:1995vs} and references therein, except that $\eta_1$
and $\eta_3$ have been recently calculated to NNLO in
Refs.~\cite{Brod:2011ty} and \cite{Brod:2010mj}, respectively.  We
follow the same conventions for the RG equations as in
Ref.~\cite{Buchalla:1995vs}. Thus the Callan-Symanzik function and the
anomalous dimension $\gamma(\gbar)$ of $Q^{\Delta S=2}$ are defined by
\be
\dfrac{d \gbar}{d \ln \mu} = \beta(\gbar)\,,\qquad
\dfrac{d Q^{\Delta S=2}_{\rm R}}{d \ln \mu} =
-\gamma(\gbar)\,Q^{\Delta S=2}_{\rm R} \,\,,  
\label{eq:four_quark_operator_anomalous_dimensions}
\ee
with perturbative expansions
\begin{eqnarray}
\beta(g)  &=&  -\beta_0 \dfrac{g^3}{(4\pi)^2} \,\, - \,\, \beta_1
\dfrac{g^5}{(4\pi)^4} \,\, - \,\, \cdots 
\label{eq:four_quark_operator_anomalous_dimensions_perturbative}
\\
\gamma(g)  &=&  \gamma_0 \dfrac{g^2}{(4\pi)^2} \,\, + \,\,
\gamma_1 \dfrac{g^4}{(4\pi)^4} \,\, + \,\, \cdots \,.\nn
\end{eqnarray}
We stress that $\beta_0, \beta_1$ and $\gamma_0$ are universal,
i.e. scheme independent. $K^0-\bar K^0$ mixing is usually considered
in the naive dimensional regularization (NDR) scheme of $\msbar$, and
below we specify the perturbative coefficient $\gamma_1$ in that
scheme:
\begin{eqnarray}
& &\beta_0 = 
         \left\{\frac{11}{3}N-\frac{2}{3}\Nf\right\}, \qquad
   \beta_1 = 
         \left\{\frac{34}{3}N^2-\Nf\left(\frac{13}{3}N-\frac{1}{N}
         \right)\right\}, \label{eq:RG-coefficients}\\[0.3ex]
& &\gamma_0 = \frac{6(N-1)}{N}, \qquad
         \gamma_1 = \frac{N-1}{2N} 
         \left\{-21 + \frac{57}{N} - \frac{19}{3}N + \frac{4}{3}\Nf
         \right\}\,.\nn
\end{eqnarray}
Note that for QCD the above expressions must be evaluated for $N=3$
colours, while $\Nf$ denotes the number of active quark flavours. As
already stated, Eq.~(\ref{eq:Heff}) is valid at scales below the charm
threshold, after all heavier flavours have been integrated out,
i.e. $\Nf = 3$.

In Eq.~(\ref{eq:Heff}), the terms proportional to $\eta_1,\,\eta_2$
and $\eta_3$, multiplied by the contributions containing
$\gbar(\mu)^2$, correspond to the Wilson coefficient of the OPE,
computed in perturbation theory. Its dependence on the renormalization
scheme and scale $\mu$ is canceled by that of the weak matrix element
$\langle \bar K^0 \vert Q^{\Delta S=2}_{\rm R} (\mu) \vert K^0
\rangle$. The latter corresponds to the long-distance effects of the
effective Hamiltonian and must be computed nonperturbatively. For
historical, as well as technical reasons, it is convenient to express
it in terms of the $B$ parameter $B_{\rm{K}}$, defined as
\be
   B_{\rm{K}}(\mu)= \frac{{\left\langle\bar{K}^0\left|
         Q^{\Delta S=2}_{\rm R}(\mu)\right|K^0\right\rangle} }{
         {\frac{8}{3}\fK^2\mK^2}} \,\, .
\ee
The four-quark operator $Q^{\Delta S=2}(\mu)$ is renormalized at scale $\mu$
in some regularization scheme, for instance, NDR-$\msbar$. Assuming that
$B_{\rm{K}}(\mu)$ and the anomalous dimension $\gamma(g)$ are both known in
that scheme, the renormalization group independent (RGI) $B$ parameter
$\hat{B}_{\rm K}$ is related to $B_{\rm{K}}(\mu)$ by the exact formula
\be
  \hat{B}_{\rm{K}} = 
  \left(\frac{\gbar(\mu)^2}{4\pi}\right)^{-\gamma_0/(2\beta_0)}
  \exp\bigg\{ \int_0^{\gbar(\mu)} \, dg \, \bigg(
  \frac{\gamma(g)}{\beta(g)} \, + \, \frac{\gamma_0}{\beta_0g} \bigg)
  \bigg\} 
\, B_{\rm{K}}(\mu) \,\,\, .
\ee
At NLO in perturbation theory the above reduces to
\be
   \hat{B}_{\rm{K}} =
   \left(\frac{\gbar(\mu)^2}{4\pi}\right)^{- \gamma_0/(2\beta_0)}
   \left\{ 1+\dfrac{\gbar(\mu)^2}{(4\pi)^2}\left[
   \frac{\beta_1\gamma_0-\beta_0\gamma_1}{2\beta_0^2} \right]\right\}\,
   B_{\rm{K}}(\mu) \,\,\, .
\label{eq:BKRGI_NLO}
\ee
To this order, this is the scale-independent product of all
$\mu$-dependent quantities in Eq.~(\ref{eq:Heff}).

Lattice QCD calculations provide results for $B_K(\mu)$. These
results are, however, usually obtained in intermediate schemes other
than the continuum $\msbar$ scheme used to calculate the Wilson
coefficients appearing in Eq.~(\ref{eq:Heff}). Examples of
intermediate schemes are the RI/MOM scheme \cite{Martinelli:1994ty}
(also dubbed the ``Rome-Southampton method'') and the Schr\"odinger
functional (SF) scheme \cite{Luscher:1992an}. These schemes are used
as they allow a nonperturbative renormalization of the four-fermion
operator, using an auxiliary lattice simulation.  This allows
$B_K(\mu)$ to be calculated with percent-level accuracy, as described
below.

In order to make contact with phenomenology, however, and in
particular to use the results presented above, one must convert from
the intermediate scheme to the $\msbar$ scheme or to the RGI quantity
$\hat{B}_{\rm K}$. This conversion relies on one or two-loop
perturbative matching calculations, the truncation errors in which
are, for many recent calculations, the dominant source of error in
$\hat{B}_{\rm{K}}$ (see, for instance, Refs.~\cite{Laiho:2011np,Arthur:2012opa,Bae:2014sja,Blum:2014tka,Jang:2015sla}).
While this scheme-conversion error is not, strictly speaking, an error
of the lattice calculation itself, it must be included in results for
the quantities of phenomenological interest, namely
$B_K(\msbar,2\,{\rm GeV})$ and $\hat{B}_{\rm K}$. We note that this
error can be minimized by matching between the intermediate scheme and
$\msbar$ at as large a scale $\mu$ as possible (so that the coupling
which determines the rate of convergence is
minimized). Recent calculations have pushed the matching $\mu$ up to
the range $3-3.5\,$GeV. This is possible because of the use of
nonperturbative RG running determined on the
lattice~\cite{Durr:2011ap,Arthur:2012opa,Blum:2014tka}. The
Schr\"odinger functional offers the possibility to run
nonperturbatively to scales $\mu\sim M_{\rm{W}}$ where the truncation
error can be safely neglected. However, so far this has been applied
only for two flavours of Wilson quarks~\cite{Dimopoulos:2007ht}.

Perturbative truncation errors in Eq.~(\ref{eq:Heff}) also affect the
Wilson coefficients $\eta_1$, $\eta_2$ and~$\eta_3$. It turns out that
the largest uncertainty comes from that in
$\eta_1$~\cite{Brod:2011ty}. Although it is now calculated at NNLO,
the series shows poor convergence. The net effect is that the
uncertainty in $\eta_1$ is larger than that in present lattice
calculations of $B_K$.

In the Standard Model, $\epsilon_{K}$ receives contributions from: 1)
short distance physics given by $\Delta S = 2$ ``box diagrams"
involving $W^\pm$ bosons and $u,c$ and $t$ quarks; 2) long distance
physics from light hadrons contributing to the imaginary part of the
dispersive amplitude $M_{12}$ used in the two component description of
$K^0-\bar{K}^0$ mixing; 3) the imaginary part of the absorptive
amplitude $\Gamma_{12}$ from $K^0-\bar{K}^0$ mixing; and 4)
$\text{Im}(A_0)/\text{Re}(A_0)$. The terms in this decomposition can
vary with phase conventions. It is common to represent contribution~1
by $\text{Im} M_{12}^\text{SD} = \text{Im} [ \langle \bar{K}^0 | {\cal
    H}_\text{eff}^{\Delta S = 2} | K^0 \rangle]$ and contribution~2 by
$M_{12}^\text{LD}$.  Contribution~3 can be related to
$\text{Im}(A_0)/\text{Re}(A_0)$,
yielding~\cite{Buras:1998raa,Anikeev:2001rk,Nierste:2009wg,Buras:2008nn,Buras:2010pz}
\be
\epsilon_K \,\,\, = \,\,\, \exp(i \phi_\epsilon) \, \sin(\phi_\epsilon)
  \left[
  \frac{\text{Im} [\langle\bar{K}^0|{\cal H}_\text{eff}^{\Delta S=2}|K^0 \rangle]}{\Delta M_K}
 + \frac{\text{Im}(M_{12}^\text{LD})  }{\Delta M_K}
 + \frac{\text{Im}(A_0)}{\text{Re}(A_0)}
  \right]
  \label{eq:epsK}
\ee
for $\lambda_u$ real and positive; the phase of $\epsilon_{K}$ is
given by
\be
\phi_\epsilon \,\,\, = \,\,\, \arctan \frac{\Delta M_{K}}{\Delta
  \Gamma_{K}/2} \,\,\, . 
\ee
The quantities $\Delta M_K$ and $\Delta \Gamma_K$ are the mass and
decay width differences between long- and short-lived neutral kaons,
while $A_0$ is the amplitude of the kaon decay into an isospin-0 two
pion state. Experimentally known values of the above quantities
are\,\cite{Agashe:2014kda}:
\begin{eqnarray}
\vert \epsilon_{K} \vert \,\, &=& \,\, 2.228(11) \times 10^{-3} \,\, ,
\nn \\
\phi_\epsilon \,\, &=& \,\, 43.52(5)^\circ \,\, ,
 \\
\Delta M_{K} \,\, &=& \,\, 3.4839(59) \times 10^{-12}\, {\rm MeV} \,\, ,
\nn \\
\Delta \Gamma_{K}  \,\, &=& \,\ 7.3382(33) \times 10^{-15} \,{\rm GeV}
\,\,.\nn 
\end{eqnarray}
A recent analytical estimate of the contributions of $M_{12}^\text{LD}$
(Refs.~\cite{Buras:2008nn,Buras:2010pz}) leads to
\be
    \epsilon_{K} \,\,\, = \,\,\, \exp(i \phi_\epsilon) \,\,
    \sin(\phi_\epsilon) \,\, \Big [ \frac{\text{Im} [ \langle \bar K^0 \vert
                    {\cal H}_{\rm eff}^{\Delta S = 2} \vert K^0 \rangle ]} {\Delta M_K }
                \,\,\, + \,\,\, \rho \frac{\text{Im}(A_0)}{\text{Re}(A_0)} \,\, \Big ] \,\,\, .
                \label{eq:epsK-phenom}
\ee
A phenomenological estimate for $\xi=\Im(A_0)/\Re(A_0)$ can be determined using the
experimental value of $\epsilon^\prime/\epsilon$ \cite{Buras:2010pz}
\begin{equation}
   \xi = -6.0(1.5)\cdot10^{-4}\sqrt{2}|\epsilon_K|
       = -1.9(5)\cdot10^{-4}. 
\end{equation}
A more precise result has been obtained from the ratio of amplitudes
$\Im(A_2)/\Re(A_2)$ computed in lattice QCD~\cite{Blum:2015ywa} (where
$A_2$ denotes the $\Delta{I}=3/2$ decay amplitude for $K\to\pi\pi$):
\begin{equation}
   \xi = -1.6(2)\cdot10^{-4}.
\end{equation}
The value of $\xi$ can then be combined with a ${\chi}\rm PT$-based
estimate for the long-range contribution,
i.e. $\rho=0.6(3)$~\cite{Buras:2010pz}. Overall, the combination
$\rho\xi$ leads to a suppression of $|\epsilon_K|$ by $6(2)\%$
relative to the naive estimate (i.e. the first term in square brackets
in Eq.~(\ref{eq:epsK})), regardless of whether the phenomenological or
lattice estimate for $\xi$ is used. The uncertainty in the suppression
factor is dominated by the error on $\rho$. Although this is a small
correction, we note that its contribution to the error of $\epsilon_K$
is larger than that arising from the value of $B_{\rm K}$ reported
below.

Efforts are under way to compute both the real and imaginary
long-distance contribution to the
$K_L-K_S$ mass difference in lattice
QCD\,\cite{Christ:2012se,Bai:2014cva,Christ:2015pwa}. However, the
results are not yet precise enough to improve the accuracy in the
determination of the parameter $\rho$.

\subsection{Lattice computation of $B_{\rm{K}}$}

Lattice calculations of $B_{\rm{K}}$ are affected by the same
systematic effects discussed in previous sections. However, the issue
of renormalization merits special attention. The reason is that the
multiplicative renormalizability of the relevant operator $Q^{\Delta
S=2}$ is lost once the regularized QCD action ceases to be invariant
under chiral transformations. For Wilson fermions, $Q^{\Delta S=2}$
mixes with four additional dimension-six operators, which belong to
different representations of the chiral group, with mixing
coefficients that are finite functions of the gauge coupling. This
complicated renormalization pattern was identified as the main source
of systematic error in earlier, mostly quenched calculations of
$B_{\rm{K}}$ with Wilson quarks. It can be bypassed via the
implementation of specifically designed methods, which are either
based on Ward identities~\cite{Becirevic:2000cy} or on a modification
of the Wilson quark action, known as twisted mass
QCD~\cite{Frezzotti:2000nk,Dimopoulos:2006dm}.

An advantage of staggered fermions is the presence of a remnant $U(1)$
chiral symmetry. However, at nonvanishing lattice spacing, the
symmetry among the extra unphysical degrees of freedom (tastes) is
broken. As a result, mixing with other dimension-six operators cannot
be avoided in the staggered formulation, which complicates the
determination of the $B$ parameter. The effects of the broken taste
symmetry are usually treated via an effective field theory, such as
staggered Chiral Perturbation Theory (S$\chi$PT).

Fermionic lattice actions based on the Ginsparg-Wilson
relation~\cite{Ginsparg:1981bj} are invariant under the chiral group,
and hence four-quark operators such as $Q^{\Delta S=2}$ renormalize
multiplicatively. However, depending on the particular formulation of
Ginsparg-Wilson fermions, residual chiral symmetry breaking effects
may be present in actual calculations. For instance, in the case of
domain wall fermions, the finiteness of the extra 5th dimension
implies that the decoupling of modes with different chirality is not
exact, which produces a residual nonzero quark mass in the chiral
limit. Whether or not a significant mixing with dimension-six
operators is induced as well must be investigated on a case-by-case
basis.

Recent lattice QCD calculations of $B_K$ have been performed with
$\Nf=2+1+1$ dynamical quarks\,\cite{Carrasco:2015pra}, and we want to
mention a few conceptual issues that arise in this context. As
described in section\,\ref{sec:indCP}, kaon mixing is expressed in
terms of an effective four-quark interaction $Q^{{\Delta}S=2}$,
considered below the charm threshold. When the matrix element of
$Q^{{\Delta}S=2}$ is evaluated in a theory that contains a dynamical
charm quark, the resulting estimate for $B_K$ must then be matched to
the three-flavour theory which underlies the effective four-quark
interaction.\footnote{We thank Martin L\"uscher for an interesting
  discussion on this issue.} In general, the matching of $2+1$-flavour
QCD with the theory containing $2+1+1$ flavours of sea quarks below
the charm threshold can be accomplished by adjusting the coupling and
quark masses of the $\Nf=2+1$ theory so that the two theories match at
energies $E<m_c$. The corrections associated with this matching are of
order $(E/m_c)^2$, since the subleading operators have dimension
eight \cite{Cirigliano:2000ev}.
When the kaon mixing amplitude is considered, the matching also
involves the relation between the relevant box graphs and the
effective four-quark operator. In this case, corrections of order
$(E/m_c)^2$ arise not only from the charm quarks in the sea, but also
from the valence sector, since the charm quark propagates in the box
diagrams. One expects that the sea quark effects are subdominant, as
they are suppressed by powers of $\alpha_s$. We note that the original
derivation of the effective four-quark interaction is valid up to
corrections of order $(E/m_c)^2$. While the kaon mixing amplitudes
evaluated in the $\Nf=2+1$ and $2+1+1$ theories are thus subject to
corrections of the same order in $E/m_c$ as the derivation of the
conventional four-quark interaction, the general conceptual issue
regarding the calculation of $B_K$ in QCD with $\Nf=2+1+1$ flavours
should be addressed in detail in future calculations.

Another issue in this context is how the lattice scale and the
physical values of the quark masses are determined in the $2+1$ and
$2+1+1$ flavour theories. Here it is important to consider in which
way the quantities used to fix the bare parameters are affected by a
dynamical charm quark. Apart from a brief discussion in
Ref.\,\cite{Carrasco:2015pra}, these issues have not been fully worked
out in the literature, but these kinds of mismatches were seen in
simple lattice-QCD observables as quenched calculations gave way to
$\Nf=2$ and then $2+1$ flavour results. Given the scale of the charm
quark mass relative to the scale of $B_K$, we expect these errors to
be modest, but a more quantitative understanding is needed as
statistical errors on $B_K$ are reduced. Within this review we will
not discuss this issue further.

\begin{figure}[ht]
\centering
\includegraphics[width=13cm]{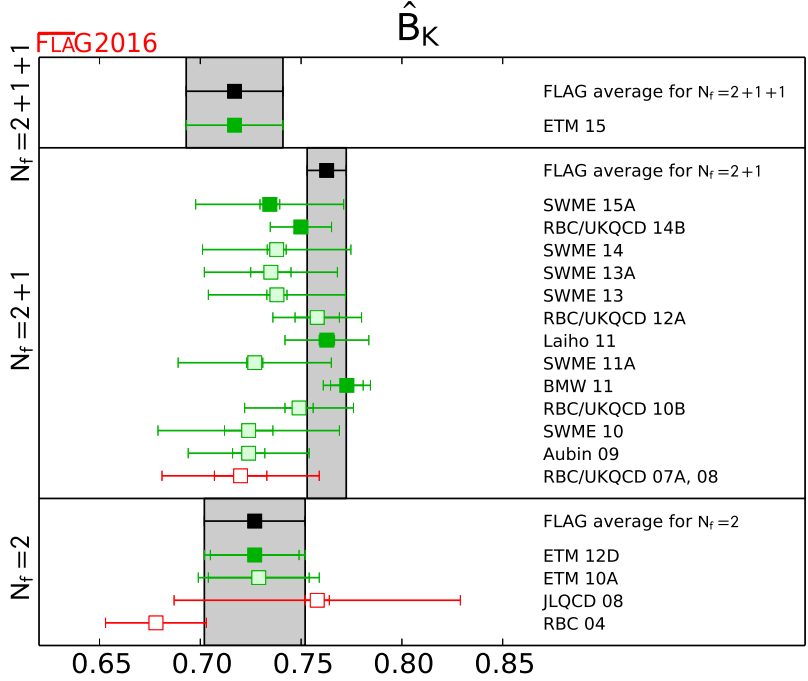}
\caption{Recent unquenched lattice results for the RGI $B$ parameter
  $\hat{B}_{\rm{K}}$. The grey bands indicate our global averages
  described in the text. For $\Nf=2+1+1$ and $\Nf=2$ the global estimate coincide
  with the results by ETM\,12D and ETM\,10A, respectively. \label{fig_BKsumm}}
\end{figure}

Below we focus on recent results for $B_{\rm{K}}$, obtained for
$\Nf=2, 2+1$ and $2+1+1$ flavours of dynamical quarks. A compilation
of results is shown in Tabs.~\ref{tab_BKsumm}
and~\ref{tab_BKsumm_nf2}, as well as Fig.~\ref{fig_BKsumm}. An
overview of the quality of systematic error studies is represented by
the colour coded entries in Tabs.~\ref{tab_BKsumm}
and~\ref{tab_BKsumm_nf2}. In Appendix~\ref{app-BK} we gather the
simulation details and results from different collaborations, the
values of the most relevant lattice parameters, and comparative tables
on the various estimates of systematic errors.

Some of the groups whose results are listed in Tabs.~\ref{tab_BKsumm}
and~\ref{tab_BKsumm_nf2} do not quote results for both
$B_{\rm{K}}(\overline{\rm MS},2\,{\rm GeV})$ -- which we denote by the
shorthand $B_{\rm{K}}$ from now on -- and $\hat{B}_{\rm{K}}$. This
concerns Refs.~\cite{Aoki:2004ht,Constantinou:2010qv,Bertone:2012cu}
for $\Nf=2$,
Refs.\cite{Laiho:2011np,Arthur:2012opa,Blum:2014tka,Jang:2015sla}
for~$2+1$ and Ref.~\cite{Carrasco:2015pra} for~$2+1+1$ flavours. In these
cases we perform the conversion ourselves by evaluating the
proportionality factor in Eq.~(\ref{eq:BKRGI_NLO}) at $\mu=2\,\gev$,
using the following procedure: For $\Nf=2+1$ we use the value
$\alpha_s(M_{\rm{Z}})=0.1185$ from the 2014 edition of the
PDG\,\cite{Agashe:2014kda} and run it across the quark thresholds at
$m_b=4.18$\,GeV and $m_c=1.275$\,GeV, and then run up in the
three-flavour theory to $\mu=2\,\gev$. All running is done using the
four-loop RG $\beta$-function. The resulting value of
$\alpha_s^{\msbar}(2\,\gev)=0.29672$ is then used to evaluate
$\hat{B}_{\rm{K}}/B_{\rm{K}}$ in perturbation theory at NLO, which
gives $\hat{B}_{\rm{K}}/B_{\rm{K}}=1.369$ in the three-flavour
theory. This value of the conversion factor has also been applied to
the result computed in QCD with $\Nf=2+1+1$ flavours of dynamical
quarks~\cite{Carrasco:2015pra}.

In two-flavour QCD one can insert the updated nonperturbative
estimate for the $\Lambda$ parameter by the ALPHA
Collaboration\,\cite{Fritzsch:2012wq},
i.e.\,$\Lambda^{(2)}=310(20)$\,MeV, into the NLO expressions for
$\alpha_s$. The resulting value of the perturbative conversion factor
$\hat{B}_K/B_K$ for $\Nf=2$ is then equal to~1.386. However, since the
running coupling in the $\msbar$ scheme enters at several stages in
the entire matching and running procedure, it is difficult to use this
estimate of $\alpha_s$ consistently without a partial reanalysis of
the data in
Refs.~\cite{Aoki:2004ht,Constantinou:2010qv,Bertone:2012cu}. We have
therefore chosen to apply the conversion factor of~1.369 not only to
results obtained for $\Nf=2+1$ flavours but also to the two-flavour
theory (in cases where only one of $\hat{B_K}$ and $B_K$ are
quoted). We note that the difference between 1.386 and 1.369 will
produce an ambiguity of the order of~1\%, which is well below the
overall uncertainties in
Refs.~\cite{Aoki:2004ht,Constantinou:2010qv}. We have indicated
explicitly in Tab.~\ref{tab_BKsumm_nf2} in which way the conversion
factor 1.369 has been applied to the results of
Refs.~\cite{Aoki:2004ht,Constantinou:2010qv,Bertone:2012cu}.

\begin{table}[t]
\begin{center}
\mbox{} \\[3.0cm]
{\footnotesize{
\vspace*{-2cm}\begin{tabular*}{\textwidth}[l]{l @{\extracolsep{\fill}} r@{\hspace{1mm}}l@{\hspace{1mm}}l@{\hspace{1mm}}l@{\hspace{1mm}}l@{\hspace{1mm}}l@{\hspace{1mm}}l@{\hspace{1mm}}l@{\hspace{1mm}}l@{\hspace{1mm}}l}
Collaboration & Ref. & $\Nf$ & 
\hspace{0.15cm}\begin{rotate}{60}{publication status}\end{rotate}\hspace{-0.15cm} &
\hspace{0.15cm}\begin{rotate}{60}{continuum extrapolation}\end{rotate}\hspace{-0.15cm} &
\hspace{0.15cm}\begin{rotate}{60}{chiral extrapolation}\end{rotate}\hspace{-0.15cm}&
\hspace{0.15cm}\begin{rotate}{60}{finite volume}\end{rotate}\hspace{-0.15cm}&
\hspace{0.15cm}\begin{rotate}{60}{renormalization}\end{rotate}\hspace{-0.15cm}  &
\hspace{0.15cm}\begin{rotate}{60}{running}\end{rotate}\hspace{-0.15cm} & 
\rule{0.3cm}{0cm}$B_{{K}}(\overline{\rm MS},2\,{\rm GeV})$ 
& \rule{0.3cm}{0cm}$\hat{B}_{{K}}$ \\
&&&&&&&&&& \\[-0.1cm]
\hline
\hline
&&&&&&&&&& \\[-0.1cm]

ETM 15 & \cite{Carrasco:2015pra} & 2+1+1 & \gA & \good & \soso & \soso
& \good&  $\,a$ &   0.524(13)(12)  & 0.717(18)(16)$^1$ \\[0.5ex]
&&&&&&&&&& \\[-0.1cm]
\hline
&&&&&&&&&& \\[-0.1cm]
SWME 15A & \cite{Jang:2015sla} & 2+1 & \gA & \good & \soso &
\good & \soso$^\ddagger$  & $-$ & 0.537(4)(26) & 0.735(5)(36)$^2$ \\[0.5ex]

RBC/UKQCD 14B
& \cite{Blum:2014tka} & 2+1 & \gA & \good & \good &
     \soso  & \good & $\,b$  & 0.5478(18)(110)$^3$ & 0.7499(24)(150) \\[0.5ex]  

SWME 14 & \cite{Bae:2014sja} & 2+1 & \gA & \good & \soso &
\good & \soso$^\ddagger$  & $-$ & 0.5388(34)(266) & 0.7379(47)(365) \\[0.5ex]

SWME 13A & \cite{Bae:2013tca} & 2+1 & \gA & \good & \soso  &
\good & \soso$^\ddagger$  & $-$ & 0.537(7)(24) & 0.735(10)(33) \\[0.5ex]

SWME 13 & \cite{Bae:2013lja} & 2+1 & \rC & \good & \soso &
\good & \soso$^\ddagger$ & $-$ & 0.539(3)(25) & 0.738(5)(34) \\[0.5ex]

RBC/UKQCD 12A
& \cite{Arthur:2012opa} & 2+1 & \gA & \soso & \good &
     \soso & \good & $\,b$ & 0.554(8)(14)$^3$ & 0.758(11)(19) \\[0.5ex]  

Laiho 11 & \cite{Laiho:2011np} & 2+1 & \rC & \good & \soso &
     \soso & \good & $-$ & 0.5572(28)(150)& 0.7628(38)(205)$^2$ \\[0.5ex]  

SWME 11A & \cite{Bae:2011ff} & 2+1 & \gA & \good & \soso &
\soso & \soso$^\ddagger$ & $-$ & 0.531(3)(27) & 0.727(4)(38) \\[0.5ex]

BMW 11 & \cite{Durr:2011ap} & 2+1 & \gA & \good & \good & \good & \good
& $\,c$ & 0.5644(59)(58) & 0.7727(81)(84) \\[0.5ex]

RBC/UKQCD 10B & \cite{Aoki:2010pe} & 2+1 & \gA & \soso & \soso & \good &
\good & $\,d$ & 0.549(5)(26) & 0.749(7)(26) \\[0.5ex] 

SWME 10 & \cite{Bae:2010ki} & 2+1 & \gA & \good & \soso & \soso & \soso
& $-$ & 0.529(9)(32) &  0.724(12)(43) \\[0.5ex] 

Aubin 09 & \cite{Aubin:2009jh} & 2+1 & \gA & \soso & \soso &
     \soso & \tbg & $-$ & 0.527(6)(21)& 0.724(8)(29) \\[0.5ex]  

RBC/UKQCD 07A, 08 \rule{1em}{0em}& \cite{Antonio:2007pb,Allton:2008pn} & 2+1 & \gA
                              & \tbr & \soso & \tbg     & \tbg & $-$ &
0.524(10)(28) & 0.720(13)(37) \\[0.5ex]  
HPQCD/UKQCD 06  & \cite{Gamiz:2006sq} & 2+1 & \gA
                              & \tbr & \soso$^\ast$ & \tbg     & \tbr &
$-$ & 0.618(18)(135)& 0.83(18) \\[0.5ex]  
&&&&&&&&&& \\[-0.1cm]
\hline
\hline\\[-0.1cm]
\end{tabular*}
}}
\begin{minipage}{\linewidth}
{\footnotesize 
\begin{itemize}
\item[$^\ddagger$] The renormalization is performed using perturbation
        theory at one loop, with a conservative estimate of the uncertainty. \\[-5mm]
\item[$^\ast$] This result has been obtained with only two ``light'' sea
        quark masses. \\[-5mm]
\item[$a$]  $B_K$ is renormalized nonperturbatively at scales $1/a \sim 2.2-3.3\,\gev$ in the $\Nf = 4$ RI/MOM scheme 
     using two different lattice momentum scale intervals, the first around $1/a$ while the second around  3.5 GeV. 
        The impact of the two ways to the final 
        result is taken into account  in the error budget. Conversion to $\msbar$ is at one-loop at 3 GeV.  \\[-5mm]
\item[$b$] $B_K$ is renormalized nonperturbatively at a scale of 1.4 GeV
        in two RI/SMOM schemes for $\Nf = 3$, and 
	then run to 3 GeV using a nonperturbatively determined step-scaling
        function. 
	Conversion to $\msbar$ is at one-loop order at 3 GeV.\\[-5mm]
\item[$c$] $B_K$ is renormalized and run nonperturbatively to a scale of
        $3.4\,\gev$ in the RI/MOM scheme.
	nonperturbative and NLO
        perturbative running agrees down to scales of $1.8\,\gev$ within
        statistical
	uncertainties of about 2\%.\\[-5mm]
\item[$d$] $B_K$ is renormalized nonperturbatively at a scale of 2\,GeV
        in two RI/SMOM schemes for $\Nf = 3$, and then 
	run to 3 GeV using a nonperturbatively determined step-scaling
        function. Conversion to $\msbar$ is at 
	one-loop order at 3 GeV.\\[-5mm]
\item[$^1$] $B_{K}(\msbar, 2\,\gev)$ and $\hat{B}_{{K}}$ are related
        using the conversion factor  1.369 {\it i.e.} the one obtained
        with $N_f=2+1$.  \\[-5mm]
\item[$^2$] $\hat{B}_{{K}}$ is obtained from the estimate for
        $B_{K}(\msbar, 2\,\gev)$ using the conversion factor 1.369. \\[-5mm]
\item[$^3$] $B_{K}(\msbar, 2\,\gev)$ is obtained from the estimate for
        $\hat{B}_{{K}}$ using the conversion factor 1.369.  
\end{itemize}
}
\end{minipage}
\caption{Results for the Kaon $B$ parameter in QCD with $\Nf=2+1+1$
  and $\Nf=2+1$ dynamical flavours, together with a summary of
  systematic errors. Any available information about nonperturbative
  running is indicated in the column ``running", with details given at
  the bottom of the table.\label{tab_BKsumm}}
\end{center}
\end{table}

\begin{table}[t]
\begin{center}
\mbox{} \\[3.0cm]
{\footnotesize{
\vspace*{-2cm}\begin{tabular*}{\textwidth}[l]{l @{\extracolsep{\fill}} r l l l l l l l l l}
Collaboration & Ref. & $\Nf$ & 
\hspace{0.15cm}\begin{rotate}{60}{publication status}\end{rotate}\hspace{-0.15cm} &
\hspace{0.15cm}\begin{rotate}{60}{continuum extrapolation}\end{rotate}\hspace{-0.15cm} &
\hspace{0.15cm}\begin{rotate}{60}{chiral extrapolation}\end{rotate}\hspace{-0.15cm}&
\hspace{0.15cm}\begin{rotate}{60}{finite volume}\end{rotate}\hspace{-0.15cm}&
\hspace{0.15cm}\begin{rotate}{60}{renormalization}\end{rotate}\hspace{-0.15cm}  &
\hspace{0.15cm}\begin{rotate}{60}{running}\end{rotate}\hspace{-0.15cm} & 
\rule{0.3cm}{0cm}$B_{{K}}(\overline{\rm MS},2\,{\rm GeV})$ 
& \rule{0.3cm}{0cm}$\hat{B}_{{K}}$ \\
&&&&&&&&&& \\[-0.1cm]
\hline
\hline
&&&&&&&&&& \\[-0.1cm]

ETM 12D & \cite{Bertone:2012cu} & 2 & \gA & \good & \soso & \soso
& \good&  $\,e$ &   0.531(16)(9)  & 0.727(22)(12)$^1$ \\[0.5ex]

ETM 10A & \cite{Constantinou:2010qv} & 2 & \gA & \good & \soso & \soso
& \good&  $\,f$ &   0.533(18)(12)$^1$  & 0.729(25)(17) \\[0.5ex]
JLQCD 08 & \cite{Aoki:2008ss} & 2 & \gA  & \tbr      & \soso      &
\tbr          &\tbg    & $-$ & 0.537(4)(40) &
0.758(6)(71)\\[0.5ex]  
RBC 04   & \cite{Aoki:2004ht} & 2 & \gA & \tbr      & \tbr      &
\tbr$^\dagger$ & \tbg      &$-$ & 0.495(18)    & 0.678(25)$^1$
\\[0.5ex]  
UKQCD 04 & \cite{Flynn:2004au} & 2  & \gA  & \tbr      & \tbr      &
\tbr$^\dagger$ & \tbr      & $-$ & 0.49(13)     & 0.68(18)
\\[0.5ex]  
&&&&&&&&&& \\[-0.1cm]
\hline
\hline\\[-0.1cm]
\end{tabular*}
}}
\begin{minipage}{\linewidth}
{\footnotesize 
\begin{itemize}
\item[$^\dagger$] These results have been obtained at  
	$(M_\pi L)_{\rm min} > 4$ in a lattice box 
        with a spatial extension  $L < 2$~fm.\\[-5mm]
\item[$e$] $B_K$ is renormalized nonperturbatively at scales $1/a \sim 2
        - 3.7\,\gev$ in the $\Nf = 2$ RI/MOM scheme. In this
        scheme, nonperturbative and NLO
        perturbative running are shown to agree from 4 GeV down to 2 GeV to
        better than 3\%
        \cite{Constantinou:2010gr,Constantinou:2010qv}.  \\[-5mm]
\item[$f$] $B_K$ is renormalized nonperturbatively at scales $1/a \sim 2
        - 3\,\gev$ in the $\Nf = 2$ RI/MOM scheme. In this
        scheme, nonperturbative and NLO
        perturbative running are shown to agree from 4 GeV down to 2 GeV to
        better than 3\%
        \cite{Constantinou:2010gr,Constantinou:2010qv}.  \\[-5mm]
\item[$^1$] $B_{K}(\msbar, 2\,\gev)$ and $\hat{B}_{{K}}$ are related using the conversion factor  1.369 {\it i.e.} the one obtained with $N_f=2+1$. 
\end{itemize}
}
\end{minipage}
\caption{Results for the Kaon $B$ parameter in QCD with $\Nf=2$
  dynamical flavours, together with a summary of systematic
  errors. Any available information about nonperturbative running is
  indicated in the column ``running", with details given at the bottom
  of the table.\label{tab_BKsumm_nf2}}
\end{center}
\end{table}

Since the last edition of the FLAG review\,\cite{Aoki:2013ldr} several
new or updated results have been reported. For QCD with $N_f=2+1+1$
there is now a published calculation from the ETM
Collaboration\,\cite{Carrasco:2015pra}; updated results for $N_f=2+1$
have been reported by several collaborations, i.e. RBC/UKQCD\,14B
\cite{Blum:2014tka}, SWME\,13A\,\cite{Bae:2013tca},
SWME\,14\,\cite{Bae:2014sja} and SWME\,15A\,\cite{Jang:2015sla}. For
$N_f=2$ we now include the result from ETMC, i.e.
ETM\,12D\,\cite{Bertone:2012cu}. We briefly discuss the main features
of the most recent calculations below.

The calculation by ETM\,15~\cite{Carrasco:2015pra} employs
Osterwalder-Seiler valence quarks on twisted-mass dynamical quark
ensembles. Both valence and sea quarks are tuned to maximal
twist. This mixed action setup guarantees that the four-fermion matrix
elements are automatically $\cO(a)$ improved and free of wrong chirality
mixing effects. The calculation has been carried out at three values
of the lattice spacing ($a \simeq 0.06 - 0.09$ fm). Light pseudoscalar
mass values are in the range $210-450$\,MeV. The spatial lattice sizes
vary between 2.1 to 2.9\,fm and correspond to $M_{\pi, {\rm min}} L
\simeq 3.2 - 3.5$.  Finite volume effects are investigated at the
coarsest lattice spacing by controlling the consistency of results
obtained at two lattice volumes at 280\,MeV for the light pseudoscalar
mass. The determination of the bag parameter is performed using
simultaneous chiral and continuum fits. The renormalization factors
have been evaluated using the RI/MOM technique for $N_f=4$ degenerate
Wilson twisted-mass dynamical quark gauge configurations generated for
this purpose. In order to gain control over discretization effects the
evaluation of the renormalization factors has been carried out
following two different methods. The uncertainty from the RI
computation is estimated at 2\%. The conversion to
$\overline{\rm{MS}}$ produces an additional 0.6\% of systematic
error. The overall uncertainty for the bag parameter is computed from
a distribution of several results, each one of them corresponding to a
variant of the analysis procedure.

The collection of results from the SWME collaboration
\cite{Bae:2010ki,Bae:2011ff,Bae:2013lja,Bae:2013tca,Bae:2014sja,Jang:2015sla}
have all been obtained using a mixed action, i.e. HYP-smeared valence
staggered quarks on the Asqtad improved, rooted staggered MILC
ensembles. For the latest set of results, labelled SWME\,14,\,15A
\cite{Bae:2014sja,Jang:2015sla} an extended set of ensembles,
comprising finer lattice spacings and a smallest pion mass of 174\,MeV
has been added to the calculation. The final estimate for $B_K$ is
obtained from a combined chiral and continuum extrapolation using the
data computed for the three finest lattice spacings. The dominant
systematic error of 4.4\% is associated with the matching factor
between the lattice and $\msbar$ schemes. It has been computed in
perturbation theory at one loop, and its error was estimated assuming
a missing two-loop matching term of size $1\times\alpha(1/a)^2$,
i.e. with no factors of $1/(4\pi)$ included. Different functional
forms for the chiral fits contribute another 2\% to the error
budget. It should also be noted that Bayesian priors are used to
constrain some of the coefficients in the chiral ansatz. The total
systematic error amounts to about 5\%. Compared to the earlier
calculations of SWME one finds that ``the overall error is only
slightly reduced, but, more importantly, the methods of estimating
errors have been improved''\,\cite{Bae:2014sja}.

The RBC and UKQCD Collaborations have updated their value for $B_K$
using $\Nf=2+1$ flavours of domain wall fermions \cite{Blum:2014tka}.
Previous results came from ensembles at three different lattice
spacings with unitary pion masses in the range of 170 to 430\,MeV. The
new work adds an ensemble with essentially physical light and strange
quark masses at two of the lattice spacings, along with a third finer
lattice with 370~MeV pion masses. This finer ensemble provides an
additional constraint on continuum extrapolations. Lattice spacings
and quark masses are determined via a combined continuum and chiral
extrapolation to all ensembles. With lattice spacings at hand,
nonperturbative renormalization and nonperturbative step scaling are
used to find the renormalized value of $B_K$ at 3~GeV in the
RI-SMOM($\gamma^\mu,\gamma^\mu)$ and RI-SMOM($\slash{q}, \slash{q})$
schemes for all of the ensembles. These $B_K$ values for each pion
mass are determined for the physical strange quark mass through
valence strange quark interpolations/extrapolations and dynamical
strange quark mass reweighting. The light quark mass dependence is
then fit to $SU(2)$ chiral perturbation theory. Because the new
ensembles have quark masses within a few percent of their physical
values, the systematic error related to the extrapolation to physical
values is neglected. The new physical point ensembles have
(5.5\,fm)$^3$ volumes, and chiral perturbation theory fits with and
without finite volume corrections differ by 10-20\% of the statistical
errors, so no finite volume error is quoted. The fits are dominated by
the physical point ensembles, which have small errors. Fits with $B_K$
normalized in both RI-SMOM schemes are done, and the difference is used
to estimate the systematic error due to nonperturbative
renormalization.

The $\Nf=2$ calculation described in ETM\,12D \cite{Bertone:2012cu}
uses a mixed action setup employing twisted-mass dynamical quarks and
Osterwalder-Seiler quarks in the valence, both tuned to maximal
twist. The work of ETM\,12D is an update of the calculation of ETM\,10A
\cite{Constantinou:2010qv}. The main addition is the inclusion of a
fourth (superfine) lattice spacing ($a \simeq 0.05$ fm). Thus, the
computation is performed at four values of the lattice spacing ($a
\simeq 0.05 - 0.1$ fm), and the lightest simulated value of the light
pseudoscalar mass is about 280 MeV. Final results are obtained with
combined chiral and continuum fits. Finite volume effects are studied
at one value of the lattice spacing ($a \simeq 0.08$ fm), and it is
found that results obtained on two lattice volumes, namely for $L=2.2$
and 2.9\,fm at $M_\pi\approx 300$\,MeV are in good agreement within
errors. The four- and two-fermion renormalization factors needed in
the bag parameter evaluation are computed nonperturbatively using the
Rome-Southampton method. The systematic error due to the matching of
RI and $\overline{\rm{MS}}$ schemes is estimated to be 2.5\%.

We now describe our procedure for obtaining global averages. The rules
of section \ref{sec:color-code} stipulate that results free
of red tags and published in a refereed journal may enter an
average. Papers that at the time of writing are still unpublished but
are obvious updates of earlier published results can also be taken
into account.

There is only one result for $N_f=2+1+1$, computed by the ETM
Collaboration\,\cite{Carrasco:2015pra}. Since it is free of red tags,
it qualifies as the currently best global estimate, i.e.
%
%FLAGRESULT BEGIN
% TAG      &BK &END
% REFS     &\cite{Carrasco:2015pra}  &END
% UNITS    & 1 &END
% FLAVOURs & 2+1+1 &END
%FLAGRESULT END
%FLAGRESULTFORMULA BEGIN
\begin{equation}
\Nf=2+1+1:\hspace{.2cm}\FLAGAVBEGIN\hat{B}_{\rm{K}} = 0.717(18)(16)\FLAGAVEND\, ,
\hspace{.2cm}B_{\rm{K}}^\msbar (2\,{\rm GeV}) = 0.524(13)(12)\hspace{.3cm}\Ref~\mbox{\cite{Carrasco:2015pra}}.
\end{equation}
%FLAGRESULTFORMULA END
%
The bulk of results for the kaon $B$ parameter has been obtained for
$\Nf=2+1$. As in the previous edition of the FLAG
review\,\cite{Aoki:2013ldr} we include the results from
SWME\,\cite{Bae:2013tca,Bae:2014sja,Jang:2015sla}, despite the fact
that nonperturbative information on the renormalization factors is
not available. Instead, the matching factor has been determined in
perturbation theory at one loop, but with a sufficiently conservative
error of 4.4\%.

Thus, for $\Nf=2+1$ our global average is based on the results of
BMW\,11~\cite{Durr:2011ap}, Laiho\,11~\cite{Laiho:2011np},
RBC/UKQCD\,14B~\cite{Blum:2014tka} and
SWME\,15A~\cite{Jang:2015sla}. The last three are the latest updates from
a series of calculations by the same collaborations. Our procedure is
as follows: in a first step statistical and systematic errors of each
individual result for the RGI $B$ parameter, $\hat{B}_{\rm{K}}$, are
combined in quadrature. Next, a weighted average is computed from the
set of results. For the final error estimate we take correlations
between different collaborations into account. To this end we note
that we consider the statistical and finite-volume errors of SWME\,15A
and Laiho\,11 to be correlated, since both groups use the Asqtad
ensembles generated by the MILC Collaboration. Laiho\,11 and
RBC/UKQCD\,14B both use domain wall quarks in the valence sector and
also employ similar procedures for the nonperturbative determination
of matching factors. Hence, we treat the quoted renormalization and
matching uncertainties by the two groups as correlated. After
constructing the global covariance matrix according to
Schmelling~\cite{Schmelling:1994pz}, we arrive at
%
%FLAGRESULT BEGIN
% TAG      &BK &END
% REFS     &\cite{Durr:2011ap,Laiho:2011np,Blum:2014tka,Jang:2015sla} &END
% UNITS    & 1 &END
% FLAVOURs & 2+1 &END
%FLAGRESULT END
%FLAGRESULTFORMULA BEGIN
\begin{equation}
  \Nf=2+1:\hspace{2.5cm}\FLAGAVBEGIN \hat{B}_{\rm{K}} = 0.7625(97)\FLAGAVEND\qquad\Refs~\mbox{\cite{Durr:2011ap,Laiho:2011np,Blum:2014tka,Jang:2015sla}},
\end{equation}
%FLAGRESULTFORMULA END
%
with $\chi^2/{\rm d.o.f.}=0.675$. After applying the NLO conversion
factor $\hat{B}_{\rm{K}}/B_{\rm{K}}^\msbar (2\,{\rm GeV})=1.369$, this
translates into
\begin{equation}
  \Nf=2+1:\hspace{1cm} B_{\rm{K}}^\msbar(2\,{\rm GeV})=0.5570(71)\qquad\Refs~\mbox{\cite{Durr:2011ap,Laiho:2011np,Blum:2014tka,Jang:2015sla}}.
\end{equation}
These values and their uncertainties are very close to the global
estimates quoted in the previous edition of the FLAG
review\,\cite{Aoki:2013ldr}. Note, however, that the statistical
errors of each calculation entering the global average have now been
reduced to a level that makes them statistically incompatible. It is
only because of the relatively large systematic errors that the
weighted average produces a value of $\cO(1)$ for the reduced $\chi^2$.

Passing over to describing the results computed for $\Nf=2$ flavours,
we note that there is only the set of results published in
ETM\,12D\,\cite{Bertone:2012cu} and
ETM\,10A~\cite{Constantinou:2010qv} that allow for an extensive
investigation of systematic uncertainties. We identify the result from
ETM\,12D\,\cite{Bertone:2012cu}, which is an update of ETM\,10A, with
the currently best global estimate for two-flavour QCD, i.e.
%
%FLAGRESULT BEGIN
% TAG      &BK &END
% REFS     &\cite{Bertone:2012cu} &END
% UNITS    & 1 &END
% FLAVOURs & 2 &END
%FLAGRESULT END
%FLAGRESULTFORMULA BEGIN
\begin{equation}
\Nf=2:\hspace{.3cm}\FLAGAVBEGIN\hat{B}_{\rm{K}} = 0.727(22)(12)\FLAGAVEND ,
\hspace{.3cm}B_{\rm{K}}^\msbar (2\,{\rm GeV}) = 0.531(16)(19)\hspace{.4cm}\Ref~\mbox{\cite{Bertone:2012cu}}.
\end{equation}
%FLAGRESULTFORMULA END
%
The result in the $\msbar$ scheme has been obtained by applying the
same conversion factor of 1.369 as in the three-flavour theory.

\subsection{Kaon BSM $B$ parameters}
\label{sec:Bi}

We now report on lattice results concerning the matrix elements of
operators that encode the effects of physics beyond the Standard Model
(BSM) to the mixing of neutral kaons. In this theoretical framework
both the SM and BSM contributions add up to reproduce the
experimentally observed value of $\epsilon_K$. Since BSM contributions
involve heavy but unobserved particles, it is natural to assume that
they are short-distance dominated. The effective Hamiltonian for
generic ${\Delta}S=2$ processes including BSM contributions reads
\begin{equation}
  {\cal H}_{\rm eff,BSM}^{\Delta S=2} = \sum_{i=1}^5
  C_i(\mu)Q_i(\mu),
\end{equation}
where $Q_1$ is the four-quark operator of Eq.~(\ref{eq:Q1def}) that
gives rise to the SM contribution to $\epsilon_K$. In the so-called
SUSY basis introduced by Gabbiani et al.~\cite{Gabbiani:1996hi} the
(parity-even) operators $Q_2,\ldots,Q_5$ read\,\footnote{Thanks to QCD
  parity invariance we can ignore three more dimension-six operators
  whose parity conserving parts coincide with the corresponding parity
  conserving contributions of the operators $Q_1, Q_2$ and $Q_3$.}
\begin{eqnarray}
 & & Q_2 = \big(\bar{s}^a(1-\gamma_5)d^a\big)
           \big(\bar{s}^b(1-\gamma_5)d^b\big), \nonumber\\
 & & Q_3 = \big(\bar{s}^a(1-\gamma_5)d^b\big)
           \big(\bar{s}^b(1-\gamma_5)d^a\big), \nonumber\\
 & & Q_4 = \big(\bar{s}^a(1-\gamma_5)d^a\big)
           \big(\bar{s}^b(1+\gamma_5)d^b\big), \nonumber\\
 & & Q_5 = \big(\bar{s}^a(1-\gamma_5)d^b\big)
           \big(\bar{s}^b(1+\gamma_5)d^a\big),
\end{eqnarray}
where $a$ and $b$ denote colour indices.  In analogy to the case of
$B_{\rm{K}}$ one then defines the $B$ parameters of $Q_2,\ldots,Q_5$
according to
\be
   B_i(\mu) = \frac{\left\langle \bar{K}^0\left| Q_i(\mu)\right|
     \right\rangle}{N_i\left\langle\bar{K}^0\left|\bar{s}\gamma_5
     d\right|0\right\rangle \left\langle0\left|\bar{s}\gamma_5
     d\right|K^0\right\rangle}, \quad i=2,\ldots,5.
\ee
The factors $\{N_2,\ldots,N_5\}$ are given by $\{-5/3, 1/3, 2, 2/3\}$,
and it is understood that $B_i(\mu)$ is specified in some
renormalization scheme, such as $\msbar$ or a variant of the
regularization-independent momentum subtraction (RI-MOM) scheme.

The SUSY basis has been adopted in
Refs.\,\cite{Boyle:2012qb,Bertone:2012cu,Carrasco:2015pra}. Alternatively,
one can employ the chiral basis of Buras, Misiak and
Urban\,\cite{Buras:2000if}. The SWME Collaboration prefers the latter,
since the anomalous dimension which enters the RG running has been
calculated to two loops in perturbation
theory\,\cite{Buras:2000if}. Results obtained in the chiral basis can
be easily converted to the SUSY basis via
\be
   B_3^{\rm SUSY}={\textstyle\frac{1}{2}}\left( 5B_2^{\rm chiral} -
   3B_3^{\rm chiral} \right).
\ee
The remaining $B$ parameters are the same in both bases. In the
following we adopt the SUSY basis and drop the superscript.

Older quenched results for the BSM $B$ parameters can be found in
Refs.~\cite{Allton:1998sm, Donini:1999nn, Babich:2006bh}. Recent
estimates for $B_2,\ldots,B_5$ have been reported for QCD with $\Nf=2$
(ETM\,12D~\cite{Bertone:2012cu}), $\Nf=2+1$
(RBC/UKQCD\,12E\,\cite{Boyle:2012qb}, SWME\,13A\,\cite{Bae:2013tca},
SWME\,14C\,\cite{Jang:2014aea}, SWME\,15A\,\cite{Jang:2015sla}) and
$\Nf=2+1+1$ (ETM\,15\,\cite{Carrasco:2015pra}) flavours of dynamical
quarks. The main features of these calculations are identical to the
case of $B_{\rm{K}}$ discussed above. We note, in particular, that
SWME perform the matching between rooted staggered quarks and the
$\msbar$ scheme using perturbation theory at one loop, while RBC/UKQCD
and ETMC employ nonperturbative renormalization for domain wall and
twisted-mass Wilson quarks, respectively. Control over systematic
uncertainties (chiral and continuum extrapolations, finite-volume
effects) in $B_2,\ldots,B_5$ is expected to be at the same level as
for $B_{\rm{K}}$, as far as the results by ETM\,12D, ETM\,15 and
SWME\,15A are concerned. The calculation by RBC/UKQCD\,12E has been
performed at a single value of the lattice spacing and a minimum pion
mass of 290\,MeV. Thus, the results do not benefit from the same
improvements regarding control over the chiral and continuum
extrapolations as in the case of $B_{\rm{K}}$\,\cite{Blum:2014tka}.
Recent progress from RBC/UKQCD using two values of the lattice spacing
have been reported in Refs.~\cite{Lytle:2013oqa} and
\cite{Hudspith:2015wev}.

\begin{figure}[ht]
\centering
\leavevmode
\includegraphics[height=10cm]{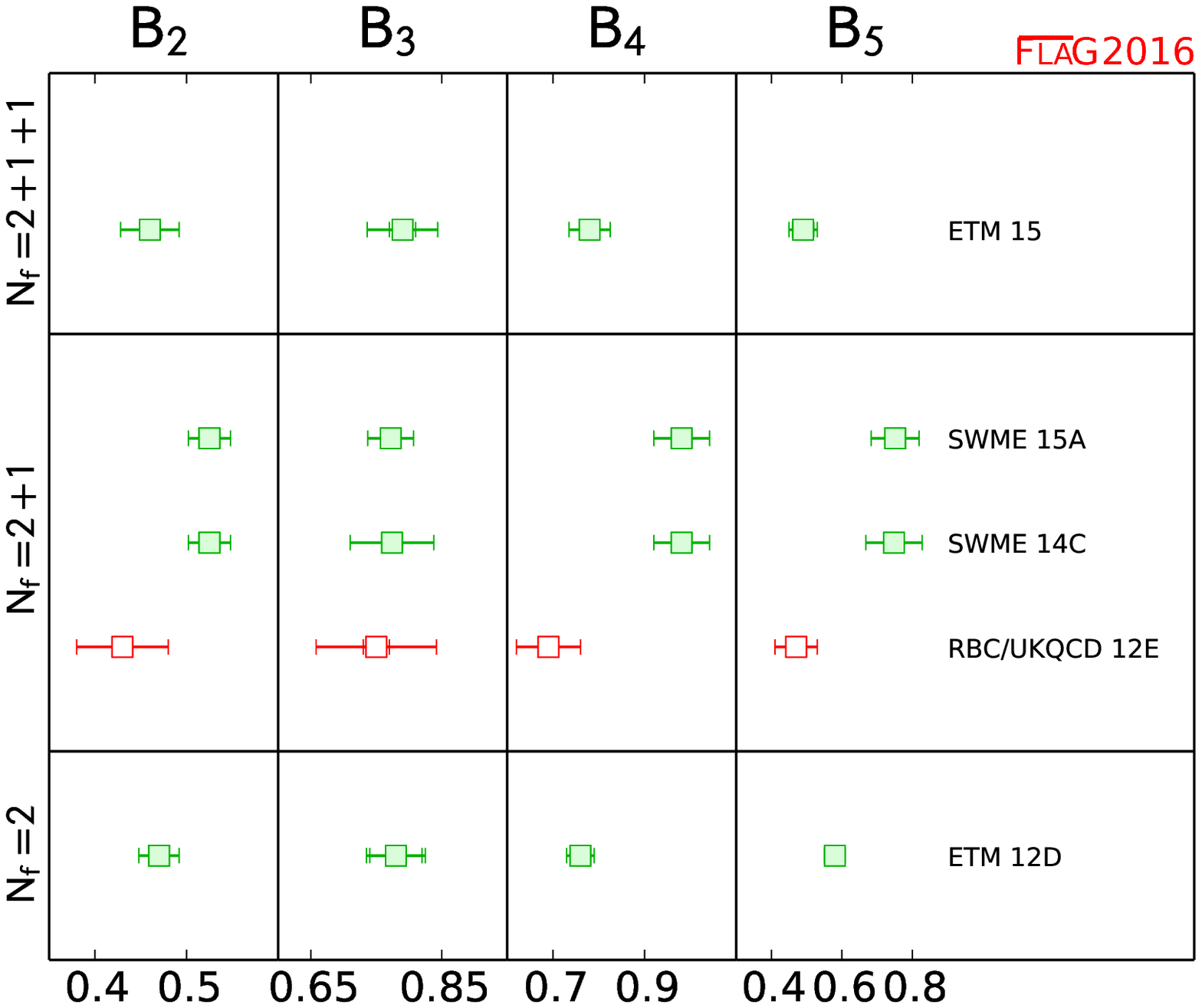}
\caption{Lattice results for the BSM $B$ parameters defined in the
  $\msbar$ scheme at a reference scale of 3\,GeV, see Tab.~\ref{tab_Bi}.
\label{fig_Bisumm}}
\end{figure}

Results for the $B$ parameters $B_2,\ldots,B_5$ computed with $\Nf=2,
2+1$ and $2+1+1$ dynamical quarks are listed and compared in
Tab.~\ref{tab_Bi} and Fig.~\ref{fig_Bisumm}. In general one finds
that the BSM $B$ parameters computed by different collaborations do
not show the same level of consistency as the SM kaon mixing parameter
$B_K$ discussed previously. In particular, the results for $B_2, B_4$
and $B_5$ from SWME\,\cite{Bae:2013tca,Jang:2014aea,Jang:2015sla},
obtained using staggered quarks and employing perturbative matching
differ significantly from those quoted by the
ETM\,\cite{Bertone:2012cu,Carrasco:2015pra} and
RBC/UKQCD\,\cite{Boyle:2012qb} Collaborations, which both determine
the matching factors nonperturbatively. A recent update of the
RBC/UKQCD calculation described in Ref.\,\cite{Hudspith:2015wev}
provides a hint that the nonperturbative determination of the
matching factors depends strongly on the details in the implementation
of the Rome-Southampton method. The use of nonexceptional momentum
configurations in the calculation of the vertex functions produces a
significant modification of the renormalization factors, which in turn
brings the results from RBC/UKQCD much closer to the estimates from
SWME.

Therefore, insufficient control over the renormalization and matching
procedure appears to be the most likely explanation for the observed
deviations. In the absence of further investigations that corroborate
this conjecture, it is difficult to quote global estimates for the BSM
$B$ parameters $B_2,\ldots,B_5$. However, we observe that for each
choice of $\Nf$ there is only one set of results that meets the
required quality criteria, i.e. ETM\,15 \cite{Carrasco:2015pra} for
$\Nf=2+1+1$, SWME\,15A \cite{Jang:2015sla} for $\Nf=2+1$, and
ETM\,12D \cite{Bertone:2012cu} for two-flavour QCD.

\begin{table}[!h]
\begin{center}
\mbox{} \\[3.0cm]
{\footnotesize{
\begin{tabular*}{\textwidth}[l]{l @{\extracolsep{\fill}}r@{\hspace{1mm}}l@{\hspace{1mm}}l@{\hspace{1mm}}l@{\hspace{1mm}}l@{\hspace{1mm}}l@{\hspace{1mm}}l@{\hspace{1mm}}l@{\hspace{1mm}}l@{\hspace{1mm}}l@{\hspace{1mm}}l@{\hspace{1mm}}l}
Collaboration & Ref. & $\Nf$ & 
\hspace{0.15cm}\begin{rotate}{60}{publication status}\end{rotate}\hspace{-0.15cm} &
\hspace{0.15cm}\begin{rotate}{60}{continuum extrapolation}\end{rotate}\hspace{-0.15cm} &
\hspace{0.15cm}\begin{rotate}{60}{chiral extrapolation}\end{rotate}\hspace{-0.15cm}&
\hspace{0.15cm}\begin{rotate}{60}{finite volume}\end{rotate}\hspace{-0.15cm}&
\hspace{0.15cm}\begin{rotate}{60}{renormalization}\end{rotate}\hspace{-0.15cm}  &
\hspace{0.15cm}\begin{rotate}{60}{running}\end{rotate}\hspace{-0.15cm} & 
$B_2$ & $B_3$ & $B_4$ & $B_5$ \\
&&&&&&&&& \\[-0.1cm]
\hline
\hline
&&&&&&&&& \\[-0.1cm]

ETM 15 & \cite{Carrasco:2015pra} & 2+1+1 & \gA & \good & \soso & \soso
& \good&  $\,a$ & 0.46(1)(3) & 0.79(2)(5) & 0.78(2)(4) & 0.49(3)(3)  \\[0.5ex]
&&&&&&&&& \\[-0.1cm]

\hline

&&&&&&&&& \\[-0.1cm]
SWME 15A & \cite{Jang:2015sla} & 2+1 & \gA & \good & \soso &
\good & \soso$^\dagger$ & $-$ & 0.525(1)(23) & 0.772(6)(35) & 0.981(3)(62) & 0.751(7)(68)  \\[0.5ex]
&&&&&&&&& \\[-0.1cm]

SWME 14C & \cite{Jang:2014aea} & 2+1 & C & \good & \soso &
\good & \soso$^\dagger$ & $-$ & 0.525(1)(23) & 0.774(6)(64) & 0.981(3)(61) & 0.748(9)(79)  \\[0.5ex]
&&&&&&&&& \\[-0.1cm]

SWME 13A$^\ddagger$ & \cite{Bae:2013tca} & 2+1 & \gA & \good & \soso  &
\good & \soso$^\dagger$ & $-$ & 0.549(3)(28)  & 0.790(30) & 1.033(6)(46) & 0.855(6)(43)   \\[0.5ex]
&&&&&&&&& \\[-0.1cm]

RBC/
& \cite{Boyle:2012qb} & 2+1 & \gA & \tbr & \soso & \good &
\good & $\,b$ & 0.43(1)(5)  & 0.75(2)(9)  & 0.69(1)(7)  & 0.47(1)(6)
\\
UKQCD 12E & & & & & & & & & & & & \\[0.5ex]  
&&&&&&&&& \\[-0.1cm]

\hline

&&&&&&&&& \\[-0.1cm]
ETM 12D & \cite{Bertone:2012cu} & 2 & \gA & \good & \soso & \soso
& \good&  $\,c$ & 0.47(2)(1)  & 0.78(4)(2)  & 0.76(2)(2)  & 0.58(2)(2)  \\[0.5ex]

&&&&&&&&& \\[-0.1cm]
\hline
\hline\\[-0.1cm]
\end{tabular*}
}}
\begin{minipage}{\linewidth}
{\footnotesize 
\begin{itemize}
\item[$^\dagger$] The renormalization is performed using perturbation
        theory at one loop, with a conservative estimate of
         the uncertainty. \\[-5mm]
\item[$a$] $B_i$ are renormalized nonperturbatively at scales $1/a
        \sim 2.2-3.3\,\gev$ in the $\Nf = 4$ RI/MOM scheme 
        using two different lattice momentum scale intervals, with
        values around $1/a$ for the first and around
        3.5~GeV for the second one. The impact of
        these two ways to the final result is taken into account
         in the error budget. Conversion to $\msbar$ is at one loop at 3~GeV.\\[-5mm]
\item[$b$] The $B$ parameters are renormalized nonperturbatively at a scale of 3~GeV. \\[-5mm]
\item[$c$] $B_i$ are renormalized nonperturbatively at scales $1/a
        \sim 2-3.7\,\gev$ in the $\Nf = 2$ RI/MOM scheme using
        two different lattice momentum scale intervals,  
        with values around $1/a$ for the first and around 3~GeV
        for the second one.\\[-5mm]
\item[$^\ddagger$] The computation of $B_4$ and $B_5$ has been
        revised in Refs. \cite{Jang:2015sla} and \cite{Jang:2014aea}. 
\end{itemize}
}
\end{minipage}
\caption{Results for the BSM $B$ parameters $B_2,\ldots,B_5$ in the
  $\msbar$ scheme at a reference scale of 3\,GeV. Any available
  information on nonperturbative running is indicated in the column
  ``running", with details given at the bottom of the
  Tab.~\label{tab_Bi}}
\end{center}
\end{table}

%% file: HQ/HQD.tex
%=================================================
\section{$D$-meson decay constants and form factors}
\label{sec:DDecays}
%=================================================

Leptonic and semileptonic decays of charmed $D$ and $D_s$ mesons occur
via charged $W$-boson exchange, and are sensitive probes of $c \to d$
and $c \to s$ quark flavour-changing transitions.  Given experimental
measurements of the branching fractions combined with sufficiently
precise theoretical calculations of the hadronic matrix elements, they
enable the determination of the CKM matrix elements $|V_{cd}|$ and
$|V_{cs}|$ (within the Standard Model) and a precise test of the
unitarity of the second row of the CKM matrix.  Here we summarize the
status of lattice-QCD calculations of the charmed leptonic decay
constants.  Significant progress has
been made in charm physics on the lattice in recent years,
largely due to the availability of gauge configurations produced using
highly-improved lattice-fermion actions that enable treating the
$c$-quark with the same action as for the $u$, $d$, and $s$-quarks.

This Section updates the corresponding one in the last FLAG review~\cite{Aoki:2013ldr} for results that
appeared after November 30, 2013.
As already done in Ref.~\cite{Aoki:2013ldr}, we limit our
review to results based on modern simulations with reasonably light
pion masses (below approximately 500~MeV). This excludes results
obtained from the earliest unquenched simulations, which typically had
two flavours in the sea, and which were limited to heavier pion masses
because of the constraints imposed by the computational resources and
methods available at that time.
Recent lattice-QCD averages for $D_{(s)}$-meson decay constants were also 
presented by the Particle Data Group in the review on 
``Leptonic Decays of Charged Pseudoscalar Mesons"~\cite{Rosner:2015wva}.  
The PDG three- and four-flavour averages for $f_D$, $f_{D_s}$, and their ratio are identical to those obtained here.
This is because both reviews include the same sets of calculations in the averages, 
and make the same assumptions about the correlations between the calculations.

Following our review of lattice-QCD calculations of $D_{(s)}$-meson
leptonic decay constants and semileptonic form factors, we then
interpret our results within the context of the Standard Model.  We
combine our best-determined values of the hadronic matrix elements
with the most recent experimentally-measured branching fractions to
obtain $|V_{cd(s)}|$ and test the unitarity of the second row of the
CKM matrix.

%\newpage
\input{HQ/HQSubsections/DL.tex}

\input{HQ/HQSubsections/DSL.tex}

\input{HQ/HQSubsections/DCKM.tex}

%% file: HQ/HQSubsections/DL.tex
\subsection{Leptonic decay constants $f_D$ and $f_{D_s}$}
\label{sec:fD}

In the Standard Model the decay constant $f_{D_{(s)}}$ of a charged
pseudoscalar $D$ or $D_s$ meson is related to the branching ratio for
leptonic decays mediated by a $W$ boson through the formula
\be
{\mathcal{B}}(D_{(s)} \to \ell\nu_\ell)= {{G_F^2|V_{cq}|^2 \tau_{D_{(s)}}}\over{8 \pi}} f_{D_{(s)}}^2 m_\ell^2 
m_{D_{(s)}} \left(1-{{m_\ell^2}\over{m_{D_{(s)}}^2}}\right)^2\;,
 \label{eq:Dtoellnu}
\ee
where $V_{cd}$ ($V_{cs}$) is the appropriate CKM matrix element for a
$D$ ($D_s$) meson.  The branching fractions have been experimentally
measured by CLEO, Belle, Babar and BES with a precision around 4-5$\%$ for
both the $D$ and the $D_s$-meson
decay modes~\cite{Rosner:2015wva}.  When
combined with lattice results for the decay constants, they allow for
determinations of $|V_{cs}|$ and $|V_{cd}|$.

\begin{figure}[tb]
\hspace{-0.8cm}\includegraphics[width=0.58\linewidth]{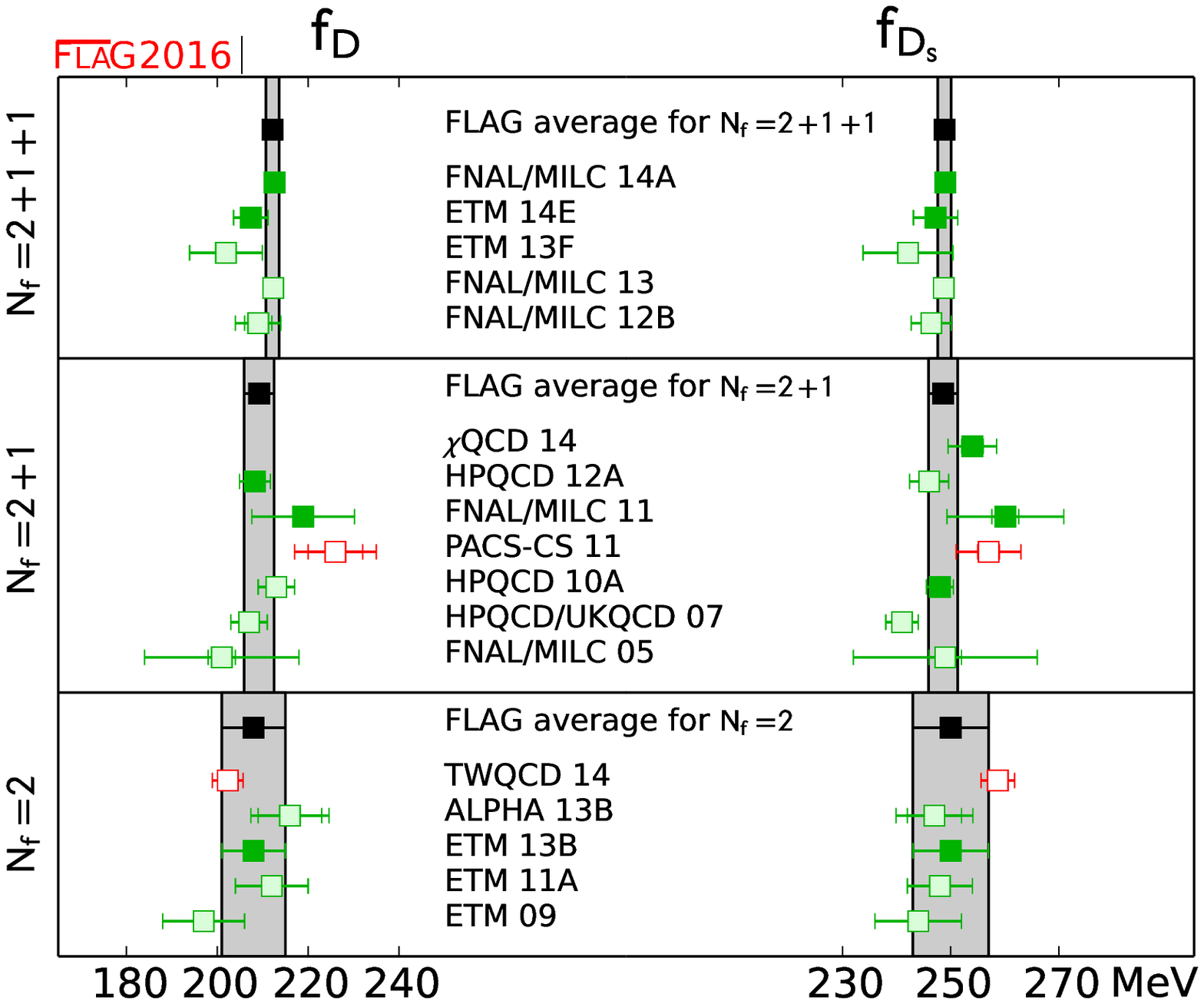} \hspace{-1cm}
\includegraphics[width=0.58\linewidth]{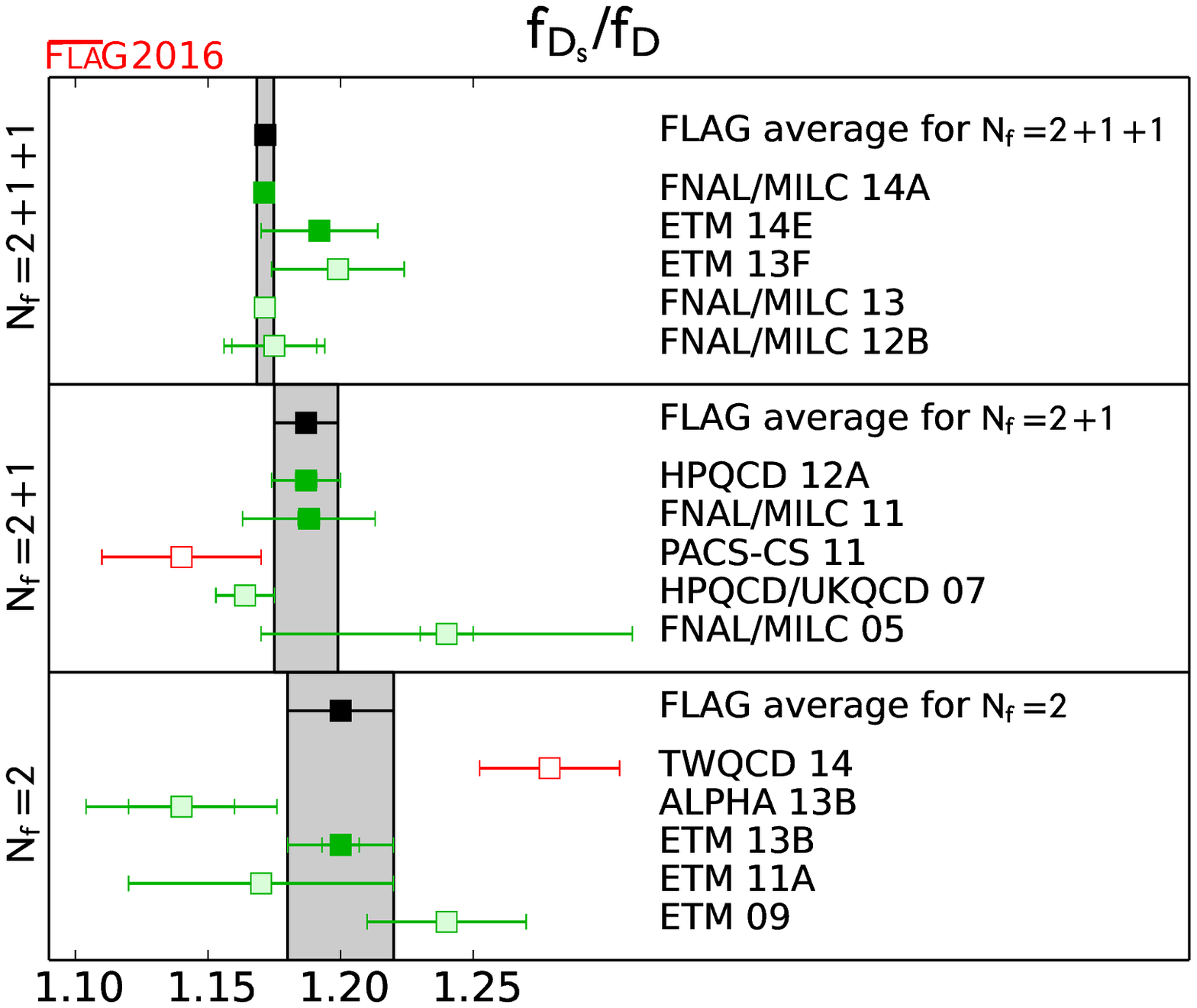}

\vspace{-2mm}
\caption{Decay constants of the $D$ and $D_s$ mesons [values in Tab.~\ref{tab_FDsummary}]. 
and Eqs.~\ref{eq:Nf2av},~\ref{eq:Nf2+1av},~\ref{eq:Nf2+1+1av}]. 
The significance of the colours is explained in Sec.~2.
The black squares and grey bands
  indicate our averages. }
\label{fig:fD}
\end{figure}

In lattice-QCD calculations the decay constants $f_{D_{(s)}}$ are extracted from 
Euclidean  matrix elements of the axial current
\be
\langle 0| A^{\mu}_{cq} | D_q(p) \rangle = i f_{D_q}\;p_{D_q}^\mu  \;,
\label{eq:dkconst}
\ee
with $q=d,s$ and $ A^{\mu}_{cq} =\bar{c}\gamma_\mu \gamma_5
q$. Results for $N_f=2,\; 2+1$ and $2+1+1$ dynamical flavours are
summarized in Tab.~\ref{tab_FDsummary} and Fig.~\ref{fig:fD}.
Since the publication of the last FLAG review, a handful of results
for  $f_D$ and $f_{D_s}$ have appeared, which we are going to briefly describe
here. We consider isospin-averaged quantities, although in a few cases
results for $f_{D^+}$ are quoted (FNAL/MILC~11 and FNAL/MILC~14A, where the difference
between $f_D$ and $f_{D^+}$ has been estimated be at the 0.5 MeV level).
%
%%%%%%%%%%%%%%%%%%%%%%%%%%%%%%%%%%%%%%%%%%%% THE TABLES %%%%%%%%%%%%%%%%%%%%%%%%%%%%%%%%%%%%%%%%%%%%%%%%%%%%%%%%%%%%%%
\begin{table}[!htb]
\begin{center}
\mbox{} \\[3.0cm]
\footnotesize
\begin{tabular*}{\textwidth}[l]{@{\extracolsep{\fill}}l@{\hspace{1mm}}r@{\hspace{1mm}}l@{\hspace{1mm}}l@{\hspace{1mm}}l@{\hspace{1mm}}l@{\hspace{1mm}}l@{\hspace{1mm}}l@{\hspace{1mm}}l@{\hspace{1mm}}l@{\hspace{1mm}}l@{\hspace{1mm}}l}
Collaboration & Ref. & $\Nf$ & 
\hspace{0.15cm}\begin{rotate}{60}{publication status}\end{rotate}\hspace{-0.15cm} &
\hspace{0.15cm}\begin{rotate}{60}{continuum extrapolation}\end{rotate}\hspace{-0.15cm} &
\hspace{0.15cm}\begin{rotate}{60}{chiral extrapolation}\end{rotate}\hspace{-0.15cm}&
\hspace{0.15cm}\begin{rotate}{60}{finite volume}\end{rotate}\hspace{-0.15cm}&
\hspace{0.15cm}\begin{rotate}{60}{renormalization/matching}\end{rotate}\hspace{-0.15cm}  &
\hspace{0.15cm}\begin{rotate}{60}{heavy-quark treatment}\end{rotate}\hspace{-0.15cm} & 
\rule{0.4cm}{0cm}$f_D$ & \rule{0.4cm}{0cm}$f_{D_s}$  & 
 \rule{0.3cm}{0cm}$f_{D_s}/f_D$ \\[0.2cm]
\hline
\hline
&&&&&&&&&&& \\[-0.1cm]
FNAL/MILC 14A$^{**}$ & \cite{Bazavov:2014wgs} & 2+1+1 & \gA & \good & \good &\good & \good & \okay & 
212.6(0.4) $+1.0 \choose -1.2$   & 249.0(0.3)$+1.1 \choose -1.5$ &  1.1712(10)$+29 \choose -32$ \\[0.5ex]

ETM 14E$^{\dagger}$ & \cite{Carrasco:2014poa} & 2+1+1 & \gA & \good & \soso  &  \soso & \good  &  \okay &
207.4(3.8)   & 247.2(4.1) &  1.192(22) \\[0.5ex]

ETM 13F & \cite{Dimopoulos:2013qfa} & 2+1+1 & \rC & \soso & \soso  &  \soso & \good  &  \okay &
202(8)   & 242(8) &  1.199(25) \\[0.5ex]

FNAL/MILC 13$^\nabla$ & \cite{Bazavov:2013nfa} & 2+1+1 & \rC & \good    & \good    & \good     
&\good & \okay  & 212.3(0.3)(1.0)   & 248.7(0.2)(1.0) & 1.1714(10)(25)\\[0.5ex]

FNAL/MILC 12B & \cite{Bazavov:2012dg} & 2+1+1 & \rC & \good    & \good    & \good     
&\good & \okay  & 209.2(3.0)(3.6)   & 246.4(0.5)(3.6) & 1.175(16)(11)\\[0.5ex]

&&&&&&&&&&& \\[-0.1cm]
\hline
&&&&&&&&&&& \\[-0.1cm]
$\chi$QCD~14 &\cite{Yang:2014sea} & 2+1 &  \gA &\soso &\soso & \soso & \good & \okay
& & 254(2)(4) & \\[0.5ex]
HPQCD 12A &\cite{Na:2012iu} & 2+1 & \gA &\soso  &\soso &\soso &\good &\okay 
& 208.3(1.0)(3.3) & 246.0(0.7)(3.5) & 1.187(4)(12)\\[0.5ex]

FNAL/MILC 11& \cite{Bazavov:2011aa} & 2+1 & \gA & \soso &\soso &\soso  & 
 \soso & \okay & 218.9(11.3) & 260.1(10.8)&   1.188(25)   \\[0.5ex]  

PACS-CS 11 & \cite{Namekawa:2011wt} & 2+1 & \gA & \tbr & \good & \tbr  & 
\soso & \okay & 226(6)(1)(5) & 257(2)(1)(5)&  1.14(3)   \\[0.5ex] 

HPQCD 10A & \cite{Davies:2010ip} & 2+1 & \gA & \good  & \soso  & 
\good & \good & \okay & 213(4)$^{*}$ & 248.0(2.5)  \\[0.5ex]

HPQCD/UKQCD 07 & \cite{Follana:2007uv} & 2+1 &  \gA & \good & \soso & 
\soso & \good  & \okay & 207(4) & 241 (3)& 1.164(11)  \\[0.5ex] 

FNAL/MILC 05 & \cite{Aubin:2005ar} & 2+1 & \gA &\soso &   \soso    &
\soso      & \soso    &  \okay       & 201(3)(17) & 249(3)(16)  & 1.24(1)(7) \\[0.5ex]

&&&&&&&&&&& \\[-0.1cm]
\hline
&&&&&&&&&&& \\[-0.1cm]
TWQCD 14$^{\square\square}$ & \cite{Chen:2014hva} & 2 & \gA & \tbr & \soso  &  \tbr & \good  &  \okay &
202.3(2.2)(2.6)   & 258.7(1.1)(2.9) &  1.2788(264) \\[0.5ex]

ALPHA 13B & \cite{Heitger:2013oaa} & 2 & \rC & \soso & \good & \good & \good & \okay &
216(7)(5)   & 247(5)(5) &  1.14(2)(3) \\[0.5ex]

ETM 13B$^\square$ & \cite{Carrasco:2013zta} & 2 & \gA & \good & \soso  &  \soso & \good  &  \okay &
208(7)   & 250(7) &  1.20(2) \\[0.5ex]

ETM 11A & \cite{Dimopoulos:2011gx} & 2 & \gA & \good & \soso  &  \soso & \good  &  \okay &
212(8)   & 248(6) &  1.17(5) \\[0.5ex]

ETM 09 & \cite{Blossier:2009bx} & 2 & \gA & \soso & \soso  &  \soso & \good  &  \okay & 
197(9)   & 244(8) &  1.24(3) \\[0.5ex]

&&&&&&&&&&& \\[-0.1cm]
\hline
\hline\\
\end{tabular*}\\[-0.2cm]
\begin{minipage}{\linewidth}
{\footnotesize 
\begin{itemize}
   \item[$^{\dagger}$] Update of ETM 13F.\\[-5mm]
   \item[$^{\nabla}$] Update of FNAL/MILC 12B.\\[-5mm]
\item[$^{*}$] This result is obtained by using the central value for $f_{D_s}/f_D$ from HPQCD/UKQCD~07 
and increasing the error to account for the effects from the change in the physical value of $r_1$. \\[-5mm]
\item[$^{\square}$] Update of ETM 11A and ETM 09. \\[-5mm]
\item[$^{\square\square}$] 1 lattice spacing $\simeq 0.1$ fm only. $M_{\pi,{\rm min}}L=1.93$.\\[-5mm]
\item[$^{**}$] 
At $\beta = 5.8$, $M_{\pi, \rm min}L=3.2$ but this ensemble is primarily used for the systematic error estimate.
\end{itemize}
}
\end{minipage}
\caption{Decay constants of the $D$ and $D_{s}$ mesons (in MeV) and their ratio.}
\label{tab_FDsummary}
\end{center}
\end{table}
%%%%%%%%%%%%%%%%%%%%%%%%%%%%%%%%%%%%%%%%%%%%%%%%%%%%%%%%%%%%%%%%%%%%%%%%%%%%%%%%%%%%%%%%%%%%%%%%%%%%%%%%%%%%%%%%%%%
%
%
%%%%%%%%%%%%%%%%%%%%%%%%%%%%%%%%%%%%%%%%%%%%%%%%%%%%%%%%%%%%%%%%%%% THE FIGURES %%%%%%%%%%%%%%%%%%%%%%%%%%%%%%%%%%%%%%%%
%
%\newpage

Two new results have appeared for $N_f=2$. The averages however remain unchanged, as we will
see in the following. In Ref.~\cite{Heitger:2013oaa}, the ALPHA collaboration
directly computed the matrix element in Eq.~(\ref{eq:dkconst})
(for $\mu=0$ and $q=d,s$) on two $N_f=2$ ensembles of nonperturbatively $\cO(a)$ improved
Wilson fermions at lattice spacings of 0.065 and 0.048 fm.
Pion masses range between 440 and 190 MeV and the condition $Lm_\pi\geq 4$ is always met.
Chiral/continuum extrapolations are performed adopting either a fit ansatz linear in $m_\pi^2$ and $a^2$
or, for $f_D$, by using a fit form  inspired by partially quenched 
Heavy Meson Chiral Perturbation Theory
(HM$\chi$PT).
Together with the scale setting, these extrapolations dominate the final systematic errors.
As the scale is set through another decay constant ($f_K$), what is actually computed is $f_{D_{(s)}}/f_K$ and
most of the uncertainty on the renormalization constant of the axial current drops out.
Since the results only appeared as a proceeding contribution to the Lattice~2013 conference, they do not enter the
final averages.

The TWQCD collaboration reported in Ref.~\cite{Chen:2014hva} about the first computation of the masses and decay constants
of pseudoscalar $D_{(s)}$ mesons in two-flavour lattice QCD with domain-wall fermions. This is a
calculation performed at one lattice spacing only ($a\approx 0.061$fm) and in a rather small volume
($24^3 \times 48$, with $M_{\pi,{\rm min}}L \approx 1.9$).
For these reasons the quoted values of the decay constants do not qualify for the averages and should be regarded as the
result of a pilot study in view of a longer and ongoing effort, in which the remaining systematics will be addressed
through computations at different volumes as well as several lattice spacings.

The $N_f=2$ averages therefore coincide with those in the previous FLAG review and are given by the values in ETM~13B, namely
%FLAGRESULT BEGIN
% TAG      & fD    & fDs	&fDsofD &END
% REFS     & \cite{Carrasco:2013zta} &\cite{Carrasco:2013zta} &\cite{Carrasco:2013zta} &END
% UNITS    & '[MeV]' & '[MeV]' & 1 &END
% NUMRESULTS & 1 & 1 & 1 &END
% FLAVOURs & 2 & 2 & 2 &END
%FLAGRESULT END
%FLAGRESULTFORMULA BEGIN
\begin{align}
         &&\FLAGAVBEGIN f_D    &=208(7)\FLAGAVEND\;{\rm MeV}         &&\Ref~\mbox{\cite{Carrasco:2013zta}},\nonumber\\[2mm]
&N_f=2:   &\FLAGAVBEGIN f_{D_s}&= 250(7) \FLAGAVEND \;{\rm MeV}      &&\Ref~\mbox{\cite{Carrasco:2013zta}},\label{eq:Nf2av}\\[2mm]
	 &&\FLAGAVBEGIN {{f_{D_s}}/{f_D}}&=1.20(2)\FLAGAVEND     &&\Ref~\mbox{\cite{Carrasco:2013zta}}.\nonumber
\end{align}
%FLAGRESULTFORMULA END

The situation is quite similar for the $N_f=2+1$ case, where only one new result, and for $f_{D_s}$ only, appeared in the last two years.
The $\chi$QCD collaboration used (valence) overlap fermions on a sea of 2+1 flavours of domain-wall fermions (corresponding to
the gauge configurations generated by RBC/UKQCD and described in Ref.~\cite{Aoki:2010dy}) to compute the charm- and the strange-quark masses
as well as $f_{D_s}$. 
The decay constant is obtained by combining the determinations
from  either an exactly conserved PCAC Ward identity or from the matrix element 
of the local axial current.
The latter needs to be renormalized and the corresponding renormalization constant has been determined nonperturbatively
in Ref.~\cite{Liu:2013yxz}.
The computation of $f_{D_s}$ has been performed at two lattice spacings ($a=0.113$ and $a=0.085$ fm) with the value of the bare charm-quark mass, in lattice
units, ranging between 0.3 and 0.75. Pion masses reach down to about 300 MeV and $M_{\pi,{\rm min}}L$ is always larger than 4.
The chiral extrapolation and lattice artifacts are responsible for the largest systematic uncertainties, both being estimated to be around
1\%, on top of a statistical error of about the same size.
The lattice spacing dependence is estimated by changing the functional form
in the chiral/continuum extrapolation by terms of $\cO(a^4)$.
As the authors point out, it will be possible to make a more accurate 
assessment of the discretization errors only once the planned 
ensembles at a finer lattice spacing are available.

The RBC/UKQCD collaboration presented intermediate results for the  
$D$ and $D_s$ decay constants with 2+1 flavours of M\"obius domain-wall fermions in Ref.~\cite{Boyle:2015kyy}.
Since the analysis has not been completed yet, no values for $f_{D_{(s)}}$ are quoted.

Summarizing the $\Nf=2+1$ case, the average for $f_D$ did not change with respect to the last review
and it is obtained from the HPQCD~12A and the FNAL/MILC~11 determinations, whereas for $f_{D_s}$ 
the value changes in order to include the result from the $\chi$QCD collaboration (together
with the values in HPQCD~10A and in FNAL/MILC~11). The updated estimates then read
%FLAGRESULT BEGIN
% TAG      & fD    & fDs	&fDsofD &END
% REFS     & \cite{Na:2012iu,Bazavov:2011aa} &\cite{Davies:2010ip,Bazavov:2011aa,Yang:2014sea} &\cite{Na:2012iu,Bazavov:2011aa} &END
% UNITS    & '[MeV]' & '[MeV]' & 1 &END
% NUMRESULTS & 2 & 3 & 2 &END
% FLAVOURs & 2+1 & 2+1 & 2+1 &END
%FLAGRESULT END
%FLAGRESULTFORMULA BEGIN
\begin{align}
        &&\FLAGAVBEGIN f_D    &=209.2(3.3)\FLAGAVEND \;{\rm MeV} &&\Refs~\mbox{\cite{Na:2012iu,Bazavov:2011aa}},\nonumber\\[2mm]
&N_f=2+1:&\FLAGAVBEGIN f_{D_s}&=249.8(2.3)\FLAGAVEND \;{\rm MeV} &&\Refs~\mbox{\cite{Davies:2010ip,Bazavov:2011aa,Yang:2014sea}},\label{eq:Nf2+1av} \\[2mm]
        &&\FLAGAVBEGIN {{f_{D_s}}/{f_D}}&=1.187(12)\FLAGAVEND&&\Refs~\mbox{\cite{Na:2012iu,Bazavov:2011aa}},\nonumber
\end{align}
%FLAGRESULTFORMULA END
where the error on the $\Nf=2+1$ average of $f_{D_s}$ has been rescaled by the factor $\sqrt{\chi^2/d.o.f.}=1.1$ (see Sec.~2).
In addition, 
the statistical errors between the results of FNAL/MILC and HPQCD have been everywhere treated as 100\% correlated since
the two collaborations use overlapping sets of configurations. The same procedure had been used in the 2013 review.

Two new determinations appeared from simulations with 2+1+1 dynamical flavours. These are FNAL/MILC~14A and ETM~14E.
The FNAL/MILC~14A results in Ref.~\cite{Bazavov:2014wgs} are obtained using the HISQ ensembles with up, down, strange and charm dynamical
quarks, generated by the MILC collaboration~\cite{Bazavov:2012xda} (see also Ref.~\cite{Gamiz:2013xxa} for the RMS pion masses)
employing HISQ sea quarks and a 1-loop tadpole improved Symanzik gauge action. The RHMC as well as the RHMD algorithms have been used in this case.
The latter is an inexact algorithm, where the accept/reject step at the end of the molecular-dynamics trajectory is skipped. 
In Ref.~\cite{Bazavov:2012xda} results for the plaquette, the bare fermion
condensates and a few meson masses, using both algorithms, are compared and found to
agree within statistical uncertainties. 
The relative scale is set through $F_{4ps}$, the decay 
constant of a fictitious meson with valence masses of $0.4 m_s$ and physical
sea-quark masses. For the absolute scale $f_\pi$ is used.
In FNAL/MILC~14A four different lattice spacings, ranging from 0.15 to 0.06 fm, have been considered with
all quark masses close to their physical values. The analysis includes additional ensembles with light sea-quark 
masses that are heavier than in nature, and where in some cases the strange sea-quark masses are lighter than in nature.
This allowed to actually perform two different analyses; the ``physical mass analysis'' and the ``chiral analysis''. 
The second analysis uses staggered chiral perturbation theory for all-staggered  heavy-light mesons in order to include
the unphysical-mass ensembles. This results in smaller statistical errors compared to the ``physical mass analysis''.
The latter is used for the central values and the former as a cross-check and as an ingredient in the systematic error analysis.
Chiral and continuum extrapolation uncertainties are estimated by considering a total of 114 different fits.
The quark-mass and lattice-spacing dependence of the decay constants are modelled in heavy-meson,
rooted, all-staggered chiral perturbation theory (HMrAS$\chi$PT) including all
NNLO and N$^3$LO mass-dependent, analytic, terms.
Fits differ in the way some of the LEC's are fixed, in the number of NNLO parameters related to discretization effects included,
in the use of priors, in whether the $a=0.15$~fm ensembles are included or not and in the inputs used for the quark masses
and the lattice spacings.
The number of parameters ranges between 23 and 28 and the number of data points varies between 314 and 366.
The maximum difference between these results and the central values is taken as an estimate of the chiral/continuum
extrapolation errors. The central fit is chosen to give results that are close to the centres of the distributions, in order
to symmetrize the errors.
FNAL/MILC also provides in Ref.~\cite{Bazavov:2014wgs} an estimate of strong isospin-breaking effects by computing the $D$ meson decay constant with
the mass of the light quark in the valence set to the physical value of the down-quark mass. The result reads $f_{D^+} - f_D=0.47(1)$$+25 \choose -6$~MeV.
This effect is of the size of the quoted errors, and the number in Tab.~\ref{tab_FDsummary} indeed corresponds to  $f_{D^+}$. 
The final accuracy on the decay constants is at the level of half-a-percent. It is therefore necessary to consider the electroweak corrections
to the decay rates  when extracting $|V_{cd}|$ and $|V_{cs}|$ from leptonic transitions of $D_{(s)}$ mesons.
The most difficult to quantify is due to electromagnetic effects that depend on the meson hadronic structure.
In Ref.~\cite{Bazavov:2014wgs} this contribution to the decay rates is estimated to be between  1.1\% and 2.8\%, by 
considering the corresponding contribution for $\pi$ and $K$ decays, as computed in $\chi$PT, and allowing for a factor 2 to 5.
After correcting the PDG data for the decay rates in Ref.~\cite{Agashe:2014kda}, by including the effects mentioned above with their corresponding uncertainty,
the FNAL/MILC collaboration uses the results for $f_D$ and $f_{D_s}$ to produce estimates for $|V_{cd}|$ and $|V_{cs}|$, 
as well as a unitarity test of the second row of the CKM matrix, which yields $1- |V_{cd}|^2 - |V_{cs}|^2 - |V_{cb}|^2 =-0.07(4)$,
indicating a slight tension with CKM unitarity.\footnote{Notice that the contribution of $|V_{cb}|^2$ to the unitarity relation
is more than one order of magnitude below the quoted error, and it can therefore be neglected.}

The ETM collaboration has also published results with $2+1+1$ dynamical
flavours in Ref.~\cite{Carrasco:2014poa}~(ETM~14E), updating the values that appeared
in the Lattice 2013 Conference proceedings~\cite{Dimopoulos:2013qfa}~(ETM~13F).
The configurations have been generated using the Iwasaki action in the
gauge and the Wilson twisted mass action for sea quarks.  The charm
and strange valence quarks are discretized as Osterwalder-Seiler
fermions~\cite{Osterwalder:1977pc}.  Three different lattice spacings
in the range $0.09 - 0.06$ fm have been considered with pion masses as
low as 210 MeV in lattices of linear spatial extent of about 2 to 3
fm (see Ref.~\cite{Carrasco:2014cwa} for details on the simulations). 
In~ETM~14E $f_{D_s}$ is obtained by extrapolating the ratio
 $f_{D_s}/m_{D_s}$, differently from ETM~13B, where $f_{D_s}r_0$ was
extrapolated. The new choice is found to be affected by smaller
discretization effects. For the chiral/continuum extrapolation
terms linear and quadratic in $m_l$ and one term linear in $a^2$ 
are included in the parameterization. Systematic uncertainties
are assessed by comparing to a linear fit in $m_l$ and by taking
the difference with the result at the finest lattice resolution.
The decay constant $f_D$ is determined by fitting the double ratio
$(f_{D_s}/f_D)/(f_K/f_\pi)$ using continuum HM$\chi$PT, as discretization
effects are not visible, within errors, for that quantity.
An alternative fit without chiral logs is used to estimate 
the systematic uncertainty associated to the chiral extrapolation.
The main systematic uncertainties are due to the continuum and chiral
extrapolations and to the error on $f_K/f_\pi$, which is
also determined in~ETM~14E.
Using the experimental averages of $f_D|V_{cd}|$ and $f_{D_s}|V_{cs}|$
available in 2014 from PDG~\cite{Agashe:2014kda}, the ETM collaboration also provides a unitarity test of the 
second row of the CKM matrix, obtaining
$1- |V_{cd}|^2 - |V_{cs}|^2 - |V_{cb}|^2 =-0.08(5)$, which is consistent with the estimate from~FNAL/MILC~14A
and with the value in the latest PDG~report~\cite{Rosner:2015wva}, which quotes $-0.063(34)$ for the same combination
of matrix elements. That indicates a slight tension with three-generation unitarity.

Finally, by combining in a weighted average the FNAL/MILC~14A and the ETM~14E results, we get the estimates
%FLAGRESULT BEGIN
% TAG      & fD    & fDs	&fDsofD &END
% REFS     & \cite{Bazavov:2014wgs,Carrasco:2014poa} &\cite{Bazavov:2014wgs,Carrasco:2014poa} &\cite{Bazavov:2014wgs,Carrasco:2014poa} &END
% UNITS    & '[MeV]' & '[MeV]' & 1 &END
% NUMRESULTS & 2 & 2 & 2 &END
% FLAVOURs & 2+1+1 & 2+1+1 & 2+1+1 &END
%FLAGRESULT END
%FLAGRESULTFORMULA BEGIN
\begin{align}
	  &&\FLAGAVBEGIN f_D&=212.15(1.45)\FLAGAVEND \;{\rm MeV}	&&\Refs~\mbox{\cite{Bazavov:2014wgs,Carrasco:2014poa}},\nonumber\\[2mm]  
&N_f=2+1+1:&\FLAGAVBEGIN f_{D_s}&= 248.83(1.27)\FLAGAVEND \;{\rm MeV}	&&\Refs~\mbox{\cite{Bazavov:2014wgs,Carrasco:2014poa}},\label{eq:Nf2+1+1av} \\[2mm]
          &&\FLAGAVBEGIN {{f_{D_s}}/{f_D}}&=1.1716(32)\FLAGAVEND	&&\Refs~\mbox{\cite{Bazavov:2014wgs,Carrasco:2014poa}},\nonumber
\end{align}
%FLAGRESULTFORMULA END
where the error on the average of $f_{D}$ has been rescaled by the factor $\sqrt{\chi^2/d.o.f.}=1.3$.
The PDG~\cite{Agashe:2014kda} produces {\it experimental} averages of the decay constants, by combining the measurements
of  $f_D|V_{cd}|$ and $f_{D_s}|V_{cs}|$ with values of $|V_{cd}|$ and $|V_{cs}|$ obtained by relating them to other
CKM elements (i.e., by assuming unitarity). Given the choices detailed in Ref.~\cite{Agashe:2014kda}, the values read
\begin{equation}
f_{D^+}^{exp}=203.7 (4.8) \; {\rm MeV}, \quad\! f_{D_s^+}^{exp}= 257.8 (4.1)\;{\rm MeV},
\end{equation}
which disagree with the $\Nf=2+1+1$ lattice averages in Eq.~(\ref{eq:Nf2+1+1av}) at the two-sigma level.

%% file: HQ/HQSubsections/DSL.tex
\subsection{Semileptonic form factors for $D\to \pi \ell \nu$ and $D\to K \ell \nu$}
 \label{sec:DtoPiK}

The form factors for semileptonic $D\to \pi \ell\nu$ and $D\to
K \ell \nu$ decays, when combined with experimental measurements of the
decay widths, enable determinations of the CKM matrix elements
$|V_{cd}|$ and $|V_{cs}|$ via:
\begin{eqnarray}
	\frac{d\Gamma(D\to P\ell\nu)}{dq^2} = \frac{G_F^2 |V_{cx}|^2}{24 \pi^3}
	\,\frac{(q^2-m_\ell^2)^2\sqrt{E_P^2-m_P^2}}{q^4m_{D}^2} \,
	\bigg[ \left(1+\frac{m_\ell^2}{2q^2}\right)m_{D}^2(E_P^2-m_P^2)|f_+(q^2)|^2 & \nonumber\\
+ \frac{3m_\ell^2}{8q^2}(m_{D}^2-m_P^2)^2|f_0(q^2)|^2 & \!\!\!\! \bigg]\,, \label{eq:DtoPiKFull}
\end{eqnarray}
where $x = d, s$ is the daughter light quark, $P= \pi, K$ is the
daughter light pseudoscalar meson, and $q = (p_D - p_P)$ is the
momentum of the outgoing lepton pair.  The vector and scalar form
factors $f_+(q^2)$ and $f_0(q^2)$ parameterize the hadronic matrix
element of the heavy-to-light quark flavour-changing vector current
$V_\mu = \overline{x} \gamma_\mu c$:
\begin{equation}
\langle P| V_\mu | D \rangle  = f_+(q^2) \left( {p_D}_\mu+ {p_P}_\mu - \frac{m_D^2 - m_P^2}{q^2}\,q_\mu \right) + f_0(q^2) \frac{m_D^2 - m_P^2}{q^2}\,q_\mu \,,
\end{equation}
and satisfy the kinematic constraint $f_+(0) = f_0(0)$.  Because the contribution to the decay width from
the scalar form factor is proportional to $m_\ell^2$, it can be
neglected for $\ell = e, \mu$, and Eq.~(\ref{eq:DtoPiKFull})
simplifies to
\begin{equation}
\frac{d\Gamma \!\left(D \to P \ell \nu\right)}{d q^2} = \frac{G_F^2}{24 \pi^3} |\vec{p}_{P}|^3 {|V_{cx}|^2 |f_+^{DP} (q^2)|^2} \,. \label{eq:DtoPiK}
\end{equation}

In practice, most lattice-QCD calculations of $D\to \pi \ell\nu$ and
$D\to K \ell \nu$ focus on providing the value of the vector form
factor at a single value of the momentum transfer, $f_+(q^2=0)$, which
is sufficient to obtain $|V_{cd}|$ and $|V_{cs}|$.  Because the decay
rate cannot be measured directly at $q^2=0$, comparison
of these lattice-QCD results with experiment requires a slight
extrapolation of the experimental measurement.  Some lattice-QCD
calculations also provide determinations of the $D\to \pi \ell\nu$ and
$D\to K \ell \nu$ form factors over the full kinematic range $0 < q^2
< q^2_{\rm max} = (m_D - m_P)^2$, thereby allowing a comparison of the
shapes of the lattice simulation and experimental data.  This
nontrivial test in the $D$ system provides a strong check of
lattice-QCD methods that are also used in the $B$-meson system.

Lattice-QCD calculations of the $D\to \pi \ell\nu$ and $D\to
K \ell \nu$ form factors typically use the same light-quark and
charm-quark actions as those of the leptonic decay constants $f_D$ and
$f_{D_s}$.  Therefore many of the same issues arise, e.g., chiral
extrapolation of the light-quark mass(es) to the physical point,
discretization errors from the charm quark, and matching the lattice
weak operator to the continuum, as discussed in the previous section.
Two strategies have been adopted to eliminate the need to renormalize
the heavy-light vector current in recent calculations of
$D\to \pi \ell\nu$ and $D\to K \ell \nu$, both of which can be applied
to simulations in which the same relativistic action is used for the
light $(u,d,s)$ and charm quarks.  The first method was proposed by
Be{\'c}irevi{\'c} and Haas in Ref.~\cite{Becirevic:2007cr}, and
introduces double-ratios of lattice three-point correlation functions
in which the vector current renormalization cancels.  Discretization
errors in the double ratio are of $\cO((am_h)^2)$ provided
that the vector-current matrix elements are $\cO(a)$
improved.  The vector and scalar form factors $f_+(q^2)$ and
$f_0(q^2)$ are obtained by taking suitable linear combinations of
these double ratios.  The second method was introduced by the HPQCD
Collaboration in Ref.~\cite{Na:2010uf}.  In this case, the quantity
$(m_{c} - m_{x} ) \langle P | S | D \rangle$, where $m_{x}$ and $m_c$
are the bare lattice quark masses and $S = \bar{x}c$ is the lattice
scalar current, does not get renormalized.  The desired form factor at
$q^2=0$ can be obtained by (i) using a Ward identity to
relate the matrix element of the vector current to that of the scalar
current, and (ii) taking advantage of the kinematic identity $f_+(0) = f_0(0)$, such that $f_+(q^2=0) = (m_{c} -
m_{x} ) \langle P | S | D \rangle / (m^2_D - m^2_P)$.

Additional complications enter for semileptonic decay matrix elements
due to the nonzero momentum of the outgoing pion or kaon.  Both
statistical errors and discretization errors increase at larger meson
momenta, so results for the lattice form factors are most precise at
$q^2_{\rm max}$.  However, because lattice calculations are performed
in a finite spatial volume, the pion or kaon three-momentum can only
take discrete values in units of $2\pi/L$ when periodic boundary
conditions are used.  For typical box sizes in recent lattice $D$- and
$B$-meson form-factor calculations, $L \sim 2.5$--3~fm; thus the
smallest nonzero momentum in most of these analyses lies in the range
$p_P \equiv |\vec{p}_P| \sim 400$--$500$~MeV.  The largest momentum in lattice
heavy-light form-factor calculations is typically restricted to
$ p_P \leq 4\pi/L$. %(2,0,0)$.  
For $D \to \pi \ell \nu$ and $D \to
K \ell \nu$, $q^2=0$ corresponds to $p_\pi \sim 940$~MeV and $p_K \sim
1$~GeV, respectively, and the full recoil-momentum region is within
the range of accessible lattice momenta.\footnote{This situation
differs from that of calculations of the $K\to\pi\ell\nu$ form factor,
where the physical pion recoil momenta are smaller than $2 \pi/L$.
For $K\to\pi\ell\nu$ it is now standard to use nonperiodic
(``twisted") boundary
conditions~\cite{Bedaque:2004kc,Sachrajda:2004mi} to simulate directly
at $q^2=0$; see Sec.~\ref{sec:Direct}.  Some collaborations have also
begun to use twisted boundary conditions for $D$
decays~\cite{DiVita:2011py,Koponen:2011ev,Koponen:2012di,Koponen:2013tua}.}  Therefore
the interpolation to $q^2=0$ is relatively insensitive to the fit
function used to parameterize the momentum dependence, and the
associated systematic uncertainty in $f_+(0)$ is small. In contrast,
determinations of the form-factor shape can depend strongly on the
parameterization of the momentum dependence, and the systematic
uncertainty due to the choice of model function is often difficult to
quantify.  This is becoming relevant for $D \to \pi \ell \nu$ and $D
\to K \ell \nu$ decays as more collaborations are beginning to present
results for $f_+(q^2)$ and $f_0(q^2)$ over the full kinematic range.
The parameterization of the form-factor shape is even more important
for semileptonic $B$ decays, for which the momentum range needed to
connect to experiment is often far from $q^2_{max}$.

A class of functions based on general field-theory properties, known
as $z$-expansions, has been introduced to allow model-independent
parameterizations of the $q^2$ dependence of semileptonic form factors
over the entire kinematic range (see, e.g.,
Refs.~\cite{Boyd:1994tt,Bourrely:2008za}).  The use of such functions
is now standard for the analysis of $B \to \pi \ell\nu$ transitions
and the determination of
$|V_{ub}|$~\cite{Bailey:2008wp,Ha:2010rf,Lees:2012vv,Amhis:2012bh}; we
therefore discuss approaches for parameterizing the $q^2$ dependence
of semileptonic form factors, including $z$-expansions, in
Sec.~\ref{sec:BtoPiK}.  Here we briefly summarize the aspects most
relevant to calculations of $D \to \pi \ell \nu$ and $D \to K \ell
\nu$.  In general, all semileptonic form factors can be expressed as a
series expansion in powers of $z$ times an overall multiplicative
function that accounts for any sub-threshold poles and branch cuts,
where the new variable $z$ is a nonlinear function of $q^2$.  The
series coefficients $a_n$ depend upon the physical process (as well as
the choice of the prefactors), and can only be determined empirically
by fits to lattice or experimental data.  Unitarity establishes strict
upper bounds on the size of the $a_n$'s, while guidance from
heavy-quark power counting provides even tighter constraints.
Some works are now using a variation of this approach,
commonly referred to as ``modified $z$-expansion," that
is used to simultaneously extrapolate their lattice simulation data
to the physical light-quark masses and the continuum limit, and to
interpolate/extrapolate their lattice data in $q^2$.
More comments on this method are also provided in~Sec.~\ref{sec:BtoPiK}.

\subsubsection{Results for $f_+(0)$}

We now review the status of lattice calculations of the $D \to \pi
\ell \nu$ and $D \to K \ell \nu$ form factors at $q^2=0$.  As in the
previous version of this review, although we also describe ongoing
calculations of the form-factor shapes, we do not rate these
calculations, since all of them are still unpublished, except for
conference proceedings that provide only partial results.\footnote{In Ref.~\cite{Aubin:2004ej},
to be discussed below, form factors are indeed computed for several values of $q^2$,
and fitted to a Be\'cirevi\'c-Kaidalov parameterization (cf. Sec.~\ref{sec:zparam})
to extract their values at $q^2=0$. However, while results for fit parameters are provided,
the values of the form factors at $q^2 \neq 0$ are not provided, which prevents
us from performing an independent analysis of their shape using model-independent
parameterizations.}

The most advanced $N_f = 2$ lattice-QCD calculation of the
$D \to \pi \ell \nu$ and $D \to K \ell \nu$ form factors is by the ETM
Collaboration~\cite{DiVita:2011py}.  This still preliminary work uses
the twisted-mass Wilson action for both the light and charm quarks,
with three lattice spacings down to $a \approx 0.068$~fm and (charged)
pion masses down to $m_\pi \approx 270$~MeV.  The calculation employs
the ratio method of Ref.~\cite{Becirevic:2007cr} to avoid the need to
renormalize the vector current, and extrapolates to the physical
light-quark masses using $SU(2)$ heavy-light meson $\chi$PT.  ETM simulate with nonperiodic boundary
conditions for the valence quarks to access arbitrary momentum values
over the full physical $q^2$ range, and interpolate to $q^2=0$ using
the Be{\'c}irevi{\'c}-Kaidalov ansatz~\cite{Becirevic:1999kt}.  The
statistical errors in $f_+^{D\pi}(0)$ and $f_+^{DK}(0)$ are 9\% and
7\%, respectively, and lead to rather large systematic uncertainties
in the fits to the light-quark mass and energy dependence (7\% and
5\%, respectively).  Another significant source of uncertainty is from
discretization errors (5\% and 3\%, respectively).  On the finest
lattice spacing used in this analysis $am_c \sim 0.17$, so $\cO((am_c)^2)$ cutoff errors are expected to be about 5\%.  This can be
reduced by including the existing $N_f = 2$ twisted-mass ensembles
with $a \approx 0.051$~fm discussed in Ref.~\cite{Baron:2009wt}. Work
is in progress by the ETM Collaboration also to compute the form
factors $f_+^{D\pi},~f_0^{D\pi}$ and $f_+^{DK},~f_0^{DK}$
for the whole kinematically available range on the $N_f = 2+1+1$ twisted-mass
Wilson lattices~\cite{Baron:2010bv}.  This calculation will include
dynamical charm-quark effects and use three lattice spacings down to
$a\approx 0.06$~fm. A BCL $z$-parameterization is being used
to describe the $q^2$ dependence. The latest progress report on this
work, which provides values of the form factors at $q^2=0$
with statistical errors only, can be found in Ref.~\cite{Carrasco:2015bhi}.

The first published $N_f = 2+1$ lattice-QCD calculation of the
$D \to \pi \ell \nu$ and $D \to K \ell \nu$ form factors is by the
Fermilab Lattice, MILC, and HPQCD Collaborations~\cite{Aubin:2004ej}.
(Because only two of the authors of this work are in HPQCD, and to
distinguish it from other more recent works on the same topic by
HPQCD, we hereafter refer to this work as ``FNAL/MILC.")  This work
uses asqtad-improved staggered sea quarks and light ($u,d,s$) valence
quarks and the Fermilab action for the charm quarks, with a single
lattice spacing of $a \approx 0.12$ fm.  At this lattice spacing, the
staggered taste splittings are still fairly large, and the minimum RMS
pion mass is $\approx 510$~MeV.  This calculation renormalizes the
vector current using a mostly nonperturbative approach, such that the
perturbative truncation error is expected to be negligible compared to
other systematics.  The Fermilab Lattice and MILC Collaborations
present results for the $D \to \pi \ell \nu$ and $D \to K \ell \nu$
semileptonic form factors over the full kinematic range, rather than
just at $q^2=0$.  In fact, the publication of this
result predated the precise measurements of the $D\to K \ell\nu$ decay
width by the FOCUS~\cite{Link:2004dh} and Belle
experiments~\cite{Abe:2005sh}, and predicted the shape of
$f_+^{DK}(q^2)$ quite accurately.  This bolsters confidence in
calculations of the $B$-meson semileptonic decay form factors using
the same methodology.  Work is in progress~\cite{Bailey:2012sa} to
reduce both the statistical and systematic errors in $f_+^{D\pi}(q^2)$
and $f_+^{DK}(q^2)$ through increasing the number of configurations
analysed, simulating with lighter pions, and adding lattice spacings
as fine as $a \approx 0.045$~fm. 
In parallel, a much more ambitious computation of $D \to \pi \ell \nu$ and $D \to K \ell \nu$
by FNAL/MILC is now ongoing, using $N_f=2+1+1$ MILC HISQ ensembles at four values
of the lattice spacing down to $a=0.042~{\rm fm}$ and pion masses down to the
physical point. The latest report on this computation, focusing on the form factors
at $q^2=0$, but without explicit values of the latter yet, can be found in Ref.~\cite{Primer:2015qpz}.

\begin{table}[h]
\begin{center}
\mbox{} \\[3.0cm]
\footnotesize
\begin{tabular*}{\textwidth}[l]{l @{\extracolsep{\fill}} r l l l l l l l c c}
Collaboration & Ref. & $\Nf$ & 
\hspace{0.15cm}\begin{rotate}{60}{publication status}\end{rotate}\hspace{-0.15cm} &
\hspace{0.15cm}\begin{rotate}{60}{continuum extrapolation}\end{rotate}\hspace{-0.15cm} &
\hspace{0.15cm}\begin{rotate}{60}{chiral extrapolation}\end{rotate}\hspace{-0.15cm}&
\hspace{0.15cm}\begin{rotate}{60}{finite volume}\end{rotate}\hspace{-0.15cm}&
\hspace{0.15cm}\begin{rotate}{60}{renormalization}\end{rotate}\hspace{-0.15cm}  &
\hspace{0.15cm}\begin{rotate}{60}{heavy-quark treatment}\end{rotate}\hspace{-0.15cm}  &
$f_+^{D\pi}(0)$ & $f_+^{DK}(0)$\\
&&&&&&&&& \\[-0.1cm]
\hline
\hline
&&&&&&&&& \\[-0.1cm]
HPQCD 11 & \cite{Na:2011mc} & 2+1 & \gA  & \soso & \soso & \soso & \good &  \okay & 0.666(29) &\\[0.5ex]
HPQCD 10B & \cite{Na:2010uf} & 2+1 & \gA  & \soso & \soso & \soso & \good &  \okay & & 0.747(19)  \\[0.5ex]
FNAL/MILC 04 & \cite{Aubin:2004ej} & 2+1 & \gA  & \tbr & \tbr & \soso & \soso & \okay & 0.64(3)(6)& 0.73(3)(7)
\\[0.5ex]
&&&&&&&&& \\[-0.1cm]
\hline\\[0.5ex]
ETM 11B & \cite{DiVita:2011py} & 2 & \rC  & \soso & \soso & \good & \good &  \okay & 0.65(6)(6) & 0.76(5)(5)\\[0.5ex]
&&&&&&&&& \\[-0.1cm]
\hline
\hline
\end{tabular*}
\caption{$D \to \pi\ell\nu$ and $D\to K\ell\nu$ semileptonic form factors at $q^2=0$.\label{tab_DtoPiKsumm2}}
\end{center}
\end{table}

The most precise published calculations of the
$D \to \pi \ell \nu$~\cite{Na:2011mc} and $D \to
K \ell \nu$~\cite{Na:2010uf} form factors are by the HPQCD
Collaboration.  These analyses also use the $N_f = 2+1$
asqtad-improved staggered MILC configurations at two lattice spacings
$a \approx 0.09$ and 0.12~fm, but use the HISQ action for the valence
$u,d,s$, and $c$ quarks.  In these mixed-action calculations, the HISQ
valence light-quark masses are tuned so that the ratio $m_l/m_s$ is
approximately the same as for the sea quarks; the minimum RMS sea-pion
mass is $\approx 390$~MeV.  They calculate the form factors at $q^2=0$
by relating them to the matrix element of the scalar
current, which is not renormalized. They use the ``modified
$z$-expansion'' to simultaneously extrapolate to the physical
light-quark masses and continuum and interpolate to $q^2 = 0$, and
allow the coefficients of the series expansion to vary with the light-
and charm-quark masses.  The form of the light-quark dependence is
inspired by $\chi$PT, and includes logarithms of the form $m_\pi^2
{\rm log} (m_\pi^2)$ as well as polynomials in the valence-, sea-, and
charm-quark masses.  Polynomials in $E_{\pi(K)}$ are also included to
parameterize momentum-dependent discretization errors. (See Ref.~~\cite{Na:2011mc}
for further technical details.)
The number of terms is increased until the result for $f_+(0)$
stabilizes, such that the quoted fit error for $f_+(0)$ includes both
statistical uncertainties and those due to most systematics.  The
largest uncertainties in these calculations are from statistics and
charm-quark discretization errors.

The HPQCD Collaboration is now extending their work on $D$-meson
semileptonic form factors to determining their shape over the full
kinematic range~\cite{Koponen:2011ev}, and recently obtained results
for the $D \to K \ell \nu$ form factors $f_+(q^2)$ and
$f_0(q^2)$~\cite{Koponen:2012di}.  This analysis uses a subset of the
ensembles included in their earlier work, with two sea-quark masses at
$a \approx 0.12$~fm and one sea-quark mass at $a \approx 0.09$~fm, but
with approximately three times more statistics on the coarser
ensembles and ten times more statistics on the finer ensemble.  As
above, the scalar current is not renormalized.  The spatial vector
current renormalization factor is obtained by requiring that
$f_+(0)^{H\to H} = 1$ for $H = D, D_s, \eta_s$, and $\eta_c$.  The
renormalization factors for the flavour-diagonal currents agree for
different momenta as well as for charm-charm and strange-strange
external mesons within a few percent, and are then used to renormalize
the flavour-changing charm-strange and charm-light currents.  The
charm-strange temporal vector current is normalized by matching to the
scalar current $f_0(q^2_{\rm max})$.  Also as above, they
simultaneously extrapolate to the physical light-quark masses and
continuum and interpolate/extrapolate in $q^2$ using the modified
$z$-expansion.  In this case, however, they only allow for light-quark
mass and lattice-spacing dependence in the series coefficients, but
not for charm-quark mass or kaon energy dependence, and constrain the
parameters with Bayesian priors.  It is not clear, however, that only
three sea-quark ensembles at two lattice spacings are sufficient to
resolve the quark-mass and lattice spacing dependence, even within the
context of constrained fitting.  The quoted error in the zero-recoil
form factor $f_+(0) = 0.745(11)$ is significantly smaller than in
their 2010 work, but we are unable to understand the sources of this
improvement with the limited information provided in
Ref.~\cite{Koponen:2012di}. The preprint does not provide an error
budget, nor any information on how the systematic uncertainties are
estimated. Thus we cannot rate this calculation, and do not include it
in the summary table and plot.

\begin{figure}[h]
\begin{center}
\includegraphics[width=0.7\linewidth]{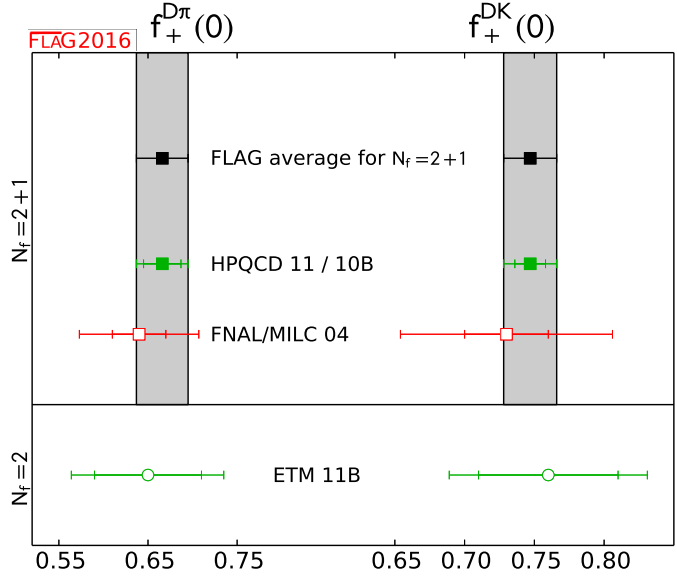}

\vspace{-2mm}
\caption{$D\to\pi \ell\nu$ and $D\to K\ell\nu$ semileptonic form
  factors at $q^2=0$. The HPQCD result for
  $f_+^{D\pi}(0)$ is from HPQCD 11, the one for $f_+^{DK}(0)$
  represents HPQCD 10B (see Table
  \ref{tab_DtoPiKsumm2}). \label{fig:DtoPiK}}
 \end{center}
\end{figure}

Table~\ref{tab_DtoPiKsumm2} summarizes the existing $N_f =2$ and $N_f
= 2+1$ calculations of the $D \to \pi \ell \nu$ and $D \to K \ell \nu$
semileptonic form factors.  The quality of the systematic error
studies is indicated by the symbols.  Additional tables in
appendix~\ref{app:DtoPi_Notes} provide further details on the
simulation parameters and comparisons of the error estimates.  Recall
that only calculations without red tags that are published in a
refereed journal are included in the FLAG average.  Of the
calculations described above, only those of HPQCD~10B,11 satisfy all
of the quality criteria.  Therefore our average of the
$D \to \pi \ell \nu$ and $D \to K \ell \nu$ semileptonic form factors
from $N_f = 2+1$ lattice QCD is
%
%FLAGRESULT BEGIN
% TAG      & fDtopi    & fDtoK	 &END
% REFS     & \cite{Na:2011mc} &\cite{Na:2010uf} &END
% UNITS    & 1 & 1 &END
% NUMRESULTS & 2 & 2 & 2 &END
% FLAVOURs & 2+1 & 2+1 &END
%FLAGRESULT END
%FLAGRESULTFORMULA BEGIN
\begin{align}
	&&\FLAGAVBEGIN f_+^{D\pi}(0)&=  0.666(29)\FLAGAVEND&&\Refs~\mbox{\cite{Na:2011mc}},\nonumber\\[-3mm]
&N_f=2+1:&\label{eq:Nf=2p1Dsemi}\\[-3mm]
        &&\FLAGAVBEGIN f_+^{DK}(0)  &= 0.747(19)\FLAGAVEND &&\Refs~\mbox{\cite{Na:2010uf}}.\nonumber
\end{align}
%FLAGRESULTFORMULA END
%
Fig.~\ref{fig:DtoPiK} displays the existing $N_f =2$ and $N_f = 2+1$
results for $f_+^{D\pi}(0)$ and $f_+^{DK}(0)$; the grey bands show our
average of these quantities.  Section~\ref{sec:Vcd} discusses the
implications of these results for determinations of the CKM matrix
elements $|V_{cd}|$ and $|V_{cs}|$ and tests of unitarity of the
second row of the CKM matrix.

%% file: HQ/HQSubsections/DCKM.tex
\subsection{Determinations of $|V_{cd}|$ and $|V_{cs}|$ and test of  second-row CKM unitarity}
\label{sec:Vcd}

We now interpret the lattice-QCD results for the $D_{(s)}$ meson decays
as determinations of the CKM
matrix elements $|V_{cd}|$ and $|V_{cs}|$ in the Standard Model.

For the leptonic decays, we use the latest experimental averages from
Rosner, Stone and Van de Water for the Particle Data Group~\cite{Rosner:2015wva}
\begin{equation}
	f_D |V_{cd}| = 45.91(1.05)~{\rm MeV} \,, \qquad f_{D_s} |V_{cs}| = 250.9(4.0)~{\rm MeV} \,.
\end{equation}
By combining these with the average values of $f_D$ and $f_{D_s}$ from
the individual $N_f = 2$, $N_f = 2+1$ and $N_f=2+1+1$ lattice-QCD 
calculations that
satisfy the FLAG criteria, we obtain the results for the CKM
matrix elements $|V_{cd}|$ and $|V_{cs}|$ in
Tab.~\ref{tab:VcdVcsIndividual}.  
%%%%%%%%%%%%%%%%%%%%%%%%%%%%%%%%%%%%%%%%%%%%%%%%%%%%%
For our preferred values we use the
averaged $N_f=2$ and $N_f = 2+1$ results for $f_D$ and $f_{D_s}$ in
Eqs.~(\ref{eq:Nf2av}),~(\ref{eq:Nf2+1av}) 
and~(\ref{eq:Nf2+1+1av}).  We obtain
\begin{align}
&{\rm leptonic~decays}, N_f=2+1+1:&|V_{cd}| &= 0.2164(14)(49)\,, &|V_{cs}| &= 1.008  (5)(16) \,, \\
&{\rm leptonic~decays}, N_f=2+1:  &|V_{cd}| &= 0.2195(35)(50)\,, &|V_{cs}| &= 1.004  (9)(16) \,, \\
&{\rm leptonic~decays}, N_f=2:    &|V_{cd}| &= 0.2207(74)(50)\,, &|V_{cs}| &= 1.004 (28)(16) \,,
\end{align}
where the errors shown are from the lattice calculation and experiment
(plus nonlattice theory), respectively.  For the $N_f = 2+1$ and the $N_f=2+1+1$
determinations, the uncertainties from the lattice-QCD calculations of
the decay constants are smaller than the
experimental uncertainties in the branching fractions.
Although the results for
$|V_{cs}|$ are slightly larger than one, they are  consistent with
unity within errors.

The leptonic determinations of these CKM matrix elements have uncertainities that are reaching the few-percent level.
However, higher-order electroweak and hadronic corrections to the rate have not been computed for the case of $D_{(s)}$ mesons,
whereas they have been estimated to be around 1--2\% for pion and kaon decays~\cite{Cirigliano:2007ga}. 
It is therefore important that such theoretical calculations are tackled soon, perhaps directly on the lattice, as proposed
in Ref.~\cite{Carrasco:2015xwa}. 
\begin{table}[tb]
\begin{center}
\noindent
%\footnotesize
\begin{tabular*}{\textwidth}[l]{@{\extracolsep{\fill}}lrlcr}
Collaboration & Ref. &$\Nf$&from&\rule{0.8cm}{0cm}$|V_{cd}|$ or $|V_{cs}|$\\
&&&& \\[-2ex]
\hline \hline &&&&\\[-2ex]
FNAL/MILC~14A & \cite{Bazavov:2014wgs} & 2+1+1 & $ f_D$ & 0.2159(12)(49) \\
ETM~14E & \cite{Carrasco:2014poa} &  2+1+1 & $ f_D$ & 0.2214(41)(51) \\
HPQCD 12A & \cite{Na:2012iu} & 2+1 & $f_{D}$  & 0.2204(36)(50) \\
HPQCD 11 & \cite{Na:2011mc} & 2+1 & $D \to \pi \ell \nu$  & 0.2140(93)(29) \\
FNAL/MILC 11  & \cite{Bazavov:2011aa} & 2+1 & $f_{D}$  &  0.2097(108)(48)  \\
ETM 13B  & \cite{Carrasco:2013zta} & 2 & $f_{D}$  &  0.2207(74)(50)  \\
&&&& \\[-2ex]
 \hline
&&&& \\[-2ex]
FNAL/MILC~14A & \cite{Bazavov:2014wgs} & 2+1+1 & $ f_{D_s}$ & 1.008(5)(16) \\
ETM~14E & \cite{Carrasco:2014poa} &  2+1+1 & $ f_{D_s}$ & 1.015(17)(16) \\
HPQCD 10A & \cite{Davies:2010ip} & 2+1 & $f_{D_s}$  & 1.012(10)(16)  \\
FNAL/MILC 11 & \cite{Bazavov:2011aa} & 2+1 & $f_{D_s}$  &  0.965(40)(16) \\
HPQCD 10B & \cite{Na:2010uf} & 2+1 & $D \to K \ell \nu$  & 0.975(25)(7) \\
$\chi$QCD~14 & \cite{Yang:2014sea} & 2+1 & $f_{D_s}$  &  0.988(17)(16) \\
ETM 13B & \cite{Carrasco:2013zta} & 2 & $f_{D_s}$  &  1.004(28)(16) \\
&&&& \\[-2ex]
 \hline \hline 
\end{tabular*}
\caption{Determinations of $|V_{cd}|$ (upper panel) and $|V_{cs}|$
  (lower panel) obtained from lattice calculations of $D$-meson
  leptonic decay constants
 and semileptonic form factors.
The errors
  shown are from the lattice calculation and experiment (plus
  nonlattice theory), respectively. \label{tab:VcdVcsIndividual}}
\end{center}
\end{table}

For the semileptonic decays, there is no update on the lattice side
from the previous version of our review. As experimental input for the
determination of $|V_{cb}|$ we use the latest experimental averages
from the Heavy Flavour Averaging Group~\cite{Amhis:2014hma}:
\begin{equation}
	f_+^{D\pi}(0) |V_{cd}| = 0.1425(19) \,, \qquad f_+^{DK}(0) |V_{cs}| = 0.728(5)  \,.
\end{equation}
For each of $f_+^{D\pi}(0)$ and $f_+^{DK}(0)$, there is only a single
$N_f = 2+1$ lattice-QCD calculation that satisfies the FLAG criteria.
Using these results, which are given in Eq.~(\ref{eq:Nf=2p1Dsemi}), we
obtain our preferred values for $|V_{cd}|$ and $|V_{cs}|$:
\begin{eqnarray}
	|V_{cd}| = 0.2140(93)(29)  \,, \quad  |V_{cs}| = 0.975(25)(7)   \,, \quad  ({\rm semileptonic~decays}, N_f=2+1) 
\end{eqnarray}
where the errors shown are from the lattice calculation and experiment
(plus nonlattice theory), respectively. These values are compared
with individual leptonic determinations in Tab.~\ref{tab:VcdVcsIndividual}.

\begin{figure}[h]

\begin{center}
\includegraphics[width=0.7\linewidth]{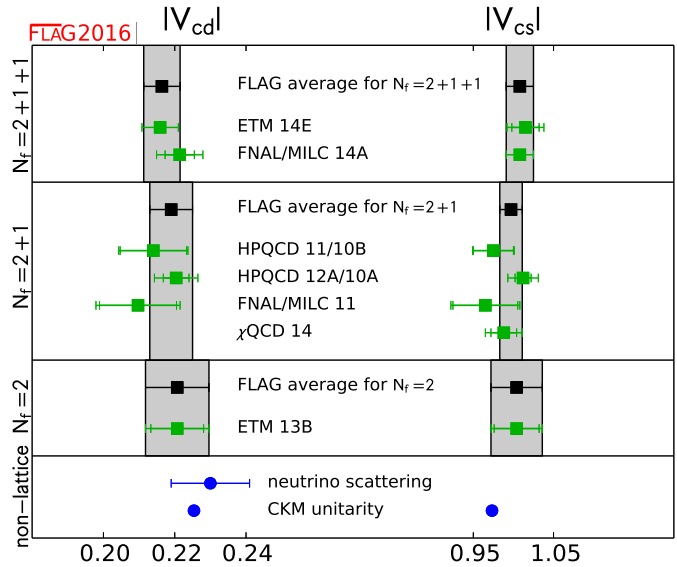}

\vspace{-2mm}
\caption{Comparison of determinations of $|V_{cd}|$ and $|V_{cs}|$
  obtained from lattice methods with nonlattice determinations and
  the Standard Model prediction based on CKM unitarity.  When two
  references are listed on a single row, the first corresponds to the
  lattice input for $|V_{cd}|$ and the second to that for $|V_{cs}|$.
  The results denoted by squares are from leptonic decays, while those
  denoted by triangles are from semileptonic
  decays.
\label{fig:VcdVcs}}
\end{center}
\end{figure}

Table~\ref{tab:VcdVcsSummary} summarizes the results for $|V_{cd}|$
and $|V_{cs}|$ from leptonic 
and semileptonic
 decays, and compares
them to determinations from neutrino scattering (for $|V_{cd}|$ only)
and CKM unitarity.  These results are also plotted in
Fig.~\ref{fig:VcdVcs}.  
For both $|V_{cd}|$ and $|V_{cs}|$, the errors in the direct determinations from
leptonic 
and semileptonic 
decays are approximately one order of magnitude larger
than the indirect determination from CKM unitarity.  
Some tensions at the 2$\sigma$ level are present between the direct and
the indirect estimates, namely in $|V_{cd}|$ using the $N_f=2+1+1$ lattice result
and in $|V_{cs}|$ using both the $N_f=2+1$ and the $N_f=2+1+1$ values.

In order to provide final estimates, for $N_f=2$ and $N_f=2+1+1$
we take the only available results coming from leptonic decays,
while for $N_f=2+1$ we average leptonic and semileptonic channels.
For this purpose, we assume that the statistical errors are 100\% correlated between 
the FNAL/MILC and HPQCD computations because they use the MILC asqtad gauge configurations.
We also assume that the heavy-quark discretization errors are 100\%
correlated between the HPQCD calculations of leptonic and semileptonic
decays because they use the same charm-quark action, and that the
scale-setting uncertainties are 100\% correlated between the HPQCD
results as well.  Finally, we include the 100\% correlation between
the experimental inputs for the two extractions of $|V_{cd(s)}|$ from
leptonic decays. We finally quote
\begin{align}
&{\rm our~average}, N_f=2+1+1:&|V_{cd}| &= 0.2164(51) \,,& |V_{cs}| &= 1.008(17) \,, \\
&{\rm our~average}, N_f=2+1:  &|V_{cd}| &= 0.2190(60) \,,& |V_{cs}| &= 0.997(14) \,, \\
&{\rm our~average}, N_f=2:    &|V_{cd}| &= 0.2207(89) \,,& |V_{cs}| &= 1.004(32) \,, 
\label{eq:Vcdsfinal}
\end{align}
where the errors include both theoretical and experimental
uncertainties.

Using the lattice determinations of $|V_{cd}|$ and $|V_{cs}|$ in
Tab.~\ref{tab:VcdVcsSummary}, we can test the unitarity of the second row
of the CKM matrix.  We obtain
\begin{align}
&N_f=2+1+1:   &|V_{cd}|^2 + |V_{cs}|^2 + |V_{cb}|^2 - 1 &= 0.06(3) \,,\\  
&N_f=2+1:     &|V_{cd}|^2 + |V_{cs}|^2 + |V_{cb}|^2 - 1 &= 0.04(3) \,,  \\
&N_f=2:       &|V_{cd}|^2 + |V_{cs}|^2 + |V_{cb}|^2 - 1 &= 0.06(7) \,.  
\end{align}
Again, tensions at the 2$\sigma$ level with CKM unitarity are visible, as also
reported in the PDG review~\cite{Rosner:2015wva}, where the value 0.063(34) is quoted for the quantity in the equations above.
Given the
current level of precision, this result does not depend on 
 $|V_{cb}|$, which is of $\cO(10^{-2})$. 
%[see Eq.~(\ref{eq:VcbNf2p1})].

\begin{table}[tb]
\begin{center}
\noindent
%\footnotesize
\begin{tabular*}{\textwidth}[l]{@{\extracolsep{\fill}}lcrcc}
& from & Ref. &\rule{0.8cm}{0cm}$|V_{cd}|$ & \rule{0.8cm}{0cm}$|V_{cs}|$\\
&& \\[-2ex]
\hline \hline &&\\[-2ex]
$N_f = 2+1+1$&  $f_D$ \& $f_{D_s}$ && 0.2164(51) & 1.008(17) \\
$N_f = 2+1$&  $f_D$ \& $f_{D_s}$ && 0.2195(61) & 1.004(18) \\
$N_f = 2$ &  $f_D$ \& $f_{D_s}$ && 0.2207(89) & 1.004(32) \\
&& \\[-2ex]
 \hline
&& \\[-2ex]
$N_f = 2+1$ & $D \to \pi \ell\nu$ and $D\to K \ell\nu$ && 0.2140(97) & 0.975(26) \\
&& \\[-2ex]
 \hline
&& \\[-2ex]
PDG & neutrino scattering & \cite{Agashe:2014kda} & 0.230(11)&  \\
Rosner 15 ({\it for the} PDG) & CKM unitarity & \cite{Rosner:2015wva} & 0.2254(7) & 0.9733(2) \\
&& \\[-2ex]
 \hline \hline 
\end{tabular*}
\caption{Comparison of determinations of $|V_{cd}|$ and $|V_{cs}|$
  obtained from lattice methods with nonlattice determinations and
  the Standard Model prediction assuming CKM
  unitarity.
\label{tab:VcdVcsSummary}}
\end{center}
\end{table}

\clearpage

%% file: HQ/HQB.tex
%=================================================
\section{$B$-meson decay constants, mixing parameters and form factors}
\label{sec:BDecays}
%=================================================

The (semi)leptonic decay and mixing processes of $B_{(s)}$ mesons have been playing
a crucial role in flavour physics.   In particular, they contain
important information for the investigation of the $b{-}d$ unitarity
triangle in the Cabibbo-Kobayashi-Maskawa (CKM) matrix, and can be
ideal probes to physics beyond the Standard Model.
The charged-current decay channels $B^{+} \rightarrow l^{+}
\nu_{l}$ and $B^{0} \rightarrow \pi^{-} l^{+} \nu_{l}$, where $l^{+}$
is a charged lepton with $\nu_{l}$ being the corresponding neutrino, are
essential in extracting the CKM matrix element $|V_{ub}|$.  Similarly,
the $B$ to $D^{(\ast)}$ semileptonic transitions can be used to
determine $|V_{cb}|$.   The flavour changing neutral current (FCNC)
processes, such as $B\to K^{(*)} \ell^+
\ell^-$ and $B_{d(s)} \to \ell^+ \ell^-$,  occur only beyond the tree level in weak interactions and are suppressed in the Standard
Model. Therefore, these processes can be sensitive to new
physics, since heavy particles can contribute to the loop diagrams.
They are also suitable channels for the extraction of the CKM matrix
elements involving the top quark which can appear in the loop. For
instance, the neutral $B_{d(s)}$-meson mixings are FCNC processes and
are dominated by the 1-loop ``box'' diagrams containing the top quark
and the $W$ bosons.  Thus, using the experimentally measured neutral $B^0_{d(s)}$-meson oscillation
frequencies, $\Delta M_{d(s)}$, and the theoretical calculations for
the relevant hadronic mixing matrix elements, one can obtain
$|V_{td}|$ and $|V_{ts}|$ in the Standard Model.\footnote{The neutral
  $B$-meson leptonic
  decays, $B_{d,s} \to \mu^{+} \mu^{-}$, were recently observed at
  the LHC experiments, and the corresponding branching fractions can
  be obtained by combining the data from the CMS and the LHCb
  collaborations~\cite{CMS:2014xfa}.  Nevertheless, the errors of these experimental
results are currently too large to enable a precise determination of
$|V_{td}|$ and $|V_{ts}|$.}

Accommodating the light quarks and the $b$ quark simultaneously in
lattice-QCD computations is a challenging endeavour. To incorporate
the pion and the $b$ hadrons with their physical masses, the simulations have to be performed using the lattice
size $\hat{L} = L/a \sim \cO(10^{2})$, where $a$ is the lattice spacing and $L$
is the physical (dimensionful) box size.   This is a few times larger
than what one can practically afford in contemporary numerical projects.
Therefore, in addition to employing Chiral Perturbation Theory for the extrapolations in the
light-quark mass, current lattice calculations for quantities involving
$b$ hadrons often make use of effective theories that allow one to
expand in inverse powers of $m_{b}$. In this regard, two general
approaches are widely adopted.  On the one hand, effective field theories
such as Heavy-Quark Effective Theory (HQET) and Nonrelativistic
QCD (NRQCD) can be directly implemented in numerical computations. On
the other hand, a relativistic quark action can be improved {\it \'a la}
Symanzik to suppress cutoff errors, and then re-interpreted in a manner
that is suitable for heavy-quark physics calculations.   
This latter strategy is often referred to as the method of the Relativistic
Heavy-Quark Action (RHQA).
The utilization of such effective theories inevitably introduces systematic
uncertainties that are not present in light-quark calculations.  These
uncertainties 
can arise from the truncation of the expansion in constructing the
effective theories (as in HQET and NRQCD),
or from more intricate
cutoff effects (as in NRQCD and RQHA).  They can also be introduced
through more complicated renormalization
procedures which often lead to significant systematic effects in
matching the lattice operators to their continuum counterparts.  For
instance, due to the use of different actions for the heavy and the
light quarks, it is more difficult to construct absolutely 
normalized bottom-light currents.

Complementary to the above ``effective theory approaches'', 
another popular method is to simulate the heavy and the light quarks
using the same (normally improved) lattice action at several values of
the heavy-quark mass, $m_{h}$, with $a m_{h} < 1$ and $m_{h} < m_{b}$.   
This enables one to employ HQET-inspired relations to extrapolate the
computed quantities to the physical $b$ mass.  When combined with
results obtained in the static heavy-quark limit, this approach can be
rendered into an interpolation, instead of extrapolation, in
$m_{h}$. The discretization errors are the main source of the
systematic effects in this method, and very small lattice spacings are
needed to keep such errors under control.

Because of the challenge described above, the efforts that have been
made to obtain reliable, accurate lattice-QCD results for physics of the $b$ quark
have been enormous.   These efforts include significant theoretical progress in
formulating QCD with heavy quarks on the lattice. This aspect is
briefly reviewed in Appendix~\ref{app:HQactions}.

In this section, we summarize the results of the $B$-meson leptonic
decay constants, the neutral $B$-mixing parameters, and the
semileptonic form factors, from lattice QCD.  To be focused on the
calculations which have strong phenomenological impact, we limit the
review to results based on modern simulations containing dynamical
fermions with reasonably light pion masses (below
approximately 500~MeV).  Compared to the progress in the light-quark
sector, heavy-quark physics on the lattice is not as
mature. Consequently, fewer collaborations have finished calculations 
for these quantities.   In addition, the existing results are often
obtained at coarser lattice spacings and heavier pions.
Therefore, for some quantities, there is only a single
lattice calculation that satisfies the criteria to be included in our
average. Nevertheless, several collaborations are currently pursuing
this line of research with various lattice $b$-quark
actions, finer lattice spacings, and lighter pions.  Thus many new
results with controlled errors are expected to appear in the near future.

Following our review of $B_{(s)}$-meson
leptonic decay constants, the neutral $B$-meson mixing parameters, and
semileptonic form factors, we then interpret our results within the
context of the Standard Model.  We combine our best-determined values
of the hadronic matrix elements with the most recent
experimentally-measured branching fractions to obtain $|V_{(u)cb}|$
and compare these results to those obtained from inclusive
semileptonic $B$ decays.

Recent lattice-QCD averages for $B^+$- and $B_s$-meson decay constants
were also presented by the Particle Data Group (PDG) in~Ref.~\cite{Rosner:2015wva}.  The PDG three-
and four-flavour averages 
for these quantities differ from those quoted here because the PDG
provides the charged-meson decay constant, $f_{B^+}$, while we present 
the isospin-averaged meson-decay constant, $f_B$.

\input{HQ/HQSubsections/BL.tex}
\input{HQ/HQSubsections/BB.tex}

\input{HQ/HQSubsections/BSL.tex}

\input{HQ/HQSubsections/BCKM.tex}

%% file: HQ/HQSubsections/BL.tex
\subsection{Leptonic decay constants $f_B$ and $f_{B_s}$}
\label{sec:fB}

The $B$ and $B_s$ meson decay constants are crucial input for
extracting information from leptonic $B$ decays. Charged $B$ mesons
can decay to the lepton-neutrino final state  through the
charged-current weak interaction.  On the other hand, neutral
$B_{d(s)}$ mesons can decay to a charged-lepton pair via a
flavour-changing neutral current (FCNC) process.

In the Standard Model the decay rate for $B^+ \to \ell^+ \nu_{\ell}$
is described by a formula identical to Eq.~(\ref{eq:Dtoellnu}), with $D_{(s)}$ replaced by $B$, and the 
relevant CKM matrix element, $V_{cq}$, substituted by $V_{ub}$,
\be
\Gamma ( B \to \ell \nu_{\ell} ) =  \frac{ m_B}{8 \pi} G_F^2  f_B^2 |V_{ub}|^2 m_{\ell}^2 
           \left(1-\frac{ m_{\ell}^2}{m_B^2} \right)^2 \;. \label{eq:B_leptonic_rate}
\ee
The only charged-current $B$ meson decay that has been observed so far is 
$B^{+} \to \tau^{+} \nu_{\tau}$, which has been measured by the Belle
and Babar collaborations~\cite{Lees:2012ju,Kronenbitter:2015kls}.
Both collaborations have reported results with errors around $20\%$. These measurements can be used to 
determine $|V_{ub}|$ when combined with lattice-QCD predictions of the corresponding
decay constant. 

Neutral $B_{d(s)}$-meson decays to a charged lepton pair, $B_{d(s)}
\rightarrow l^{+} l^{-}$ is a FCNC process, and can only occur at
1-loop in the Standard Model.  Hence these processes are expected to
be rare, and are sensitive to physics beyond the Standard Model.
The corresponding expression for the branching fraction has the  form 
\be
B(B_q \to \ell^+ \ell^-) = \frac{\tau_{B_q}}{1+y_q} \frac{G_F^2 \alpha^2 }{ 16\pi^3} m_{B_q} f_{B_q}^2 \vert V_{tb}^\ast V_{tq}\vert^2 m_\ell^2 C_{10}^{\rm SM} \sqrt{1 - \frac{4m_\ell^2}{m_{B_q}^2}} , 
\ee
where the light-quark $q=s$ or $d$, and 
the coefficient $C_{10}^{\rm SM}$ includes the NLO electro-weak and NNLO QCD
matching corrections~\cite{Bobeth:2013uxa}. The factor $1/(1+y_q)$,
with $y_q=\Delta \Gamma_{B_q}/(2 \Gamma_{B_q})$, accounts for the fact
that the measured branching fraction corresponds to a time-integrated
rate of the oscillating $B_q$ system to
$\ell^+\ell^-$~\cite{DeBruyn:2012wj}. That correction is particularly
important for the $B_s$ decays because of the relatively large
$y_s=0.06(1)$~\cite{Amhis:2014hma, Aaij:2014zsa}.
Evidence for both
$B_s \to \mu^+ \mu^-$ and $B_s \to \mu^+ \mu^-$ decays was recently observed
by the CMS and the LHCb collaborations.
Combining the data from both collaborations, the branching fractions
can be extracted to be~\cite{CMS:2014xfa},
\begin{eqnarray} 
   B(B_d \to \mu^+ \mu^-) &=& (3.9^{+1.6}_{-1.4}) \,10^{-10} , \nonumber\\
   B(B_s \to \mu^+ \mu^-) &=& (2.8^{+0.7}_{-0.6}) \,10^{-9} ,
\label{eq:B_to_mumu_CMS_LHCb_2014}
\end{eqnarray}
which are compatible with the Standard Model predictions at the $2.2\sigma$
and $1.2\sigma$ level, respectively.

The decay constants $f_{B_q}$ (with $q=u,d,s$) parameterize the matrix
elements of the corresponding axial-vector currents, $A^{\mu}_{bq}
= \bar{b}\gamma^{\mu}\gamma^5q$, analogously to the definition of
$f_{D_q}$ in Sec.~\ref{sec:fD}:
\be
\langle 0| A^{\mu} | B_q(p) \rangle = i p_B^{\mu} f_{B_q} \;.
\label{eq:fB_from_ME}
\ee
For heavy-light mesons, it is convenient to define and analyze the quantity 
\be
 \Phi_{B_q} \equiv f_{B_q} \sqrt{m_{B_q}} \;,
\ee
which approaches a constant (up to logarithmic corrections) in the
$m_B \to \infty$ limit according to HQET.
In the following discussion we denote lattice data for $\Phi$($f$)
obtained at a heavy-quark mass $m_h$ and light valence-quark mass
$m_{\ell}$ as $\Phi_{h\ell}$($f_{hl}$), to differentiate them from
the corresponding quantities at the physical $b$ and light-quark
masses.

The $SU(3)$-breaking ratio, $f_{B_s}/f_B$, is of 
interest. This is because in lattice-QCD calculations for this
quantity,  many systematic effects can be partially reduced.  
These include discretization errors, heavy-quark mass
tuning effects, and renormalization/matching errors, amongst others. 
On the other hand, 
this $SU(3)$-breaking ratio is still sensitive to the chiral
extrapolation. Given that the chiral extrapolation is under control,
one can then adopt $f_{B_s}/f_B$ as input in extracting
phenomenologically-interesting quantities. For instance, this ratio
can be used to determine $|V_{ts}/V_{td}|$. In addition, it often
happens to be easier to obtain lattice results for $f_{B_{s}}$ with
smaller errors. Therefore, one can combine the $B_{s}$-meson
decay constant with the $SU(3)$-breaking ratio to calculate $f_{B}$. Such
strategy can lead to better precision in the computation of the
$B$-meson decay constant, and has been adopted by the ETM~\cite{Carrasco:2013zta} and the
HPQCD collaborations~\cite{Na:2012sp}.  

It is clear that the decay constants for charged and neutral $B$
mesons play different roles in flavour physics phenomenology.  As
already mentioned above, the knowledge of the $B^{+}$-meson decay constant,
$f_{B^{+}}$, is essential for extracting $|V_{ub}|$ from
leptonic $B^{+}$ decays. The neutral $B$-meson decay constants,
$f_{B^{0}}$ and $f_{B_{s}}$, are inputs for obtaining $|V_{td}|$
using information from the $B$-meson mixing processes. In
view of this, it is desirable to include isospin-breaking effects in
lattice computations for these quantities, and have results for
$f_{B^{+}}$ and $f_{B^{0}}$.  Nevertheless, as will be discussed in
detail in this section, such effects are small compared to the current
errors of the decay constants calculated using lattice QCD.  In this review, we
will then concentrate on the isospin-averaged result, $f_{B}$, and the
$B_{s}$-meson decay constant, as well as the $SU(3)$-breaking ratio,
$f_{B_{s}}/f_{B}$.   For the world average for the lattice
determination of $f_{B^{+}}$ and $f_{B_{s}}/f_{B^{+}}$,
we refer the reader to the latest work from the Particle Data
Group (PDG)~\cite{Rosner:2015wva}.   Notice that the lattice results used in
Ref.~\cite{Rosner:2015wva} and the current review are identical.  We
will discuss this in further detail at the end of this subsection.

\begin{figure}[!htb]
\hspace{-0.8cm}\includegraphics[width=0.58\linewidth]{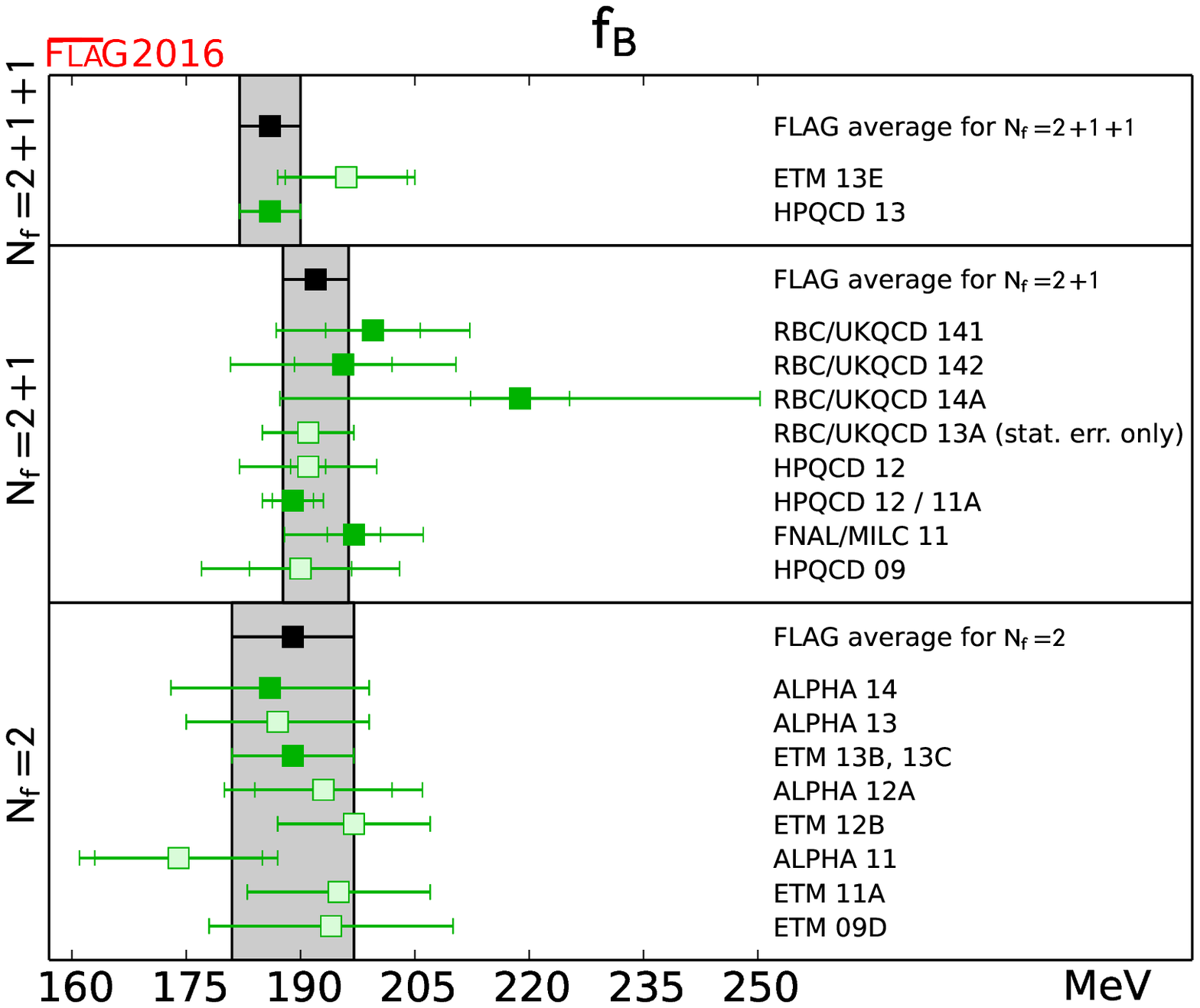}\hspace{-0.95cm}
\includegraphics[width=0.58\linewidth]{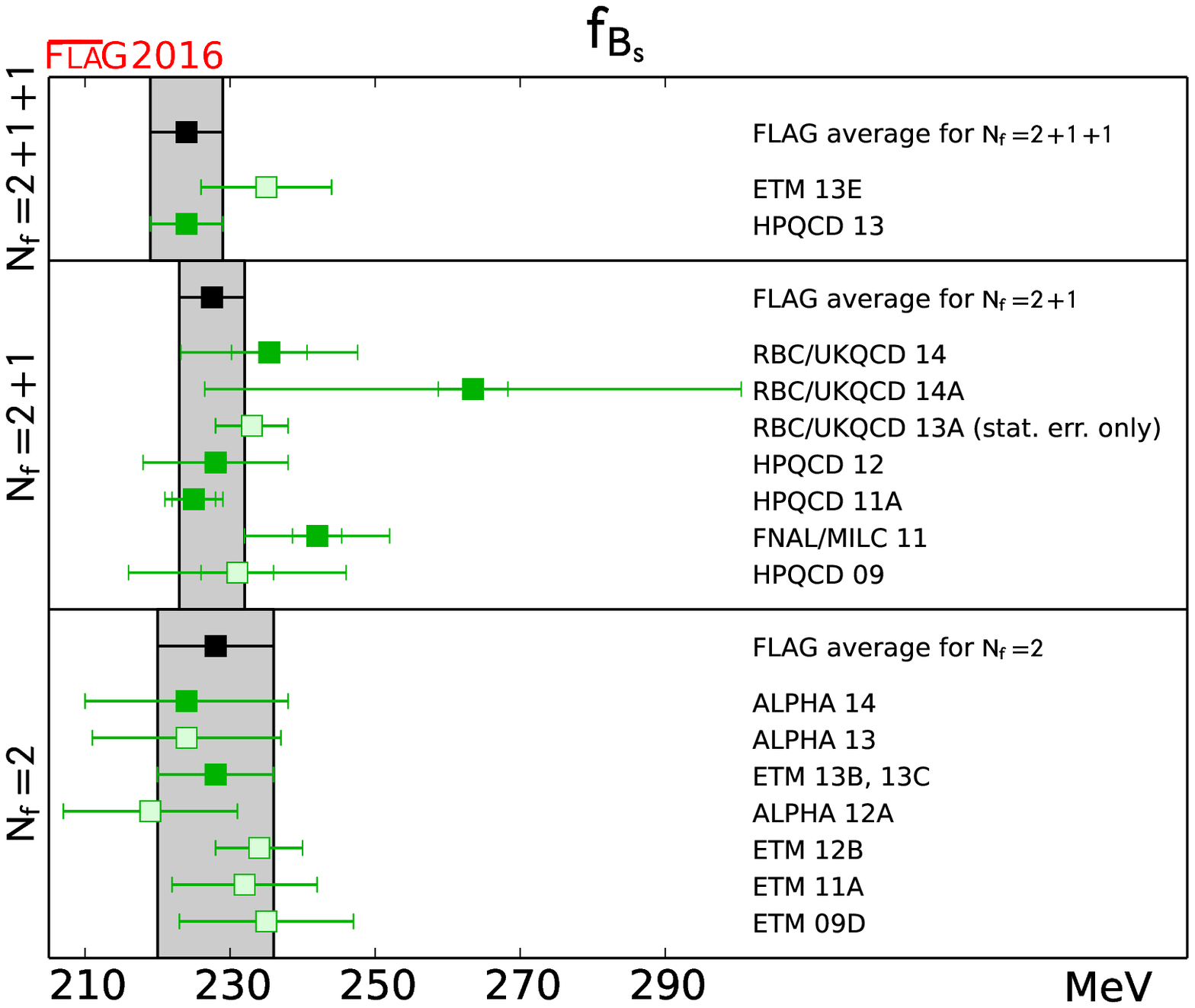}
 \vspace{-2mm}
\caption{Decay constants of the $B$ and $B_s$ mesons. The values are taken from Tab.~\ref{tab:FBssumm} 
(the $f_B$ entry for FNAL/MILC 11 represents $f_{B^+}$). The
significance of the colours is explained in
Sec.~2.
The black squares and grey bands indicate
our averages in Eqs.~(\ref{eq:fbav2}), (\ref{eq:fbav21}) and
(\ref{eq:fbav211}).}
\label{fig:fB}
\end{figure}

\begin{figure}[!htb]
\begin{center}
\includegraphics[width=0.58\linewidth]{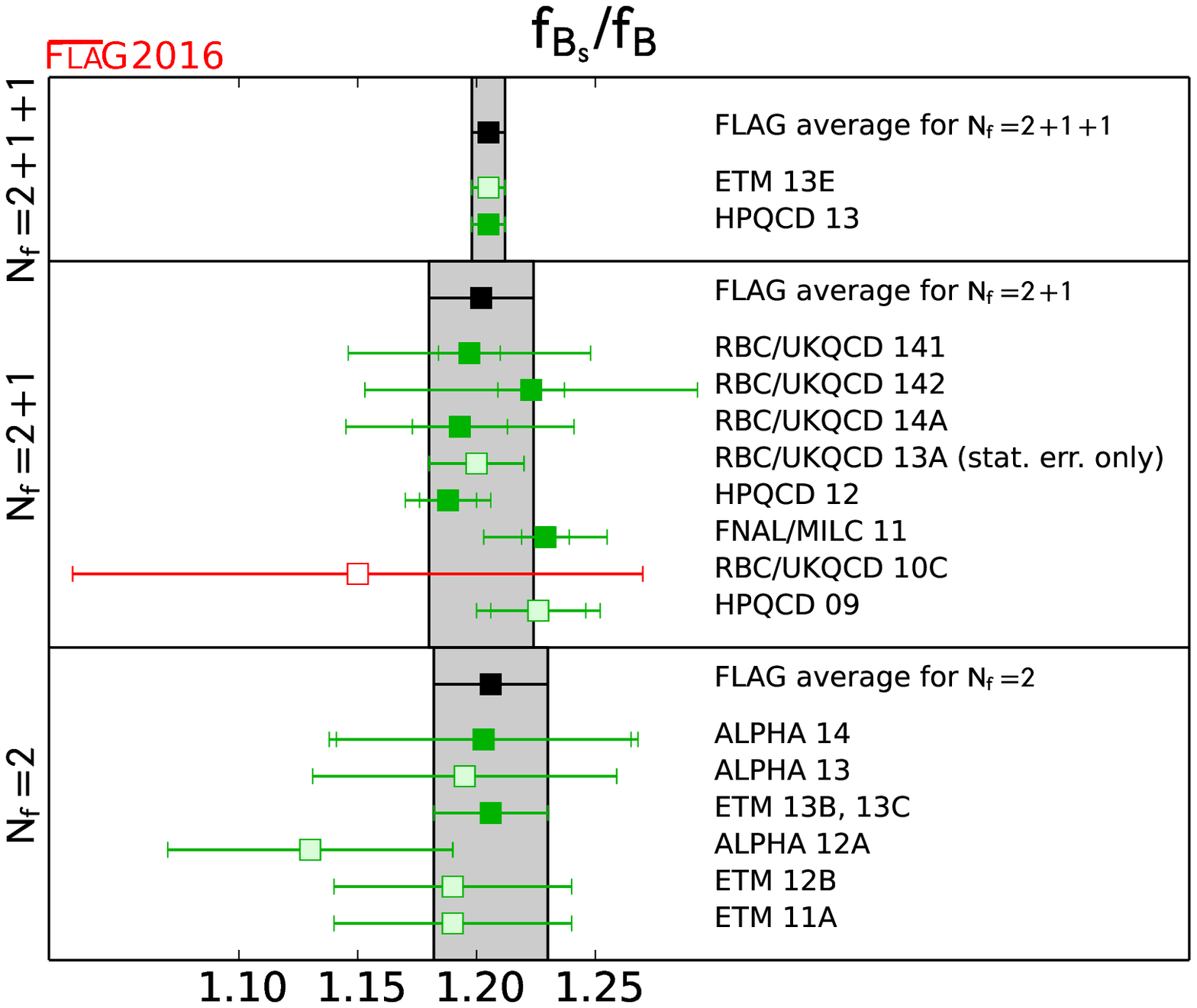}
\vspace{-2mm}
\caption{Ratio of the decay constants of the $B$ and $B_s$ mesons. The values are taken from Tab.~\ref{tab:FBratsumm} 
(the $f_B$ entry for FNAL/MILC 11 represents $f_{B^+}$). The
significance of the colours is explained in
Sec.~2.
The black squares and grey bands indicate
our averages in Eqs.~(\ref{eq:fbav2}), (\ref{eq:fbav21}) and
(\ref{eq:fbav211}).}
\label{fig:fBratio}
\end{center}
\end{figure}

The status of lattice-QCD computations for $B$-meson decay constants
and the $SU(3)$-breaking ratio, using gauge-field ensembles
with light dynamical fermions, is summarized in Tabs.~\ref{tab:FBssumm}
and~\ref{tab:FBratsumm}.  Figs.~\ref{fig:fB} and~\ref{fig:fBratio} contain the graphic
presentation of the collected results and our averages.  Many results
in these tables and plots were
already reviewed in detail in the previous FLAG
report~\cite{Aoki:2013ldr}.  Below we will describe the new results
that appeared after December 2013.  In addition, we will comment on
our updated strategies in performing the averaging.
\begin{table}[!htb]
\mbox{} \\[3.0cm]
\footnotesize
\begin{tabular*}{\textwidth}[l]{@{\extracolsep{\fill}}l@{\hspace{1mm}}r@{\hspace{1mm}}l@{\hspace{1mm}}l@{\hspace{1mm}}l@{\hspace{1mm}}l@{\hspace{1mm}}l@{\hspace{1mm}}l@{\hspace{1mm}}l@{\hspace{5mm}}l@{\hspace{1mm}}l@{\hspace{1mm}}l@{\hspace{1mm}}l@{\hspace{1mm}}l}
Collaboration & Ref. & $\Nf$ & 
\hspace{0.15cm}\begin{rotate}{60}{publication status}\end{rotate}\hspace{-0.15cm} &
\hspace{0.15cm}\begin{rotate}{60}{continuum extrapolation}\end{rotate}\hspace{-0.15cm} &
\hspace{0.15cm}\begin{rotate}{60}{chiral extrapolation}\end{rotate}\hspace{-0.15cm}&
\hspace{0.15cm}\begin{rotate}{60}{finite volume}\end{rotate}\hspace{-0.15cm}&
\hspace{0.15cm}\begin{rotate}{60}{renormalization/matching}\end{rotate}\hspace{-0.15cm}  &
\hspace{0.15cm}\begin{rotate}{60}{heavy-quark treatment}\end{rotate}\hspace{-0.15cm} & 
 $f_{B^+}$ & $f_{B^0}$   & $f_{B}$ & $f_{B_s}$  \\
&&&&&&&&&&&&\\[-0.1cm]
\hline
\hline
&&&&&&&&&&&& \\[-0.1cm]
ETM 13E & \cite{Carrasco:2013naa} & 2+1+1 & \rC & \soso & \soso & \soso 
& \soso &  \okay &  $-$ & $-$ & 196(9) & 235(9) \\[0.5ex]

HPQCD 13 & \cite{Dowdall:2013tga} & 2+1+1 & \gA & \good & \good & \good & \soso
& \okay &  184(4) & 188(4) &186(4) & 224(5)  \\[0.5ex]

&&&&&&&&&& \\[-0.1cm]
\hline
&&&&&&&&&& \\[-0.1cm]

RBC/UKQCD 14 & \cite{Christ:2014uea} & 2+1 & \gA & \soso & \soso & \soso 
  & \soso & \okay & 195.6(14.9) & 199.5(12.6) & $-$ & 235.4(12.2) \\[0.5ex]

RBC/UKQCD 14A & \cite{Aoki:2014nga} & 2+1 & \gA & \soso & \soso & \soso 
  & \soso & \okay & $-$ & $-$ & 219(31) & 264(37) \\[0.5ex]

RBC/UKQCD 13A & \cite{Witzel:2013sla} & 2+1 & \rC & \soso & \soso & \soso 
  & \soso & \okay & $-$ & $-$ &  191(6)$_{\rm stat}^\diamond$ & 233(5)$_{\rm stat}^\diamond$ \\[0.5ex]

HPQCD 12 & \cite{Na:2012sp} & 2+1 & \gA & \soso & \soso & \soso & \soso
& \okay & $-$ & $-$ & 191(9) & 228(10)  \\[0.5ex]

HPQCD 12 & \cite{Na:2012sp} & 2+1 & \gA & \soso & \soso & \soso & \soso
& \okay & $-$ & $-$ & 189(4)$^\triangle$ &  $-$  \\[0.5ex]

HPQCD 11A & \cite{McNeile:2011ng} & 2+1 & \gA & \good & \soso &
 \good & \good & \okay & $-$ & $-$ & $-$ & 225(4)$^\nabla$ \\[0.5ex] 

FNAL/MILC 11 & \cite{Bazavov:2011aa} & 2+1 & \gA & \soso & \soso &
     \good & \soso & \okay & 197(9) & $-$ & $-$ & 242(10) &  \\[0.5ex]  

HPQCD 09 & \cite{Gamiz:2009ku} & 2+1 & \gA & \soso & \soso & \soso &
\soso & \okay & $-$ & $-$ & 190(13)$^\bullet$ & 231(15)$^\bullet$  \\[0.5ex] 

&&&&&&&&&& \\[-0.1cm]
\hline
&&&&&&&&&& \\[-0.1cm]

ALPHA 14 & \cite{Bernardoni:2014fva} & 2 & \gA & \good & \good &\good 
& \good & \okay &  $-$ & $-$ & 186(13) & 224(14) \\[0.5ex]

ALPHA 13 & \cite{Bernardoni:2013oda} & 2 & \rC  & \good   & \good   &
\good    &\good  & \okay   & $-$ & $-$ & 187(12)(2) &  224(13) &  \\[0.5ex] 

ETM 13B, 13C$^\dagger$ & \cite{Carrasco:2013zta,Carrasco:2013iba} & 2 & \gA & \good & \soso & \good
& \soso &  \okay &  $-$ & $-$ & 189(8) & 228(8) \\[0.5ex]

ALPHA 12A& \cite{Bernardoni:2012ti} & 2 & \rC  & \good      & \good      &
\good          &\good  & \okay   & $-$ & $-$ & 193(9)(4) &  219(12) &  \\[0.5ex] 

ETM 12B & \cite{Carrasco:2012de} & 2 & \rC & \good & \soso & \good
& \soso &  \okay &  $-$ & $-$ & 197(10) & 234(6) \\[0.5ex]

ALPHA 11& \cite{Blossier:2011dk} & 2 & \rC  & \good      & \soso      &
\good          &\good  & \okay  & $-$ & $-$ & 174(11)(2) &  $-$ &  \\[0.5ex]  

ETM 11A & \cite{Dimopoulos:2011gx} & 2 & \gA & \soso & \soso & \good
& \soso &  \okay & $-$ & $-$ & 195(12) & 232(10) \\[0.5ex]

ETM 09D & \cite{Blossier:2009hg} & 2 & \gA & \soso & \soso & \soso
& \soso &  \okay & $-$ & $-$ & 194(16) & 235(12) \\[0.5ex]
&&&&&&&&&& \\[-0.1cm]
\hline
\hline\\
\end{tabular*}\\[-0.2cm]
\begin{minipage}{\linewidth}
{\footnotesize 
\begin{itemize}
   \item[$^\diamond$] Statistical errors only. \\[-5mm]
   \item[$^\triangle$] Obtained by combining $f_{B_s}$ from HPQCD~11A with $f_{B_s}/f_B$ calculated in this work.\\[-5mm]
   \item[$^\nabla$] This result uses one ensemble per lattice spacing with light to strange sea-quark mass 
        ratio $m_{\ell}/m_s \approx 0.2$. \\[-5mm]
   \item[$^\bullet$] This result uses an old determination of  $r_1=0.321(5)$ fm from Ref.~\cite{Gray:2005ur} that 
        has since been superseded. \\[-5mm]
   \item[$^\dagger$] Update of ETM~11A and 12B. 
\end{itemize}
}
\end{minipage}
\caption{Decay constants of the $B$, $B^+$, $B^0$ and $B_{s}$ mesons
  (in MeV). Here $f_B$ stands for the mean value of $f_{B^+}$ and
  $f_{B^0}$, extrapolated (or interpolated) in the mass of the light
  valence-quark to the physical value of $m_{ud}$.}
\label{tab:FBssumm}
\end{table}

\begin{table}[!htb]
\begin{center}
\mbox{} \\[3.0cm]
\footnotesize
\begin{tabular*}{\textwidth}[l]{@{\extracolsep{\fill}}l@{\hspace{1mm}}r@{\hspace{1mm}}l@{\hspace{1mm}}l@{\hspace{1mm}}l@{\hspace{1mm}}l@{\hspace{1mm}}l@{\hspace{1mm}}l@{\hspace{1mm}}l@{\hspace{5mm}}l@{\hspace{1mm}}l@{\hspace{1mm}}l@{\hspace{1mm}}l}
Collaboration & Ref. & $\Nf$ & 
\hspace{0.15cm}\begin{rotate}{60}{publication status}\end{rotate}\hspace{-0.15cm} &
\hspace{0.15cm}\begin{rotate}{60}{continuum extrapolation}\end{rotate}\hspace{-0.15cm} &
\hspace{0.15cm}\begin{rotate}{60}{chiral extrapolation}\end{rotate}\hspace{-0.15cm}&
\hspace{0.15cm}\begin{rotate}{60}{finite volume}\end{rotate}\hspace{-0.15cm}&
\hspace{0.15cm}\begin{rotate}{60}{renormalization/matching}\end{rotate}\hspace{-0.15cm}  &
\hspace{0.15cm}\begin{rotate}{60}{heavy-quark treatment}\end{rotate}\hspace{-0.15cm} & 
 $f_{B_s}/f_{B^+}$  & $f_{B_s}/f_{B^0}$  & $f_{B_s}/f_{B}$  \\
&&&&&&&&&& \\[-0.1cm]
\hline
\hline
&&&&&&&&&& \\[-0.1cm]

ETM 13E & \cite{Carrasco:2013naa} & 2+1+1 & \rC & \good & \soso & \soso
& \soso &  \okay &  $-$ & $-$ & 1.201(25) \\[0.5ex]

HPQCD 13 & \cite{Dowdall:2013tga} & 2+1+1 & \gA & \good & \good & \good & \soso
& \okay & 1.217(8) & 1.194(7) & 1.205(7)  \\[0.5ex]

&&&&&&&&&& \\[-0.1cm]
\hline
&&&&&&&&&& \\[-0.1cm]

RBC/UKQCD 14 & \cite{Christ:2014uea} & 2+1 & \gA & \soso & \soso & \soso 
  & \soso & \okay & 1.223(71) & 1.197(50) & $-$ \\[0.5ex]

RBC/UKQCD 14A & \cite{Aoki:2014nga} & 2+1 & \gA & \soso & \soso & \soso 
  & \soso & \okay & $-$ & $-$ & 1.193(48) \\[0.5ex]

RBC/UKQCD 13A & \cite{Witzel:2013sla} & 2+1 & \rC & \soso & \soso & \soso 
  & \soso & \okay & $-$ & $-$ &  1.20(2)$_{\rm stat}^\diamond$ \\[0.5ex]

HPQCD 12 & \cite{Na:2012sp} & 2+1 & \gA & \soso & \soso & \soso & \soso
& \okay & $-$ & $-$ & 1.188(18) \\[0.5ex]

FNAL/MILC 11 & \cite{Bazavov:2011aa} & 2+1 & \gA & \soso & \soso &
     \good& \soso & \okay & 1.229(26) & $-$ & $-$ \\[0.5ex]  
     
RBC/UKQCD 10C & \cite{Albertus:2010nm} & 2+1 & \gA & \tbr & \tbr & \tbr 
  & \soso & \okay & $-$ & $-$ & 1.15(12) \\[0.5ex]

HPQCD 09 & \cite{Gamiz:2009ku} & 2+1 & \gA & \soso & \soso & \soso &
\soso & \okay & $-$ & $-$ & 1.226(26)  \\[0.5ex] 

&&&&&&&&&& \\[-0.1cm]
\hline
&&&&&&&&&& \\[-0.1cm]
ALPHA 14 \al \cite{Bernardoni:2014fva} & 2 & \gA & \good & \good & \good 
& \good &  \okay &  $-$ \al $-$ & 1.203(65)\\[0.5ex]

ALPHA 13 & \cite{Bernardoni:2013oda} & 2 & \rC  & \good  & \good  &
\good   &\good  & \okay   & $-$ & $-$ & 1.195(61)(20)  &  \\[0.5ex] 

ETM 13B, 13C$^\dagger$ & \cite{Carrasco:2013zta,Carrasco:2013iba} & 2 & \gA & \good & \soso & \good
& \soso &  \okay &  $-$ & $-$ & 1.206(24)  \\[0.5ex]

ALPHA 12A & \cite{Bernardoni:2012ti} & 2 & \rC & \good & \good & \good
& \good &  \okay & $-$ & $-$ & 1.13(6)  \\ [0.5ex]

ETM 12B & \cite{Carrasco:2012de} & 2 & \rC & \good & \soso & \good
& \soso &  \okay & $-$ & $-$ & 1.19(5) \\ [0.5ex]

ETM 11A & \cite{Dimopoulos:2011gx} & 2 & \gA & \soso & \soso & \good
& \soso &  \okay & $-$ & $-$ & 1.19(5) \\ [0.5ex]
&&&&&&&&&& \\[-0.1cm]
\hline
\hline\\
\end{tabular*}\\[-0.2cm]
\begin{minipage}{\linewidth}
{\footnotesize 
\begin{itemize}
   \item[$^\diamond$] Statistical errors only. \\[-5mm]
   \item[$^\dagger$] Update of ETM 11A and 12B. 
\end{itemize}
}
\end{minipage}
\caption{Ratios of decay constants of the $B$ and $B_s$ mesons (for details see Tab.~\ref{tab:FBssumm}).}
\label{tab:FBratsumm}
\end{center}
\end{table}
Only one new $N_{f}=2$ project for computing $f_{B}$, $f_{B_{s}}$ and
$f_{B_{s}}/f_{B}$ was completed after the publication of the previous
FLAG review.  
This was carried out by the ALPHA
collaboration~\cite{Bernardoni:2014fva} (ALPHA~14 in Tabs.~\ref{tab:FBssumm}
and~\ref{tab:FBratsumm}), on the CLS (Coordinated
Lattice Simulations) gauge-field ensembles which were generated using
the Wilson plaquette action and $N_{f} = 2$ non-perturbatively
$\cO(a)$-improved 
Wilson fermions with the
DD-HMC~\cite{Luscher:2003qa,Luscher:2005rx,Luscher:2007es} or the
MP-HMC~\cite{Marinkovic:2010eg} algorithm.  There are
three choices of lattice spacing, 0.048, 0.065 and 0.075 fm, in
these ensembles.   At each lattice spacing, three to four lattice
sizes are adopted in the simulations.  The hyper-cubic boxes are of
the shape $L^{3}\times T$, with the temporal extent $T=2L$.  The
smallest box used in ALPHA~14 is $L \approx 2$ fm.
On each of these lattice sizes,
one sea-quark mass is employed in the computation, and the condition
$M_{\pi} L > 4$ is always ensured.   This leads to subpercentage-level
finite-size effects~\cite{Arndt:2004bg}.
The corresponding lightest pions composed of the
sea quarks for these three values of the lattice spacing are 270, 190,
and 280 MeV, respectively.    In this work, the
lattice-regularized HQET action and the axial current to the order of $1/m_{B}$, as tuned in 
Refs.~\cite{Heitger:2003nj,DellaMorte:2006cb,Blossier:2010jk,Blossier:2012qu,Bernardoni:2013xba}
with non-perturbative matching to QCD, are used
to compute the heavy-light meson decay constant.    This matching
procedure removes both the logarithmic and the power divergences in
the effective theory regularized on the lattice.   The valence light (up and
down) quarks are implemented with the unitary setup, such that
the valence and the sea pions have identical masses.   On the
other hand, the valence strange-quark mass is tuned on the CLS
gauge-field ensembles employing the kaon decay
constant~\cite{Fritzsch:2012wq}.    The static-light axial current in
this work is also $\cO(a)$-improved to 1-loop order.
Using the lattice data, the
ground-state contributions to the relevant correlators are obtained
through the method of the generalized eigenvalue problem (GEVP), 
as detailed in Ref.~\cite{Blossier:2009kd}.  With this GEVP approach in
ALPHA 14, the systematic errors arising from the excited-state contamination are
typically less than one third of the statistical errors in the
extracted decay constants.  Combined chiral-continuum extrapolations,
adopting the NLO HM$\chi$PT predictions, 
are then performed to determine the decay constants in the limit of
physical pion mass and vanishing lattice spacing.   The errors of the
final results in ALPHA~14 include statistical uncertainties, the
discrepancy to the static-limit results, the effects of the
lattice spacing, the uncertainties from the HQET parameters in the
matching procedure, and the systematic effects in the chiral
extrapolations as estimated by comparing with fits to formulae without
the chiral logarithms.  
Since the fits to
the predictions of finite-volume HM$\chi$PT~\cite{Arndt:2004bg} have not been implemented,
systematic effects resulting from the finite lattice size are not
included in the analysis.   Nevertheless, given that the condition $M_{\pi} L > 4$
is always satisfied in ALPHA~14, these effects should be at the
subpercentage level according to the 1-loop formulae in Ref.~\cite{Arndt:2004bg}.

The new result, ALPHA~14, satisfies all our criteria for being
included in the averaging process.  Therefore, in the current edition of the FLAG
report, two $N_{f}=2$ calculations for the $B$-meson decay constants
and the $SU(3)$-breaking ratio contribute to our averages.  The other
determination of these quantities (ETM~13B, 13C in Tabs.~\ref{tab:FBssumm}
and~\ref{tab:FBratsumm}) was
already reviewed in detail in the previous FLAG publication.  These two
projects are based on completely different lattice simulations, and
there is no correlation between the errors quoted in them.   This
gives our estimate,
%
%FLAGRESULT BEGIN
% TAG      & fB    & fBs	&fBsofB &END
% REFS     & \cite{Bernardoni:2014fva,Carrasco:2013zta,Carrasco:2013iba} &\cite{Bernardoni:2014fva,Carrasco:2013zta,Carrasco:2013iba} &\cite{Bernardoni:2014fva,Carrasco:2013zta,Carrasco:2013iba} &END
% UNITS    & '[MeV]' & '[MeV]' & 1 &END
% NUMRESULTS & 2 & 2 & 2 &END
% FLAVOURs & 2 & 2 & 2 &END
%FLAGRESULT END
%FLAGRESULTFORMULA BEGIN
\begin{align}
      &&\FLAGAVBEGIN f_B&=188(7)\FLAGAVEND \;{\rm MeV} 	        &&\Refs~\mbox{\cite{Bernardoni:2014fva,Carrasco:2013zta,Carrasco:2013iba}},\nonumber\\
&N_f=2:&\FLAGAVBEGIN f_{B_s}&= 227(7)\FLAGAVEND \;{\rm MeV}     &&\Refs~\mbox{\cite{Bernardoni:2014fva,Carrasco:2013zta,Carrasco:2013iba}},\label{eq:fbav2}\\
      &&\FLAGAVBEGIN {{f_{B_s}}/{f_B}}&=1.206(23)\FLAGAVEND &&\Refs~\mbox{\cite{Bernardoni:2014fva,Carrasco:2013zta,Carrasco:2013iba}}.\nonumber
\end{align}
%FLAGRESULTFORMULA END
%
Two groups of authors (RBC/UKQCD~14~\cite{Christ:2014uea} and RBC/UKQCD~14A~\cite{Aoki:2014nga}  in Tabs.~\ref{tab:FBssumm}
and~\ref{tab:FBratsumm}) presented their $N_{f} = 2+1$ results for
$f_{B}$, $f_{B_{s}}$ and $f_{B_{s}}/f_{B}$ after the publication of
the previous FLAG report in 2013.   Both groups belong to
the RBC/UKQCD collaboration.  They use the same gauge-field
ensembles generated by this collaboration, with the Iwasaki gauge
action and domain-wall dynamical quarks~\cite{Aoki:2010dy}, adopting
the ``RHMC II'' algorithm~\cite{Allton:2008pn}.  Two values of the lattice
spacing, 0.11 and 0.086 fm, are used in the simulations, with the
corresponding lattice sizes being $24^{3}\times 64$ and $32^{3}\times
64$, respectively.  This fixes the spatial size $L \approx 2.7$ fm in all
the data sets.  For the coarse lattice, two choices of the sea-quark
masses, with $M_{\pi} \approx 328$ and 420 MeV, are implemented in the
simulations.  On the other hand, three values of the sea-quark masses
($M_{\pi} \approx 289$, 344, 394 MeV) are used on the fine lattice.
This makes certain that $M_{\pi} L > 4$ is always satisfied.  
At each value of the lattice spacing, only one sea strange-quark mass
is implemented, which is about $10\%$ higher than its physical value. 

In RBC/UKQCD 14, the heavy-quark is described by the relativistic
lattice action proposed in Ref.~\cite{Christ:2006us}.    The three
parameters of this 
relativistic heavy-quark (RHQ) action are tuned non-perturbatively in 
Ref. ~\cite{Aoki:2012xaa}  by requiring that the spin-averaged
$B_{s}$-meson mass, $\overline{M}_{B_{s}} = (M_{B_s} + 3M_{B^{\ast}_{s}} )/4$, and the hyperfine
splitting, $\Delta_{M_{B_s}} = M_{B_{s}^{\ast}} - M_{B_s}$ equal the PDG values, and
that the lattice rest and kinetic meson masses are equal.  Statistical 
uncertainties in the tuned parameters are propagated to the decay 
constants via jackknife resampling.  Simulations with different
values of the RHQ parameters are used to estimate the remaining
uncertainties in the decay constants from the tuning procedure.
Regarding
valence light- and strange-quarks, the authors of RBC/UKQCD~14 adopt
exactly the same domain-wall discretization as that in the sea-quark
sector.  For each lattice spacing, such valence domain-wall fermion
propagators at six choices of the mass parameter are generated.  These
six values straddle between the lightest and strange sea-quark
masses in the gauge-field ensembles, and several of them correspond to
the unitary points.  With the above lattice setting, the heavy-meson decay constants are
obtained, employing an axial current that is $\cO(a)$-improved  to 1-loop level.
The renormalization of the axial current is carried out
with a mostly nonperturbative procedure proposed in
Ref.~\cite{ElKhadra:2001rv}.    Linear interpolations for the
heavy-quark action parameters, as well as the valence strange-quark
mass are then performed on these heavy-meson decay constants.  As for
the chiral extrapolation for the light-quark mass, it is implemented
together with the continuum extrapolation (linear in $a^{2}$)
adopting $SU(2)$-HM$\chi$PT at NLO.\footnote{The authors of RBC/UKQCD~14
claim that using the NLO $SU(3)$-HM$\chi$PT extrapolation
formulae, acceptable fits for the decay constants can be found.  On
the other hand, no reasonable fit for the ratio, $f_{B_{s}}/f_{B}$,
can result from this procedure, because this ratio has
smaller statistical errors.  The NLO $SU(3)$-HM$\chi$PT predictions
are then used as a means to estimate the systematic effects arising from the
chiral-continuum extrapolation.} The decay constants, $f_{B^{+}}$ and
$f_{B^{0}}$, are determined by chirally extrapolating to the physical
$u$- and $d$-quark masses, respectively, and their isospin-averaged
counterpart, $f_{B}$, is not reported.  
Notice that only the unitary points in
the light-quark mass are used in the central procedure for the chiral extrapolation.  This
extrapolation serves as the method to confirm that finite-size effects
are at the subpercentage level by
comparing with the prediction of finite-volume HM$\chi$PT~\cite{Arndt:2004bg}.   
Furthermore, since there is no observed sea-quark dependence in
$f_{B_{s}}$, it is extrapolated to the continuum limit straight after
the interpolation of the valence strange-quark mass.   The authors of
RBC/UKQCD~14 provided a comprehensive list of systematic errors in their
work.  The dominant effect is from the chiral-continuum
extrapolation.  This was investigated using several alternative
procedures by varying the fit ans\"{a}tze and omitting the data points
at the heaviest pion mass.   The error arising from the continuum
extrapolation of $f_{B_{s}}$ is estimated by taking the result on the
finer lattice as the alternative.  One other important source of the
systematic errors is the heavy-quark discretization effect, which is
estimated using a power-counting argument in the improvement programme.

In the other newly completed $B$-meson decay constants project,
RBC/UKQCD~14A, the static heavy-quark action is implemented with the
HYP smearing~\cite{Hasenfratz:2001hp} that reduces the power
divergences.   As for the valence light- and strange-quarks, the same domain-wall
discretization as adopted for the sea quarks is used.  The masses of
the valence light quarks are chosen to be at the unitary points.  On
the other hand,  for each lattice spacing, two values of the valence
strange-quark mass are utilized, with one of them identical to that of
its sea-quark counterpart, and the other slightly smaller than the physical
strange-quark mass.  Employing the propagators of these valence quarks
computed on the RBC/UKQCD gauge-field ensembles, the relevant matrix elements of
the axial current are calculated to extract the decay constant.  
Notice that the source and sink smearings are applied on the valence
light- and strange-quark propagators, in order to obtain better overlap
with the ground state.  The axial current is $\cO(a)$-improved to
1-loop order, and its
renormalization/matching is performed in a two-step
fashion.  Namely, it is first matched from the lattice-regularized
HQET to the same effective theory in the continuum at the inverse
lattice spacing, $a^{-1}$, and then matched to QCD at the physical
$b$-quark mass, $m_{b}$.  At each of these two steps, the matching is
carried out at 1-loop level, and the 2-loop running between $a^{-1}$ and
$m_{b}$ is implemented accordingly.   Regarding the extrapolation to
the physical light-quark mass, it is achieved using $SU(2)$-HM$\chi$PT,
after linearly interpolating the decay constants to the physical
strange-quark mass in the valence sector.  Unlike RBC/UKQCD~14,
here the isospin-averaged $f_{B}$, instead of the individual
$f_{B^{+}}$ and $f_{B^{0}}$, is reported in RBC/UKQCD~14A.
This chiral fit is combined with the continuum extrapolation by
including a term proportional to $a^{2}$ in the HM$\chi$PT formulae.
In addition, finite-size effects are also estimated by replacing the 
1-loop integrals with sums in HM$\chi$PT~\cite{Arndt:2004bg}.  The
predominant systematic error in $f_{B_{s}}$ and $f_{B}$ is from the
1-loop renormalization/matching procedure.  This error is
accounted for by employing a power-counting method, and is evaluated
to be around $6\%$.  Obviously, it is small for $f_{B_{s}}/f_{B}$.
Another significant systematic effect (about $2\sim 3\%$ in all relevant
quantities) results from the chiral-continuum extrapolation.  This
effect is estimated by omitting the chiral logarithms in the fitting procedure.
Finally, based upon a power-counting argument, the authors of RBC/UKQCD~14A
include a $10\%$ error on $f_{B_{(s)}}$, and a $2.2\%$ error on $f_{B_{s}}/f_{B}$,
to account for the use of the static heavy quarks in their work.

Both new computations from the RBC/UKQCD collaboration satisfy the criteria
for being considered in our averages of the relevant quantities.
Since they are based on exactly the same gauge-field configurations,
we treat the statistical errors in these two results as $100\%$
correlated.   It also has to be pointed out that only $f_{B^{+}}$ and
$f_{B^{0}}$ are reported in RBC/UKQCD~14, while we are
concentrating on the isospin-averaged $f_{B}$ in our current work.
For this purpose, we regard both $f_{B^{+}}$ and $f_{B^{0}}$ in
RBC/UKQCD~14 as $f_{B}$, and completely correlate all the errors.

In addition to RBC/UKQCD~14 and RBC/UKQCD~14A, a few
other results in Tabs.~\ref{tab:FBssumm} and~\ref{tab:FBratsumm} are
also in our averaging procedure.   These include HPQCD~12, HPQCD~11A,
and FNAL/MILC~11.
Notice that there are two results of $f_{B}$ from
HPQCD~12 in Tab.~\ref{tab:FBssumm}.   Both of these were in the
averaging procedure in the last edition of the FLAG report.  However,
for our current work, we only include the one with smaller error.
This result is obtained by taking $f_{B_{s}}/f_{B}$ computed with
the NRQCD description of the $b$ quark in HPQCD~12, 
and multiplying it by $f_{B_{s}}$ calculated employing the HISQ
discretization for the heavy quarks in HPQCD~11A.  This strategy
significantly reduces the systematic effect arising from the
renormalization of the axial current in Eq.~(\ref{eq:fB_from_ME}), as
compared to the ``direct''  determination of $f_{B}$ using NRQCD heavy
quarks in HPQCD~12.   Since the calculations performed in 
FNAL/MILC~11, HPQCD~12 and HPQCD~11A all involve the gauge-field
ensembles generated by the MILC collaboration, we treat their
statistical errors as $100\%$ correlated.  Following the above discussion,
our procedure leads to the averages,
%
%FLAGRESULT BEGIN
% TAG      & fB    & fBs	&fBsofB &END
% REFS     & \cite{Christ:2014uea,Aoki:2014nga,Na:2012sp,McNeile:2011ng,Bazavov:2011aa} &\cite{Christ:2014uea,Aoki:2014nga,Na:2012sp,McNeile:2011ng,Bazavov:2011aa} &\cite{Christ:2014uea,Aoki:2014nga,Na:2012sp,McNeile:2011ng,Bazavov:2011aa} &END
% UNITS    & '[MeV]' & '[MeV]' & 1 &END
% NUMRESULTS & 5 & 5 & 5 &END
% FLAVOURs & 2+1 & 2+1 & 2+1 &END
%FLAGRESULT END
%FLAGRESULTFORMULA BEGIN
\begin{align}
        &&\FLAGAVBEGIN f_B&=192.0(4.3)\FLAGAVEND \;{\rm MeV}	  &&\Refs~\mbox{\cite{Christ:2014uea,Aoki:2014nga,Na:2012sp,McNeile:2011ng,Bazavov:2011aa}},\nonumber\\
&N_f=2+1:&\FLAGAVBEGIN f_{B_s}&=228.4(3.7)\FLAGAVEND \;{\rm MeV} &&\Refs~\mbox{\cite{Christ:2014uea,Aoki:2014nga,Na:2012sp,McNeile:2011ng,Bazavov:2011aa}}, \label{eq:fbav21}\\
        &&\FLAGAVBEGIN  {{f_{B_s}}/{f_B}}&=1.201(16)\FLAGAVEND&&\Refs~\mbox{\cite{Christ:2014uea,Aoki:2014nga,Na:2012sp,McNeile:2011ng,Bazavov:2011aa}}.\nonumber
\end{align}
%FLAGRESULTFORMULA END
%

There have been no new $N_{f} = 2+1+1$ results for the $B$-meson decay
constants and the $SU(3)$-breaking ratio since the release of the
previous FLAG publication.\footnote{At the Lattice 2015 conference, the Fermilab Lattice and MILC
  collaborations reported their on-going project for computing the
  $B$-meson decay constants in $N_{f}=2+1+1$
  QCD~\cite{Lattice:2015nee}.  However, no result has been shown
  yet.}  Therefore, our averages remain the same as those in the
previous FLAG report,
%
%FLAGRESULT BEGIN
% TAG      & fB    & fBs	&fBsofB &END
% REFS     & \cite{Dowdall:2013tga} &\cite{Dowdall:2013tga} &\cite{Dowdall:2013tga} &END
% UNITS    & '[MeV]' & '[MeV]' & 1 &END
% NUMRESULTS & 1 & 1 & 1 &END
% FLAVOURs & 2+1+1 & 2+1+1 & 2+1+1 &END
%FLAGRESULT END
%FLAGRESULTFORMULA BEGIN
\begin{align}
          &&\FLAGAVBEGIN f_B&=186(4)\FLAGAVEND  \;{\rm MeV} 	     &&\Refs~\mbox{\cite{Dowdall:2013tga}},\nonumber\\
&N_f=2+1+1:&\FLAGAVBEGIN f_{B_s}&=224(5)\FLAGAVEND \;{\rm MeV}       &&\Refs~\mbox{\cite{Dowdall:2013tga}},\label{eq:fbav211}\\
          &&\FLAGAVBEGIN {{f_{B_s}}/{f_B}}&=1.205(7) \FLAGAVEND  &&\Refs~\mbox{\cite{Dowdall:2013tga}}.\nonumber
\end{align}
%FLAGRESULTFORMULA END
%

The PDG recently presented their averages for the
$N_{f}=2+1$ and $N_{f}=2+1+1$ lattice-QCD determinations of
$f_{B^{+}}$, $f_{B_{s}}$ and
$f_{B_{s}}/f_{B^{+}}$~\cite{Rosner:2015wva}\footnote{We thank Ruth Van
de Water for communication and discussion regarding the comparison of
the averaging strategies.}.  The lattice-computation results used in
Ref.~\cite{Rosner:2015wva} are identical to those included in our
current work.   Regarding our isospin-averaged $f_{B}$ as the
representative for $f_{B^{+}}$, then the results from current FLAG and
PDG estimations for these quantities are well compatible.  In the PDG
work, they ``corrected'' the isospin-averaged $f_{B}$, as reported by
various lattice collaborations, using the $N_{f}=2+1+1$ strong isospin-breaking effect
computed in HPQCD~13~\cite{Dowdall:2013tga} (see 
Tab.~\ref{tab:FBssumm} in this subsection).  This only accounts for the contribution from the
valence-quark masses.   However, since the isospin-breaking effects
from the sea-quark masses appear in the form $(m^{({\rm sea})}_{u} -
m^{({\rm sea})}_{d})^{2}$, the valence sector is the
predominant source of strong isospin breaking~\cite{WalkerLoud:2009nf}.\footnote{We thank Ruth
  Van de Water and Andre Walker-Loud for helpful discussion on this
point.}

%% file: HQ/HQSubsections/BB.tex
\subsection{Neutral $B$-meson mixing matrix elements}
\label{sec:BMix}

Neutral $B$-meson mixing is induced in the Standard Model through
1-loop box diagrams to lowest order in the electroweak theory,
similar to those for short-distance effects in neutral kaon mixing. The effective Hamiltonian
is given by
\begin{equation}
  {\cal H}_{\rm eff}^{\Delta B = 2, {\rm SM}} \,\, = \,\,
  \frac{G_F^2 M_{\rm{W}}^2}{16\pi^2} ({\cal F}^0_d {\cal Q}^d_1 + {\cal F}^0_s {\cal Q}^s_1)\,\, +
   \,\, {\rm h.c.} \,\,,
   \label{eq:HeffB}
\end{equation}
with
\begin{equation}
 {\cal Q}^q_1 =
   \left[\bar{b}\gamma_\mu(1-\gamma_5)q\right]
   \left[\bar{b}\gamma_\mu(1-\gamma_5)q\right],
   \label{eq:Q1}
\end{equation}
where $q=d$ or $s$. The short-distance function ${\cal F}^0_q$ in
Eq.~(\ref{eq:HeffB}) is much simpler compared to the kaon mixing case
due to the hierarchy in the CKM matrix elements. Here, only one term
is relevant,
\begin{equation}
 {\cal F}^0_q = \lambda_{tq}^2 S_0(x_t)
\end{equation}
where
\begin{equation}
 \lambda_{tq} = V^*_{tq}V_{tb},
\end{equation}
and where $S_0(x_t)$ is an Inami-Lim function with $x_t=m_t^2/M_W^2$,
which describes the basic electroweak loop contributions without QCD
\cite{Inami:1980fz}. The transition amplitude for $B_q^0$ with $q=d$
or $s$ can be written as
\begin{eqnarray}
\label{eq:BmixHeff}
&&\langle \bar B^0_q \vert {\cal H}_{\rm eff}^{\Delta B = 2} \vert B^0_q\rangle  \,\, = \,\, \frac{G_F^2 M_{\rm{W}}^2}{16 \pi^2}  
\Big [ \lambda_{tq}^2 S_0(x_t) \eta_{2B} \Big ]  \nn \\ 
&&\times 
  \left(\frac{\gbar(\mu)^2}{4\pi}\right)^{-\gamma_0/(2\beta_0)}
  \exp\bigg\{ \int_0^{\gbar(\mu)} \, dg \, \bigg(
  \frac{\gamma(g)}{\beta(g)} \, + \, \frac{\gamma_0}{\beta_0g} \bigg)
  \bigg\} 
   \langle \bar B^0_q \vert  Q^q_{\rm R} (\mu) \vert B^0_q
   \rangle \,\, + \,\, {\rm h.c.} \,\, ,
   \label{eq:BBME}
\end{eqnarray}
where $Q^q_{\rm R} (\mu)$ is the renormalized four-fermion operator
(usually in the NDR scheme of $\msbar$). The running coupling
($\gbar$), the $\beta$-function ($\beta(g)$), and the anomalous
dimension of the four-quark operator ($\gamma(g)$) are defined in
Eqs.~(\ref{eq:four_quark_operator_anomalous_dimensions})~and~(\ref{eq:four_quark_operator_anomalous_dimensions_perturbative}).
The product of $\mu$ dependent terms on the second line of
Eq.~(\ref{eq:BBME}) is, of course, $\mu$-independent (up to truncation
errors arising from the use of perturbation theory). The explicit expression for
the short-distance QCD correction factor $\eta_{2B}$ (calculated to
NLO) can be found in Ref.~\cite{Buchalla:1995vs}.

For historical reasons the $B$-meson mixing matrix elements are often
parameterized in terms of bag parameters defined as
\begin{equation}
 B_{B_q}(\mu)= \frac{{\left\langle\bar{B}^0_q\left|
   Q^q_{\rm R}(\mu)\right|B^0_q\right\rangle} }{
         {\frac{8}{3}f_{B_q}^2\mB^2}} \,\, .
         \label{eq:bagdef}
\end{equation}
The RGI $B$ parameter $\hat{B}$ is defined, as in the case of the kaon,
and expressed to 2-loop order as
\begin{equation}
 \hat{B}_{B_q} = 
   \left(\frac{\gbar(\mu)^2}{4\pi}\right)^{- \gamma_0/(2\beta_0)}
   \left\{ 1+\dfrac{\gbar(\mu)^2}{(4\pi)^2}\left[
   \frac{\beta_1\gamma_0-\beta_0\gamma_1}{2\beta_0^2} \right]\right\}\,
   B_{B_q}(\mu) \,\,\, ,
\label{eq:BBRGI_NLO}
\end{equation}
with $\beta_0$, $\beta_1$, $\gamma_0$, and $\gamma_1$ defined in
Eq.~(\ref{eq:RG-coefficients}). Note, as Eq.~(\ref{eq:BBME}) is
evaluated above the bottom threshold ($m_b<\mu<m_t$), the active number
of flavours here is $N_f=5$.

Nonzero transition amplitudes result in a mass difference between the
$CP$ eigenstates of the neutral $B$-meson system. Writing the mass
difference for a $B_q^0$ meson as $\Delta m_q$, its Standard Model
prediction is
\begin{equation}
 \Delta m_q = \frac{G^2_Fm^2_W m_{B_q}}{6\pi^2} \,
  |\lambda_{tq}|^2 S_0(x_t) \eta_{2B} f_{B_q}^2 \hat{B}_{B_q}.
\end{equation}
Experimentally the mass difference is measured as oscillation
frequency of the $CP$ eigenstates. The frequencies are measured
precisely with an error of less than a percent. Many different
experiments have measured $\Delta m_d$, but the current average
\cite{Agashe:2014kda} is based on measurements from the
$B$-factory experiments Belle and Babar, and from the LHC experiment
LHC$b$. For $\Delta m_s$ the experimental average is dominated by results
from LHC$b$
\cite{Agashe:2014kda}.  With these experimental results and
lattice-QCD calculations of $f_{B_q}^2\hat{B}_{B_q}$ at hand,
$\lambda_{tq}$ can be determined.  In lattice-QCD calculations the
flavour $SU(3)$-breaking ratio
\begin{equation}
 \xi^2 = \frac{f_{B_s}^2B_{B_s}}{f_{B_d}^2B_{B_d}}
 \label{eq:xidef}
\end{equation} 
can be obtained more precisely than the individual $B_q$-mixing matrix
elements because statistical and systematic errors cancel in part.
With this the ratio $|V_{td}/V_{ts}|$ can be determined, which can be used
to constrain the apex of the CKM triangle.

Neutral $B$-meson mixing, being loop-induced in the Standard Model is
also a sensitive probe of new physics. The most general $\Delta B=2$
effective Hamiltonian that describes contributions to $B$-meson mixing
in the Standard Model and beyond is given in terms of five local
four-fermion operators:
\be
  {\cal H}_{\rm eff, BSM}^{\Delta B = 2} = \sum_{q=d,s}\sum_{i=1}^5 {\cal C}_i {\cal Q}^q_i \;,
\ee
where ${\cal Q}_1$ is defined in Eq.~(\ref{eq:Q1}) and where
\bd
{\cal Q}^q_2 =  \left[\bar{b}(1-\gamma_5)q\right]
   \left[\bar{b}(1-\gamma_5)q\right], \qquad
{\cal Q}^q_3 =  \left[\bar{b}^{\alpha}(1-\gamma_5)q^{\beta}\right]
   \left[\bar{b}^{\beta}(1-\gamma_5)q^{\alpha}\right],
 \ed
  \be
{\cal Q}^q_4 =  \left[\bar{b}(1-\gamma_5)q\right]
   \left[\bar{b}(1+\gamma_5)q\right], \qquad
{\cal Q}^q_5 =  \left[\bar{b}^{\alpha}(1-\gamma_5)q^{\beta}\right]
   \left[\bar{b}^{\beta}(1+\gamma_5)q^{\alpha}\right], 
   \label{eq:Q25}
\ee 
with the superscripts $\alpha,\beta$ denoting colour indices, which
are shown only when they are contracted across the two bilinears.
There are three other basis operators in the $\Delta
B=2$ effective Hamiltonian. When evaluated in QCD, however, 
they give identical matrix elements to the ones already listed due to
parity invariance in QCD.
The short-distance Wilson coefficients ${\cal C}_i$ depend on the
underlying theory and can be calculated perturbatively.  In the
Standard Model only matrix elements of ${\cal Q}^q_1$ contribute to
$\Delta m_q$, while all operators do for example for general SUSY
extensions of the Standard Model~\cite{Gabbiani:1996hi}.
The matrix elements or bag parameters for the non-SM operators are also 
useful to estimate the width difference in the Standard Model,
where combinations of matrix elements of ${\cal Q}^q_1$,
${\cal Q}^q_2$, and ${\cal Q}^q_3$ contribute to $\Delta \Gamma_q$ 
at $\cO(1/m_b)$~\cite{Lenz:2006hd,Beneke:1996gn}.  

\begin{figure}[!htb]
\hspace{-0.8cm}\includegraphics[width=0.57\linewidth]{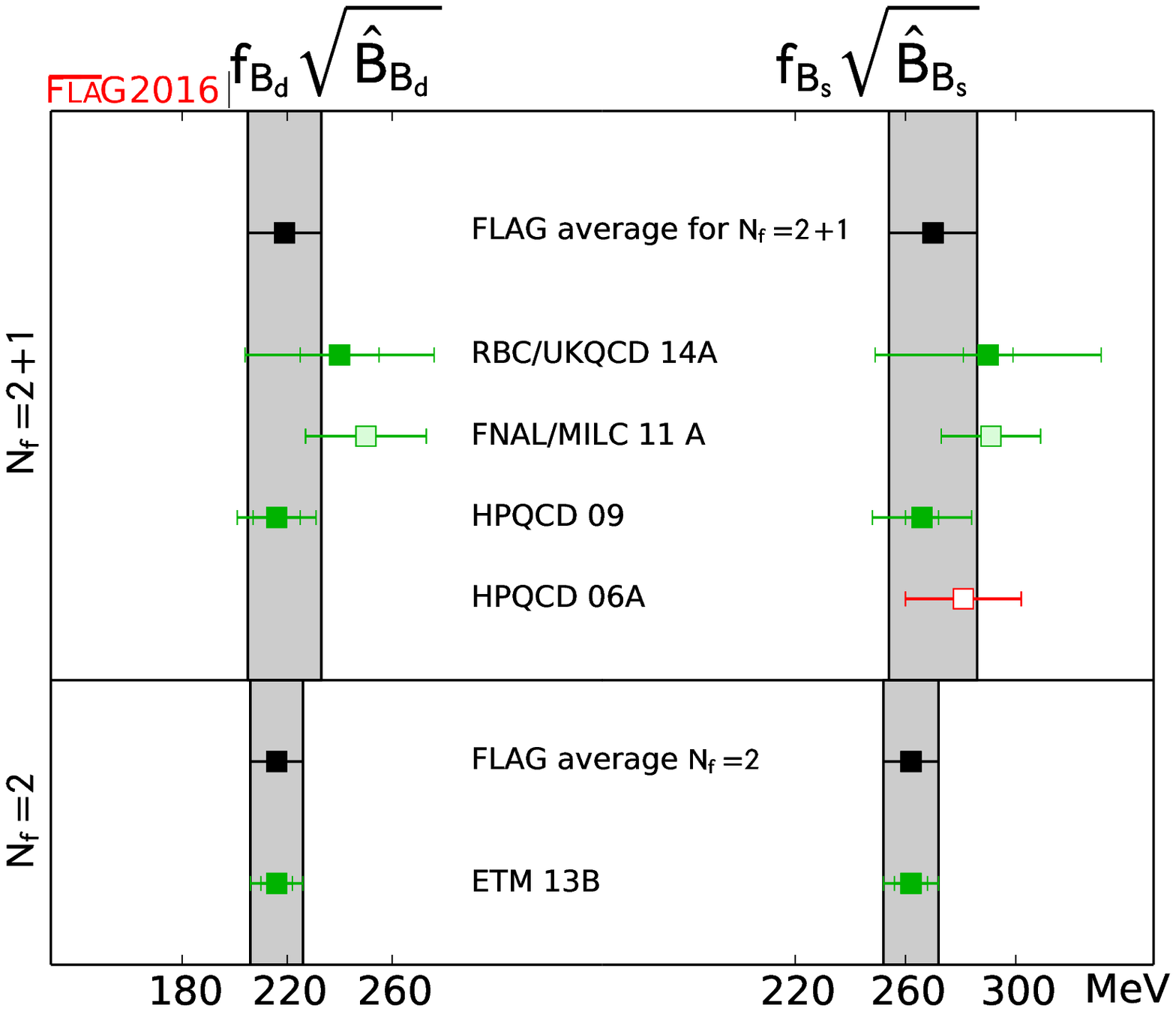}\hspace{-0.8cm}
\includegraphics[width=0.57\linewidth]{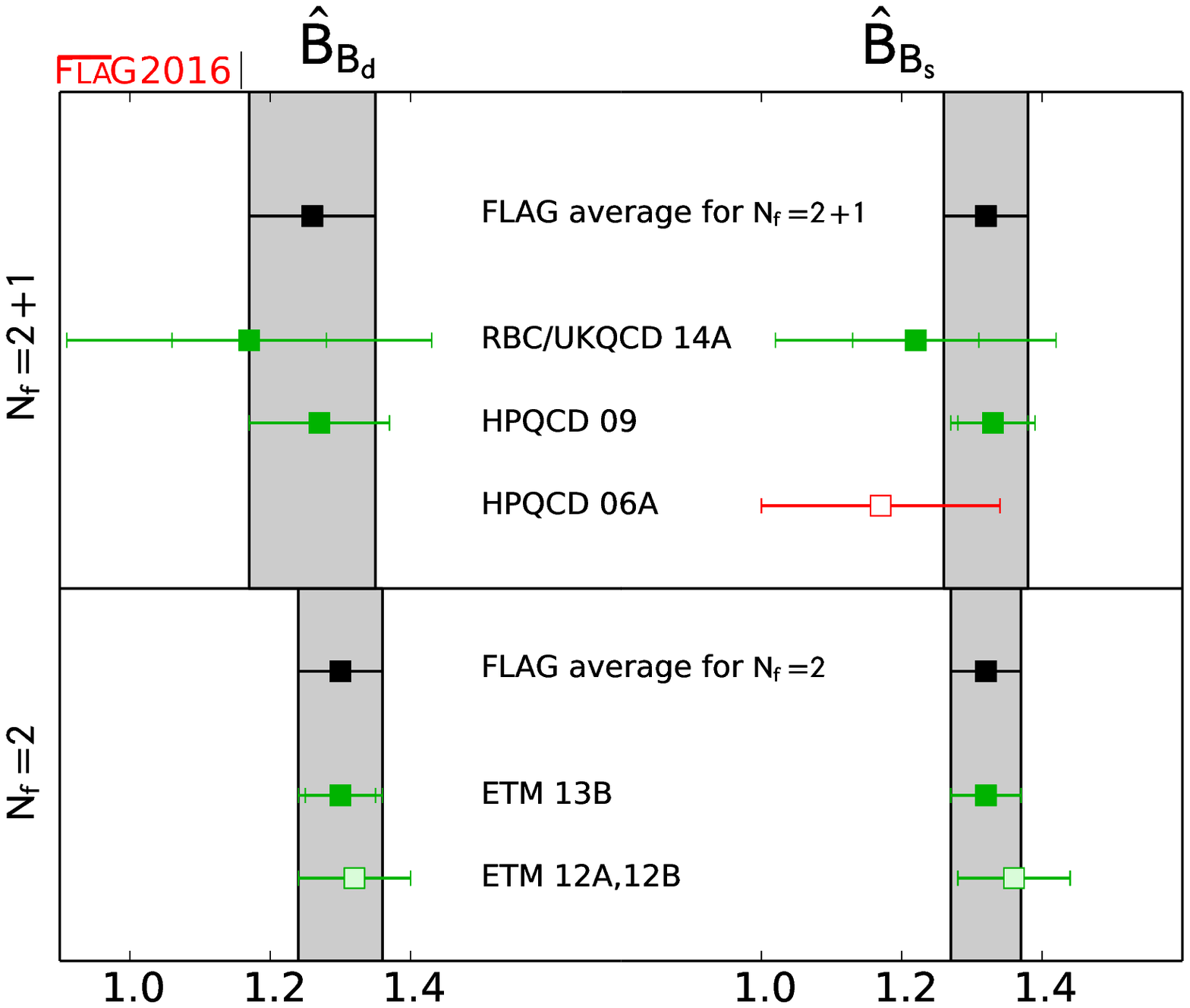}

\vspace{-5mm}
\caption{Neutral $B$- and $B_{\rm s}$-meson mixing matrix
 elements and bag parameters [values in Tab.~\ref{tab_BBssumm} and
 Eqs.~(\ref{eq:avfBB2}), (\ref{eq:avfBB}), (\ref{eq:avBB2}), (\ref{eq:avBB})].
 \label{fig:fBsqrtBB2}}
\end{figure}

In this section we report on results from lattice-QCD calculations for
the neutral $B$-meson mixing parameters $\hat{B}_{B_d}$,
$\hat{B}_{B_s}$, $f_{B_d}\sqrt{\hat{B}_{B_d}}$,
$f_{B_s}\sqrt{\hat{B}_{B_s}}$ and the $SU(3)$-breaking ratios
$B_{B_s}/B_{B_d}$ and $\xi$ defined in Eqs.~(\ref{eq:bagdef}),
(\ref{eq:BBRGI_NLO}), and (\ref{eq:xidef}).  The results are
summarized in Tabs.~\ref{tab_BBssumm} and \ref{tab_BBratsumm} and in
Figs.~\ref{fig:fBsqrtBB2} and \ref{fig:xi}. Additional details about
the underlying simulations and systematic error estimates are given in
Appendix~\ref{app:BMix_Notes}.  Some collaborations do not provide the
RGI quantities $\hat{B}_{B_q}$ but quote instead
$B_B(\mu)^{\overline{MS},NDR}$. In such cases we convert the results
to the RGI quantities quoted in Tab.~\ref{tab_BBssumm} using
Eq.~(\ref{eq:BBRGI_NLO}). More details on the conversion factors are
provided below in the descriptions of the individual results.
We do not provide the $B$-meson matrix elements of the other operators
${\cal Q}_{2-5}$ in this report. They have been calculated in
Ref.~\cite{Carrasco:2013zta} for the $N_f=2$ case and 
in Ref.~\cite{Bouchard:2011xj}, which is a conference proceedings article.

\begin{table}[!htb]
\begin{center}
\mbox{} \\[3.0cm]
\footnotesize
\begin{tabular*}{\textwidth}[l]{l@{\extracolsep{\fill}}@{\hspace{1mm}}r@{\hspace{1mm}}l@{\hspace{1mm}}l@{\hspace{1mm}}l@{\hspace{1mm}}l@{\hspace{1mm}}l@{\hspace{1mm}}l@{\hspace{1mm}}l@{\hspace{5mm}}l@{\hspace{1mm}}l@{\hspace{1mm}}l@{\hspace{1mm}}l@{\hspace{1mm}}l}
Collaboration \al Ref. \al $\Nf$ \al
\hspace{0.15cm}\begin{rotate}{60}{publication status}\end{rotate}\hspace{-0.15cm} \al
\hspace{0.15cm}\begin{rotate}{60}{continuum extrapolation}\end{rotate}\hspace{-0.15cm} \al
\hspace{0.15cm}\begin{rotate}{60}{chiral extrapolation}\end{rotate}\hspace{-0.15cm}\al
\hspace{0.15cm}\begin{rotate}{60}{finite volume}\end{rotate}\hspace{-0.15cm}\al
\hspace{0.15cm}\begin{rotate}{60}{renormalization/matching}\end{rotate}\hspace{-0.15cm}  \al
\hspace{0.15cm}\begin{rotate}{60}{heavy-quark treatment}\end{rotate}\hspace{-0.15cm} \al 
\rule{0.12cm}{0cm}
\parbox[b]{1.2cm}{$f_{\rm B_d}\sqrt{\hat{B}_{\rm B_d}}$} \al
\rule{0.12cm}{0cm}
\parbox[b]{1.2cm}{$f_{\rm B_s}\sqrt{\hat{B}_{\rm B_s}}$} \al
\rule{0.12cm}{0cm}
$\hat{B}_{\rm B_d}$ \al 
\rule{0.12cm}{0cm}
$\hat{B}_{\rm B_{\rm s}}$ \\
&&&&&&&&&& \\[-0.1cm]
\hline
\hline
&&&&&&&&&& \\[-0.1cm]

RBC/UKQCD 14A \al \cite{Aoki:2014nga} \al 2+1 \al \gA \al \soso \al \soso \al
     \soso \al \soso
	\al \okay & 240(15)(33) \al 290(09)(40) \al 1.17(11)(24) \al 1.22(06)(19)\\[0.5ex]

FNAL/MILC 11A \al \cite{Bouchard:2011xj} \al 2+1 \al \rC \al \good \al \soso \al
     \good \al \soso
	\al \okay & 250(23)$^\dagger$ \al 291(18)$^\dagger$ \al $-$ \al $-$\\[0.5ex]

HPQCD 09 \al \cite{Gamiz:2009ku} \al 2+1 \al \gA \al \soso \al \soso$^\nabla$ \al \soso \al
\soso 
\al \okay & 216(15)$^\ast$ \al 266(18)$^\ast$ \al 1.27(10)$^\ast$ \al 1.33(6)$^\ast$ \\[0.5ex] 

HPQCD 06A \al \cite{Dalgic:2006gp} \al 2+1 \al \gA \al \tbr \al \tbr \al \good \al 
\soso 
	\al \okay & $-$ \al  281(21) \al $-$ \al 1.17(17) \\
&&&&&&&&&& \\[-0.1cm]
\hline
&&&&&&&&&& \\[-0.1cm]
ETM 13B \al \cite{Carrasco:2013zta} \al 2 \al \gA \al \good \al \soso \al \soso \al
    \good \al \okay & 216(6)(8) \al 262(6)(8) \al  1.30(5)(3) \al 1.32(5)(2) \\[0.5ex]

ETM 12A, 12B \al \cite{Carrasco:2012dd,Carrasco:2012de} \al 2 \al \rC \al \good \al \soso \al \soso \al
    \good \al \okay & $-$ \al $-$ \al  1.32(8)$^\diamond$ \al 1.36(8)$^\diamond$ \\[0.5ex]
&&&&&&&&&& \\[-0.1cm]
\hline
\hline\\
\end{tabular*}\\[-0.2cm]
\begin{minipage}{\linewidth}
{\footnotesize 
\begin{itemize}
   \item[$^\dagger$] Reported $f_B^2B$ at $\mu=m_b$ is converted to RGI by
	multiplying the 2-loop factor
	1.517.\\[-5mm]
   \item[$^\nabla$] Wrong-spin contributions are not included in the rS$\chi$PT fits. \\[-5mm]
        \item[$^\ast$] This result uses an old determination of  $r_1=0.321(5)$~fm from Ref.~\cite{Gray:2005ur} that has since been superseded. \\[-5mm]
        \item[$^\diamond$] Reported $B$ at $\mu=m_b=4.35$ GeV is converted to
     RGI by multiplying the 2-loop factor 1.521.
\end{itemize}
}
\end{minipage}
\caption{Neutral $B$- and $B_{\rm s}$-meson mixing matrix
 elements (in MeV) and bag parameters.}
\label{tab_BBssumm}
\end{center}
\end{table}

\begin{table}[!htb]
\begin{center}
\mbox{} \\[3.0cm]
\footnotesize
\begin{tabular*}{\textwidth}[l]{l @{\extracolsep{\fill}} r l l l l l l l l l}
Collaboration & Ref. & $\Nf$ & 
\hspace{0.15cm}\begin{rotate}{60}{publication status}\end{rotate}\hspace{-0.15cm} &
\hspace{0.15cm}\begin{rotate}{60}{continuum extrapolation}\end{rotate}\hspace{-0.15cm} &
\hspace{0.15cm}\begin{rotate}{60}{chiral extrapolation}\end{rotate}\hspace{-0.15cm}&
\hspace{0.15cm}\begin{rotate}{60}{finite volume}\end{rotate}\hspace{-0.15cm}&
\hspace{0.15cm}\begin{rotate}{60}{renormalization/matching}\end{rotate}\hspace{-0.15cm}  &
\hspace{0.15cm}\begin{rotate}{60}{heavy-quark treatment}\end{rotate}\hspace{-0.15cm} & 
\rule{0.12cm}{0cm}$\xi$ &
 \rule{0.12cm}{0cm}$B_{\rm B_{\rm s}}/B_{\rm B_d}$ \\
&&&&&&&&&& \\[-0.1cm]
\hline
\hline
&&&&&&&&&& \\[-0.1cm]

RBC/UKQCD 14A & \cite{Aoki:2014nga} & 2+1 & \gA & \soso & \soso &
     \soso & \soso & \okay & 1.208(41)(52) & 1.028(60)(49) \\[0.5ex]

FNAL/MILC 12 & \cite{Bazavov:2012zs} & 2+1 & \gA & \soso & \soso &
     \good & \soso & \okay & 1.268(63) & 1.06(11) \\[0.5ex]

RBC/UKQCD 10C
 & \cite{Albertus:2010nm} & 2+1 & \gA & \tbr & \tbr & \tbr
  & \soso & \okay & 1.13(12) & $-$ \\[0.5ex]

HPQCD 09 & \cite{Gamiz:2009ku} & 2+1 & \gA & \soso & \soso$^\nabla$ & \soso &
\soso & \okay & 1.258(33) & 1.05(7) \\[0.5ex] 

&&&&&&&&&& \\[-0.1cm]

\hline

&&&&&&&&&& \\[-0.1cm]

ETM 13B & \cite{Carrasco:2013zta} & 2 & \gA & \good & \soso & \soso & \good
			     & \okay & 1.225(16)(14)(22) & 1.007(15)(14) \\

ETM 12A, 12B & \cite{Carrasco:2012dd,Carrasco:2012de} & 2 & \rC & \good & \soso & \soso & \good
			     & \okay & 1.21(6) & 1.03(2) \\
&&&&&&&&&& \\[-0.1cm]
\hline
\hline\\
\end{tabular*}\\[-0.2cm]
\begin{minipage}{\linewidth}
{\footnotesize 
\begin{itemize}
   \item[$^\nabla$] Wrong-spin contributions are not included in the rS$\chi$PT fits. 
\end{itemize}
}
\end{minipage}
\caption{Results for $SU(3)$-breaking ratios of neutral $B_{d}$- and 
 $B_{s}$-meson mixing matrix elements and bag parameters.}
\label{tab_BBratsumm}
\end{center}
\end{table}

\begin{figure}[!htb]
\begin{center}
\includegraphics[width=0.57\linewidth]{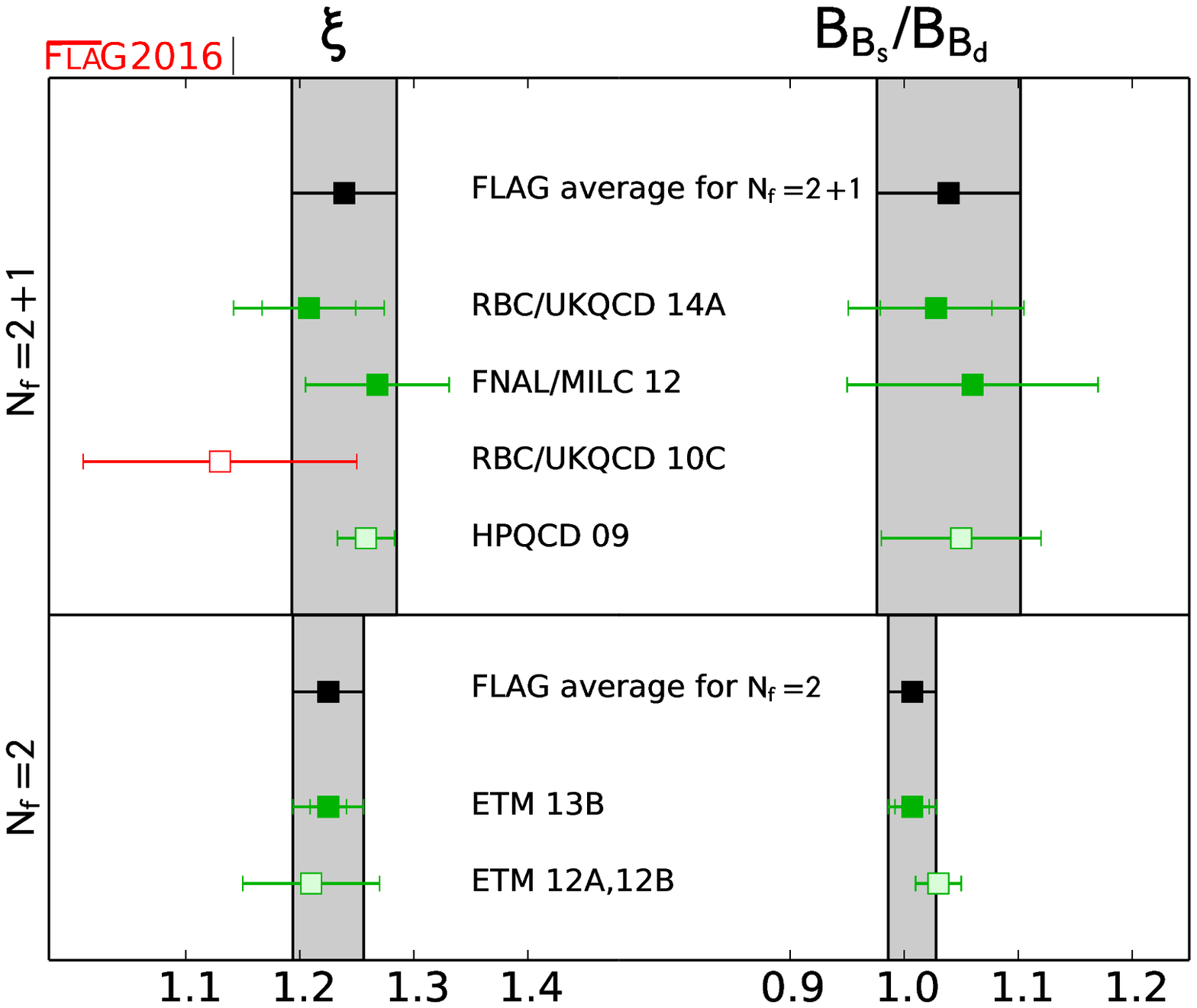}

\vspace{-2mm}
\caption{The $SU(3)$-breaking quantities $\xi$ and $B_{B_s}/B_{B_d}$
 [values in Tab.~\ref{tab_BBratsumm} and Eqs.~(\ref{eq:avxiBB2}) and (\ref{eq:avxiBB})].}\label{fig:xi} 
\end{center}
\end{figure}

There are no new results for $N_f=2$ reported after the previous FLAG
review. However the paper by the ETM collaboration (ETM~13B)~\cite{Carrasco:2013zta}, which was a preprint,
has been published in a journal, thus, it is now eligible to enter the
averages. Because this is the only result that passes the quality
criteria for $N_f=2$, we quote their values as our averages
in this version:
%FLAGRESULT BEGIN
% TAG      & fBsqrtBB    & fBssqrtBBs & BB  &BBs & xi  &BBsoBB	 &END
% REFS     & \cite{Carrasco:2013zta} & \cite{Carrasco:2013zta} & \cite{Carrasco:2013zta} & \cite{Carrasco:2013zta} & \cite{Carrasco:2013zta} & \cite{Carrasco:2013zta} &END
% UNITS    & '[MeV]' & '[MeV]' & 1 & 1 & 1 & 1 &END
% NUMRESULTS & 1 & 1 & 1& 1& 1& 1 &END
% FLAVOURs & 2& 2& 2& 2& 2 & 2 &END
%FLAGRESULT END
%FLAGRESULTFORMULA BEGIN
\begin{align}
      &&  \FLAGAVBEGIN f_{B_d}\sqrt{\hat{B}_{b_d}}&= 216(10)\FLAGAVEND\;\; {\rm MeV}
         &\FLAGAVBEGIN f_{B_s}\sqrt{\hat{B}_{B_s}}&= 262(10)\FLAGAVEND\;\; {\rm MeV}
         &\Ref~\mbox{\cite{Carrasco:2013zta}},  \label{eq:avfBB2}\\
N_f=2:&&\FLAGAVBEGIN \hat{B}_{B_d}&= 1.30(6)\FLAGAVEND 
         &\FLAGAVBEGIN \hat{B}_{B_s}&= 1.32(5)\FLAGAVEND 
	 &\Ref~\mbox{\cite{Carrasco:2013zta}},  \label{eq:avBB2}\\
      &&  \FLAGAVBEGIN \xi &=  1.225(31)\FLAGAVEND  
  	& \FLAGAVBEGIN B_{B_s}/B_{B_d} & =  1.007(21)\FLAGAVEND
 	&\Ref~\mbox{\cite{Carrasco:2013zta}}. \label{eq:avxiBB2}
\end{align}
%FLAGRESULTFORMULA END

For the $N_f=2+1$ case there is a new report
(RBC/UKQCD~14A)~\cite{Aoki:2014nga} by the RBC/UKQCD collaboration 
on the neutral $B$-meson mixing parameter, using domain-wall fermions
for the light quarks and the static approximation for the $b$ quark.
Used gauge configuration ensembles are the $N_f=2+1$ domain-wall
fermion and Iwasaki gauge actions with two lattice spacings
($a\approx 0.09, 0.11$~fm) and a minimum pion mass of about 290 MeV. 
Two different static-quark actions, smeared with 
HYP1~\cite{Hasenfratz:2001hp} and 
HYP2~\cite{DellaMorte:2005yc} are used
to further constrain the continuum limit. The operators used are 
1-loop $\cO(a)$-improved with the tadpole improved perturbation theory.
Two different types of chiral formulae are adopted for the 
combined continuum and chiral extrapolation:
$SU(2)$ NLO HM$\chi$PT and first order polynomial in quark masses with
linear $\cO(a^2)$ terms.
The central values are determined as the average of the results with two
different chiral formulae. The systematic error is estimated as half
of the full difference of the two, with an exception for the quantity only
involving $B_s^0$, where the NLO $\chi$PT is identical to the first
order polynomial. In such cases, the fit excluding the heaviest $ud$
mass point is used for the estimate of the systematic error.
The systematic error due to the static approximation is estimated
by the simple power counting: the size of $\Lambda_{QCD}/m_b$, where 
$\Lambda_{QCD}=0.5$ GeV and $m_b(\mu=m_b)^{\overline{\rm MS}}=4.18$ GeV
(PDG) leads to 12\%. This is the dominant systematic error for individual
$f_B\sqrt{B_B}$ or $B_B$. Due to this large error, the effect of the
inclusion in the FLAG averages of these quantities is small.
The dominant systematic error for the $SU(3)$-breaking error, instead, comes from
the combined continuum and chiral extrapolation, while the statistical 
uncertainty is a bit larger than that.

Due to the addition of this new result, the values for $N_f=2+1$ are 
now averages from multiple results by multiple collaborations,
rather than being given by the values from a single computation, as it  was done
in the previous FLAG report. Our averages are:
%FLAGRESULT BEGIN
% TAG      & fBsqrtBB    & fBssqrtBBs & BB  &BBs & xi  &BBsoBB	 &END
% REFS     & \cite{Gamiz:2009ku,Aoki:2014nga}& \cite{Gamiz:2009ku,Aoki:2014nga}& \cite{Gamiz:2009ku,Aoki:2014nga}& \cite{Gamiz:2009ku,Aoki:2014nga} &\cite{Bazavov:2012zs,Aoki:2014nga}&\cite{Bazavov:2012zs,Aoki:2014nga} &END
% UNITS    & '[MeV]' & '[MeV]' & 1 & 1 & 1 & 1 &END
% NUMRESULTS & 2 & 2 & 2& 2& 2& 2 &END
% FLAVOURs & 2+1& 2+1& 2+1& 2+1& 2+1 & 2+1 &END
%FLAGRESULT END
%FLAGRESULTFORMULA BEGIN
\begin{align}
        && \FLAGAVBEGIN f_{B_d}\sqrt{\hat{B}_{B_d}} &=  219(14)\FLAGAVEND \, {\rm MeV}
          &\FLAGAVBEGIN f_{B_s}\sqrt{\hat{B}_{B_s}} &=  270(16)\FLAGAVEND \, {\rm MeV}
	  &\Refs~\mbox{\cite{Gamiz:2009ku,Aoki:2014nga}},  \label{eq:avfBB}\\
&N_f=2+1:& \FLAGAVBEGIN \hat{B}_{B_d}  &= 1.26(9)\FLAGAVEND 
          & \FLAGAVBEGIN \hat{B}_{B_s} &=  1.32(6)\FLAGAVEND 
          &\Refs~\mbox{\cite{Gamiz:2009ku,Aoki:2014nga}}, \label{eq:avBB}\\
        && \FLAGAVBEGIN \xi  &=  1.239(46)\FLAGAVEND 
          &\FLAGAVBEGIN B_{B_s}/B_{B_d}  &=  1.039(63)\FLAGAVEND
          &\Refs~\mbox{\cite{Bazavov:2012zs,Aoki:2014nga}}. \label{eq:avxiBB}
\end{align}
%FLAGRESULTFORMULA END
Here Eqs.~(\ref{eq:avfBB}) and (\ref{eq:avBB}) are averages from 
HPQCD~09~\cite{Gamiz:2009ku} and RBC/UKQCD~14A~\cite{Aoki:2014nga},
while Eq.~(\ref{eq:avxiBB}) is from FNAL/MILC~12~\cite{Bazavov:2012zs}
and RBC/UKQCD~14A~\cite{Aoki:2014nga}.

As discussed in detail in the previous FLAG review~\cite{Aoki:2013ldr}
HPQCD~09 does not include wrong-spin contributions, which are staggered
fermion artifacts, to the chiral extrapolation analysis. 
It is possible that the effect is significant for $\xi$ and
$B_{B_s}/B_{B_d}$, since the chiral extrapolation error is a dominant one
for these $SU(3)$ flavour breaking ratios.
Indeed, a test done by FNAL/MILC~12~\cite{Bazavov:2012zs} indicates
that the omission of the wrong spin contribution in the chiral analysis
may be a significant source of error.
We therefore took the conservative
choice to exclude $\xi$ and $B_{B_s}/B_{B_d}$ by HPQCD~09 from from our
average and we follow the same strategy in this report as well.

We note that the above results are all correlated with each other:  the
numbers in Eqs.~(\ref{eq:avfBB}) and (\ref{eq:avBB}) are dominated by those from
HPQCD~09~\cite{Gamiz:2009ku}, while those in Eq.~(\ref{eq:avxiBB})   
involve FNAL/MILC~12~\cite{Bazavov:2012zs} --
the same Asqtad MILC ensembles are used in these 
simulations. The results are also correlated with the averages obtained in 
Sec.~\ref{sec:fB} and shown in
Eq.~(\ref{eq:fbav21}), because the calculations of $B$-meson decay constants and  
mixing quantities 
are performed on the same (or on similar) sets of ensembles, and results obtained by a 
given collaboration 
use the same actions and setups. These correlations must be considered when 
using our averages as inputs to UT fits. In the future, as more independent 
calculations enter the averages, correlations between the lattice-QCD inputs to the UT 
fit will become less significant.

%% file: HQ/HQSubsections/BSL.tex
\subsection{Semileptonic form factors for $B$ decays to light flavours}
\label{sec:BtoPiK}

The Standard Model differential rate for the decay $B_{(s)}\to
P\ell\nu$ involving a quark-level $b\to u$ transition is given, at
leading order in the weak interaction, by a formula identical to the
one for $D$ decays in Eq.~(\ref{eq:DtoPiKFull}) but with $D \to
B_{(s)}$ and the relevant CKM matrix element $|V_{cq}| \to |V_{ub}|$:
\begin{eqnarray}
	\frac{d\Gamma(B_{(s)}\to P\ell\nu)}{dq^2} = \frac{G_F^2 |V_{ub}|^2}{24 \pi^3}
	\,\frac{(q^2-m_\ell^2)^2\sqrt{E_P^2-m_P^2}}{q^4m_{B_{(s)}}^2}
	\bigg[& \!\!\!\!\!\!\!\!\!\!\! \left(1+\frac{m_\ell^2}{2q^2}\right)m_{B_{(s)}}^2(E_P^2-m_P^2)|f_+(q^2)|^2 \nonumber\\
&~~~~\,+\,\frac{3m_\ell^2}{8q^2}(m_{B_{(s)}}^2-m_P^2)^2|f_0(q^2)|^2
\bigg]\,. \label{eq:B_semileptonic_rate}
\end{eqnarray}
Again, for $\ell=e,\mu$ the contribution from the scalar form factor
$f_0$ can be neglected, and one has a similar expression to
Eq.~(\ref{eq:DtoPiK}), which in principle allows for a direct
extraction of $|V_{ub}|$ by matching theoretical predictions to
experimental data.  However, while for $D$ (or $K$) decays the entire
physical range $0 \leq q^2 \leq q^2_{\rm max}$ can be covered with
moderate momenta accessible to lattice simulations, in
$B \to \pi \ell\nu$ decays one has $q^2_{\rm max} \sim 26~{\rm GeV}^2$
and only part of the full kinematic range is reachable.
As a consequence, obtaining $|V_{ub}|$ from $B\to\pi\ell\nu$ is more
complicated than obtaining $|V_{cd(s)}|$ from semileptonic $D$-meson
decays.

In practice, lattice computations are restricted
to small values of the momentum transfer (see Sec.~\ref{sec:DtoPiK})
where statistical and momentum-dependent discretization errors can be
controlled,\footnote{The variance of hadron correlation functions at
nonzero three-momentum is dominated at large Euclidean times by
zero-momentum multiparticle states~\cite{DellaMorte:2012xc}; therefore
the noise-to-signal grows more rapidly than for the vanishing three-momentum
case.} which in existing calculations roughly cover the upper third of
the kinematically allowed $q^2$ range.
Since, on the other hand, the decay rate is
suppressed by phase space at large $q^2$, most of the semileptonic $B\to
\pi$ events are selected in experiment at lower values of $q^2$, leading
to more accurate experimental results for the binned differential rate
in that region.\footnote{Upcoming data from Belle~II are expected to
significantly improve the precision of experimental results,
in particular, for larger values of $q^2$.}
It is therefore a challenge to find a window of
intermediate values of $q^2$ at which both the experimental and
lattice results can be reliably evaluated.

In current practice, the extraction of CKM matrix elements requires
that both experimental and lattice data for the $q^2$
dependence be parameterized by fitting data to a specific
ansatz. Before the generalization of the sophisticated ans\"{a}tze that
will be discussed below, the most common procedure to overcome this difficulty involved matching
the theoretical prediction and the experimental result for the
integrated decay rate over some finite interval in $q^2$,
\begin{gather}\label{eq:Deltazeta}
	\Delta \zeta = \frac{1}{|V_{ub}|^2} \int_{q^2_{1}}^{q^2_{2}} \left( \frac{d \Gamma}{d q^2} \right) dq^2\,.
\end{gather}
In the most recent literature, it has become customary to perform a joint fit to lattice
and experimental results, keeping the relative normalization
$|V_{ub}|^2$ as a free parameter. In either case, good control of the systematic
uncertainty induced by the choice of parameterization is crucial
to obtain a precise determination of $|V_{ub}|$.

\subsubsection{Parameterizations of semileptonic form factors}
\label{sec:zparam}

In this section, we discuss the description of the $q^2$ dependence of
form factors, using the vector form factor $f_+$ of $B\to\pi\ell\nu$ decays
as a benchmark case. Since in this channel the parameterization of the
$q^2$ dependence is crucial for the extraction of $|V_{ub}|$ from the existing
measurements (involving decays to light leptons), as explained
above, it has been studied in great detail in the literature. Some comments
about the generalization of the techniques involved will follow.

\paragraph{The vector form factor for $B\to\pi\ell\nu$}

All form factors are analytic functions of $q^2$ outside physical
poles and inelastic threshold branch points; in the case of
$B\to\pi\ell\nu$, the only pole expected below the $B\pi$ production
region, starting at $q^2 = t_+ = (m_B+m_\pi)^2$, is the $B^*$.  A
simple ansatz for the $q^2$ dependence of the $B\to\pi\ell\nu$
semileptonic form factors that incorporates vector-meson dominance is
the Be\'cirevi\'c-Kaidalov (BK)
parameterization~\cite{Becirevic:1999kt},
which for the vector form factor reads:
\begin{gather}
f_+(q^2) = \frac{f(0)}{(1-q^2/m_{B^*}^2)(1-\alpha q^2/m_{B^*}^2)}\,.
\label{eq:BKparam}
\end{gather}
Because the BK ansatz has few free parameters, it has been used
extensively to parameterize the shape of experimental
branching-fraction measurements and theoretical form-factor
calculations.  A variant of this parameterization proposed by Ball and
Zwicky (BZ) adds extra pole factors to the expressions in
Eq.~(\ref{eq:BKparam}) in order to mimic the effect of multiparticle
states~\cite{Ball:2004ye}. A similar idea, extending the use of effective
poles also to $D\to\pi\ell\nu$ decays, is explored in Ref.~\cite{Becirevic:2014kaa}.
Finally, yet another variant (RH) has been proposed by
Hill in Ref.~\cite{Hill:2005ju}. Although all of these parameterizations
capture some known properties of form factors, they do not manifestly
satisfy others.  For example,
perturbative QCD scaling constrains the
behaviour of $f_+$ in the deep Euclidean region~\cite{Lepage:1980fj,Akhoury:1993uw,Lellouch:1995yv}, and
angular momentum conservation constrains the asymptotic behaviour near
thresholds --- e.g., ${\rm Im}\,f_+(q^2) \sim (q^2-t_+)^{3/2}$ (see, e.g., Ref.~\cite{Bourrely:2008za}).  Most importantly, these parameterizations do not allow for an easy
quantification of systematic uncertainties.

A more systematic approach that improves upon the use of simple models
for the $q^2$ behaviour exploits the positivity and analyticity
properties of two-point functions of vector currents to obtain optimal
parameterizations of form
factors~\cite{Bourrely:1980gp,Boyd:1994tt,Lellouch:1995yv,Boyd:1997qw,Arnesen:2005ez,Becher:2005bg}.
Any form factor $f$ can be shown to admit a series expansion of the
form
\begin{gather}
f(q^2) = \frac{1}{B(q^2)\phi(q^2,t_0)}\,\sum_{n=0}^\infty a_n(t_0)\,z(q^2,t_0)^n\,,
\end{gather}
where the squared momentum transfer is replaced by the variable
\begin{gather}
z(q^2,t_0) = \frac{\sqrt{t_+-q^2}-\sqrt{t_+-t_0}}{\sqrt{t_+-q^2}+\sqrt{t_+-t_0}}\,.
\end{gather}
This is a conformal transformation, depending on an arbitrary real
parameter $t_0<t_+$, that maps the $q^2$ plane cut for $q^2 \geq t_+$
onto the disk $|z(q^2,t_0)|<1$ in the $z$ complex plane. The function
$B(q^2)$ is called the {\it Blaschke factor}, and contains poles and
cuts below $t_+$ --- for instance, in the case of $B\to\pi$ decays,
\begin{gather}
B(q^2)=\frac{z(q^2,t_0)-z(m_{B^*}^2,t_0)}{1-z(q^2,t_0)z(m_{B^*}^2,t_0)}=z(q^2,m_{B^*}^2)\,.
\end{gather}
Finally, the quantity $\phi(q^2,t_0)$, called the {\em outer
function}, is some otherwise arbitrary function that does not introduce further
poles or branch cuts.  The crucial property of this series expansion
is that the sum of the squares of the coefficients
\begin{gather}
\sum_{n=0}^\infty a_n^2 = \frac{1}{2\pi i}\oint \frac{dz}{z}\,|B(z)\phi(z)f(z)|^2\,,
\end{gather}
is a finite quantity. Therefore, by using this parameterization an
absolute bound to the uncertainty induced by truncating the series can
be obtained.  The aim in choosing $\phi$ is to obtain
a bound that is useful in practice, while
(ideally) preserving the correct behaviour of the form factor at high
$q^2$ and around thresholds.

The simplest form of the bound would correspond to $\sum_{n=0}^\infty
a_n^2=1$.  {\it Imposing} this bound yields the following ``standard''
choice for the outer function
\begin{gather}
\label{eq:comp_of}
\begin{split}
\phi(q^2,t_0)=&\sqrt{\frac{1}{32\pi\chi_{1^-}(0)}}\,
\left(\sqrt{t_+-q^2}+\sqrt{t_+-t_0}\right)\\
&\times\,\left(\sqrt{t_+-q^2}+\sqrt{t_+-t_-}\right)^{3/2}
\left(\sqrt{t_+-q^2}+\sqrt{t_+}\right)^{-5}
\,\frac{t_+-q^2}{(t_+-t_0)^{1/4}}\,,
\end{split}
\end{gather}
where $t_-=(m_B-m_\pi)^2$, and $\chi_{1^-}(0)$ is the derivative of the transverse component of
the polarization function (i.e., the Fourier transform of the vector
two-point function) $\Pi_{\mu\nu}(q)$ at Euclidian momentum
$Q^2=-q^2=0$. It is computed perturbatively, using operator product
expansion techniques, by relating the $B\to\pi\ell\nu$ decay amplitude
to $\ell\nu\to B\pi$ inelastic scattering via crossing symmetry and
reproducing the correct value of the inclusive $\ell\nu\to X_b$ amplitude.
We will refer to the series parameterization with the outer function
in Eq.~(\ref{eq:comp_of}) as Boyd, Grinstein, and Lebed (BGL).  The
perturbative and OPE truncations imply that the bound is not strict,
and one should take it as
\begin{gather}
\sum_{n=0}^N a_n^2 \lesssim 1\,,
\end{gather}
where this holds for any choice of $N$.  Since the values of $|z|$ in
the kinematical region of interest are well below~1 for judicious
choices of $t_0$, this provides a very stringent bound on systematic
uncertainties related to truncation for $N\geq 2$. On the other hand,
the outer function in Eq.~(\ref{eq:comp_of}) is somewhat unwieldy and,
more relevantly, spoils the correct large $q^2$ behaviour and induces
an unphysical singularity at the $B\pi$ threshold.

A simpler choice of outer function has been proposed by Bourrely,
Caprini and Lellouch (BCL) in Ref.~\cite{Bourrely:2008za}, which leads to a
parameterization of the form
\begin{gather}
\label{eq:bcl}
f_+(q^2)=\frac{1}{1-q^2/m_{B^*}^2}\,\sum_{n=0}^N a_n(t_0) z(q^2,t_0)^n\,.
\end{gather}
This satisfies all the basic properties of the form factor, at the price
of changing the expression for the bound to
\begin{gather}
\sum_{j,k=0}^N B_{jk}(t_0)a_j(t_0)a_k(t_0) \leq 1\,.
\end{gather}
The constants $B_{jk}$ can be computed and shown to be
$|B_{jk}|\lesssim \cO(10^{-2})$ for judicious choices of
$t_0$; therefore, one again finds that truncating at $N\geq 2$
provides sufficiently stringent bounds for the current level of
experimental and theoretical precision.  It is actually possible to
optimize the properties of the expansion by taking
\begin{gather}
t_0 = t_{\rm opt} = (m_B+m_\pi)(\sqrt{m_B}-\sqrt{m_\pi})^2\,,
\end{gather}
which for physical values of the masses results in the semileptonic
domain being mapped onto the symmetric interval $|z| \ltapprox 0.279$
(where this range differs slightly for the $B^{\pm}$ and $B^0$ decay
channels), minimizing the maximum truncation error.  If one also
imposes that the asymptotic behaviour ${\rm Im}\,f_+(q^2) \sim
(q^2-t_+)^{3/2}$ near threshold is satisfied, then the highest-order
coefficient is further constrained as
\begin{gather}
\label{eq:red_coeff}
a_N=-\,\frac{(-1)^N}{N}\,\sum_{n=0}^{N-1}(-1)^n\,n\,a_n\,.
\end{gather}
Substituting the above constraint on $a_N$ into Eq.~(\ref{eq:bcl})
leads to the constrained BCL parameterization
\begin{gather}
\label{eq:bcl_c}
f_+(q^2)=\frac{1}{1-q^2/m_{B^*}^2}\,\sum_{n=0}^{N-1} a_n\left[z^n-(-1)^{n-N}\,\frac{n}{N}\,z^N\right]\,,
\end{gather}
which is the standard implementation of the BCL parameterization used
in the literature.

Parameterizations of the BGL and BCL kind, to which we will refer
collectively as ``$z$-parameterizations'', have already been adopted
by the BaBar and Belle collaborations to report their results, and
also by the Heavy Flavour Averaging Group (HFAG). Some lattice
collaborations, such as FNAL/MILC and ALPHA, have already started to
report their results for form factors in this way.  The emerging trend
is to use the BCL parameterization as a standard way of presenting
results for the $q^2$ dependence of semileptonic form factors. Our
policy will be to quote results for $z$-parameterizations when the
latter are provided in the paper (including the covariance matrix of
the fits); when this is not the case, but the published form factors
include the full correlation matrix for values at different $q^2$, we
will perform our own fit to the constrained BCL ansatz
in Eq.~(\ref{eq:bcl_c}); otherwise no fit will be quoted.
We however stress the importance of providing, apart from parameterization
coefficients, values for the form factors themselves (in the continuum limit
and at physical quark masses) for a number of values of $q^2$, so that
the results can be independently parameterized by the readers if so wished.

\paragraph{Extension to other form factors}

The discussion above largely extends to the scalar form factor in $B\to\pi\ell\nu$ decays,
as well as to form factors for other semileptonic transitions (e.g., $B_s\to K$ and $B_{(s)} \to D^{(*)}_{(s)}$,
and semileptonic $D$ and $K$ decays).
As a matter of fact, after the publication of our previous review $z$-parameterizations
have been applied in several such cases, as discussed in the relevant sections.

A general discussion of semileptonic meson decay in this context can be found,
e.g., in Ref.~\cite{Hill:2006ub}. Extending what has been discussed above for
$B\to\pi$, the form factors for a generic $H \to L$
transition will display a cut starting at the production threshold $t_+$, and the optimal
value of $t_0$ required in $z$-parameterizations is $t_0=t_+(1-\sqrt{1-t_-/t_+})$
(where $t_\pm=(m_H\pm m_L)^2$).
For unitarity bounds to apply, the Blaschke factor has to include all sub-threshold
poles with the quantum numbers of the hadronic current --- i.e., vector (resp. scalar) resonances
in $B\pi$ scattering for the vector (resp. scalar) form factors of $B\to\pi$, $B_s\to K$,
or $\Lambda_b \to p$; and vector (resp. scalar) resonances
in $B_c\pi$ scattering for the vector (resp. scalar) form factors of $B\to D$
or $\Lambda_b \to \Lambda_c$.\footnote{A more complicated analytic structure
may arise in other cases, such as channels with vector mesons in the final state.
We will however not discuss form-factor parameterizations for any such process.}
Thus, as emphasized above, the control over systematic uncertainties brought in by using
$z$-parameterizations strongly depends on implementation details.
This has practical consequences, in particular, when the resonance spectrum
in a given channel is not sufficiently well-known. Caveats may also
apply for channels where resonances with a nonnegligible width appear.
A further issue is whether $t_+=(m_H+m_L)^2$ is the proper choice for the start of the cut in cases such as $B_s\to K\ell\nu$ and $B\to D\ell\nu$, where there are lighter two-particle states that project on the current ($B$,$\pi$ and $B_c$,$\pi$ for the two processes, respectively).\footnote{We are grateful
to G.~Herdo\'{\i}za, R.J.~Hill, A.~Kronfeld and A.~Szczepaniak for illuminating discussions
on this issue.}
In any such
situation, it is not clear a priori that a given $z$-parameterization will
satisfy strict bounds, as has been seen, e.g., in determinations of the proton charge radius
from electron-proton scattering~\cite{Hill:2010yb,Hill:2011wy,Epstein:2014zua}.

The HPQCD Collaboration pioneered a variation on the $z$-parameterization
approach, which they refer to as a ``modified $z$-expansion," that
is used to simultaneously extrapolate their lattice simulation data
to the physical light-quark masses and the continuum limit, and to
interpolate/extrapolate their lattice data in $q^2$.  This entails
allowing the coefficients $a_n$ to depend on the light-quark masses,
squared lattice spacing, and, in some cases the charm-quark mass and
pion or kaon energy.  Because the modified $z$-expansion is not
derived from an underlying effective field theory, there are several
potential concerns with this approach that have yet to be studied.
The most significant is that there is no theoretical
derivation relating the coefficients of the modified $z$-expansion to
those of the physical coefficients measured in experiment; it
therefore introduces an unquantified model dependence in the
form-factor shape. As a result, the applicability of unitarity bounds has to be examined carefully.
Related to this, $z$-parameterization coefficients implicitly depend on quark masses,
and particular care should be taken in the event that some state can move
across the inelastic threshold as quark masses are changed (which would
in turn also affect the form of the Blaschke factor). Also, the lattice
spacing dependence of form factors provided by Symanzik effective theory
techniques may not extend trivially to $z$-parameterization coefficients.
The modified $z$-expansion is now being utilized by collaborations
other than HPQCD and for quantities other than $D \to \pi \ell \nu$
and $D \to K \ell \nu$, where it was originally employed.
We advise treating results that utilize the modified $z$-expansion to
obtain form-factor shapes and CKM matrix elements with caution,
however, since the systematics of this approach warrant further study.

\subsubsection{Form factors for $B\to\pi\ell\nu$}
\label{sec:BtoPi}

The semileptonic decay processes $B\to\pi\ell\nu$ enable determinations of the CKM matrixelement $|V_{ub}|$
within the Standard Model via Eq.~(\ref{eq:B_semileptonic_rate}).
At the time of our previous review, the only available results for
$B\to\pi\ell\nu$ form factors came from the HPQCD~\cite{Dalgic:2006dt}
and FNAL/MILC~\cite{Bailey:2008wp} Collaborations.
Only HPQCD provided results for the scalar form factor $f_0$.
The last two years, however, have witnessed significant progress: FNAL/MILC
have significantly upgraded their $B\to\pi\ell\nu$ results~\cite{Lattice:2015tia},\footnote{Since the new FNAL/MILC results supersede Ref.~\cite{Bailey:2008wp}, we will not
discuss this latter work in the present version of the review.}
while a completely new computation has been provided by RBC/UKQCD~\cite{Flynn:2015mha}.
All the above computations employ $N_f=2+1$ dynamical configurations,
and provide values for both form factors $f_+$ and $f_0$.
Finally, HPQCD have recently published the first $N_f=2+1+1$ results for the $B\to\pi\ell\nu$ scalar
form factor, working at zero recoil and pion masses down to the physical value~\cite{Colquhoun:2015mfa};
this adds to previous reports on ongoing work to upgrade their 2006
computation~\cite{Bouchard:2012tb,Bouchard:2013zda}. Since this latter
result has no immediate impact on current $|V_{ub}|$ determinations,
which come from the vector-form-factor-dominated decay channels into light leptons,
we will from now on concentrate on the $N_f=2+1$ determinations of the
$q^2$ dependence of $B\to\pi$ form factors.

Both the HPQCD and the FNAL/MILC computations of $B\to\pi\ell\nu$
amplitudes use ensembles of gauge configurations with $N_f=2+1$
flavours of rooted staggered quarks produced by the MILC Collaboration;
however, the latest FNAL/MILC work makes a much more extensive
use of the currently available ensembles, both in terms of
lattice spacings and light-quark masses.
HPQCD have results at two values of the lattice spacing
($a\sim0.12,~0.09~{\rm fm}$), while FNAL/MILC employs four values
($a\sim0.12,~0.09,~0.06,~0.045~{\rm fm}$).
Lattice-discretization
effects are estimated within HMrS$\chi$PT in the FNAL/MILC
computation, while HPQCD quotes the results at $a\sim 0.12~{\rm fm}$
as central values and uses the $a\sim 0.09~{\rm fm}$ results to quote
an uncertainty.
The relative scale is fixed in both cases through $r_1/a$.
HPQCD set the absolute scale through the $\Upsilon$ $2S$--$1S$ splitting,
while FNAL/MILC uses a combination of $f_\pi$ and the same $\Upsilon$
splitting, as described in Ref.~\cite{Bazavov:2011aa}.
The spatial extent of the lattices employed by HPQCD is $L\simeq 2.4~{\rm fm}$,
save for the lightest mass point (at $a\sim 0.09~{\rm fm}$) for which $L\simeq 2.9~{\rm fm}$.
FNAL/MILC, on the other hand, uses extents up to $L \simeq 5.8~{\rm fm}$, in order
to allow for light pion masses while keeping finite volume effects under
control. Indeed, while in the 2006 HPQCD work the lightest RMS pion mass is $400~{\rm MeV}$,
the latest FNAL/MILC work includes pions as light as $165~{\rm MeV}$ --- in both cases
the bound $m_\pi L \gtrsim 3.8$ is kept.
Other than the qualitatively different range of MILC ensembles used
in the two computations, the main difference between HPQCD and FNAL/MILC lies in the treatment of
heavy quarks. HPQCD uses the NRQCD formalism, with a 1-loop matching
of the relevant currents to the ones in the relativistic
theory. FNAL/MILC employs the clover action with the Fermilab
interpretation, with a mostly nonperturbative renormalization of the
relevant currents, within which light-light and heavy-heavy currents
are renormalized nonperturbatively and 1-loop perturbation theory is
used for the relative normalization.  (See Tab.~\ref{tab_BtoPisumm2};
full details about the computations are provided in tables in
Appendix~\ref{app:BtoPi_Notes}.)

The RBC/UKQCD computation is based on $N_f=2+1$ DWF ensembles at two
values of the lattice spacing ($a\sim0.12,~0.09~{\rm fm}$), and pion masses
in a narrow interval ranging from slightly above $400~{\rm MeV}$ to slightly below $300~{\rm MeV}$,
keeping $m_\pi L \gtrsim 4$.
The scale is set using the $\Omega^-$ baryon mass. Discretization effects
coming from the light sector
are estimated in the $1\%$ ballpark using HM$\chi$PT supplemented with effective higher-order
interactions to describe cutoff effects.
The $b$ quark is treated using the Columbia RHQ action, with
a mostly nonperturbative renormalization of the relevant currents. Discretization
effects coming from the heavy sector are estimated with power-counting
arguments to be below $2\%$.

Given the large kinematical range available in the $B\to\pi$ transition,
chiral extrapolations are an important source of systematic uncertainty:
apart from the eventual need to reach physical pion masses in the extrapolation,
the applicability of $\chi$PT is not guaranteed for large values of the pion energy $E_\pi$.
Indeed, in all computations $E_\pi$ reaches values in the $1~{\rm GeV}$ ballpark,
and chiral extrapolation systematics is the dominant source of errors.
FNAL/MILC uses $SU(2)$ NLO HMrS$\chi$PT for the continuum-chiral extrapolation,
supplemented by NNLO analytic terms
and hard-pion $\chi$PT terms~\cite{Bijnens:2010ws};\footnote{Note that issues are known to exist with hard-pion $\chi$PT, cf. Ref.~\cite{Procura:2013dsa}.} systematic uncertainties
are estimated through an extensive study of the effects of varying the
specific fit ansatz and/or data range. RBC/UKQCD uses
$SU(2)$ hard-pion HM$\chi$PT to perform its combined continuum-chiral
extrapolation, and obtains sizeable estimates for systematic uncertainties
by varying the ans\"{a}tze and ranges used in fits. HPQCD performs chiral
extrapolations using HMrS$\chi$PT formulae, and estimates systematic
uncertainties by comparing the result with the ones from fits to a
linear behaviour in the light-quark mass, continuum HM$\chi$PT, and
partially quenched HMrS$\chi$PT formulae (including also data with
different sea and valence light-quark masses).

FNAL/MILC and RBC/UKQCD describe the $q^2$ dependence of
$f_+$ and $f_0$ by applying a BCL parameterization to
the form factors extrapolated to the continuum
limit, within the range of values of $q^2$ covered by data.
RBC/UKQCD generate synthetic data for the form factors at some values
of $q^2$ (evenly spaced in $z$) from the continuous function of $q^2$ obtained
from the joint chiral-continuum extrapolation,
which are then used as input for the fits. After having checked that the
kinematical constraint $f_+(0)=f_0(0)$ is satisfied within errors by the extrapolation
to $q^2=0$ of the results of separate fits, this constraint is imposed
to improve fit quality. In the case of FNAL/MILC, rather than producing
synthetic data a functional method is used to extract the $z$-parameterization
directly from the fit functions employed in the continuum-chiral extrapolation.
The resulting preferred fits for both works are quoted in Tab.~\ref{tab_BtoPisumm2}.
In the case of HPQCD, the parameterization of the $q^2$ dependence of form factors is
somewhat intertwined with chiral extrapolations: a set of fiducial
values $\{E_\pi^{(n)}\}$ is fixed for each value of the light-quark
mass, and $f_{+,0}$ are interpolated to each of the $E_\pi^{(n)}$;
chiral extrapolations are then performed at fixed $E_\pi$
(i.e. $m_\pi$ and $q^2$ are varied subject to $E_\pi$=constant). The
interpolation is performed using a BZ ansatz.  The $q^2$ dependence of
the resulting form factors in the chiral limit is then described by
means of a BZ ansatz, which is cross-checked against BK, RH, and BGL
parameterizations. Unfortunately, the correlation matrix for the values
of the form factors at different $q^2$ is not provided, which severely
limits the possibilities of combining them with other computations into
a global $z$-parameterization.

Based on the parameterized form factors, HPQCD and RBC/UKQCD provide
values for integrated decay rates $\Delta \zeta^{B\pi}$, as defined
in Eq.~(\ref{eq:Deltazeta}); they are quoted in Tab.~\ref{tab_BtoPisumm2}.
The latest FNAL/MILC work, on the other hand, does not quote a value
for the integrated ratio. Furthermore, as mentioned above, the field has recently moved forward
to determine CKM matrix elements from direct joint fits of experimental
results and theoretical form factors, rather than a matching through
$\Delta \zeta^{B\pi}$. Thus, we will not provide here a FLAG average for the integrated rate,
and focus on averaging lattice results for the form factors themselves.

\begin{table}[t]
\begin{center}
\mbox{} \\[3.0cm]
\footnotesize
\begin{tabular*}{\textwidth}[l]{l @{\extracolsep{\fill}} c @{\hspace{2mm}} c l l l l l l l @{\hspace{2mm}} c }
Collaboration & Ref. & $\Nf$ & 
\hspace{0.15cm}\begin{rotate}{60}{publication status}\end{rotate}\hspace{-0.15cm} &
\hspace{0.15cm}\begin{rotate}{60}{continuum extrapolation}\end{rotate}\hspace{-0.15cm} &
\hspace{0.15cm}\begin{rotate}{60}{chiral extrapolation}\end{rotate}\hspace{-0.15cm}&
\hspace{0.15cm}\begin{rotate}{60}{finite volume}\end{rotate}\hspace{-0.15cm}&
\hspace{0.15cm}\begin{rotate}{60}{renormalization}\end{rotate}\hspace{-0.15cm}  &
\hspace{0.15cm}\begin{rotate}{60}{heavy-quark treatment}\end{rotate}\hspace{-0.15cm}  &
\hspace{0.15cm}\begin{rotate}{60}{$z$-parameterization}\end{rotate}\hspace{-0.15cm} &
\rule{0.3cm}{0cm}$\Delta \zeta^{B\pi} $ \\%
&&&&&&&&&& \\[-0.0cm]
\hline
\hline
&&&&&&&&&& \\[-0.0cm]
\SLfnalmilcBpi & \cite{Lattice:2015tia} & 2+1 & \gA  & \good & \soso & \good & \soso & \okay &
 BCL & n/a\\[-0.0cm]
\SLrbcukqcdBpi & \cite{Flynn:2015mha} & 2+1 & \gA  & \soso & \soso & \soso & \soso & \okay &
 BCL & $\quad 1.77(34)$ \\[-0.0cm]
HPQCD 06 & \cite{Dalgic:2006dt} & 2+1 & \gA  & \soso & \soso & \soso
& \soso & \okay &
 n/a & $\quad $2.07(41)(39) \\[-0.0cm]
&&&&&&&&&& \\[-0.0cm]
\hline
\hline
\end{tabular*}
\caption{Results for the $B \to \pi\ell\nu$ semileptonic form factor.  The quantity $\Delta\zeta$ is defined in Eq.~(\ref{eq:Deltazeta}); the quoted values correspond to $q_1=4$~GeV, $q_2=q_{max}$, and are given in $\mbox{ps}^{-1}$.
\label{tab_BtoPisumm2}}
\end{center}
\end{table}

In our previous review, we averaged the results for $f_+(q^2)$ in HPQCD~06
and the superseded FNAL/MILC~2008 determination~\cite{Bailey:2008wp},
fitting them jointly to our preferred BCL $z$-parameterization,
Eq.~(\ref{eq:bcl_c}). The new results do not, however, allow for an
update of such a joint fit: RBC/UKQCD only provides synthetic
values of $f_+$ and $f_0$ at a few values of $q^2$ as an illustration
of their results, and FNAL/MILC does not quote synthetic values at all.
In both cases, full results for BCL $z$-parameterizations defined by
Eq.~(\ref{eq:bcl_c}) are quoted.
In the case of HPQCD~06, unfortunately,
a fit to a BCL $z$-parameterization is not possible, as discussed above.
Our procedure to take into account all the available computations will thus
proceed as follows: we will take as input the synthetic values for $f_+(q^2)$
quoted by RBC/UKQCD, as well as the HPQCD datum at $q^2=17.34~\GeV^2$; and will
add to these synthetic values for $f_+(q^2)$ that we generate using FNAL/MILC's
preferred $z$-parameterization. The RBC/UKQCD data are treated as fully uncorrelated
to the other determinations, while a conservative $100\%$ correlation of statistical
errors is introduced for FNAL/MILC and HPQCD, which are both based on MILC $N_f=2+1$
ensembles.

\begin{figure}[tbp]
\begin{center}
\setbox1=\hbox{\includegraphics[width=0.55\textwidth]{HQ/Figures/fp_Bpi_latt.eps}}
\begin{minipage}{0.55\textwidth}
\includegraphics[width=\textwidth]{HQ/Figures/fp_Bpi_latt.eps}
\end{minipage}
\begin{minipage}{0.55\textwidth}
   \llap{\makebox[\wd1][l]{\raisebox{66mm}{
   \hspace{6mm}\includegraphics[width=0.15\textwidth]{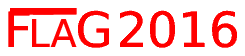}
   }}}
\end{minipage}
\vspace{-2mm}
\caption{The form factor $(1 - q^2/m_{B^*}^2) f_+(q^2)$ for $B \to \pi\ell\nu$ plotted versus $z$.
(See text for a discussion of the dataset.)
The grey band displays our preferred three-parameter BCL fit to the plotted data with errors
(see Eq.~(\ref{eq:BtoPiLatBCLFit})).}\label{fig:LQCDzfit}
\end{center}
\end{figure}

The resulting dataset is then fitted to the BCL parameterization
in Eq.~(\ref{eq:bcl_c}). We assess the systematic
uncertainty due to truncating the series expansion by considering fits
to different orders in $z$.  Fig.~\ref{fig:LQCDzfit} plots the
FNAL/MILC, RBC/UKQCD, and HPQCD data points for $(1-q^2/m_{B^*}^2) f_+(q^2)$
versus $z$; the data is highly linear, and only a simple two-parameter
fit is needed for a good $\chi^2/{\rm d.o.f.}$. (Note that a fit to
the constrained BCL form in Eq.~(\ref{eq:bcl_c}) with two free
parameters corresponds to a polynomial through $\cO(z^2)$,
etc.)  Further, we cannot constrain the coefficients of the
$z$-expansion beyond this order, as evidenced by the error on the
coefficient $a_2$ being significantly greater than 100\% for a
three-parameter fit. 
We quote as our preferred result the outcome of the three-parameter
$\cO(z^3)$ BCL fit:
 \begin{gather}
N_f=2+1:  \qquad 
a_0 = 0.421(13)\,,~~~~
a_1 = -0.35(10)\,,~~~~
a_2 = -0.41(64)\,;\label{eq:BtoPiLatBCLFit}\\[1.0ex]
\nonumber \qquad\qquad
{\rm corr}(a_i,a_j)=\left(\begin{array}{rrr}
 1.000 &  0.306 &  0.084 \\
 0.306 &  1.000 &  0.856 \\
 0.084 &  0.856 &  1.000
\end{array}\right)\,.
\end{gather}
The uncertainties on $a_0$ and $a_1$ encompass the central values
obtained from $\cO(z^2)$ and $\cO(z^4)$ fits,
and thus adequately reflect the systematic uncertainty on those
series coefficients. This can be
used as the averaged FLAG result for the lattice-computed form factor
$f_+(q^2)$. The coefficient $a_3$ can be obtained from the values for
$a_0$--$a_2$ using Eq.~(\ref{eq:red_coeff}). The fit is illustrated
in Fig.~\ref{fig:LQCDzfit}.

It is worth stressing that, with respect to our average in the previous
edition of the FLAG report, the relative error on $a_0$, which dominates
the theory contribution to the determination of $|V_{ub}|$, has
decreased from $7.3\%$ to $3.1\%$. The dominant factor in this
remarkable improvement is the new FNAL/MILC determination of $f_+$.
We emphasize that future
lattice-QCD calculations of semileptonic form factors should publish
their full statistical and systematic correlation matrices to enable
others to use the data. It is also preferable to have direct
access to values of the form factors, since this allows for an independent
analysis that avoids further assumptions about the compatibility of
the procedures to arrive at a given $z$-parameterization.

The above procedure could in principle be extended to $f_0$.  However,
in this case significant inconsistencies among published data exist:
while the results of RBC/UKQCD and FNAL/MILC for $f_0$ are compatible
within the quoted errors, the results in HPQCD~06 are incompatible
with the less precise RBC/UKQCD determination at the $\sim 2\sigma$
level, and with the similarly precise FNAL/MILC determination at more
than $3\sigma$, cf. the right panel of Fig.~18
in Ref.~\cite{Lattice:2015tia}.  This situation is very different with
respect to the one for $f_+$, where the three determinations are
consistent within errors, cf. the left panel of Fig.~18
in Ref.~\cite{Lattice:2015tia}.  On the other hand, $f_0$ does not enter
current determinations of $|V_{ub}|$ in $B\to\pi$ transitions, due to
the lack of an experimental measurement of $B\to\pi\tau\nu$ processes.
As a consequence, we will refrain from providing an average of results
for $f_0$ until the reasons of the aforementioned discrepancy are
clarified.\footnote{One consistency check of the form-factor
  computation consists in verifying whether the kinematical constraint
  $f_+(0)=f_0(0)$ is satisfied. This is however of limited value in
  the existing computations, since the $q^2=0$ point is reached by a
  long extrapolation from the covered region in momentum transfer.
  RBC/UKQCD~15 and FNAL/MILC~15 do check that $f_+(0)=f_0(0)$ within
  the resulting large uncertainties, while HPQCD~06 make use of the
  $f_+(0)=f_0(0)$ as part of their fits for the $q^2$ dependence.}

\subsubsection{Form factors for $B_s\to K\ell\nu$}
\label{sec:BtoK}

Similar to $B\to\pi\ell\nu$, measurements of $B_s\to K\ell\nu$ enable determinations
of the CKM matrix element $|V_{ub}|$
within the Standard Model via Eq.~(\ref{eq:B_semileptonic_rate}).
From the lattice point of view the two channels are very similar ---
as a matter of fact, $B_s\to K\ell\nu$ is actually somewhat simpler,
in that the fact that the kaon mass region is easily accessed by all simulations
makes the systematic uncertainties related to chiral extrapolation
smaller. On the other hand, $B_s\to K\ell\nu$ channels have not been measured
experimentally yet, and therefore lattice results provide SM predictions
for the relevant rates.

At the time of our previous review, only preliminary
results existed for $B_s \to K\ell\nu$ form factors.
However, as with $B \to \pi\ell\nu$, great progress has been made
during the last year, and first full results for $B_s\to K\ell\nu$ form factors have been
provided by HPQCD~\cite{Bouchard:2014ypa} and RBC/UKQCD~\cite{Lattice:2015tia}
for both form factors $f_+$ and $f_0$, in both cases using $N_f=2+1$ dynamical configurations.
Finally, the ALPHA Collaboration determination of $B_s\to K\ell\nu$
form factors with $N_f=2$ is also well underway~\cite{Bahr:2014iqa};
however, since the latter is so far described only in
conference proceedings which do not provide quotable results, it
will not be discussed here.

The RBC/UKQCD computation has been published together with the $B\to\pi\ell\nu$
computation discussed in Sec.~\ref{sec:BtoPi}, all technical details being
practically identical. The main difference is that errors are significantly smaller,
mostly due to the reduction of systematic uncertainties due to the chiral extrapolation;
detailed information is provided in tables in Appendix~\ref{app:BtoPi_Notes}.
The HPQCD computation uses ensembles of gauge configurations with $N_f=2+1$
flavours of rooted staggered quarks produced by the MILC Collaboration
at two values of the lattice spacing ($a\sim0.12,~0.09~{\rm fm}$), for three
and two different sea-pion masses, respectively, down to a value of $260~{\rm MeV}$.
The $b$ quark is treated within the NRQCD formalism, with a 1-loop matching
of the relevant currents to the ones in the relativistic theory, omitting terms
of $\cO(\alpha_s\Lambda_{\rm QCD}/m_b)$. A HISQ action
is used for the valence $s$ quark. The continuum-chiral extrapolation
is combined with the description of the $q^2$ dependence of the form factors
into a modified $z$-expansion (cf.~Sec.~\ref{sec:zparam}) that formally coincides
in the continuum with the BCL ansatz. The dependence of
form factors on the pion energy and quark masses is fitted to a 1-loop ansatz
inspired by hard-pion $\chi$PT~\cite{Bijnens:2010ws},
that factorizes out the chiral logarithms describing soft physics.
See Tab.~\ref{tab_BstoKsumm} and the tables in Appendix~\ref{app:BtoPi_Notes} for full details.

\begin{table}[t]
\begin{center}
\mbox{} \\[3.0cm]
\footnotesize
\begin{tabular*}{\textwidth}[l]{l @{\extracolsep{\fill}} c @{\hspace{2mm}} c l l l l l l l l }
Collaboration & Ref. & $\Nf$ & 
\hspace{0.15cm}\begin{rotate}{60}{publication status}\end{rotate}\hspace{-0.15cm} &
\hspace{0.15cm}\begin{rotate}{60}{continuum extrapolation}\end{rotate}\hspace{-0.15cm} &
\hspace{0.15cm}\begin{rotate}{60}{chiral extrapolation}\end{rotate}\hspace{-0.15cm}&
\hspace{0.15cm}\begin{rotate}{60}{finite volume}\end{rotate}\hspace{-0.15cm}&
\hspace{0.15cm}\begin{rotate}{60}{renormalization}\end{rotate}\hspace{-0.15cm}  &
\hspace{0.15cm}\begin{rotate}{60}{heavy-quark treatment}\end{rotate}\hspace{-0.15cm}  &
\hspace{0.15cm}\begin{rotate}{60}{$z$-parameterization}\end{rotate}\hspace{-0.15cm}\\%
&&&&&&&&& \\[-0.0cm]
\hline
\hline
&&&&&&&&& \\[-0.0cm]
\SLrbcukqcdBpi & \cite{Flynn:2015mha} & 2+1 & \gA  & \soso & \soso & \soso & \soso & \okay &
BCL \\[-0.0cm]
\SLhpqcdBsK & \cite{Bouchard:2014ypa} & 2+1 & \gA  & \soso & \soso & \soso & \soso & \okay &
BCL$^\dagger$   \\[-0.0cm]
&&&&&&&&& \\[-0.0cm]
\hline
\hline\\
\end{tabular*}\\[-0.2cm]
\begin{minipage}{\linewidth}
{\footnotesize 
\begin{itemize}
   \item[$^\dagger$] Results from modified $z$-expansion. 
\end{itemize}
}
\end{minipage}
\caption{Results for the $B_s \to K\ell\nu$ semileptonic form factor. 
\label{tab_BstoKsumm}}
\end{center}
\end{table}

Both RBC/UKQCD and HPQCD quote values for integrated differential decay rates over
the full kinematically available region. However,
since the absence of experiment makes the relevant integration interval subject to change,
we will not discuss them here, and focus on averages of form factors. In order to proceed
to combine the results from the two collaborations, we will follow a similar approach to
the one adopted above for $B\to\pi\ell\nu$: we will take as direct input the synthetic values
of the form factors provided by RBC/UKQCD, use the preferred HPQCD parameterization
to produce synthetic values, and perform a joint fit to the two datasets.
Note that, contrary to $B\to\pi\ell\nu$, there are no significant discrepancies
that prevent a meaningful averaging; this allows us to provide results that are relevant
for predictions concerning the channel $B_s\to K\tau\nu$.
The fits will be fully independent for $f_+$ and $f_0$ --- in particular, we will not impose
kinematical constraints at $q^2=0$ --- and we will treat the two
datasets as completely uncorrelated. Whenever our averages for $f_+$ and $f_0$
are used together, we recommend to conservatively take a 100\% correlation between
the corresponding errors.

\begin{figure}[tbp]
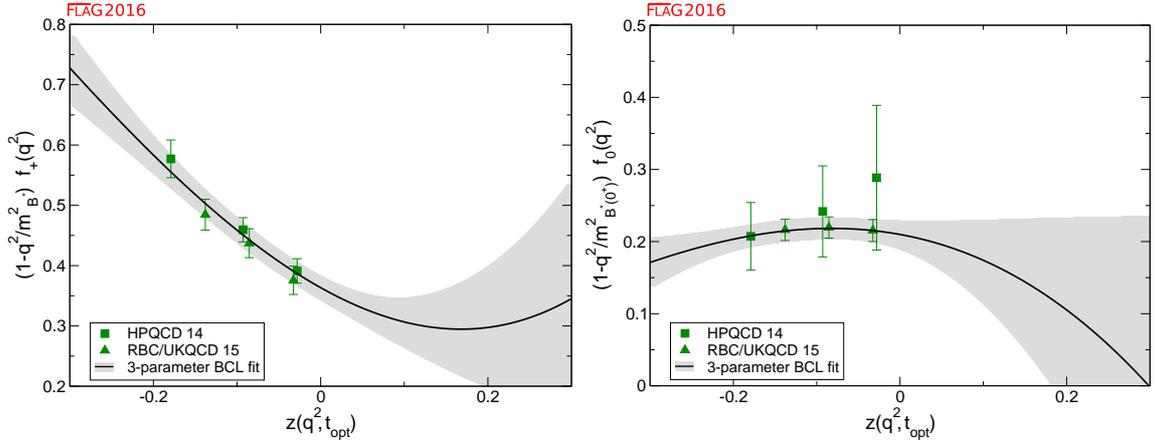

\begin{center}
\setbox1=\hbox{\includegraphics[width=0.48\textwidth]{HQ/Figures/fp_BsK_latt.eps}}
\setbox2=\hbox{\includegraphics[width=0.48\textwidth]{HQ/Figures/f0_BsK_latt.eps}}
\begin{minipage}{0.48\textwidth}
\includegraphics[width=\textwidth]{HQ/Figures/fp_BsK_latt.eps}
\end{minipage}
\begin{minipage}{0.48\textwidth}
\includegraphics[width=\textwidth]{HQ/Figures/f0_BsK_latt.eps}
\end{minipage}
\begin{minipage}{0.48\textwidth}
   \llap{\makebox[\wd1][l]{\raisebox{58mm}{
   \hspace{-71mm}\includegraphics[width=0.15\textwidth]{HQ/Figures/FLAG_Logo.eps}
   }}}
\end{minipage}
\begin{minipage}{0.48\textwidth}
   \llap{\makebox[\wd2][l]{\raisebox{58mm}{
   \hspace{4mm}\includegraphics[width=0.15\textwidth]{HQ/Figures/FLAG_Logo.eps}
   }}}
\end{minipage}
\vspace{-3mm}
\caption{The form factors $(1 - q^2/m_{B^*}^2) f_+(q^2)$
(left) and $(1 - q^2/m_{B^*(0+)}^2) f_0(q^2)$ (right) for $B_s \to K\ell\nu$ plotted versus $z$.
(See text for a discussion of the datasets.)
The grey band displays our preferred three-parameter BCL fit to the plotted data with errors
(see Eqs.~(\ref{eq:BstoKLatBCLFitvector},\ref{eq:BstoKLatBCLFitscalar})).}
\label{fig:LQCDzfitBsK}
\end{center}
\end{figure}

For the fits we employ a BCL ansatz with $t_+=(M_{B_s}+M_{K^\pm})^2 \simeq 34.35~\GeV^2$ and
$t_0=(M_{B_s}+M_{K^\pm})(\sqrt{M_{B_s}}-\sqrt{M_{K^\pm}})^2 \simeq 15.27~\GeV^2$.
Our pole factors will contain a single pole in both the vector and scalar
channels, for which we take the mass values $M_{B^*}=5.325~\GeV$
and $M_{B^*(0^+)}=5.65~\GeV$.\footnote{The values of the scalar resonance mass
in $B\pi$ scattering taken by HPQCD and RBC/UKQCD are $M_{B^*(0+)}=5.6794(10)~\GeV$
and $M_{B^*(0+)}=5.63~\GeV$, respectively. We use an average of the two values,
and have checked that changing it by $\sim 1\%$ has a negligible impact on the fit results.}
We quote as our preferred result the outcome of the three-parameter
$\cO(z^3)$ BCL fit:
\begin{gather}
N_f=2+1:  \qquad 
a_0^{(+)} = 0.363(16)\,,~~~~
a_1^{(+)} = -0.78(19)\,,~~~~
a_2^{(+)} = 1.9(1.3)\,;\label{eq:BstoKLatBCLFitvector}\\[1.0ex]
\nonumber \qquad\qquad
{\rm corr}(a_i^{(+)},a_j^{(+)})=\left(\begin{array}{rrr}
 1.000 &  0.343 &  0.220 \\
 0.343 &  1.000 &  0.874 \\
 0.220 &  0.874 &  1.000
\end{array}\right)
\end{gather}
for the vector form factor, and
\begin{gather}
N_f=2+1:  \qquad 
a_0^{(0)} = 0.210(18)\,,~~~~
a_1^{(0)} = -0.21(28)\,,~~~~
a_2^{(0)} = -1.4(1.3)\,;\label{eq:BstoKLatBCLFitscalar}\\[1.0ex]
\nonumber \qquad\qquad
{\rm corr}(a_i^{(0)},a_j^{(0)})=\left(\begin{array}{rrr}
 1.000 &  0.306 &  0.084 \\
 0.306 &  1.000 &  0.856 \\
 0.084 &  0.856 &  1.000
\end{array}\right)\,,
\end{gather}
where the uncertainties on $a_0$ and $a_1$ encompass the central values
obtained from $\cO(z^2)$ fits,
and thus adequately reflect the systematic uncertainty on those
series coefficients.\footnote{In this case, $\cO(z^4)$ fits with just
two degrees of freedom, are significantly less stable. Still,
the results for $a_0$ and $a_1$ are always compatible with the ones
at $\cO(z^2)$ and $\cO(z^3)$ within one standard deviation.}
These can be
used as the averaged FLAG results for the lattice-computed form factors
$f_+(q^2)$ and $f_0(q^2)$. The coefficient $a_3$ can be obtained from the values for
$a_0$--$a_2$ using Eq.~(\ref{eq:red_coeff}). The fit is illustrated
in Fig.~\ref{fig:LQCDzfitBsK}.

\subsubsection{Form factors for rare and radiative $B$-semileptonic decays to light flavours}

Lattice-QCD input is also available for some exclusive semileptonic
decay channels involving neutral-current $b\to q$ transitions at the
quark level, where $q=d,s$. Being forbidden at tree level in the SM, these processes
allow for stringent tests of potential new physics; simple examples
are $B\to K^*\gamma$, $B\to K^{(*)}\ell^+\ell^-$, or $B\to\pi\ell^+\ell^-$ where the $B$
meson (and therefore the light meson in the final state) can be either neutral or charged.

The corresponding SM effective
weak Hamiltonian is considerably more complicated than the one for the
tree-level processes discussed above: after neglecting top-quark
effects, as many as ten dimension-six operators formed by the product
of two hadronic currents or one hadronic and one leptonic current
appear.\footnote{See, e.g., Ref.~\cite{Antonelli:2009ws} and references
therein.}  Three of the latter, coming from penguin and box diagrams,
dominate at short distances; supplementing this with a combination of high-energy
OPE arguments and results from Soft Collinear Effective Theory relevant
at intermediate energies, it is possible to argue that their contributions
are still dominant when long-distance physics is also taken into account.
Within this approximation, the relevant long-distance contribution thus
consists of matrix elements of current operators (vector,
tensor, and axial-vector) between one-hadron states, which in turn can
be parameterized in terms of a number of form factors
(see Ref.~\cite{Liu:2009dj} for a complete description).
On top of the aforementioned approximations, the
lattice computation of the relevant form factors in channels with a
vector meson in the final state faces extra challenges on top of those
already present in the case of a pseudoscalar meson: the state
is unstable, and the extraction of the relevant matrix element from
correlation functions is significantly more complicated; $\chi$PT
cannot be used as a guide to extrapolate results at unphysically heavy
pion masses to the chiral limit. While the field theory procedures to take
resonance effects into account are available~\cite{Luscher:1986pf,Luscher:1990ux,Luscher:1991cf,Lage:2009zv,Bernard:2010fp,Doring:2011vk,Hansen:2012tf,Briceno:2012yi,Dudek:2014qha}, they have not yet
been implemented in the existing preliminary computations, which therefore
suffer from uncontrolled systematic errors in
calculations of weak decay form factors into unstable vector meson
final states, such as the $K^*$ or $\rho$ mesons.\footnote{In cases such as $B\to D^*$
transitions, that will be discussed below, this is much less of a practical
problem due to the very narrow nature of the resonance.}

As a consequence of the complexity of the problem, the level of maturity
of these computations is significantly below the one present in the SM tree-level
decays discussed in other sections of this review. Therefore, we will only
provide below a short guide to the existing results, without attempting to obtain
averages for the known long-distance contributions to the relevant amplitudes.

In channels with pseudoscalar mesons in the final state, there are results
for the vector, scalar, and tensor form factors for $B_s\to K\ell^+\ell^-$ decays
by HPQCD~\cite{Bouchard:2013pna}, and (very recent) results for both $B\to\pi\ell^+\ell^-$~\cite{Bailey:2015nbd}
and $B_s\to K\ell^+\ell^-$~\cite{Bailey:2015dka} from FNAL/MILC.
Both computations
employ MILC $N_f=2+1$ asqtad ensembles. HPQCD has also a companion
paper~\cite{Bouchard:2013mia} in which they calculate the
Standard Model predictions for the differential branching fractions
and other observables and compare to experiment.
The HPQCD computation employs NRQCD $b$ quarks and
HISQ valence light quarks, and parameterizes the form
factors over the full kinematic range using a model-independent
$z$-expansion as in Sec.~\ref{sec:zparam}, including the covariance matrix of the fit coefficients.
In the case of the (separate) FNAL/MILC computations, both of them
use Fermilab $b$ quarks and asqtad light quarks, and a BCL $z$-parameterization of the form factors.

Concerning channels with vector mesons in the final state, Horgan {\it et al.}
have obtained the seven form factors governing $B \to K^* \ell^+
\ell^-$ (as well as those for $B_s \to \phi\, \ell^+ \ell^-$) in
Ref.~\cite{Horgan:2013hoa} using NRQCD $b$ quarks and asqtad staggered
light quarks.  In this work, they use a modified $z$-expansion to
simultaneously extrapolate to the physical light-quark masses and
continuum and extrapolate in $q^2$ to the full kinematic range.  As
discussed in Sec.~\ref{sec:DtoPiK}, the modified $z$-expansion is
not based on an underlying effective theory, and the associated
uncertainties have yet to be fully studied.  Horgan {\it et al.} use
their form-factor results to calculate the differential branching
fractions and angular distributions and discuss the implications for
phenomenology in a companion paper~\cite{Horgan:2013pva}. Finally,
ongoing work on $B\to K^*\ell^+\ell^-$ and $B_s\to \phi\ell^+\ell^-$
by RBC/UKQCD, including first results, have recently been reported in Ref.~\cite{Flynn:2015ynk}.

\subsection{Semileptonic form factors for $B \to D \ell \nu$, $B \to D^*  \ell \nu$, and $B \to D \tau \nu$}
\label{sec:BtoD}

The semileptonic processes $ B \rightarrow D \ell \nu$ and
$B \rightarrow D^* \ell \nu$ have been studied
extensively by experimentalists and theorists over the years.  They
allow for the determination of the CKM matrix element $|V_{cb}|$, an
extremely important parameter of the Standard Model. $|V_{cb}|$
appears in many quantities that serve as inputs into CKM Unitarity
Triangle analyses and reducing its uncertainties is of paramount
importance.  For example, when $\epsilon_K$, the measure of indirect
$CP$ violation in the neutral kaon system, is written in terms of the
parameters $\rho$ and $\eta$ that specify the apex of the unitarity
triangle, a factor of $|V_{cb}|^4$ multiplies the dominant term.  As a
result, the errors coming from $|V_{cb}|$ (and not those from $B_K$)
are now the dominant uncertainty in the Standard Model (SM) prediction
for this quantity.

The decay rates for $B \rightarrow D^{(*)}\ell\nu$ can be parameterized in terms of
vector and scalar form factors in the same way as, e.g., $B\to\pi\ell\nu$, see Sec.~\ref{sec:BtoPiK}.
Traditionally, the light channels $\ell=e,~\mu$ have however been dealt with using a somewhat
different notation, viz.
\begin{eqnarray}
    \frac{d\Gamma_{B^-\to D^{0} \ell^-\bar{\nu}}}{dw} & = &
        \frac{G^2_{\rm F} m^3_{D}}{48\pi^3}(m_B+m_{D})^2(w^2-1)^{3/2}  |\eta_\mathrm{EW}|^2|V_{cb}|^2 |\mathcal{G}(w)|^2,
    \label{eq:vxb:BtoD} \\
    \frac{d\Gamma_{B^-\to D^{0*}\ell^-\bar{\nu}}}{dw} & = &
        \frac{G^2_{\rm F} m^3_{D^*}}{4\pi^3}(m_B-m_{D^*})^2(w^2-1)^{1/2}  |\eta_\mathrm{EW}|^2|V_{cb}|^2\chi(w)|\mathcal{F}(w)|^2 ,
    \label{eq:vxb:BtoDstar}
\end{eqnarray}
where $w \equiv v_B \cdot v_{D^{(*)}}$, $v_P=p_P/m_P$ are the
four-velocities of the mesons, and $\eta_\mathrm{EW}=1.0066$
 is the 1-loop electroweak correction~\cite{Sirlin:1981ie}. The
 function $\chi(w)$ in Eq.~(\ref{eq:vxb:BtoDstar}) depends upon the
 recoil $w$ and the meson masses, and reduces to unity at zero
 recoil~\cite{Antonelli:2009ws}.  These formulas do not include terms
 that are proportional to the lepton mass squared, which can be
 neglected for $\ell = e, \mu$.
Until recently, most unquenched lattice calculations for $B \rightarrow D^* \ell \nu$ and
$B \rightarrow D \ell \nu$ decays focused on the form
factors at zero recoil ${\cal F}^{B \rightarrow D^*}(1)$ and ${\cal G}^{B \rightarrow D}(1)$;
these can then be combined with experimental input to extract $|V_{cb}|$.
The main reasons for concentrating on the zero recoil point are that
(i) the decay rate then depends on a single form factor, and (ii) for
$B \rightarrow D^*\ell\nu$, there are no $\cO(\Lambda_{QCD}/m_Q)$
contributions due to Luke's theorem~\cite{Luke:1990eg}. Further, the zero recoil form
factor can be computed via a double ratio in which most of the current
renormalization cancels and heavy-quark discretization errors are
suppressed by an additional power of $\Lambda_{QCD}/m_Q$.
Recent work on $B \rightarrow D^{(*)}\ell\nu$ transitions
has started to explore the dependence of the relevant form factors on the
momentum transfer, using a similar methodology to the one employed
in $B\to\pi\ell\nu$ transitions; we refer the reader to Sec.~\ref{sec:BtoPiK}
for a detailed discussion.

At the time of the previous version of this review, there were no published complete
computations of the form factors for $B \rightarrow D\ell\nu$ decays: $N_f=2+1$
results by FNAL/MILC for ${\cal G}^{B \rightarrow D}(1)$ had only appeared
in proceedings form~\cite{Okamoto:2004xg,Qiu:2013ofa}, while the (now published) $N_f=2$ study
by Atoui {\it et al.}~\cite{Atoui:2013zza}, that in addition to providing ${\cal G}^{B \rightarrow D}(1)$
explores the $w>1$ region, was still in preprint form. This latter work also 
provided the first results for $B_s \rightarrow D_s\ell\nu$
amplitudes, again including information about the momentum transfer dependence;
this will allow for an independent determination of $|V_{cb}|$ as soon as
experimental data are available for these transitions.
Meanwhile, the only fully published unquenched results for ${\cal F}^{B \rightarrow D^*}(1)$,
obtained by FNAL/MILC, dated from 2008~\cite{Bernard:2008dn}.
In the last two years, however, significant progress has been attained in $N_f=2+1$ computations:
the FNAL/MILC value for ${\cal F}^{B \rightarrow D^*}(1)$ has been updated
in Ref.~\cite{Bailey:2014tva}, and full results for $B \rightarrow D\ell\nu$
at $w \geq 1$ have been published by FNAL/MILC~\cite{Lattice:2015rga} and HPQCD~\cite{Na:2015kha}.
These works also provide full results for the scalar form factor, allowing us to analyze the decay in the $\tau$ channel.
In the discussion below, we will only refer to this latest generation
of results, which supersedes previous $N_f=2+1$ determinations and allows
for an extraction of $|V_{cb}|$ that incorporates information about the $q^2$ dependence
of the decay rate (cf.~Sec.~\ref{sec:Vcb}).

\subsubsection{ $B_{(s)} \rightarrow D_{(s)}$ decays}

We will first discuss the $N_f=2+1$ computations of $B \rightarrow D \ell \nu$
by FNAL/MILC and HPQCD mentioned above, both based on MILC asqtad ensembles.
Full details about all the computations are provided in Tab.~\ref{tab_BtoDStarsumm2}
and in the tables in App.~\ref{app:BtoD_Notes}.

The FNAL/MILC study~\cite{Lattice:2015rga} employs ensembles at four values of the lattice
spacing ranging between approximately $0.045~{\rm fm}$
and $0.12~{\rm fm}$, and several values of the light-quark mass corresponding to pions
with RMS masses ranging between $260~{\rm MeV}$ and $670~{\rm MeV}$ (with just
one ensemble with $M_\pi^{\rm RMS} \simeq 330~{\rm MeV}$ at the finest lattice spacing).
The $b$ and $c$ quarks are treated using the Fermilab approach.
The quantities directly studied are the form factors $h_\pm$
defined by
\begin{equation}
\frac{\langle D(p_D)| i\bar c \gamma_\mu b| B(p_B)\rangle}{\sqrt{m_D m_B}} =
h_+(w)(v_B+v_D)_\mu\,+\,h_-(w)(v_B-v_D)_\mu\,,
\end{equation}
which are related to the standard vector and scalar form factors by
\begin{equation}
f_+(q^2) = \frac{1}{2\sqrt{r}}\,\left[(1+r)h_+(w)-(1-r)h_-(w)\right]\,,~~~~
f_0(q^2) = \sqrt{r}\left[\frac{1+w}{1+r}\,h_+(w)\,+\,\frac{1-w}{1-r}\,h_-(w)\right]\,,
\end{equation}
with $r=m_D/m_B$. (Recall that
$q^2=(p_B-p_D)^2=m_B^2+m_D^2-2wm_Bm_D$.)  The hadronic form factor
relevant for experiment, $\mathcal{G}(w)$, is then obtained from the
relation $\mathcal{G}(w)=4rf_+(q^2)/(1+r)$. The form factors are
obtained from double ratios of three-point functions in which the
flavour-conserving current renormalization factors cancel. The
remaining matching factor $\rho_{V^\mu_{cb}}$ is estimated with
1-loop lattice perturbation theory.
In order to obtain $h_\pm(w)$, a joint continuum-chiral fit is performed
to an ansatz that
contains the light-quark mass and lattice spacing dependence predicted
by next-to-leading order HMrS$\chi$PT,
and the leading dependence on $m_c$
predicted by the heavy-quark expansion ($1/m_c^2$ for $h_+$ and
$1/m_c$ for $h_-$). The $w$-dependence, which allows for an
interpolation in $w$, is given by analytic terms up to $(1-w)^2$, as
well as a contribution from the log proportional to $g^2_{D^*D\pi}$.
The total resulting systematic error is $1.2\%$ for $f_+$ and $1.1\%$ for $f_0$.
This dominates the final error budget for the form factors.
After $f_+$ and $f_0$ have been determined as functions of $w$ within the interval
of values of $q^2$ covered by the computation, synthetic data points are
generated to be subsequently fitted to a $z$-expansion of the BGL form, cf.~Sec.~\ref{sec:BtoPiK},
with pole factors set to unity.
This in turn enables one to determine $|V_{cb}|$ from a joint fit of this $z$-expansion
and experimental data. The value of the zero-recoil form factor resulting
from the $z$-expansion is
\begin{equation}
{\cal G}^{B \rightarrow D}(1)=1.1054(4)_{\rm stat}(8)_{\rm sys}\,.
\end{equation}

The HPQCD computation~\cite{Na:2015kha} considers ensembles at two values of the lattice
spacing, $a=0.09,~0.12~{\rm fm}$, and two and three values of light-quark masses, respectively.
The $b$ quark is treated using NRQCD, while for the $c$ quark the HISQ action is used.
The form factors studied, extracted from suitable three-point functions, are
\begin{equation}
\langle D(p_D)| V^0 | B\rangle = \sqrt{2M_B}f_\parallel\,,~~~~~~~~
\langle D(p_D)| V^k | B\rangle = \sqrt{2M_B}p^k_D f_\perp\,,
\end{equation}
where $V_\mu$ is the relevant vector current and the $B$ rest frame is assumed.
The standard vector and scalar form factors are retrieved as
\begin{equation}
f_+ = \frac{1}{\sqrt{2M_B}}f_\parallel \,+\, \frac{1}{\sqrt{2M_B}}(M_B-E_D)f_\perp\,,~~~~
f_0 = \frac{\sqrt{2M_B}}{M_B^2-M_D^2}\left[(M_B-E_D)f_\parallel+(M_B^2-E_D^2)f_\perp\right]\,.
\end{equation}
The currents in the effective theory are matched at 1-loop to their continuum
counterparts. Results for the form factors are then fitted to a modified BCL $z$-expansion
ansatz, that takes into account simultaneously the lattice spacing, light-quark masses,
and $q^2$ dependence. For the mass dependence NLO chiral logs are included, in the
form obtained in hard-pion $\chi$PT. As in the case of the FNAL/MILC computation,
once $f_+$ and $f_0$ have been determined as functions of $q^2$, $|V_{cb}|$ can
be determined from a joint fit of this $z$-expansion and experimental data.
The work quotes for the zero-recoil vector form factor the result
\begin{equation}
{\cal G}^{B \rightarrow D}(1)=1.035(40)\,.
\end{equation}
This value is 1.8$\sigma$ smaller than the FNAL/MILC result and significantly less precise.
The dominant source of errors in the $|V_{cb}|$ determination by HPQCD are discretization
effects and the systematic uncertainty associated with the perturbative matching.

In order to combine the form factors determinations of HPQCD and FNAL/MILC
into a lattice average, we proceed in a similar way as with $B\to\pi\ell\nu$
and $B_s\to K\ell\nu$ above. FNAL/MILC quotes synthetic values for the
form factors at three values of $w$ (or, alternatively, $q^2$) with a full
correlation matrix, which we take directly as input. In the case of HPQCD,
we use their preferred modified $z$-expansion parameterization to produce
synthetic values of the form factors at two different values of $q^2$.
This leaves us with a total of five data points in the kinematical
range $w\in[1.00,1.11]$. As in the case of $B\to\pi\ell\nu$, we conservatively
assume a 100\% correlation of statistical uncertainties between HPQCD
and FNAL/MILC. We then fit this dataset to a BCL ansatz, using 
$t_+=(M_{B^0}+M_{D^\pm})^2 \simeq 51.12~\GeV^2$ and
$t_0=(M_{B^0}+M_{D^\pm})(\sqrt{M_{B^0}}-\sqrt{M_{D^\pm}})^2 \simeq 6.19~\GeV^2$.
In our fits, pole factors have been set to unity --- i.e., we do not
take into account the effect of sub-threshold poles, which is then
implicitly absorbed into the series coefficients. The reason for this
is our imperfect knowledge of the relevant resonance spectrum in this channel,
which does not allow us to decide the precise number of poles needed.\footnote{As noted
above, this is the same approach adopted by FNAL/MILC in their fits to a BGL
ansatz. HPQCD, meanwhile, uses one single pole in the pole factors that
enter their modified $z$-expansion, using their spectral studies to fix
the value of the relevant resonance masses.}
This in turn implies that unitarity bounds do not rigorously apply,
which has to be taken into account when interpreting the results (cf. Sec.~\ref{sec:zparam}).

\begin{figure}[tbp]
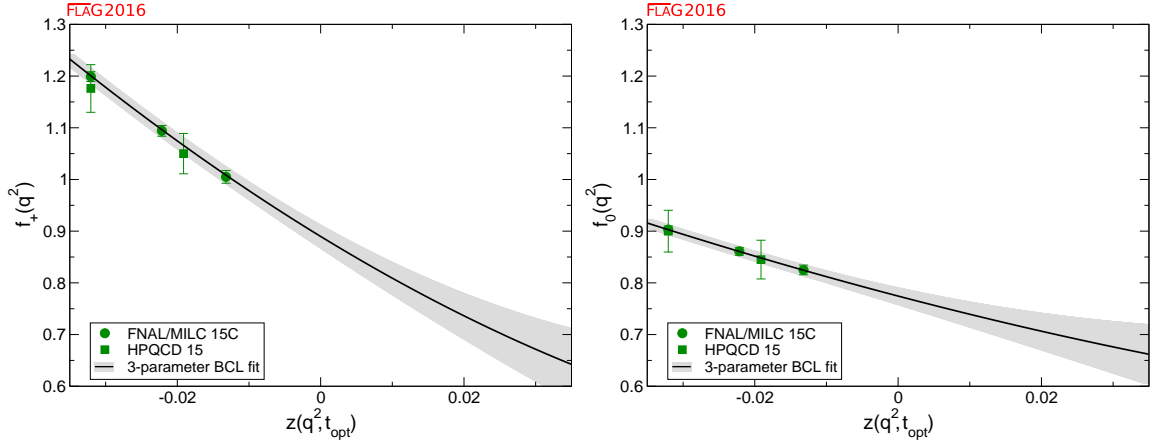

\begin{center}
\setbox1=\hbox{\includegraphics[width=0.48\textwidth]{HQ/Figures/fp_BD_latt.eps}}
\setbox2=\hbox{\includegraphics[width=0.48\textwidth]{HQ/Figures/f0_BD_latt.eps}}
\begin{minipage}{0.48\textwidth}
\includegraphics[width=\textwidth]{HQ/Figures/fp_BD_latt.eps}
\end{minipage}
\begin{minipage}{0.48\textwidth}
\includegraphics[width=\textwidth]{HQ/Figures/f0_BD_latt.eps}
\end{minipage}
\begin{minipage}{0.48\textwidth}
   \llap{\makebox[\wd1][l]{\raisebox{58mm}{
   \hspace{-71mm}\includegraphics[width=0.15\textwidth]{HQ/Figures/FLAG_Logo.eps}
   }}}
\end{minipage}
\begin{minipage}{0.48\textwidth}
   \llap{\makebox[\wd2][l]{\raisebox{58mm}{
   \hspace{4mm}\includegraphics[width=0.15\textwidth]{HQ/Figures/FLAG_Logo.eps}
   }}}
\end{minipage}
\vspace{-3mm}
\caption{The form factors $f_+(q^2)$
(left) and $f_0(q^2)$ (right) for $B \to D\ell\nu$ plotted versus $z$.
(See text for a discussion of the datasets.)
The grey band displays our preferred three-parameter BCL fit to the plotted data with errors
(see Eqs.~(\ref{eq:BtoDLatBCLFitvector},\ref{eq:BtoDLatBCLFitscalar})).}
\label{fig:LQCDzfitBD}
\end{center}
\end{figure}

The fits to $\cO(z^2)$ and $\cO(z^3)$ are always
well-behaved, and there are no qualitative differences between the vector
and the scalar channels. We conservatively take the $\cO(z^3)$
fit as our best result and quote for the vector form factor
\begin{gather}
N_f=2+1:  \qquad 
a_0^{(+)} = 0.890(18)\,,~~~~
a_1^{(+)} = -8.47(93)\,,~~~~
a_2^{(+)} = 39(16)\,;\label{eq:BtoDLatBCLFitvector}\\[1.0ex]
\nonumber \qquad\qquad
{\rm corr}(a_i^{(+)},a_j^{(+)})=\left(\begin{array}{rrr}
 1.000 &  0.806 &  0.711 \\
 0.806 &  1.000 &  0.971 \\
 0.711 &  0.971 &  1.000
\end{array}\right)\,,
\end{gather}
whereas for the scalar form factor we quote
\begin{gather}
N_f=2+1:  \qquad 
a_0^{(0)} = 0.774(14)\,,~~~~
a_1^{(0)} = -3.64(77)\,,~~~~
a_2^{(0)} = -12(14)\,;\label{eq:BtoDLatBCLFitscalar}\\[1.0ex]
\nonumber \qquad\qquad
{\rm corr}(a_i^{(0)},a_j^{(0)})=\left(\begin{array}{rrr}
 1.000 &  0.848 &  0.777 \\
 0.848 &  1.000 &  0.974 \\
 0.777 &  0.974 &  1.000
\end{array}\right)\,.
\end{gather}
(The large values (bearing large errors) of the higher-order coefficients
are likely due to the effect of unaccounted-for resonance poles, cf. the
discussion above.)
These can be
used as the averaged FLAG results for the lattice-computed form factors
$f_+(q^2)$ and $f_0(q^2)$. The coefficient $a_3$ can be obtained from the values for
$a_0$--$a_2$ using Eq.~(\ref{eq:red_coeff}). The fit is illustrated
in Fig.~\ref{fig:LQCDzfitBD}.

Ref.~\cite{Atoui:2013zza} is the only existing $N_f=2$ work on $B \rightarrow D\ell\nu$
transitions, that furthermore provides the only available results
for $B_s \rightarrow D_s\ell\nu$.
This computation uses the publicly available ETM configurations
obtained with the twisted-mass QCD action at maximal twist.  Four
values of the lattice spacing, ranging between $0.054~{\rm fm}$ and
$0.098~{\rm fm}$, are considered, with physical box lengths ranging
between $1.7~{\rm fm}$ and $2.7~{\rm fm}$.  At two values of the
lattice spacing two different physical volumes are available.
Charged-pion masses range between $\approx 270~{\rm MeV}$ and $\approx
490~{\rm MeV}$, with two or three masses available per lattice spacing
and volume, save for the $a \approx 0.054~{\rm fm}$ point at which
only one light mass is available for each of the two volumes. The
strange and heavy valence quarks are also treated with maximally
twisted-mass QCD.

The quantities of interest are again the form factors $h_\pm$ defined above.
In order to control discretization effects from the heavy quarks, a strategy
similar to the one employed by the ETM Collaboration in their studies of
$B$-meson decay constants (cf. Sec.~\ref{sec:fB}) is employed: the value of
${\cal G}(w)$ is computed at a fixed value of $m_c$ and several values of
a heavier quark mass $m_h^{(k)}=\lambda^k m_c$, where $\lambda$ is a fixed
scaling parameter, and step-scaling functions are built as
\begin{equation}
\Sigma_k(w) = \frac{{\cal G}(w,\lambda^{k+1} m_c,m_c,a^2)}{{\cal G}(w,\lambda^k m_c,m_c,a^2)}\,.
\end{equation}
Each ratio is extrapolated to the continuum limit,
$\sigma_k(w)=\lim_{a \to 0}\Sigma_k(w)$.  One then exploits the fact
that the $m_h \to \infty$ limit of the step-scaling is fixed --- in
particular, it is easy to find from the heavy-quark expansion that
$\lim_{m_h\to\infty}\sigma(1)=1$. In this way, the physical result at
the $b$-quark mass can be reached by interpolating $\sigma(w)$ between
the charm region (where the computation can be carried out with
controlled systematics) and the known static limit value.

In practice, the values of $m_c$ and $m_s$ are fixed at each value of
the lattice spacing such that the experimental kaon and $D_s$ masses
are reached at the physical point, as determined
in Ref.~\cite{Blossier:2010cr}.  For the scaling parameter, $\lambda=1.176$
is chosen, and eight scaling steps are performed, reaching
$m_h/m_c=1.176^9\simeq 4.30$, approximately corresponding to the ratio
of the physical $b$- and $c$-masses in the $\overline{\rm MS}$ scheme
at $2~{\rm GeV}$.  All observables are obtained from ratios that do
not require (re)normalization.  The ansatz for the continuum and
chiral extrapolation of $\Sigma_k$ contains a constant and linear
terms in $m_{\rm sea}$ and $a^2$.  Twisted boundary conditions in
space are used for valence-quark fields for better momentum
resolution.  Applying this strategy the form factors are finally
obtained at four reference values of $w$ between $1.004$ and $1.062$,
and, after a slight extrapolation to $w=1$, the result is quoted
\begin{equation}
{\cal G}^{B_s \rightarrow D_s}(1) = 1.052(46)\,.
\end{equation}

The authors also provide values for the form factor relevant for the
meson states with light valence quarks, obtained from a similar
analysis to the one described above for the $B_s\rightarrow D_s$ case.
Values are quoted from fits with and without a linear $m_{\rm
  sea}/m_s$ term in the chiral extrapolation. The result in the former
case, which safely covers systematic uncertainties, is
\begin{equation}
{\cal G}^{B \rightarrow D}(1)=1.033(95)\,.
\label{eq:avBDnf2}
\end{equation}
Given the identical strategy, and the small sensitivity of the ratios
used in their method to the light valence- and sea-quark masses, we
assign this result the same ratings in Tab.~\ref{tab_BtoDStarsumm2}
as those for their calculation of ${\cal G}^{B_s \rightarrow D_s}(1)$.
Currently the precision of this calculation is not competitive with
that of $N_f=2+1$ works, but this is due largely to the small number of
configurations analysed by Atoui {\it et al.}  The viability of their method
has been clearly demonstrated, however, which leaves significant room
for improvement on the errors of both the $B \to D$ and $B_s \to D_s$
form factors with this approach by including either additional
two-flavour data or analysing more recent ensembles with $N_f>2$.

Finally, Atoui {\it et al.} also study the scalar and tensor form factors, as well as the
momentum transfer dependence of $f_{+,0}$. The value of the ratio $f_0(q^2)/f_+(q^2)$
is provided at a reference value of $q^2$ as a proxy for the slope of ${\cal G}(w)$
around the zero-recoil limit.

\subsubsection{Ratios of $B\to D\ell\nu$ form factors}

The availability of results for the scalar form factor $f_0$ in the latest
generation of results for $B\to D\ell\nu$ amplitudes allows us to study
interesting observables that involve the decay in the $\tau$ channel.
One such quantity is the ratio
\begin{equation}
R(D) = {\cal B}(B \rightarrow D \tau \nu) /
{\cal B}(B \rightarrow D \ell \nu)\hspace{1cm}\mbox{with}\;\ell=e,\mu\,,
\end{equation}
which is sensitive to $f_0$, and can be accurately determined by experiment.\footnote{A
similar ratio $R(D^*)$ can be considered for $B \rightarrow D^*$ transitions ---
as a matter of fact, the experimental value of $R(D^*)$ is significantly more accurate
than the one of $R(D)$. However, the absence of lattice results for the $B\to D^*$
scalar form factor, and indeed of results at nonzero recoil (see below), takes
$R(D^*)$ out of our current scope.}
Indeed, the recent availability of experimental results for $R(D)$ has made
this quantity particularly relevant in the search for possible physics beyond
the Standard Model.
Both FNAL/MILC and HPQCD provide values for $R(D)$ from their recent form
factor computations, discussed above. In the FNAL/MILC case, this result
supersedes their 2012 determination, which was discussed in the previous
version of this review. The quoted values by FNAL/MILC and HPQCD are
\begin{equation}
R(D) = 0.299(11)\,\Ref~\mbox{~\cite{Lattice:2015rga}}\,,~~~~~
R(D) = 0.300(8)\,\Ref~\mbox{~\cite{Na:2015kha}}\,.
\end{equation}
These results are in excellent agreement, and can be averaged (using the same
considerations for the correlation between the two computations as we did
in the averaging of form factors) into
\begin{equation}
R(D) = 0.300(8)\,,~~~~\mbox{our average.}
\end{equation}
This result is about $1.6\sigma$ lower than the current experimental average
for this quantity. It has to be stressed that achieving this level of precision
critically depends on the reliability with which the low-$q^2$ region is
controlled by the parameterizations of the form factors.

Another area of immediate interest in searches for physics
beyond the Standard Model is the measurement of $B_s \rightarrow \mu^+ \mu^-$ decays,
recently achieved by LHCb.\footnote{See Ref.~\cite{CMS:2014xfa} for the latest
results, obtained from a joint analysis of CMS and LHCb data.}
In addition to the $B_s$ decay constant (see Sec.~\ref{sec:fB}),
one of the hadronic inputs required by the LHCb analysis is the ratios
of $B_q$ meson ($q = d,s$) fragmentation fractions, $f_s / f_d$.
A dedicated $N_f=2+1$ study by FNAL/MILC\footnote{This work also provided
a value for $R(D)$, now superseded by Ref.~\cite{Lattice:2015rga}.} Ref.~\cite{Bailey:2012rr} addresses the
ratios of scalar form factors $f_0^{(q)}(q^2)$, and quotes:
\begin{equation}
f_0^{(s)}(M_\pi^2) / f_0^{(d)}(M_K^2) = 1.046(44)(15),  
\qquad 
f_0^{(s)}(M_\pi^2) / f_0^{(d)}(M_\pi^2) = 1.054(47)(17),
\end{equation}
where the first error is statistical and the second systematic.  These
results lead to fragmentation fraction ratios $f_s/f_d$ that are
consistent with LHCb's measurements via other methods~\cite{Aaij:2011hi}.

\subsubsection{$B \rightarrow D^*$ decays}

The most precise computation of the zero-recoil form
factors needed for the determination of $|V_{cb}|$ from exclusive $B$
semileptonic decays comes from the $B \rightarrow D^* \ell \nu$ form
factor at zero recoil, ${\cal F}^{B \rightarrow D^*}(1)$, calculated
by the FNAL/MILC Collaboration. The original computation, published
in Ref.~\cite{Bernard:2008dn}, has now been updated~\cite{Bailey:2014tva}
by employing a much more extensive set of gauge ensembles and
increasing the statistics of the ensembles originally considered, while
preserving the analysis strategy. There is currently no unquenched
computation of the relevant form factors at nonzero recoil.

This work uses
the MILC $N_f = 2 + 1$ ensembles.  The bottom and charm quarks are
simulated using the clover action with the Fermilab interpretation and
light quarks are treated via the asqtad staggered fermion action.  At
zero recoil ${\cal F}^{B \rightarrow D^*}(1)$ reduces to a single form
factor $h_{A_1}(1)$ coming from the axial-vector current
\begin{equation}
\langle D^*(v,\epsilon^\prime)| {\cal A}_\mu | \overline{B}(v) \rangle = i \sqrt{2m_B 2 m_{D^*}} \; {\epsilon^\prime_\mu}^\ast h_{A_1}(1),
\end{equation}
where $\epsilon^\prime$ is the polarization of the $D^*$.
The form factor is accessed through a ratio of three-point
correlators, viz.
\begin{equation}
{\cal R}_{A_1} = \frac{\langle D^*|\bar{c} \gamma_j \gamma_5 b | \overline{B} 
\rangle \; \langle \overline{B}| \bar{b} \gamma_j \gamma_5 c | D^* \rangle}
{\langle D^*|\bar{c} \gamma_4 c | D^* 
\rangle \; \langle \overline{B}| \bar{b} \gamma_4 b | \overline{B} \rangle} 
= |h_{A_1}(1)|^2.
\end{equation}
Simulation data are obtained on
MILC ensembles with five lattice spacings, ranging from $a \approx 0.15~{\rm fm}$
to $a \approx 0.045~{\rm fm}$, and as many as five values of the light-quark masses
per ensemble (though just one at the finest lattice spacing).
Results are then extrapolated to the physical, continuum/chiral, limit
employing staggered $\chi$PT.

The $D^*$ meson is not a stable particle in QCD and decays
 predominantly into a $D$ plus a pion.  Nevertheless, heavy-light
 meson $\chi$PT can be applied to extrapolate lattice simulation
 results for the $B\to D^*\ell\nu$ form factor to the physical
 light-quark mass.  The $D^*$ width is quite narrow, 0.096 MeV for the
 $D^{*\pm}(2010)$ and less than 2.1MeV for the $D^{*0}(2007)$, making
 this system much more stable and long lived than the $\rho$ or the
 $K^*$ systems. The fact that the $D^* - D$ mass difference is close
 to the pion mass leads to the well known ``cusp'' in ${\cal
 R}_{A_1}$ just above the physical pion
 mass~\cite{Randall:1993qg,Savage:2001jw,Hashimoto:2001nb}. This cusp
 makes the chiral extrapolation sensitive to values used in the
 $\chi$PT formulas for the $D^*D \pi$ coupling $g_{D^*D\pi}$.  The
 error budget in Ref.~\cite{Bailey:2014tva} includes a separate
 error of 0.3\% coming from the uncertainty in $g_{D^*D \pi}$ in
 addition to general chiral extrapolation errors in order to take this
 sensitivity into account.

The final updated value presented in Ref.~\cite{Bailey:2014tva},
that we quote as our average for this quantity, is
\begin{equation}
{\cal F}^{B \rightarrow D^*}(1) = h_{A_1}(1) = 0.906(4)(12)\,,
\label{eq:avBDstar}
\end{equation}
where the first error is statistical, and the second the sum of systematic errors
added in quadrature, making up a total error of $1.4$\% (down from the original
$2.6$\% of Ref.~\cite{Bernard:2008dn}). The largest systematic
uncertainty comes from discretization errors followed by effects of
higher-order corrections in the chiral perturbation theory ansatz.

\begin{table}[h] 
\begin{center}
\mbox{} \\[3.0cm]
\footnotesize\hspace{-0.2cm}
\begin{tabular*}{\textwidth}[l]{l @{\extracolsep{\fill}} r l l l l l l l c l}
Collaboration & Ref. & $\Nf$ & 
\hspace{0.15cm}\begin{rotate}{60}{publication status}\end{rotate}\hspace{-0.15cm} &
\hspace{0.15cm}\begin{rotate}{60}{continuum extrapolation}\end{rotate}\hspace{-0.15cm} &
\hspace{0.15cm}\begin{rotate}{60}{chiral extrapolation}\end{rotate}\hspace{-0.15cm}&
\hspace{0.15cm}\begin{rotate}{60}{finite volume}\end{rotate}\hspace{-0.15cm}&
\hspace{0.15cm}\begin{rotate}{60}{renormalization}\end{rotate}\hspace{-0.15cm}  &
\hspace{0.15cm}\begin{rotate}{60}{heavy-quark treatment}\end{rotate}\hspace{-0.15cm}  & \multicolumn{2}{l}{\hspace{5mm} $w=1$ form factor / ratio}\\
&&&&&&&&&& \\[-0.1cm]
\hline
\hline
&&&&&&&&&& \\[-0.1cm]
\SLfnalmilcBDstar & \cite{Bailey:2014tva} & 2+1 & \gA & \good &  \soso &  \good & \soso & \okay&${\mathcal F}^{B\to D^*}(1)$   & 0.906(4)(12) \\[0.5ex]  
&&&&&&&&& \\[-0.1cm]
\hline
&&&&&&&&& \\[-0.1cm]
\SLhpqcdBD & \cite{Na:2015kha} & 2+1 & \gA & \soso &  \soso &  \soso & \soso & \okay & ${\mathcal G}^{B\to D}(1)$ & 1.035(40) \\[0.5ex]  
\SLfnalmilcBD & \cite{Lattice:2015rga} & 2+1 & \gA & \good &  \soso &  \good & \soso & \okay & ${\mathcal G}^{B\to D}(1)$  & $1.1054(4)(8)$ \\[0.5ex]  
&&&&&&&&& \\[-0.1cm]
\hline
&&&&&&&&& \\[-0.1cm]
\SLhpqcdBD & \cite{Na:2015kha} & 2+1 & \gA & \soso &  \soso &  \soso & \soso & \okay & $R(D)$ & 0.300(8) \\[0.5ex]  
\SLfnalmilcBD & \cite{Lattice:2015rga} & 2+1 & \gA & \good &  \soso &  \good & \soso & \okay & $R(D)$  & 0.299(11) \\[0.5ex]  
&&&&&&&&& \\[-0.1cm]
\hline
&&&&&&&&& \\[-0.1cm]
Atoui 13 & \cite{Atoui:2013zza} & 2 & \gA & \good & \soso & \good & --- & \okay & ${\mathcal G}^{B\to D}(1)$  & 1.033(95) \\[0.5ex]
&&&&&&&&& \\[-0.1cm]
\hline
&&&&&&&&& \\[-0.1cm]
Atoui 13 & \cite{Atoui:2013zza} & 2 & \gA & \good & \soso & \good & --- & \okay & ${\mathcal G}^{B_s\to D_s}(1)$  & 1.052(46) \\
&&&&&&&&& \\[-0.1cm]
\hline
\hline
\end{tabular*}
\caption{Lattice results for the $B \to D^* \ell\nu$, $B\to D\ell\nu$, and $B_s \to D_s \ell \nu$ semileptonic form factors and $R(D)$. \label{tab_BtoDStarsumm2}}
\end{center}
\end{table}

\subsection{Semileptonic form factors for $\Lambda_b\to p\ell\nu$ and $\Lambda_b\to \Lambda_c\ell\nu$}
\label{sec:Lambdab}

A recent new development in Lattice QCD computations for heavy-quark physics is
the study of semileptonic decays of the $\Lambda_b$ baryon, with first unquenched results
provided in a work by Detmold, Lehner and Meinel~\cite{Detmold:2015aaa}. The importance of this
result is that, together with a recent analysis by LHCb of the ratio of
decay rates $\Gamma(\Lambda_b\to p\ell\nu)/\Gamma(\Lambda_b\to \Lambda_c\ell\nu)$~\cite{Aaij:2015bfa},
it allows for an exclusive determination of the ratio $|V_{ub}|/|V_{cb}|$ largely
independent from the outcome of different exclusive channels, thus contributing
a very interesting piece of information to the existing tensions in the determination
of third-column CKM matrix elements (cf.~Secs.~\ref{sec:Vub},\ref{sec:Vcb}).
For that reason, we will discuss these results briefly, notwithstanding the fact
that baryon physics is in general out of the scope of the present review.

The amplitudes of the decays $\Lambda_b\to p\ell\nu$ and $\Lambda_b\to \Lambda_c\ell\nu$
receive contributions from both the vector and the axial components of the current
in the matrix elements $\langle p|\bar q\gamma^\mu(\mathbf{1}-\gamma_5)b|\Lambda_b\rangle$
and $\langle \Lambda_c|\bar q\gamma^\mu(\mathbf{1}-\gamma_5)b|\Lambda_b\rangle$,
and can be parameterized in terms of six different form factors --- see, e.g., Ref.~\cite{Feldmann:2011xf}
for a complete description. They split into three form factors $f_+$, $f_0$, $f_\perp$ in the
parity-even sector, mediated by the vector component of the current, and another three form factors
$g_+,g_0,g_\perp$ in the parity-odd sector, mediated by the axial component. All
of them provide contributions that are parametrically comparable.

The computation of Detmold {\it et al.} uses RBC/UKQCD $N_f=2+1$ DWF ensembles,
and treats the $b$ and $c$ quarks within the Columbia RHQ approach.
Two values of the lattice spacing ($a\sim0.112,~0.085~{\rm fm}$) are considered,
with the absolute scale set from the $\Upsilon(2S)$--$\Upsilon(1S)$ splitting.
Sea pion masses lie in a narrow interval ranging from slightly above
$400~{\rm MeV}$ to slightly below $300~{\rm MeV}$, keeping $m_\pi L \gtrsim 4$;
however, lighter pion masses are considered in the valence DWF action
for the $u,d$ quarks, leading to partial quenching effects in the chiral
extrapolation. More importantly, this also leads to values of $M_{\pi,{\rm min}}L$ close to $3.0$
(cf. App~\ref{app:BtoPi_Notes} for details);
compounded with the fact that there is only one lattice volume in the computation,
an application of the FLAG criteria would lead to a $\bad$ rating for finite volume effects.
It has to be stressed, however, that our criteria have been developed in the context
of meson physics, and their application to the baryon sector is not straightforward;
as a consequence, we will refrain from providing a conclusive rating of this computation
for the time being.

Results for the form factors are obtained from suitable three-point functions,
and fitted to a modified $z$-expansion ansatz that combines the $q^2$ dependence
with the chiral and continuum extrapolations. The main results of the paper are
the predictions (errors are statistical and systematic, respectively)
\begin{gather}
\begin{split}
\frac{1}{|V_{ub}|^2}\int_{15~{\rm GeV}^2}^{q^2_{\rm max}}\frac{{\rm d}\Gamma(\Lambda_b\to p\mu^-\bar\nu_\mu)}{{\rm d}q^2}\,{\rm d}q^2 &= 12.32(93)(80)~{\rm ps}^{-1}\,,\\
\frac{1}{|V_{cb}|^2}\int_{15~{\rm GeV}^2}^{q^2_{\rm max}}\frac{{\rm d}\Gamma(\Lambda_b\to \Lambda_c\mu^-\bar\nu_\mu)}{{\rm d}q^2}\,{\rm d}q^2 &= 8.39(18)(32)~{\rm ps}^{-1}\,,
\end{split}
\end{gather}
which are the input for the LHCb analysis. Prediction for the total rates in all possible
lepton channels, as well as for ratios similar to $R(D)$ (cf. Sec.~\ref{sec:BtoD}) between the $\tau$
and light lepton channels are also available.

%% file: HQ/HQSubsections/BCKM.tex
\subsection{Determination of $|V_{ub}|$}
\label{sec:Vub}

We now use the lattice-determined Standard Model transition amplitudes
for leptonic (Sec.~\ref{sec:fB}) and semileptonic
(Sec.~\ref{sec:BtoPiK}) $B$-meson decays to obtain exclusive
determinations of the CKM matrix element $|V_{ub}|$.
In this section, we describe the aspect of our work
that involves experimental input for the relevant charged-current
exclusive decay processes.
The relevant
formulae are Eqs.~(\ref{eq:B_leptonic_rate})
and~(\ref{eq:B_semileptonic_rate}). Among leptonic channels the only
input comes from $B\to\tau\nu_\tau$, since the rates for decays to $e$
and $\mu$ have not yet been measured.  In the semileptonic case we
only consider $B\to\pi\ell\nu$ transitions (experimentally
measured for $\ell=e,\mu$). As discussed in Secs.~\ref{sec:BtoPiK} and~\ref{sec:Lambdab},
there are now lattice predictions for the rates of the decays $B_s\to K\ell\nu$
and $\Lambda_b\to p\ell\nu$; however, in the former case the process has not been
experimentally measured yet, while in the latter case the only existing lattice computation
does not meet FLAG requirements for controlled systematics.

We first investigate the determination of $|V_{ub}|$ through the
$B\to\tau\nu_\tau$ transition.  This is the only experimentally
measured leptonic decay channel of the charged $B$-meson.
After the publication of the previous FLAG report~\cite{Aoki:2013ldr}
in 2013, the experimental measurements of the branching fraction of
this channel, $B(B^{-} \to \tau^{-} \bar{\nu})$, were updated.
While the results from the BaBar collaboration remain the same as
those reported before the end of 2013, the Belle collaboration reanalysed the
data and reported that the value of $B(B^{-} \to \tau^{-} \bar{\nu})$ obtained
with semileptonic tags changed from 
$1.54^{+0.38_0.29}_{-0.37-0.31} \times 10^{-4}$ to $1.25 \pm 0.28 \pm
0.27 \times 10^{-4}$~\cite{Kronenbitter:2015kls}. Table~\ref{tab:leptonic_B_decay_exp} summarizes the
current status of experimental results for this branching fraction.
\begin{table}[h]
\begin{center}
\noindent
\begin{tabular*}{\textwidth}[l]{@{\extracolsep{\fill}}lll}
Collaboration & Tagging method  & $B(B^{-}\to \tau^{-}\bar{\nu})
                                  \times 10^{4}$\\
&& \\[-2ex]
\hline \hline &&\\[-2ex]
Belle~\cite{Adachi:2012mm} &  Hadronic  & $0.72^{+0.27}_{-0.25}\pm 0.11$ \\
Belle~\cite{Kronenbitter:2015kls} &  Semileptonic & $1.25 \pm 0.28 \pm
                                                    0.27$ \\
&& \\[-2ex]
 \hline
&& \\[-2ex]
BaBar~\cite{Lees:2012ju} & Hadronic & $1.83^{+0.53}_{-0.49}\pm 0.24$\\
BaBar~\cite{Aubert:2009wt} & Semileptonic  & $1.7 \pm 0.8 \pm 0.2$\\
&& \\[-2ex]
\hline \hline && \\[-2ex]
\end{tabular*}
\caption{Experimental measurements for $B(B^{-}\to \tau^{-}\bar{\nu})$.
  The first error on each result is statistical, while the second
  error is systematic.}
\label{tab:leptonic_B_decay_exp}
\end{center}
\end{table}

It is obvious that all the measurements listed in
Tab.~\ref{tab:leptonic_B_decay_exp} have significance less than
$5\sigma$, and the uncertainties are dominated by statistical errors. These measurements lead to the averages
of experimental measurements for $B(B^{-}\to \tau
\bar{\nu})$~\cite{Kronenbitter:2015kls,Lees:2012ju},
\begin{eqnarray}
 B(B^{-}\to \tau \bar{\nu} ) &=& 0.91 \pm 0.22 \mbox{ }{\rm from}\mbox{ } {\rm Belle,}\nonumber\\
                                          &=& 1.79 \pm 0.48 \mbox{
                                          }{\rm from }\mbox{ }
                                              {\rm BaBar.}
\label{eq:leptonic_B_decay_exp_ave}
\end{eqnarray}
We notice that minor tension between results from the two collaborations can be
observed, even in the presence of large errors. Despite this situation,
in Ref.~\cite{Rosner:2015wva} the Particle Data Group performed a global
average of $B(B^{-}\to \tau \bar{\nu} )$ employing all the information in
Tab.~\ref{tab:leptonic_B_decay_exp}. Here we choose to proceed
with the strategy of quoting different values of $|V_{ub}|$ as
determined using inputs from the Belle and the
BaBar experiments shown in Eq.~(\ref{eq:leptonic_B_decay_exp_ave}),
respectively.

Combining the results in Eq.~(\ref{eq:leptonic_B_decay_exp_ave}) with
the experimental measurements of the mass of the $\tau$-lepton and the
$B$-meson lifetime and mass, the Particle Data Group presented~\cite{Rosner:2015wva}
\begin{eqnarray}
 |V_{ub}| f_{B} &=& 0.72 \pm 0.09 \mbox{ }{\rm MeV}\mbox{ }{\rm from}\mbox{ } {\rm Belle,}\nonumber\\
                                          &=& 1.01 \pm 0.14 \mbox{
                                              }{\rm MeV}\mbox{
                                          }{\rm from }\mbox{ }
                                              {\rm BaBar,}
\label{eq:Vub_fB}
\end{eqnarray}
which can be used to extract $|V_{ub}|$.  
\begin{align}
&N_f=2    &\mbox{Belle}~B\to\tau\nu_\tau:   && |V_{ub}| &= 3.83(48)(15) \times 10^{-3}  \,,\nonumber\\
&N_f=2+1  &\mbox{Belle}~B\to\tau\nu_\tau:   && |V_{ub}| &= 3.75(47)(9) \times 10^{-3}   \,,\nonumber\\
&N_f=2+1+1&\mbox{Belle}~B\to\tau\nu_\tau:   && |V_{ub}| &= 3.87(48)(9) \times 10^{-3}   \,;\nonumber\\
&         & \\                                           
&N_f=2    &\mbox{Babar}~B\to\tau\nu_\tau:   && |V_{ub}| &=  5.37(74)(21) \times 10^{-3} \,,\nonumber\\
&N_f=2+1  &\mbox{Babar}~B\to\tau\nu_\tau:   && |V_{ub}| &=  5.26(73)(12) \times 10^{-3} \,,\nonumber\\
&N_f=2+1+1&\mbox{Babar}~B\to\tau\nu_\tau:   && |V_{ub}| &= 5.43(75)(12) \times 10^{-3}  \,.\nonumber
\end{align}
where the first error comes from experiment and the second comes from
the uncertainty in $f_B$.  

Let us now turn our attention to semileptonic decays. The experimental
value of $|V_{ub}|f_+(q^2)$ can be extracted from the measured branching
fractions for $B^0\to\pi^\pm\ell\nu$ and/or $B^\pm\to\pi^0\ell\nu$
applying Eq.~(\ref{eq:B_semileptonic_rate});\footnote{Since $\ell=e,\mu$ the contribution
from the scalar form factor in Eq.~(\ref{eq:B_semileptonic_rate}) is negligible.}
$|V_{ub}|$ can then be determined by performing fits to the constrained BCL $z$ parameterization
of the form factor $f_+(q^2)$ given in Eq.~(\ref{eq:bcl_c}). This can be done in two ways:
one option is to perform separate fits to lattice and experimental results, and extract
the value of $|V_{ub}|$ from the ratio of the respective $a_0$ coefficients; a second option
is to perform a simultaneous fit to lattice and experimental data, leaving their relative
normalization $|V_{ub}|$ as a free parameter. We adopt the second strategy, because it combines
the lattice and experimental input in a more efficient way, leading to a smaller uncertainty
on $|V_{ub}|$.

The available state-of-the-art experimental input, as employed, e.g., by HFAG, consists of five datasets:
three untagged measurements by BaBar (6-bin~\cite{delAmoSanchez:2010af} and 12-bin~\cite{Lees:2012vv}) and Belle~\cite{Ha:2010rf}, all of which assume isospin symmetry and provide combined
$B^0\to\pi^-$ and $B^+\to\pi^0$ data; and the two tagged Belle measurements of
$\bar B^0\to\pi^+$ (13-bin) and $B^-\to\pi^0$ (7-bin )~\cite{Sibidanov:2013rkk}.
In the previous version of the FLAG review~\cite{Aoki:2013ldr} we only used the
13-bin Belle and 12-bin BaBar datasets, and performed separate
fits to them due to the lack of information on systematic correlations between them.
Now however we will follow established practice, and perform a combined fit to all the
experimental data. This is based on the existence of new information about cross-correlations,
that allows us to obtain a meaningful final error estimate.\footnote{See, e.g., Sec.~V.D of~~\cite{Lattice:2015tia}
for a detailed discussion.} The lattice input dataset will be the same discussed in Sec.~\ref{sec:BtoPiK}.

\begin{figure}[tbp]
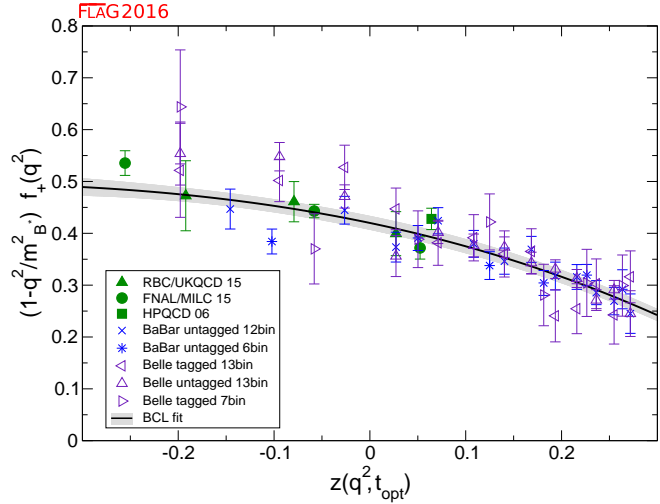

\begin{center}
\setbox1=\hbox{\includegraphics[width=0.55\textwidth]{HQ/Figures/fp_Bpi_latt+exp.eps}}
\begin{minipage}{0.55\textwidth}
\includegraphics[width=\textwidth]{HQ/Figures/fp_Bpi_latt+exp.eps}
\end{minipage}
\begin{minipage}{0.55\textwidth}
   \llap{\makebox[\wd1][l]{\raisebox{66mm}{
   \hspace{6mm}\includegraphics[width=0.15\textwidth]{HQ/Figures/FLAG_Logo.eps}
   }}}
\end{minipage}
\vspace{-2mm}
\caption{
Lattice and experimental data for $(1-q^2/m_{B^*}^2)f_+^{B\to\pi}(q^2)$ versus $z$.
The filled green symbols denote lattice-QCD points included in the fit, while blue and indigo
points show experimental data divided by the value of $|V_{ub}|$ obtained from the fit. The grey band shows the preferred three-parameter BCL fit to the lattice-QCD and experimental data with errors.
}
\label{fig:Vub_SL_fit}
\end{center}
\end{figure}

A simple three-parameter constrained BCL fit (i.e., through $\cO(z^2)$ plus $|V_{ub}|$) is
enough to describe the combined datasets satisfactorily; however, the inclusion of experimental points
allows for a better determination of the higher orders in the BCL parameterization with respect
to the lattice-only fit. In order to address the potential systematic uncertainty due
to truncating the series in $z$, we continue to add terms to the fit until the result for $|V_{ub}|$
stabilizes, i.e., the central value settles and the errors stop increasing. We find that this happens
at $\cO(z^3)$, and take the value of $|V_{ub}|$ from the combined fit through this order
as our estimate,
\begin{align}
&N_f=2+1&B\to\pi\ell\nu: && |V_{ub}| &= 3.62(14) \times 10^{-3}\,.
\end{align}
Fig.~\ref{fig:Vub_SL_fit} shows both the lattice and experimental data for
$(1-q^2/m_{B^*}^2)f_+(q^2)$ as a function of $z(q^2)$, together with our preferred fit;
experimental data have been rescaled by the resulting value for $|V_{ub}|^2$.
It is worth noting the good consistency between the form factor shapes from
lattice and experimental data. This can be quantified, e.g., by computing the ratio of the
two leading coefficients in the constrained BCL parameterization: the fit to lattice form
factors yields $a_1/a_0=-0.83(25)$ (cf. Eq.~(\ref{eq:BtoPiLatBCLFit})),
while the above lattice+experiment fit yields $a_1/a_0=-0.921(88)$.

We plot the values of $|V_{ub}|$ we have obtained in
Fig.~\ref{fig:Vxbsummary}, where the determination through inclusive decays
by the Heavy Flavour Averaging Group (HFAG)~\cite{Amhis:2014hma},
yielding $|V_{ub}|=4.62(20)(29) \times 10^{-3}$, is also shown for comparison.
In this plot the
tension between the BaBar and the Belle measurements of $B(B^{-} \to
\tau^{-} \bar{\nu})$ is manifest. As discussed above, it is for this
reason that we do not extract $|V_{ub}|$ through the average of
results for this branching fraction from these two collaborations. In
fact this means that a reliable determination of $|V_{ub}|$ using
information from leptonic $B$-meson decays is still absent;
the situation will only clearly improve with the more precise experimental
data expected from Belle~II.
The value for $|V_{ub}|$ obtained from semileptonic $B$ decays for $N_f=2+1$, on the other hand,
is significantly more precise than both the leptonic and the inclusive determinations,
and exhibits the well-known $\sim 3\sigma$ tension with the latter.

\subsection{Determination of $|V_{cb}|$}
\label{sec:Vcb}

We will now use the lattice QCD results for the $B \to D^{(*)}\ell\nu$ form factors
in order to obtain determinations of the CKM matrix element $|V_{cb}|$ in the Standard Model.
The relevant formulae are given in~Eq.~(\ref{eq:vxb:BtoDstar}).

Let us summarize the lattice input that satisfies FLAG requirements for the control
of systematic uncertainties, discussed in Sec.~\ref{sec:BtoD}.
In the (experimentally more precise) $B\to D^*\ell\nu$ channel, there is only one
$N_f=2+1$ lattice computation of the relevant form factor $\mathcal{F}^{B\to D^*}$ at zero recoil.
Concerning the $B \to D\ell\nu$ channel, for $N_f=2$ there is one determination
of the relevant form factor $\mathcal{G}^{B\to D}$ at zero recoil\footnote{The same work
provides $\mathcal{G}^{B_s\to D_s}$, for which there are, however, no experimental data.}; while
for $N_f=2+1$ there are two determinations of the $B \to D$ form factor as a function
of the recoil parameter in roughly the lowest third of the kinematically allowed region.
In this latter case, it is possible to replicate the analysis carried out for $|V_{ub}|$
in~Sec.~\ref{sec:Vub}, and perform a joint fit to lattice and experimental data;
in the former, the value of $|V_{cb}|$ has to be extracted by matching to the
experimental value for $\mathcal{F}^{B\to D^*}(1)\eta_{\rm EW}|V_{cb}|$ and
$\mathcal{G}^{B\to D}(1)\eta_{\rm EW}|V_{cb}|$.

The latest experimental average by HFAG~\cite{Amhis:2014hma} for the $B\to D^*$ form factor
at zero recoil is
\begin{gather}
\mathcal{F}^{B\to D^*}(1)\eta_{\rm EW}|V_{cb}| = 35.81(0.45)\times 10^{-3}\,.
\end{gather}
By using $\eta_{\rm EW}=1.00662$ and the lattice value for $\mathcal{F}^{B\to D^*}(1)$
in~Eq.~(\ref{eq:avBDstar}), we thus extract our average
\begin{align}
&N_f=2+1&B\to D^*\ell\nu: && |V_{cb}| &= 39.27(56)(49) \times 10^{-3} \,,
\end{align}
where the first uncertainty comes from the lattice computation and the second from the
experimental input.
For the zero-recoil $B \to D$ form factor, HFAG quotes
\begin{gather}
\mbox{HFAG:} \qquad \mathcal{G}^{B\to D}(1)\eta_{\rm EW}|V_{cb}| = 42.65(1.53)\times 10^{-3}\,.
\label{eq:BDHFAG}
\end{gather}
This average is strongly dominated by the BaBar input.
The set of experimental results for $B \to D\ell\nu$ has however been significantly improved
by the recent publication of a new Belle measurement~\cite{Glattauer:2015teq},
which quotes
\begin{gather}
\mbox{Belle 2016:} \qquad \mathcal{G}^{B\to D}(1)\eta_{\rm EW}|V_{cb}| = 42.29(1.37)\times 10^{-3}\,.
\end{gather}
Given the difficulties to include this latter number in a global average replicating the
procedure followed by HFAG, and the fact that the final uncertainty will be completely dominated
by the error of the lattice input in Eq.~(\ref{eq:avBDnf2}), we will conservatively
use the value in Eq.~(\ref{eq:BDHFAG}) to provide an average for $N_f=2$, and quote
\begin{align}
&N_f=2&B\to D\ell\nu: && |V_{cb}| &= 41.0(3.8)(1.5) \times 10^{-3} \,.
\end{align}

Finally, for $N_f=2+1$ we will perform, as discussed above, a joint fit to the available
lattice data, discussed in Sec.~\ref{sec:BtoD}, and state-of-the-art experimental determinations.
In this case we will combine the aforementioned recent Belle measurement~\cite{Glattauer:2015teq},
which provides partial integrated decay rates in 10 bins in the recoil parameter $w$,
with the 2010 BaBar dataset in Ref.~\cite{Aubert:2009ac}, which quotes the value of
$\mathcal{G}^{B\to D}(w)\eta_{\rm EW}|V_{cb}|$ for four values of $w$.
The fit is dominated by the more precise Belle data; given this, and the fact that only partial
correlations among systematic uncertainties are to be expected, we will treat both datasets
are uncorrelated.\footnote{We have checked that results using just one experimental dataset
are compatible within $1\sigma$. In the case of BaBar, we have
taken into account the introduction of some EW corrections in the data.}
A constrained BCL fit through $\cO(z^3)$, using the same ansatz as for lattice-only data
in Sec.~\ref{sec:BtoD}, yields our average
\begin{align}
&N_f=2+1&B\to D\ell\nu: &&|V_{cb}| &= 40.85(98) \times 10^{-3} \,,
\end{align}
where the error combines the lattice and experimental uncertainties.
The fit is illustrated in Fig.~\ref{fig:Vcb_SL_fit}.

\begin{figure}[!h]
\begin{center}
\begin{minipage}{0.49\textwidth}
\includegraphics[width=1.1\linewidth]{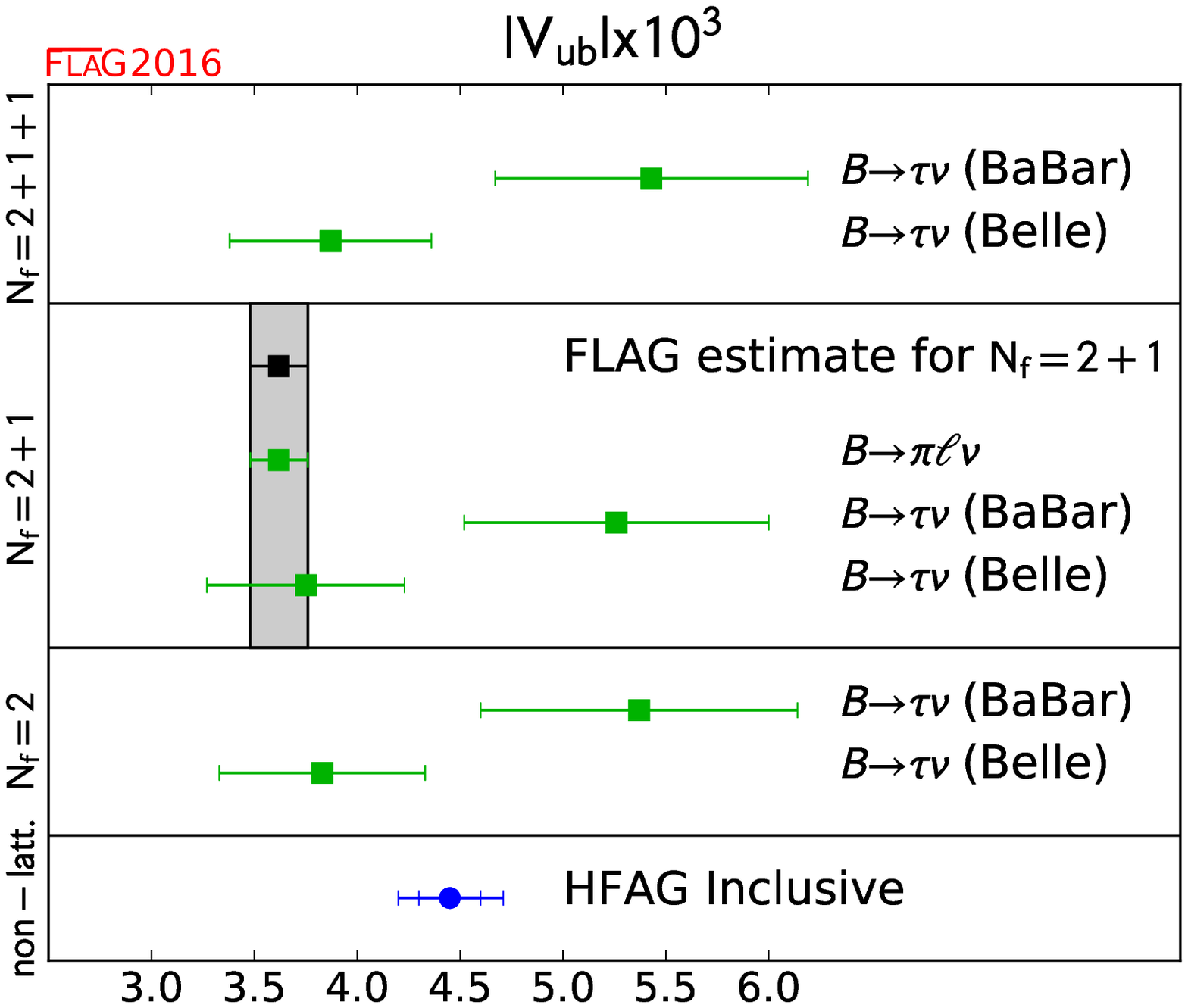}
\end{minipage}
\begin{minipage}{0.49\textwidth}
\includegraphics[width=1.1\linewidth]{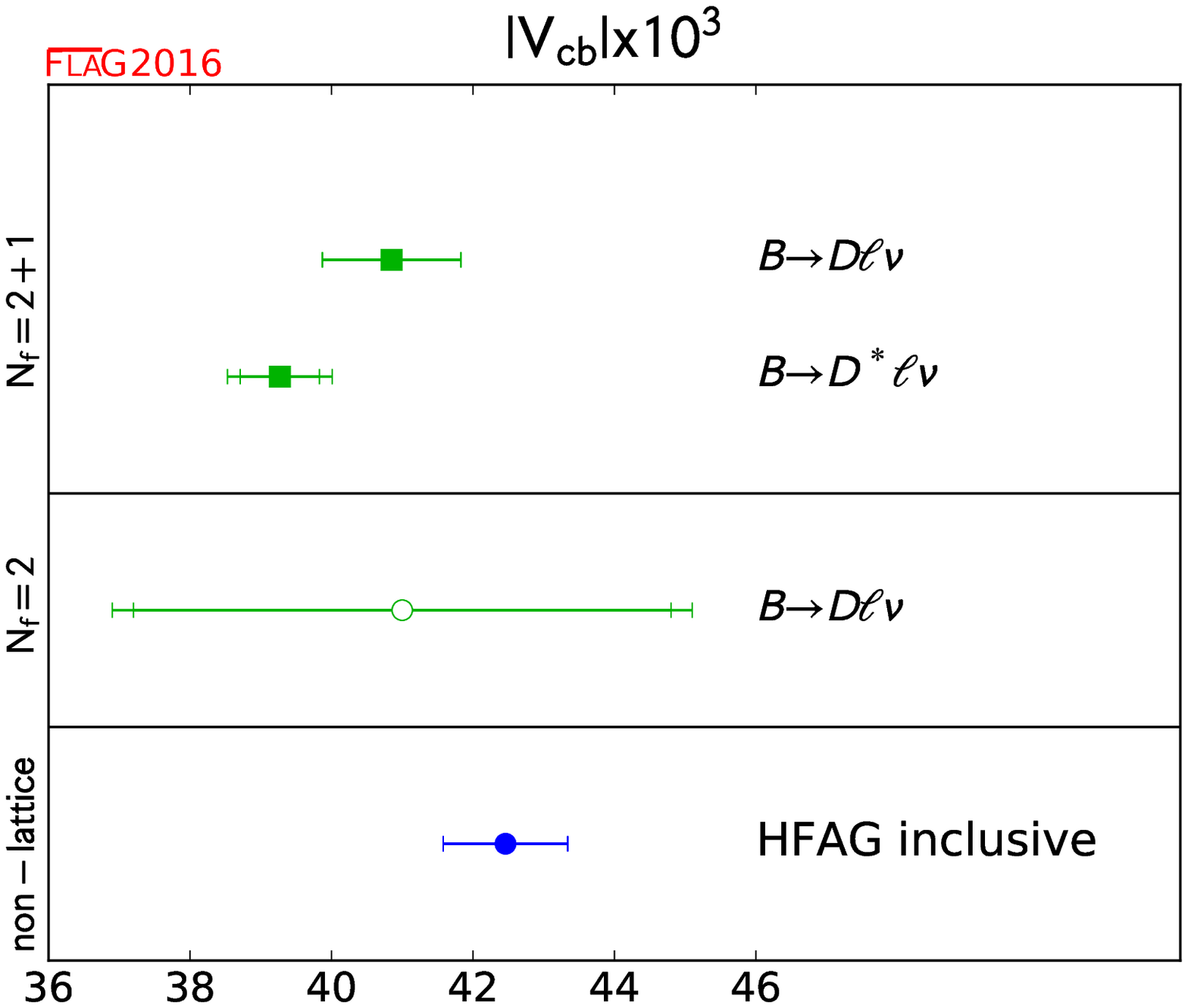}
\end{minipage}
\vspace{-2mm}
\caption{Left: Summary of $|V_{ub}|$ determined using: i) the $B$-meson leptonic
decay branching fraction, $B(B^{-} \to \tau^{-} \bar{\nu})$, measured
at the Belle and BaBar experiments, and our averages for $f_{B}$ from
lattice QCD; and ii) the various measurements of the $B\to\pi\ell\nu$
decay rates by Belle and BaBar, and our averages for lattice determinations
of the relevant vector form factor $f_+(q^2)$.
Right: Same for determinations of $|V_{cb}|$ using semileptonic decays.
The HFAG inclusive results are from Ref.~\cite{Amhis:2014hma}.}
\label{fig:Vxbsummary}
\end{center}
\end{figure}

Our results
are summarized in Tab.~\ref{tab:Vcbsummary}, which also shows the HFAG inclusive
determination of $|V_{cb}|$ for comparison, and illustrated in Fig.~\ref{fig:Vxbsummary}.
The $N_f=2+1$ results coming from $B\to D^*\ell\nu$ and $B\to D\ell\nu$ could in principle
be averaged; we will however not do so, due to the difficulties of properly taking into
account experimental correlations. We will thus leave them as separate exclusive estimates,
which show good mutual consistence, and the well-known tension with the inclusive determination.

\begin{table}[!t]
\begin{center}
\noindent
\begin{tabular*}{\textwidth}[l]{@{\extracolsep{\fill}}lcc}
 & from  & $|V_{cb}| \times 10^3$\\
&& \\[-2ex]
\hline \hline &&\\[-2ex]
our average for $N_f=2+1$ & $B \to D^*\ell\nu$ & $39.27(56)(49)$ \\
our average for $N_f=2+1$ & $B \to D\ell\nu$ & $40.85(98)$ \\
&& \\[-2ex]
 \hline
our average for $N_f=2$ & $B \to D\ell\nu$ & $41.0(3.8)(1.5)$ \\
&& \\[-2ex]
 \hline
HFAG inclusive average & $B \to X_c\ell\nu$ & $42.46(88)$ \\
&& \\[-2ex]
\hline \hline && \\[-2ex]
\end{tabular*}
\caption{Results for $|V_{cb}|$. When two errors are quoted in
our averages, the first one comes from the lattice form factor, and the
second from the experimental measurement. The HFAG inclusive average
obtained in the kinetic scheme
from Ref.~\cite{Amhis:2014hma} is shown for comparison.}
\label{tab:Vcbsummary}
\end{center}
\end{table}
\begin{figure}[!t]
\begin{center}
\setbox1=\hbox{\includegraphics[width=0.55\textwidth]{HQ/Figures/fp_BD_latt+exp.eps}}
\begin{minipage}{0.55\textwidth}
\includegraphics[width=\textwidth]{HQ/Figures/fp_BD_latt+exp.eps}
\end{minipage}
\begin{minipage}{0.55\textwidth}
   \llap{\makebox[\wd1][l]{\raisebox{66mm}{
   \hspace{6mm}\includegraphics[width=0.15\textwidth]{HQ/Figures/FLAG_Logo.eps}
   }}}
\end{minipage}
\vspace{-2mm}
\caption{
Lattice and experimental data for $f_+^{B\to D}(q^2)$ versus $z$.
The filled green symbols denote lattice-QCD points included in the fit, while blue and indigo
points show experimental data divided by the value of $|V_{cb}|$ obtained from the fit. The grey band shows the preferred three-parameter BCL fit to the lattice-QCD and experimental data with errors.
}
\label{fig:Vcb_SL_fit}
\end{center}
\end{figure}

%% file: Alpha_s/alpha.tex
\section{The strong coupling $\alpha_{\rm s}$}
\label{sec:alpha_s}

\input{Alpha_s/s12.tex} % general
\input{Alpha_s/s3.tex} % $\alpha_s$ from the Schr\"odinger Functional \label{s:SF}
\input{Alpha_s/s4.tex} % {$\alpha_s$ from the potential at short distances} \label{s:qq}
\input{Alpha_s/s5.tex} % from the vacuum polarization at short distances \label{s:vac}
\input{Alpha_s/s6.tex} % from observables at the lattice spacing scale \label{s:WL}
\input{Alpha_s/s7.tex} % from current two-point functions \label{s:curr}
\input{Alpha_s/s8.tex} % from QCD vertices} \label{s:glu}

\input{Alpha_s/s9.tex} % Summary \label{s:alpsumm}

%% file: Alpha_s/s12.tex
\subsection{Introduction}

% ----------------------------------------------------------------------

\label{introduction}

% ----------------------------------------------------------------------

The strong coupling $\gbar(\mu)$ defined at scale $\mu$, plays a key
role in the understanding of QCD and in its application for collider
physics. For example, the parametric uncertainty from $\alpha_s$ is one of
the dominant sources of uncertainty in the Standard Model prediction for
the $H \to b\bar{b}$ partial width, and the largest source of uncertainty
for $H \to gg$.  Thus higher precision determinations of $\alpha_s$ are
needed to maximize the potential of experimental measurements at the LHC,
and for high-precision Higgs studies at future
colliders~\cite{Heinemeyer:2013tqa,Adams:2013qkq,Dawson:2013bba}. The
value of $\alpha_s$ also yields one of the essential boundary conditions
for completions of the standard model at high energies. 

In order to determine the running coupling at scale $\mu$
\begin{eqnarray}
   \alpha_s(\mu) = { \gbar^2(\mu) \over 4\pi} \,,
\end{eqnarray}
we should first ``measure'' a short-distance quantity ${\oO}$ at scale
$\mu$ either experimentally or by lattice calculations and then 
match it with a perturbative expansion in terms of a running coupling,
conventionally taken as $\alpha_{\overline{\rm MS}}(\mu)$,
\begin{eqnarray}
   {\oO}(\mu) = c_1 \alpha_{\overline{\rm MS}}(\mu)
              +  c_2 \alpha_{\overline{\rm MS}}(\mu)^2 + \cdots \,.
\label{eq:alpha_MSbar}
\end{eqnarray}
The essential difference between continuum determinations of
$\alpha_s$ and lattice determinations is the origin of the values of
$\oO$ in \eq{eq:alpha_MSbar}.

The basis of continuum determinations are 
experimentally measurable cross sections from which $\oO$ is
defined. These cross sections have to be sufficiently inclusive 
and at sufficiently high scales such that perturbation theory 
can be applied. Often hadronization corrections have to be used
to connect the observed hadronic cross sections to the perturbative
ones. Experimental data at high $\mu$, where perturbation theory
is progressively more precise, usually have increasing experimental errors, 
and it is  not easy to find processes which allow one
to follow the $\mu$ dependence of a single $\oO(\mu)$ over
a range where $\alpha_s(\mu)$ changes significantly and precision is 
maintained.

In contrast, in lattice gauge theory, one can design $\oO(\mu)$ as
Euclidean short-distance quantities which are not directly related to
experimental observables. This allows us to follow the $\mu$
dependence until the perturbative regime is reached and
nonperturbative ``corrections'' are negligible.  The only
experimental input for lattice computations of $\alpha_s$ is the
hadron spectrum which fixes the overall energy scale of the theory and
the quark masses. Therefore experimental errors are completely
negligible and issues such as hadronization do not occur.  We can
construct many short-distance quantities that are easy to calculate
nonperturbatively in lattice simulations with small statistical
uncertainties.  We can also simulate at parameter values that do not
exist in nature (for example with unphysical quark masses between
bottom and charm) to help control systematic uncertainties.  These
features mean that precise results for $\alpha_s$ can be achieved
with lattice gauge theory computations.  Further, as in the continuum,
the different methods available to determine $\alpha_s$ in
lattice calculations with different associated systematic
uncertainties enable valuable cross-checks.  Practical limitations are
discussed in the next section, but a simple one is worth mentioning
here. Experimental results (and therefore the continuum
determinations) of course have all quarks present, while in lattice
gauge theories only the light ones are included and one then is forced
to use the matching at thresholds, as discussed in the following
subsection.

It is important to keep in mind that the dominant source of uncertainty
in most present day lattice-QCD calculations of $\alpha_s$ are from
the truncation of continuum/lattice perturbation theory and from
discretization errors. Perturbative truncation errors are of
a different nature than most other lattice (or continuum)
systematics, in that they often cannot easily be estimated
from studying the data itself. Further, the size
of higher-order coefficients in the perturbative series can sometimes
turn out to be larger than naive expectations based on power counting
from the behaviour of lower-order terms.  

The various phenomenological approaches to determining the running
coupling, $\alpha^{(5)}_{\overline{\rm MS}}(M_Z)$ are summarized by the
Particle Data Group \cite{Agashe:2014kda}. 
The PDG review lists $4$ categories of phenomenological results
used to obtain the running coupling using hadronic
$\tau$ decays, hadronic final states of $e^+e^-$ annihilation,
deep inelastic lepton--nucleon scattering and electroweak precision data.
Excluding lattice results, the PDG quotes a
weighted average of
\begin{eqnarray}
   \alpha^{(5)}_{\overline{\rm MS}}(M_Z) = 0.1175(17) \,,
\label{PDG_nolat}
\end{eqnarray}
compared to $
   \alpha^{(5)}_{\overline{\rm MS}}(M_Z) = 0.1183(12) 
$
of the previous review \cite{Beringer:1900zz}.
For a general overview of the various phenomenological
and lattice approaches see e.g.~Ref.~\cite{Bethke:2011tr}.  We note
that perturbative truncation errors are also the dominant
source of uncertainty in several of the phenomenological
determinations of $\alpha_s$.  In particular, the extraction of
$\alpha_s$ from $\tau$ data, which is the most precise and has the
largest impact on the nonlattice average in Eq.~(\ref{PDG_nolat}) is
especially sensitive to the treatment of higher-order perturbative
terms.  This is important to keep in mind when comparing our chosen
range for $\alpha^{(5)}_{\overline{\rm MS}}(M_Z)$ from lattice
determinations in Eq.~(\ref{eq:alpmz}) with the nonlattice average
from the PDG.

\subsubsection{Scheme and scale dependence of $\alpha_s$ and $\Lambda_{\rm QCD}$}

Despite the fact that the notion of the QCD coupling is 
initially a perturbative concept, the associated $\Lambda$ parameter
is nonperturbatively defined
\begin{eqnarray}
   \Lambda 
      \equiv \mu\,(b_0\gbar^2(\mu))^{-b_1/(2b_0^2)} 
              e^{-1/(2b_0\gbar^2(\mu))}
             \exp\left[ -\int_0^{\gbar(\mu)}\,dx 
                        \left( {1\over \beta(x)} + {1 \over b_0x^3} 
                                                 - {b_1 \over b_0^2x}
                        \right) \right] \,,
\label{eq:Lambda}
\end{eqnarray}
where $\beta$ is the full renormalization group function in the scheme
which defines $\gbar$, and $b_0$ and $b_1$ are the first two
scheme-independent coefficients of the perturbative expansion
\begin{eqnarray}
   \beta(x) \sim -b_0 x^3 -b_1 x^5 + \ldots \,,
\end{eqnarray}
with
\begin{eqnarray}
   b_0 = {1\over (4\pi)^2}
           \left( 11 - {2\over 3}N_f \right) \,, \qquad
   b_1 = {1\over (4\pi)^4}
           \left( 102 - {38 \over 3} N_f \right) \,.
\label{b0+b1}
\end{eqnarray}
Thus the $\Lambda$ parameter is renormalization-scheme-dependent but in an
exactly computable way, and lattice gauge theory is an ideal method to
relate it to the low-energy properties of QCD.

The change in the coupling from one scheme, $S$, to another (taken here
to be the $\overline{\rm MS}$ scheme) is perturbative,
\begin{eqnarray}
   g_{\overline{\rm MS}}^2(\mu) 
      = g_{\rm S}^2(\mu) (1 + c^{(1)}_g g_{\rm S}^2(\mu) + \ldots ) \,,
\label{eq:g_conversion}
\end{eqnarray}
where $c^{(i)}_g$ are the finite renormalization coefficients.  The
scale $\mu$ must be taken high enough for the error in keeping only
the first few terms in the expansion to be small.
On the other hand, the conversion to the $\Lambda$ parameter
in the $\overline{\rm MS}$ scheme is given exactly by
\begin{eqnarray}
   \Lambda_{\overline{\rm MS}} 
      = \Lambda_{\rm S} \exp\left[ c_g^{(1)}/(2b_0)\right] \,.
\end{eqnarray}

By convention $\alpha_\msbar$ is usually quoted at a scale $\mu=M_Z$
where the appropriate effective coupling is the one in the
5-flavour theory: $\alpha^{(5)}_{\overline{\rm MS}}(M_Z)$.  In
order to obtain it from a result with fewer flavours, one connects effective
theories with different number of flavours as discussed by Bernreuther
and Wetzel~\cite{Bernreuther:1981sg}.  For example one considers the
$\msbar$ scheme, matches the 3-flavour theory to the 4-flavour
theory at a scale given by the charm-quark mass, runs with the
4-loop $\beta$-function of the 4-flavour theory to a scale given by
the $b$-quark mass and there matches to the 5-flavour theory, after
which one runs up to $\mu=M_Z$.  For the matching relation at a given
quark threshold we use the mass $m_\star$ which satisfies $m_\star=
\overline{m}_\msbar(m_\star)$, where $\overline{m}$ is the running
mass (analogous to the running coupling). Then
\begin{eqnarray}
\label{e:convnfm1}
 \gbar^2_{N_f-1}(m_\star) =  \gbar^2_{N_f}(m_\star)\times 
       [1+t_2\,\gbar^4_{N_f}(m_\star)+t_3\,\gbar^6_{N_f}(m_\star)+ \ldots]
\label{e:grelation}
\end{eqnarray}
with \cite{Chetyrkin:2005ia}
\begin{eqnarray}
  t_2 &=&  {1 \over (4\pi^2)^2} {11\over72}\\
  t_3 &=&  {1 \over (4\pi^2)^3} \left[- {82043\over27648}\zeta_3 + 
                     {564731\over124416}-{2633\over31104}(N_f-1)\right]\, 
\end{eqnarray}
(where $\zeta_3$ is the Riemann zeta-function) provides the matching
at the thresholds in the $\msbar$ scheme.  While $t_2$, $t_3$ are
numerically small coefficients, the charm threshold scale is also
relatively low and so there are nonperturbative
uncertainties in the matching procedure, which are difficult to
estimate but which we assume here to be negligible.
Obviously there is no perturbative matching formula across
the strange ``threshold''; here matching is entirely nonperturbative.
Model dependent extrapolations of $\gbar^2_{N_f}$ from $N_f=0,2$ to
$N_f=3$ were done in the early days of lattice gauge theory. We will
include these in our listings of results but not in our estimates,
since such extrapolations are based on untestable assumptions.

\subsubsection{Overview of the review of $\alpha_s$}

We begin by explaining lattice-specific difficulties in \sect{s:crit}
and the FLAG criteria designed to assess whether the
associated systematic uncertainties can be controlled and estimated in
a reasonable manner.  We then discuss, in \sect{s:SF} -- \sect{s:glu},
the various lattice approaches. For completeness, we present results
from calculations with $N_f = 0, 2, 3$, and 4 flavours.  Finally, in
Sec.~\ref{s:alpsumm}, we present averages together with our best
estimates for $\alpha_{\overline{\rm MS}}^{(5)}$. These are determined
from 3- and 4-flavour QCD simulations. The earlier $N_f = 0, 2$
works obtained results for $N_f = 3$ by extrapolation in
$N_f$. Because this is not a theoretically controlled procedure, we do
not include these results in our averages.  For the $\Lambda$
parameter, we also give results for other number of flavours,
including $N_f=0$. Even though the latter numbers should not be used
for phenomenology, they represent valuable nonperturbative
information concerning field theories with variable numbers of quarks.

\subsubsection{Differences compared to the FLAG 13 report}

For the benefit of the readers which are familiar with our previous 
report, we list here where changes and additions can be found which
go beyond slight improvements of the presentation.

Our criteria are unchanged as far as the explicit ratings on
renormalization scale, perturbative behaviour and continuum
extrapolation are concerned. However, where we discuss the criteria,
we emphasize that it is also important whether finite-size effects and
topology sampling are under control. In a few cases, this influences
our decision on which computations enter our ranges and averages.

New computations which are reviewed here are 
\begin{itemize}
\item[]
    Karbstein 14 \cite{Karbstein:2014bsa}
    and Bazavov 14 \cite{Bazavov:2014soa}
    based on the static-quark potential (\sect{s:qq}),
\item[]
    FlowQCD 15 \cite{Asakawa:2015vta} based on a tadpole-improved
    bare coupling (\sect{s:WL}),
\item[]    
    HPQCD 14A  \cite{Chakraborty:2014aca} based on heavy-quark current 
    two-point functions (\sect{s:curr}).
\end{itemize}
They influence the final ranges marginally.

% ----------------------------------------------------------------------

\subsection{Discussion of criteria for computations entering the averages}

% ----------------------------------------------------------------------

\label{s:crit}

% ----------------------------------------------------------------------

As in the PDG review, we only use calculations of $\alpha_s$ published
in peer-reviewed journals, and that use NNLO or higher-order
perturbative expansions, to obtain our final range in
Sec.~\ref{s:alpsumm}.  We also, however, introduce further
criteria designed to assess the ability to control important
systematics which we describe here.  Some of these criteria, 
  e.g.~that for the continuum extrapolation, are associated with
lattice-specific systematics and have no continuum analogue.  Other
criteria, e.g.~that for the renormalization scale, could in
principle be applied to nonlattice determinations.
Expecting that lattice calculations
will continue to improve significantly in the near future, our goal in
reviewing the state of the art here is to be conservative and avoid
prematurely choosing an overly small range.

In lattice calculations, we generally take ${\oO}$ to be some
combination of physical amplitudes or Euclidean correlation functions
which are free from UV and IR divergences and have a well-defined
continuum limit.  Examples include the force between static quarks and
$2$-point functions of quark bilinear currents.

In comparison to values of observables ${\oO}$ determined
experimentally, those from lattice calculations require two more
steps.  The first step concerns setting the scale $\mu$ in \mbox{GeV},
where one needs to use some experimentally measurable low-energy scale
as input. Ideally one employs a hadron mass. Alternatively convenient
intermediate scales such as $\sqrt{t_0}$, $w_0$, $r_0$, $r_1$,
\cite{Luscher:2010iy,Borsanyi:2012zs,Sommer:1993ce,Bernard:2000gd} can
be used if their relation to an experimental dimensionful observable
is established. The low-energy scale needs to be computed at the same
bare parameters where ${\oO}$ is determined, at least as long as
one does not use the step-scaling method (see below).  This induces a
practical difficulty given present computing resources.  In the
determination of the low-energy reference scale the volume needs to be
large enough to avoid finite-size effects. On the other hand, in order
for the perturbative expansion of Eq.~(\ref{eq:alpha_MSbar}) to be
reliable, one has to reach sufficiently high values of $\mu$,
i.e.\ short enough distances. To avoid uncontrollable discretization
effects the lattice spacing $a$ has to be accordingly small.  This
means
\begin{eqnarray}
   L \gg \mbox{hadron size}\sim \Lambda_{\rm QCD}^{-1}\quad 
   \mbox{and} \quad  1/a \gg \mu \,,
   \label{eq:scaleproblem}
\end{eqnarray}
(where $L$ is the box size) and therefore
\begin{eqnarray} 
   L/a \ggg \mu/\Lambda_{\rm QCD} \,.
   \label{eq:scaleproblem2}
\end{eqnarray}
The currently available computer power, however, limits $L/a$, 
typically to
$L/a = 20-64$. 
Unless one accepts compromises in controlling  discretization errors
or finite-size effects, this means one needs to set 
the scale $\mu$ according to
\begin{eqnarray}
   \mu \lll L/a \times \Lambda_{\rm QCD} & \sim 5-20\, \mbox{GeV} \,.
\end{eqnarray}
Therefore, $\mu$ can be $1-3\, \mbox{GeV}$ at most.
This raises the concern whether the asymptotic perturbative expansion
truncated at $1$-loop, $2$-loop, or $3$-loop in Eq.~(\ref{eq:alpha_MSbar})
is sufficiently accurate. There is a finite-size scaling method,
usually called step-scaling method, which solves this problem by identifying 
$\mu=1/L$ in the definition of ${\oO}(\mu)$, see \sect{s:SF}. 

For the second step after setting the scale $\mu$ in physical units
($\mbox{GeV}$), one should compute ${\oO}$ on the lattice,
${\oO}_{\rm lat}(a,\mu)$ for several lattice spacings and take the continuum
limit to obtain the left hand side of Eq.~(\ref{eq:alpha_MSbar}) as
\begin{eqnarray}
   {\oO}(\mu) \equiv \lim_{a\rightarrow 0} {\oO}_{\rm lat}(a,\mu) 
              \mbox{  with $\mu$ fixed}\,.
\end{eqnarray}
This is necessary to remove the discretization error.

Here it is assumed that the quantity ${\oO}$ has a continuum limit,
which is regularization-independent up to discretization errors.
The method discussed in \sect{s:WL}, which is based on the perturbative
expansion of a lattice-regulated, divergent short-distance quantity
$W_{\rm lat}(a)$ differs in this respect and must be
treated separately.

In summary, a controlled determination of $\alpha_s$ 
needs to satisfy the following:
\begin{enumerate}

   \item The determination of $\alpha_s$ is based on a
         comparison of a
         short-distance quantity ${\oO}$ at scale $\mu$ with a well--defined
         continuum limit without UV and IR divergences to a perturbative
         expansion formula in Eq.~(\ref{eq:alpha_MSbar}).

   \item The scale $\mu$ is large enough so that the perturbative expansion
         in Eq.~(\ref{eq:alpha_MSbar}) is precise 
         to the order at which it is truncated,
         i.e. it has good {\em asymptotic} convergence.
         \label{pt_converg}

   \item If ${\oO}$ is defined by physical quantities in infinite volume,  
         one needs to satisfy \eq{eq:scaleproblem2}.
         \label{constraints}

   \item[] Nonuniversal quantities need a separate discussion, see
        \sect{s:WL}.

\end{enumerate}

Conditions \ref{pt_converg}. and \ref{constraints}. give approximate lower and
upper bounds for $\mu$ respectively. It is important to see whether there is a
window to satisfy \ref{pt_converg}. and \ref{constraints}. at the same time.
If it exists, it remains to examine whether a particular lattice
calculation is done inside the window or not. 

Obviously, an important issue for the reliability of a calculation is
whether the scale $\mu$ that can be reached lies in a regime where
perturbation theory can be applied with confidence. However, the value
of $\mu$ does not provide an unambiguous criterion. For instance, the
Schr\"odinger Functional, or SF-coupling (Sec.~\ref{s:SF}) is
conventionally taken at the scale $\mu=1/L$, but one could also choose
$\mu=2/L$. Instead of $\mu$ we therefore define an effective
$\alpha_{\rm eff}$.  For schemes such as SF (see Sec.~\ref{s:SF}) or
$qq$ (see Sec.~\ref{s:qq}) this is directly the coupling  of
the scheme. For other schemes such as the vacuum polarization we use
the perturbative expansion \eq{eq:alpha_MSbar} for the observable
${\oO}$ to define
\begin{eqnarray}
   \alpha_{\rm eff} =  {\oO}/c_1 \,.
   \label{eq:alpeff}
\end{eqnarray}
If there is an $\alpha_s$-independent term it should first be subtracted.
Note that this is nothing but defining an effective,
regularization-independent coupling,
a physical renormalization scheme.

Let us now comment further on the use of the perturbative series.
Since it is only an asymptotic expansion, the remainder $R_n({\oO})={\oO}-\sum_{i\leq n}c_i \alpha_s^i$ of a truncated
perturbative expression ${\oO}\sim\sum_{i\leq n}c_i \alpha_s^i$
cannot just be estimated as a perturbative error $k\,\alpha_s^{n+1}$.
The error is nonperturbative. Often one speaks of ``nonperturbative
contributions'', but nonperturbative and perturbative cannot be
strictly separated due to the asymptotic nature of the series (see
e.g.~Ref.~\cite{Martinelli:1996pk}).

Still, we do have some general ideas concerning the 
size of nonperturbative effects. The known ones such as instantons
or renormalons decay for large $\mu$ like inverse powers of $\mu$
and are thus roughly of the form 
\begin{eqnarray}
   \exp(-\gamma/\alpha_s) \,,
\end{eqnarray}
with some positive constant $\gamma$. Thus we have,
loosely speaking,
\begin{eqnarray}
   {\oO} = c_1 \alpha_s + c_2 \alpha_s^2 + \ldots + c_n\alpha_s^n
                  + \cO(\alpha_s^{n+1}) 
                  + \cO(\exp(-\gamma/\alpha_s)) \,.
   \label{eq:Owitherr}
\end{eqnarray}
For small $\alpha_s$, the $\exp(-\gamma/\alpha_s)$ is negligible.
Similarly the perturbative estimate for the magnitude of
relative errors in \eq{eq:Owitherr} is small; as an
illustration for $n=3$ and $\alpha_s = 0.2$ the relative error
is $\sim 0.8\%$ (assuming coefficients $|c_{n+1} /c_1 | \sim 1$).

For larger values of $\alpha_s$ nonperturbative effects can become
significant in Eq.~(\ref{eq:Owitherr}). An instructive example comes
from the values obtained from $\tau$
decays, for which $\alpha_s\approx 0.3$. Here, different applications
of perturbation theory (fixed order, FOPT, and contour improved, CIPT)
each look reasonably asymptotically convergent but the difference does
not seem to decrease much with the order (see, e.g., the contribution
of Pich in Ref.~\cite{Bethke:2011tr}). In addition nonperturbative terms
in the spectral function may be nonnegligible even after the
integration up to $m_\tau$ (see, e.g., Ref.~\cite{Boito:2014sta}, Golterman
in Ref.~\cite{Bethke:2011tr}). All of this is because $\alpha_s$ is not
really small.

Since the size of the nonperturbative effects is very hard to
estimate one should try to avoid such regions of the coupling.  In a
fully controlled computation one would like to verify the perturbative
behaviour by changing $\alpha_s$ over a significant range instead of
estimating the errors as $\sim \alpha_s^{n+1}$ .  Some computations
try to take nonperturbative power `corrections' to the perturbative
series into account by including such terms in a fit to the $\mu$
dependence. We note that this is a delicate procedure, both because
the separation of nonperturbative and perturbative is theoretically
not well defined and because in practice a term like, e.g.,
$\alpha_s(\mu)^3$ is hard to distinguish from a $1/\mu^2$ term when
the $\mu$-range is restricted and statistical and systematic errors
are present. We consider it safer to restrict the fit range to the
region where the power corrections are negligible compared to the
estimated perturbative error.

The above considerations lead us to the following special
criteria for the determination of $\alpha_s$. 

%\eject
\begin{itemize}
   \item Renormalization scale         
         \begin{itemize}
            \item[\good] all points relevant in the analysis have
             $\alpha_{\rm eff} < 0.2$
            \item[\soso] all points have $\alpha_{\rm eff} < 0.4$
                         and at least one $\alpha_{\rm eff} \le 0.25$
            \item[\bad]  otherwise                                   
         \end{itemize}

   \item Perturbative behaviour 
        \begin{itemize}
           \item[\good] verified over a range of a factor $4$ change
                        in $\alpha_{\rm eff}^{n_\mathrm{l}}$ without power
                        corrections  or alternatively 
                        $\alpha_{\rm eff}^{n_\mathrm{l}}=0.01$ is reached
           \item[\soso] agreement with perturbation theory 
                        over a range of a factor
                        $2.25$ in $\alpha_{\rm eff}^{n_\mathrm{l}}$ 
                        possibly fitting with power corrections or
                        alternatively $\alpha_{\rm eff}^{n_\mathrm{l}}=0.02$
                        is reached
           \item[\bad]  otherwise
       \end{itemize}
        Here $n_\mathrm{l}$ is the loop order to which the 
        connection of $\alpha_{\rm eff}$ to the $\msbar$ scheme is known.
        The $\beta$-function of $\alpha_{\rm eff}$ is then known to 
        $n_\mathrm{l}+1$ loop order.%
        \footnote{Once one is in the perturbative region with 
        $\alpha_{\rm eff}$, the error in 
        extracting the $\Lambda$ parameter due to the truncation of 
        perturbation theory scales like  $\alpha_{\rm eff}^{n_\mathrm{l}}$,
        as seen e.g.\ in Eq.~(\ref{eq:Lambda}). In order to well
        detect/control such corrections, one needs to change
        the correction term significantly; 
        we require a factor of four for a $\good$ and a factor 2.25
        for a $\soso$. In comparison to FLAG 13, where $n_l=2$ was taken
        as the default, we have made the $n_l$ dependence explicit and
        list it in Tabs.~\ref{tab_Nf=0_renormalization_a} --
        \ref{tab_Nf=4_renormalization}. 
        An exception to the above is the situation 
        where the correction terms are small anyway, i.e.  
        $\alpha_{\rm eff}^{n_\mathrm{l}} \approx 0.02$ is reached.}
         
   \item Continuum extrapolation 
        
        At a reference point of $\alpha_{\rm eff} = 0.3$ (or less) we require
         \begin{itemize}
            \item[\good] three lattice spacings with
                         $\mu a < 1/2$ and full $\cO(a)$
                         improvement, \\
                         or three lattice spacings with
                         $\mu a \leq 1/4$ and $2$-loop $\cO(a)$
                         improvement, \\
                         or $\mu a \leq 1/8$ and $1$-loop $\cO(a)$
                         improvement 
            \item[\soso] three lattice spacings with $\mu a < 1.5$
                         reaching down to $\mu a =1$ and full
                         $\cO(a)$ improvement, \\
                         or three lattice spacings with
                         $\mu a \leq 1/4$ and 1-loop $\cO(a)$
                         improvement        
            \item[\bad]  otherwise 
         \end{itemize}

    \item Finite-size effects \\[1ex]
    These are a less serious issue for the determination 
    of $\alpha_s$ since one looks at short-distance observables 
    where such effects are expected to be suppressed. 
    We therefore have no special criterion in our tables, but
    do check that volumes are not too small and in particular
    the scale is determined in large enough volume.% 
    \footnote{
    Note also that the determination of the scale does not need
    to be very precise, since using the lowest-order $\beta$-function
    shows that a 3\% error in the scale determination corresponds to a
    $\sim 0.5\%$ error in $\alpha_s(M_Z)$.  So as long as systematic
    errors from chiral extrapolation and finite-volume effects are below
    3\% we do not need to be concerned about those. This covers
    most cases.}  Remarks 
    are added in the text when appropriate.
    \item Topology sampling \\[1ex]
    In principle a good way to improve the quality 
    of determinations of $\alpha_s$ is to push to very small lattice 
    spacings thus enabling large $\mu$. It is known  
    that the sampling of field space becomes very difficult for the 
    HMC algorithm when the lattice spacing is small and one has the
    standard periodic boundary conditions. In practice, for all known
    discretizations the topological charge slows down dramatically for
    $a\approx 0.05\,\fm$ and smaller \cite{DelDebbio:2002xa,Bernard:2003gq,Schaefer:2010hu,Chowdhury:2013mea,Brower:2014bqa,Bazavov:2014pvz,Fukaya:2015ara}. Open boundary conditions solve the problem 
    \cite{Luscher:2011kk} but are rarely used. Since the effect of
    the freezing is generally not known, we also do need to pay
    attention to this issue. Remarks are added in the text when appropriate.

\end{itemize}  
       
We assume that quark-mass effects
of light quarks (including strange) are negligible in the effective
coupling itself where large, perturbative, $\mu$ is considered.

We also need to specify what is meant by $\mu$. Here are our choices:
\begin{eqnarray}
   \text{Schr\"odinger Functional} &:& \mu=1/L\,,
   \nonumber  \\
   \text{heavy quark-antiquark potential} &:& \mu=2/r\,,
   \nonumber  \\
   \text{observables in momentum space} &:& \mu =q \,,
   \nonumber   \\ 
    \text{moments of heavy-quark currents} 
                                        &:& \mu=2\bar{m}_\mathrm{h}
\label{mu_def}
\end{eqnarray}
where $q$ is the magnitude of the momentum and $\bar{m}_\mathrm{h}$
the heavy-quark mass.  We note again that the above criteria cannot
be applied when regularization dependent quantities
$W_\mathrm{lat}(a)$ are used instead of ${\cO}(\mu)$. These cases
are specifically discussed in \sect{s:WL}.

A popular scale choice is the intermediate $r_0$ scale, although one
should also bear in mind that its determination from physical
observables has also to be taken into account.  The phenomenological
value of $r_0$ was originally determined as $r_0 \approx
0.49\,\mbox{fm}$ through potential models describing quarkonia
\cite{Sommer:1993ce}. Recent determinations from 2-flavour QCD are
$r_0 = 0.420(14) - 0.450(14)\,\mbox{fm}$ by the ETM collaboration
\cite{Baron:2009wt,Blossier:2009bx}, using as input $f_\pi$ and $f_K$
and carrying out various continuum extrapolations. On the other hand,
the ALPHA collaboration \cite{Fritzsch:2012wq} determined $r_0 =
0.503(10)\,\mbox{fm}$ with input from $f_K$, and the QCDSF
Collaboration \cite{Bali:2012qs} cites $0.501(10)(11)\,\mbox{fm}$ from
the mass of the nucleon (no continuum limit).  Recent determinations
from 3-flavour QCD are consistent with $r_1 = 0.313(3)\,\mbox{fm}$
and $r_0 = 0.472(5)\,\mbox{fm}$
\cite{Davies:2009tsa,Bazavov:2010hj,Bazavov:2011nk}. Due to the
uncertainty in these estimates, and as many results are based directly
on $r_0$ to set the scale, we shall often give both the dimensionless
number $r_0 \Lambda_{\overline{\rm MS}}$, as well as $\Lambda_{\overline{\rm MS}}$.
In the cases where no physical $r_0$ scale is given in
the original papers or we convert to
the $r_0$ scale, we use the value $r_0=0.472\,\fm$. In case
$r_1 \Lambda_{\overline{\rm MS}}$ is given in the publications,
we use $r_0 /r_1 = 1.508$ \cite{Bazavov:2011nk} to convert,
neglecting the error on this ratio. In some, mostly early,
computations the string tension, $\sqrt{\sigma}$ was used.
We convert to $r_0$ using $r_0^2\sigma = 1.65-\pi/12$,
which has been shown to be an excellent approximation 
in the relevant pure gauge theory \cite{Necco:2001xg,Luscher:2002qv}.
The new scales $t_0,w_0$ based on the Wilson flow are very attractive
alternatives to $r_0$ but have not yet been used as much and their
discretization errors are still under discussion 
\cite{Ramos:2014kka,Fodor:2014cpa,Bazavov:2015yea,Bornyakov:2015eaa}.
We remain with $r_0$ as our main reference scale for now.

The attentive reader will have noticed that bounds such as $\mu a <
1.5$ or at least one value of $\alpha_\mathrm{eff}\leq 0.25$
which we require for a $\soso$ are
not very stringent. There is a considerable difference between
$\soso$ and $\good$. We have chosen the above bounds, unchanged as compared 
to \flagold, since not too many
computations would satisfy more stringent ones at present.
Nevertheless, we believe that the \soso\ criteria already give
reasonable bases for estimates of systematic errors. In the future, we
expect that we will be able to tighten our criteria for inclusion in
the average, and that many more computations will reach the present
\good\ rating in one or more categories. 

In principle one should also
account for electro-weak radiative corrections. However, both in the
determination of $\alpha_\mathrm{s}$ at intermediate scales $\mu$ and
in the running to high scales, we expect electro-weak effects to be
much smaller than the presently reached precision. Such effects are
therefore not further discussed.

% ----------------------------------------------------------------------

%% file: Alpha_s/s3.tex
\subsection{$\alpha_s$ from the Schr\"odinger Functional}
\label{s:SF}
% ----------------------------------------------------------------------
\subsubsection{General considerations}

% --------------------------------------------------------------------

The method of step-scaling functions avoids the scale problem,
\eq{eq:scaleproblem}. It is in principle independent of the particular
boundary conditions used and was first developed with periodic
boundary conditions in a two-dimensional
model~\cite{Luscher:1991wu}. However, at present most applications in
QCD use Schr\"odinger functional boundary
conditions~\cite{Luscher:1992an,Sint:1993un}.  An important reason is
that these boundary conditions avoid zero modes for the quark fields
and quartic modes \cite{Coste:1985mn} in the perturbative expansion in
the gauge fields.  Furthermore the corresponding renormalization
scheme is well studied in perturbation
theory~\cite{Luscher:1993gh,Sint:1995ch,Bode:1999sm} with the
3-loop $\beta$-function and 2-loop cutoff effects (for the
standard Wilson regularization) known.

Let us first briefly review the step-scaling strategy.  The essential
idea is to split the determination of the running coupling at large
$\mu$ and of a hadronic scale into two lattice calculations and
connect them by `step scaling'.  In the former part, we determine the
running coupling constant in a finite-volume scheme, in practice a
`Schr\"odinger Functional (SF) scheme' in which the renormalization
scale is set by the inverse lattice size $\mu = 1/L$. In this
calculation, one takes a high renormalization scale while keeping the
lattice spacing sufficiently small as
\begin{eqnarray}
   \mu \equiv 1/L \sim 10\,\ldots\, 100\,\mbox{GeV}\,, \qquad a/L \ll 1 \,.
\end{eqnarray}
In the latter part, one chooses a certain 
$\gbar^2_\mathrm{max}=\gbar^2(1/L_\mathrm{max})$, 
typically such that $L_\mathrm{max}$ is around $0.5\,\fm$. With a 
common discretization, one then determines $L_\mathrm{max}/a$ and
(in a large volume $L \ge 2 - 3\,\mbox{fm} $)
a hadronic scale such as a hadron mass, $\sqrt{t_0}/a$ or $r_0/a$
at the same bare parameters. In this way one gets numbers for
$L_\mathrm{max}/r_0$ and by changing the lattice spacing $a$ 
carries out a continuum limit extrapolation of that ratio. 
 
In order to connect $\gbar^2(1/L_\mathrm{max})$ to $\gbar^2(\mu)$ at
high $\mu$, one determines the change of the coupling in the continuum
limit when the scale changes from $L$ to $L/2$, starting from
$L=L_{\rm max}$ and arriving at $\mu = 2^k /L_{\rm max}$. This part of
the strategy is called step scaling. Combining these results yields
$\gbar^2(\mu)$ at $\mu = 2^k {r_0 \over L_\mathrm{max}} r_0^{-1}$,
where $r_0$ stands for the particular chosen hadronic scale.

In order to have a perturbatively well-defined scheme,
the SF scheme uses Dirichlet boundary condition at time 
$t = 0$ and $t = T$. These break translation invariance and permit
${\cO}(a)$ counter terms at the boundary through quantum corrections. 
Therefore, the leading discretization error is ${\cO}(a)$.
Improving the lattice action is achieved by adding
counter terms at the boundaries whose coefficients are denoted
as $c_t,\tilde c_t$. In practice, these coefficients are computed
with $1$-loop or $2$-loop perturbative accuracy.
A better precision in this step yields a better 
control over discretization errors, which is important, as can be
seen, e.g., in Refs.~\cite{Takeda:2004xha,Necco:2001xg}.
The finite $c^{(i)}_g$, \eq{eq:g_conversion}, are 
known for $i=1,2$ 
\cite{Sint:1995ch,Bode:1999sm}.

Also computations with Dirichlet boundary conditions do in principle
suffer from the insufficient change of topology in the HMC algorithm
at small lattice spacing. However, in a small volume the weight of
nonzero charge sectors in the path integral is exponentially
suppressed~\cite{Luscher:1981zf}~\footnote{We simplify here and assume
  that the classical solution associated with the used boundary
  conditions has charge zero.  In practice this is the case.} and one
practically should not sample any nontrivial topology. Considering
the suppression quantitatively Ref.~\cite{Fritzsch:2013yxa} finds a
strong suppression below $L\approx 0.8\,\fm$. Therefore the lack of
topology change of the HMC is not a real issue in the computations
discussed here.  A mix of Dirichlet and open boundary conditions is
expected to remove this worry \cite{Luscher:2014kea} and may be
considered in the future.

% --------------------------------------------------------------------

\subsubsection{Discussion of computations}

% --------------------------------------------------------------------

In Tab.~\ref{tab_SF3} we give results from various determinations
\begin{table}[!htb]
   \vspace{3.0cm}
   \footnotesize
   \begin{tabular*}{\textwidth}[l]{l@{\extracolsep{\fill}}rlllllllll}
      Collaboration & Ref. & $\Nf$ &
      \hspace{0.15cm}\begin{rotate}{60}{publication status}\end{rotate}
                                                       \hspace{-0.15cm} &
      \hspace{0.15cm}\begin{rotate}{60}{renormalization scale}\end{rotate}
                                                       \hspace{-0.15cm} &
      \hspace{0.15cm}\begin{rotate}{60}{perturbative behaviour}\end{rotate}
                                                       \hspace{-0.15cm} &
      \hspace{0.15cm}\begin{rotate}{60}{continuum extrapolation}\end{rotate}
                               \hspace{-0.25cm} & %\rule{0.2cm}{0cm} 
                         scale & $\Lambda_\msbar[\MeV]$ & $r_0\Lambda_\msbar$ \\
      & & & & & & & & \\[-0.1cm]
      \hline
      \hline
      & & & & & & & & \\
      ALPHA 10A & \cite{Tekin:2010mm} & 4 & \gA &\good & \good &\good 
                    & \multicolumn{3}{l}{only running of $\alpha_s$ in Fig.~4}
                    \\  
      Perez 10 & \cite{PerezRubio:2010ke} & 4 & \oP &\good & \good &\soso  
                    & \multicolumn{3}{l}{only step-scaling function in Fig.~4}
                    \\           
      & & & & & & & & & \\[-0.1cm]
      \hline
      & & & & & & & & & \\[-0.1cm]
      PACS-CS 09A& \cite{Aoki:2009tf} & 2+1 
                    & \gA &\good &\good &\soso
                    & $m_\rho$ & $371(13)(8)(^{+0}_{-27})$$^{\#}$
                    & $0.888(30)(18)(^{+0}_{-65})$$^\dagger$
                    \\ % RS 22.6.13, 1.11.13
                    &&&\gA &\good &\good &\soso 
                    & $m_\rho$  & $345(59)$$^{\#\#}$
                    & $0.824(141)$$^\dagger$
                    \\ % RS 22.6.13
      & & & & & & & & \\[-0.1cm]
      \hline  \\[-1.0ex]
      & & & & & & & & \\[-0.1cm]
      ALPHA 12$^*$  & \cite{Fritzsch:2012wq} & 2 
                    & \gA &\good &\good &\good
                    &  $f_{\rm K}$ & $310(20)$ &  $0.789(52)$
                    \\
      ALPHA 04 & \cite{DellaMorte:2004bc} & 2 
                    & \gA &\bad &\good &\good
                    & $r_0 = 0.5\,\mbox{fm}$$^\S$  & $245(16)(16)^\S$ & $0.62(2)(2)^\S$
                    \\
      ALPHA 01A & \cite{Bode:2001jv} & 2 
                    &\gA & \good & \good & \good 
                    &\multicolumn{3}{l}{only running of $\alpha_s$  in Fig.~5}
                    \\
      & & & & & & & & \\[-0.1cm]
      \hline  \\[-1.0ex]
      & & & & & & & & \\[-0.1cm]
      CP-PACS 04$^\&$  & \cite{Takeda:2004xha} & 0 
                    & \gA & \good & \good & \soso  
                    & \multicolumn{3}{l}{only tables of $g^2_{\rm SF}$}
                    \\
      ALPHA 98$^{\dagger\dagger}$ & \cite{Capitani:1998mq} & 0 
                    & \gA & \good & \good & \soso 
                    &  $r_0=0.5\fm$ & $238(19)$ & 0.602(48) 
                    \\
      L\"uscher 93  & \cite{Luscher:1993gh} & 0 
                    & \gA & \good & \soso & \soso
                    & $r_0=0.5\fm$ & 233(23)  & 0.590(60)$^{\S\S}$ 
                    \\
      &&&&&&& \\[-0.1cm]
      \hline
      \hline\\
\end{tabular*}\\[-0.2cm]
\begin{minipage}{\linewidth}
{\footnotesize 
\begin{itemize}
\item[$^{\#}$]    Result with a constant (in $a$) continuum extrapolation
         of the combination $L_\mathrm{max}m_\rho$. \\[-5mm]
\item[$^\dagger$] In conversion to $r_0\Lambda_{\overline{\rm MS}}$, $r_0$ is
         taken to be $0.472\,\mbox{fm}$. \\[-5mm]
\item[$^{\#\#}$]     Result with a linear continuum extrapolation
         in $a$ of the combination $L_\mathrm{max}m_\rho$. \\[-5mm]
\item[$^*$]      Supersedes ALPHA 04. \\[-5mm]
\item[$^\S$]     The $N_f=2$ results were based on values for $r_0/a$
         which have later been found to be too small by
          \cite{Fritzsch:2012wq}. The effect will be of the order of
          10--15\%, presumably an increase in $\Lambda r_0$.
          We have taken this into account by a $\bad$ in the 
          renormalization scale. \\[-5mm]
\item[$^\&$]     This investigation was a precursor for PACS-CS 09A
          and confirmed two step-scaling functions as well as the
          scale setting of ALPHA~98. \\[-5mm]
\item[$^{\dagger\dagger}$]  Uses data of L\"uscher~93 and therefore supersedes it. \\[-5mm]
\item[$^{\S\S}$] Converted from $\alpha_\msbar(37r_0^{-1})=0.1108(25)$.
\end{itemize}
}
\end{minipage}
\caption{Results for the $\Lambda$ parameter from computations using 
         step scaling of the SF-coupling. Entries without values for $\Lambda$
         computed the running and established perturbative behaviour
         at large $\mu$. 
         }
\label{tab_SF3}
\end{table}
of the $\Lambda$ parameter. For a clear assessment of the $N_f$
dependence, the last column also shows results that refer to a common
hadronic scale, $r_0$. As discussed above, the renormalization scale
can be chosen large enough such that $\alpha_s < 0.2$ and the
perturbative behaviour can be verified.  Consequently only $\good$ is
present for these criteria except for early work
where the $n_l=2$ loop connection to $\msbar$ was not yet known.
With dynamical fermions, results for the
step-scaling functions are always available for at least $a/L = \mu a
=1/4,1/6, 1/8$.  All calculations have a nonperturbatively
$\cO(a)$ improved action in the bulk. For the discussed
boundary $\cO(a)$ terms this is not so. In most recent
calculations 2-loop $\cO(a)$ improvement is employed together
with at least three lattice spacings.\footnote{With 2-loop
  $\cO(a)$ improvement we here mean $c_\mathrm{t}$ including
  the $g_0^4$ term and $\tilde c_\mathrm{t}$ with the $g_0^2$
  term. For gluonic observables such as the running coupling this is
  sufficient for cutoff effects being suppressed to $\cO(g^6
  a)$.} This means a \good\ for the continuum extrapolation.  In the
other contributions only 1-loop $c_t$ was available and we arrive at \soso. We
note that the discretization errors in the step-scaling functions are
usually found to be very small, at the percent level or
below. However, the overall desired precision is very high as well,
and the results in CP-PACS 04~\cite{Takeda:2004xha} show that
discretization errors at the below percent level cannot be taken for
granted.  In particular with staggered fermions (unimproved except for
boundary terms) few percent effects are seen in
Perez~10~\cite{PerezRubio:2010ke}.

In the work by PACS-CS 09A~\cite{Aoki:2009tf}, the continuum
extrapolation in the scale setting is performed using a constant
function in $a$ and with a linear function.
Potentially the former leaves a considerable residual discretization 
error. We here use, as discussed with the collaboration, 
the continuum extrapolation linear in $a$,
as given in the second line of PACS-CS 09A \cite{Aoki:2009tf}
results in Tab.~\ref{tab_SF3}.

A single computation, PACS-CS 09A~\cite{Aoki:2009tf}, quotes also
$\alpha_\msbar(M_Z)$. 
We take the linear continuum extrapolation as discussed above:
\begin{eqnarray}
 \alpha_\msbar^{(5)}(M_Z)=0.118(3)\,, 
\end{eqnarray}
where the conversion from a 3-flavour result to 5-flavours
was done perturbatively (see \sect{s:crit}).
Other results do not have a sufficient number of quark flavours
(ALPHA~10A \cite{Tekin:2010mm}, Perez~10 \cite{PerezRubio:2010ke})
or do not yet contain the conversion of the
scale to physical units. Thus no value for $\alpha_\msbar^{(5)}(M_Z)$ is quoted.
 
More results for $\alpha_\msbar^{(5)}(M_Z)$ using step-scaling functions 
can be expected soon. Their precision is likely to be much better than what
we were able to report on here. A major reason is the use of the gradient flow
\cite{Luscher:2010iy} in definitions of finite volume schemes 
\cite{Fodor:2012td,Fritzsch:2013je}.

% ----------------------------------------------------------------------

%% file: Alpha_s/s4.tex
\subsection{$\alpha_s$ from the potential at short distances}
\label{s:qq}

% ----------------------------------------------------------------------

\subsubsection{General considerations}

% --------------------------------------------------------------------

The basic method was introduced in Ref.~\cite{Michael:1992nj} and developed in
Ref.~\cite{Booth:1992bm}. The force or potential between an infinitely
massive quark and antiquark pair defines an effective coupling
constant via
\begin{eqnarray}
   F(r) = {d V(r) \over dr} 
        = C_F {\alpha_\mathrm{qq}(r) \over r^2} \,.
\label{force_alpha}
\end{eqnarray}
The coupling can be evaluated nonperturbatively from the potential
through a numerical differentiation, see below. In perturbation theory
one also defines couplings in different schemes $\alpha_{\bar{V}}$,
$\alpha_V$ via 
\begin{eqnarray}
   V(r) = - C_F {\alpha_{\bar{V}}(r) \over r} \,, 
   \qquad \mbox{or} \quad
   \tilde{V}(Q) = - C_F {\alpha_V(Q) \over Q^2} \,,
\label{pot_alpha}
\end{eqnarray}
where one fixes the unphysical constant in the potential
by $\lim_{r\to\infty}V(r)=0$ and $\tilde{V}(Q)$ is the
Fourier transform of $V(r)$. Nonperturbatively, the subtraction
of a constant in the potential introduces an additional 
renormalization constant, the value of $V(r_\mathrm{ref})$ at some 
distance $r_\mathrm{ref}$.  Perturbatively, it is believed to entail a 
renormalon ambiguity. In perturbation theory, these definitions
are all simply related to each other, and their perturbative
expansions are known including the $\alpha_s^4$ 
and $\alpha_s^5 \log\alpha_s$  terms
\cite{Fischler:1977yf,Billoire:1979ih,Peter:1997me,Schroder:1998vy,
Brambilla:1999qa,Smirnov:2009fh,Anzai:2009tm,Brambilla:2009bi}.
 
The potential $V(r)$ is determined from ratios of Wilson loops,
$W(r,t)$, which behave as
\begin{eqnarray}
   \langle W(r, t) \rangle 
      = |c_0|^2 e^{-V(r)t} + \sum_{n\not= 0} |c_n|^2 e^{-V_n(r)t} \,,
      \label{e:vfromw}
\end{eqnarray}
where $t$ is taken as the temporal extension of the loop, $r$ is the
spatial one and $V_n$ are excited-state potentials.  To improve the
overlap with the ground state, and to suppress the effects of excited
states, $t$ is taken large. Also various additional techniques are
used, such as a variational basis of operators (spatial paths) to help
in projecting out the ground state.  Furthermore some
lattice-discretization effects can be reduced by averaging over Wilson
loops related by rotational symmetry in the continuum.

In order to reduce discretization errors it is of advantage 
to define the numerical derivative giving the force as
\begin{eqnarray}
   F(r_\mathrm{I}) = { V(r) - V(r-a) \over a } \,,
\end{eqnarray}
where $r_\mathrm{I}$ is chosen so that at tree level the force is the
continuum force. $F(r_\mathrm{I})$ is then a `tree-level improved' quantity
and similarly the tree-level improved potential can be defined
\cite{Necco:2001gh}.

Lattice potential results are in position space,
while perturbation theory is naturally computed in momentum space at
large momentum.
Usually, the Fourier transform is then taken of the perturbation
expansion to match to the lattice data.

Finally, as was noted in Sec.~\ref{s:crit}, a determination
of the force can also be used to determine the $r_0$ scale,
by defining it from the static force by
\begin{eqnarray}
   r_0^2 F(r_0) = {1.65} \,,
\end{eqnarray}
and with $r_1^2 F(r_1) = 1$ the $r_1$ scale.

% --------------------------------------------------------------------

\subsubsection{Discussion of computations}
\label{short_dist_discuss}

% --------------------------------------------------------------------

In Tab.~\ref{tab_short_dist}, we list results of determinations
\begin{table}[htb]
   \vspace{3.0cm}
   \footnotesize
   \begin{tabular*}{\textwidth}[l]{l@{\extracolsep{\fill}}rlllllllll}
      Collaboration & Ref. & $N_f$ &
      \hspace{0.15cm}\begin{rotate}{60}{publication status}\end{rotate}
                                                       \hspace{-0.15cm} &
      \hspace{0.15cm}\begin{rotate}{60}{renormalization scale}\end{rotate}
                                                       \hspace{-0.15cm} &
      \hspace{0.15cm}\begin{rotate}{60}{perturbative behaviour}\end{rotate}
                                                       \hspace{-0.15cm} &
      \hspace{0.15cm}\begin{rotate}{60}{continuum extrapolation}\end{rotate}
                               \hspace{-0.25cm} & %\rule{0.2cm}{0cm} 
                         scale & $\Lambda_\msbar[\MeV]$ & $r_0\Lambda_\msbar$ \\
      & & & & & & & & & \\[-0.1cm]
      \hline
      \hline
% \good \soso  \bad
      & & & & & & & & & \\[-0.1cm]

      {Bazavov 14}
                   & \cite{Bazavov:2014soa}  & 2+1       & \gA & \soso
                   & \good  & \soso
                   & $r_1 = 0.3106(17)\,\mbox{fm}^a$
                   & $315(^{+18}_{-12})^b$
                   & $0.746(^{+42}_{-27})$                              \\

      {Bazavov 12}
                   & \cite{Bazavov:2012ka}   & 2+1       & \gA & \soso$^\dagger$
                   & \soso   & \soso$^\#$
                   & $r_0 = 0.468\,\mbox{fm}$ 
                   & $295(30)$\,$^\star$ 
                   & $0.70(7)$$^{\star\star}$                                   \\
      & & & & & & & & & \\[-0.1cm]
      \hline
      & & & & & & & & & \\[-0.1cm]

     Karbstein 14 
                   & \cite{Karbstein:2014bsa} & 2        & \gA & \soso
                   & \soso & \soso
                   & $r_0 = 0.42\,\mbox{fm}$
                   & $331(21)$
                   & 0.692(31)                                         \\

      ETM 11C      & \cite{Jansen:2011vv}    & 2         & \gA & \soso  
                   & \soso  & \soso
                   & $r_0 = 0.42\,\mbox{fm}$
                   & $315(30)$$^\S$ 
                   & $0.658(55)$                                        \\
      & & & & & & & & & \\[-0.1cm]
      \hline
      & & & & & & & & & \\[-0.1cm]
      Brambilla 10 & \cite{Brambilla:2010pp} & 0         & \gA & \soso 
                   & \good\ & \soso$^{\dagger\dagger}$ &  & $266(13)$$^{+}$&
                   $0.637(^{+32}_{-30})$$^{\dagger\dagger}$                  \\
      UKQCD 92     & \cite{Booth:1992bm}    & 0         & \gA & \good 
                                  & \soso$^{++}$   & \bad   
                                  & $\sqrt{\sigma}=0.44\,\GeV$ 
                                  & $256(20)$
                                  & 0.686(54)                             \\
      Bali 92     & \cite{Bali:1992ru}    & 0         & \gA & \good 
                                  & \soso$^{++}$   & \bad 
                                  & $\sqrt{\sigma}=0.44\,\GeV$
                                  & $247(10)$                             
                                  & 0.661(27)                             \\
      & & & & & & & & & \\[-0.1cm]
      \hline
      \hline\\
\end{tabular*}\\[-0.2cm]
\begin{minipage}{\linewidth}
{\footnotesize 
\begin{itemize}
   \item[$^a$]
   Determination on lattices with $m_\pi L=2.2 - 2.6$. 
   About 10 changes of topological charge on the finest lattice 
   \cite{Bazavov:2014pvz}. 
   Scale from $r_1$ \cite{Bazavov:2014pvz}
   as determined from  $f_\pi$ in Ref.~\cite{Bazavov:2010hj}.      \\[-5mm]
   \item[$^b$]
         $\alpha^{(3)}_{\overline{\rm MS}}(1.5\,\mbox{GeV}) = 0.336(^{+12}_{-8})$, 
         $\alpha^{(5)}_{\overline{\rm MS}}(M_Z) = 0.1166(^{+12}_{-8})$.
         \\[-5mm]
   \item[$^\dagger$]
   Since values of $\alpha_\mathrm{eff}$ within our designated range are used,
   we assign a \soso\ despite
   values of $\alpha_\mathrm{eff}$ up to $\alpha_\mathrm{eff}=0.5$ being used.  
   \\[-5mm]
   \item[$^\#$]     Since values of $2a/r$ within our designated range are used,
   we assign a \soso\ although
   only values of $2a/r\geq1.14$ are used at $\alpha_\mathrm{eff}=0.3$.
   \\[-5mm]
   \item[$^\star$] Using results from Ref.~\cite{Bazavov:2011nk}.  \\[-5mm]
   \item[$^{\star\star}$]
         $\alpha^{(3)}_{\overline{\rm MS}}(1.5\,\mbox{GeV}) = 0.326(19)$, 
         $\alpha^{(5)}_{\overline{\rm MS}}(M_Z) = 0.1156(^{+21}_{-22})$.  \\[-5mm]
   \item[$^\S$] Both potential and $r_0/a$ are determined on a small 
   ($L=3.2r_0$) lattice.   \\[-5mm]
   \item[$^{\dagger\dagger}$] Uses lattice results of Ref.~\cite{Necco:2001xg}, 
   some of which have very small lattice spacings where 
   according to more recent investigations a bias due to the freezing of
   topology may be present.  \\[-5mm] 
   \item[$^+$] Only $r_0\Lambda_{\overline{\rm MS}}$ is given,
        our conversion using $r_0 = 0.472\,\mbox{fm}$.   \\[-5mm]
   \item[$^{++}$] We give a $\soso$ because only a NLO formula is used and
       the error bars are very large; our criterion does not apply 
       well to these very early calculations.           
\end{itemize}
}
\end{minipage}
\normalsize
\caption{Short-distance potential results.}
\label{tab_short_dist}
\end{table}
of $r_0\Lambda_{\msbar}$ (together with $\Lambda_{\msbar}$
using the scale determination of the authors). 
Since the last review, FLAG 13, there have been
two new computations, Karbstein 14 \cite{Karbstein:2014bsa}
and Bazavov 14 \cite{Bazavov:2014soa}.

The first determinations in the three-colour Yang Mills theory are by
UKQCD 92 \cite{Booth:1992bm} and Bali 92 \cite{Bali:1992ru} who used
$\alpha_\mathrm{qq}$ as explained above, but not in the tree-level
improved form. Rather a phenomenologically determined lattice artifact
correction was subtracted from the lattice potentials.  The comparison
with perturbation theory was on a more qualitative level on the basis
of a 2-loop $\beta$-function ($n_l=1$) and a continuum extrapolation
could not be performed as yet. A much more precise computation of
$\alpha_\mathrm{qq}$ with continuum extrapolation was performed in
Refs.~\cite{Necco:2001xg,Necco:2001gh}. Satisfactory agreement with
perturbation theory was found \cite{Necco:2001gh} but the stability of
the perturbative prediction was not considered sufficient to be able
to extract a $\Lambda$ parameter.

In Brambilla 10 \cite{Brambilla:2010pp} the same quenched lattice
results of Ref.~\cite{Necco:2001gh} were used and a fit was performed to
the continuum potential, instead of the force. Perturbation theory to
$n_l=3$ loop
was used including a resummation of terms $\alpha_s^3 (\alpha_s \ln\alpha_s)^n $ 
and $\alpha_s^4 (\alpha_s \ln\alpha_s)^n $. Close
agreement with perturbation theory was found when a renormalon
subtraction was performed. Note that the renormalon subtraction
introduces a second scale into the perturbative formula which is
absent when the force is considered.

Bazavov 14 \cite{Bazavov:2014soa} is an update of
Bazavov 12 \cite{Bazavov:2012ka} and modify this procedure
somewhat. They consider the well-defined perturbative expansion
for the force, where renormalon problems disappear. They set $\mu = 1/r$
to eliminate logarithms and then integrate the force to obtain an
expression for the potential. 
The resulting integration constant is fixed by requiring
the perturbative potential to be equal to the nonperturbative 
one exactly at a reference distance $r_{\rm ref}$ and the two are then
compared at other values of $r$. As a further check,
the force is also used directly.

For the quenched calculation Brambilla 10 \cite{Brambilla:2010pp}
very small lattice spacings were available,
$a \sim 0.025\,\mbox{fm}$, \cite{Necco:2001gh}.
For ETM 11C \cite{Jansen:2011vv}, Bazavov 12 \cite{Bazavov:2012ka},
Karbstein 14 \cite{Karbstein:2014bsa}
and Bazavov 14 \cite{Bazavov:2014soa} using dynamical
fermions such small lattice spacings are not yet realized 
(Bazavov 14 reaches down to $a \sim 0.041\,\mbox{fm}$). They
all use the tree-level improved potential as described above. 
We note that the value of $\Lambda_\msbar$ in physical units by
ETM 11C \cite{Jansen:2011vv} is based on a value of $r_0=0.42$~fm. 
This is at least 10\% smaller than the large majority of
other values of $r_0$. Also the value of $r_0/a$ or $r_1/a$
on the finest lattices in ETM 11C \cite{Jansen:2011vv}
and Bazavov 14 \cite{Bazavov:2014soa} come from
rather small lattices with $m_\pi L \approx 2.4$, $2.2$ respectively.

Instead of the procedure discussed previously, Karbstein 14 
\cite{Karbstein:2014bsa} reanalyzes the data of ETM 11C 
\cite{Jansen:2011vv} by first estimating
the Fourier transform $\tilde V(p)$ of $V(r)$ and then fits 
the perturbative expansion of $\tilde V(p)$ in terms of 
$\alpha_\msbar(p)$. Of course, the Fourier transform cannot 
be computed without modelling the $r$-dependence of $V(r)$
at short and at large distances. The authors fit a linearly rising
potential at large distances together with string-like
corrections of order $r^{-n}$ and define the potential at large 
distances by this fit.\footnote{Note that at large distances,
where string breaking is known to occur, this is not 
any more the ground state potential defined by \eq{e:vfromw}.}
Recall that for observables in momentum space
we take the renormalization scale entering our criteria as $\mu=p$,
Eq.~(\ref{mu_def}). The analysis (as in ETM 11C \cite{Jansen:2011vv})
is dominated by the data at the smallest lattice spacing, where
a controlled determination of the overall scale  is difficult due to 
possible finite-size effects.

One of the main issues for all these computations is whether the
perturbative running of the coupling constant
has been reached. While for quenched or $N_f=0$ fermions this seems
to be the case at the smallest distances, for dynamical fermions at present 
there is no consensus. Brambilla 10
\cite{Brambilla:2010pp}, Bazavov 12 \cite{Bazavov:2012ka}
and Bazavov 14 \cite{Bazavov:2014soa} report good agreement with perturbation
theory after the renormalon is subtracted or eliminated, but
Ref.~\cite{Knechtli:2011pz} uses the force directly,
where no renormalon contributes, and finds that far shorter distances
are needed than are presently accessible for dynamical fermion
simulations in order to match to perturbation theory.  
Further work is needed to clarify this point. 

A second issue is the coverage of configuration space in some of the
simulations, which use very small lattice spacings with periodic
boundary conditions. Affected are the smallest two lattice spacings
of Bazavov 14 \cite{Bazavov:2014soa} where very few tunnelings of
the topological charge occur \cite{Bazavov:2014pvz}.
With present knowledge, it also seems  possible that the older data
by Refs.~\cite{Necco:2001xg,Necco:2001gh} used by Brambilla 10 
\cite{Brambilla:2010pp} are partially done in (close to) frozen topology.

% ----------------------------------------------------------------------

%% file: Alpha_s/s5.tex
\subsection{$\alpha_s$ from the vacuum polarization at short distances}

% ----------------------------------------------------------------------

\label{s:vac}

% ----------------------------------------------------------------------

\subsubsection{General considerations}

% ----------------------------------------------------------------------

The vacuum polarization function for the flavour nonsinglet 
currents $J^a_\mu$ ($a=1,2,3$) in the momentum representation is
parameterized as 
\begin{eqnarray}
   \langle J^a_\mu J^b_\nu \rangle 
      =\delta^{ab} [(\delta_{\mu\nu}Q^2 - Q_\mu Q_\nu) \Pi^{(1)}(Q) 
                                     - Q_\mu Q_\nu\Pi^{(0)}(Q)] \,,
\end{eqnarray}
where $Q_\mu$ is a space like momentum and $J_\mu\equiv V_\mu$
for a vector current and $J_\mu\equiv A_\mu$ for an axial-vector current. 
Defining $\Pi_J(Q)\equiv \Pi_J^{(0)}(Q)+\Pi_J^{(1)}(Q)$,
the operator product expansion (OPE) of the vacuum polarization
function $\Pi_{V+A}(Q)=\Pi_V(Q)+\Pi_A(Q)$ is given by
\begin{eqnarray}
   \lefteqn{\Pi_{V+A}|_{\rm OPE}(Q^2,\alpha_s)}
      & &                                             \nonumber  \\
      &=& c + C_1(Q^2) + C_m^{V+A}(Q^2)
                       \frac{\bar{m}^2(Q)}{Q^2}
            + \sum_{q=u,d,s}C_{\bar{q}q}^{V+A}(Q^2)
                        \frac{\langle m_q\bar{q}q \rangle}{Q^4}
                                                      \nonumber  \\
      & &   + C_{GG}(Q^2) 
                \frac{\langle \alpha_s GG\rangle}{Q^4}+{\cO}(Q^{-6}) \,,
\label{eq:vacpol}
\end{eqnarray}
for large
$Q^2$. $C_X^{V+A}(Q^2)=\sum_{i\geq0}\left( C_X^{V+A}\right)^{(i)}\alpha_s^i(Q^2)$
are the perturbative coefficient functions for the
operators $X$ ($X=1$, $\bar{q}q$, $GG$) and $\bar m$ is the running 
mass of the mass-degenerate up and down quarks.
$C_1$ is known including $\alpha_s^4$
in a continuum renormalization scheme such as the
$\overline{\rm MS}$ scheme
\cite{Surguladze:1990tg,Gorishnii:1990vf,Baikov:2008jh}.
Nonperturbatively, there are terms in $C_X$ which do not have a 
series expansion in $\alpha_s$. For an example for the unit
operator see Ref.~\cite{Balitsky:1993ki}.
The term $c$ is $Q$--independent and divergent in the limit of infinite
ultraviolet cutoff. However the Adler function defined as 
\begin{eqnarray}
   D(Q^2) \equiv - Q^2 { d\Pi(Q^2) \over dQ^2} \,,
\end{eqnarray}
is a scheme-independent finite quantity. Therefore one can determine
the running coupling constant in the $\overline{\rm MS}$ scheme
from the vacuum polarization function computed by a lattice-QCD
simulation. 
In more detail, the lattice data of the vacuum polarization is fitted with the 
perturbative formula Eq.~(\ref{eq:vacpol}) with fit parameter 
$\Lambda_{\overline{\rm MS}}$ parameterizing the running coupling 
$\alpha_{\overline{\rm MS}}(Q^2)$.  

While there is no problem in discussing the OPE at the
nonperturbative level, the `condensates' such as ${\langle \alpha_s
  GG\rangle}$ are ambiguous, since they mix with lower-dimensional
operators including the unity operator.  Therefore one should work in
the high-$Q^2$ regime where power corrections are negligible within
the given accuracy. Thus setting the renormalization scale as
$\mu\equiv \sqrt{Q^2}$, one should seek, as always, the window
$\Lambda_{\rm QCD} \ll \mu \ll a^{-1}$.

% --------------------------------------------------------------------

\subsubsection{Discussion of computations}

% --------------------------------------------------------------------

Results using this method are, to date, only available using
overlap fermions. These are collected in Tab.~\ref{tab_vac} for
\begin{table}[!htb]
   \vspace{3.0cm}
   \footnotesize
   \begin{tabular*}{\textwidth}[l]{l@{\extracolsep{\fill}}rllllllll}
   Collaboration & Ref. & $\Nf$ &
   \hspace{0.15cm}\begin{rotate}{60}{publication status}\end{rotate}
                                                    \hspace{-0.15cm} &
   \hspace{0.15cm}\begin{rotate}{60}{renormalization scale}\end{rotate}
                                                    \hspace{-0.15cm} &
   \hspace{0.15cm}\begin{rotate}{60}{perturbative behaviour}\end{rotate}
                                                    \hspace{-0.15cm} &
   \hspace{0.15cm}\begin{rotate}{60}{continuum extrapolation}\end{rotate}
      \hspace{-0.25cm} & %\rule{0.2cm}{0cm} 
                         scale & $\Lambda_\msbar[\MeV]$ & $r_0\Lambda_\msbar$ \\
   & & & & & & & & & \\[-0.1cm]
   \hline
   \hline
   & & & & & & & & & \\[-0.1cm]
   JLQCD 10 & \cite{Shintani:2010ph} & 2+1 &\gA & \bad 
            & \bad & \bad
            & $r_0 = 0.472\,\mbox{fm}$
            & $247(5)$$^\dagger$
            & $0.591(12)$              \\
   & & & & & & & & & \\[-0.1cm]
   \hline
   & & & & & & & & & \\[-0.1cm]
   JLQCD/TWQCD 08C & \cite{Shintani:2008ga} & 2 & \gA & \soso 
            & \bad & \bad
            & $r_0 = 0.49\,\mbox{fm}$
            & $234(9)(^{+16}_{-0})$
            & $0.581(22)(^{+40}_{-0})$    \\
            
   & & & & & & & & & \\[-0.1cm]
   \hline
   \hline
\end{tabular*}
\begin{tabular*}{\textwidth}[l]{l@{\extracolsep{\fill}}llllllll}
\multicolumn{8}{l}{\vbox{\begin{flushleft}
   $^\dagger$  $\alpha_\msbar^{(5)}(M_Z)=0.1118(3)(^{+16}_{-17})$. \\
\end{flushleft}}}
\end{tabular*}
\vspace{-0.3cm}
\normalsize
\caption{Vacuum polarization results.}
\label{tab_vac}
\end{table}
$N_f=2$, JLQCD/TWQCD 08C \cite{Shintani:2008ga} and for $N_f = 2+1$, JLQCD 10
\cite{Shintani:2010ph}.
At present, only one lattice spacing $a \approx 0.11\,\mbox{fm}$ 
has been simulated.

The fit to \eq{eq:vacpol} is done with the 4-loop relation between
the running coupling and $\lms$.  It is found that without introducing
condensate contributions, the momentum scale where the perturbative
formula gives good agreement with the lattice results is very narrow,
$aQ \simeq 0.8-1.0$.  When condensate contributions are included the
perturbative formula gives good agreement with the lattice results for
the extended range $aQ \simeq 0.6-1.0$. Since there is only a single
lattice spacing there is a \bad\ for the continuum limit.  The
renormalization scale $\mu$ is in the range of $Q=1.6-2\,\mbox{GeV}$.
Approximating $\alpha_{\rm eff}\approx \alpha_{\overline{\rm MS}}(Q)$,
we estimate that $\alpha_{\rm eff}=0.25-0.30$ for $N_f=2$
and $\alpha_{\rm  eff}=0.29-0.33$ for $N_f=2+1$. Thus we give a 
\soso\ and \bad\ for $\Nf=2$ and $\Nf=2+1$ respectively for
the renormalization scale and a \bad\ for the perturbative behaviour.

We note that more investigations of this method are in progress
\cite{Hudspith:2015xoa}.

% ----------------------------------------------------------------------

%% file: Alpha_s/s6.tex
\subsection{$\alpha_s$ from observables at the lattice spacing scale}

% ----------------------------------------------------------------------

\label{s:WL}

% ----------------------------------------------------------------------

% ----------------------------------------------------------------------

\subsubsection{General considerations}

% ----------------------------------------------------------------------

The general method is to evaluate a short-distance quantity ${\oO}$
at the scale of the lattice spacing $\sim 1/a$ and then determine
its relationship to $\alpha_{\overline{\rm MS}}$ via a power series expansion.

This is epitomized by the strategy of the HPQCD collaboration
\cite{Mason:2005zx,Davies:2008sw}, discussed here for illustration,
which computes and then fits to a variety of short-distance quantities, $Y$,
\begin{eqnarray}
   Y = \sum_{n=1}^{n_{\rm max}} c_n \alphah^n(q^*) \,.
\label{Ydef}
\end{eqnarray}
$Y$ is taken as the logarithm of small Wilson loops (including some
nonplanar ones), Creutz ratios, `tadpole-improved' Wilson loops and
the tadpole-improved or `boosted' bare coupling ($\cO(20)$ quantities in
total). $c_n$ are perturbative coefficients (each depending on the
choice of $Y$) known to $n = 3$ with additional coefficients up to
$n_{\rm max}$ being numerically fitted.  $\alphah$ is the running
coupling constant related to $\alphav$ from the static-quark potential
(see Sec.~\ref{s:qq}).\footnote{ $\alphah$ is defined by
  $\Lambda_\mathrm{V'}=\Lambda_\mathrm{V}$ and
  $b_i^\mathrm{V'}=b_i^\mathrm{V}$ for $i=0,1,2$ but $b_i^\mathrm{V'}=0$ for
  $i\geq3$. }

 The coupling
constant is fixed at a scale $q^* = d/a$.
This is chosen as the mean value of $\ln q$ with the one gluon loop
as measure
\cite{Lepage:1992xa,Hornbostel:2002af}. (Thus a different result
for $d$ is found for every short-distance quantity.)
A rough estimate yields $d \approx \pi$, and in general the
renormalization scale is always found to lie in this region.

For example for the Wilson loop $W_{mn} \equiv \langle W(ma,na) \rangle$
we have
\begin{eqnarray}
   \ln\left( \frac{W_{mn}}{u_0^{2(m+n)}}\right)
      = c_1 \alphah(q^*) +  c_2 \alphah^2(q^*)  + c_3 \alphah^3(q^*)
        + \cdots \,,
\label{short-cut}
\end{eqnarray}
for the tadpole-improved version, where $c_1$, $c_2\,, \ldots$
are the appropriate perturbative coefficients and $u_0 = W_{11}^{1/4}$.
Substituting the nonperturbative simulation value in the left hand side,
we can determine $\alphah(q^*)$, at the scale $q^*$.
Note that one finds empirically that perturbation theory for these
tadpole-improved quantities have smaller $c_n$ coefficients and so
the series has a faster apparent convergence.

Using the $\beta$-function in the $\rm V'$ scheme,
results can be run to a reference value, chosen as
$\alpha_0 \equiv \alphah(q_0)$, $q_0 = 7.5\,\mbox{GeV}$.
This is then converted perturbatively to the continuum
$\msbar$ scheme
\begin{eqnarray}
   \alpha_{\overline{\rm MS}}(q_0)
      = \alpha_0 + d_1 \alpha_0^2 + d_2 \alpha_0^3 + \cdots \,,
\end{eqnarray}
where $d_1, d_2$ are known one and two loop coefficients.

Other collaborations have focused more on the bare `boosted'
coupling constant and directly determined its relationship to
$\alpha_{\overline{\rm MS}}$. Specifically, the boosted coupling is
defined by 
\begin{eqnarray}
   \alphap(1/a) = {1\over 4\pi} {g_0^2 \over u_0^4} \,,
\end{eqnarray}
again determined at a scale $\sim 1/a$. As discussed previously
since the plaquette expectation value in the boosted coupling
contains the tadpole diagram contributions to all orders, which
are dominant contributions in perturbation theory,
there is an expectation that the perturbation theory using
the boosted coupling has 
smaller perturbative coefficients \cite{Lepage:1992xa}, and hence smaller 
perturbative errors.
 
% ----------------------------------------------------------------------

\subsubsection{Continuum limit}

% ----------------------------------------------------------------------

Lattice results always come along with discretization errors,
which one needs to remove by a continuum extrapolation.
As mentioned previously, in this respect the present
method differs in principle from those in which $\alpha_s$ is determined
from physical observables. In the general case, the numerical
results of the lattice simulations at a value of $\mu$ fixed in physical 
units can be extrapolated to the continuum limit, and the result can be 
analyzed as to whether it shows perturbative running as a function of 
$\mu$ in the continuum. For observables at the cutoff-scale ($q^*=d/a$),  
discretization effects cannot easily be separated out
from perturbation theory, as the scale for the coupling
comes from the lattice spacing. 
Therefore the restriction  $a\mu  \ll 1$ (the `continuum extrapolation'
criterion) is not applicable here. Discretization errors of 
order $a^2$ are, however, present. Since 
$a\sim \exp(-1/(2b_0 g_0^2)) \sim \exp(-1/(8\pi b_0 \alpha(q^*))$, 
these errors now appear as power corrections to the perturbative 
running, and have to be taken into account in the study of the 
perturbative behaviour, which is to be verified by changing $a$. 
One thus usually fits with power corrections in this method.

In order to keep a symmetry with the `continuum extrapolation' 
criterion for physical observables and to remember that discretization 
errors are, of course, relevant, 
we replace it here by one for the lattice spacings used:
\begin{itemize}
   \item Lattice spacings
         \begin{itemize}
            \item[\good] 
               3 or more lattice spacings, at least 2 points below
               $a = 0.1\,\mbox{fm}$
            \item[\soso]
               2 lattice spacings, at least 1 point below
               $a = 0.1\,\mbox{fm}$
            \item[\bad]
               otherwise 
         \end{itemize}
\end{itemize}

% ----------------------------------------------------------------------

\subsubsection{Discussion of computations}

% ----------------------------------------------------------------------

Note that due to $\mu \sim 1/a$ being relatively large the
results easily have a $\good$ or $\soso$ in the rating on 
renormalization scale.

The work of El-Khadra 92 \cite{ElKhadra:1992vn} employs a 1-loop
formula to relate $\alpha^{(0)}_{\overline{\rm MS}}(\pi/a)$
to the boosted coupling for three lattice spacings
$a^{-1} = 1.15$, $1.78$, $2.43\,\mbox{GeV}$. (The lattice spacing
is determined from the charmonium 1S-1P splitting.) They obtain
$\Lambda^{(0)}_{\overline{\rm MS}}=234\,\mbox{MeV}$, corresponding
to $\alpha_{\rm eff} = \alpha^{(0)}_{\overline{\rm MS}}(\pi/a)
\approx 0.15$ - $0.2$. The work of Aoki 94 \cite{Aoki:1994pc}
calculates $\alpha^{(2)}_V$ and $\alpha^{(2)}_{\overline{\rm MS}}$
for a single lattice spacing $a^{-1}\sim 2\,\mbox{GeV}$ again
determined from charmonium 1S-1P splitting in 2-flavour QCD.
Using 1-loop perturbation theory with boosted coupling,
they obtain $\alpha^{(2)}_V=0.169$ and $\alpha^{(2)}_{\overline{\rm MS}}=0.142$.
Davies 94 \cite{Davies:1994ei} gives a determination of $\alphav$
from the expansion 
\begin{equation}
   -\ln W_{11} \equiv \frac{4\pi}{3}\alphav^{(N_f)}(3.41/a)
        \times [1 - (1.185+0.070N_f)\alphav^{(N_f)} ]\,,
\end{equation}
neglecting higher-order terms.  They compute the $\Upsilon$ spectrum
in $N_f=0$, $2$ QCD for single lattice spacings at $a^{-1} = 2.57$,
$2.47\,\mbox{GeV}$ and obtain $\alphav(3.41/a)\simeq 0.1$5, $0.18$
respectively.  Extrapolating the inverse coupling linearly in $N_f$, a
value of $\alphav^{(3)}(8.3\,\mbox{GeV}) = 0.196(3)$ is obtained.
SESAM 99 \cite{Spitz:1999tu} follows a similar strategy, again for a
single lattice spacing. They linearly extrapolated results for
$1/\alphav^{(0)}$, $1/\alphav^{(2)}$ at a fixed scale of
$9\,\mbox{GeV}$ to give $\alphav^{(3)}$, which is then perturbatively
converted to $\alpha_{\overline{\rm MS}}^{(3)}$. This finally gave
$\alpha_{\overline{\rm MS}}^{(5)}(M_Z) = 0.1118(17)$.  Wingate 95
\cite{Wingate:1995fd} also follow this method.  With the scale
determined from the charmonium 1S-1P splitting for single lattice
spacings in $N_f = 0$, $2$ giving $a^{-1}\simeq 1.80\,\mbox{GeV}$ for
$N_f=0$ and $a^{-1}\simeq 1.66\,\mbox{GeV}$ for $N_f=2$ they obtain
$\alphav^{(0)}(3.41/a)\simeq 0.15$ and $\alphav^{(2)}\simeq 0.18$
respectively. Extrapolating the coupling linearly in $N_f$, they
obtain $\alphav^{(3)}(6.48\,\mbox{GeV})=0.194(17)$.

\begin{table}[!h]
   \vspace{3.0cm}
   \footnotesize
   \begin{tabular*}{\textwidth}[l]{l@{\extracolsep{\fill}}rllllllll}
   Collaboration & Ref. & $N_f$ &
   \hspace{0.15cm}\begin{rotate}{60}{publication status}\end{rotate}
                                                    \hspace{-0.15cm} &
   \hspace{0.15cm}\begin{rotate}{60}{renormalization scale}\end{rotate}
                                                    \hspace{-0.15cm} &
   \hspace{0.15cm}\begin{rotate}{60}{perturbative behaviour}\end{rotate}
                                                    \hspace{-0.15cm} &
   \hspace{0.15cm}\begin{rotate}{60}{lattice spacings}\end{rotate}
      \hspace{-0.25cm} & %\rule{0.2cm}{0cm} 
                         scale & $\Lambda_\msbar[\MeV]$ & $r_0\Lambda_\msbar$ \\
   & & & & & & & & \\[-0.1cm]
   \hline
   \hline
   & & & & & & & & \\[-0.1cm]
   HPQCD 10$^a$$^\S$& \cite{McNeile:2010ji}& 2+1 & \gA & \soso
            & \good & \good
            & $r_1 = 0.3133(23)\, \mbox{fm}$
            & 340(9) 
            & 0.812(22)                                   \\ 
%            &                     &     &     &
   HPQCD 08A$^a$& \cite{Davies:2008sw} & 2+1 & \gA & \soso
            & \good & \good
            & $r_1 = 0.321(5)\,\mbox{fm}$$^{\dagger\dagger}$
            & 338(12)$^\star$
            & 0.809(29)                                   \\
%            &                     &     &     &
   Maltman 08$^a$& \cite{Maltman:2008bx}& 2+1 & \gA & \soso
            & \soso & \good
            & $r_1 = 0.318\, \mbox{fm}$
            & 352(17)$^\dagger$
            & 0.841(40)                                   \\ 
%            &                     &     &     &
   HPQCD 05A$^a$ & \cite{Mason:2005zx} & 2+1 & \gA & \soso
            & \soso & \soso
            & $r_1$$^{\dagger\dagger}$
            & 319(17)$^{\star\star}$
            & 0.763(42)                                   \\
%            &                     &     &     &
   & & & & & & & & &  \\[-0.1cm]
   \hline
   & & & & & & & & &  \\[-0.1cm]
   QCDSF/UKQCD 05 & \cite{Gockeler:2005rv}  & 2 & \gA & \good
            & \bad  & \good
            & $r_0 = 0.467(33)\,\mbox{fm}$
            & 261(17)(26)
            & 0.617(40)(21)$^b$                           \\
%            &                     &     &     &
   SESAM 99$^c$ & \cite{Spitz:1999tu} & 2 & \gA & \soso
            & \bad  & \bad
            & $c\bar{c}$(1S-1P)
            & 
            &                                             \\
   Wingate 95$^d$ & \cite{Wingate:1995fd} & 2 & \gA & \good
            & \bad  & \bad
            & $c\bar{c}$(1S-1P)
            & 
            &                                             \\
   Davies 94$^e$ & \cite{Davies:1994ei} & 2 & \gA & \good
            & \bad & \bad
            & $\Upsilon$
            & 
            &                                             \\
   Aoki 94$^f$ & \cite{Aoki:1994pc} & 2 & \gA & \good
            & \bad & \bad
            & $c\bar{c}$(1S-1P)
            & 
            &                                             \\
   & & & & & & & & &  \\[-0.1cm]
   \hline
   & & & & & & & & &  \\[-0.1cm]

   FlowQCD 15
            & \cite{Asakawa:2015vta}        & 0 & \oP 
            & \good  & \good   & \good
            & $w_{0.4} = 0.193(3)\,\mbox{fm}$$^i$
            & $258(6)$$^i$
            & 0.618(11)$^i$                             \\

   QCDSF/UKQCD 05 & \cite{Gockeler:2005rv}  & 0 & \gA & \good
            & \soso & \good
            & $r_0 = 0.467(33)\,\mbox{fm}$
            & 259(1)(20)
            & 0.614(2)(5)$^b$                              \\
%            &                     &     &     &
   SESAM 99$^c$ & \cite{Spitz:1999tu} & 0 & \gA & \good
            & \bad  & \bad
            & $c\bar{c}$(1S-1P)
            & 
            &                                             \\
   Wingate 95$^d$ & \cite{Wingate:1995fd} & 0 & \gA & \good
            & \bad  & \bad
            & $c\bar{c}$(1S-1P)
            & 
            &                                             \\
   Davies 94$^e$ & \cite{Davies:1994ei}  & 0 & \gA & \good
            & \bad & \bad
            & $\Upsilon$
            & 
            &                                             \\
   El-Khadra 92$^g$ & \cite{ElKhadra:1992vn} & 0 & \gA & \good
            & \bad    & \soso
            & $c\bar{c}$(1S-1P)
            & 234(10)
            & 0.560(24)$^h$                               \\
   & & & & & & & & &  \\[-0.1cm]
   \hline
   \hline\\
\end{tabular*}\\[-0.2cm]
\begin{minipage}{\linewidth}
{\footnotesize 
\begin{itemize}
   \item[$^a$]       The numbers for $\Lambda$ have been converted from the values for 
              $\alpha_s^{(5)}(M_Z)$. \\[-5mm]
   \item[$^{\S}$]     $\alpha_{\overline{\rm MS}}^{(3)}(5\ \mbox{GeV})=0.2034(21)$,
              $\alpha^{(5)}_{\overline{\rm MS}}(M_Z)=0.1184(6)$,
              only update of intermediate scale and $c$-, $b$-quark masses,
              supersedes HPQCD 08A.\\[-5mm]
   \item[$^\dagger$] $\alpha^{(5)}_{\overline{\rm MS}}(M_Z)=0.1192(11)$. \\[-4mm]
   \item[$^\star$]    $\alpha_V^{(3)}(7.5\,\mbox{GeV})=0.2120(28)$, 
              $\alpha^{(5)}_{\overline{\rm MS}}(M_Z)=0.1183(8)$,
              supersedes HPQCD 05. \\[-5mm]
   \item[$^{\dagger\dagger}$] Scale is originally determined from $\Upsilon$
              mass splitting. $r_1$ is used as an intermediate scale.
              In conversion to $r_0\Lambda_{\overline{\rm MS}}$, $r_0$ is
              taken to be $0.472\,\mbox{fm}$. \\[-5mm]
   \item[$^{\star\star}$] $\alpha_V^{(3)}(7.5\,\mbox{GeV})=0.2082(40)$,
              $\alpha^{(5)}_{\overline{\rm MS}}(M_Z)=0.1170(12)$. \\[-5mm]
   \item[$^b$]       This supersedes 
              Refs.~\cite{Gockeler:2004ad,Booth:2001uy,Booth:2001qp}.
              $\alpha^{(5)}_{\overline{\rm MS}}(M_Z)=0.112(1)(2)$.
              The $N_f=2$ results were based on values for $r_0 /a$
              which have later been found to be too 
              small~\cite{Fritzsch:2012wq}. The effect will  
              be of the order of 10--15\%, presumably an increase in 
              $\Lambda r_0$. \\[-5mm]
   \item[$^c$]       $\alpha^{(5)}_{\overline{\rm MS}}(M_Z)=0.1118(17)$. \\[-4mm]
   \item[$^d$]    
   $\alpha_V^{(3)}(6.48\,\mbox{GeV})=0.194(7)$ extrapolated from $\Nf=0,2$.
              $\alpha^{(5)}_{\overline{\rm MS}}(M_Z)=0.107(5)$.   \\[-4mm]
   \item[$^e$]  
              $\alpha_P^{(3)}(8.2\,\mbox{GeV})=0.1959(34)$ extrapolated
              from $N_f=0,2$. $\alpha^{(5)}_{\overline{\rm MS}}(M_Z)=0.115(2)$.
              \\[-5mm]
   \item[ $^f$]       Estimated $\alpha^{(5)}_{\overline{\rm MS}}(M_Z)=0.108(5)(4)$. \\[-5mm]
   \item[$^g$]       This early computation violates our requirement that
              scheme conversions are done at the 2-loop level.
              $\Lambda_{\overline{\rm MS}}^{(4)}=160(^{+47}_{-37})\mbox{MeV}$, 
              $\alpha^{(4)}_{\overline{\rm MS}}(5\mbox{GeV})=0.174(12)$.
              We converted this number to give
              $\alpha^{(5)}_{\overline{\rm MS}}(M_Z)=0.106(4)$.  \\[-5mm]
   \item[$^h$]       We used $r_0=0.472\,\mbox{fm}$ to convert to $r_0 \lms$. \\[-5mm]
   \item[$^i$]       Reference scale $w_{0.4}$ where $w_x$ is defined 
              by $\left. t\partial_t[t^2 \langle E(t)\rangle]\right|_{t=w_x^2}=x$
              in terms of the action density $E(t)$ at positive flow time $t$ 
              \cite{Asakawa:2015vta}. Our conversion to $r_0$ scale
              using \cite{Asakawa:2015vta} $r_0/w_{0.4}=2.587(45)$ and
              $r_0=0.472\,\mbox{fm}$. 
\end{itemize}
}
\end{minipage}
\normalsize
\caption{Wilson loop results. }
\label{tab_wloops}
\end{table}

%%%%%%%%%%%%%%%

The QCDSF/UKQCD collaborations, QCDSF/UKQCD 05
\cite{Gockeler:2005rv}, \cite{Gockeler:2004ad,Booth:2001uy,Booth:2001qp},
use the 2-loop relation (re-written here in terms of $\alpha$)
\begin{eqnarray}
   {1 \over \alpha_{\overline{\rm MS}}(\mu)} 
      = {1 \over \alphap(1/a)} 
        + 4\pi(2b_0\ln a\mu - t_1^P) 
        + (4\pi)^2(2b_1\ln a\mu - t_2^P)\alphap(1/a) \,,
\label{gPtoMSbar}
\end{eqnarray}
where $t_1^P$ and $t_2^P$ are known. (A 2-loop relation corresponds
to a 3-loop lattice $\beta$-function.)  This was used to
directly compute $\alpha_{\rm \overline{\rm MS}}$, and the scale was
chosen so that the $\cO(\alphap^0)$ term vanishes, i.e.\
\begin{eqnarray}
   \mu^* = {1 \over a} \exp{[t_1^P/(2b_0)] } 
        \approx \left\{ \begin{array}{cc}
                           2.63/a  & N_f = 0 \\
                           1.4/a   & N_f = 2 \\
                        \end{array}
                 \right. \,.
\label{amustar}
\end{eqnarray}
The method is to first compute $\alphap(1/a)$ and from this using
Eq.~(\ref{gPtoMSbar}) to find $\alpha_{\overline{\rm MS}}(\mu^*)$.
The RG equation, Eq.~(\ref{eq:Lambda}), then determines
$\mu^*/\Lambda_{\overline{\rm MS}}$ and hence using
Eq.~(\ref{amustar}) leads to the result for $r_0\Lambda_{\overline{\rm MS}}$.
This avoids giving the scale in $\mbox{MeV}$ until the end.
In the $\Nf=0$ case $7$ lattice spacings were used
\cite{Necco:2001xg}, giving a range $\mu^*/\Lambda_{\overline{\rm MS}}
\approx 24$ - $72$ (or $a^{-1} \approx 2$ - $7\,\mbox{GeV}$) and
$\alpha_{\rm eff} = \alpha_{\overline{\rm MS}}(\mu^*) \approx 0.15$ -
$ 0.10$. Neglecting higher-order perturbative terms (see discussion
after Eq.~(\ref{qcdsf:ouruncert}) below) in Eq.~(\ref{gPtoMSbar}) this
is sufficient to allow a continuum extrapolation of
$r_0\Lambda_{\overline{\rm MS}}$.
A similar computation for $N_f = 2$ by QCDSF/UKQCD~05 \cite{Gockeler:2005rv}
gave $\mu^*/\Lambda_{\overline{\rm MS}} \approx 12$ - $17$
(or roughly $a^{-1} \approx 2$ - $3\,\mbox{GeV}$) 
and $\alpha_{\rm eff} = \alpha_{\overline{\rm MS}}(\mu^*)
\approx 0.20$ - $ 0.18$.
The $N_f=2$ results of QCDSF/UKQCD~05 \cite{Gockeler:2005rv} are affected by an 
uncertainty which was not known at the time of publication: 
It has been realized that the values of $r_0/a$ of Ref.~\cite{Gockeler:2005rv}
were significantly too low~\cite{Fritzsch:2012wq}. 
As this effect is expected to depend on $a$, it
influences the perturbative behaviour leading us to assign 
a \bad\ for that criterion. 

Since FLAG 13, there has been one new result for $N_f = 0$ 
by FlowQCD 15 \cite{Asakawa:2015vta}. They also use the techniques
as described in Eqs.~(\ref{gPtoMSbar}), (\ref{amustar}), but together
with the gradient flow scale $w_0$ (rather than the $r_0$ scale).
The continuum limit is estimated by extrapolating the data at $9$
lattice spacings linearly in $a^2$. The data range used is
$\mu^*/\Lambda_{\overline{\rm MS}} \approx 40$ - $120$ (or 
$a^{-1} \approx 3$ - $11\,\mbox{GeV}$) and
$\alpha_{\overline{\rm MS}}(\mu^*) \approx 0.12$ - $0.09$.
Since a very small value of $\alpha_\msbar$ is reached, there is a $\good$ 
in the perturbative behaviour. Note that our conversion to the common
$r_0$ scale leads to a significant increase of the error of the
$\Lambda$ parameter compared to%
\footnote{The scale $w_{0.4}$ used in FlowQCD 15 
\cite{Asakawa:2015vta} is a modified $w_0$ Wilson flow scale.
With this notation $w_0 \equiv w_{0.3}$.}
$w_{0.4} \Lambda_\msbar=0.2388(5)(13)$.

The work of HPQCD 05A \cite{Mason:2005zx} (which supersedes
the original work \cite{Davies:2003ik}) uses three lattice spacings
$a^{-1} \approx 1.2$, $1.6$, $2.3\,\mbox{GeV}$ for $2+1$
flavour QCD. Typically the renormalization scale
$q \approx \pi/a \approx 3.50 - 7.10\,\mbox{GeV}$, corresponding to
$\alpha_\mathrm{V'}
 \approx 0.22-0.28$. 

In the later update HPQCD 08A \cite{Davies:2008sw} twelve data sets
(with six lattice spacings) are now used reaching up to $a^{-1}
\approx 4.4\,\mbox{GeV}$ corresponding to $\alpha_\mathrm{V'}\approx
0.18$. The values used for the scale $r_1$ were further updated in
HPQCD 10 \cite{McNeile:2010ji}. Maltman 08 \cite{Maltman:2008bx}
uses most of the same lattice ensembles as HPQCD
08A~\cite{Davies:2008sw}, but considers a much smaller set of
quantities (three versus 22) that are less sensitive to condensates.
They also use different strategies for evaluating the condensates and
for the perturbative expansion, and a slightly different value for the
scale $r_1$. The central values of the final results from 
Maltman 08 \cite{Maltman:2008bx} and HPQCD 08A \cite{Davies:2008sw}
differ by 0.0009 (which would be decreased to 0.0007
taking into account a reduction of 0.0002 in the value of the $r_1$
scale used by Maltman 08 \cite{Maltman:2008bx}).
 
As mentioned before, the perturbative coefficients are computed
through $3$-loop order~\cite{Mason:2004zt}, while the higher-order
perturbative coefficients $c_n$ with $ n_{\rm max} \ge n > 3$ (with
$n_{\rm max} = 10$) are numerically fitted using the
lattice-simulation data for the lattice spacings with the help of
Bayesian methods.  It turns out that corrections in \eq{short-cut} are
of order $|c_i/c_1|\alpha^i=$ 5--15\% and 3--10\% for $i=2,3$,
respectively.  The inclusion of a fourth-order term is necessary to
obtain a good fit to the data, and leads to a shift of the result by
$1$ -- $2$ sigma. For all but one of the 22 quantities, central values
of $|c_4/c_1|\approx 2-4$ were found, with errors from the fits of
$\approx 2$.

An important source of uncertainty is the truncation 
of perturbation theory. In HPQCD 08A \cite{Davies:2008sw}, 10
\cite{McNeile:2010ji} it
is estimated to be about $0.4$\% of $\alpha_\msbar(M_Z)$.  In \flagold\
we included a rather detailed discussion of the issue with the result
that we prefer for the time being a more conservative error
based on the above estimate $|c_4/c_1| = 2$. 
From Eq.~(\ref{Ydef}) this gives an estimate of the uncertainty
in $\alpha_{\rm eff}$ of
\begin{eqnarray}
  \Delta \alpha_{\rm eff}(\mu_1) = 
          \left|{c_4 \over c_1}\right|\alpha_{\rm eff}^4(\mu_1) \,,
\label{qcdsf:ouruncert}
\end{eqnarray}
at the scale $\mu_1$ where $\alpha_{\rm eff}$ is computed from
the Wilson loops. This can be used with a variation
in $\Lambda$ at lowest order of perturbation theory and also
applied to $\alpha_s$ evolved to a different scale $\mu_2$%
\footnote{From Eq.~(\ref{e:grelation}) we see that
$\alpha_s$ is continuous and differentiable across
the mass thresholds (at the same scale). Therefore 
to leading order $\alpha_s$ and $\Delta \alpha_s$
are independent of $N_f$.},
\begin{eqnarray}
   {\Delta\Lambda \over \Lambda} 
      = {1\over 8\pi b_0 \alpha_s} 
                  {\Delta \alpha_s \over \alpha_s}
                                                         \,, \qquad
   {\Delta \alpha_s(\mu_2) \over \Delta \alpha_s(\mu_1)}
      = {\alpha_s^2(\mu_2) \over \alpha_s^2(\mu_1)} \,.
   \label{e:dLL}   
\end{eqnarray}
We shall later use this with $\mu_2 = M_Z$
and $\alpha_s(\mu_1)=0.2$ as a typical value extracted 
from Wilson loops in HPQCD 10 \cite{McNeile:2010ji}, HPQCD 08A
\cite{Davies:2008sw}.

Again we note that the results of QCDSF/UKQCD 05
\cite{Gockeler:2005rv} ($N_f = 0$) and FlowQCD 15
\cite{Asakawa:2015vta} may be affected by frozen topology as they have
lattice spacings significantly below $a = 0.05\,\mbox{fm}$.
The associated additional systematic error is presently unknown.  

Tab.~\ref{tab_wloops} summarizes the results.

% ----------------------------------------------------------------------

%% file: Alpha_s/s7.tex
\subsection{$\alpha_s$ from current two-point functions}

% ----------------------------------------------------------------------

\label{s:curr}

% ----------------------------------------------------------------------

\subsubsection{General considerations}

% ----------------------------------------------------------------------

The method has been introduced in Ref.~\cite{Allison:2008xk}
and updated in Ref.~\cite{McNeile:2010ji}, see also
Ref.~\cite{Bochkarev:1995ai}. Since \flagold\ a new application,
HPQCD 14A  \cite{Chakraborty:2014aca}, with 2+1+1 flavours has appeared.
There the definition for larger-$n$ moments is somewhat simplified
and we describe it here. The previously used one can be found in \flagold.

The basic observable is constructed from a current 
\begin{eqnarray}
  J(x) = i m_{0h}\overline\psi_h(x)\gamma_5\psi_{h'}(x)
  \label{e:Jx}
\end{eqnarray}
of two mass-degenerate heavy-valence quarks, $h$, $h^\prime$.
The pre-factor $m_{0h}$ denotes the bare mass of the quark.
With a residual chiral symmetry, $J(x)$ is a renormalization group
invariant local field, i.e.\ it requires no renormalization.
Staggered fermions and twisted mass fermions have such a residual
chiral symmetry. The (Euclidean) time-slice correlation function
\begin{eqnarray}
   G(x_0) = a^3 \sum_{\vec{x}} \langle J^\dagger(x) J(0) \rangle \,,
\end{eqnarray}
($J^\dagger(x) = im_{0h}\overline\psi_{h'}(x)\gamma_5\psi_{h}(x)$)
has a $\sim x_0^{-3}$  singularity at short distances and moments
\begin{eqnarray}
   G_n = a \sum_{t=-(T/2-a)}^{T/2-a} t^n \,G(t) \,,
\label{Gn_smu}
\end{eqnarray}
are nonvanishing for even $n$ and furthermore finite for $n \ge 4$. 
Here $T$ is the time extent of the lattice.
The moments are dominated by contributions at $t$ of order $1/m_{0h}$.
For large mass $m_{0h}$ these are short distances and the moments
become increasingly perturbative for decreasing $n$.
Denoting the lowest-order perturbation theory moments by $G_n^{(0)}$,
one defines the normalized moments
\begin{eqnarray}
   \tilde R_n = \left\{ \begin{array}{cc}
                    G_4/G_4^{(0)}          & \mbox{for $n=4$} \,, \\[0.5em]
                    { G_n^{1/(n-4)} \over
                     m_{0c}\left( G_n^{(0)} \right)^{1/(n-4)} }
                                        & \mbox{for $n \ge 6$} \,, \\
                 \end{array}
         \right.
\label{ratio_GG}
\end{eqnarray}
of even order $n$. Note that \eq{e:Jx} contains the variable (bare) heavy-quark mass $m_{0h}$, while \eq{ratio_GG} is defined with the charm-quark mass,
tuned to its physical value. The normalization
$m_{0c}\left( G_n^{(0)} \right)^{1/(n-4)}$ in \eq{ratio_GG} ensures that
$\tilde R_n$ remains renormalization group invariant, %
but introduces a mass scale.
In the continuum limit
the normalized moments can then be parameterized in terms of functions
\begin{eqnarray}
   \tilde R_n \equiv \left\{ \begin{array}{cc}
                         r_4(\alpha_s(\mu))
                                        & \mbox{for $n=4$} \,,     \\[0.5em]
                         {r_n(\alpha_s(\mu)) \over \bar{m}_c(\mu)}
                                        & \mbox{for $n \ge 6$} \,, \\
                      \end{array}
              \right.
              \label{e:Rn}
\end{eqnarray}
with $\bar{m}_c(\mu)$ being the renormalized charm-quark mass.
The reduced moments
$r_n$ have a perturbative expansion
\begin{eqnarray}
   r_n = 1 + r_{n,1}\alpha_s + r_{n,2}\alpha_s^2 + r_{n,3}\alpha_s^3 + \ldots\,,
\label{rn_expan}
\end{eqnarray}
where the written terms $r_{n,i}(\mu/\bar{m}_h(\mu))$, $i \le 3$ are known
for low $n$ from Refs.~\cite{Chetyrkin:2006xg,Boughezal:2006px,Maier:2008he,
Maier:2009fz,Kiyo:2009gb}. In practice, the expansion is performed in
the $\overline{\rm MS}$ scheme. Matching nonperturbative lattice results
for the moments to the perturbative expansion, one determines an
approximation to $\alpha_{\overline{\rm MS}}(\mu)$ as well as $\bar m_c(\mu)$.
With the lattice spacing (scale) determined from some extra physical input,
this calibrates $\mu$.  As usual suitable pseudoscalar masses
determine the bare quark masses, here in particular the charm mass, 
and then through \eq{e:Rn} the renormalized charm-quark mass.

A difficulty with this approach is that large masses are needed to enter
the perturbative domain. Lattice artefacts can then be sizeable and
have a complicated form. The ratios in Eq.~(\ref{ratio_GG}) use the
tree-level lattice results in the usual way for normalization.
This results in unity as the leading term in Eq.~(\ref{rn_expan}),
suppressing some of the kinematical lattice artefacts.
We note that in contrast to e.g.\ the definition of $\alpha_\mathrm{qq}$,
here the cutoff effects are of order $a^k\alpha_s$, while there the
tree-level term defines $\alpha_s$ and therefore the cutoff effects
after tree-level improvement are of order $a^k\alpha_s^2$.

Finite-size effects (FSE) due to the omission of
$|t| > T /2$ in Eq.~(\ref{Gn_smu}) grow with $n$ as 
$(m_\mathrm{p}T/2)^n\, \exp{(-m_\mathrm{p} T/2)}$. 
In practice, however, since the (lower) moments
are short-distance dominated, the FSE are expected to be irrelevant
at the present level of precision.  

Moments of correlation functions of the quark's electromagnetic
current can also be obtained from experimental data for $e^+e^-$
annihilation~\cite{Kuhn:2007vp,Chetyrkin:2009fv}.  This enables a
nonlattice determination of $\alpha_s$ using a similar analysis
method.  In particular, the same continuum perturbation theory
computation enters both the lattice and the phenomenological determinations.

% ----------------------------------------------------------------------

\subsubsection{Discussion of computations}

% ----------------------------------------------------------------------

The method has originally been applied in HPQCD 08B \cite{Allison:2008xk}
and in HPQCD 10 \cite{McNeile:2010ji}, based on the MILC ensembles with
$2 + 1$ flavours of Asqtad staggered quarks and HISQ valence quarks. The
scale was set using $r_1 = 0.321(5)\,\mbox{fm}$ in HPQCD 08B 
\cite{Allison:2008xk} and the updated value 
$r_1 = 0.3133(23)\,\mbox{fm}$ in HPQCD 10 \cite{McNeile:2010ji}. The
effective range of couplings used is here given for $n = 4$, which is
the moment most dominated by short (perturbative) distances and
important in the determination of $\alpha_s$. The range is similar for
other ratios. With $r_{4,1} = 0.7427$ and $R_4 = 1.28$ determined
in the continuum limit at the charm mass in Ref.~\cite{Allison:2008xk}, we
have $\alpha_{\rm eff} = 0.38$ at the charm-quark mass, which is the
mass value where HPQCD 08B \cite{Allison:2008xk} carries out the analysis.
In HPQCD 10 \cite{McNeile:2010ji} a set of masses is used,
with $R_4 \in [1.09, 1.29]$ which corresponds 
to $\alpha_{\rm eff} \in [0.12, 0.40]$.

\begin{figure}[!htb]
\hspace{-0.4cm} 
   \includegraphics[width=0.56\textwidth]{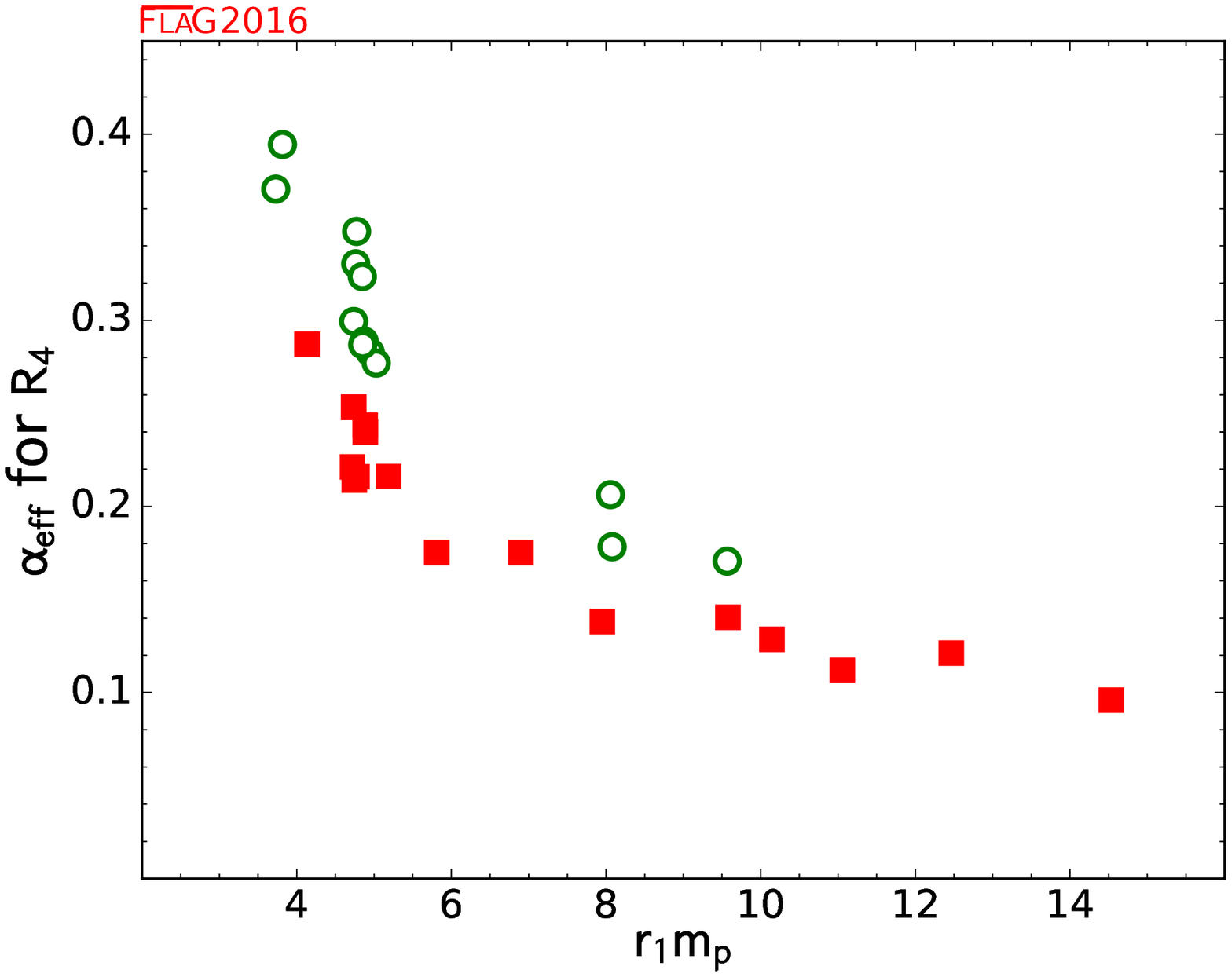}\hspace{-0.5cm} 
   \includegraphics[width=0.56\textwidth]{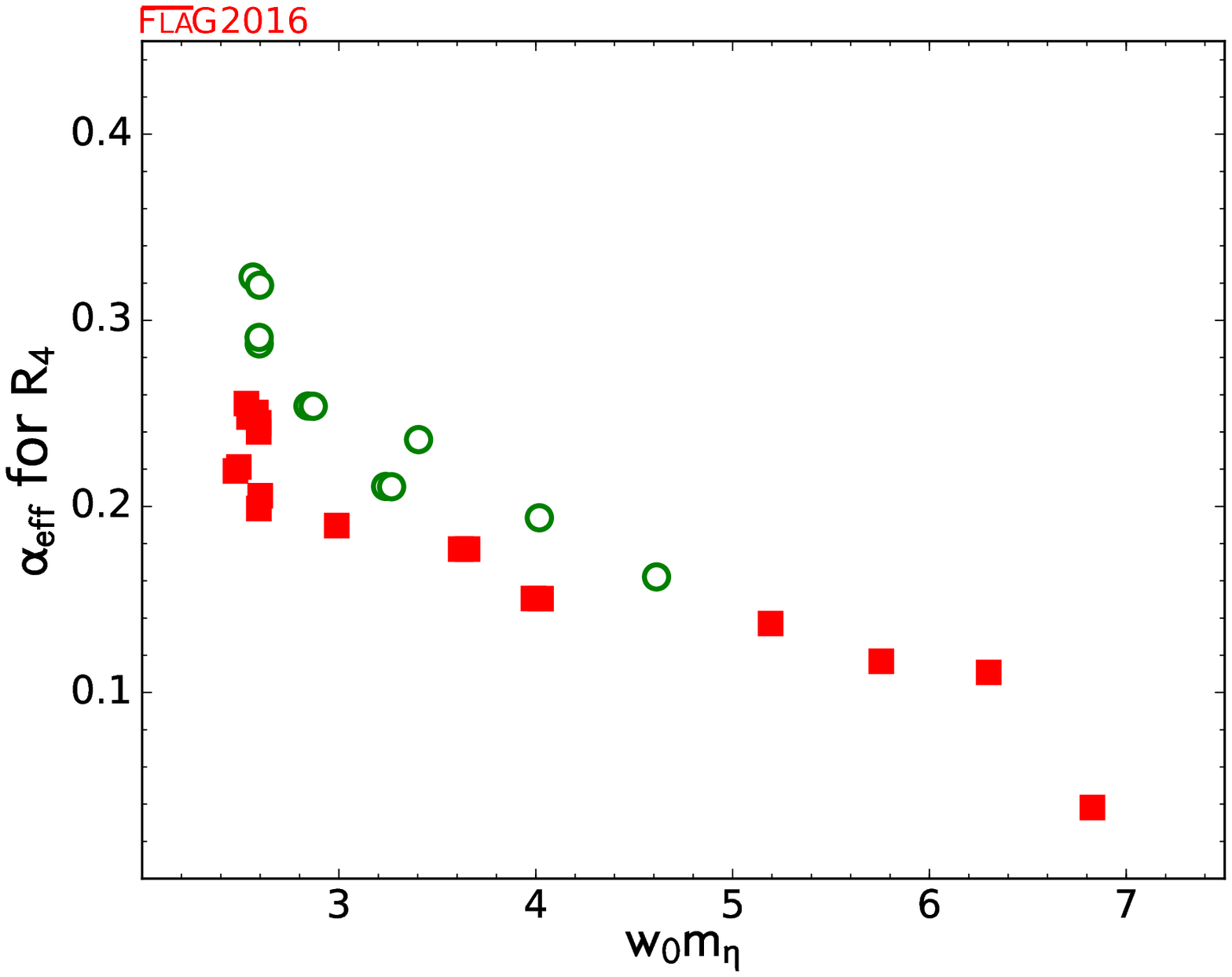}
\caption{$\alpha_{\rm eff}$ for $R_4$ from HPQCD 10 data (left) and
         from HPQCD 14A  (right). A similar graph for 
         $R_6/R_8$ is shown in \flagold.
         Symbols correspond to \protect\soso\ for data with
         $1\leq a\mu\leq 1.5$ and \protect\bad\ for $a\mu>1.5$,
         while \protect\good\ ($a\mu < 1/2$) is not present.
         This corresponds exactly to the $a\mu$ part of our continuum
         limit criterion, but does not consider how many lattice
         spacings are present.
         Note that mistunings in the
         quark masses have not been accounted for, but, estimated as
         in HPQCD 14A  \protect\cite{Chakraborty:2014aca}, they are
         smaller than the size of the symbols 
         in the graphs.}
\label{hpqcd_alpha_eff}
\end{figure}

The available data of HPQCD 10 \cite{McNeile:2010ji} is summarized in 
the left panel of Fig.~\ref{hpqcd_alpha_eff}
where we plot $\alpha_\mathrm{eff}$ against $m_\mathrm{p} r_1$.
For the continuum limit criterion, we choose the scale $\mu = 2\bar m_h
\approx m_\mathrm{p}/1.1$, where we have taken $\bar m_h$ in the $\msbar$
scheme at scale $\bar m_h$ and the numerical value $1.1$ was determined in
HPQCD 10B \cite{Na:2010uf}.

The data in Fig.~\ref{hpqcd_alpha_eff} are
grouped according to the range of $a\mu$ that they cover.
The vertical spread of the results for $\alpha_{\rm eff}$ 
at fixed $r_1m_\mathrm{p}$
in the figure measures the discretization errors seen: 
in the continuum we would expect all the points 
to lie on one universal curve. The plots illustrate the selection applied by
our criterion for the continuum limit with our choices for
$\mu$. Fig.~\ref{hpqcd_alpha_eff} gives reason for concern, since it shows
that the discretization errors that need to be removed in the continuum
extrapolation are not small.

With our choices for $\mu$, the continuum limit criterion is satisfied
for 3 lattice spacings when $\alpha_\mathrm{eff} \leq 0.3$ and $n=4$.
Larger-$n$ moments are more influenced by nonperturbative effects.
For the $n$ values considered, adding a gluon condensate term only changed error
bars slightly in HPQCD's analysis.
We note that HPQCD in their papers perform a global fit to all data
using a joint expansion in powers of $\alpha_s^n$, $\left(
\Lambda/(m_\mathrm{p}/2) \right)^j$ to parameterize the heavy-quark
mass dependence, and $\left( am_\mathrm{p}/2 \right)^{2i}$ to
parameterize the lattice-spacing dependence.  To obtain a good fit,
they must exclude data with $am_\mathrm{p} > 1.95$ and include
lattice-spacing terms $a^{2i}$ with $i$ greater than $10$.  Because
  these fits include many more fit parameters than data points, HPQCD
  uses their expectations for the sizes of coefficients as Bayesean
  priors.  The fits include data with masses as large as $am_{\textrm
    p}/2 \sim0.86$, so there is only minimal suppression of the many
  high-order contributions for the heavier masses.  It is not clear,
  however, how sensitive the final results are to the larger
  $am_{\textrm p}/2$ values in the data.  The continuum limit of the
  fit is in agreement with a perturbative scale dependence (a
  5-loop running $\alpha_{\overline{\rm MS}}$ with a fitted
  5-loop coefficient in the $\beta$-function is used).  Indeed, Fig.~2
  of Ref.~\cite{McNeile:2010ji} suggests that HPQCD's fit describes
  the data well.

The new computation, HPQCD 14A  \cite{Chakraborty:2014aca},
is based on MILC's 2+1+1 HISQ staggered ensembles. Compared to 
HPQCD 10 \cite{McNeile:2010ji} valence- and 
sea-quarks now use the same discretization and the scale is set 
through the gradient flow scale $w_0$, determined to 
$w_0=0.1715(9)\,\fm$ in Ref.~\cite{Dowdall:2012ab}.

We again show the values of $\alpha_\mathrm{eff}$ as a function
of the physical scale. Discretization errors are noticeable.
A number of data points, satisfy our continuum limit criterion
$a\mu < 1.5$, at two different lattice 
spacings. This does not by itself lead to a \soso\ but the next-larger 
lattice spacing does not miss the criterion by much, see 
Tab.~\ref{tab_Nf=4_continuumlimit}.
We therefore assign a \soso\ in that criterion.

The other details of the analysis by 
HPQCD 10 \cite{McNeile:2010ji} are very similar to the
ones described above, with one noteworthy exception. 
The new definition of 
the moments does not involve the pseudoscalar $h \bar h$ mass anymore.
Therefore its relation to the quark mass does not need to be modeled
in the fit. Since it is now replaced by the renormalized 
charm-quark mass, the analysis produces a result for 
$\alpha_s$ and the charm-quark mass at the same time. Here we only discuss
the result for $\alpha_s$. 

In Tab.~\ref{tab_current_2pt} we list the current
two-point function results.
\begin{table}[!htb]
   \vspace{3.0cm}
   \footnotesize
   \begin{tabular*}{\textwidth}[l]{l@{\extracolsep{\fill}}rlllllllll}
      Collaboration & Ref. & $N_f$ &
      \hspace{0.15cm}\begin{rotate}{60}{publication status}\end{rotate}
                                                       \hspace{-0.15cm} &
      \hspace{0.15cm}\begin{rotate}{60}{renormalization scale}\end{rotate}
                                                       \hspace{-0.15cm} &
      \hspace{0.15cm}\begin{rotate}{60}{perturbative behaviour}\end{rotate}
                                                       \hspace{-0.15cm} &
      \hspace{0.15cm}\begin{rotate}{60}{continuum extrapolation}\end{rotate}
      \hspace{-0.25cm} & %\rule{0.2cm}{0cm} 
                         scale & $\Lambda_\msbar[\MeV]$ & $r_0\Lambda_\msbar$ \\
      &&&&&&&& \\[-0.1cm]
      \hline
      \hline
      &&&&&&&& \\[-0.1cm]

      HPQCD 14A   &  \cite{Chakraborty:2014aca} 
                                              & 2+1+1   & \gA & \soso
                   & \good      & \soso
                   & $w_0=0.1715(9)\,\mbox{fm}^a$
                   & 294(11)$^{bc}$
                   & 0.703(26)                                    \\

      &&&&&&&& \\[-0.1cm]
      \hline
      &&&&&&&& \\[-0.1cm]

      HPQCD 10     & \cite{McNeile:2010ji}  & 2+1       & \gA & \soso
                   & \good   & \soso           
                   & $r_1 = 0.3133(23)\, \mbox{fm}$$^\dagger$
                   & 338(10)$^\star$           &  0.809(25)            \\
      HPQCD 08B    & \cite{Allison:2008xk}  & 2+1       & \gA & \bad 
                   & \bad  & \bad           
                   & $r_1 = 0.321(5)\,\mbox{fm}$$^\dagger$  
                   & 325(18)$^+$             &  0.777(42)  \\
      &&&&&&&& \\[-0.1cm]
      \hline
      \hline\\
\end{tabular*}\\[-0.2cm]
\begin{minipage}{\linewidth}
{\footnotesize 
\begin{itemize}
   \item[$^a$]  Scale determined in \cite{Dowdall:2013rya} using $f_\pi$. \\[-5mm]
   \item[$^b$]  $\alpha^{(4)}_{\overline{\rm MS}}(5\,\mbox{GeV}) = 0.2128(25)$, 
         $\alpha^{(5)}_{\overline{\rm MS}}(M_Z) = 0.11822(74)$.         \\[-5mm]
   \item[$^c$] Our conversion for $\Lambda_{\overline{\rm MS}}$ for $N_f = 4$.
         We also used $r_0 = 0.472\,\mbox{fm}$.\\[-5mm]
   \item[$^\dagger$] Scale is determined from $\Upsilon$ mass splitting.    \\[-5mm]
   \item[$^\star$]  $\alpha^{(3)}_{\overline{\rm MS}}(5\,\mbox{GeV}) = 0.2034(21)$,
            $\alpha^{(5)}_{\overline{\rm MS}}(M_Z) = 0.1183(7)$.         \\[-4mm]
   \item[$^+$]     $\alpha^{(4)}_{\overline{\rm MS}}(3\,\mbox{GeV}) = 0.251(6)$,
            $\alpha^{(5)}_{\overline{\rm MS}}(M_Z) = 0.1174(12)$.        
\end{itemize}
}
\end{minipage}
\normalsize
\caption{Current two-point function results. }
\label{tab_current_2pt}
\end{table}
Thus far, only one group has used this approach,
which models complicated and potentially large cutoff effects together
with a perturbative coefficient. We therefore are waiting to see confirmation
by other collaborations of the small systematic errors obtained
(cf.\ discussion in Sec.~\ref{subsubsect:Our range}). (We note that more investigations of
this method are in progress \cite{Nakayama:2015hrn}.)
We do, however, include the values of $\alpha_{\overline{\rm MS}}(M_Z)$
and $\Lambda_{\overline{\rm MS}}$ of HPQCD 10 \cite{McNeile:2010ji}
and HPQCD 14A  \cite{Chakraborty:2014aca} in our final range.

% ----------------------------------------------------------------------

%% file: Alpha_s/s8.tex
\subsection{$\alpha_s$ from QCD vertices}

% ----------------------------------------------------------------------
\label{s:glu}

% ----------------------------------------------------------------------

\subsubsection{General considerations}

% --------------------------------------------------------------------

The most intuitive and in principle direct way to determine the
coupling constant in QCD is to compute the appropriate
three- or four-point gluon vertices
or alternatively the
quark-quark-gluon vertex or ghost-ghost-gluon vertex (i.e.\ $
q\overline{q}A$ or $c\overline{c}A$ vertex respectively).  A suitable
combination of renormalization constants then leads to the relation
between the bare (lattice) and renormalized coupling constant. This
procedure requires the implementation of a nonperturbative
renormalization condition and the fixing of the gauge. For the study
of nonperturbative gauge fixing and the associated Gribov ambiguity,
we refer to Refs.~\cite{Cucchieri:1997dx,Giusti:2001xf,Maas:2009ph} and
references therein.
In practice the Landau gauge is used and the
renormalization constants are defined 
by requiring that the vertex is equal to the tree level
value at a certain momentum configuration.
The resulting renormalization
schemes are called `MOM' scheme (symmetric momentum configuration)
or `$\rm \widetilde{MOM}$' (one momentum vanishes), 
which are then converted perturbatively
to the $\overline{\rm MS}$ scheme.

A pioneering work to determine the three-gluon vertex in the $N_f = 0$
theory is Alles~96~\cite{Alles:1996ka} (which was followed by
Ref.~\cite{Boucaud:2001qz} for two flavour QCD); a more recent $N_f = 0$
computation was Ref.~\cite{Boucaud:2005gg} in which the three-gluon vertex
as well as the ghost-ghost-gluon vertex was considered.  (This
requires in general a computation of the propagator of the
Faddeev--Popov ghost on the lattice.) The latter paper concluded that
the resulting $\Lambda_{\overline{\rm MS}}$ depended strongly on the
scheme used, the order of perturbation theory used in the matching and
also on nonperturbative corrections \cite{Boucaud:2005xn}.

Subsequently in Refs.~\cite{Sternbeck:2007br,Boucaud:2008gn} a specific
$\widetilde{\rm MOM}$ scheme with zero ghost momentum for the
ghost-ghost-gluon vertex was used. In this scheme, dubbed
the `MM' (Minimal MOM) or `Taylor' (T) scheme, the vertex
is not renormalized, and so the renormalized coupling reduces to
\begin{eqnarray}
   \alpha_{\rm T}(\mu) 
      = D^{\rm gluon}_{\rm lat}(\mu, a) D^{\rm ghost}_{\rm lat}(\mu, a)^2 \,
                      {g_0^2(a) \over 4\pi} \,,
\end{eqnarray}
where $D^{\rm ghost}_{\rm lat}$ and $D^{\rm gluon}_{\rm lat}$ are the
(bare lattice) dressed ghost and gluon `form factors' of these
propagator functions in the Landau gauge,
\begin{eqnarray}
   D^{ab}(p) = - \delta^{ab}\, {D^{\rm ghost}(p) \over p^2}\,, \qquad
   D_{\mu\nu}^{ab}(p) 
      = \delta^{ab} \left( \delta_{\mu\nu} - {p_\mu p_\nu \over p^2} \right) \,
        {D^{\rm gluon}(p) \over p^2 } \,,
\end{eqnarray}
and we have written the formula in the continuum with 
$D^{\rm ghost/gluon}(p)=D^{\rm ghost/gluon}_{\rm lat}(p, 0)$.
Thus there is now no need to compute the ghost-ghost-gluon vertex,
just the ghost and gluon propagators.

% --------------------------------------------------------------------

\subsubsection{Discussion of computations}
\label{s:glu_discuss}

% --------------------------------------------------------------------

\begin{table}[!h]
   \vspace{3.0cm}
   \footnotesize
   \begin{tabular*}{\textwidth}[l]{l@{\extracolsep{\fill}}rllllllll}
   Collaboration & Ref. & $\Nf$ &
   \hspace{0.15cm}\begin{rotate}{60}{publication status}\end{rotate}
                                                    \hspace{-0.15cm} &
   \hspace{0.15cm}\begin{rotate}{60}{renormalization scale}\end{rotate}
                                                    \hspace{-0.15cm} &
   \hspace{0.15cm}\begin{rotate}{60}{perturbative behaviour}\end{rotate}
                                                    \hspace{-0.15cm} &
   \hspace{0.15cm}\begin{rotate}{60}{continuum extrapolation}\end{rotate}
      \hspace{-0.25cm} & 
                         scale & $\Lambda_\msbar[\MeV]$ & $r_0\Lambda_\msbar$ \\
      & & & & & & & & \\[-0.1cm]
      \hline
      \hline
      & & & & & & & & \\[-0.1cm]
       ETM 13D      & \cite{Blossier:2013ioa}   & {2+1+1} & {\gA}
                    & \soso & \soso  & \bad  
                    & $f_\pi$
                    & $314(7)(14)(10)$$^\S$
                    & $0.752(18)(34)(81)$$^\dagger$                        \\
       ETM 12C        & \cite{Blossier:2012ef}   & 2+1+1 & \gA 
                    & \soso & \soso  & \bad  
                    & $f_\pi$
                    & $324(17)$$^\S$
                    & $0.775(41)$$^\dagger$                                \\
      ETM 11D       & \cite{Blossier:2011tf}   & 2+1+1 & \gA 
                    & \soso & \soso  & \bad  
                    & $f_\pi$
                    & $316(13)(8)(^{+0}_{-9})$$^\star$
                    & $0.756(31)(19)(^{+0}_{-22})$$^\dagger$                 \\
      & & & & & & & & \\[-0.1cm]
      \hline
      & & & & & & & & \\[-0.1cm]
      Sternbeck 12  & \cite{Sternbeck:2012qs}  & 2+1  & \rC
                    &     &        & 
                    & \multicolumn{3}{l}{only running of 
                                         $\alpha_s$ in Fig.~4}            \\
      & & & & & & & & \\[-0.1cm]
      \hline
      & & & & & & & & \\[-0.1cm]
      Sternbeck 12  & \cite{Sternbeck:2012qs}  & 2  & \rC
                    &  &  & 
                    & \multicolumn{3}{l}{Agreement with $r_0\Lambda_\msbar$ 
                                         value of \cite{Fritzsch:2012wq} } \\
      Sternbeck 10  & \cite{Sternbeck:2010xu}  & 2  & \rC 
                    & \soso  & \good & \bad
                    &
                    & $251(15)$$^\#$
                    & $0.60(3)(2)$                                       \\
      ETM 10F       & \cite{Blossier:2010ky}   & 2  & \gA 
                    & \soso  & \soso  & \soso 
                    & $f_\pi$
                    & $330(23)(22)(^{+0}_{-33})$\hspace{-2mm}
                    & $0.72(5)$$^+$                                       \\
      Boucaud 01B    & \cite{Boucaud:2001qz}    & 2 & \gA 
                    & \soso & \soso  & \bad
                    & $K^{\ast}-K$
                    & $264(27)$$^{\star\star}$
                    & 0.669(69)                              \\
      & & & & & & & & \\[-0.1cm]
      \hline
      & & & & & & & & \\[-0.1cm]
      Sternbeck 12  & \cite{Sternbeck:2012qs}   & 0 & \rC 
                    &  &  &
                    &  \multicolumn{3}{l}{Agreement with $r_0\Lambda_\msbar$
                                          value of \cite{Brambilla:2010pp}} \\
      Sternbeck 10  & \cite{Sternbeck:2010xu}   & 0 & \rC
                    & \good & \good & \bad
                    &
                    & $259(4)$$^\#$
                    & $0.62(1)$                                            \\
      Ilgenfritz 10 & \cite{Ilgenfritz:2010gu}  & 0 & \gA
                    &    \good    &  \good      & \bad 
                    & \multicolumn{2}{l}{only running of
                                         $\alpha_s$ in Fig.~13}           \\
{Boucaud 08}    & \cite{Boucaud:2008gn}       & 0         &\gA  
                    & \soso & \good   & \bad 
                    & $\sqrt{\sigma} = 445\,\mbox{MeV}$
                    & $224(3)(^{+8}_{-5})$
                    & $0.59(1)(^{+2}_{-1})$         
 \\
{Boucaud 05}    & \cite{Boucaud:2005gg}       & 0       &\gA  
                    & \bad & \good   & \bad 
                    & $\sqrt{\sigma} = 445\,\mbox{MeV}$
                    & 320(32)
                    & 0.85(9)           
 \\
   Soto 01        & \cite{DeSoto:2001qx}        & 0         & \gA  
                    & \soso & \soso  & \soso
                    & $\sqrt{\sigma} = 445\,\mbox{MeV}$
                    & 260(18)
                    & 0.69(5)          
 \\
{Boucaud 01A}    & \cite{Boucaud:2001st}      & 0         &\gA  
                    & \soso & \soso  & \soso
                    & $\sqrt{\sigma} = 445\,\mbox{MeV}$
                    & 233(28)~MeV
                    & 0.62(7)       
 \\
{Boucaud 00B}   & \cite{Boucaud:2000nd}      & 0         &\gA  
                    & \soso & \soso  & \soso
                    & 
                    & \multicolumn{2}{l}{only running of
                                         $\alpha_s$}
 \\
{Boucaud 00A}     &\cite{Boucaud:2000ey}     &  0    &\gA  
                    & \soso & \soso  & \soso
                    & $\sqrt{\sigma} = 445\,\mbox{MeV}$
                    & $237(3)(^{+~0}_{-10})$
                    & $0.63(1)(^{+0}_{-3})$            
 \\
{Becirevic 99B}  & \cite{Becirevic:1999hj} & 0 &\gA  
                    & \soso & \soso  & \bad 
                    & $\sqrt{\sigma} = 445\,\mbox{MeV}$
                    & $319(14)(^{+10}_{-20})$
                    & $0.84(4)(^{+3}_{-5})$   
 \\
{Becirevic 99A}  & \cite{Becirevic:1999uc} & 0 &\gA  
                    & \soso & \soso  & \bad 
                    & $\sqrt{\sigma} = 445\,\mbox{MeV}$
                    & $\lesssim 353(2)(^{+25}_{-15})$
                    & $\lesssim 0.93 (^{+7}_{-4})$         
 \\
{Boucaud 98B}  & \cite{Boucaud:1998xi} & 0 &\gA  
                    & \bad  & \soso  & \bad 
                    & $\sqrt{\sigma} = 445\,\mbox{MeV}$
                    & 295(5)(15)
                    & 0.78(4)           
 \\
{Boucaud 98A}    & \cite{Boucaud:1998bq} & 0 &\gA  
                    & \bad  & \soso  & \bad 
                    & $\sqrt{\sigma} = 445\,\mbox{MeV}$
                    & 300(5)
                    & 0.79(1)         
\\
{Alles 96}    & \cite{Alles:1996ka} & 0 &\gA  
                    & \bad  & \bad   & \bad 
                    & $\sqrt{\sigma} = 440\,\mbox{MeV}$\hspace{0.3mm}$^{++}$\hspace{-0.3cm}        
                    & 340(50)
                    & 0.91(13)   
\\
      & & & & & & & & \\[-0.1cm]
      \hline
      \hline\\
\end{tabular*}\\[-0.2cm]
\begin{minipage}{\linewidth}
{\footnotesize 
\begin{itemize}
\item[$^\dagger$] We use the 2+1 value $r_0=0.472$~fm.                        \\[-5mm]
   \item[$^\S$] $\alpha_{\overline{\rm MS}}^{(5)}(M_Z)=0.1200(14)$.                   \\[-5mm]
   \item[$^\star$] First error is statistical; second is due to the lattice
           spacing and third is due to the chiral extrapolation.
           $\alpha_{\overline{\rm MS}}^{(5)}(M_Z)=0.1198(9)(5)(^{+0}_{-5})$.    \\[-5mm]
   \item[$^\#$] In the paper only $r_0\Lambda_{\overline{\rm MS}}$ is given,
         we converted to $\MeV$ with $r_0=0.472$~fm.                    \\[-5mm]
   \item[$^+$] The determination of $r_0$
        from the $f_\pi$ scale is found in Ref.~\cite{Baron:2009wt}.          \\[-5mm]
   \item[$^{\star\star}$]  $\alpha_{\overline{\rm MS}}^{(5)}(M_Z)=0.113(3)(4)$.         \\[-5mm]
   \item[$^{++}$]  The scale is taken from the string tension computation
           of Ref.~\cite{Bali:1992ru}.
\end{itemize}
}
\end{minipage}
\normalsize
\caption{Results for the gluon--ghost vertex.}
\label{tab_vertex}
\end{table}

For the calculations considered here, to match to perturbative
scaling, it was first necessary to reduce lattice artifacts by an
$H(4)$ extrapolation procedure (addressing $O(4)$ rotational
invariance), e.g.\ ETM 10F \cite{Blossier:2010ky} or by lattice
perturbation theory, e.g.\ Sternbeck 12 \cite{Sternbeck:2012qs}.  To
match to perturbation theory, collaborations vary in their approach.
In ETM 10F \cite{Blossier:2010ky} it was necessary to include the
operator $A^2$ in the OPE of the ghost and gluon propagators, while in
{Sternbeck 12 \cite{Sternbeck:2012qs}} very large momenta are used and
$a^2p^2$ and $a^4p^4$ terms are included in their fit to the momentum
dependence. A further later refinement was the introduction of
higher nonperturbative OPE power corrections in ETM 11D
\cite{Blossier:2011tf} and ETM 12C \cite{Blossier:2012ef}.
Although
the expected leading power correction, $1/p^4$, was tried, ETM finds
good agreement with their data only when they fit with the
next-to-leading-order term, $1/p^6$.  The update ETM 13D
\cite{Blossier:2013ioa} investigates this point in more detail, using
better data with reduced statistical errors.  They find that after
again including the $1/p^6$ term they can describe their data over a
large momentum range from about 1.75~GeV to 7~GeV.

In all calculations except for Sternbeck 10 \cite{Sternbeck:2010xu},
Sternbeck 12 \cite{Sternbeck:2012qs} ,
the matching with the perturbative formula is performed including
power corrections in the form of 
condensates, in particular $\langle A^2 \rangle$. 
Three lattice spacings are present in almost all 
calculations with $N_f=0$, $2$, but the scales $ap$ are rather large.
This mostly results in a $\bad$ on the continuum extrapolation
(Sternbeck 10 \cite{Sternbeck:2010xu},
  Boucaud 01B \cite{Boucaud:2001qz} for $N_f=2$.
 Ilgenfritz 10 \cite{Ilgenfritz:2010gu},   
 Boucaud 08 \cite{Boucaud:2008gn},
 Boucaud 05 \cite{Boucaud:2005gg}, 
 Becirevic 99B \cite{Becirevic:1999hj},
  Becirevic 99A \cite{Becirevic:1999uc},
 Boucaud 98B \cite{Boucaud:1998xi},
 Boucaud 98A \cite{Boucaud:1998bq},
 Alles 96 \cite{Alles:1996ka} for $N_f=0$).
A \soso\ is reached in the $\Nf=0$ computations 
Boucaud 00A \cite{Boucaud:2000ey}, 00B \cite{Boucaud:2000nd},
01A \cite{Boucaud:2001st}, Soto 01 \cite{DeSoto:2001qx} due to
a rather small lattice spacing,  but this is done on a lattice
of a small physical size. 
The $N_f=2+1+1$ calculation, fitting with condensates, 
is carried out for two lattice spacings
and with $ap>1.5$, giving $\bad$
for the continuum extrapolation as well. 
In ETM 10F \cite{Blossier:2010ky} we have
$0.25 < \alpha_{\rm eff} < 0.4$, while in ETM 11D \cite{Blossier:2011tf}, 
ETM 12C \cite{Blossier:2012ef} (and ETM 13 \cite{Cichy:2013gja})
we find $0.24 < \alpha_{\rm eff} < 0.38$ which gives a green circle
in these cases for the renormalization scale.
In ETM 10F \cite{Blossier:2010ky} the values of $ap$ violate our criterion
for a continuum limit only slightly, and 
we give a \soso.

In {Sternbeck 10 \cite{Sternbeck:2010xu}}, the coupling ranges over
$0.07 \leq \alpha_{\rm eff} \leq 0.32$ for $N_f=0$ and $0.19 \leq
\alpha_{\rm eff} \leq 0.38$ for $N_f=2$ giving $\good$ and $\soso$ for
the renormalization scale respectively.  The fit with the perturbative
formula is carried out without condensates, giving a satisfactory
description of the data.  In {Boucaud 01A \cite{Boucaud:2001st}},
depending on $a$, a large range of $\alpha_{\rm eff}$ is used which
goes down to $0.2$ giving a $\soso$ for the renormalization scale and
perturbative behaviour, and several lattice spacings are used leading
to $\soso$ in the continuum extrapolation.  The $\Nf=2$ computation
Boucaud 01B \cite{Boucaud:2001st}, fails the continuum limit criterion
because both $a\mu$ is too large and an unimproved Wilson fermion
action is used.  Finally in the conference proceedings
Sternbeck 12 \cite{Sternbeck:2012qs}, the $N_f=0,2,3$ coupling
$\alpha_\mathrm{T}$ is studied.  Subtracting 1-loop lattice artefacts
and subsequently fitting with $a^2p^2$ and $a^4p^4$ additional lattice
artefacts, agreement with the perturbative running is found for large
momenta ($r_0^2p^2 > 600$) without the need for power corrections.  In
these comparisons, the values of $r_0\Lambda_\msbar$ from other
collaborations are used. As no numbers are given, we have not
introduced ratings for this study.

In Tab.~\ref{tab_vertex} we summarize the results. Presently there
are no $N_f \geq 3$ calculations of $\alpha_s$ from QCD vertices that
satisfy the FLAG criteria to be included in the range.

% ----------------------------------------------------------------------

%% file: Alpha_s/s9.tex
\subsection{Summary}
\label{s:alpsumm}

% --------------------------------------------------------------------

\begin{table}[!htb]
   \vspace{3.0cm}
   \footnotesize
   \begin{tabular*}{\textwidth}[l]{l@{\extracolsep{\fill}}rlllllllr}
   Collaboration & Ref. & $N_f$ &
   \hspace{0.15cm}\begin{rotate}{60}{publication status}\end{rotate}
                                                    \hspace{-0.15cm} &
   \hspace{0.15cm}\begin{rotate}{60}{renormalization scale}\end{rotate}
                                                    \hspace{-0.15cm} &
   \hspace{0.15cm}\begin{rotate}{60}{perturbative behaviour}\end{rotate}
                                                    \hspace{-0.15cm} &
   \hspace{0.15cm}\begin{rotate}{60}{continuum extrapolation}\end{rotate}
      \hspace{-0.25cm} & %\rule{0.2cm}{0cm} 
       $\alpha_\msbar(M_\mathrm{Z})$ & Method  & Table \\
   & & & & & & & & \\[-0.1cm]
   \hline
   \hline
   & & & & & & & & \\[-0.1cm]
    HPQCD 14A 
                    &  \cite{Chakraborty:2014aca} & 2+1+1 & \gA 
                    & \soso & \good   & \soso
                    & 0.11822(74)
                    & current two points
                    & \ref{tab_current_2pt}                    \\
   ETM 13D    &  \cite{Blossier:2013ioa}   & 2+1+1& \gA
                    & \soso & \soso  & \bad 
%                    & $f_\pi$
                    & 0.1196(4)(8)(16)
                    & gluon-ghost vertex
                    & \ref{tab_vertex}                         \\
   ETM 12C    & \cite{Blossier:2012ef}   & 2+1+1 & \gA 
                    & \soso & \soso  & \bad  
%                    & $f_\pi$
                    & 0.1200(14)
		 & gluon-ghost vertex
                    & \ref{tab_vertex}                         \\
   ETM 11D   & \cite{Blossier:2011tf}   & 2+1+1 & \gA 
             & \soso & \soso & \bad  
%                   & $f_\pi$
                    & $0.1198(9)(5)(^{+0}_{-5})$
                    & gluon-ghost vertex
                    & \ref{tab_vertex}                         \\
   & & & & & & & & &  \\[-0.1cm]
   \hline
   & & & & & & & & & \\[-0.1cm]

   {Bazavov 14}
            & \cite{Bazavov:2014soa}    & 2+1       & \gA & \soso
            & \good   & \soso
            & $0.1166(^{+12}_{-8})$
            & $Q$-$\bar{Q}$ potential
            & \ref{tab_short_dist}                            \\

   {Bazavov 12}
            & \cite{Bazavov:2012ka}   & 2+1       & \gA & \soso
            & \soso  & \soso
%         & $r_0 = 0.468\,\mbox{fm}$ 
            & $0.1156(^{+21}_{-22})$ 
            & $Q$-$\bar{Q}$ potential
            & \ref{tab_short_dist}                            \\
   HPQCD 10   & \cite{McNeile:2010ji}  & 2+1       & \gA & \soso
             & \good  & \soso          
%           & $r_1 = 0.3133(23)\, \mbox{fm}$
             & 0.1183(7)          
             & current two points
             & \ref{tab_current_2pt}                              \\
   HPQCD 10& \cite{McNeile:2010ji}& 2+1 & \gA & \soso
            & \good & \good
%            & $r_1 = 0.3133(23)\, \mbox{fm}$
            & 0.1184(6)    
            & Wilson loops
            & \ref{tab_wloops}  
            \\
   JLQCD 10 & \cite{Shintani:2010ph} & 2+1 &\gA & \bad 
            & \bad & \bad
            & $0.1118(3)(^{+16}_{-17})$    
            & vacuum polarization  
            & \ref{tab_vac} \\
  PACS-CS 09A& \cite{Aoki:2009tf} & 2+1 
            & \gA &\good &\good &\soso
%            & $m_\rho$ 
            & $0.118(3)$$^\#$
            & Schr{\"o}dinger functional\hspace{-0.5cm}
            & \ref{tab_SF3}                                        \\
   Maltman 08& \cite{Maltman:2008bx}& 2+1 & \gA & \soso
            & \soso & \good
%            & $r_1 = 0.318\, \mbox{fm}$
            & $0.1192(11)$
            & Wilson loops
            & \ref{tab_wloops}                               \\ 
   HPQCD 08B  & \cite{Allison:2008xk}  & 2+1       & \gA & \bad
             & \bad  & \bad
%          & $r_1 = 0.321(5)\,\mbox{fm}$
             & 0.1174(12) 
             & current two points
             & \ref{tab_current_2pt}                               \\
   HPQCD 08A& \cite{Davies:2008sw} & 2+1 & \gA & \soso
            & \good & \good
%            & $r_1 = 0.321(5)\,\mbox{fm}$
            & 0.1183(8)
             & Wilson loops
            & \ref{tab_wloops}                                      \\
   HPQCD 05A & \cite{Mason:2005zx} & 2+1 & \gA & \soso
            & \soso & \soso
 %           & $r_1$
            & 0.1170(12)
            & Wilson loops
            & \ref{tab_wloops}                                       \\
   & & & & & & & & \\[-0.1cm]
   \hline
   & & & & & & & & \\[-0.1cm]
   QCDSF/UKQCD 05 & \cite{Gockeler:2005rv}  & $0,2 \to 3$ & \gA & \good 
            & \bad  & \good
%            & $r_0 = 0.467(33)\,\mbox{fm}$
            & 0.112(1)(2)
	   & Wilson loops
            & \ref{tab_wloops}                                       \\
   Boucaud 01B  & \cite{Boucaud:2001qz}    & $2\to 3$ & \gA 
            & \soso & \soso  & \bad
%            & $K^{\ast}-K$
            & 0.113(3)(4)
	   & gluon-ghost vertex
            & \ref{tab_vertex}                                       \\
   SESAM 99 & \cite{Spitz:1999tu} & $0,2\to3$ & \gA & \good
            & \bad  & \bad
%            & $c\bar{c}$(1S-1P)
            & 0.1118(17)
            & Wilson loops
            & \ref{tab_wloops}                                       \\
   Wingate 95 & \cite{Wingate:1995fd} & $0,2\to3$  & \gA & \good
            & \bad  & \bad
%            & $c\bar{c}$(1S-1P)
            & 0.107(5)
            & Wilson loops
            & \ref{tab_wloops}                                       \\
   Davies 94& \cite{Davies:1994ei} & $0,2\to3$  & \gA & \good
            & \bad & \bad
%            & $\Upsilon$
            & 0.115(2)
            & Wilson loops
            & \ref{tab_wloops}                                       \\
   Aoki 94  & \cite{Aoki:1994pc} &  $2\to3$  & \gA & \good
            & \bad & \bad
%            & $c\bar{c}$(1S-1P)
            & 0.108(5)(4)
	   & Wilson loops
            & \ref{tab_wloops}                                       \\
   El-Khadra 92 & \cite{ElKhadra:1992vn} & $0\to3$ & \gA & \good
            & \bad  & \soso
%            & $c\bar{c}$(1S-1P)
            & 0.106(4)
            & Wilson loops
            & \ref{tab_wloops}                                      \\
   & & & & & & & & &  \\[-0.1cm]
   \hline
   \hline
\end{tabular*}
\begin{tabular*}{\textwidth}[l]{l@{\extracolsep{\fill}}lllllll}
\multicolumn{8}{l}{\vbox{\begin{flushleft} 
   $^\#$  Result with a linear continuum extrapolation in $a$.
\end{flushleft}}}
\end{tabular*}
\vspace{-0.3cm}
\caption{Results for $\alpha_\msbar(M_\mathrm{Z})$. 
         $N_f = 3$ results are matched at the
         charm and bottom thresholds and scaled to $M_Z$ to obtain
         the $N_f =5$ result.
         The arrows in the $N_f$ column indicates which $N_f$
         ($N_f = 0$, $2$ or a combination of both)
         were used to first extrapolate to $N_f = 3$ or estimate
         the $N_f = 3$ value through a model/assumption.
         The exact procedures used vary and are given in the
         various papers.}
\label{tab_alphmsbar}
\end{table}

% ----------------------------------------------------------------------

\subsubsection{The present situation}

\begin{figure}[!htb]\hspace{-2cm}\begin{center}
      \includegraphics[width=14.0cm]{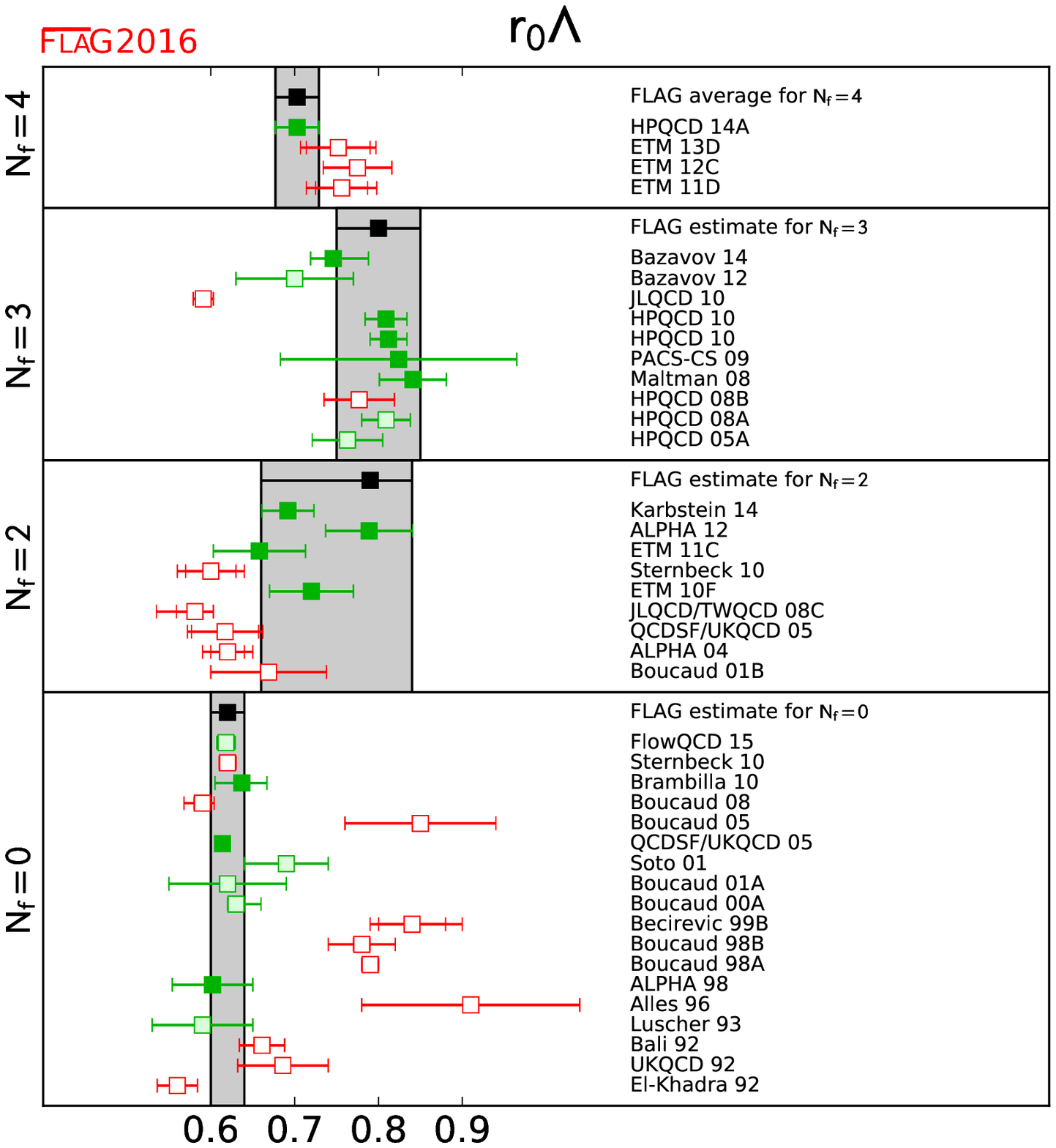}
      \end{center}
\vspace{-1cm}
\caption{$r_0\Lambda_{\overline{\rm MS}}$ estimates for
         $N_f = 0$, $2$, $3$, $4$ flavours.
         Full green squares are used in our final
         ranges, pale green squares also indicate that there are no
         red squares in the colour coding but the computations were
         superseded by later more complete ones or not
         published, while red open squares mean that there is at
         least one red square in the colour coding.}
\label{r0LamMSbar15}
\end{figure}

We first summarize the status of lattice-QCD calculations of the QCD
scale $\Lambda_\msbar$.  Fig.~\ref{r0LamMSbar15} shows all results for
$r_0\Lambda_{\overline{\rm MS}}$ discussed in the previous sections.

Many of the numbers are the ones given directly in the papers. 
However, when only $\Lambda_{\overline{\rm MS}}$ in physical units
($\mbox{MeV}$) is available, we have converted them by multiplying
with the value of $r_0$ in physical units. The notation used
is full green squares for results used in our final average,
while a lightly shaded green square indicates that there are no
red squares in the previous colour coding but the computation does
not enter the ranges because either it has been superseded by an update
or it is not published. Red open squares mean that there is at least
one red square in the colour coding.

For $N_f=0$ there is relatively little spread in the more recent
numbers, even in those which do not satisfy our criteria. 

When two flavours of quarks are included, the numbers extracted 
by the various groups show a considerable spread, as in particular
older computations did not yet control the systematics sufficiently. 
This illustrates the difficulty of the problem and emphasizes the 
need for strict criteria.  
The agreement among the more modern calculations with three or more flavours, 
however, is quite good.

\begin{figure}[!htb]
   \begin{center}
      \includegraphics[width=14.0cm]{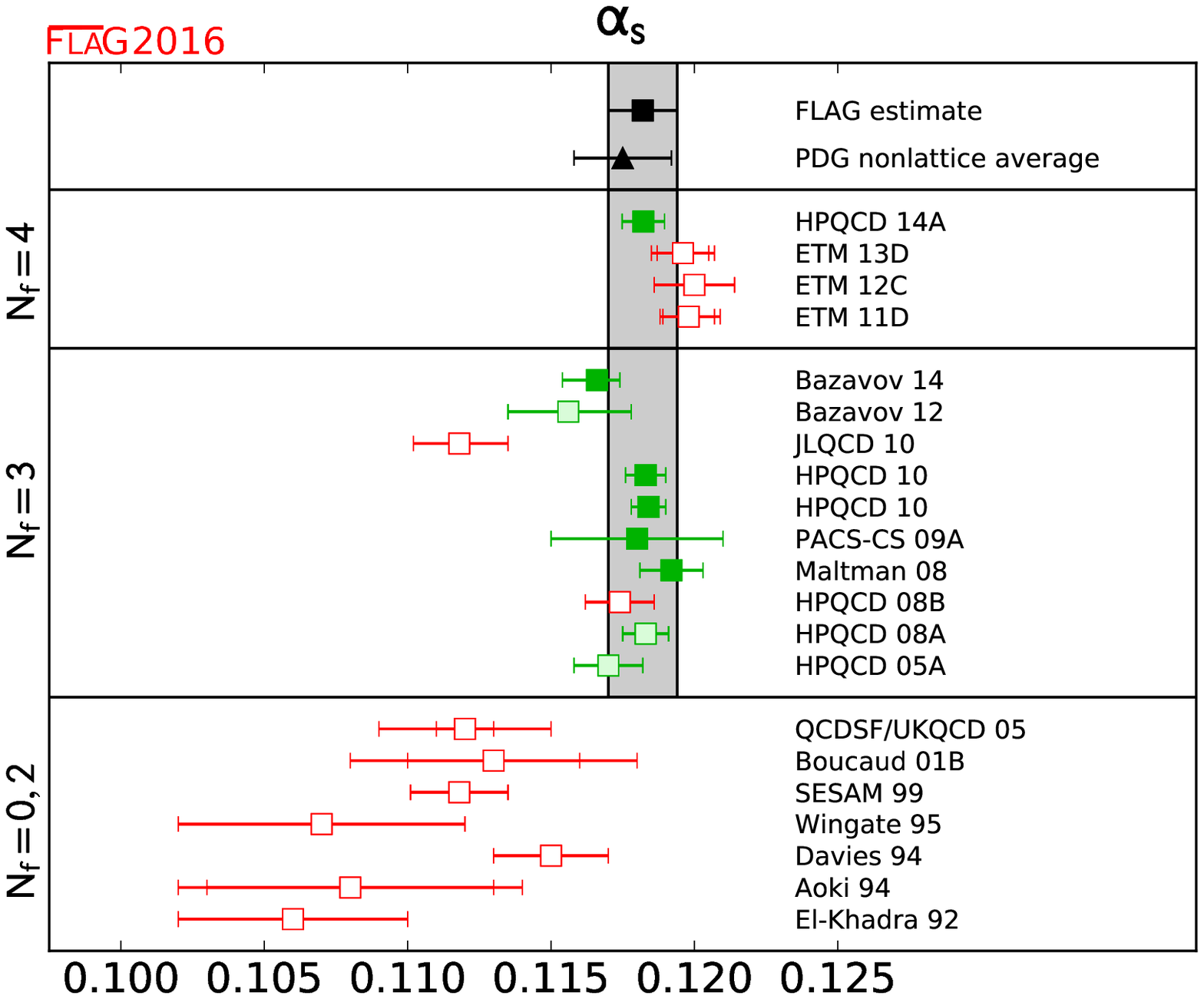}
   \end{center}
\caption{$\alpha_{\overline{\rm MS}}^{(5)}(M_Z)$, the coupling
  constant in the $\overline{\rm MS}$ scheme at the $Z$ mass. The
  results labeled $N_f=0,2$ use estimates for $N_f=3$ obtained by
  first extrapolating in $N_f$ from $N_f=0,2$ results. Since this is
  not a theoretically justified procedure, these are not included in
  our final estimate and are thus given a red symbol. However, they
  are shown to indicate the progress made since these early
  calculations. The PDG entry indicates the outcome of their analysis
  excluding lattice results (see section
  \ref{subsubsec:alpha_s_Conclusions}).}
\label{alphasMSbarZ}
\end{figure}

We now turn to the status of the essential result for phenomenology,
$\alpha_{\overline{\rm MS}}^{(5)}(M_Z)$.  In Tab.~\ref{tab_alphmsbar}
and Fig.~\ref{alphasMSbarZ} we show all the results for
$\alpha_{\overline{\rm MS}}^{(5)}(M_Z)$ (i.e.\ $\alpha_{\overline{\rm MS}}$ at
the $Z$ mass) obtained from $N_f=2+1$ and $N_f = 2+1+1$
simulations. For comparison, we also include results from $N_f = 0$, $2$
simulations, which are not relevant for phenomenology. For the $N_f
\geq 3$ simulations, the conversion from $\Nf = 3$ or $\Nf = 4$
to $\Nf = 5$ is made by matching the coupling constant at the charm and
bottom quark thresholds and using the scale as determined or used by the
authors. For $N_f = 0$, $2$ the results for $\alpha_\msbar$ in the
summary table come from evaluations of $\alpha_\msbar$ at a relatively
low scale and are extrapolated in $\Nf$ to $\Nf = 3$.

As can be seen from the tables and figures, at present there are
several computations satisfying the criteria to be included in
the FLAG average. Since \flagold\ two new computations of
$\alpha_{\overline{\rm MS}}^{(5)}(M_Z)$, Bazavov 14 \cite{Bazavov:2014soa}
and HPQCD 14A  \cite{Chakraborty:2014aca},
pass all our criteria with a \soso. 
We note that none of those calculations of
$\alpha_{\overline{\rm MS}}^{(5)}(M_Z)$ satisfy all of our more
stringent criteria: a $\good$ for the renormalization scale,
perturbative behaviour and continuum extrapolation.  The results,
however, are obtained from four different methods that have different
associated systematics, and agree quite well within the stated
uncertainties.

\subsubsection{Our range for $\alpha_{\overline{\rm MS}}^{(5)}$}
\label{subsubsect:Our range}
We now explain the determination of our range.  We only include those
results without a red tag and that are published in a refereed journal.
We also do not include any numbers which were obtained by
extrapolating from theories with less than three flavours.  There is
no real basis for such extrapolations; rather they use ad hoc
assumptions on the low-energy behaviour of the theories. One also
notices from the published results that the estimated numbers are
quite significantly below those with at least 2+1 flavours.

A general issue with most recent determinations of $\alpha_\msbar$,
both lattice and nonlattice, is that they are dominated by
perturbative truncation errors, which are difficult to estimate.
Further, all results discussed here except for
those of Secs.~\ref{s:SF},~\ref{s:WL} are based on extractions of
$\alpha_\msbar$ that are largely influenced by data with
$\alpha_\mathrm{eff}\geq 0.3$.  At smaller $\alpha_s$ the momentum scale
$\mu$ quickly is at or above $a^{-1}$. We have included computations
using $a\mu$ up to $1.5$ and $\alpha_\mathrm{eff}$ up to 0.4, but one
would ideally like to be significantly below that. Accordingly we
choose at this stage to estimate the error ranges in a conservative
manner, and not simply perform weighted averages with the individual
errors estimated by each group.

Many of the methods have thus far only been applied by a single
collaboration, and with simulation parameters that could still be
improved.  We therefore think that the following aspects of the
individual calculations are important to keep in mind, and look
forward to additional clarification and/or corroboration in the
future.

\vspace{0.2em}
\noindent $\bullet\,$ The potential computations 
Brambilla 10 \cite{Brambilla:2010pp},
ETM 11C \cite{Jansen:2011vv} and Bazavov 12 \cite{Bazavov:2012ka}  give
evidence that they have reached distances where perturbation theory
can be used. However, in addition to $\Lambda$, a scale
is introduced into the perturbative prediction by the process of
subtracting the renormalon contribution. 
This subtraction is avoided in
Bazavov 14 \cite{Bazavov:2014soa} by using the force and again 
agreement with perturbative running is reported.
The extractions of $\Lambda$
are dominated by data with $\alpha_\mathrm{eff}\geq 0.3$. In contrast,
Ref.~\cite{Knechtli:2011pz}, which studies the force instead of the
potential and therefore does not need a renormalon subtraction, finds
that significantly smaller lattice spacings would be needed in order
for perturbation theory to be reliable in a region of $\mu=1/r$ where 
discretization errors are controlled.  
Further study is still needed to clarify the situation.

\vspace{0.2em}
\noindent $\bullet\,$ In the determination of $\alpha_s$ from
observables at the lattice spacing scale, there is an interplay
of higher-order perturbative terms and lattice artefacts.
In HPQCD 05A \cite{Mason:2005zx}, HPQCD 08A \cite{Davies:2008sw}
and Maltman 08 \cite{Maltman:2008bx} both lattice artifacts (which are
power corrections in this approach) and higher-order perturbative
terms are fitted.  We note that, Maltman 08~\cite{Maltman:2008bx} and
HPQCD 08A~\cite{Davies:2008sw} analyze largely the same data set but
use different versions of the perturbative expansion and treatments of
nonperturbative terms.  After adjusting for the slightly different
lattice scales used, the values of $\alpha_\msbar(M_Z)$ differ by
$0.0004$ to $0.0008$ for the three quantities considered.  In fact the
largest of these differences ($0.0008$) comes from a tadpole-improved
loop, which is expected to be best behaved perturbatively.

\vspace{0.2em}
\noindent $\bullet\,$ Other computations with very small errors are
HPQCD 10 \cite{McNeile:2010ji} and HPQCD~14A~\cite{Chakraborty:2014aca},
where correlation functions of heavy quarks are
used to construct short-distance quantities. Due to the large quark
masses needed to reach the region of small coupling, considerable
discretization errors are present, see \fig{hpqcd_alpha_eff}. These
are treated by fits to the perturbative running (a 5-loop running
$\alpha_{\overline{\rm MS}}$ with a fitted 5-loop coefficient in
the $\beta$-function is used) with high-order terms in a double expansion
in $a^2\Lambda^2$ and $a^2 m_\mathrm{h}^2$ supplemented by priors
which limit the size of the coefficients.  The priors play an
especially important role in these fits given the much larger number
of fit parameters than data points.  We note, however, that the size
of the coefficients does not prevent high-order terms from
contributing significantly, since the data includes values of
$am_{\textrm p}/2$ that are rather close to 1.  

As previously mentioned $\alpha_{\overline{\rm MS}}^{(5)}(M_Z)$ is
summarized in Tab.~\ref{tab_alphmsbar} and Fig.~\ref{alphasMSbarZ}.
A number of calculations that include at least the
effect of the strange quark make up our final estimate. These are
Bazavov 14 \cite{Bazavov:2014soa}, HPQCD 14A  
\cite{Chakraborty:2014aca}, HPQCD 10 \cite{McNeile:2010ji}
(Wilson loops and current two-point correlators),
PACS-CS 09A \cite{Aoki:2009tf}, Maltman 08 \cite{Maltman:2008bx}
while HPQCD~08A/05A \cite{Davies:2008sw,Mason:2005zx}
and Bazavov 12 \cite{Bazavov:2012ka}
have been superseded by more recent calculations. 
We obtain the central value for our range,
\begin{eqnarray}
  \alpha_{\overline{\rm MS}}^{(5)}(M_Z) = 0.1182(12) \,, 
\label{eq:alpmz}
\end{eqnarray}
from the weighted average of the six results.\footnote{We have
symmetrized the asymmetric error bars of 
Bazavov 14 \cite{Bazavov:2014soa} to $0.1166(10)$ in taking the average.}
Of the results that enter our range, those from Wilson loops (HPQCD~10
\cite{McNeile:2010ji}, and Maltman 08 \cite{Maltman:2008bx})
and current two-point correlators (HPQCD~10 \cite{McNeile:2010ji})
presently have the smallest quoted errors. We have just listed reasons
to be careful in estimating the present overall uncertainty. We therefore take
a larger range for $\alpha_{\overline{\rm MS}}^{(5)}(M_Z)$ than one
would obtain from the weighted average, or even from the most precise
individual calculation.  We arrive at its value as follows.
We make a conservative estimate of the perturbative uncertainty
in the calculation of $\alpha_s$ from small Wilson loops. 
One approach for making such
an estimate would be to take the largest of the differences between
the calculations of Maltman 08 \cite{Maltman:2008bx} and HPQCD
08A \cite{Davies:2008sw}, $0.0008$, which comes from the quantity
computed by both groups that is expected to be best behaved
perturbatively.  This is somewhat larger than some of the estimates in
the individual papers. Our choice is instead to take an estimate of
the perturbative truncation error as the overall uncertainty. 
As explained in \sect{s:WL} the first unknown coefficient
in the perturbative series was estimated in the fits to be
$|c_4/c_1|\approx 2$. Using it in Eqs.~(\ref{qcdsf:ouruncert},\ref{e:dLL})
\footnote{More precisely, we use $\alpha_{\overline{\rm MS}}^{(3)}(5\GeV)=0.203$
corresponding to \eq{e:lms3} and
$\alpha_{\overline{\rm MS}}^{(5)}(M_\mathrm{Z})=0.1182$
in Eqs.~(\ref{qcdsf:ouruncert},\ref{e:dLL}).} yields 
$\Delta \alpha^{(5)}_{\overline{\rm MS}}(M_Z) = 0.0012$.
This is larger than the estimate of $0.0008$ above and is what we 
adopt as the uncertainty of the Wilson loop results. 
The second number with small errors entering the average
comes from the analysis of moments of heavy quark correlators. Here an 
independent estimate of the uncertainty
due to the fit to the $a$-dependence (see \fig{hpqcd_alpha_eff})
is much more difficult to make; as discussed
above, and in the absence of confirmation by other groups, we are not
yet ready to use the result of HPQCD 10 \cite{McNeile:2010ji}
from the analysis of moments to reduce the size of our range.
Thus the overall size of the range is determined by our estimate
of the uncertainty of $\alpha_{\overline{\rm MS}}^{(5)}(M_Z)$ from 
Wilson loops. It is further reassuring
to see that almost all central values that qualify for averaging
are within the so-determined range.

The range for $\alpha_{\overline{\rm MS}}^{(5)}(M_Z)$ presented here
is based on results with rather different systematics (apart from the
matching across the charm threshold). We therefore believe that the
true value is quite likely to lie within this range.
 
We emphasize once more that all computations which enter
this range rely on a perturbative inclusion of the charm and beauty
quarks.  While perturbation theory for the matching of $\gbar^2_{N_f}$
and $\gbar^2_{N_f-1}$ looks very well behaved even at the mass of the
charm, this scale is rather low and we have no accurate information
about the precision of perturbation theory. Nonperturbative studies
are not yet precise enough \cite{Bruno:2014ufa}. However, it seems unlikely
that the associated uncertainty is comparable with the present
errors. With future improved precision, this will become a relevant
issue. Note that this uncertainty is also present in some of the
phenomenological determinations, in particular from $\tau$ decays.

\subsubsection{Ranges for $[r_0 \Lambda]^{(\Nf)}$
and $\lms$ }

In the present situation, we give ranges for $[r_0 \Lambda]^{(\Nf)}$
and $\lms$, discussing their determination case by case.  We include
results with $\Nf<3$ because it is interesting to see the
$\Nf$-dependence of the connection of low- and high-energy QCD.  This
aids our understanding of the field theory and helps in finding
possible ways to tackle it beyond the lattice approach. It is also of
interest in providing an impression on the size of the vacuum
polarization effects of quarks, in particular with an eye on the still
difficult-to-treat heavier charm and beauty quarks. Even if this
information is rather qualitative, it may be valuable, given that it
is of a completely nonperturbative nature.
We emphasize that results for $[r_0 \Lambda]^{(0)}$
and $[r_0 \Lambda]^{(2)}$ are {\em not}\/ meant to be used
in phenomenology. 

For $\Nf=2+1+1$, we presently do not quote a range
as there is a single result: HPQCD 14A 
\cite{Chakraborty:2014aca} found $[r_0 \Lambda]^{(4)} = 0.70(3)$.

For $\Nf=2+1$, we take as a central value the weighted average of
Bazavov 14 \cite{Bazavov:2014soa}, HPQCD~10 \cite{McNeile:2010ji}
(Wilson loops and current two-point correlators),
PACS-CS~09A \cite{Aoki:2009tf} and Maltman~08 \cite{Maltman:2008bx}.
Since the uncertainty in $r_0$ is small compared to that of $\Lambda$,
we can directly propagate the error from \eq{eq:alpmz} and arrive at 
\begin{eqnarray}
   [r_0 \lms]^{(3)} = 0.80(5) \,.
\label{eq:lms3}
\end{eqnarray}
It is in good agreement with all 2+1 results without red tags. 
In physical units, using $r_0=0.472$~fm and neglecting
its error, this means
\begin{eqnarray}
   \lms^{(3)} = 336(19)\,\mbox{MeV}\,.
\label{e:lms3}
\end{eqnarray}

For $N_f=2$, at present there is one computation with a \good\ rating
for all criteria, ALPHA 12 \cite{Fritzsch:2012wq}. We adopt it as our
central value and enlarge the error to cover the central values of the
other three results with filled green boxes. This results in an
asymmetric error. Our range is unchanged as compared to \flagold,
\begin{eqnarray}
   [r_0 \lms]^{(2)} = 0.79(^{+~5}_{-{13}}) \,, \quad
   \label{eq:lms2}
\end{eqnarray}
and in physical units, using $r_0=0.472$fm, 
\begin{eqnarray}
   \lms^{(2)} = 330(^{+21}_{-{54}}) \mbox{MeV}\,.  \quad 
\end{eqnarray}
A weighted average of the four eligible numbers would yield 
$[r_0 \lms]^{(2)} = 0.709(22)$, not covering the best result and in
particular leading to a smaller error than we feel is justified, given
the issues discussed previously in
Sec.~\ref{short_dist_discuss} (Karbstein 14 \cite{Karbstein:2014bsa},
ETM 11C \cite{Jansen:2011vv}) and 
Sec.~\ref{s:glu_discuss} (ETM 10F \cite{Blossier:2010ky}).
Thus we believe that our estimate is a conservative choice; the low value of 
ETM 11C \cite{Jansen:2011vv} leads to a large downward error.
We hope that future work will improve the situation.

For $N_f=0$ we take into account 
ALPHA~98 \cite{Capitani:1998mq}, QCDSF/UKQCD~05 \cite{Gockeler:2005rv},
and Brambilla~10 \cite{Brambilla:2010pp} for forming a range.
We exclude the older estimates shown in the graph which
have a limited control of the systematic errors due to power law
corrections and discretization errors.\footnote{We have assigned a
  \soso\ for the continuum limit, in Boucaud 00A \cite{Boucaud:2000ey},
  00B \cite{Boucaud:2000nd}, 01A \cite{Boucaud:2001st},
  Soto~01 \cite{DeSoto:2001qx} but these results are from lattices of a
  very small physical size with finite-size effects that are not
  easily quantified.}  
None of the computations have a full set of \good\ 
and has P for publication status. Taking a weighted average of the 
three numbers, we obtain $[r_0 \lms]^{(0)} = 0.615(5)$,
dominated by the QCDSF/UKQCD~05 \cite{Gockeler:2005rv}
result.

Since we are not yet convinced that such a
small uncertainty has been reached, we prefer to presently take a
range which encompasses all four central values and whose uncertainty
comes close to our estimate of the perturbative error
in QCDSF/UKQCD~05 \cite{Gockeler:2005rv}:
based on $|c_4/c_1| \approx 2$ as before, we find 
$\Delta [r_0 \lms]^{(0)} = 0.018$. We then have
\begin{eqnarray}
   [r_0 \lms]^{(0)} =  0.62(2) \,.  \quad
   \label{eq:lms0}
\end{eqnarray}
Converting to physical units, again using $r_0=0.472\,\mbox{fm}$ yields
\begin{eqnarray}
   \lms^{(0)} =  260(7)\,\mbox{MeV}\,. \quad
\end{eqnarray}
While the conversion of the $\Lambda$ parameter to physical units is
quite unambiguous for $\Nf=2+1$, our choice of $r_0=0.472$~fm also for
smaller numbers of flavour amounts to a convention, in particular for
$\Nf=0$. Indeed, in the Tabs.~\ref{tab_SF3}--\ref{tab_vertex}
somewhat different numbers in MeV are found.

How sure are we about our ranges for $[r_0 \lms]^{(N_f)}$? In one case we
have a result, \eq{eq:lms2} which easily passes our criteria, in
another one (\eq{eq:lms0}) we have three compatible results which are
close to that quality and agree. For $\Nf=2+1$ the range
(\eq{eq:lms3}) takes account of results with rather different
systematics. We therefore find it difficult to imagine that the ranges
could be violated by much.

\subsubsection{Conclusions}
\label{subsubsec:alpha_s_Conclusions}
With the present results our range for the strong coupling is
%FLAGRESULT BEGIN
% TAG      & alphas&LambdaQCD &END
% REFS     & \cite{Bazavov:2014soa,Chakraborty:2014aca,McNeile:2010ji,Aoki:2009tf,Maltman:2008bx} & \cite{Bazavov:2014soa,Chakraborty:2014aca,McNeile:2010ji,Aoki:2009tf,Maltman:2008bx}&END
% UNITS    & 1 &'[MeV]'&END
% FLAVOURs & 5 & 5 &END
%FLAGRESULT END
%FLAGRESULTFORMULA BEGIN
(repeating Eq.~(\ref{eq:alpmz}))
\begin{eqnarray*}
 \FLAGAVBEGIN \alpha_{\overline{\rm MS}}^{(5)}(M_Z) = 0.1182(12)\FLAGAVEND\qquad\Refs~\mbox{\cite{Bazavov:2014soa,Chakraborty:2014aca,McNeile:2010ji,Aoki:2009tf,Maltman:2008bx}}, 
\end{eqnarray*}
and the associated $\Lambda$ parameter
\begin{eqnarray}
  \FLAGAVBEGIN \Lambda_{\overline{\rm MS}}^{(5)} = 211(14)\FLAGAVEND\,\MeV\hspace{5mm}\qquad\Refs~\mbox{\cite{Bazavov:2014soa,Chakraborty:2014aca,McNeile:2010ji,Aoki:2009tf,Maltman:2008bx}}. 
\end{eqnarray} 
%FLAGRESULTFORMULA END
These have changed little compared to the previous FLAG review. 
As can be seen from Fig.~\ref{alphasMSbarZ}, when surveying the green
data points, the individual lattice results agree within their quoted
errors. Furthermore those points are based on different methods for
determining $\alpha_s$, each with its own difficulties and
limitations.  Thus the overall consistency of the lattice $\alpha_s$
results engenders confidence in our range.

It is interesting to compare to the new Particle Data Group world average,
which appeared in February 2016 \cite{Agashe:2014kda}. 
The PDG performs their averages, both of lattice determinations and 
of different categories of phenomenological determinations of $\alpha_s$, 
in a way differing significantly from how we determine our range. 
They perform an unweighted average of the mean values. As its error they use
the average of the quoted errors of the different 
determinations that went into the average. 
This procedure leads to larger final uncertainties than the one
used in the previous edition \cite{Beringer:1900zz}.
When one applies this method to
the numbers entering Eq.~(\ref{eq:alpmz}), i.e. the ones satisfying
our criteria, one obtains
$
  \alpha_{\overline{\rm MS}}^{(5)}(M_Z) = 0.1181(12)\,.
$
This number is close to our result
Eq.~(\ref{eq:alpmz}). It differs a little from the value 
quoted by the PDG since in a couple of cases we used updated 
results and because not all determinations entering the PDG average
satisfy our citeria. For comparison, the 
PDG number for lattice results is $0.1187(12)$,
and their average of all phenomenological results is 
$0.1175(17)$.

Our range for the lattice determination of $\alpha_{\overline{\rm MS}}(M_Z)$
in Eq.~(\ref{eq:alpmz}) is in excellent agreement with
the PDG nonlattice average  Eq.~(\ref{PDG_nolat}). This is an excellent check for the 
subtle interplay of theory, phenomenology and experiments in the
nonlattice determinations. The work done on the lattice provides an
entirely independent determination, with negligible experimental 
uncertainty,  which reaches a better
precision even with our conservative estimate of its uncertainty.

We finish by commenting on perspectives for the future.  In the next
few years we anticipate that a growing number of lattice calculations
of $\alpha_s$ from different quantities and by different
collaborations will enable increasingly precise determinations,
coupled with stringent cross-checks.  The determination of $\alpha_s$
from observables at the lattice spacing scale may improve due to a
further reduction of the lattice spacing. This reduces
$\alpha_\mathrm{eff}$ and thus the dominating error in
$\alpha_\msbar$ as long as perturbative results for the simulated 
action are available to high order.  Schr\"odinger functional methods for $N_f=2+1$ will
certainly reach the precision of the present $N_f=2$ results soon, as
this just requires an application of the presently known
techniques. Furthermore, we may expect a significant reduction of
errors due to new definitions of running couplings
\cite{Fodor:2012td,Fritzsch:2013je} using the Yang Mills gradient flow
\cite{Luscher:2010iy}. Factors of two and more in precision are
certainly possible. At this point it will then also be necessary to
include the charm quark in the computations such that the perturbative
matching of $N_f=2+1$ and $2+1+1$ theories at the charm quark
threshold is avoided. First generation $N_f=2+1+1$ simulations are presently being carried out.

%% file: acknowledgment.tex
\section*{Acknowledgments}
\addcontentsline{toc}{section}{Acknowledgments}
FLAG wishes to thank Peter Boyle for early participation. We also wish to thank S.~Bethke, P.~Boucaud, S.~Descotes-Genon,
G.~Dissertori, A.~X.~El~Khadra, W.~Lee, K.~Maltmann, G.~P.~Salam,
R.~S.~Van de Water, and A.~Walker-Loud for discussions.

The kick-off meeting for the present review was held in Bern and was supported by the Albert Einstein Center for Fundamental Physics of the University of Bern. Its hospitality and financial support are gratefully acknowledged. 

G.C., S.D., M.D.M., P.D., R.H., A.J., V.L., A.V., and H.W. are grateful to the Mainz Institute for Theoretical Physics (MITP) for hospitality and partial support during the completion stage of this work.

Members of FLAG were supported by funding agencies; in particular:
\begin{itemize}
\item S.A.~acknowledges partial support from the Grant-in-Aid of the Japanese Ministry of Education, Sciences and Technology, Sports and Culture (MEXT) for Scientific Research (No.~25287046 and 25800147), by MEXT Strategic Program for Innovative Research (SPIRE) Field 5, by a priority issue (Elucidation of the fundamental laws and evolution of the universe) to be tackled by using Post-K Computer, by the Joint Institute for Computational Fundamental Science (JICFuS); 

\item Y.A.~acknowledges support from JSPS KAKENHI Grant No.~22224003;

\item S.H.~and T.K.~acknowledge support from Grants No.~JP26247043 and JP26400259, and by the Post-K supercomputer project through JICFuS;

\item M.D.M.~acknowledges  support from the Danish National Research Foundation DNRF:90 grant and from a Lundbeck Foundation Fellowship grant;

\item S.D.~acknowledges the DFG for partial funding through the SFB/TRR-55 program;

\item H.F.~acknowledges a Grant-in-Aid of the Japanese Ministry of Education (No.~25800147);

\item A.J.~acknowledges support from the European Research Council under the European Community's Seventh Framework Programme (FP7/2007-2013); EU PITN-GA-2009-238353 (STRONGnet) and ERC grant agreement No.~279757;

\item L.L.~and C.-J.D.L.~acknowledge partial support from the OCEVU Labex
  (ANR-11-LABX-0060) and the A*MIDEX project (ANR-11-IDEX-0001-02)
  which are funded by the ``Investissements d'Avenir'' French
  government programme and managed by the ``Agence nationale de la
  recherche'' (ANR); C.-J.D.L.~also acknowledges support from the Taiwanese MoST grant number
  102-2112- M-009-002-MY3;

\item C.P. acknowledges support from the EU PITN-GA-2009-238353 (STRONGnet), Spanish MICINN and MINECO grants FPA2012-31686, FPA2012-31880, and FPA2015-68541-P (MINECO/FEDER), and MINECO's Centro de Excelencia Severo Ochoa Programme under grant SEV-2012-0249;

\item This work was partially supported by the US Department of Energy
  under grant numbers DE-FG02-91ER40628 (for C.B.), DE-FG02-92ER40716
  (for T.B.), DE-FG03-92ER40711 (for M.G.), DE-SC0010120 (for S.G.), DE-SC0011941 (for R.M.), and DE-SC0011637 (S.R.S.).
\end{itemize}

%% file: Glossary/gloss_app.tex
\subsection{Lattice actions}\label{sec_lattice_actions}
In this appendix we give brief descriptions of the lattice actions
used in the simulations and summarize their main features.

\subsubsection{Gauge actions \label{sec_gauge_actions}}

The simplest and most widely used discretization of the Yang-Mills
part of the QCD action is the Wilson plaquette action\,\cite{Wilson:1974sk}:
\be
 S_{\rm G} = \beta\sum_{x} \sum_{\mu<\nu}\Big(
  1-\frac{1}{3}{\rm Re\,\Tr}\,W_{\mu\nu}^{1\times1}(x)\Big),
\label{eq_plaquette}
\ee
where $\beta \equiv 6/g_0^2$ (with $g_0$ the bare gauge coupling) and
the plaquette $W_{\mu\nu}^{1\times1}(x)$ is the product of
link variables around an elementary square of the lattice, i.e.
\be
  W_{\mu\nu}^{1\times1}(x) \equiv U_\mu(x)U_\nu(x+a\hat{\mu})
   U_\mu(x+a\hat{\nu})^{-1} U_\nu(x)^{-1}.
\ee
This expression reproduces the Euclidean Yang-Mills action in the
continuum up to corrections of order~$a^2$.  There is a general
formalism, known as the ``Symanzik improvement programme''
\cite{Symanzik:1983dc,Symanzik:1983gh}, which is designed to cancel
the leading lattice artifacts, such that observables have an
accelerated rate of convergence to the continuum limit.  The
improvement programme is implemented by adding higher-dimensional
operators, whose coefficients must be tuned appropriately in order to
cancel the leading lattice artifacts. The effectiveness of this
procedure depends largely on the method with which the coefficients
are determined. The most widely applied methods (in ascending order of
effectiveness) include perturbation theory, tadpole-improved
(partially resummed) perturbation theory, renormalization group
methods, and the nonperturbative evaluation of improvement
conditions.

In the case of Yang-Mills theory, the simplest version of an improved
lattice action is obtained by adding rectangular $1\times2$ loops to
the plaquette action, i.e.
\be
   S_{\rm G}^{\rm imp} = \beta\sum_{x}\left\{ c_0\sum_{\mu<\nu}\Big(
  1-\frac{1}{3}{\rm Re\,\Tr}\,W_{\mu\nu}^{1\times1}(x)\Big) +
   c_1\sum_{\mu,\nu} \Big(
  1-\frac{1}{3}{\rm Re\,\Tr}\,W_{\mu\nu}^{1\times2}(x)\Big) \right\},
\label{eq_Sym}
\ee
where the coefficients $c_0, c_1$ satisfy the normalization condition
$c_0+8c_1=1$. The {\sl Symanzik-improved} \cite{Luscher:1984xn},
{\sl Iwasaki} \cite{Iwasaki:1985we}, and {\sl DBW2}
\cite{Takaishi:1996xj,deForcrand:1999bi} actions are all defined
through \eq{eq_Sym} via particular choices for $c_0, c_1$. Details are
listed in Table\,\ref{tab_gaugeactions} together with the
abbreviations used in the summary tables. Another widely used variant is the {\sl tadpole Symanzik-improved} \cite{Lepage:1992xa,Alford:1995hw} action which is obtained by adding additional 6-link parallelogram loops $W_{\mu\nu\sigma}^{1\times 1\times 1}(x)$ to the action in Eq.~(\ref{eq_Sym}), i.e.
\be
S_{\rm G}^{\rm tadSym} = S_{\rm G}^{\rm imp} + \beta \sum_x c_2 \sum_{\mu<\nu<\sigma}\Big(1-\frac{1}{3} {\rm Re\,\Tr}\,W_{\mu\nu\sigma}^{1\times1\times1}(x)\Big),
\ee
where
\be
  W_{\mu\nu\sigma}^{1\times1\times1}(x) \equiv U_\mu(x)U_\nu(x+a\hat{\mu})U_\sigma(x+a\hat\mu+a\hat\nu)
   U_\mu(x+a\hat\sigma+a\hat{\nu})^{-1} U_\nu(x+a\hat\sigma)^{-1} U_\sigma(x)^{-1}
\ee
allows for one-loop improvement \cite{Luscher:1984xn}.

\vspace{-0.07cm}
\begin{table}[!h]
\begin{center}
{\footnotesize
\begin{tabular*}{\textwidth}[l]{l @{\extracolsep{\fill}} c l}
\hline\hline \\[-1.0ex]
Abbrev. & $c_1$ & Description 
\\[1.0ex] \hline \hline \\[-1.0ex]
Wilson    & 0 & Wilson plaquette action \\[1.0ex] \hline \\[-1.0ex]
tlSym   & $-1/12$ & tree-level Symanzik-improved gauge action \\[1.0ex] \hline \\[-1.0ex]
tadSym  & variable & tadpole Symanzik-improved gauge action
 \\[1.0ex] \hline \\[-1.0ex]
Iwasaki & $-0.331$ & Renormalization group improved (``Iwasaki'')
action \\[1.0ex] \hline \\[-1.0ex]
DBW2 & $-1.4088$ & Renormalization group improved (``DBW2'') action 
\\ [1.0ex] 
\hline\hline
\end{tabular*}
}
\caption{Summary of lattice gauge actions. The leading lattice
 artifacts are $\cO(a^2)$ or better for all
  discretizations. \label{tab_gaugeactions}} 
\end{center}
\end{table}

%\clearpage

\subsubsection{Light-quark actions \label{sec_quark_actions}}

If one attempts to discretize the quark action, one is faced with the
fermion doubling problem: the naive lattice transcription produces a
16-fold degeneracy of the fermion spectrum. \\

\noindent
{\it Wilson fermions}\\
\noindent

Wilson's solution to the fermion doubling problem is based on adding a
dimension-5 (irrelevant) operator to the lattice action. The
Wilson-Dirac operator for the massless case reads
\cite{Wilson:1974sk,Wilson:1975id}
\be
     D_{\rm w} = \half\gamma_\mu(\nabla_\mu+\nabla_\mu^*)
   +a\nabla_\mu^*\nabla_\mu,
\ee
where $\nabla_\mu,\,\nabla_\mu^*$ denote the covariant forward and
backward lattice derivatives, respectively.  The addition of the
Wilson term $a\nabla_\mu^*\nabla_\mu$, results in fermion doublers
acquiring a mass proportional to the inverse lattice spacing; close to
the continuum limit these extra degrees of freedom are removed from
the low-energy spectrum. However, the Wilson term also results in an
explicit breaking of chiral symmetry even at zero bare quark mass.
Consequently, it also generates divergences proportional to the UV
cutoff (inverse lattice spacing), besides the usual logarithmic
ones. Therefore the chiral limit of the regularized theory is not
defined simply by the vanishing of the bare quark mass but must be
appropriately tuned. As a consequence quark-mass renormalization
requires a power subtraction on top of the standard multiplicative
logarithmic renormalization.  The breaking of chiral symmetry also
implies that the nonrenormalization theorem has to be applied with
care~\cite{Karsten:1980wd,Bochicchio:1985xa}, resulting in a
normalization factor for the axial current which is a regular function
of the bare coupling.  On the other hand, vector symmetry is
unaffected by the Wilson term and thus a lattice (point split) vector
current is conserved and obeys the usual nonrenormalization theorem
with a trivial (unity) normalization factor. Thus, compared to lattice
fermion actions which preserve chiral symmetry, or a subgroup of it,
the Wilson regularization typically results in more complicated
renormalization patterns.

Furthermore, the leading-order lattice artifacts are of order~$a$.
With the help of the Symanzik improvement programme, the leading
artifacts can be cancelled in the action by adding the so-called
``Clover'' or Sheikholeslami-Wohlert (SW) term~\cite{Luscher:1996sc}.
The resulting expression in the massless case reads
\be
   D_{\rm sw} = D_{\rm w}
   +\frac{ia}{4}\,\csw\sigma_{\mu\nu}\widehat{F}_{\mu\nu},
\label{eq_DSW}
\ee
where $\sigma_{\mu\nu}=\frac{i}{2}[\gamma_\mu,\gamma_\nu]$, and
$\widehat{F}_{\mu\nu}$ is a lattice transcription of the gluon field
strength tensor $F_{\mu\nu}$. The coefficient $\csw$ can be determined
perturbatively at tree-level ($\csw = 1$; tree-level improvement or
tlSW for short), via a mean field approach \cite{Lepage:1992xa}
(mean-field improvement or mfSW) or via a nonperturbative approach
\cite{Luscher:1996ug} (nonperturbatively improved or npSW).
Hadron masses, computed using $D_{\rm sw}$, with the coefficient
$\csw$ determined nonperturbatively, will approach the continuum
limit with a rate proportional to~$a^2$; with tlSW for $\csw$ the rate
is proportional to~$g_0^2 a$.

Other observables require additional improvement
coefficients~\cite{Luscher:1996sc}.  A common example consists in the
computation of the matrix element $\langle \alpha \vert Q \vert \beta
\rangle$ of a composite field $Q$ of dimension-$d$ with external
states $\vert \alpha \rangle$ and $\vert \beta \rangle$. In the
simplest cases, the above bare matrix element diverges logarithmically
and a single renormalization parameter $Z_Q$ is adequate to render it
finite. It then approaches the continuum limit with a rate
proportional to the lattice spacing $a$, even when the lattice action
contains the Clover term. In order to reduce discretization errors to
${\cO}(a^2)$, the lattice definition of the composite operator $Q$
must be modified (or ``improved"), by the addition of all
dimension-$(d+1)$ operators with the same lattice symmetries as $Q$.
Each of these terms is accompanied by a coefficient which must be
tuned in a way analogous to that of $\csw$. Once these coefficients
are determined nonperturbatively, the renormalized matrix element of
the improved operator, computed with a npSW action, converges to the
continuum limit with a rate proportional to~$a^2$. A tlSW improvement
of these coefficients and $\csw$ will result in a rate proportional
to~$g_0^2 a$.

It is important to stress that the improvement procedure does not
affect the chiral properties of Wilson fermions; chiral symmetry
remains broken.

Finally, we mention ``twisted-mass QCD'' as a method which was
originally designed to address another problem of Wilson's
discretization: the Wilson-Dirac operator is not protected against the
occurrence of unphysical zero modes, which manifest themselves as
``exceptional'' configurations. They occur with a certain frequency in
numerical simulations with Wilson quarks and can lead to strong
statistical fluctuations. The problem can be cured by introducing a
so-called ``chirally twisted'' mass term. The most common formulation
applies to a flavour doublet $\bar \psi = ( u \quad d)$ of
mass-degenerate quarks, with the fermionic part of the QCD action in
the continuum assuming the form \cite{Frezzotti:2000nk}
\be
   S_{\rm F}^{\rm tm;cont} = \int d^4{x}\, \psibar(x)(\gamma_\mu
   D_\mu +
   m + i\mu_{\rm q}\gamma_5\tau^3)\psi(x).
\ee
Here, $\mu_{\rm q}$ is the twisted-mass parameter, and $\tau^3$ is a
Pauli matrix in flavour space. The standard action in the continuum
can be recovered via a global chiral field rotation. The physical
quark mass is obtained as a function of the two mass parameters $m$
and $\mu_{\rm q}$. The corresponding lattice regularization of twisted-mass QCD (tmWil) for $\Nf=2$ flavours is defined through the fermion
matrix
\be
   D_{\rm w}+m_0+i\mu_{\rm q}\gamma_5\tau^3 \,\, .
\label{eq_tmQCD}
\ee
Although this formulation breaks physical parity and flavour
symmetries, resulting in nondegenerate neutral and charged pions,
is has a number of advantages over standard Wilson
fermions. Firstly, the presence of the twisted-mass parameter
$\mu_{\rm q}$ protects the discretized theory against unphysical zero
modes. A second attractive feature of twisted-mass lattice QCD is the
fact that, once the bare mass parameter $m_0$ is tuned to its ``critical value"
(corresponding to massless pions in the standard Wilson formulation),
the leading lattice artifacts are of order $a^2$ without the
need to add the Sheikholeslami-Wohlert term in the action, or other
improving coefficients~\cite{Frezzotti:2003ni}. A third important advantage
is that, although the problem of explicit chiral
symmetry breaking remains, quantities computed with twisted fermions
with a suitable tuning of the mass parameter $\mu_{\rm q}$,
are subject to renormalization patterns which are simpler than the ones with
standard Wilson fermions. Well known examples are the pseudoscalar decay
constant  and $B_{\rm K}$.\\

\noindent
{\it Staggered Fermions}\\
\noindent

An alternative procedure to deal with the doubling problem is based on so-called ``staggered''
or Kogut-Susskind fermions \cite{Kogut:1974ag,Banks:1975gq, Banks:1976ia, Susskind:1976jm}.
Here the degeneracy is only lifted partially, from 16 down to 4.  It has become customary
to refer to these residual doublers as ``tastes'' in order to distinguish them from physical
flavours.  Taste changing interactions 
can occur via the exchange of gluons with one or more components
  of momentum near the cutoff $\pi/a$.  This leads to the breaking of the $SU(4)$ vector symmetry among 
  tastes, thereby generating order $a^2$ lattice artifacts.

The residual doubling of staggered quarks (four tastes per
flavour) is removed by taking a fractional power of the fermion determinant \cite{Marinari:1981qf} --- the ``fourth-root 
procedure,'' or, sometimes, the ``fourth root trick.''  
This procedure would be unproblematic if
the action had full $SU(4)$ taste symmetry, which would give a
Dirac operator that was block-diagonal in taste space.  
However, the breaking of taste symmetry at nonzero lattice spacing leads to a
variety of problems. In fact, the fourth root of the determinant is not equivalent
to the determinant of any local lattice Dirac operator \cite{Bernard:2006ee}.
This in turn leads 
to violations of unitarity 
on the lattice \cite{Prelovsek:2005rf,Bernard:2006zw,Bernard:2007qf,Aubin:2008wk}.

According to standard renormalization group lore, the taste
violations, which are associated with lattice operators of dimension
greater than four, might be expected go away in the continuum limit,
resulting in the restoration of locality and unitarity.  However,
there is a problem with applying the standard lore to this nonstandard
situation: the usual renormalization group reasoning assumes that the
lattice action is local.  Nevertheless, Shamir
\cite{Shamir:2004zc,Shamir:2006nj} shows that one may apply the
renormalization group to a ``nearby'' local theory, and thereby gives
a strong argument that that the desired local, unitary theory of QCD
is reproduced by the rooted staggered lattice theory in the continuum
limit.

A version of chiral perturbation that includes the lattice artifacts
due to taste violations and rooting (``rooted staggered chiral
perturbation theory'') can also be worked out
\cite{Lee:1999zxa,Aubin:2003mg,Sharpe:2004is} and shown to correctly
describe the unitarity-violating lattice artifacts in the pion sector
\cite{Bernard:2006zw,Bernard:2007ma}.  This provides additional
evidence that the desired continuum limit can be obtained. Further, it
gives a practical method for removing the lattice artifacts from
simulation results. Versions of rooted staggered chiral perturbation
theory exist for heavy-light mesons with staggered light quarks but
nonstaggered heavy quarks \cite{Aubin:2005aq}, heavy-light mesons with
staggered light and heavy quarks
\cite{Komijani:2012fq,Bernard:2013qwa}, staggered baryons
\cite{Bailey:2007iq}, and mixed actions with a staggered sea
\cite{Bar:2005tu,Bae:2010ki}, as well as the pion-only version
referenced above.

There is also considerable numerical evidence that the rooting
procedure works as desired.  This includes investigations in the
Schwinger model \cite{Durr:2003xs,Durr:2004ta,Durr:2006ze}, studies of
the eigenvalues of the Dirac operator in QCD
\cite{Follana:2004sz,Durr:2004as,Wong:2004nk,Donald:2011if}, and
evidence for taste restoration in the pion spectrum as $a\to0$
\cite{Aubin:2004fs,Bazavov:2009bb}.

Issues with the rooting procedure have led Creutz
\cite{Creutz:2006ys,Creutz:2006wv,Creutz:2007yg,Creutz:2007pr,Creutz:2007rk,Creutz:2008kb,Creutz:2008nk}
to argue that the continuum limit of the rooted staggered theory
cannot be QCD.  These objections have however been answered in
Refs.~\cite{Bernard:2006vv,Sharpe:2006re,Bernard:2007eh,Kronfeld:2007ek,Bernard:2008gr,Adams:2008db,Golterman:2008gt,
  Donald:2011if}. In particular, a claim that the continuum 't Hooft
vertex \cite{'tHooft:1976up,'tHooft:1976fv} could not be properly
reproduced by the rooted theory has been refuted
\cite{Bernard:2007eh,Donald:2011if}.

Overall, despite the lack of rigorous proof of the correctness of the
rooting procedure, we think the evidence is strong enough to consider staggered
QCD simulations on a par with simulations using other actions.
See the following reviews for further evidence and discussion:
Refs.~\cite{Durr:2005ax,Sharpe:2006re,Kronfeld:2007ek,Golterman:2008gt,Bazavov:2009bb}.
\\

\noindent
{\it Improved Staggered Fermions}\\
\noindent

An improvement program can be used to suppress taste-changing
interactions, leading to ``improved staggered fermions,'' with the
so-called ``Asqtad'' \cite{Orginos:1999cr}, ``HISQ''
\cite{Follana:2006rc}, ``Stout-smeared'' \cite{Aoki:2005vt}, and
``HYP'' \cite{Hasenfratz:2001hp} actions as the most common versions.
All these actions smear the gauge links in order to reduce the
coupling of high-momentum gluons to the quarks, with the main goal of
decreasing taste-violating interactions. In the Asqtad case, this is
accomplished by replacing the gluon links in the derivatives by
averages over 1-, 3-, 5-, and 7-link paths.  The other actions reduce
taste changing even further by smearing more.  In addition to the
smearing, the Asqtad and HISQ actions include a three-hop term in the
action (the ``Naik term'' \cite{Naik:1986bn}) to remove order $a^2$
errors in the dispersion relation, as well as a ``Lepage term''
\cite{Lepage:1998vj} to cancel other order $a^2$ artifacts introduced
by the smearing.  In both the Asqtad and HISQ actions, the leading
taste violations are of order $\alpha_S^2 a^2$, and ``generic''
lattices artifacts (those associated with discretization errors other
than taste violations) are of order $\alpha_S a^2$.  The overall
coefficients of these errors are, however, significantly smaller with
HISQ than with Asqtad.  With the Stout-smeared and HYP actions, the
errors are formally larger (order $\alpha_S a^2$ for taste violations
and order $a^2$ for generic lattices artifacts).  Nevertheless, the
smearing seems to be very efficient, and the actual size of errors at
accessible lattice spacings appears to be at least as small as with
HISQ.

Although logically distinct from the light-quark improvement program
for these actions, it is customary with the HISQ action to include an
additional correction designed to reduce discretization errors for
heavy quarks (in practice, usually charm quarks)
\cite{Follana:2006rc}. The Naik term is adjusted to remove leading
$(am_c)^4$ and $\alpha_S(am_c)^2$ errors, where $m_c$ is the
charm-quark mass and ``leading'' in this context means leading in
powers of the heavy-quark velocity $v$ ($v/c\sim 1/3$ for $D_s$).
With these improvements, the claim is that one can use the staggered
action for charm quarks, although it must be emphasized that it is not
obvious {\it a priori}\/ how large a value of $am_c$ may be tolerated
for a given desired accuracy, and this must be studied in the
simulations.  \\

\noindent
{\it Ginsparg-Wilson fermions}\\
\noindent

Fermionic lattice actions, which do not suffer from the doubling
problem whilst preserving chiral symmetry go under the name of
``Ginsparg-Wilson fermions''. In the continuum the massless Dirac
operator ($D$) anti-commutes with $\gamma_5$. At nonzero lattice spacing a 
chiral symmetry can be realized if this condition is relaxed
to \cite{Hasenfratz:1998jp,Hasenfratz:1998ri,Luscher:1998pqa}
\be
   \left\{D,\gamma_5\right\} = aD\gamma_5 D,
\label{eq_GWrelation}
\ee
which is now known as the Ginsparg-Wilson relation
\cite{Ginsparg:1981bj}. The Nielsen-Ninomiya
theorem~\cite{Nielsen:1981hk}, which states that any lattice
formulation for which $D$ anticommutes with $\gamma_5$ necessarily has
doubler fermions, is circumvented since $\{D,\gamma_5\}\neq 0$.

A lattice Dirac operator which satisfies \eq{eq_GWrelation} can be
constructed in several ways. The so-called ``overlap'' or
Neuberger-Dirac operator~\cite{Neuberger:1997fp} acts in four
space-time dimensions and is, in its simplest form, defined by
\be
   D_{\rm N} = \frac{1}{\abar} \left( 1-\epsilon(A)
   \right),\quad\mathrm{where}\quad\epsilon(A)\equiv A (A^\dagger A)^{-1/2}, \quad A=1+s-aD_{\rm w},\quad \abar=\frac{a}{1+s},
\label{eq_overlap}
\ee
$D_{\rm w}$ is the massless Wilson-Dirac operator and $|s|<1$
is a tunable parameter. The overlap operator $D_{\rm N}$ removes all
doublers from the spectrum, and can readily be shown to satisfy the
Ginsparg-Wilson relation. The occurrence of the sign function $\epsilon(A)$ in
$D_{\rm N}$ renders the application of $D_{\rm N}$ in a computer
program potentially very costly, since it must be implemented using,
for instance, a polynomial approximation.

The most widely used approach to satisfying the Ginsparg-Wilson
relation \eq{eq_GWrelation} in large-scale numerical simulations is
provided by \textit{Domain Wall Fermions}
(DWF)~\cite{Kaplan:1992bt,Shamir:1993zy,Furman:1994ky} and we
therefore describe this in some more detail. Following early
exploratory studies~\cite{Blum:1996jf}. this approach has been
developed into a practical formulation of lattice QCD with good chiral
and flavour symmetries leading to results which contribute
significantly to this review. In this formulation, the fermion fields
$\psi(x,s)$ depend on a discrete fifth coordinate $s=1,\ldots,N$ as well as
the physical 4-dimensional space-time coordinates $x_\mu,\,\mu=1\cdots
4$ (the gluon fields do not depend on $s$). The lattice on which the
simulations are performed, is therefore a five-dimensional one of size
$L^3\times T\times N$, where $L,\,T$ and $N$ represent the number of
points in the spatial, temporal and fifth dimensions respectively.
The remarkable feature of DWF is that for each flavour there exists a
physical light mode corresponding to the field $q(x)$:
\begin{eqnarray}
q(x)&=&\frac{1+\gamma^5}{2}\psi(x,1)+\frac{1-\gamma^5}{2}\psi(x,N)\\
\bar{q}(x)&=&\overline{\psi}(x,N)\frac{1+\gamma^5}{2} + \overline{\psi}(x,1)\frac{1-\gamma^5}{2}\,.
\end{eqnarray}
The left and right-handed modes of the physical field are located on
opposite boundaries in the 5th dimensional space which, for
$N\to\infty$, allows for independent transformations of the left and
right components of the quark fields, that is for chiral
transformations. Unlike Wilson fermions, where for each flavour the
quark-mass parameter in the action is fine-tuned requiring a
subtraction of contributions of $\cO(1/a)$ where $a$ is the lattice
spacing, with DWF no such subtraction is necessary for the physical
modes, whereas the unphysical modes have masses of $\cO(1/a)$ and
decouple.

In actual simulations $N$ is finite and there are small violations of
chiral symmetry which must be accounted for. The theoretical framework
for the study of the residual breaking of chiral symmetry has been a
subject of intensive investigation (for a review and references to the
original literature see e.g.~\cite{Sharpe:2007yd}). The breaking
requires one or more \emph{crossings} of the fifth dimension to couple
the left and right-handed modes; the more crossings that are required
the smaller the effect.  For many physical quantities the leading
effects of chiral symmetry breaking due to finite $N$ are parameterized
by a \emph{residual} mass, $m_{\mathrm{res}}$.  For example, the PCAC
relation (for degenerate quarks of mass $m$) $\partial_\mu A_\mu(x) =
2m P(x)$, where $A_\mu$ and $P$ represent the axial current and
pseudoscalar density respectively, is satisfied with
$m=m^\mathrm{DWF}+m_\mathrm{res}$, where $m^\mathrm{DWF}$ is the bare
mass in the DWF action. The mixing of operators which transform under
different representations of chiral symmetry is found to be negligibly
small in current simulations. The important thing to note is that the
chiral symmetry breaking effects are small and that there are
techniques to mitigate their consequences.

The main price which has to be paid for the good chiral symmetry is
that the simulations are performed in 5 dimensions, requiring
approximately a factor of N in computing resources and resulting in
practice in ensembles at fewer values of the lattice spacing and quark
masses than is possible with other formulations. The current
generation of DWF simulations is being performed at physical quark
masses so that ensembles with good chiral and flavour symmetries are
being generated and analysed~\cite{Arthur:2012opa}. For a discussion
of the equivalence of DWF and overlap fermions
see Refs.~\cite{Borici:1999zw,Borici:1999da}.

A third example of an operator which satisfies the Ginsparg-Wilson
relation is the so-called fixed-point action
\cite{Bietenholz:1995cy,Hasenfratz:2000xz,Hasenfratz:2002rp}. This
construction proceeds via a renormalization group approach. A related
formalism are the so-called ``chirally improved'' fermions
\cite{Gattringer:2000js}.\\

\begin{table}
\begin{center}
{\footnotesize
\begin{tabular*}{\textwidth}[l]{l @{\extracolsep{\fill}} l l l l}
\hline \hline  \\[-1.0ex]
\parbox[t]{1.5cm}{Abbrev.} & Discretization & \parbox[t]{2.2cm}{Leading lattice \\artifacts} & Chiral symmetry &  Remarks
\\[4.0ex] \hline \hline \\[-1.0ex]
Wilson     & Wilson & $\cO(a)$ & broken & 
\\[1.0ex] \hline \\[-1.0ex]
tmWil   & twisted-mass Wilson &  \parbox[t]{2.2cm}{$\cO(a^2)$
at\\ maximal twist} & broken & \parbox[t]{5cm}{flavour-symmetry breaking:\\ $(M_\text{PS}^{0})^2-(M_\text{PS}^\pm)^2\sim \cO(a^2)$}
\\[4.0ex] \hline \\[-1.0ex]
tlSW      & Sheikholeslami-Wohlert & $\cO(g^2 a)$ & broken & tree-level
impr., $\csw=1$
\\[1.0ex] \hline \\[-1.0ex]
\parbox[t]{1.0cm}{n-HYP tlSW}      & Sheikholeslami-Wohlert & $\cO(g^2 a)$ & broken & \parbox[t]{5cm}{tree-level
impr., $\csw=1$,\\
n-HYP smeared gauge links
}
\\[4.0ex] \hline \\[-1.0ex]
\parbox[t]{1.2cm}{stout tlSW}      & Sheikholeslami-Wohlert & $\cO(g^2 a)$ & broken & \parbox[t]{5cm}{tree-level
impr., $\csw=1$,\\
stout smeared gauge links
}
\\[4.0ex] \hline \\[-1.0ex]
\parbox[t]{1.2cm}{HEX tlSW}      & Sheikholeslami-Wohlert & $\cO(g^2 a)$ & broken & \parbox[t]{5cm}{tree-level
impr., $\csw=1$,\\
HEX smeared gauge links
}
\\[4.0ex] \hline \\[-1.0ex]
mfSW      & Sheikholeslami-Wohlert & $\cO(g^2 a)$ & broken & mean-field impr.
\\[1.0ex] \hline \\[-1.0ex]
npSW      & Sheikholeslami-Wohlert & $\cO(a^2)$ & broken & nonperturbatively impr.
\\[1.0ex] \hline \\[-1.0ex]
KS      & Staggered & $\cO(a^2)$ & \parbox[t]{3cm}{$\rm
  U(1)\times U(1)$ subgr.\\ unbroken} & rooting for $\Nf<4$
\\[4.0ex] \hline \\[-1.0ex]
Asqtad  & Staggered & $\cO(a^2)$ & \parbox[t]{3cm}{$\rm
  U(1)\times U(1)$ subgr.\\ unbroken}  & \parbox[t]{5cm}{Asqtad
  smeared gauge links, \\rooting for $\Nf<4$}  
\\[4.0ex] \hline \\[-1.0ex]
HISQ  & Staggered & $\cO(a^2)$ & \parbox[t]{3cm}{$\rm
  U(1)\times U(1)$ subgr.\\ unbroken}  & \parbox[t]{5cm}{HISQ
  smeared gauge links, \\rooting for $\Nf<4$}  
\\[4.0ex] \hline \\[-1.0ex]
DW      & Domain Wall & \parbox[t]{2.2cm}{asymptotically \\$\cO(a^2)$} & \parbox[t]{3cm}{remnant
  breaking \\exponentially suppr.} & \parbox[t]{5cm}{exact chiral symmetry and\\$\cO(a)$ impr. only in the limit \\
 $L_s\rightarrow \infty$}
\\[7.0ex] \hline \\[-1.0ex]
oDW      & optimal Domain Wall & \parbox[t]{2.2cm}{asymptotically \\$\cO(a^2)$} & \parbox[t]{3cm}{remnant
  breaking \\exponentially suppr.} & \parbox[t]{5cm}{exact chiral symmetry and\\$\cO(a)$ impr. only in the limit \\
 $L_s\rightarrow \infty$}
\\[7.0ex] \hline \\[-1.0ex]
M-DW      & Moebius Domain Wall & \parbox[t]{2.2cm}{asymptotically \\$\cO(a^2)$} & \parbox[t]{3cm}{remnant
  breaking \\exponentially suppr.} & \parbox[t]{5cm}{exact chiral symmetry and\\$\cO(a)$ impr. only in the limit \\
 $L_s\rightarrow \infty$}
\\[7.0ex] \hline \\[-1.0ex]
overlap    & Neuberger & $\cO(a^2)$ & exact
\\[1.0ex] 
\hline\hline
\end{tabular*}
}
\caption{The most widely used discretizations of the quark action
  and some of their properties. Note that in order to maintain the
  leading lattice artifacts of the action in nonspectral observables
  (like operator matrix elements)
  the corresponding nonspectral operators need to be improved as well. 
\label{tab_quarkactions}}
\end{center}
\end{table}

\noindent
{\it Smearing}\\
\noindent

A simple modification which can help improve the action as well as the
computational performance is the use of smeared gauge fields in the
covariant derivatives of the fermionic action. Any smearing procedure
is acceptable as long as it consists of only adding irrelevant (local)
operators. Moreover, it can be combined with any discretization of the
quark action.  The ``Asqtad'' staggered quark action mentioned above
\cite{Orginos:1999cr} is an example which makes use of so-called
``Asqtad'' smeared (or ``fat'') links. Another example is the use of
n-HYP smeared \cite{Hasenfratz:2001hp,Hasenfratz:2007rf}, stout smeared
\cite{Morningstar:2003gk,Durr:2008rw} or HEX (hypercubic stout) smeared \cite{Capitani:2006ni} gauge links in the tree-level clover improved
discretization of the quark action, denoted by ``n-HYP tlSW'',
``stout tlSW'' and ``HEX tlSW'' in the following.\\

\noindent
In Table \ref{tab_quarkactions} we summarize the most widely used
discretizations of the quark action and their main properties together
with the abbreviations used in the summary tables. Note that in order
to maintain the leading lattice artifacts of the actions as given in
the table in nonspectral observables (like operator matrix elements)
the corresponding nonspectral operators need to be improved as well.

%\clearpage
%=================================================
\subsubsection{Heavy-quark actions}
\label{app:HQactions}
%=================================================

Charm and bottom quarks are often simulated with different
lattice-quark actions than up, down, and strange quarks because their
masses are large relative to typical lattice spacings in current
simulations; for example, $a m_c \sim 0.4$ and $am_b \sim 1.3$ at
$a=0.06$~fm.  Therefore, for the actions described in the previous
section, using a sufficiently small lattice spacing to control generic
$(am_h)^n$ discretization errors is computationally costly, and in
fact prohibitive at the physical $b$-quark mass.

One approach for lattice heavy quarks is direct application of
effective theory.  In this case the lattice heavy-quark action only
correctly describes phenomena in a specific kinematic regime, such as
Heavy-Quark Effective Theory
(HQET)~\cite{Isgur:1989vq,Eichten:1989zv,Isgur:1989ed} or
Nonrelativistic QCD (NRQCD)~\cite{Caswell:1985ui,Bodwin:1994jh}.  One
can discretize the effective Lagrangian to obtain, for example,
Lattice HQET~\cite{Heitger:2003nj} or Lattice
NRQCD~\cite{Thacker:1990bm,Lepage:1992tx}, and then simulate the
effective theory numerically.  The coefficients of the operators in
the lattice-HQET and lattice-NRQCD actions are free parameters that
must be determined by matching to the underlying theory (QCD) through
the chosen order in $1/m_h$ or $v_h^2$, where $m_h$ is the heavy-quark
mass and $v_h$ is the heavy-quark velocity in the the heavy-light
meson rest frame.

Another approach is to interpret a relativistic quark action such as
those described in the previous section in a manner suitable for heavy
quarks.  One can extend the standard Symanzik improvement program,
which allows one to systematically remove lattice cutoff effects by
adding higher-dimension operators to the action, by allowing the
coefficients of the dimension 4 and higher operators to depend
explicitly upon the heavy-quark mass.  Different prescriptions for
tuning the parameters correspond to different implementations: those
in common use are often called the Fermilab
action~\cite{ElKhadra:1996mp}, the relativistic heavy-quark action
(RHQ)~\cite{Christ:2006us}, and the Tsukuba
formulation~\cite{Aoki:2001ra}.  In the Fermilab approach, HQET is
used to match the lattice theory to continuum QCD at the desired order
in $1/m_h$.

More generally, effective theory can be used to estimate the size of
cutoff errors from the various lattice heavy-quark actions.  The power
counting for the sizes of operators with heavy quarks depends on the
typical momenta of the heavy quarks in the system.  Bound-state
dynamics differ considerably between heavy-heavy and heavy-light
systems.  In heavy-light systems, the heavy quark provides an
approximately static source for the attractive binding force, like the
proton in a hydrogen atom.  The typical heavy-quark momentum in the
bound-state rest frame is $|\vec{p}_h| \sim \Lambda_{\rm QCD}$, and
heavy-light operators scale as powers of $(\Lambda_{\rm QCD}/m_h)^n$.
This is often called ``HQET power-counting", although it applies to
heavy-light operators in HQET, NRQCD, and even relativistic
heavy-quark actions described below.  Heavy-heavy systems are similar
to positronium or the deuteron, with the typical heavy-quark momentum
$|\vec{p}_h| \sim \alpha_S m_h$.  Therefore motion of the heavy quarks
in the bound state rest frame cannot be neglected.  Heavy-heavy
operators have complicated power counting rules in terms of
$v_h^2$~\cite{Lepage:1992tx}; this is often called ``NRQCD power
counting."

Alternatively, one can simulate bottom or charm quarks with the same
action as up, down, and strange quarks provided that (1) the action is
sufficiently improved, and (2) the lattice spacing is sufficiently
fine.  These qualitative criteria do not specify precisely how large a
numerical value of $am_h$ can be allowed while obtaining a given
precision for physical quantities; this must be established
empirically in numerical simulations.  At present, both the HISQ and
twisted-mass Wilson actions discussed previously are being used to
simulate charm quarks.
Simulations with HISQ quarks have employed heavier-quark masses than
those with twisted-mass Wilson quarks because the action is more
highly improved, but neither action can be used to simulate at the
physical $am_b$ for current lattice spacings.  Therefore calculations
of heavy-light decay constants with these actions still rely on
effective theory to reach the $b$-quark mass: the ETM Collaboration
interpolates between twisted-mass Wilson data generated near $am_c$
and the static point~\cite{Dimopoulos:2011gx}, while the HPQCD
Collaboration extrapolates HISQ data generated below $am_b$ up to the
physical point using an HQET-inspired series expansion in
$(1/m_h)^n$~\cite{McNeile:2011ng}.
\\

%=================================================
%Heavy-quark effective theory
%=================================================

\noindent
{\it Heavy-quark effective theory}\\
\noindent

HQET was introduced by Eichten and Hill in
Ref.~\cite{Eichten:1989zv}. It provides the correct asymptotic
description of QCD correlation functions in the static limit
$m_{h}/|\vec{p}_h| \!\to\! \infty$. Subleading effects are described
by higher dimensional operators whose coupling constants are formally
of ${\cO}((1/m_{h})^n)$.  The HQET expansion works well for
heavy-light systems in which the heavy-quark momentum is small
compared to the mass.

The HQET Lagrangian density at the leading (static) order in the rest
frame of the heavy quark is given by
\be
{\mathcal L}^{\rm stat}(x) = \overline{\psi}_{h}(x) \,D_0\, \psi_{h}(x)\;,
\ee
with
\be
P_+ \psi_{h} = \psi_{h} \; , \quad\quad \overline{\psi}_{h} P_+=\overline{\psi}_{h} \;,  
\quad\quad P_+={{1+\gamma_0}\over{2}} \;.
\ee
A bare quark mass $m_{\rm bare}^{\rm stat}$ has to be added to the energy  
levels $E^{\rm stat}$ computed with this Lagrangian to obtain the physical ones.
 For example, the mass of the $B$ meson in the static approximation is given by
\be
m_{B} = E^{\rm stat} + m_{\rm bare}^{\rm stat} \;.
\ee
At tree-level $m_{\rm bare}^{\rm stat}$ is simply the (static approximation of
the) $b$-quark mass, but in the quantized lattice formulation it has
to further compensate a divergence linear in the inverse lattice spacing.
Weak composite fields  are also rewritten in terms of the static fields, e.g.
\begin{equation}
A_0(x)^{\rm stat}=Z_{\rm A}^{\rm stat} \left( \overline{\psi}(x) \gamma_0\gamma_5\psi_h(x)\right)\;,
\end{equation}
where the renormalization factor of the axial current in the static
theory $Z_{\rm A}^{\rm stat}$ is scale-dependent.  Recent lattice-QCD
calculations using static $b$ quarks and dynamical light
quarks \cite{Albertus:2010nm,Dimopoulos:2011gx} perform the operator
matching at one-loop in mean-field improved lattice perturbation
theory~\cite{Ishikawa:2011dd,Blossier:2011dg}.  Therefore the
heavy-quark discretization, truncation, and matching errors in these
results are of ${\cO}(a^2 \Lambda_{\rm QCD}^2)$, ${\cO}
(\Lambda_{\rm QCD}/m_h)$, and ${\cO}(\alpha_s^2, \alpha_s^2
a \Lambda_{\rm QCD})$.

In order to reduce heavy-quark truncation errors in $B$-meson masses
and matrix elements to the few-percent level, state-of-the-art
lattice-HQET computations now include corrections of ${\cO}(1/m_h)$.  Adding the $1/m_{h}$ terms, the HQET Lagrangian reads
\begin{eqnarray}
{\mathcal L}^{\rm HQET}(x) &=&  {\mathcal L}^{\rm stat}(x) - \omegakin{\mathcal{O}}_{\rm kin}(x)
        - \omegaspin{\mathcal{O}}_{\rm spin}(x)  \,, \\[2.0ex]
  \mathcal{O}_{\rm kin}(x) &=& \overline{\psi}_{h}(x){\bf D}^2\psi_{h}(x) \,,\quad
  \mathcal{O}_{\rm spin}(x) = \overline{\psi}_{h}(x){\boldsymbol\sigma}\!\cdot\!{\bf B}\psi_{h}(x)\,.
\end{eqnarray}
At this order, two other parameters appear in the Lagrangian,
$\omegakin$ and $\omegaspin$. The normalization is such that the
tree-level values of the coefficients are
$\omegakin=\omegaspin=1/(2m_{h})$.  Similarly the operators are
formally expanded in inverse powers of the heavy-quark mass.  The time
component of the axial current, relevant for the computation of
mesonic decay constants is given by
\begin{eqnarray}
A_0^{\rm HQET}(x) &=& Z_{\rm A}^{\rm HQET}\left(A_0^{\rm stat}(x) +\sum_{i=1}^2 c_{\rm A}^{(i)} A_0^{(i)}(x)\right)\;, \\
A_0^{(1)}(x)&=&\overline{\psi}\frac{1}{2}\gamma_5 \gamma_k  (\nabla_k-\overleftarrow{\nabla}_k)\psi_h(x), \qquad k=1,2,3\\
A_0^{(2)} &=& -\partial_kA_k^{\rm stat}(x)\;, \quad A_k^{\rm stat}=\overline{\psi}(x) \gamma_k\gamma_5\psi_h(x)\;,
\end{eqnarray}
and depends on two additional parameters $c_{\rm A}^{(1)}$ and $c_{\rm A}^{(2)}$.

A framework for nonperturbative HQET on the lattice has been
introduced in Refs.~\cite{Heitger:2003nj,Blossier:2010jk}.  As pointed out
in Refs.~\cite{Sommer:2006sj,DellaMorte:2007ny}, since $\alpha_s(m_h)$
decreases logarithmically with $m_h$, whereas corrections in the
effective theory are power-like in $\Lambda/m_h$, it is possible that
the leading errors in a calculation will be due to the perturbative
matching of the action and the currents at a given order
$(\Lambda/m_h)^l$ rather than to the missing ${\cO}((\Lambda/m_h)^{l+1})$ terms.  Thus, in order to keep matching
errors below the uncertainty due to truncating the HQET expansion, the
matching is performed nonperturbatively beyond leading order in
$1/m_{h}$. The asymptotic convergence of HQET in the limit
$m_h \to \infty$ indeed holds only in that case.

The higher dimensional interaction terms in the effective Lagrangian
are treated as space-time volume insertions into static correlation
functions.  For correlators of some multi-local fields ${\oO}$
and up to the $1/m_h$ corrections to the operator, this means
\begin{equation}
\langle {\oO} \rangle =\langle {\oO} \rangle_{\rm stat} +\omegakin a^4 \sum_x
\langle {\oO\mathcal{O}}_{\rm kin}(x) \rangle_{\rm stat} + \omegaspin a^4 \sum_x
\langle {\oO\mathcal{O}}_{\rm spin}(x) \rangle_{\rm stat} \;, 
\end{equation}
where $\langle {\oO} \rangle_{\rm stat}$ denotes the static
expectation value with ${\mathcal{L}}^{\rm stat}(x)
+{\mathcal{L}}^{\rm light}(x)$.  Nonperturbative renormalization of
these correlators guarantees the existence of a well-defined continuum
limit to any order in $1/m_h$.  The parameters of the effective action
and operators are then determined by matching a suitable number of
observables calculated in HQET (to a given order in $1/m_{h}$) and in
QCD in a small volume (typically with $L\simeq 0.5$ fm), where the
full relativistic dynamics of the $b$-quark can be simulated and the
parameters can be computed with good accuracy.
In Refs.~\cite{Blossier:2010jk,Blossier:2012qu} the Schr\"odinger Functional
(SF) setup has been adopted to define a set of quantities, given by
the small volume equivalent of decay constants, pseudoscalar-vector
splittings, effective masses and ratio of correlation functions for
different kinematics, that can be used to implement the matching
conditions.  The kinematical conditions are usually modified by
changing the periodicity in space of the fermions, i.e. by directly
exploiting a finite-volume effect.  The new scale $L$, which is
introduced in this way, is chosen such that higher orders in $1/m_hL$
and in $\Lambda_{\rm QCD}/m_h$ are of about the same size. At the end
of the matching step the parameters are known at lattice spacings
which are of the order of $0.01$ fm, significantly smaller than the
resolutions used for large volume, phenomenological, applications. For
this reason a set of SF-step scaling functions is introduced in the
effective theory to evolve the parameters to larger lattice spacings.
The whole procedure yields the nonperturbative parameters with an
accuracy which allows to compute phenomenological quantities with a
precision of a few percent
(see Refs.~\cite{Blossier:2010mk,Bernardoni:2012ti} for the case of the
$B_{(s)}$ decay constants).  Such an accuracy can not be achieved by
performing the nonperturbative matching in large volume against
experimental measurements, which in addition would reduce the
predictivity of the theory.  For the lattice-HQET action matched
nonperturbatively through ${\cO}(1/m_h)$, discretization and
truncation errors are of ${\cO}(a \Lambda^2_{\rm QCD}/m_h,
a^2 \Lambda^2_{\rm QCD})$ and ${\cO}((\Lambda_{\rm QCD}/m_h )^2)$.

The noise-to-signal ratio of static-light correlation functions grows
exponentially in Euclidean time, $\propto e^{\mu x_0}$ . The rate
$\mu$ is nonuniversal but diverges as $1/a$ as one approaches the
continuum limit. By changing the discretization of the covariant
derivative in the static action one may achieve an exponential
reduction of the noise to signal ratio. Such a strategy led to the
introduction of the $S^{\rm stat}_{\rm HYP1,2}$
actions~\cite{DellaMorte:2005yc}, where the thin links in $D_0$ are
replaced by HYP-smeared links~\cite{Hasenfratz:2001hp}.  These actions
are now used in all lattice applications of HQET.
\\

%=================================================
%Nonrelativistic QCD
%=================================================

\noindent
{\it Nonrelativistic QCD}\\
\noindent

Nonrelativistic QCD (NRQCD) \cite{Thacker:1990bm,Lepage:1992tx} is an
 effective theory that can be matched to full QCD order by order in
 the heavy-quark velocity $v_h^2$ (for heavy-heavy systems) or in
 $\Lambda_{\rm QCD}/m_h$ (for heavy-light systems) and in powers of
 $\alpha_s$.  Relativistic corrections appear as higher-dimensional
 operators in the Hamiltonian.
 
 As an effective field theory, NRQCD is only useful with an
 ultraviolet cutoff of order $m_h$ or less. On the lattice this means
 that it can be used only for $am_h>1$, which means that $\cO(a^n)$
 errors cannot be removed by taking $a\to0$ at fixed $m_h$. Instead
 heavy-quark discretization errors are systematically removed by
 adding additional operators to the lattice Hamiltonian.  Thus, while
 strictly speaking no continuum limit exists at fixed $m_h$, continuum
 physics can be obtained at finite lattice spacing to arbitrarily high
 precision provided enough terms are included, and provided that the
 coefficients of these terms are calculated with sufficient accuracy.
 Residual discretization errors can be parameterized as corrections to
 the coefficients in the nonrelativistic expansion, as shown in
 Eq.~(\ref{deltaH}).  Typically they are of the form
 $(a|\vec{p}_h|)^n$ multiplied by a function of $am_h$ that is smooth
 over the limited range of heavy-quark masses (with $am_h > 1$) used
 in simulations, and can therefore can be represented by a low-order
 polynomial in $am_h$ by Taylor's theorem (see
 Ref.~\cite{Gregory:2010gm} for further discussion).  Power-counting
 estimates of these effects can be compared to the observed lattice
 spacing dependence in simulations. Provided that these effects are
 small, such comparisons can be used to estimate and correct the
 residual discretization effects.

An important feature of the NRQCD approach is that the same action can
be applied to both heavy-heavy and heavy-light systems. This allows,
for instance, the bare $b$-quark mass to be fixed via experimental
input from $\Upsilon$ so that simulations carried out in the $B$ or
$B_s$ systems have no adjustable parameters left.  Precision
calculations of the $B_s$-meson mass (or of the mass splitting
$M_{B_s} - M_\Upsilon/2$) can then be used to test the reliability of
the method before turning to quantities one is trying to predict, such
as decay constants $f_B$ and $f_{B_s}$, semileptonic form factors or
neutral $B$ mixing parameters.

Given the same lattice-NRQCD heavy-quark action, simulation results
will not be as accurate for charm quarks as for bottom ($1/m_b <
1/m_c$, and $v_b < v_c$ in heavy-heavy systems).  For charm, however,
a more serious concern is the restriction that $am_h$ must be greater
than one.  This limits lattice-NRQCD simulations at the physical
$am_c$ to relatively coarse lattice spacings for which light-quark and
gluon discretization errors could be large.  Thus recent lattice-NRQCD
simulations have focused on bottom quarks because $am_b > 1$ in the
range of typical lattice spacings between $\approx$ 0.06 and 0.15~fm.

In most simulations with NRQCD $b$-quarks during the past decade one
has worked with an NRQCD action that includes tree-level relativistic
corrections through ${\cO}(v_h^4)$ and discretization corrections
through ${\cO}(a^2)$,
 \begin{eqnarray}
 \label{nrqcdact}
&&  S_{\rm NRQCD}  =
a^4 \sum_x \Bigg\{  {\Psi}^\dagger_t \Psi_t -
 {\Psi}^\dagger_t
\left(1 \!-\!\frac{a \delta H}{2}\right)_t
 \left(1\!-\!\frac{aH_0}{2n}\right)^{n}_t \nonumber \\
& \times &
 U^\dagger_t(t-a)
 \left(1\!-\!\frac{aH_0}{2n}\right)^{n}_{t-a}
\left(1\!-\!\frac{a\delta H}{2}\right)_{t-a} \Psi_{t-a} \Bigg\} \, ,
 \end{eqnarray}
where the subscripts $``t"$ and $``t-a"$ denote that the heavy-quark, gauge, $\bf{E}$,  and $\bf{B}$-fields are on time slices $t$ or $t-a$, respectively.
 $H_0$ is the nonrelativistic kinetic energy operator,
 \be
 H_0 = - {\delsq\over2m_h} \, ,
 \ee
and $\delta H$ includes relativistic and finite-lattice-spacing
corrections,
 \begin{eqnarray}
\delta H
&=& - c_1\,\frac{(\delsq)^2}{8m_h^3}
+ c_2\,\frac{i g}{8m_h^2}\left(\delv\cdot\Ev - \Ev\cdot\delv\right) \nl
& &
 - c_3\,\frac{g}{8m_h^2} \sigmav\cdot(\delvt\times\Ev - \Ev\times\delvt)\nl
& & - c_4\,\frac{g}{2m_h}\,\sigmav\cdot\Bv
  + c_5\,\frac{a^2\delfour}{24m_h}  - c_6\,\frac{a(\delsq)^2}
{16nm_h^2} \, .
\label{deltaH}
\end{eqnarray}
 $m_h$ is the bare heavy-quark mass, $\delsq$ the lattice Laplacian,
$\delv$ the symmetric lattice derivative and $\delfour$ the lattice
discretization of the continuum $\sum_i D^4_i$.  $\delvt$ is the
improved symmetric lattice derivative and the $\Ev$ and $\Bv$ fields
have been improved beyond the usual clover leaf construction. The
stability parameter $n$ is discussed in Ref.~\cite{Lepage:1992tx}.  In most
cases the $c_i$'s have been set equal to their tree-level values $c_i
= 1$.  With this implementation of the NRQCD action, errors in
heavy-light-meson masses and splittings are of ${\cO}(\alpha_S \Lambda_{\rm QCD}/m_h )$, ${\cO}(\alpha_S (\Lambda_{\rm
QCD}/m_h)^2 )$, ${\cO}((\Lambda_{\rm QCD}/m_h )^3)$, and ${\cO}(\alpha_s a^2 \Lambda_{\rm QCD}^2)$, with coefficients that are
functions of $am_h$.  One-loop corrections to many of the coefficients
in Eq.~(\ref{deltaH}) have now been calculated, and are starting to be
included in
simulations \cite{Morningstar:1994qe,Hammant:2011bt,Dowdall:2011wh}.

Most of the operator matchings involving heavy-light currents or
four-fermion operators with NRQCD $b$-quarks and AsqTad or HISQ light
quarks have been carried out at one-loop order in lattice perturbation
theory.  In calculations published to date of electroweak matrix
elements, heavy-light currents with massless light quarks have been
matched through ${\cO}(\alpha_s, \Lambda_{\rm QCD}/m_h, \alpha_s/(a
m_h),
\alpha_s \Lambda_{\rm QCD}/m_h)$, and four-fermion operators through \\
 ${\cO}(\alpha_s, \Lambda_{\rm QCD}/m_h, 
\alpha_s/(a m_h))$.
NRQCD/HISQ currents with massive HISQ quarks are also of interest,
e.g.  for the bottom-charm currents in $B \rightarrow D^{(*)}, l \nu$
semileptonic decays and the relevant matching calculations have been
performed at one-loop order in Ref.~\cite{Monahan:2012dq}.  Taking all
the above into account, the most significant systematic error in
electroweak matrix elements published to date with NRQCD $b$-quarks is
the ${\cO}(\alpha_s^2)$ perturbative matching uncertainty.  Work is
therefore underway to use current-current correlator methods combined
with very high order continuum perturbation theory to do current
matchings nonperturbatively~\cite{Koponen:2010jy}.
\\

%=================================================
%Relativistic heavy quarks
%=================================================

\noindent
{\it Relativistic heavy quarks}\\
\noindent

An approach for relativistic heavy-quark lattice formulations was
first introduced by El-Khadra, Kronfeld, and Mackenzie in
Ref.~\cite{ElKhadra:1996mp}.  Here they showed that, for a general
lattice action with massive quarks and non-Abelian gauge fields,
discretization errors can be factorized into the form $f(m_h
a)(a|\vec{p}_h|)^n$, and that the function $f(m_h a)$ is bounded to be
of ${\cO}(1)$ or less for all values of the quark mass $m_h$.
Therefore cutoff effects are of ${\cO}(a \Lambda_{\rm QCD})^n$
and ${\cO}((a|\vec{p}_h|)^n)$, even for $am_h \gtapprox 1$, and
can be controlled using a Symanzik-like procedure.  As in the standard
Symanzik improvement program, cutoff effects are systematically
removed by introducing higher-dimension operators to the lattice
action and suitably tuning their coefficients.  In the relativistic
heavy-quark approach, however, the operator coefficients are allowed
to depend explicitly on the quark mass.  By including lattice
operators through dimension $n$ and adjusting their coefficients
$c_{n,i}(m_h a)$ correctly, one enforces that matrix elements in the
lattice theory are equal to the analogous matrix elements in continuum
QCD through $(a|\vec{p}_h|)^n$, such that residual heavy-quark
discretization errors are of ${\cO}(a|\vec{p}_h|)^{n+1}$.

The relativistic heavy-quark approach can be used to compute the
matrix elements of states containing heavy quarks for which the
heavy-quark spatial momentum $|\vec{p}_h|$ is small compared to the
lattice spacing.  Thus it is suitable to describe bottom and charm
quarks in both heavy-light and heavy-heavy systems.  Calculations of
bottomonium and charmonium spectra serve as nontrivial tests of the
method and its accuracy.

At fixed lattice spacing, relativistic heavy-quark formulations
recover the massless limit when $(am_h) \ll 1$, recover the static
limit when $(am_h) \gg 1$, and smoothy interpolate between the two;
thus they can be used for any value of the quark mass, and, in
particular, for both charm and bottom.  Discretization errors for
relativistic heavy-quark formulations are generically of the form
$\alpha_s^k f(am_h)(a |\vec{p}_h|)^n$, where $k$ reflects the order of
the perturbative matching for operators of ${\cO}((a
|\vec{p}_h|)^n)$.  For each $n$, such errors are removed completely if
the operator matching is nonperturbative. When $(am_h) \sim 1$, this
gives rise to nontrivial lattice-spacing dependence in physical
quantities, and it is prudent to compare estimates based on
power-counting with a direct study of scaling behaviour using a range
of lattice spacings.
At fixed quark mass, relativistic heavy-quark actions possess a smooth
continuum limit without power-divergences.  Of course, as
$m_h \to \infty$ at fixed lattice spacing, the power divergences of
the static limit are recovered (see, e.g. Ref.~\cite{Harada:2001fi}).

The relativistic heavy-quark formulations in use all begin with the
anisotropic Sheikholeslami-Wohlert (``clover")
action~\cite{Sheikholeslami:1985ij}:
\begin{equation}
S_\textrm{lat} = a^4 \sum_{x,x'} \bar{\psi}(x') \left( m_0 + \gamma_0 D_0 + \zeta \vec{\gamma} \cdot \vec{D} - \frac{a}{2} (D^0)^2 - \frac{a}{2} \zeta (\vec{D})^2+ \sum_{\mu,\nu} \frac{ia}{4} c_{\rm SW} \sigma_{\mu\nu} F_{\mu\nu} \right)_{x' x} \psi(x) \,,
\label{eq:HQAct}
\end{equation}
where $D_\mu$ is the lattice covariant derivative and $F_{\mu\nu}$ is
the lattice field-strength tensor.  Here we show the form of the
action given in Ref.~\cite{Christ:2006us}.  The introduction of a
space-time anisotropy, parameterized by $\zeta$ in
Eq.~(\ref{eq:HQAct}), is convenient for heavy-quark systems because
the characteristic heavy-quark four-momenta do not respect space-time
axis exchange ($\vec{p}_h < m_h$ in the bound-state rest frame).
Further, the Sheikoleslami-Wohlert action respects the continuum
heavy-quark spin and flavour symmetries, so HQET can be used to
interpret and estimate lattice discretization
effects~\cite{Kronfeld:2000ck,Harada:2001fi,Harada:2001fj}.  We
discuss three different prescriptions for tuning the parameters of the
action in common use below.  In particular, we focus on aspects of the
action and operator improvement and matching relevant for evaluating
the quality of the calculations discussed in the main text.

The meson energy-momentum dispersion relation plays an important role
in relativistic heavy-quark formulations:
\begin{equation}
	E(\vec{p}) = M_1 + \frac{\vec{p}^2}{2M_2} + {\cO}(\vec{p}^4) \,,
\end{equation}
where $M_1$ and $M_2$ are known as the rest and kinetic masses,
respectively.  Because the lattice breaks Lorentz invariance, there
are corrections proportional to powers of the momentum.  Further, the
lattice rest masses and kinetic masses are not equal ($M_1 \neq M_2$),
and only become equal in the continuum limit.

The Fermilab interpretation~\cite{ElKhadra:1996mp} is suitable for
calculations of mass splittings and matrix elements of systems with
heavy quarks.  The Fermilab action is based on the hopping-parameter
form of the Wilson action, in which $\kappa_h$ parameterizes the
heavy-quark mass.  In practice, $\kappa_h$ is tuned such that the the
kinetic meson mass equals the experimentally-measured heavy-strange
meson mass ($m_{B_s}$ for bottom and $m_{D_s}$ for charm).  In
principle, one could also tune the anisotropy parameter such that $M_1
= M_2$.  This is not necessary, however, to obtain mass splittings and
matrix elements, which are not affected by
$M_1$~\cite{Kronfeld:2000ck}.  Therefore in the Fermilab action the
anisotropy parameter is set equal to unity.  The clover coefficient in
the Fermilab action is fixed to the value $c_{\rm SW} = 1/u_0^3$ from
mean-field improved lattice perturbation theory~\cite{Lepage:1992xa}.
With this prescription, discretization effects are of ${\cO}(\alpha_sa|\vec{p}_h|, (a|\vec{p}_h|)^2)$.  Calculations of
electroweak matrix elements also require improving the lattice current
and four-fermion operators to the same order, and matching them to the
continuum.  Calculations with the Fermilab action remove tree-level
${\cO}(a)$ errors in electroweak operators by rotating the
heavy-quark field used in the matrix element and setting the rotation
coefficient to its tadpole-improved tree-level value (see e.g.
Eqs.~(7.8) and (7.10) of Ref.~\cite{ElKhadra:1996mp}).  Finally,
electroweak operators are typically renormalized using a mostly
nonperturbative approach in which the flavour-conserving light-light
and heavy-heavy current renormalization factors $Z_V^{ll}$ and
$Z_V^{hh}$ are computed nonperturbatively~\cite{ElKhadra:2001rv}.  The
flavour-conserving factors account for most of the heavy-light current
renormalization.  The remaining correction is expected to be close to
unity due to the cancellation of most of the radiative corrections
including tadpole graphs~\cite{Harada:2001fi}; therefore it can be
reliably computed at one-loop in mean-field improved lattice
perturbation theory with truncation errors at the percent to
few-percent level.

The relativistic heavy-quark (RHQ) formulation developed by Li, Lin,
and Christ builds upon the Fermilab approach, but tunes all the
parameters of the action in Eq.~(\ref{eq:HQAct})
nonperturbatively~\cite{Christ:2006us}.  In practice, the three
parameters $\{m_0a, c_{\rm SW}, \zeta\}$ are fixed to reproduce the
experimentally-measured $B_s$ meson mass and hyperfine splitting
($m_{B_s^*}-m_{B_s}$), and to make the kinetic and rest masses of the
lattice $B_s$ meson equal~\cite{Aoki:2012xaa}.  This is done by
computing the heavy-strange meson mass, hyperfine splitting, and ratio
$M_1/M_2$ for several sets of bare parameters $\{m_0a, c_{\rm
SW}, \zeta\}$ and interpolating linearly to the physical $B_s$ point.
By fixing the $B_s$-meson hyperfine splitting, one loses a potential
experimental prediction with respect to the Fermilab formulation.
However, by requiring that $M_1 = M_2$, one gains the ability to use
the meson rest masses, which are generally more precise than the
kinetic masses, in the RHQ approach.  The nonperturbative
parameter-tuning procedure eliminates ${\cO}(a)$ errors from
the RHQ action, such that discretization errors are of ${\cO}((a|\vec{p}_h|)^2)$.  Calculations of $B$-meson decay constants and
semileptonic form factors with the RHQ action are in
progress~\cite{Witzel:2012pr,Kawanai:2012id}, as is the corresponding
one-loop mean-field improved lattice perturbation
theory~\cite{Lehner:2012bt}.  For these works, cutoff effects in the
electroweak vector and axial-vector currents will be removed through
${\cO}(\alpha_s a)$, such that the remaining discretization
errors are of ${\cO}(\alpha_s^2a|\vec{p}_h|,
(a|\vec{p}_h|)^2)$.  Matching the lattice operators to the continuum
will be done following the mostly nonperturbative approach described
above.

The Tsukuba heavy-quark action is also based on the
Sheikholeslami-Wohlert action in Eq.~(\ref{eq:HQAct}), but allows for
further anisotropies and hence has additional parameters: specifically
the clover coefficients in the spatial $(c_B)$ and temporal $(c_E)$
directions differ, as do the anisotropy coefficients of the $\vec{D}$
and $\vec{D}^2$ operators~\cite{Aoki:2001ra}.  In practice, the
contribution to the clover coefficient in the massless limit is
computed nonperturbatively~\cite{Aoki:2005et}, while the
mass-dependent contributions, which differ for $c_B$ and $c_E$, are
calculated at one-loop in mean-field improved lattice perturbation
theory~\cite{Aoki:2003dg}.  The hopping parameter is fixed
nonperturbatively to reproduce the experimentally-measured
spin-averaged $1S$ charmonium mass~\cite{Namekawa:2011wt}.  One of the
anisotropy parameters ($r_t$ in Ref.~\cite{Namekawa:2011wt}) is also
set to its one-loop perturbative value, while the other ($\nu$ in
Ref.~\cite{Namekawa:2011wt}) is fixed noperturbatively to obtain the
continuum dispersion relation for the spin-averaged charmonium $1S$
states (such that $M_1 = M_2$).  For the renormalization and
improvement coefficients of weak current operators, the contributions
in the chiral limit are obtained
nonperturbatively~\cite{Kaneko:2007wh,Aoki:2010wm}, while the
mass-dependent contributions are estimated using one-loop lattice
perturbation theory~\cite{Aoki:2004th}.  With these choices, lattice
cutoff effects from the action and operators are of ${\cO}(\alpha_s^2 a|\vec{p}|, (a|\vec{p}_h|)^2)$.
\\

%=================================================
%Light quark actions + HQET
%=================================================

\noindent
{\it Light-quark actions combined with HQET}\\
\noindent

The heavy-quark formulations discussed in the previous sections use
effective field theory to avoid the occurence of discretization errors
of the form $(am_h)^n$.  In this section we describe methods that use
improved actions that were originally designed for light-quark systems
for $B$ physics calculations. Such actions unavoidably contain
discretization errors that grow as a power of the heavy-quark mass. In
order to use them for heavy-quark physics, they must be improved to at
least ${\cO}(am_h)^2$.  However, since $am_b > 1$ at the smallest
lattice spacings available in current simulations, these methods also
require input from HQET to guide the simulation results to the
physical $b$-quark mass.

The ETM collaboration has developed two methods, the ``ratio
method'' \cite{Blossier:2009hg} and the ``interpolation
method'' \cite{Guazzini:2006bn,Blossier:2009gd}. They use these
methods together with simulations with twisted-mass Wilson fermions,
which have discretization errors of $\cO(am_h)^2$.  In the interpolation
method $\Phi_{hs}$ and $\Phi_{h\ell}$ (or $\Phi_{hs}/\Phi_{h\ell}$)
are calculated for a range of heavy-quark masses in the charm region
and above, while roughly keeping $am_h \ltsim 0.5 $. The relativistic
results are combined with a separate calculation of the decay
constants in the static limit, and then interpolated to the physical
$b$ quark mass. In ETM's implementation of this method, the heavy
Wilson decay constants are matched to HQET using NLO in continuum
perturbation theory. The static limit result is renormalized using
one-loop mean-field improved lattice perturbation theory, while for
the relativistic data PCAC is used to calculate absolutely normalized
matrix elements. Both, the relativistic and static limit data are then
run to the common reference scale $\mu_b = 4.5 \GeV$ at NLO in
continuum perturbation theory.  In the ratio method, one constructs
physical quantities $P(m_h)$ from the relativistic data that have a
well-defined static limit ($P(m_h) \to$ const.~for $m_h \to \infty$)
and evaluates them at the heavy-quark masses used in the simulations.
Ratios of these quantities are then formed at a fixed ratio of heavy
quark masses, $z = P(m_h) / P(m_h/\lambda)$ (where $1 < \lambda \lsim
1.3$), which ensures that $z$ is equal to unity in the static limit.
Hence, a separate static limit calculation is not needed with this
method.  In ETM's implementation of the ratio method for the $B$-meson
decay constant, $P(m_h)$ is constructed from the decay constants and
the heavy-quark pole mass as $P(m_h) = f_{h\ell}(m_h) \cdot (m^{\rm
pole}_h)^{1/2}$. The corresponding $z$-ratio therefore also includes
ratios of perturbative matching factors for the pole mass to $\msbar$
conversion.  For the interpolation to the physical $b$-quark mass,
ratios of perturbative matching factors converting the data from QCD
to HQET are also included. The QCD-to-HQET matching factors improve
the approach to the static limit by removing the leading logarithmic
corrections. In ETM's implementation of this method (ETM 11 and 12)
both conversion factors are evaluated at NLO in continuum perturbation
theory. The ratios are then simply fit to a polynomial in $1/m_h$ and
interpolated to the physical $b$-quark mass.  The ratios constructed
from $f_{h\ell}$ ($f_{hs}$) are called $z$ ($z_s$).  In order to
obtain the $B$ meson decay constants, the ratios are combined with
relativistic decay constant data evaluated at the smallest reference
mass.

The HPQCD collaboration has introduced a method in
Ref.~\cite{McNeile:2011ng} which we shall refer to as the ``heavy
HISQ'' method.  The first key ingredient is the use of the HISQ action
for the heavy and light valence quarks, which has leading
discretization errors of ${\cO} \left(\alpha_s (v/c) (am_h)^2,
(v/c)^2 (am_h)^4\right)$.  With the same action for the heavy and
light valence quarks it is possible to use PCAC to avoid
renormalization uncertainties.  Another key ingredient is the
availability of gauge ensembles over a large range of lattice
spacings, in this case in the form of the library of $N_f = 2+1$
asqtad ensembles made public by the MILC collaboration which includes
lattice spacings as small as $a \approx 0.045$~fm.  Since the HISQ
action is so highly improved and with lattice spacings as small as
$0.045$~fm, HPQCD is able to use a large range of heavy-quark masses,
from below the charm region to almost up to the physical $b$ quark
mass with $am_h \ltsim 0.85$. They then fit their data in a combined
continuum and HQET fit (i.e. using a fit function that is motivated by
HQET) to a polynomial in $1/m_H$ (the heavy pseudo scalar meson mass
of a meson containing a heavy ($h$) quark).

\bigskip

In Table~\ref{tab_heavy_quarkactions} we list the discretizations of
the quark action most widely used for heavy $c$ and $b$ quarks
together with the abbreviations used in the summary tables.  We also
summarize the main properties of these actions and the leading lattice
discretization errors for calculations of heavy-light meson matrix
quantities with them.  Note that in order to maintain the leading
lattice artifacts of the actions as given in the table in nonspectral
observables (like operator matrix elements) the corresponding
nonspectral operators need to be improved as well.

\begin{table}
\begin{center}
{\footnotesize
\begin{tabular*}{\textwidth}[l]{l @{\extracolsep{\fill}} l l l}
\hline \hline  \\[-1.0ex]
\parbox[t]{1.5cm}{Abbrev.} & Discretization & 
\parbox[t]{4cm}{Leading lattice artifacts\\and truncation errors\\for heavy-light mesons} &  
Remarks
\\[7.0ex] \hline \hline \\[-1.0ex]
tmWil   & twisted-mass Wilson &  ${\cO}\big((am_h)^2\big)$ & \parbox[t]{4.cm}{PCAC relation for axial-vector current}  
\\[3.0ex] \hline \\[-1.0ex]
HISQ  & Staggered & \parbox[t]{4cm}{${\cO}\big (\alpha_S (am_h)^2 (v/c), \\(am_h)^4 (v/c)^2 \big)$}  & \parbox[t]{4.cm}{PCAC relation for axial-vector current; Ward identity for vector current}  
\\[6.0ex] \hline \\[-1.0ex]
static  & static effective action &  \parbox[t]{4cm}{${\cO}\big( a^2 \Lambda_{\rm QCD}^2, \Lambda_{\rm QCD}/m_h, \\ \alpha_s^2, \alpha_s^2 a \Lambda_{\rm QCD} \big)$}  & \parbox[t]{4.5cm}{implementations use APE, HYP1, and HYP2 smearing}  
\\[4.0ex] \hline \\[-1.0ex]
HQET  & Heavy-Quark Effective Theory &  \parbox[t]{4cm}{${\cO}\big( a \Lambda^2_{\rm QCD}/m_h,  a^2 \Lambda^2_{\rm QCD},\\
 (\Lambda_{\rm QCD}/m_h)^2 \big)$}  & \parbox[t]{4.5cm}{Nonperturbative matching through ${\cO}(1/m_h)$}  
\\[4.0ex] \hline \\[-1.0ex]
NRQCD  & Nonrelativistic QCD & \parbox[t]{4cm}{${\cO}\big(\alpha_S \Lambda_{\rm QCD}/m_h, \\ \alpha_S (\Lambda_{\rm QCD}/m_h)^2 , \\ (\Lambda_{\rm QCD}/m_h )^3,  \alpha_s a^2 \Lambda_{\rm QCD}^2 \big)$}  & \parbox[t]{4.5cm}{Tree-level relativistic corrections through 
${\cO}(v_h^4)$ and discretization corrections through ${\cO}(a^2)$}  
\\[9.5ex] \hline \\[-1.0ex] 
Fermilab  & Sheikholeslami-Wohlert & ${\cO}\big(\alpha_sa\Lambda_{\rm QCD}, (a\Lambda_{\rm QCD})^2\big)$  & \parbox[t]{4.5cm}{Hopping parameter tuned nonperturbatively; clover coefficient computed at tree-level in mean-field-improved lattice perturbation theory}  
\\[12.0ex] \hline \\[-1.0ex] 

RHQ       & Sheikholeslami-Wohlert & ${\cO}\big( \alpha_s^2 a\Lambda_{\rm QCD}, (a\Lambda_{\rm QCD})^2 \big)$  & \parbox[t]{4.5cm}{Hopping parameter, anisoptropy and clover coefficient tuned nonperturbatively by fixing the $B_s$-meson hyperfine splitting} \\[12.0ex] \hline \\[-1.0ex] 

Tsukuba  & Sheikholeslami-Wohlert & ${\cO}\big( \alpha_s^2 a\Lambda_{\rm QCD}, (a\Lambda_{\rm QCD})^2 \big)$  & \parbox[t]{4.5cm}{NP clover coefficient at $ma=0$ plus mass-dependent corrections calculated at one-loop in lattice perturbation theory; $\nu$ calculated NP from dispersion relation; $r_s$ calculated at one-loop in lattice perturbation theory}  
\\[20.0ex]
\hline\hline
\end{tabular*}
}
\caption{Discretizations of the quark action most widely used for heavy $c$ and $b$ quarks  and some of their properties.
\label{tab_heavy_quarkactions}}
\end{center}
\end{table}

%\clearpage
\subsection{Setting the scale \label{sec_scale}}

In simulations of lattice QCD quantities such as hadron masses and
decay constants are obtained in ``lattice units'' i.e.~as
dimensionless numbers. In order to convert them into physical units
they must be expressed in terms of some experimentally known,
dimensionful reference quantity $Q$. This procedure is called
``setting the scale''. It amounts to computing the nonperturbative
relation between the bare gauge coupling $g_0$ (which is an input
parameter in any lattice simulation) and the lattice spacing~$a$
expressed in physical units. To this end one chooses a value for $g_0$
and computes the value of the reference quantity in a simulation: This
yields the dimensionless combination, $(aQ)|_{g_0}$, at the chosen
value of $g_0$. The calibration of the lattice spacing is then
achieved via
\be
 a^{-1}\,[{\rm MeV}] = \frac{Q|_{\rm{exp}}\,[{\rm MeV}]}{(aQ)|_{g_0}},
\ee
where $Q|_{\rm{exp}}$ denotes the experimentally known value of the
reference quantity. Common choices for $Q$ are the mass of the
nucleon, the $\Omega$ baryon or the decay constants of the pion and
the kaon. Vector mesons, such as the $\rho$ or $K^\ast$-meson, are
unstable and therefore their masses are not very well suited for
setting the scale, despite the fact that they have been used over many
years for that purpose.

Another widely used quantity to set the scale is the hadronic radius
$r_0$, which can be determined from the force between static quarks
via the relation \cite{Sommer:1993ce}
\be
   F(r_0)r_0^2 = 1.65.
\ee
If the force is derived from potential models describing heavy
quarkonia, the above relation determines the value of $r_0$ as
$r_0\approx0.5$\,fm. A variant of this procedure is obtained
\cite{Bernard:2000gd} by using the definition $F(r_1)r_1^2=1.00$,
which yields $r_1\approx0.32$\,fm. It is important to realize that
both $r_0$ and $r_1$ are not directly accessible in experiment, so
that their values derived from phenomenological potentials are
necessarily model-dependent. Inspite of the inherent ambiguity
whenever hadronic radii are used to calibrate the lattice spacing,
they are very useful quantities for performing scaling tests and
continuum extrapolations of lattice data. Furthermore, they can be
easily computed with good statistical accuracy in lattice simulations.

\subsection{Matching and running \label{sec_match}}

The lattice formulation of QCD amounts to introducing a particular
regularization scheme. Thus, in order to be useful for phenomenology,
hadronic matrix elements computed in lattice simulations must be
related to some continuum reference scheme, such as the
$\msbar$-scheme of dimensional regularization. The matching to the
continuum scheme usually involves running to some reference scale
using the renormalization group. 

In principle, the matching factors which relate lattice matrix
elements to the $\msbar$-scheme, can be computed in perturbation
theory formulated in terms of the bare coupling. It has been known for
a long time, though, that the perturbative expansion is not under good 
control. Several techniques have been developed which allow for a
nonperturbative matching between lattice regularization and continuum
schemes, and are briefly introduced here.\\

%\newpage

\noindent
{\sl Regularization-independent Momentum Subtraction}\\
\noindent

In the {\sl Regularization-independent Momentum Subtraction}
(``RI/MOM'' or ``RI'') scheme \cite{Martinelli:1994ty} a
nonperturbative renormalization condition is formulated in terms of
Green functions involving quark states in a fixed gauge (usually
Landau gauge) at nonzero virtuality. In this way one relates
operators in lattice regularization nonperturbatively to the RI
scheme. In a second step one matches the operator in the RI scheme to
its counterpart in the $\msbar$-scheme. The advantage of this
procedure is that the latter relation involves perturbation theory
formulated in the continuum theory. The uncontrolled use of lattice
perturbation theory can thus be avoided. A technical complication is
associated with the accessible momentum scales (i.e. virtualities),
which must be large enough (typically several $\gev$) in order for the
perturbative relation to $\msbar$ to be reliable. The momentum scales
in simulations must stay well below the cutoff scale (i.e. $2\pi$ over
the lattice spacing), since otherwise large lattice artifacts are
incurred. Thus, the applicability of the RI scheme traditionally relies on the
existence of a ``window'' of momentum scales, which satisfy
\be
   \Lambda_{\rm QCD} \;\lesssim\; p \;\lesssim\; 2\pi a^{-1}.
\ee
However, solutions for mitigating this limitation, which involve
continuum limit, nonperturbative running to higher scales in the
RI/MOM scheme, have recently been proposed and implemented
\cite{Arthur:2010ht,Durr:2010vn,Durr:2010aw,Aoki:2010pe}.\\

\noindent
{\it Schr\"odinger functional}\\
\noindent

Another example of a nonperturbative matching procedure is provided
by the Schr\"odinger functional (SF) scheme \cite{Luscher:1992an}. It
is based on the formulation of QCD in a finite volume. If all quark
masses are set to zero the box length remains the only scale in the
theory, such that observables like the coupling constant run with the
box size~$L$. The great advantage is that the RG running of
scale-dependent quantities can be computed nonperturbatively using
recursive finite-size scaling techniques. It is thus possible to run
nonperturbatively up to scales of, say, $100\,\gev$, where one is
sure that the perturbative relation between the SF and
$\msbar$-schemes is controlled.\\

\noindent
{\sl Perturbation theory}\\
\noindent

The third matching procedure is based on perturbation theory in which
higher order are effectively resummed \cite{Lepage:1992xa}. Although
this procedure is easier to implement, it is hard to estimate the
uncertainty associated with it.\\

\noindent
{\sl Mostly nonperturbative renormalization}\\
\noindent

Some calculations of heavy-light and heavy-heavy matrix elements adopt a mostly nonperturbative matching approach.  Let us consider a weak 
decay process mediated by a current with quark flavours $h$ and $q$, where $h$ is the initial heavy quark (either bottom or charm) and 
$q$ can be a light ($\ell = u,d$), strange, or charm quark. The matrix elements of lattice current  $J_{hq}$ are matched to the 
corresponding continuum matrix elements with continuum current ${\cal J}_{hq}$ by calculating the renormalization factor $Z_{J_{hq}}$. 
The mostly nonperturbative renormalization method takes advantage of rewriting the current renormalization factor as the following product:
\begin{align}
Z_{J_{hq}} = \rho_{J_{hq}} \sqrt{Z_{V^4_{hh}}Z_{V^4_{qq}}} \,
\label{eq:Zvbl}
\end{align}
The flavour-conserving renormalization factors $Z_{V^4_{hh}}$ and $Z_{V^4_{qq}}$ can be obtained nonperturbatively from standard heavy-light 
and light-light meson charge normalization conditions.  $Z_{V^4_{hh}}$ and $Z_{V^4_{qq}}$ account for  the bulk of the renormalization. The remaining 
correction $\rho_{J_{hq}}$ is expected to be close to unity because most of the radiative corrections, including self-energy corrections and 
contributions from tadpole graphs, cancel in the ratio~\cite{Harada:2001fj,Harada:2001fi}.  The one-loop coefficients of $\rho_{J_{hq}}$  have been calculated for
heavy-light and heavy-heavy currents for Fermilab heavy and both (improved) Wilson light \cite{Harada:2001fj,Harada:2001fi} and 
asqtad light  \cite{ElKhadra:2007qe} quarks. In all cases the one-loop coefficients are found to be very small, yielding sub-percent to few percent level corrections.

\bigskip
\noindent
In Table~\ref{tab_match} we list the abbreviations used in the
compilation of results together with a short description.

\begin{table}[ht]
%\begin{center}
{\footnotesize
\begin{tabular*}{\textwidth}[l]{l @{\extracolsep{\fill}} l}
\hline \hline \\[-1.0ex]
Abbrev. & Description
\\[1.0ex] \hline \hline \\[-1.0ex]
RI  &  regularization-independent momentum subtraction scheme 
\\[1.0ex] \hline \\[-1.0ex]
SF  &  Schr\"odinger functional scheme
\\[1.0ex] \hline \\[-1.0ex]
PT1$\ell$ & matching/running computed in perturbation theory at one loop
\\[1.0ex] \hline \\[-1.0ex]
PT2$\ell$ & matching/running computed in perturbation theory at two loops 
\\[1.0ex] \hline \\[-1.0ex]
mNPR & mostly nonperturbative renormalization 
%\\[1.0ex] \hline \\[-1.0ex]
%
\\[1.0ex]
\hline\hline
\end{tabular*}
}
\caption{The most widely used matching and running
  techniques. \label{tab_match} 
}
%\end{center}
\end{table}

\subsection{Chiral extrapolation\label{sec_ChiPT}}
As mentioned in the introduction, Symanzik's framework can be combined 
with Chiral Perturbation Theory. The well-known terms occurring in the
chiral effective Lagrangian are then supplemented by contributions 
proportional to powers of the lattice spacing $a$. The additional terms are 
constrained by the symmetries of the lattice action and therefore 
depend on the specific choice of the discretization. 
The resulting effective theory can be used to analyse the $a$-dependence of 
the various quantities of interest -- provided the quark masses and the momenta
considered are in the range where the truncated chiral perturbation series yields 
an adequate approximation. Understanding the dependence on the lattice spacing 
is of central importance for a controlled extrapolation to the continuum limit.
 
For staggered fermions, this program has first been carried out for a
single staggered flavour (a single staggered field) \cite{Lee:1999zxa}
at $\cO(a^2)$. In the following, this effective theory is denoted by
S{\Ch}PT. It was later generalized to an arbitrary number of flavours
\cite{Aubin:2003mg,Aubin:2003uc}, and to next-to-leading order
\cite{Sharpe:2004is}. The corresponding theory is commonly called
Rooted Staggered chiral perturbation theory and is denoted by
RS{\Ch}PT.

For Wilson fermions, the effective theory has been developed in
\cite{Sharpe:1998xm,Rupak:2002sm,Aoki:2003yv}
and is called W{\Ch}PT, while the theory for Wilson twisted-mass
fermions \cite{Sharpe:2004ny,Aoki:2004ta,Bar:2010jk} is termed tmW{\Ch}PT.

Another important approach is to consider theories in which the
valence and sea quark masses are chosen to be different. These
theories are called {\it partially quenched}. The acronym for the
corresponding chiral effective theory is PQ{\Ch}PT
\cite{Bernard:1993sv,Golterman:1997st,Sharpe:1997by,Sharpe:2000bc}.

Finally, one can also consider theories where the fermion
discretizations used for the sea and the valence quarks are different. The
effective chiral theories for these ``mixed action'' theories are
referred to as MA{\Ch}PT  \cite{Bar:2002nr,Bar:2003mh,Bar:2005tu,Golterman:2005xa,Chen:2006wf,Chen:2007ug,Chen:2009su}.\\

\noindent
{\sl Finite-Volume Regimes of QCD}\\
\noindent

Once QCD with $\Nf$ nondegenerate flavours is regulated both in the UV and
in the IR, there are $3+\Nf$ scales in play: The scale
$\Lambda_\mathrm{QCD}$ that reflects ``dimensional transmutation''
(alternatively, one could use the pion decay constant or the nucleon mass,
in the chiral limit), the inverse lattice spacing $1/a$, the inverse box
size $1/L$, as well as $\Nf$ meson masses (or functions of meson masses)
that are sensitive to the $\Nf$ quark masses, e.g.\ $\Mpi^2$,
$2\Mka^2-\Mpi^2$ and the spin-averaged masses of ${}^1S$ states of
quarkonia.

Ultimately, we are interested in results with the two regulators
removed, i.e.\ physical quantities for which the limits $a \to 0$ and
$L \to \infty$ have been carried out. In both cases there is an
effective field theory (EFT) which guides the extrapolation.  For the
$a \to 0$ limit, this is a version of the Symanzik EFT which depends,
in its details, on the lattice action that is used, as outlined in
Sec.~\ref{sec_lattice_actions}.  The finite-volume effects are
dominated by the lightest particles, the pions.  Therefore, a chiral
EFT, also known as {\Ch}PT, is appropriate to parameterize the
finite-volume effects, i.e.\ the deviation of masses and other
observables, such as matrix elements, in a finite-volume from their
infinite volume, physical values.  Most simulations of
phenomenological interest are carried out in boxes of size $L \gg
1/Mpi$, that is in boxes whose diameter is large compared to the
Compton wavelength that the pion would have, at the given quark mass,
in infinite volume. In this situation the finite-volume corrections
are small, and in many cases the ratio $M_\mathrm{had}(L) /
M_\mathrm{had}$ or $f(L) / f$, where $f$ denotes some generic matrix
element, can be calculated in {\Ch}PT, such that the leading
finite-volume effects can be taken out analytically. In the
terminology of {\Ch}PT this setting is referred to as the $p$-regime,
as the typical contributing momenta $p \sim M_\pi \gg 1/L$.  A
peculiar situation occurs if the condition $L \gg 1/\Mpi$ is violated
(while $L\Lambda_\mathrm{QCD}\gg1$ still holds), in other words if the
quark mass is taken so light that the Compton wavelength that the pion
would have (at the given $m_q$) in infinite volume, is as large or
even larger than the actual box size. Then the pion zero-momentum mode
dominates and needs to be treated separately. While this setup is
unlikely to be useful for standard phenomenological computations, the
low-energy constants of {\Ch}PT can still be calculated, by matching
to a re-ordered version of the chiral series, and following the
details of the reordering such an extreme regime is called the
$\epsilon$- or $\delta$-regime, respectively.  Accordingly, further
particulars of these regimes are discussed in subsection
\ref{sec:chPT} of this report.

\subsection{Summary of simulated lattice actions}
In the following tables we summarize the gauge and quark actions used
in the various calculations with $N_f=2, 2+1$ and $2+1+1$ quark
flavours. The calculations with $N_f=0$ quark flavours mentioned in
Sec.~\ref{sec:alpha_s} all used the Wilson gauge action and are not
listed. Abbreviations are explained in Secs.~\ref{sec_gauge_actions}, \ref{sec_quark_actions} and
\ref{app:HQactions}, and summarized in Tabs.~\ref{tab_gaugeactions},
\ref{tab_quarkactions} and \ref{tab_heavy_quarkactions}.
\hspace{-1.5cm}
\begin{table}[h]
{\footnotesize
\begin{tabular*}{\textwidth}[l]{l @{\extracolsep{\fill}} c c c c}
\hline \hline \\[-1.0ex]
Collab. & Ref. & $\Nf$ & \parbox{1cm}{gauge\\action} & \parbox{1cm}{quark\\action} 
\\[2.0ex] \hline \hline \\[-1.0ex]
ALPHA 01A, 04, 05, 12, 13A & \cite{Bode:2001jv,DellaMorte:2004bc,DellaMorte:2005kg,Fritzsch:2012wq,Lottini:2013rfa} & 2 & Wilson & npSW \\
[2.0ex] \hline \\[-1.0ex]
Aoki 94 & \cite{Aoki:1994pc} & 2 & Wilson &  KS \\
[2.0ex] \hline \\[-1.0ex]
Bernardoni 10 & \cite{Bernardoni:2010nf} & 2 & Wilson & npSW ${}^\dagger$ \\
[2.0ex] \hline \\[-1.0ex]
Bernardoni 11 & \cite{Bernardoni:2011kd} & 2 & Wilson & npSW \\
[2.0ex] \hline \\[-1.0ex]
Brandt 13 & \cite{Brandt:2013dua} & 2 & Wilson & npSW \\
[2.0ex] \hline \\[-1.0ex]
Boucaud 01B & \cite{Boucaud:2001qz} & 2 & Wilson &  Wilson \\
[2.0ex] \hline \\[-1.0ex]
CERN-TOV 06 & \cite{DelDebbio:2006cn} & 2 & Wilson & Wilson/npSW \\
[2.0ex] \hline \\[-1.0ex]
CERN 08 & \cite{Giusti:2008vb} & 2 & Wilson & npSW \\
[2.0ex] \hline \\[-1.0ex]
{CP-PACS 01} & \cite{AliKhan:2001tx} & 2 & Iwasaki & mfSW  \\
[2.0ex] \hline \\[-1.0ex]
Davies 94  & \cite{Davies:1994ei} & 2 & Wilson  & KS \\
[2.0ex] \hline \\[-1.0ex]
D\"urr 11 & \cite{Durr:2011ed} & 2 & Wilson & npSW \\
[2.0ex] \hline \\[-1.0ex]
Engel 14 & \cite{Engel:2014eea} & 2 & Wilson & npSW \\
[2.0ex] 
\hline\hline \\
\end{tabular*}\\[-0.2cm]
\begin{minipage}{\linewidth}
{\footnotesize 
\begin{itemize}
   \item[${}^\dagger$] The calculation uses overlap fermions in the valence quark sector.\\[-5mm]
\end{itemize}
}
\end{minipage}
}
\caption{Summary of simulated lattice actions with $\Nf=2$ quark
  flavours.}
\end{table}

\begin{table}[h]
\addtocounter{table}{-1}
{\footnotesize
\begin{tabular*}{\textwidth}[l]{l @{\extracolsep{\fill}} c c c c}
\hline \hline \\[-1.0ex]
Collab. & Ref. & $\Nf$ & \parbox{1cm}{gauge\\action} & \parbox{1cm}{quark\\action} 
\\[2.0ex] \hline \hline \\[-1.0ex]
\parbox[t]{4.0cm}{ETM 07, 07A, 08, 09, 09A-D, 10B, 10D, 10F, 11C, 12, 13, 13A} & \parbox[t]{2.5cm}{
\cite{Blossier:2007vv,Boucaud:2007uk,Frezzotti:2008dr,Blossier:2009bx,Lubicz:2009ht,Jansen:2009tt,Baron:2009wt,Blossier:2009hg,Blossier:2010cr,Lubicz:2010bv,Blossier:2010ky,Jansen:2011vv,Burger:2012ti,Cichy:2013gja,Herdoiza:2013sla}} & 2 &  tlSym & tmWil \\
[9.0ex] \hline \\[-1.0ex]
{ETM 10A, 12D} & \cite{Constantinou:2010qv,Bertone:2012cu} & 2 &  tlSym & tmWil ${}^*$ \\
[2.0ex] \hline \\[-1.0ex]
ETMC 14D, 15A & \cite{Abdel-Rehim:2014nka,Abdel-Rehim:2015pwa} & 2 &  Iwasaki & tmWil with npSW \\
\end{tabular*}\\[-0.2cm]
\begin{minipage}{\linewidth}
{\footnotesize 
\begin{itemize}
   \item[${}^*$] The calculation uses Osterwalder-Seiler fermions \cite{Osterwalder:1977pc} in the valence quark sector.
\end{itemize}
}
\end{minipage}
}
\caption{(cntd.) Summary of simulated lattice actions with $\Nf=2$ quark
  flavours.}
\end{table}
%
%
%%%%%%%%%%%%%%%%%%%%%%%%%%%%%%%%%%%%%%%%%%%%%%%%%%%%%%%%%%
\begin{table}[h]
{\footnotesize
\begin{tabular*}{\textwidth}[l]{l @{\extracolsep{\fill}} c c c c}
\hline \hline \\[-1.0ex]
Collab. & Ref. & $\Nf$ & \parbox{1cm}{gauge\\action} & \parbox{1cm}{quark\\action} 
\\[2.0ex] \hline \hline \\[-1.0ex]
Aubin 08, 09 & \cite{Aubin:2008ie,Aubin:2009jh} & $2+1$ & tadSym & Asqtad ${}^\dagger$\\
[2.0ex] \hline \\[-1.0ex]
Blum 10 & \cite{Blum:2010ym} & $2+1$ & Iwasaki & DW\\
[2.0ex] \hline \\[-1.0ex]
%
%BMW 13 & \cite{Durr:2013goa} & $2+1$ &  tlSym & $\cO(a)$-improved Wilson \\
%[2.0ex] \hline \\[-1.0ex]
%
BMW 10A-C, 11, 13 & \cite{Durr:2010vn,Durr:2010aw,Portelli:2010yn,Durr:2011ap,Durr:2013goa} & $2+1$ &  tlSym & 2-level HEX tlSW \\
\end{tabular*}\\[-0.2cm]
\begin{minipage}{\linewidth}
{\footnotesize 
\begin{itemize}
   \item[${}^\dagger$] The calculation uses domain wall fermions in the valence-quark sector.\\[-5mm]
\item[${}^*$] The calculation uses HISQ staggered fermions in the valence-quark sector.
\end{itemize}
}
\end{minipage}
}
\caption{Summary of simulated lattice actions with $\Nf=2+1$ or $\Nf=3$ quark flavours.}
\end{table}

\begin{table}[h]
\addtocounter{table}{-1}
{\footnotesize
\begin{tabular*}{\textwidth}[l]{l @{\extracolsep{\fill}} c c c c}
\hline \hline \\[-1.0ex]
Collab. & Ref. & $\Nf$ & \parbox{1cm}{gauge\\action} & \parbox{1cm}{quark\\action} 
\\[2.0ex] \hline \hline \\[-1.0ex]
Laiho 11 & \cite{Laiho:2011np} & $2+1$ & tadSym & Asqtad ${}^\dagger$\\
[2.0ex] \hline \\[-1.0ex]
LHP 04 & \cite{Bonnet:2004fr} & $2+1$ & tadSym & Asqtad $^\dagger$ \\
[2.0ex] \hline \\[-1.0ex]
Maltman 08 & \cite{Maltman:2008bx} & $2+1$ & tadSym &  Asqtad \\
\end{tabular*}\\[-0.2cm]
\begin{minipage}{\linewidth}
{\footnotesize 
\begin{itemize}
   \item[${}^\dagger$] The calculation uses domain wall fermions in the valence-quark sector.\\[-5mm]
\item[${}^+$] The calculation uses HYP smeared improved staggered fermions in the valence-quark sector.
\end{itemize}
}
\end{minipage}
}
\caption{(cntd.) Summary of simulated lattice actions with $\Nf=2+1$ or $\Nf=3$ quark flavours.}
\end{table}

\begin{table}[h]
{\footnotesize
\begin{tabular*}{\textwidth}[l]{l @{\extracolsep{\fill}} c c c c}
\hline \hline \\[-1.0ex]
Collab. & Ref. & $\Nf$ & \parbox{1cm}{gauge\\action} & \parbox{1cm}{quark\\action} 
\\[2.0ex] \hline \hline \\[-1.0ex]
ALPHA 10A & \cite{Tekin:2010mm} & $4$ & Wilson &  npSW \\
[2.0ex] \hline \\[-1.0ex]
Bazavov 12 & \cite{Bazavov:2012ka} & $2+1+1$ & tlSym & HISQ \\
[2.0ex] \hline \\[-1.0ex] 
\parbox[t]{4.0cm}{ETM 10, 10E, 11, 11D, 12C, 13, 13A, 13D} &  \parbox[t]{2.5cm}{\cite{Baron:2010bv,Farchioni:2010tb,Baron:2011sf,Blossier:2011tf,Blossier:2012ef,Cichy:2013gja,Blossier:2013ioa,Herdoiza:2013sla}} & $2+1+1$ & Iwasaki & tmWil \\
[4.0ex] \hline \\[-1.0ex]
\parbox[t]{4.0cm}{ETM 14A, 14B, 15, 15C} &  \cite{Alexandrou:2014sha,Bussone:2014cha,Carrasco:2015pra,Carrasco:2015wzu} & $2+1+1$ & Iwasaki & tmWil ${}^+$ \\
[2.0ex] \hline \\[-1.0ex]
FNAL/MILC 12B, 13, 13C, 13E, 14A & \cite{Bazavov:2012dg,Bazavov:2013nfa,Gamiz:2013xxa,Bazavov:2013maa,Bazavov:2014wgs} & $2+1+1$ & tadSym & HISQ \\
[2.0ex] \hline \\[-1.0ex] 
HPQCD 14A, 15B  & \cite{Chakraborty:2014aca,Koponen:2015tkr} & $2+1+1$ & tadSym & HISQ \\
[2.0ex] \hline \\[-1.0ex] 
MILC 13A & \cite{Bazavov:2013cp} & $2+1+1$ & tadSym & HISQ \\
[2.0ex] \hline \\[-1.0ex] 
Perez 10 & \cite{PerezRubio:2010ke} & $4$ & Wilson &  npSW \\
[2.0ex] \hline\hline \\
\end{tabular*}\\[-0.2cm]
\begin{minipage}{\linewidth}
{\footnotesize 
\begin{itemize}
   \item[${}^+$] The calculation uses Osterwalder-Seiler fermions \cite{Osterwalder:1977pc} in the valence-quark sector.
\end{itemize}
}
\end{minipage}
}
\caption{Summary of simulated lattice actions with $\Nf=4$ or $\Nf=2+1+1$ quark flavours.}
\end{table}

\begin{table}[!ht]
{\footnotesize
\begin{tabular*}{\textwidth}[l]{l @{\extracolsep{\fill}} c c c c c c}
\hline\hline \\[-1.0ex]
Collab. & Ref. & $\Nf$ & Gauge & \multicolumn{3}{c}{Quark actions}  \\
& & & action & sea & light valence & heavy \\[1.0ex] \hline \hline \\[-1.0ex]
\parbox[t]{3.5cm}{ALPHA 11, 12A, 13, 14, 14B} & \parbox[t]{2.5cm}{\cite{Blossier:2011dk,Bernardoni:2012ti,Bernardoni:2013oda,Bernardoni:2014fva,Bahr:2014iqa}} & 2 &  plaquette & npSW  & npSW & HQET
\\[4.0ex] \hline \\[-1.0ex]
ALPHA 13C & \cite{Heitger:2013oaa} & 2 &  plaquette & npSW  & npSW & npSW
\\[2.0ex] \hline \\[-1.0ex]
\parbox[t]{2.5cm}{Atoui 13} & \cite{Atoui:2013zza} & 2 &  tlSym & tmWil & tmWil & tmWil
\\[2.0ex] \hline \\[-1.0ex]
\parbox[t]{3.5cm}{ETM 09, 09D, 11B, 12A, 12B, 13B, 13C} & \parbox[t]{2.5cm}{\cite{Blossier:2009bx,Blossier:2009hg,DiVita:2011py,Carrasco:2012dd,Carrasco:2012de,Carrasco:2013zta,Carrasco:2013iba}} & 2 &  tlSym & tmWil & tmWil & tmWil
\\[4.0ex] \hline \\[-1.0ex]
ETM  11A & \cite{Dimopoulos:2011gx} & 2 &  tlSym & tmWil & tmWil & tmWil, static
\\[4.0ex] \hline \\[-1.0ex]
TWQCD 14 & \cite{Chen:2014hva} & 2 & plaquette & oDW & oDW & oDW \\[2.0ex] \hline\hline
\end{tabular*}
\caption{Summary of lattice simulations $N_f=2$ sea quark flavours and with $b$ and $c$ valence quarks.}
}
\end{table}

\begin{table}[!ht]
{\footnotesize
\begin{tabular*}{\textwidth}[l]{l @{\extracolsep{\fill}} c c c c c c}
\hline\hline \\[-1.0ex]
Collab. & Ref. & $\Nf$ & Gauge & \multicolumn{3}{c}{Quark actions}  \\
& & & action & sea & light valence & heavy \\[1.0ex] \hline \hline \\[-1.0ex]
$\chi$QCD 14 & \cite{Yang:2014sea} & 2+1 & Iwasaki & DW & overlap & overlap \\[2.0ex] \hline \\[-1.0ex]
\parbox[t]{3.5cm}{FNAL/MILC 04, 04A, 05, 08, 08A, 10, 11, 11A, 12, 13B} &  \parbox[t]{2.5cm}{\cite{Aubin:2004ej,Okamoto:2004xg,Aubin:2005ar,Bernard:2008dn,Bailey:2008wp,Bailey:2010gb,Bazavov:2011aa,Bouchard:2011xj,Bazavov:2012zs,Qiu:2013ofa}}  & 2+1 & tadSym & Asqtad & Asqtad & Fermilab
\\[7.0ex] \hline \\
\end{tabular*}\\[-0.2cm]
\begin{minipage}{\linewidth}
{\footnotesize 
\begin{itemize}
   \item[$^*$] Asqtad for $u$, $d$ and $s$ quark; Fermilab for $b$ and $c$ quark.\\[-5mm]
   \item[$^\dagger$] HISQ for $u$, $d$, $s$  and $c$ quark; NRQCD for $b$ quark.
\end{itemize}
}
\end{minipage}
\caption{Summary of lattice simulations with $N_f=2+1$ or $N_f=2+1+1$ sea quark flavours and $b$ and $c$ valence quarks.  
}
}
\end{table}

%% file: qmass/qmass_app.tex
\subsection{Notes to section \ref{sec:qmass} on quark masses}

\begin{table}[!ht]
{\footnotesize
\begin{tabular*}{\textwidth}[l]{l @{\extracolsep{\fill}} c c c l}
\hline\hline \\[-1.0ex]
Collab. & Ref. & $\Nf$ & $a$ [fm] & Description
\\[1.0ex] \hline \hline \\[-1.0ex]
HPQCD 14A & \cite{Chakraborty:2014aca} & 2+1+1 & 0.15, 0.12, 0.09, 0.06  & \parbox[t]{4.5cm}{Scale set through the Wilson flow parameter $w_0$.} \\
[5.0ex] \hline \\[-1.0ex]
{FNAL/MILC 14A}& \cite{Bazavov:2014wgs} & 2+1+1 & 0.06,\,0.09,\,0.12,\,0.15  &  
	\parbox[t]{5.7cm}{ 
	HISQ action for both valence and sea quarks. Absolute scale though $f_\pi$.}\\
[5.0ex] \hline \\[-1.0ex]
{ETM 14}& \cite{Carrasco:2014cwa} & 2+1+1 & 0.062, 0.082, 0.089  &
		    \parbox[t]{5.7cm}{ 
		    Scale set through $f_\pi$. Automatic $\cO(a)$ improvement, flavour symmetry
		    breaking: $(M_{PS}^{0})^2-(M_{PS}^\pm)^2\sim \cO(a^2)$. Discretization and volume effects
		    due to the $\pi^0 - \pi^\pm$ mass splitting are taken into account through {\Ch}PT for
		    twisted mass fermions.} \\
[18.0ex] \hline \hline \\[-1.0ex]
\end{tabular*}
\caption{Continuum extrapolations/estimation of lattice artifacts in
  determinations of $m_{ud}$, $m_s$ and, in some cases $m_u$ and
  $m_d$, with $N_f=2+1+1$ quark flavours.}
}
\end{table}

\begin{table}[!ht]
{\footnotesize
\begin{tabular*}{\textwidth}[l]{l @{\extracolsep{\fill}} c c c l}
\hline\hline \\[-1.0ex]
Collab. & Ref. & $\Nf$ & $a$ [fm] & Description
\\[1.0ex] \hline \hline \\[-1.0ex]
{QCDSF/UKQCD 15}& \cite{Horsley:2015eaa} & 2+1 & 0.07 &  \parbox[t]{4.5cm}{Scale set through the gradient flow parameter $w_0$.}\\
[4.0ex] \hline \\[-1.0ex]
{RBC/UKQCD 14B}& \cite{Blum:2014tka} & 2+1 & 0.063, 0.084,\, 0.114,\, 0.144  &  \parbox[t]{4.5cm}{Scale set through $M_\Omega$.}\\
[2.0ex] \hline \\[-1.0ex]
{RBC/UKQCD 12}& \cite{Arthur:2012opa} & 2+1 & 0.085,\, 0.113,\, 0.144  &  \parbox[t]{4.5cm}{Scale set through $M_\Omega$.}\\
[2.0ex] \hline \\[-1.0ex]
PACS-CS 12 & \cite{Aoki:2012st} & 1+1+1 & 0.09
& \parbox[t]{4.5cm}{Reweighting of PACS-CS 08 $N_f=2+1$ QCD
  configurations with e.m. and $m_u\ne m_d$.} \\
[7.0ex] \hline \\[-1.0ex]
Laiho 11 & \cite{Laiho:2011np} & 2+1 & 0.06, 0.09, 0.15
& \parbox[t]{4.5cm}{MILC staggered ensembles~\cite{Bazavov:2010yq}, scale set using
  $r_1$ determined by HPQCD with $\Upsilon$ splittings, pseudoscalar
  decay constants, through $r_1$~\cite{Davies:2009tsa}.} \\ [13.0ex]
\hline \hline
\\[-1.0ex]
\end{tabular*}
\caption{Continuum extrapolations/estimation of lattice artifacts in
  determinations of $m_{ud}$, $m_s$ and, in some cases $m_u$ and
  $m_d$, with $N_f=2+1$ quark flavours.}
}
\end{table}

\begin{table}[!ht]
\addtocounter{table}{-1}
{\footnotesize
 \begin{tabular*}{\textwidth}[l]{l @{\extracolsep{\fill}} c c c l}
 \hline\hline \\[-1.0ex]
 Collab. & Ref. & $\Nf$ & $a$ [fm] & Description
 \\[1.0ex] \hline \hline \\[-1.0ex]
PACS-CS 10 & \cite{Aoki:2010wm} &  2+1 &  0.09
& \parbox[t]{4.5cm}{cf.~PACS-CS 08} \\
[1.0ex] \hline \\[-1.0ex]
{MILC 10A}& \cite{Bazavov:2010yq} & 2+1 & & cf.~MILC 09, 09A \\
[1.0ex] \hline \\[-1.0ex]
{BMW 10A, 10B}& \cite{Durr:2010vn,Durr:2010aw}
& 2+1 & \parbox[t]{2.3cm}{0.054, 0.065, 0.077, 0.093, 0.116} &  Scale set via
$M_\pi, M_K, M_\Omega$.\\
\end{tabular*}
 \caption{(cntd.) Continuum extrapolations/estimation of lattice artifacts in
   determinations of $m_{ud}$, $m_s$ and, in some cases $m_u$ and
   $m_d$, with $N_f=2+1$ quark flavours.}
}
\end{table}

\begin{table}[!ht]
{\footnotesize
\begin{tabular*}{\textwidth}[l]{l @{\extracolsep{\fill}} c c c l}
\hline\hline \\[-1.0ex]
Collab. & Ref. & $\Nf$ & $a$ [fm] & Description
\\[1.0ex] \hline \hline \\[-1.0ex]
ETM 14D		&\cite{Abdel-Rehim:2014nka}		&2  & 0.094 &\parbox[t]{5.7cm}{Scale set through $F_\pi$, $r_0$, $t_0$ and $w_0$. Twisted Wilson fermions plus clover term. Automatic $\cO(a)$ improvement.}\\
[7.0ex] \hline \\[-1.0ex]
RM123 13 & \cite{deDivitiis:2013xla} & 2 & \parbox[t]{2.3cm}{0.098,
  0.085,\\ 0.067, 0.054} & cf. ETM 10B\\
[4.0ex] \hline \\[-1.0ex]
ALPHA 12 & \cite{Fritzsch:2012wq} & 2 & \parbox[t]{2.3cm}{0.076, 0.066,\\ 0.049} & Scale set through $F_K$.\\
\hline\hline
\end{tabular*}
\caption{Continuum extrapolations/estimation of lattice artifacts in
  determinations of $m_{ud}$, $m_s$ and, in some cases $m_u$ and
  $m_d$, with $N_f=2$ quark flavours.}
}
\end{table}

\newpage

\begin{table}[!ht]
{\footnotesize
\begin{tabular*}{\textwidth}[l]{l @{\extracolsep{\fill}} c c c l}
\hline \hline  \\[-1.0ex]
Collab. & Ref. & $\Nf$ & $M_{\pi,\text{min}}$ [MeV] & Description
\\[1.0ex] \hline \hline \\[-1.0ex]
HPQCD 14A & \cite{Chakraborty:2014aca} & 2+1+1 & $128_{\pi, 5}$ ($173_{\rm RMS}$)&
 \parbox[t]{6.2cm}{Sea quark masses linearly extrapolated/interpolated to physical values. $m_s$ determined from physical $m_s/m_c$ and $m_c$.} \\
[9.0ex] \hline \\[-1.0ex]
{FNAL/MILC 14A}& \cite{Bazavov:2014wgs} & 2+1+1 & $128_{\pi,5}$ ($143_{\rm RMS}$)  &  
	\parbox[t]{6.2cm}{ 
	Linear interpolation to physical point. The lightest RMS mass is from the 
	$a=0.06$~fm ensemble and the lightest Nambu-Goldstone mass is from the
	$a=0.09$~fm ensemble.}\\ 
[9.0ex]\hline \\[-1.0ex]
{ETM 14}& \cite{Carrasco:2014cwa} & 2+1+1 & $180_{\rm \pi^0}(220_{\pi^\pm})$  & 
          \parbox[t]{6.2cm}{Chiral extrapolation performed through $SU(2)$ {\Ch}PT or polynomial fit.}\\
[4.0ex] \hline \hline \\[-1.0ex]
\end{tabular*}
\caption{Chiral extrapolation/minimum pion mass in determinations of
  $m_{ud}$, $m_s$ and, in some cases, $m_u$ and
  $m_d$,  with $N_f=2+1+1$ quark flavours.}
}
\end{table}

\begin{table}[!ht]
{\footnotesize
\begin{tabular*}{\textwidth}[l]{l @{\extracolsep{\fill}} c c c l}
\hline \hline  \\[-1.0ex]
Collab. & Ref. & $\Nf$ & $M_{\pi,\text{min}}$ [MeV] & Description
\\[1.0ex] \hline \hline \\[-1.0ex]
{QCDSF/UKQCD 15}& \cite{Horsley:2015eaa} & 2+1 & 205 (val.) &  \parbox[t]{5.7cm}{Expansion around the symmetric point $m_u = m_d = m_s$.}\\
[4.0ex] \hline \\[-1.0ex]
{RBC/UKQCD 14B}& \cite{Blum:2014tka} & 2+1 & 139 &\parbox[t]{5.7cm}{NLO PQ $SU(2)$ $\chi$PT as well as analytic ans\"atze.}\\
[4.0ex] \hline \\[-1.0ex]
{RBC/UKQCD 12}& \cite{Arthur:2012opa} & 2+1 & 170 & \parbox[t]{5.7cm}{Combined fit to Iwasaki and Iwasaki+DSDR gauge action ensembles.} \\
[4.0ex] \hline \\[-1.0ex]
PACS-CS 12 & \cite{Aoki:2012st} &  1+1+1 &  & cf.~PACS-CS 08 \\
[1.0ex] \hline \\[-1.0ex]
Laiho 11 & \cite{Laiho:2011np} & 2+1 & \parbox[t]{3cm}{210 (val.)\\ 280 (sea-RMS)} &
\parbox[t]{5.7cm}{NLO $SU(3)$, mixed-action $\chi$PT \cite{Bar:2005tu}, with N$^2$LO-N$^4$LO
  analytic terms.} \\ [4.0ex] \hline \\[-1.0ex]
PACS-CS 10 & \cite{Aoki:2010wm} &  2+1 & & cf.~PACS-CS 08 \\
[1.0ex] \hline \\[-1.0ex]
{MILC 10A}& \cite{Bazavov:2010yq} & 2+1 & & \parbox[t]{5.7cm}{NLO $SU(2)$
S{\Ch}PT. cf.~also MILC 09A, 09.} \\
[4.0ex] \hline \\[-1.0ex]
{BMW 10A, 10B}& \cite{Durr:2010vn,Durr:2010aw}
& 2+1 & 135 & \parbox[t]{6.cm}{Interpolation to the physical point.} \\
[1.0ex] \hline \\[-1.0ex]
{RBC/UKQCD 10A}& \cite{Aoki:2010dy} & 2+1 & 290 & \parbox[t]{5.7cm}{NLO PQ $SU(2)$ $\chi$PT as well as analytic ans\"atze.}\\
[4.0ex] \hline \\[-1.0ex]
{Blum 10}& \cite{Blum:2010ym,Allton:2008pn}
& 2+1 & \parbox[t]{2cm}{242 (valence), 330 (sea)} &
\parbox[t]{5.7cm}{Extrapolation done on the basis of PQ$\chi$PT formulae with
virtual photons.} \\
[4.0ex] \hline \hline\\[-1.0ex]
\end{tabular*}
\caption{Chiral extrapolation/minimum pion mass in determinations of
  $m_{ud}$, $m_s$ and, in some cases $m_u$ and
  $m_d$,  with $N_f=2+1$ quark flavours.}
}
\end{table}

\begin{table}[!ht]
\addtocounter{table}{-1}
{\footnotesize
\begin{tabular*}{\textwidth}[l]{l @{\extracolsep{\fill}} c c c l}
\hline \hline  \\[-1.0ex]
Collab. & Ref. & $\Nf$ & $M_{\pi,\text{min}}$ [MeV] & Description
\\[1.0ex] \hline \hline \\[-1.0ex]

{PACS-CS 09}& \cite{Aoki:2009ix}& 2+1 & 135 & \parbox[t]{5.5cm}{Physical point reached by
reweighting technique, no chiral extrapolation needed.}\\
[7.0ex] \hline \\[-1.0ex]
{HPQCD 09A, 10}&  \cite{Davies:2009ih,McNeile:2010ji}& 2+1 &  &  cf.~MILC 09\\
[1.0ex] \hline \\[-1.0ex]
{MILC 09A, 09} & \cite{Bazavov:2009fk,Bazavov:2009bb} & 2+1 & 177, 224 & \parbox[t]{5.5cm}{
		NLO $SU(3)$ RS{\Ch}PT, continuum {\Ch}PT at NNLO and NNNLO and NNNNLO 
		analytic terms. The lightest Nambu-Goldstone mass
                is 177 MeV (09A) and 224 MeV (09)
		(at $a=$0.09fm) and the lightest RMS mass is 258MeV (at
                $a=$0.06fm).} \\ 
[18.0ex] \hline \\[-1.0ex]
PACS-CS 08 &\cite{Aoki:2008sm}& 2+1 &156& 
		\parbox[t]{5.5cm}{NLO $SU(2)$ {\Ch}PT and $SU(3)$
                  (Wilson){\Ch}PT.}\\
[4.0ex] \hline \\[-1.0ex]
RBC/UKQCD 08	&\cite{Allton:2008pn}	&2+1& \parbox[t]{2cm}{242 (valence), 330 (sea)} & \parbox[t]{5.5cm}{$SU(3)$ PQ{\Ch}PT 
and heavy kaon NLO $SU(2)$ PQ{\Ch}PT fits.} \\
[4.0ex] \hline \\[-1.0ex]
{CP-PACS/JLQCD 07}&  \cite{Ishikawa:2007nn} & 2+1 & 620 & \parbox[t]{5.5cm}{NLO Wilson {\Ch}PT
fits to meson masses.} \\
[1.0ex] \hline \\[-1.0ex]
{HPQCD 05}& \cite{Mason:2005bj}& 2+1 & 240 & \parbox[t]{5.5cm}{PQ RS{\Ch}PT fits.} \\
[1.0ex] \hline \\[-1.0ex]
\parbox[t]{4.5cm}{HPQCD/MILC/UKQCD 04,\\ MILC 04}
& \cite{Aubin:2004ck,Aubin:2004fs}& 2+1 & 240 & \parbox[t]{5.5cm}{PQ RS{\Ch}PT fits.} \\
[4.0ex] \hline \hline \\[-1.0ex]
\end{tabular*}
\caption{(cntd.) Chiral extrapolation/minimum pion mass in determinations of
  $m_{ud}$, $m_s$ and, in some cases $m_u$ and
  $m_d$,  with $N_f=2+1$ quark flavours.}
}
\end{table}

\begin{table}[!ht]
{\footnotesize
\begin{tabular*}{\textwidth}[l]{l @{\extracolsep{\fill}} c c c l}
\hline \hline  \\[-1.0ex]
Collab. & Ref. & $\Nf$ & $M_{\pi,\text{min}}$ [MeV] & Description
\\[1.0ex] \hline \hline \\[-1.0ex]
ETM 14D         & \cite{Abdel-Rehim:2014nka}  &2  & 140 & \parbox[t]{6.2cm}{Charged/neutral pion mass breaking, $M_{\pi^pm}^2-M_{\pi^0}^2 \sim \cO(a^2)$, estimated to be $\simeq 20$ MeV.} \\
[7.0ex] \hline \\[-1.0ex]
RM123 13 & \cite{deDivitiis:2013xla} & 2 & 270 & \parbox[t]{6.2cm}{Fits based on NLO {\Ch}PT and Symanzik expansion up to $\cO(a^2)$. $\cO(\alpha)$ e.m. effects included.}\\
[7.0ex] \hline \\[-1.0ex]
ALPHA 12 & \cite{Fritzsch:2012wq} & 2 & 270 & \parbox[t]{6.2cm}{NLO
  $SU(2)$ and $SU(3)$ {\Ch}PT and $\cO(a^2)$ on LO LEC.}\\
\hline\hline
\end{tabular*}
\caption{Chiral extrapolation/minimum pion mass in determinations of
  $m_{ud}$, $m_s$ and, in some cases $m_u$ and
  $m_d$,  with $N_f=2$ quark flavours.}
}
\end{table}

\newpage

\begin{table}
{\footnotesize
\begin{tabular*}{\textwidth}[l]{l @{\extracolsep{\fill}} c c c c l}
\hline\hline  \\[-1.0ex]
Collab. & Ref. & $\Nf$ & $L$ [fm] & $M_{\pi,\text{min}} L$ & Description
\\[1.0ex] \hline \hline \\[-1.0ex]
HPQCD 14A & \cite{Chakraborty:2014aca} & 2+1+1 & 2.5-5.8  & 3.7 & \\
[4.0ex] \hline \\[-1.0ex]
{FNAL/MILC 14A}& \cite{Bazavov:2014wgs} & 2+1+1 &2.8-5.8 & $3.9_{\rm RMS}(3.7_{\pi,5}) $ &  
	\parbox[t]{4.7cm}{ Includes error estimate from NNLO S$\chi$PT. }\\
[4.0ex] \hline \\[-1.0ex]
ETM 14& \cite{Carrasco:2014cwa} & 2+1+1 & 2.0 - 3.0  & $2.7_{\pi^0}$ ($3.3_{\pi^\pm}$) & 
        \parbox[t]{4.7cm}{FV effect for the pion is corrected through resummed NNLO {\Ch}PT for twisted mass fermions,
        which takes into account the effects due to the $\pi^0 - \pi^\pm$ mass splitting.} \\
[13.0ex] \hline \hline \\[-1.0ex]
\end{tabular*}
\caption{Finite volume effects in determinations of $m_{ud}$, $m_s$
  and, in some cases $m_u$ and $m_d$, with $N_f=2+1+1$ quark flavours.}
}
\end{table}

\begin{table}
{\footnotesize
\begin{tabular*}{\textwidth}[l]{l @{\extracolsep{\fill}} c c c c l}
\hline\hline  \\[-1.0ex]
Collab. & Ref. & $\Nf$ & $L$ [fm] & $M_{\pi,\text{min}} L$ & Description
\\[1.0ex] \hline \hline \\[-1.0ex]
{QCDSF/UKQCD 15}& \cite{Horsley:2015eaa} & 2+1 & 1.7,\,2.2,\,3.4& & \parbox[t]{3.7cm}{Effective field theory used to extrapolate to infinite volume.} \\
[7.0ex] \hline \\[-1.0ex]
{RBC/UKQCD 14B}& \cite{Blum:2014tka} & 2+1 & 2.0,\,2.7,\,4.6,\,5.4 &  $3.8$ &  
\parbox[t]{3.7cm}{Uses FV chiral perturbation theory to estimate the error, which is deemed negligible and omitted.}\\
[10.0ex] \hline \\[-1.0ex]
{RBC/UKQCD 12}& \cite{Arthur:2012opa} & 2+1 & 2.7, 4.6 &  $\gtrsim 4.0$
& \parbox[t]{3.7cm}{Uses FV chiral perturbation theory to estimate the error.} \\
[7.0ex] \hline \\[-1.0ex]
PACS-CS 12 & \cite{Aoki:2012st} &  1+1+1 & & & cf.~PACS-CS 08 \\
[1.0ex] \hline \\[-1.0ex]
Laiho 11 & \cite{Laiho:2011np} & 2+1 &
\parbox[t]{2cm}{2.5, 2.9, 3.0,\\ 3.6, 3.8, 4.8}&\parbox[t]{1.5cm}{4.1 (val.)\\
4.1 (sea)}&  \parbox[t]{3.7cm}{Data corrected using NLO $SU(3)$ $\chi$PT finite-V formulae.} \\
[7.0ex] \hline \hline\\[-1.0ex]
\end{tabular*}
\caption{Finite volume effects in determinations of $m_{ud}$, $m_s$
  and, in some cases $m_u$ and $m_d$, with $N_f=2+1$ quark flavours.}
}
\end{table}

\begin{table}
\addtocounter{table}{-1}
{\footnotesize
\begin{tabular*}{\textwidth}[l]{l @{\extracolsep{\fill}} c c c c l}
\hline\hline  \\[-1.0ex]
Collab. & Ref. & $\Nf$ & $L$ [fm] & $M_{\pi,\text{min}} L$ & Description
\\[1.0ex] \hline \hline \\[-1.0ex]
PACS-CS 10 & \cite{Aoki:2010wm} &  2+1 & & & cf.~PACS-CS 08 \\
[1.0ex] \hline \\[-1.0ex]
{MILC 10A}& \cite{Bazavov:2010yq} & 2+1 & & & cf.~MILC 09A, 09 \\
[1.0ex] \hline \\[-1.0ex]
{BMW 10A, 10B}& \cite{Durr:2010vn,Durr:2010aw}
& 2+1 &  $\gtrsim 5.0$ & $\gtrsim 4.0$ & \parbox[t]{3.7cm}{FV corrections
  below 5 per mil on the largest lattices.}\\
\end{tabular*}
\caption{(cntd.) Finite volume effects in determinations of $m_{ud}$, $m_s$
  and, in some cases $m_u$ and $m_d$, with $N_f=2+1$ quark flavours.}
}
\end{table}

\begin{table}
{\footnotesize
\begin{tabular*}{\textwidth}[l]{l @{\extracolsep{\fill}} c c c c l}
\hline\hline  \\[-1.0ex]
Collab. & Ref. & $\Nf$ & $L$ [fm] & $M_{\pi,\text{min}} L$ & Description
\\[1.0ex] \hline \hline \\[-1.0ex]
ETM 14D		&\cite{Abdel-Rehim:2014nka}		&2  & 2.2,\,4.5& 3.2 & \\
[1.0ex] \hline \\[-1.0ex]
RM123 13 & \cite{deDivitiis:2013xla} & 2 & $\gtrsim 2.0$ & 3.5 & \parbox[t]{4.2cm}{One volume
$L=1.7$ fm at $m_\pi=495, \,a=0.054$fm.} \\
[4.0ex] \hline \\[-1.0ex]
ALPHA 12 & \cite{Fritzsch:2012wq} & 2 & 2.1--3.2 & 4.2
& \parbox[t]{4.2cm}{Roughly 2 distinct volumes; no analysis of FV effects.} \\
\hline\hline
\end{tabular*}
\caption{Finite volume effects in determinations of $m_{ud}$, $m_s$
  and, in some cases $m_u$ and $m_d$, 
   with $N_f=2$ quark flavours.}
}
\end{table}

\newpage

\begin{table}[!ht]
{\footnotesize
\begin{tabular*}{\textwidth}[l]{l @{\extracolsep{\fill}} c c l}
\hline\hline \\[-1.0ex]
Collab. & Ref. & $\Nf$  & Description
\\[1.0ex] \hline \hline \\[-1.0ex]
HPQCD 14A & \cite{Chakraborty:2014aca} & 2+1+1 & \parbox[t]{4.7cm}{Renormalization not required through the use of the ratio $m_c/m_s$.} \\
[7.0ex] \hline \\[-1.0ex]
{FNAL/MILC 14A}& \cite{Bazavov:2014wgs} & 2+1+1 & \parbox[t]{4.7cm}{Renormalization not required for $m_s/m_{ud}$.} \\
[4.0ex] \hline \\[-1.0ex]
ETM 14& \cite{Carrasco:2014cwa} & 2+1+1 & \parbox[t]{4.7cm}{Non-perturbative renormalization (RI/MOM).} \\
[4.0ex] \hline  \hline \\[-1.0ex]
\end{tabular*}
\caption{Renormalization in determinations of $m_{ud}$, $m_s$
  and, in some cases $m_u$ and $m_d$, with $N_f=2+1+1$ quark flavours.}
}
\end{table}

\begin{table}[!ht]
{\footnotesize
\begin{tabular*}{\textwidth}[l]{l @{\extracolsep{\fill}} c c l}
\hline\hline \\[-1.0ex]
Collab. & Ref. & $\Nf$  & Description
\\[1.0ex] \hline \hline \\[-1.0ex]
{QCDSF/UKQCD 15}& \cite{Horsley:2015eaa} & 2+1 & \parbox[t]{7cm}{Non-perturbative renormalization (RI/MOM).} \\
[1.0ex] \hline \\[-1.0ex]
{RBC/UKQCD 14B}& \cite{Blum:2014tka}  & 2+1
& \parbox[t]{7cm}{Non-perturbative renormalization (RI/SMOM).}\\
[1.0ex] \hline \\[-1.0ex]
{RBC/UKQCD 12}& \cite{Arthur:2012opa} & 2+1
& \parbox[t]{7cm}{Non-perturbative renormalization (RI/SMOM).}\\
\end{tabular*}
\caption{Renormalization in determinations of $m_{ud}$, $m_s$
  and, in some cases $m_u$ and $m_d$, with $N_f=2+1$ quark flavours.}
}
\end{table}

\begin{table}[!ht]
{\footnotesize
\begin{tabular*}{\textwidth}[l]{l @{\extracolsep{\fill}} c c l}
\hline\hline \\[-1.0ex]
Collab. & Ref. & $\Nf$  & Description
\\[1.0ex] \hline \hline \\[-1.0ex]
ETM 14D		&\cite{Abdel-Rehim:2014nka}		&2  & \parbox[t]{7cm}{Renormalization not required for $m_s/m_{ud}$.} \\
[1.0ex] \hline \\[-1.0ex]
RM123 13 & \cite{deDivitiis:2013xla} & 2 &\parbox[t]{7cm}{Non-perturbative
renormalization.} \\
[1.0ex] \hline \\[-1.0ex]
ALPHA 12 & \cite{Fritzsch:2012wq} & 2 &\parbox[t]{7cm}{Non-perturbative
renormalization.} \\
\end{tabular*}
\caption{Renormalization in determinations of $m_{ud}$, $m_s$
  and, in some cases $m_u$ and $m_d$, with $N_f=2$ quark flavours.}
}
\end{table}

%%% Local Variables: 
%%% mode: latex
%%% TeX-master: "FLAG_master"
%%% End: 

%% file: qmass/qmass_mc_app.tex
\begin{table}[!ht]
{\footnotesize
\begin{tabular*}{\textwidth}[l]{l @{\extracolsep{\fill}} c c c l}
\hline\hline \\[-1.0ex]
Collab. & Ref. & $\Nf$ & $a$ [fm] & Description
\\[1.0ex] \hline \hline \\[-1.0ex]
HPQCD 14A & \cite{Chakraborty:2014aca} & 2+1+1 & 0.06, 0.09, 0.12, 0.15  & \parbox[t]{4.5cm}{Scale set through the Wilson flow parameter $w_0$.} \\
[4.0ex] \hline \\[-1.0ex]
ETM 14& \cite{Carrasco:2014cwa} & 2+1+1 & 0.062, 0.082, 0.089  & \parbox[t]{4.5cm}{Scale set through $F_\pi$.}\\
[1.0ex] \hline \\[-1.0ex]
ETM 14A & \cite{Alexandrou:2014sha} & 2+1+1 & 0.062, 0.082, 0.089 & \parbox[t]{4.5cm}{Scale set through the nucleon mass $M_N$.} \\
[4.0ex] \hline \\[-1.0ex]
JLQCD 15B  & \cite{Nakayama:2015hrn} & 2+1 & 0.044, 0.055. 0.083 & \parbox[t]{4.5cm}{M\"obius domain wall fermions.} \\ [1.0ex] \hline
\\[-1.0ex]
$\chi$QCD 14 & \cite{Yang:2014sea} &  2+1 & 0.087, 0.11 & \parbox[t]{4.5cm}{Overlap valence fermions on domain-wall sea quarks from \cite{Aoki:2010dy}.
The lattice scale is set together with the strange and charm quark masses using the experimental values of the $D_s$, $D_s^*$ and $J/\psi$ meson masses.} \\
[18.0ex] \hline \\[-1.0ex]
HPQCD 10 & \cite{McNeile:2010ji} & 2+1 & 0.044, 0.059, 0.085, 0.12, 0.15 & \parbox[t]{4.5cm}{Scale set through the static-quark potential parameter $r_1$.} \\
[4.0ex] \hline \\[-1.0ex]
HPQCD 08B & \cite{Allison:2008xk} & 2+1 & 0.06, 0.09, 0.12, 0.15 & \parbox[t]{4.5cm}{Scale set through the static-quark potential parameter $r_1$.} \\
[4.0ex] \hline \\[-1.0ex]
ALPHA 13B& \cite{Heitger:2013oaa} & 2 & 0.048, 0.065 & \parbox[t]{4.5cm}{Scale set through $F_K$.} \\
[1.0ex] \hline \\[-1.0ex]
ETM 11F & \cite{Jansen:2011vr} & 2 & & cf.~ETM 10B \\
[1.0ex] \hline \\[-1.0ex]
ETM 10B & \cite{Blossier:2010cr} & 2 & 0.054, 0.067, 0.085, 0.098 & \parbox[t]{4.5cm}{Scale set through $F_\pi$.} \\
[1.0ex] \hline \hline \\[-1.0ex]
\end{tabular*}
\caption{Continuum extrapolations/estimation of lattice artifacts in the determinations of $m_c$.}
}
\end{table}

\begin{table}[!ht]
{\footnotesize
\begin{tabular*}{\textwidth}[l]{l @{\extracolsep{\fill}} c c c l}
\hline \hline  \\[-1.0ex]
Collab. & Ref. & $\Nf$ & $M_{\pi,\text{min}}$ [MeV] & Description
\\[1.0ex] \hline \hline \\[-1.0ex]
HPQCD 14A & \cite{Chakraborty:2014aca} & 2+1+1 & $128_{\pi, 5}$ ($173_{RMS}$)& \\
[1.0ex] \hline \\[-1.0ex]
ETM 14& \cite{Carrasco:2014cwa} & 2+1+1 & $180_{\pi^0}$ ($220_{\pi^\pm}$)  & \\
[1.0ex] \hline \\[-1.0ex]
ETM 14A & \cite{Alexandrou:2014sha} & 2+1+1 & 210 & cf.~ETM 14 \\
[1.0ex] \hline \\[-1.0ex]
JLQCD 15B  & \cite{Nakayama:2015hrn} & 2+1 & &  \\
[1.0ex] \hline \\[-1.0ex]
$\chi$QCD 14 & \cite{Yang:2014sea} & 2+1 & 290 & \\
[3.0ex] \hline \\[-1.0ex]
HPQCD 10 & \cite{McNeile:2010ji} & 2+1 & 260 & \\
[1.0ex] \hline \\[-1.0ex]
HPQCD 08B & \cite{Allison:2008xk} & 2+1 &  & \\
[1.0ex] \hline \\[-1.0ex]
ALPHA 13B& \cite{Heitger:2013oaa} & 2 & 190 & \\
[4.0ex] \hline \\[-1.0ex]
ETM 11F & \cite{Jansen:2011vr} & 2 & & cf.~ETM 10B \\
[1.0ex] \hline \\[-1.0ex]
ETM 10B & \cite{Blossier:2010cr} & 2 & 270 & \\
[1.0ex] \hline  \hline \\[-1.0ex]
\end{tabular*}
\caption{Chiral extrapolation/minimum pion mass in the determinations of $m_c$.}
}
\end{table}

\newpage

\begin{table}
{\footnotesize
\begin{tabular*}{\textwidth}[l]{l @{\extracolsep{\fill}} c c c c l}
\hline\hline  \\[-1.0ex]
Collab. & Ref. & $\Nf$ & $L$ [fm] & $M_{\pi,\text{min}} L$ & Description
\\[1.0ex] \hline \hline \\[-1.0ex]
HPQCD 14A & \cite{Chakraborty:2014aca} & 2+1+1 & 2.5 - 5.8  & 3.7 & \\
[3.0ex] \hline \\[-1.0ex]
ETM 14& \cite{Carrasco:2014cwa} & 2+1+1 & 2.0 - 3.0  & $2.7_{\pi^0}$ ($3.3_{\pi^\pm}$) & \\
[3.0ex] \hline \\[-1.0ex]
ETM 14A & \cite{Alexandrou:2014sha} & 2+1+1 & 2.0 - 3.0 & $2.7_{\pi^0}$ ($3.3_{\pi^\pm}$) & \\
[3.0ex] \hline \\[-1.0ex]
JLQCD 15B  & \cite{Nakayama:2015hrn} & 2+1 & 2.7 & & \\ 
[1.0ex] \hline \\[-1.0ex]
$\chi$QCD 14 & \cite{Yang:2014sea} &  2+1 &  2.8 & 4.1 & \\
[1.0ex] \hline \\[-1.0ex]
HPQCD 10 & \cite{McNeile:2010ji} & 2+1 & 2.3 - 3.4 & 3.8 & \\
[1.0ex] \hline \\[-1.0ex]
HPQCD 08B & \cite{Allison:2008xk} & 2+1 & & & \\
[1.0ex] \hline \\[-1.0ex]
ALPHA 13B& \cite{Heitger:2013oaa} & 2 & 4.2 & 4.0 &\\
[1.0ex] \hline \\[-1.0ex]
ETM 11F & \cite{Jansen:2011vr} & 2 & & & cf.~ETM 10B \\
[1.0ex] \hline \\[-1.0ex]
ETM 10B & \cite{Blossier:2010cr} & 2 & $\gtrsim 2.0$ & 3.5 & \\
[1.0ex] \hline  \hline \\[-1.0ex]
\end{tabular*}
\caption{Finite volume effects in the determinations of $m_c$.}
}
\end{table}

\begin{table}[!ht]
{\footnotesize
\begin{tabular*}{\textwidth}[l]{l @{\extracolsep{\fill}} c c l}
\hline\hline \\[-1.0ex]
Collab. & Ref. & $\Nf$  & Description
\\[1.0ex] \hline \hline \\[-1.0ex]
HPQCD 14A & \cite{Chakraborty:2014aca} & 2+1+1 & Renormalization not required. \\
[1.0ex] \hline \\[-1.0ex]
ETM 14& \cite{Carrasco:2014cwa} & 2+1+1 & Non-perturbative renormalization (RI/MOM). \\
[1.0ex] \hline \\[-1.0ex]
ETM 14A & \cite{Alexandrou:2014sha} & 2+1+1 & Non-perturbative renormalization (RI/MOM). \\
[1.0ex] \hline \\[-1.0ex]
JLQCD 15B  & \cite{Nakayama:2015hrn} & 2+1 & Renormalization not required. \\ [1.0ex] \hline
\\[-1.0ex]
$\chi$QCD 14 & \cite{Yang:2014sea} &  2+1 & Non-perturbative renormalization (RI/MOM). \\
[1.0ex] \hline \\[-1.0ex]
HPQCD 10 & \cite{McNeile:2010ji} & 2+1 &  Renormalization not required. \\
[1.0ex] \hline \\[-1.0ex]
HPQCD 08B & \cite{Allison:2008xk} & 2+1 & Renormalization not required. \\
[1.0ex] \hline \\[-1.0ex]
ALPHA 13B& \cite{Heitger:2013oaa} & 2 & \parbox[t]{6.5cm}{Non-perturbative renormalization (RI/MOM) plus 1-loop PT estimate for the improvement b-coefficients.} \\
[7.0ex] \hline \\[-1.0ex]
ETM 11F & \cite{Jansen:2011vr} & 2 & Renormalization not required. \\
[1.0ex] \hline \\[-1.0ex]
ETM 10B & \cite{Blossier:2010cr} & 2 & Non-perturbative renormalization (RI/MOM). \\
[1.0ex] \hline  \hline \\[-1.0ex]
\end{tabular*}
\caption{Renormalization in the determinations of $m_c$.}
}
\end{table}

%%% Local Variables: 
%%% mode: latex
%%% TeX-master: "FLAG_master"
%%% End: 

%% file: qmass/qmass_mb_app.tex
\clearpage

\begin{table}[!ht]
{\footnotesize
\begin{tabular*}{\textwidth}[l]{l @{\extracolsep{\fill}} c c c l}
\hline\hline \\[-1.0ex]
Collab. & Ref. & $\Nf$ & $a$ [fm] & Description
\\[1.0ex] \hline \hline \\[-1.0ex]
% 2+1+1
HPQCD 14B  & \cite{Colquhoun:2014ica} & 2+1+1 & 0.09, 0.12, 0.15 & \parbox[t]{4.5cm}{Scale set through the $\Upsilon^\prime-\Upsilon$ mass splitting.} \\
[4.0ex] \hline \\[-1.0ex]
ETM 14B & \cite{Bussone:2014cha} & 2+1+1 & 0.062, 0.082, 0.089 & \parbox[t]{4.5cm}{Scale set through $F_\pi$.}\\
[1.0ex] \hline \\[-1.0ex]
HPQCD 14A  & \cite{Chakraborty:2014aca} & 2+1+1 & 0.06, 0.09, 0.12, 0.15  & \parbox[t]{4.5cm}{Scale set through the Wilson flow parameter $w_0$.} \\
[4.0ex] \hline \\[-1.0ex]
%2+1
HPQCD 13B  & \cite{Lee:2013mla} & 2+1 & 0.084, 0.12 & \parbox[t]{4.5cm}{Scale set through the static-quark potential parameter $r_1$.} \\
[4.0ex] \hline \\[-1.0ex]
HPQCD 10 & \cite{McNeile:2010ji} & 2+1 & 0.044, 0.059, 0.084, 0.12, 0.15 & \parbox[t]{4.5cm}{Scale set through the static-quark potential parameter $r_1$.} \\
[4.0ex] \hline \\[-1.0ex]
% 2
ETM 13B & \cite{Carrasco:2013zta} & 2 & 0.054, 0.067, 0.085, 0.098 & \parbox[t]{4.5cm}{Scale set through the static-quark potential parameter $r_0$.} \\
[4.0ex] \hline \\[-1.0ex]
ALPHA 13C & \cite{Bernardoni:2013xba} & 2 & 0.048, 0.065, 0.075 & \parbox[t]{4.5cm}{Scale set through $F_K$.} \\
[1.0ex] \hline \\[-1.0ex]
ETM 11A & \cite{Dimopoulos:2011gx} & 2 & 0.054, 0.067, 0.085, 0.098 & \parbox[t]{4.5cm}{Scale set through $F_\pi$.} \\
[1.0ex] \hline \hline \\[-1.0ex]
\end{tabular*}
\caption{Continuum extrapolations/estimation of lattice artifacts in the determinations of $m_b$.}
}
\end{table}

\begin{table}[!ht]
{\footnotesize
\begin{tabular*}{\textwidth}[l]{l @{\extracolsep{\fill}} c c c}
\hline \hline  \\[-1.0ex]
Collab. & Ref. & $\Nf$ & $M_{\pi,\text{min}}$ [MeV] 
\\[1.0ex] \hline \hline \\[-1.0ex]
% 2+1+1
HPQCD 14B  & \cite{Colquhoun:2014ica} & 2+1+1 & 306,  128 \\
[1.0ex] \hline \\[-1.0ex]
ETM 14B & \cite{Bussone:2014cha} & 2+1+1 &  210  \\
[1.0ex] \hline \\[-1.0ex]
HPQCD 14A  & \cite{Chakraborty:2014aca} & 2+1+1 & $128_{\pi, 5}$ ($173_{RMS}$) \\
[1.0ex] \hline \\[-1.0ex]
% 2+1
HPQCD 13B  & \cite{Lee:2013mla} & 2+1 & 345  \\
[1.0ex] \hline \\[-1.0ex]
HPQCD 10 & \cite{McNeile:2010ji} & 2+1 & 260  \\
[1.0ex] \hline \\[-1.0ex]
% 2
ETM 13B & \cite{Carrasco:2013zta} & 2 & 280  \\
[1.0ex] \hline \\[-1.0ex]
ALPHA 13C & \cite{Bernardoni:2013xba} & 2 &  190 \\
[1.0ex] \hline \\[-1.0ex]
ETM 11A & \cite{Dimopoulos:2011gx} & 2 & 280   \\
[1.0ex] \hline \hline \\[-1.0ex]
\end{tabular*}
\caption{Chiral extrapolation/minimum pion mass in the determinations of $m_b$.}
}
\end{table}

\newpage

\begin{table}[!ht]
{\footnotesize
\begin{tabular*}{\textwidth}[l]{l @{\extracolsep{\fill}} c c c c}
\hline\hline  \\[-1.0ex]
Collab. & Ref. & $\Nf$ & $L$ [fm] & $M_{\pi,\text{min}} L$ 
\\[1.0ex] \hline \hline \\[-1.0ex]
% 2+1+1
HPQCD 14B  & \cite{Colquhoun:2014ica} & 2+1+1 & 2.4-7.8 & 3.0-3.8 \\
[1.0ex] \hline \\[-1.0ex]
ETM 14B & \cite{Bussone:2014cha} & 2+1+1 &  1.9-2.8 & 3.0-5.8   \\
[1.0ex] \hline \\[-1.0ex]
HPQCD 14A  & \cite{Chakraborty:2014aca} & 2+1+1 & 2.5-5.8  & 3.7  \\
[1.0ex] \hline \\[-1.0ex]
% 2+1
HPQCD 13B  & \cite{Lee:2013mla} & 2+1 & 2.4, 3.4 & 4.1  \\
[1.0ex] \hline \\[-1.0ex]
HPQCD 10 & \cite{McNeile:2010ji} & 2+1 & 2.3 - 3.4 & 3.8  \\
[1.0ex] \hline \\[-1.0ex]
% 2
ETM 13B & \cite{Carrasco:2013zta} & 2 & $\gtrsim 2.0$ & 3.5   \\
[1.0ex] \hline \\[-1.0ex]
ALPHA 13C& \cite{Bernardoni:2013xba} & 2 & 2.3-3.6 & 4.1 \\
[1.0ex] \hline \\[-1.0ex]
ETM 11A & \cite{Dimopoulos:2011gx} & 2 & $\gtrsim 2.0$ & 3.5  \\
[1.0ex] \hline \hline \\[-1.0ex]
\end{tabular*}
\caption{Finite volume effects in the determinations of $m_b$.}
}
\end{table}
 
\clearpage

\begin{table}[!ht]
{\footnotesize
\begin{tabular*}{\textwidth}[l]{l @{\extracolsep{\fill}} c c l}
\hline\hline \\[-1.0ex]
Collab. & Ref. & $\Nf$  & Description
\\[1.0ex] \hline \hline \\[-1.0ex]
%
% 2+1+1
HPQCD 14B & \cite{Colquhoun:2014ica} & 2+1+1 & \parbox[t]{4.5cm}{Renormalization not required.}\\
[1.0ex] \hline \\[-1.0ex]
ETM 14B& \cite{Bussone:2014cha} & 2+1+1 & \parbox[t]{4.5cm}{Non-perturbative renormalization (RI/MOM). }\\
[4.0ex] \hline \\[-1.0ex]
HPQCD 14A & \cite{Chakraborty:2014aca} & 2+1+1 & \parbox[t]{4.5cm}{Renormalization not required. }\\
[1.0ex] \hline \\[-1.0ex]
% 2+1
HPQCD 13B  & \cite{Lee:2013mla} & 2+1 & \parbox[t]{4.5cm}{Renormalization not required.} \\
[1.0ex] \hline \\[-1.0ex]
HPQCD 10 & \cite{McNeile:2010ji} & 2+1 &  \parbox[t]{4.5cm}{Renormalization not required.} \\
[1.0ex] \hline \\[-1.0ex]
% 2
ETM 13B & \cite{Carrasco:2013zta} & 2 &  \parbox[t]{4.5cm}{Non-perturbative renormalization (RI/MOM).}  \\
[4.0ex] \hline \\[-1.0ex]
ALPHA 13C& \cite{Bernardoni:2013xba} & 2 & \parbox[t]{4.5cm}{Non-perturbatively matched and renormalized HQET.} \\
[4.0ex] \hline \\[-1.0ex]
ETM 11A & \cite{Dimopoulos:2011gx} & 2 & \parbox[t]{4.5cm}{Renormalization not required.}\\
[1.0ex] \hline \hline \\[-1.0ex]
\end{tabular*}
\caption{Lattice renormalization in the determinations of $m_b$.}
}
\end{table}

%%% Local Variables: 
%%% mode: latex
%%% TeX-master: "FLAG_master"
%%% End: 

%% file: Vudus/VudVus_app.tex
%\documentclass[a4paper,color]{article}

%\begin{document}

\subsection{Notes to section \ref{sec:vusvud} on $|V_{ud}|$ and  $|V_{us}|$}
\label{app:VusVud}

\begin{table}[!h]
{\footnotesize
\begin{tabular*}{\textwidth}[l]{l @{\extracolsep{\fill}} c c c l}
\hline\hline \\[-1.0ex]
Collab. & Ref. & $\Nf$ & $a$ [fm] & Description
\\[1.0ex] \hline \hline \\[-2.5ex]
{ETM 15C}& \cite{Carrasco:2015wzu} & 2+1+1 & 0.062, 0.082, 0.089  &
		    \parbox[t]{5.9cm}{ 
		    Scale set through $f_\pi$. Automatic $\cO(a)$ improvement.} \\
[3.0ex] \hline  \\[-2.5ex]
FNAL/MILC 13E &\cite{Bazavov:2013maa}			&2+1+1&0.06, 0.09, 0.12, 0.15&
	\parbox[t]{5.9cm}{HISQ action for both sea and valence quarks. Relative scale through $r_1$, physical scale from 
	         pseudoscalar decay constants calculated with Asqtad fermions. The ensemble with $a \simeq 0.06$~fm 
	         is used only for cross-checking discretization effects.}\\
[17.0ex] \hline \\[-2.5ex]
FNAL/MILC 13C &\cite{Gamiz:2013xxa}			&2+1+1&0.09, 0.12, 0.15&
	\parbox[t]{5.9cm}{Relative scale through $r_1$, physical scale from 
	         $f_\pi$ calculated by MILC 09A at $N_f = 2 + 1$.}\\
\hline\hline
\end{tabular*}
\caption{Continuum extrapolations/estimation of lattice artifacts in the
  determinations of $f_+(0)$.}
}
\end{table}

\noindent
\begin{table}[!h]
{\footnotesize
\begin{tabular*}{\textwidth}[l]{l @{\extracolsep{\fill}} c c c l}
\hline \hline  \\[-1.0ex]
Collab. & Ref. & $\Nf$ & $M_{\pi,\text{min}}$ [MeV] & Description
\\[1.0ex] \hline \hline \\[-2.5ex]
{ETM 15C}& \cite{Carrasco:2015wzu} & 2+1+1 & $180_{\rm \pi^0}(220_{\pi^\pm})$  & 
          \parbox[t]{6.3cm}{Chiral extrapolation performed through $SU(2)$ or $SU(3)$ {\Ch}PT.}\\
[3.0ex] \hline  \\[-2.5ex]
FNAL/MILC 13E &\cite{Bazavov:2013maa}			&2+1+1&$173_{\rm RMS}(128_{\rm \pi,5})$&
		\parbox[t]{6.3cm}{NLO $SU(3)$ PQ staggered {\Ch}PT with
		        continuum {\Ch}PT at NNLO. Lightest Nambu-Goldstone 
		        mass is 128~MeV and lightest RMS mass is 173~MeV
			for the same gauge ensemble with $a \simeq 0.09$~fm.
			}\\
[12.0ex] \hline \\[-2.5ex]
FNAL/MILC 13C &\cite{Gamiz:2013xxa}			&2+1+1&$173_{\rm RMS}(128_{\rm \pi,5})$&
		\parbox[t]{6.3cm}{NLO $SU(3)$ PQ staggered {\Ch}PT with
		        continuum {\Ch}PT at NNLO. Lightest Nambu-Goldstone 
		        mass is 128~MeV and lightest RMS mass is 173~MeV
			for the same gauge ensemble with $a \simeq 0.09$~fm.
			}\\
\hline\hline
\end{tabular*}
\caption{Chiral extrapolation/minimum pion mass in determinations of $f_+(0)$.
The subscripts RMS and $\pi,5$ in the case of staggered fermions indicate
the root-mean-square mass and the Nambu-Goldstone boson mass, respectively. 
In the case
of twisted-mass fermions $\pi^0$ and $\pi^\pm$ indicate the neutral and
charged pion mass where applicable.}
}
\end{table}

\noindent
\begin{table}[!h]
{\footnotesize
\begin{tabular*}{\textwidth}[l]{l @{\extracolsep{\fill}} c c c c l}
\hline \hline  \\[-1.0ex]
Collab. & Ref. & $\Nf$ &$L$ [fm]& $M_{\pi,\text{min}}L$ & Description
\\[1.0ex] \hline \hline \\[-2.5ex]
{ETM 15C}& \cite{Carrasco:2015wzu} & 2+1+1 & 2.0--3.0 & $2.7_{\pi^0}(3.3_{\pi^\pm})$ &
        \parbox[t]{5.2cm}{FSE observed only in the slopes of the vector and scalar form factors.} \\
[3.0ex]\hline  \\[-2.5ex]
FNAL/MILC 13E &\cite{Bazavov:2013maa} 		&2+1+1&2.9--5.8&$4.9_{\rm RMS}(3.6_{\pi,5})$& \parbox[t]{5cm}{The values correspond to $M_{\pi,\rm RMS}=173$~MeV and 
		$M_{\pi,5}=128$~MeV, respectively.}
		\parbox[t]{5.2cm}{}\\
[6.0ex] \hline \\[-2.5ex]
FNAL/MILC 13C &\cite{Gamiz:2013xxa} 		&2+1+1&2.9--5.8&$4.9_{\rm RMS}(3.6_{\pi,5})$& \parbox[t]{5cm}{The values correspond to $M_{\pi,\rm RMS}=173$~MeV and 
		$M_{\pi,5}=128$~MeV, respectively.}
		\parbox[t]{5.2cm}{}\\
\hline\hline
\end{tabular*}
\caption{Finite volume effects in determinations of $f_+(0)$. The subscripts RMS and $\pi,5$ in the case of staggered fermions indicate
the root-mean-square mass and the Nambu-Goldstone boson mass, respectively. 
In the case
of twisted-mass fermions $\pi^0$ and $\pi^\pm$ indicate the neutral and
charged pion mass where applicable.}
}
\end{table}

\begin{table}[!h]
{\footnotesize
\begin{tabular*}{\textwidth}[l]{l @{\extracolsep{\fill}} c c c l}
\hline \hline  \\[-1.0ex]
Collab. & Ref. & $\Nf$ & $a$ [fm] & Description
\\[1ex] \hline \hline \\[-2.5ex]
{ETM 14E}& \cite{Carrasco:2014poa} & 2+1+1 & 0.062, 0.082, 0.089  &
		    \parbox[t]{5.7cm}{ 
		    Scale set through $f_\pi$. Automatic $\cO(a)$ improvement, flavour symmetry
		    breaking: $(M_{PS}^{0})^2-(M_{PS}^\pm)^2\sim \cO(a^2)$. Discretization and volume effects
		    due to the $\pi^0 - \pi^\pm$ mass splitting are taken into account through {\Ch}PT for
		    twisted-mass fermions.} \\
[17.0ex] \hline  \\[-2.5ex]
{FNAL/MILC 14A}& \cite{Bazavov:2014wgs} & 2+1+1 & 0.06,\,0.09,\,0.12,\,0.15  &  
	\parbox[t]{5.7cm}{ 
	HISQ action for both valence and sea quarks. Absolute scale though $f_\pi$.}\\
[3.0ex]  \hline  \\[-2.5ex]
{HPQCD 13A}& \cite{Dowdall:2013rya} & 2+1+1 & 0.09,\,0.12,\,0.15  &  
	\parbox[t]{5.7cm}{ 
	Relative scale through Wilson flow and absolute scale through $f_\pi$.}\\
\hline \hline 
\end{tabular*}
\caption{Continuum extrapolations/estimation of lattice artifacts in
  determinations of $f_K/f\pi$ for $N_f=2+1+1$ simulations.}
}
\end{table}

\begin{table}[!h]
{\footnotesize
\begin{tabular*}{\textwidth}[l]{l @{\extracolsep{\fill}} c c c l}
\hline \hline  \\[-1.0ex]
Collab. & Ref. & $\Nf$ & $a$ [fm] & Description
\\[1ex] \hline \hline \\[-2.5ex]
{RBC/UKQCD 14B}& \cite{Blum:2014tka} & 2+1 & 0.063,\, 0.085,\, 0.114  &  
	\parbox[t]{5.7cm}{ 
	Scale set through $m_\Omega$.}\\
\hline  \\[-2.5ex]
{RBC/UKQCD 12}& \cite{Arthur:2012opa} & 2+1 & 0.09,\, 0.11,\, 0.14  &  
	\parbox[t]{5.7cm}{ 
	Scale set through $m_\Omega$.}\\
\hline  \\[-2.5ex]
{Laiho 11}& \cite{Laiho:2011np} & 2+1 & 0.06, 0.09, 0.125 &  
	\parbox[t]{5.7cm}{Scale set through $r_1$ and $\Upsilon$ and 
	continuum extrapolation based on MA{\Ch}PT.}\\
\hline\hline
\end{tabular*}
\caption{Continuum extrapolations/estimation of lattice artifacts in
  determinations of $f_K/f\pi$ for $N_f=2+1$ simulations.}
}
\end{table}

\begin{table}[!h]
{\footnotesize
\begin{tabular*}{\textwidth}[l]{l @{\extracolsep{\fill}} c c c l}
\hline \hline  \\[-1.0ex]
Collab. & Ref. & $\Nf$ & $a$ [fm] & Description
\\[1ex] \hline \hline \\[-2.5ex]
ETM 14D		&\cite{Abdel-Rehim:2014nka}		&2  & 0.094 &\parbox[t]{5.7cm}{Scale set through $F_\pi$, $r_0$, $t_0$ and $w_0$. Twisted Wilson fermions plus clover term. Automatic $\cO(a)$ improvement.}\\
[6.0ex] \hline \\[-2.5ex]
ALPHA 13A		&\cite{Lottini:2013rfa}		&2  & 	\parbox[t]{2.5cm}{\centering 0.05, 0.065, 0.075} &\parbox[t]{5.7cm}{Scale set through $F_\pi$.	$\cO(a)$-improved Wilson action.}\\
[3.0ex] \hline \\[-2.5ex]
BGR 11        &\cite{Engel:2011aa}                &2  & 0.135 &
                   \parbox[t]{5.7cm}{Scale set through $r_0 = 0.48$ fm.
Chirally improved Dirac operator.}\\
[3.0ex] \hline \\[-2.5ex]
ETM 10D		&\cite{Lubicz:2010bv}		&2  & 	\parbox[t]{2.5cm}{\centering 0.05, 0.07,\\ 0.09, 0.10} &\parbox[t]{5.7cm}{Scale set through $F_\pi$.	Automatic $\cO(a)$ impr., flavour symmetry breaking: $(M_{PS}^{0})^2-(M_{PS}^\pm)^2\sim \cO(a^2)$.}\\
[6.0ex] \hline \\[-2.5ex]
ETM 09		&\cite{Blossier:2009bx}		&2  & 0.07, 0.09, 0.10 & 
		\parbox[t]{5.7cm}{Scale set through $F_\pi$.
		Automatic $\cO(a)$ impr., flavour symmetry
		breaking: $(M_{PS}^{0})^2-(M_{PS}^\pm)^2\sim \cO(a^2)$. 
		  }\\
[6.0ex] \hline \\[-2.5ex]
QCDSF/UKQCD 07 	&\cite{QCDSFUKQCD}	&2  & 0.06, 0.07 & 
		\parbox[t]{5.7cm}{Scale set through $F_\pi$. Non-perturbative
		$\cO(a)$-improvement.}\\
[3.0ex]
\hline\hline
\end{tabular*}
\caption{Continuum extrapolations/estimation of lattice artifacts in
  determinations of $f_K/f\pi$ for $N_f=2$ simulations.}
}
\end{table}

\vfill
\noindent
\begin{table}[!h]
{\footnotesize
\begin{tabular*}{\textwidth}[l]{l @{\extracolsep{\fill}} c c c l}
\hline \hline  \\[-1.0ex]
Collab. & Ref. & $\Nf$ & $M_{\pi,\text{min}}$ [MeV] & Description
\\[1ex] \hline \hline \\[-2.5ex]
{ETM 14E}& \cite{Carrasco:2014poa} & 2+1+1 & $180_{\rm \pi^0}(220_{\pi^\pm})$  & 
          \parbox[t]{6.2cm}{Chiral extrapolation performed through $SU(2)$ {\Ch}PT or polynomial fit.}\\
[3.0ex]\hline  \\[-2.5ex]
{FNAL/MILC 14A}& \cite{Bazavov:2014wgs} & 2+1+1 & $143_{\rm RMS}(128_{\pi,5})$  &  
	\parbox[t]{6.2cm}{ 
	Linear interpolation to physical point. The lightest RMS mass is from the 
	$a=0.06$~fm ensemble and the lightest Nambu-Goldstone mass is from the
	$a=0.09$~fm ensemble.}\\ 
[9.0ex]\hline  \\[-2.5ex]
{HPQCD 13A}& \cite{Dowdall:2013rya} & 2+1+1 &$173_{\rm RMS}(128_{\pi,5})$  &  
	\parbox[t]{6.2cm}{ 
	NLO $\chi$PT supplemented by model for NNLO. Both the lightest RMS and 
	the lightest
	Nambu-Goldstone mass are from the $a=0.09$~fm ensemble.}\\
[9.0ex] \hline  \\[-2.5ex]
{MILC 13A}& \cite{Bazavov:2013cp} & 2+1+1 & $143_{\rm RMS}(128_{\pi,5})$  &  
	\parbox[t]{6.2cm}{ 
	Linear interpolation to physical point. The lightest RMS mass is from the 
	$a=0.06$~fm ensemble and the lightest Nambu-Goldstone mass is from the
	$a=0.09$~fm ensemble.}\\ 
[9.0ex]\hline  \\[-2.5ex]
{ETM 13F}& \cite{Dimopoulos:2013qfa} & 2+1+1 & $180_{\rm \pi^0}(220_{\pi^\pm})$  & 
          \parbox[t]{6.2cm}{Chiral extrapolation performed through $SU(2)$ {\Ch}PT or polynomial fit.}\\
[3.0ex]\hline  \\[-2.5ex]
{ETM 10E}& \cite{Farchioni:2010tb} & 2+1+1 & $215_{\rm \pi^0}(265_{\pi^\pm})$  & \parbox[t]{6.2cm}{}\\
\hline  \\[-2.5ex]
{MILC 11}& \cite{Bazavov:2011fh} & 2+1+1 &$173_{\rm RMS}(128_{\pi,5})$ &\parbox[t]{6.2cm}{
		Quoted result from polynomial interpolation to the physical 
		point. The lightest RMS mass is from the 
	$a=0.06$~fm ensemble and lightest the Nambu-Goldstone mass is from the
	$a=0.09$~fm ensemble.}\\[12.0ex]
\hline\hline
\end{tabular*}
\caption{Chiral extrapolation/minimum pion mass in determinations of $f_K / f\pi$ for $N_f=2+1+1$ simulations. 
The subscripts RMS and $\pi,5$ in the case of staggered fermions indicate
the root-mean-square mass and the Nambu-Goldstone boson mass. In the case
of twisted-mass fermions $\pi^0$ and $\pi^\pm$ indicate the neutral and charged pion mass 
and, where applicable, ``val'' and ``sea'' indicate valence and sea pion masses.}
}
\end{table}

\noindent
\vspace{-3cm}
\begin{table}[!h]
{\footnotesize
\begin{tabular*}{\textwidth}[l]{l @{\extracolsep{\fill}} c c c l}
\hline \hline  \\[-1.0ex]
Collab. & Ref. & $\Nf$ & $M_{\pi,\text{min}}$ [MeV] & Description
\\[1ex] \hline \hline \\[-2.5ex]
{RBC/UKQCD 14B}& \cite{Blum:2014tka} & 2+1 & 139 &\parbox[t]{6.2cm}{
		NLO PQ $SU(2)$ $\chi$PT as well as analytic ans\"atze.}\\
[3.0ex] \hline \\[-2.5ex]
{RBC/UKQCD 12}& \cite{Arthur:2012opa} & 2+1 & 171$_{\rm sea}$, 143$_{\rm val}$ &\parbox[t]{6.2cm}{
		NLO PQ $SU(2)$ $\chi$PT as well as analytic ans\"atze.}\\
[3.0ex] \hline \\[-2.5ex]
{Laiho 11}& \cite{Laiho:2011np} & 2+1 & 250$_{\rm RMS}(220_{\pi,5})$ &
	\parbox[t]{6.2cm}{NLO MA{\Ch}PT.}\\
\end{tabular*}
\vskip -0.5cm
\caption{Chiral extrapolation/minimum pion mass in determinations of $f_K/f\pi$ for $N_f=2+1$ simulations. The subscripts RMS and $\pi,5$ in the case of staggered fermions indicate
the root-mean-square mass and the Nambu-Goldstone boson mass. In the case
of twisted-mass fermions $\pi^0$ and $\pi^\pm$ indicate the neutral and
charged pion mass and where applicable, ``val'' and ``sea'' indicate valence
and sea pion masses.}
}
\end{table}

\noindent
\vspace{-3cm}
\begin{table}[!h]
\addtocounter{table}{-1}
{\footnotesize
\begin{tabular*}{\textwidth}[l]{l @{\extracolsep{\fill}} c c c l}
\hline \hline  \\[-1.0ex]
Collab. & Ref. & $\Nf$ & $M_{\pi,\text{min}}$ [MeV] & Description
\\[1ex] \hline \hline \\[-2.5ex]
Aubin 08	&\cite{Aubin:2008ie}	&2+1&$329_{\rm RMS}(246_{\pi,5})$& 
		\parbox[t]{6.2cm}{NLO MA{\Ch}PT. According to \cite{Bazavov:2009fk}
		the lightest sea Nambu-Goldstone mass is 246\,MeV (at $a=0.09$ fm)
		and the lightest RMS mass is 329\,MeV (at $a=0.09$ fm).}\\
[9.0ex] \hline \\[-2.5ex]
PACS-CS 08, 08A &\cite{Aoki:2008sm,Kuramashi:2008tb}& 2+1 &156& 
		\parbox[t]{6.2cm}{NLO $SU(2)$ {\Ch}PT and $SU(3)$ (Wilson){\Ch}PT.}\\
\hline \\[-2.5ex]
HPQCD/UKQCD 07	&\cite{Follana:2007uv}	&2+1&$375_{\rm RMS}(263_{\pi,5})$& 
		\parbox[t]{6.2cm}{NLO $SU(3)$ chiral perturbation theory
		with NNLO and NNNLO analytic terms. 
		The lightest RMS mass is from the $a=0.09$~fm ensemble and 
		the lightest Nambu-Goldstone mass is from the 
		$a=0.12$~fm ensemble.}\\
[12.0ex] \hline \\[-2.5ex]
RBC/UKQCD 08	&\cite{Allton:2008pn}	&2+1&$330_{\rm sea}$, $242_{\rm val}$& 
		\parbox[t]{6.2cm}{While $SU(3)$ PQ{\Ch}PT fits were studied,
		final results are based on 
		heavy kaon NLO $SU(2)$ PQ{\Ch}PT. }\\
[6.0ex] \hline \\[-2.5ex]
NPLQCD 06	&\cite{Beane:2006kx}	&2+1&300& 
		\parbox[t]{6.2cm}{NLO $SU(3)$ {\Ch}PT and some NNLO terms. 
		The sea RMS mass for the employed lattices is heavier.}\\
[6.0ex] \hline \\[-2.5ex]
MILC 04	&\cite{Aubin:2004fs}	 &2+1& $400_{\rm RMS}(260_{\pi,5})$ & \parbox[t]{6.2cm}{PQ RS{\Ch}PT fits. The lightest sea Nambu-Goldstone mass is 260\,MeV
		(at $a=0.12$ fm) and the lightest RMS mass is 400\,MeV (at $a=0.09$ fm).} \\
[9ex]\hline\hline\\
\end{tabular*}
\vskip -0.5cm
\caption{(cntd.) Chiral extrapolation/minimum pion mass in determinations of $f_K/f\pi$ for $N_f=2+1$ simulations. The subscripts RMS and $\pi,5$ in the case of staggered fermions indicate
the root-mean-square mass and the Nambu-Goldstone boson mass. In the case
of twisted-mass fermions $\pi^0$ and $\pi^\pm$ indicate the neutral and
charged pion mass and where applicable, ``val'' and ``sea'' indicate valence
and sea pion masses.}
}
\end{table}

\begin{table}[!h]
{\footnotesize
\begin{tabular*}{\textwidth}[l]{l @{\extracolsep{\fill}} c c c l}
\hline \hline  \\[-1.0ex]
Collab. & Ref. & $\Nf$ & $M_{\pi,\text{min}}$ [MeV] & Description
\\[1ex] \hline \hline \\[-2.5ex]
ETM 14D         & \cite{Abdel-Rehim:2014nka}  &2  & 140 & \parbox[t]{6.2cm}{Charged/neutral pion mass breaking, $M_{\pi^\pm}^2-M_{\pi^0}^2 \sim \cO(a^2)$, estimated to be $\simeq 20$ MeV.}
\\
[6ex] \hline \\[-2.5ex]
ALPHA 13A		&\cite{Lottini:2013rfa}	&2  &190& 
                 \parbox[t]{6.2cm}{NLO $SU(3)$ {\Ch}PT and phenomenological ansatz for higher orders.}\\
[3.0ex] \hline \\[-2.5ex]
BGR 11        &\cite{Engel:2011aa}        &2  &250&
                  \parbox[t]{6.2cm}{NLO $SU(2)$ {\Ch}PT. Strange quark
mass fixed by reproducing the $\Omega$ mass.}\\
[3.0ex] \hline \\[-2.5ex]
ETM 10D         & \cite{Lubicz:2010bv}  &2  &$210_{\pi^0}(260_{\pi^\pm})$&\parbox[t]{6.2cm}{
		NLO $SU(3)$ {\Ch}PT and phenomenological ansatz for higher orders.
		Joint $f_+(0)$-$f_K/f_\pi$-fit.}\\
[6ex] \hline \\[-2.5ex]
ETM 09		&\cite{Blossier:2009bx}		&2  &$210_{\pi^0}(260_{pi^\pm})$& 
		\parbox[t]{6.2cm}{NLO heavy meson $SU(2)$ {\Ch}PT and NLO
		$SU(3)$ {\Ch}PT.}\\
[3.0ex] \hline \\[-2.5ex]
QCDSF/UKQCD 07 	&\cite{QCDSFUKQCD}	&2  &300& 
		\parbox[t]{6.2cm}{Linear extrapolation of lattice data.}\\
\hline\hline\\
\end{tabular*}
\vskip -0.5cm
\caption{Chiral extrapolation/minimum pion mass in determinations of $f_K/f\pi$ for $N_f=2$ simulations. The subscripts RMS and $\pi,5$ in the case of staggered fermions indicate
the root-mean-square mass and the Nambu-Goldstone boson mass. In the case
of twisted-mass fermions $\pi^0$ and $\pi^\pm$ indicate the neutral and
charged pion mass and where applicable, ``val'' and ``sea'' indicate valence
and sea pion masses.}
}
\end{table}

\noindent
\begin{table}[!h]
{\footnotesize
\begin{tabular*}{\textwidth}[l]{l @{\extracolsep{\fill}} c c c c l}
\hline \hline  \\[-1.0ex]
Collab. & Ref. & $\Nf$ &$L$ [fm]& $M_{\pi,\text{min}}L$ & Description
\\[1ex] \hline \hline \\[-2.5ex]
{ETM 14E}& \cite{Carrasco:2014poa} & 2+1+1 & 2.0 - 3.0 & $2.7_{\pi^0}(3.3_{\pi^\pm})$ &
        \parbox[t]{4.7cm}{FSE for the pion is corrected through resummed NNLO {\Ch}PT for twisted-mass fermions,
        which takes into account the effects due to the $\pi^0 - \pi^\pm$ mass splitting.} \\
[12.0ex] \hline \\[-2.5ex]
{FNAL/MILC 14A}& \cite{Bazavov:2014wgs} & 2+1+1 &2.8-5.8 & $3.9_{\rm RMS}(3.7_{\pi,5}) $ &  
	\parbox[t]{4.7cm}{ }\\
[0.0ex] \hline  \\[-2.5ex]
{HPQCD 13A}& \cite{Dowdall:2013rya} & 2+1+1 &2.5-5.8& $4.9_{\rm RMS}(3.7_{\pi,5})$  &  
	\parbox[t]{4.7cm}{ }\\
[0.0ex] \hline  \\[-2.5ex]
{MILC 13A}& \cite{Bazavov:2013cp} & 2+1+1 &2.8-5.8 & $3.9_{\rm RMS}(3.7_{\pi,5}) $ &  
	\parbox[t]{4.7cm}{ }\\
[0.0ex] \hline  \\[-2.5ex]
{ETM 13F}& \cite{Dimopoulos:2013qfa} & 2+1+1 & 2.0 - 3.0 & $2.7_{\pi^0}(3.3_{\pi^\pm})$ &
        \parbox[t]{4.7cm}{FSE for the pion is corrected through resummed NNLO {\Ch}PT for twisted-mass fermions,
        which takes into account the effects due to the $\pi^0 - \pi^\pm$ mass splitting.} \\
[12.0ex] \hline \\[-2.5ex]
{ETM 10E}& \cite{Farchioni:2010tb} & 2+1+1 & 1.9 - 2.9 &  $
3.1_{\pi^0}(3.9_{\pi^\pm})$ &\parbox[t]{4.7cm}{Simulation parameters from \cite{Baron:2011sf,
Herdoiza_private}.} \\
[3.0ex] \hline \\[-2.5ex]
{MILC 11}& \cite{Bazavov:2011fh} & 2+1+1 & 5.6,\,5.7 &  $4.9_{\rm RMS}(3.7_{\pi,5})$ & \\
[1.0ex] \hline \hline\\[-2.5ex]
\end{tabular*}
\caption{Finite volume effects in determinations of $f_K/f\pi$ for $N_f=2+1+1$.
The subscripts RMS and $\pi,5$ in the case of staggered fermions indicate
the root-mean-square mass and the Nambu-Goldstone boson mass. In the case
of twisted-mass fermions $\pi^0$ and $\pi^\pm$ indicate the neutral and
charged pion mass and where applicable, ``val'' and ``sea'' indicate valence
and sea pion masses.}
}
\end{table}

\begin{table}[!h]
{\footnotesize
\begin{tabular*}{\textwidth}[l]{l @{\extracolsep{\fill}} c c c c l}
\hline \hline  \\[-1.0ex]
Collab. & Ref. & $\Nf$ &$L$ [fm]& $M_{\pi,\text{min}}L$ & Description
\\[1ex] \hline \hline \\[-2.5ex]
{RBC/UKQCD 14B}& \cite{Blum:2014tka} & 2+1 & 2.0,\,2.7,\,4.6,\,5.4 &  $3.8$ & 
\\
[0.0ex] \hline \\[-2.5ex]
{RBC/UKQCD 12}& \cite{Arthur:2012opa} & 2+1 & 2.7,\,4.6 &  $3.3$ & 
\parbox[t]{4.5cm}{For partially quenched $M_\pi=143$MeV, $M_\pi L=3.3$ and for
unitary $M_\pi=171$MeV, $M_\pi L=4.0$.}\\
[9.0ex] \hline \\[-2.5ex]
{Laiho 11}& \cite{Laiho:2011np} & 2+1 & {2.5--4.0} & $4.9_{\rm RMS}(4.3_{\pi,5})$ &\parbox[t]{4.7cm}{} \\
\hline\hline
\end{tabular*}
\caption{Finite volume effects in determinations of $f_K/f\pi$ for
  $N_f=2+1$ and $N_f=2$.
The subscripts RMS and $\pi,5$ in the case of staggered fermions indicate
the root-mean-square mass and the Nambu-Goldstone boson mass. In the case
of twisted-mass fermions $\pi^0$ and $\pi^\pm$ indicate the neutral and
charged pion mass and where applicable, ``val'' and ``sea'' indicate valence
and sea pion masses.}
}
\end{table}

%\end{document}

%% file: LEC/LECs_app.tex
\newpage
\subsection{Notes to section \ref{sec:LECs} on Low-Energy Constants}

\enlargethispage{20ex}

\begin{table}[!htp]
{\footnotesize
\begin{tabular*}{\textwidth}[l]{l @{\extracolsep{\fill}} c c l l}
\hline\hline
\\[-1.2ex]
Collab. & Ref. & $\Nf$ & $a$ [fm]& Description
\\[1.0ex]
\hline \hline
\\[-1.2ex]
%HPQCD 13A, 15B  & \cite{Koponen:2015tkr} & 2+1+1 & 0.09--0.15 & \parbox[t]{7.5cm}{Configurations are shared with HPQCD13A.}\\
%[1.0ex] \hline \\[-1.2ex]%
%
HPQCD 13A, 15B  & \cite{Dowdall:2013rya,Koponen:2015tkr} & 2+1+1 & 0.09--0.15 & \parbox[t]{5.5cm}{Configurations are shared with MILC.}\\
[1.0ex] \hline \\[-1.2ex]%
%
%ETM 13 & \cite{Cichy:2013gja} & 2+1+1 & 0.0607--0.0863 & \parbox[t]{7.5cm}{Configurations are shared with ETM 11.}\\
%[1.0ex] \hline \\[-1.2ex]%
%
ETM 11, 13 & \cite{Baron:2011sf,Cichy:2013gja} & 2+1+1 & 0.0607--0.0863 & \parbox[t]{5.5cm}{3 lattice spacings fixed through $F_\pi/M_\pi$.}\\
[1.0ex] \hline\\[-1.2ex]
ETM 10 & \cite{Baron:2010bv} & 2+1+1 & 0.078, 0.086 & \parbox[t]{5.5cm}{Fixed through $F_\pi/M_\pi$.}
\\
\end{tabular*}
\caption{Continuum extrapolations/estimation of lattice artifacts in
  $\Nf=2+1+1$ and $2+1$ determinations of the Low-Energy Constants.}
}
\end{table}

\clearpage

\begin{table}[!htp]
{\footnotesize
\begin{tabular*}{\textwidth}[l]{l @{\extracolsep{\fill}} c c l l}
\hline\hline \\[-1.2ex]
Collab. & Ref. & $\Nf$ & $a$ [fm]& Description
\\[1.0ex] \hline \hline \\[-1.2ex]
ETMC 15A& \cite{Abdel-Rehim:2015pwa} & 2 & 0.0914(3)(15) & \parbox[t]{5.5cm}{Weighted average using $m_\pi$, $f_\pi$, $f_K$, $m_N$.
}\\
[4.0ex] \hline \\[-1.2ex]%
G\"ulpers 15& \cite{Gulpers:2015bba} & 2 & 0.050, 0.063, 0.079 & \parbox[t]{5.5cm}{Scale fixed through $m_{\Omega}$.
}\\
[1.0ex] \hline \\[-1.2ex]%
Engel 14& \cite{Engel:2014eea} & 2 & \parbox[t]{3.0cm}{0.0483(4), 0.0652(6), 0.0749(8)}& \parbox[t]{5.5cm}{Scale fixed through $F_K$.
}\\
\end{tabular*}
\caption{Continuum extrapolations/estimation of lattice artifacts in $\Nf=2$
  determinations of the Low-Energy Constants.}
}
\end{table}

\clearpage

%%%%%%%%%%%%%%%%%%%%%%%%%%%%%%%%%%%%%%%%%%%%%%%%%%%%%%%%%%%%%%%%%%%%%%%%%%%%%%%%

\begin{table}[!htp]
{\footnotesize
\begin{tabular*}{\textwidth}[l]{l @{\extracolsep{\fill}} c c c l}
\hline\hline \\[-1.2ex]
Collab. & Ref. & $\Nf$ & $M_{\pi,\text{min}}$ [MeV] & Description
\\[1.0ex] \hline \hline \\[-1.2ex]
HPQCD 15B  & \cite{Koponen:2015tkr} & 2+1+1 & 175 & \parbox[t]{5.2cm}{Simulated at physical point.
}\\
[1.0ex] \hline \\[-1.2ex]%
HPQCD 13A& \cite{Dowdall:2013rya} & 2+1+1 & 175 & \parbox[t]{5.2cm}{NLO chiral fit.
}\\
[1.0ex] \hline \\[-1.2ex]%
ETM 13& \cite{Cichy:2013gja} & 2+1+1 & 270 & \parbox[t]{5.2cm}{Linear fit in the quark mass.
}\\
[1.0ex] \hline \\[-1.2ex]%
ETM 11 & \cite{Baron:2011sf} & 2+1+1 & 270 & \parbox[t]{5.2cm}{NLO $SU(2)$ chiral fit.}\\
[1.0ex] \hline\\[-1.2ex]
ETM 10 & \cite{Baron:2010bv} & 2+1+1 &  $270$ & \parbox[t]{5.2cm}{$SU(2)$ NLO and NNLO fits. }\\
[1.0ex] \hline \hline
\end{tabular*}
\caption{Chiral extrapolation/minimum pion mass in $\Nf=2+1+1$ 
  determinations of the Low-Energy Constants.}
}
\end{table}

\begin{table}[!htp]
{\footnotesize
\begin{tabular*}{\textwidth}[l]{l @{\extracolsep{\fill}} c c c l}
\hline\hline \\[-1.2ex]
Collab. & Ref. & $\Nf$ & $M_{\pi,\text{min}}$ [MeV] & Description
\\[1.0ex] \hline \hline \\[-1.2ex]
RBC/UKQCD 15E & \cite{Boyle:2015exm} & 2+1 & 117.3(4.4)  & \parbox[t]{5.2cm}{GMOR for $\Sigma$, NNLO PQ $SU(2)$ $\chi$PT.}\\
[4.0ex] \hline  \\[-1.2ex]
JLQCD 15A  & \cite{Aoki:2015pba} & 2+1 & 290  & \parbox[t]{5.9cm}{Dynamical overlap, NNLO $SU(3)$.}\\
[1.0ex] \hline  \\[-1.2ex]
RBC/UKQCD 14B & \cite{Blum:2014tka} & 2+1 & 139.2  & \parbox[t]{5.9cm}{GMOR for $\Sigma$, global cont./chiral fit.}\\
\end{tabular*}
\caption{Chiral extrapolation/minimum pion mass in $2+1$
  determinations of the Low-Energy Constants.}
}
\end{table}

\clearpage

\begin{table}[!htp]
{\footnotesize
\begin{tabular*}{\textwidth}[l]{l @{\extracolsep{\fill}} c c c l}
\hline\hline
\\[-1.2ex]
Collab. & Ref. & $\Nf$ & $M_{\pi,\text{min}}$ [MeV] & Description \\
[1.0ex] \hline \hline \\[-1.2ex]
ETMC 15A& \cite{Abdel-Rehim:2015pwa} & 2 & 134 & \parbox[t]{7.5cm}{Simulation at physical point.
}\\
[1.0ex] \hline \\[-1.2ex]%
G\"ulpers 15& \cite{Gulpers:2015bba} & 2 & 193 & \parbox[t]{7.5cm}{NLO $SU(2)$ fit.
}\\
[1.0ex] \hline \\[-1.2ex]%
Engel 14& \cite{Engel:2014eea} & 2 & 193 & \parbox[t]{5.5cm}{NLO $SU(2)$ fit, Dirac op. and GMOR for $\Sigma$.
}\\
\end{tabular*}
\caption{Chiral extrapolation/minimum pion mass in $\Nf=2$ determinations of the Low-Energy Constants.}
}
\end{table}

\clearpage

%%%%%%%%%%%%%%%%%%%%%%%%%%%%%%%%%%%%%%%%%%%%%%%%%%%%%%%%%%%%%%%%%%%%%%%%%%%%%%%%

\begin{table}[!htp]
{\footnotesize
\begin{tabular*}{\textwidth}[l]{l @{\extracolsep{\fill}} c c c c l}
\hline\hline \\[-1.2ex]
Collab.\ & Ref.\ & $\Nf$ & $L$ [fm] & $M_{\pi,\text{min}}L$ & Description
\\[1.0ex] \hline \hline \\[-1.2ex]
HPQCD 15B  & \cite{Koponen:2015tkr} & 2+1+1 & & 4.8 & \parbox[t]{7.5cm}{
}\\
[1.0ex] \hline \\[-1.2ex]%
HPQCD 13A& \cite{Dowdall:2013rya} & 2+1+1 & 4.8--5.5 & 3.3 & \parbox[t]{7.5cm}{
3 volumes are compared.
}\\
[1.0ex] \hline \\[-1.2ex]%
ETM 13& \cite{Cichy:2013gja} & 2+1+1 & 1.9--2.8 & 3.0 &\parbox[t]{7.5cm}{
4 volumes compared.
}\\
\end{tabular*}
\caption{Finite volume effects in $\Nf=2+1+1$ and $2+1$ determinations of the Low-Energy Constants.}
}
\end{table}

\clearpage

\begin{table}[!htp]
{\footnotesize
\begin{tabular*}{\textwidth}[l]{l @{\extracolsep{\fill}} c c c c l}
\hline\hline \\[-1.2ex]
Collab.\ & Ref.\ & $\Nf$ & $L$ [fm] & $M_{\pi,\text{min}}L$ & Description
\\[1.0ex] \hline \hline \\[-1.2ex]
ETMC 15A& \cite{Abdel-Rehim:2015pwa} & 2 & & 4.39 & \parbox[t]{7.5cm}{ 2 volumes.
}\\
[1.0ex] \hline \\[-1.2ex]%
G\"ulpers 15& \cite{Gulpers:2015bba} & 2 & &4.09 & \parbox[t]{7.5cm}{  3 volumes, CLS confs.
}\\
[1.0ex] \hline \\[-1.2ex]%
Engel 14& \cite{Engel:2014eea} & 2 & & 4.2 & \parbox[t]{7.5cm}{ 3 volumes, CLS confs.
}\\
\end{tabular*}
\caption{Finite volume effects in $\Nf=2$ determinations of the Low-Energy Constants.}
}
\end{table}

\clearpage

%%%%%%%%%%%%%%%%%%%%%%%%%%%%%%%%%%%%%%%%%%%%%%%%%%%%%%%%%%%%%%%%%%%%%%%%%%%%%%%%

\begin{table}[!htp]
{\footnotesize
\begin{tabular*}{\textwidth}[l]{l @{\extracolsep{\fill}} c c l}
\hline\hline \\[-1.2ex]
Collab. & Ref. & $\Nf$ &  Description
\\[1.0ex] \hline \hline \\[-1.2ex]
HPQCD 15B  & \cite{Koponen:2015tkr} & 2+1+1 & ---\\
[1.0ex] \hline \\[-1.2ex]%
HPQCD 13A& \cite{Dowdall:2013rya} & 2+1+1 & ---  \\
[1.0ex] \hline \\[-1.2ex]%
ETM 10,11, 13& \cite{Baron:2010bv, Baron:2011sf, Cichy:2013gja} & 2+1+1 & Non-perturbative\\
\end{tabular*}
\caption{Renormalization in determinations of the Low-Energy Constants.}
}
\end{table}

\clearpage

%% file: BK/BK_app.tex
\subsection{Notes to section \ref{sec:BK} on Kaon mixing}
\label{app-BK}

%In the following, we summarize the characteristics (lattice actions,
%pion masses, lattice spacings, etc.) of the recent $\Nf=2+1+1$, $\Nf = 2+1$ and
%$\Nf = 2$ runs. We also provide brief descriptions of how systematic
%errors are estimated by the various authors.

\subsubsection{Kaon $B$-parameter $B_K$}
\label{app:BKSM}

%%%% a -lat. spacing %%%%%%%%%%%%%%%

\begin{table}[!ht]

{\footnotesize
\begin{tabular*}{\textwidth}{l @{\extracolsep{\fill}} c c c l}
\hline\hline \\[-1.0ex]
Collab. & Ref. & $\Nf$ & $a$ [fm] & Description 
\\[1.0ex] \hline \hline \\[-1.0ex]
%
%%% SS

ETM 15      & \cite{Carrasco:2015pra} & 2+1+1 & 0.09, 0.08, 0.06 & 
 \parbox[t]{5.7cm}{Combined chiral and continuum extrapolation. Systematic error of 2.0\% is obtained from the distribution of results over analyses which differ by  
$\cO(a^2)$ effects.}
 \\[12.0ex] \hline \\[-1.0ex] 

SWME 15A & \cite{Jang:2015sla} & 2+1 & 0.12, 0.09, 0.06, 0.045  &
\parbox[t]{5.7cm}{The three finest lattice spacings are used for the combined chiral and continnum extrapolation.  
Residual combined discretization, sea-quark extrapolation
and $\alpha_s$ matching error of 4.4\% from difference between linear
  fit in $a^2$, $m_{\rm sea}$ and a fit where $\alpha_s$ dependence is added.}
\\[18.0ex] \hline \\[-1.0ex]

RBC/UKQCD 14B & 
\cite{Blum:2014tka} & 2+1 & 
\parbox[t]{2.0cm}{0.111, 0.083, 0.063,  0.114, 0.084}  &
\parbox[t]{5.7cm}{The three first lattice spacings use different action from the 
last two ones. Combined continuum and chiral fits.}
\\[7.0ex] \hline \\[-1.0ex]

SWME 14 & \cite{Bae:2014sja} & 2+1 & 0.082, 0.059, 0.044  &
\parbox[t]{5.7cm}{Residual combined discretization and sea-quark extrapolationg
error of 0.9\% from difference between linear
  fit in $a^2$, $m_{\rm sea}$  and a constrained nine-parameter extrapolation.}
\\[12.0ex] \hline \\[-1.0ex]

SWME 13A & \cite{Bae:2013tca} & 2+1 & 0.09, 0.06, 0.045  &
\parbox[t]{5.7cm}{Residual combined discretization, sea-quark extrapolation
and $\alpha_s$ matching error of 4.4\% from difference between linear
  fit in $a^2$, $m_{\rm sea}$ and a fit where $\alpha_s$ dependence is added.}
\\[12.0ex] \hline \\[-1.0ex]

SWME 13 & \cite{Bae:2013lja} & 2+1 & 0.12, 0.09, 0.06, 0.045  &
\parbox[t]{5.7cm}{Continuum extrapolation with the coarsest lattice spacing
  omitted; residual combined discretization and sea-quark extrapolationg
error of 1.1\% from difference between linear
  fit in $a^2$, $m_{\rm sea}$ and a constrained nine-parameter extrapolation.}
\\[18.0ex] \hline \\[-1.0ex]
RBC/UKQCD 12A & 
\cite{Arthur:2012opa} & 2+1 & 0.146, 0.114, 0.087  &
\parbox[t]{5.7cm}{Coarsest lattice spacing uses different action.
Combined continuum and chiral fits.}
\\[7.0ex] \hline \hline \\[-1.0ex]
\end{tabular*}
\caption{Continuum extrapolations/estimation of lattice artifacts in
  determinations of $B_K$.}
}
\end{table}

\begin{table}[!t]

{\footnotesize
\begin{tabular*}{\textwidth}{l c c c l}
\hline\hline \\[-1.0ex]
Collab. & Ref. & $\Nf$ & $a$ [fm] & Description 
\\[1.0ex] \hline \hline \\[-1.0ex]
%
%%% SS
%
Laiho 11 & \cite{Laiho:2011np} & 2+1 & 0.12, 0.09, 0.06  &
\parbox[t]{5.7cm}{Combined continuum and chiral extrapolation based on $SU(3)$
  mixed-action partially quenched {\Ch}PT.}
\\[7.0ex] \hline \\[-1.0ex]
SWME 11, 11A & \parbox[t]{1.0cm}{\cite{Kim:2011qg,Bae:2011ff}} & 2+1 & 0.12, 0.09, 0.06, 0.045  &
\parbox[t]{5.7cm}{Continuum extrapolation with the coarsest lattice spacing
  omitted; residual discretization error of 1.9\% from difference between
  fit to a constant and a constrained five-parameter extrapolation.}
\\[12.0ex] \hline \\[-1.0ex]
BMW 11 & \cite{Durr:2011ap} & 2+1 & \parbox[t]{2.5cm}{0.093, 0.077, \\0.065, 0.054}  &
\parbox[t]{5.7cm}{Combined continuum and chiral extrapolation;
  discretization error of 0.1\% from comparison of $\cO({\alpha_s}a)$
  and $\cO(a^2)$ extrapolations.} 
\\[9.0ex] \hline \\[-1.0ex]
RBC/UKQCD 10B & \cite{Aoki:2010pe} & 2+1 & 0.114, 0.087  &
\parbox[t]{5.7cm}{Two lattice spacings. Combined chiral and continuum
fits.} 
\\[4.0ex]  \hline \\[-1.0ex]

SWME 10   & \cite{Bae:2010ki} & 2+1 &  0.12, 0.09, 0.06 &
\parbox[t]{5.7cm}{Continuum extrapolation of results obtained at four
lattice spacings; residual discretization error of 0.21\% from
difference to result at smallest lattice spacing.}   
\\[9.0ex] \hline \\[-1.0ex]

Aubin 09 & \cite{Aubin:2009jh} & 2+1 & 0.12, 0.09 & 
\parbox[t]{5.7cm}{Two
   lattice spacings; quote 0.3\% discretization error, estimated from
   various $a^2$-terms in fit function}
\\[6.0ex] \hline \\[-1.0ex]

RBC/UKQCD 07A, 08 & \parbox[t]{1.0cm}{\cite{Antonio:2007pb,Allton:2008pn}} & 2+1 & 0.114(2) & 
\parbox[t]{5.7cm}{Single lattice spacing; quote 4\%
  discretization error, estimated from the difference between
  computed and experimental values of $f_\pi$.}
\\ [9.0ex] \hline \\[-1.0ex]
HPQCD/UKQCD  06 & \cite{Gamiz:2006sq} & {2+1} & 0.12 & \parbox[t]{5.7cm}{Single
  lattice spacing; 3\% discretization error quoted without providing
  details.}
\\[4.0ex] \hline \\[-1.0ex]

ETM 12D      & \cite{Bertone:2012cu} & 2 & 0.1, 0.09, 0.07, 0.05 & 
 \parbox[t]{5.7cm}{Four lattice spacings; systematic quoted obtained from the 
difference between the finest lattice spacing and the continuum limit and comparing 
results using two evaluations of the RCs that differ by $\cO(a^2)$ effects.}
 \\[15.0ex] \hline \\[-1.0ex] 

ETM 10A      & \cite{Constantinou:2010qv} & 2 & 0.1, 0.09, 0.07 & 
 \parbox[t]{5.7cm}{Three lattice spacings; 1.2\% error quoted.}
 \\[4.0ex] \hline \\[-1.0ex] 
JLQCD 08 & \cite{Aoki:2008ss} & 2 & 0.118(1) & 
\parbox[t]{5.7cm}{Single
  lattice spacing; no error quoted.}
\\[1.0ex] \hline \\[-1.0ex]
RBC 04 & \cite{Aoki:2004ht} & 2 & 0.117(4) & 
\parbox[t]{5.7cm}{Single
   lattice spacing; no error quoted.}
\\[1.0ex] \hline \\[-1.0ex]
UKQCD 04 & \cite{Flynn:2004au} & 2 & 0.10 & \parbox[t]{5.7cm}{Single
   lattice spacing; no error quoted.}
\\[1.0ex]

\hline\hline
\end{tabular*}
\caption{(cntd.) Continuum extrapolations/estimation of lattice artifacts in
  determinations of $B_K$.}
}
\end{table}

%%%%%  M_pi MIN  %%%%%%%%%%%%%%%%%%%%%%%%%

\begin{table}[!ht]
{\footnotesize
\begin{tabular*}{\textwidth}{l @{\extracolsep{\fill}} c c c l}
\hline\hline \\[-1.0ex]
{Collab.} & {Ref.} & {$\Nf$} & {$M_{\pi,\rm min}\,[\mev]$} & {Description}  
\\[1.0ex] \hline \hline \\[-1.0ex]
%
%%% SS 
ETM 15 & \cite{Carrasco:2015pra} & 2+1+1 & \parbox[t]{1.8cm}{245,\,239,\,211} & 
  \parbox[t]{6cm}{Each $M_{\pi,\rm min}$ entry corresponds to a
  different lattice spacing. Simultaneous chiral \& continuum
  extrapolations, based on polynomial and {\Ch}PT at NLO, are carried out leads to systematic
  error of 0.8\% .}
  \\ [12.0ex] \hline \\[-1.0ex]

SWME 15A & \cite{Jang:2015sla} & 2+1 & 
\parbox[t]{1.8cm}{222/372,\, 206/174,\, 195/222, \, 206/316}
 & \parbox[t]{6cm}{Valence/sea RMS $M_{\pi,\rm min}$ entries correspond to the
four lattice spacings (the last three are used for the chiral-continuum extrapolation). Chiral extrapolations based on $SU(2)$ staggered
{\Ch}PT at NNLO (with some coefficients fixed by Bayesian priors), and
also including one analytic NNNLO term. Residual error of 0.05\%  from
changing the Bayesian priors and fit method.}
\\[23.0ex] \hline \\[-1.0ex]

RBC/UKQCD 14B & \cite{Blum:2014tka} & 2+1 & 
\parbox[t]{1.8cm}{337,\,302,\, 371,\,139,\, 139}  &
\parbox[t]{6.0cm}{$M_{\pi,\rm min}$ entries correspond to the five lattice
spacings. Combined chiral \& continuum extrapolation, using
$M_\pi<260\;$MeV and $M_\pi<370\;$MeV.}
\\[9.0ex] \hline \\[-1.0ex]

SWME 14 & \cite{Bae:2014sja} & 2+1 & 
\parbox[t]{1.8cm}{206/174,\, 195/222,\, 207/316}
 & \parbox[t]{6cm}{Valence/sea RMS $M_{\pi,\rm min}$ entries correspond to the
three lattice spacings. Chiral extrapolations based on $SU(2)$ staggered
{\Ch}PT at NNLO (with some coefficients fixed by Bayesian priors), and
also including one analytic NNNLO term. Residual error of 0.1\% error from
doubling the widths of Bayesian priors.}
\\[20.0ex] \hline \\[-1.0ex]

SWME 13A & \cite{Bae:2013tca} & 2+1 & 
\parbox[t]{1.8cm}{207/243,\, 196/262,\, 207/316}
 & \parbox[t]{6cm}{Valence/sea RMS $M_{\pi,\rm min}$ entries correspond to the
three lattice spacings. Chiral extrapolations based on $SU(2)$ staggered
{\Ch}PT at NNLO (with some coefficients fixed by Bayesian priors), and
also including one analytic NNNLO term. Residual error of 0.1\%  from
doubling the widths of Bayesian priors.}
\\[20.0ex] \hline \\[-1.0ex]

SWME 13 & \cite{Bae:2013lja} & 2+1 & 
\parbox[t]{1.8cm}{442/445,\, 299/273,\, 237/256,\, 222/334}
 & \parbox[t]{6cm}{Valence/sea RMS $M_{\pi,\rm min}$ entries correspond to the
four lattice spacings. Chiral extrapolations based on $SU(2)$ staggered
{\Ch}PT at NNLO (with some coefficients fixed by Bayesian priors), and
also including one analytic NNNLO term. Residual error of 0.33\% error from
doubling the widths of Bayesian priors.}
\\[20.0ex] \hline \\[-1.0ex]
RBC/UKQCD 12A & \cite{Arthur:2012opa} & 2+1 & 
\parbox[t]{1.8cm}{140/170,\, 240/330,\, 220/290}  &
\parbox[t]{6.0cm}{Valence/sea
$M_{\pi,\rm min}$ entries correspond to the three lattice
spacings. Combined chiral \& continuum extrapolation, using
$M_\pi<350\;$MeV.}
\\[1.0ex] \\[-1.0ex]

\hline\hline
\end{tabular*}
\caption{Chiral extrapolation/minimum pion mass in
  determinations of $B_K$.}
}
\end{table}

\begin{table}[!ht]
\addtocounter{table}{-1}
{\footnotesize
\begin{tabular*}{\textwidth}{l @{\extracolsep{\fill}} c c c l}
\hline\hline \\[-1.0ex]
{Collab.} & {Ref.} & {$\Nf$} & {$M_{\pi,\rm min}\,[\mev]$} & {Description}  
\\[1.0ex] \hline \hline \\[-1.0ex]

Laiho 11 & \cite{Laiho:2011np} & 2+1 & \parbox[t]{1.8cm}{210/280}
 & \parbox[t]{6cm}{$M_{\pi,\rm min}$ entries correspond to the smallest
  valence/sea quark masses. Chiral \& continuum fits based on NLO mixed action
  {\Ch}PT, including a subset of NNLO terms. Systematic error estimated
  from spread arising from variations in the fit function.}
\\[15.0ex] \hline \\[-1.0ex]

SWME 11, 11A & \cite{Kim:2011qg,Bae:2011ff} & 2+1 & 
\parbox[t]{1.8cm}{442/445,\, 299/325,\, 237/340,\, 222/334}
 & \parbox[t]{6cm}{Valence/sea RMS $M_{\pi,\rm min}$ entries correspond to the
four lattice spacings. Chiral extrapolations based on $SU(2)$ staggered
{\Ch}PT at NNLO (with some coefficients fixed by Bayesian priors), and
also including one analytic NNNLO term. Residual error of 0.33\% error from
doubling the widths of Bayesian priors.}
\\[20.0ex] \hline \\[-1.0ex]

BMW 11 & \cite{Durr:2011ap} & 2+1 & 
\parbox[t]{1.8cm}{219,\, 182,\\ 120,\, 131}
 & \parbox[t]{6cm}{$M_{\pi,\rm min}$ entries correspond to the
four lattice spacings used in the final result. Combined fit to the chiral and continuum
behaviour. Systematics investigated by applying cuts to the maximum pion
mass used in fits. Uncertainty of 0.1\% assigned to chiral fit.} 
\\[18.0ex] \hline \\[-1.0ex]
RBC/UKQCD 10B & \cite{Aoki:2010pe} & 2+1 & 
\parbox[t]{1.8cm}{240/330,\, 220/290}  &
\parbox[t]{6cm}{Valence/sea $M_{\pi,\rm min}$ entries correspond to the two
lattice spacings. Combined chiral and continuum extrapolations.}
\\[7.0ex] \hline \\[-1.0ex]
SWME 10   & \cite{Bae:2010ki} & 2+1 &  
\parbox[t]{1.8cm}{442/445,\, 299/325,\, 237/340} &
\parbox[t]{6cm}{Valence/sea $M_{\pi,\rm min}$ entries correspond to the three
lattice spacings. Chiral extrapolations based on $SU(2)$ staggered {\Ch}PT
at NLO, including some analytic NNLO terms. $SU(3)$ staggered {\Ch}PT as
cross-check. Combined 1.1\% error from various different variations in
the fit procedure.}
\\[18.0ex] \hline \\[-1.0ex]
Aubin 09 & \cite{Aubin:2009jh} & 2+1 & \parbox[t]{1.8cm}{240/370} &
  \parbox[t]{6cm}{ $M_{\pi,\rm min}$ entries correspond to the
  smallest valence/sea quark masses. Chiral \& continuum fits based on
  NLO mixed action {\Ch}PT at NLO, including a subset of NNLO
  terms. Systematic error estimated from spread arising from
  variations in the fit function.}
\\[18.0ex] \hline \\[-1.0ex]
RBC/UKQCD 07A, 08 & {\cite{Antonio:2007pb,Allton:2008pn}} & {2+1} & 
 \parbox[t]{1.8cm}{330} & {\parbox[t]{6cm}{Fits based on $SU(2)$ PQ{\Ch}PT
 at NLO. Effect of neglecting higher orders estimated at 6\% via
 difference between fits based on LO and NLO expressions.}} 
\\[8.0ex]  \\[-1.0ex]
\hline\hline
\end{tabular*}
\caption{(cntd.) Chiral extrapolation/minimum pion mass in
  determinations of $B_K$.}
}
\end{table}

\begin{table}[!ht]
\addtocounter{table}{-1}
{\footnotesize
\begin{tabular*}{\textwidth}{l @{\extracolsep{\fill}} c c c l}
\hline\hline \\[-1.0ex]
{Collab.} & {Ref.} & {$\Nf$} & {$M_{\pi,\rm min}\,[\mev]$} & {Description}  
\\[1.0ex] \hline \hline \\[-1.0ex]
HPQCD/UKQCD 06 & {\cite{Gamiz:2006sq}} & {2+1} & \parbox[t]{1.8cm}{360}
    & {\parbox[t]{6cm}{3\% uncertainty from chiral extrapolation 
    quoted, without giving further details.}}
\\[4.0ex] \hline \\[-1.0ex]
ETM 12D & \cite{Bertone:2012cu} & 2 & \parbox[t]{1.8cm}{400,
  280,\\ 300, 280} & 
  \parbox[t]{6cm}{Each $M_{\pi,\rm min}$ entry corresponds to a
  different lattice spacing. Simultaneous chiral \& continuum
  extrapolations, based on polynomial and {\Ch}PT at NLO, are carried out. Systematic
  error from several sources, including lattice calibration, quark
  mass calibration, chiral and continuum extrapolation etc., estimated
  at 3.0\%.} 
\\ [21.0ex] \hline \\[-1.0ex]

ETM 10A & \cite{Constantinou:2010qv} & 2 & \parbox[t]{1.8cm}{400,
  280,\\ 300} & 
  \parbox[t]{6cm}{Each $M_{\pi,\rm min}$ entry corresponds to a
  different lattice spacing. Simultaneous chiral \& continuum
  extrapolations, based on {\Ch}PT at NLO, are carried out. Systematic
  error from several sources, including lattice calibration, quark
  mass calibration, chiral and continuum extrapolation etc., estimated
  at 3.1\%.} 
\\ [21.0ex] \hline \\[-1.0ex]

JLQCD 08 & \cite{Aoki:2008ss} & 2 & 290 & \parbox[t]{6cm}{Fits based on
  NLO PQ{\Ch}PT. Range of validity investigated. Fit error included in
  statistical uncertainty.}  
\\[7.0ex] \hline \\[-1.0ex]
RBC 04 & \cite{Aoki:2004ht} & 2 & 490 & \parbox[t]{6cm}{Fits based on NLO
     PQ{\Ch}PT. Fit error included in statistical uncertainty.} 
\\[4.0ex] \hline \\[-1.0ex]
UKQCD 04 & \cite{Flynn:2004au} & 2 & 780 & \parbox[t]{6cm}{Fits to
  continuum chiral behaviour at fixed sea quark mass. Separate
  extrapolation in sea quark mass. Fit error included in overall
  uncertainty.} 
\\[10.0ex]
\hline\hline
\end{tabular*}
\caption{(cntd.) Chiral extrapolation/minimum pion mass in
  determinations of $B_K$.}
}
\end{table}

%%%% Mpi * L %%%%%%%%%%%%%%%%%%%%%%%%%

\begin{table}[!ht]
{\footnotesize
\begin{tabular*}{\textwidth}{l @{\extracolsep{\fill}} c c c c l}
\hline\hline \\[-1.0ex]
Collab. & Ref. & $\Nf$ & $L$ [fm] & ${M_{\pi,\rm min}}L$ & Description 
\\[1.0ex] \hline \hline \\[-1.0ex]
%
%%% SS

ETM 15 & \cite{Carrasco:2015pra} & 2+1+1 & \parbox[t]{1.8cm}{2.1--2.8, 
  2.6, 3.0}& \parbox[t]{1.8cm}{3.5,\,3.2,\,3.2} &  
  \parbox[t]{5cm}{Each $L$ entry corresponds to a different
  lattice spacing, with two volumes at the coarsest lattice
  spacing; results from these two volumes at $M_{\pi} \sim 280$ MeV are
  compatible.} 
\\ [12.0ex] \hline \\[-1.0ex]

SWME 15A & \cite{Jang:2015sla} & 2+1
& \parbox[t]{1.8cm}{2.4--3.4, 2.5--5.8, \,2.9--3.9,\,2.9} & 
$\gtrsim 3.8$ 
& \parbox[t]{5cm}{$L$ entries correspond to the four lattice spacings,
with several volumes in most cases. Finite-volume effects
estimated using NLO $SU(2)$ S{\Ch}PT.}
\\[9.0ex] \hline \\[-1.0ex]

RBC/UKQCD 14B & \cite{Blum:2014tka} 
& 2+1 & \parbox[t]{1.9cm}{2.7,\,2.7, 2.0,\,5.5, 5.3} &
$\gtrsim 3.8$ % 
& \parbox[t]{5cm}{
$L$ entries correspond to the five lattice
spacings. Finite volume effects estimated using NLO {\Ch}PT; 
negligible with comparison to the statistical error.}
\\[12.0ex] \hline \\[-1.0ex]

SWME 14 & \cite{Bae:2014sja} & 2+1
& \parbox[t]{1.8cm}{2.8--5.4,\, 2.8--3.8, 2.8} & 
5.6,\,3.7,\,2.9
& \parbox[t]{5cm}{$L$ entries correspond to the three lattice spacings,
with several volumes in most cases. Finite-volume effects
estimated using NLO {\Ch}PT.}
\\[9.0ex] \hline \\[-1.0ex]

SWME 13A & \cite{Bae:2013tca} & 2+1
& \parbox[t]{1.8cm}{2.4--3.4, 2.8--3.8, 2.8} & 
3.5,\,3.3,\,2.9
& \parbox[t]{5cm}{$L$ entries correspond to the three lattice spacings,
with several volumes in most cases. Finite-volume effects
estimated using NLO {\Ch}PT.}
\\[9.0ex] \hline \\[-1.0ex]

SWME 13 & \cite{Bae:2013lja} & 2+1
& \parbox[t]{1.8cm}{2.4--3.3, 2.4--5.5, 2.8--3.8, 2.8} & 
$\gtrsim 3.2$ % for valence; sea is $\gtrsim 3.9$ 
& \parbox[t]{5cm}{$L$ entries correspond to the four lattice spacings,
with several volumes in most cases. Finite-volume effects
estimated using NLO {\Ch}PT.}
\\[9.0ex] \hline \\[-1.0ex]
RBC/UKQCD 12A & \cite{Arthur:2012opa} 
& 2+1 & \parbox[t]{1.8cm}{4.6,\,2.7,\,2.8} &
$\gtrsim 3.2$ % for valence; $\gtrsim 3.9$ for sea
& \parbox[t]{5cm}{
$L$ entries correspond to the three lattice
spacings. Finite volume effects estimated using NLO {\Ch}PT.}
\\[7.0ex] \hline \\[-1.0ex]
Laiho 11 & \cite{Laiho:2011np} & 2+1 & \parbox[t]{1.8cm}{2.4,\,3.4,\,3.8} &
$\gtrsim 3.5$
 & \parbox[t]{5cm}{$L$ entries correspond to the three lattice spacings.
Finite volume effects estimated using NLO {\Ch}PT.}
\\[7.0ex] \hline \\[-1.0ex]
SWME 11, 11A & \parbox[t]{1.0cm}{\cite{Kim:2011qg,Bae:2011ff}} & 2+1
& \parbox[t]{1.8cm}{2.4/3.3, 2.4, 2.8, 2.8} & 
$\gtrsim 3.2$ % for valence; sea is $\gtrsim 3.9$ 
& \parbox[t]{5cm}{$L$ entries correspond to the four lattice spacings,
with two volumes at the coarsest lattice. Finite-volume effects
estimated using NLO {\Ch}PT.}
\\[9.0ex] \hline \\[-1.0ex]
BMW 11 & \cite{Durr:2011ap} & 2+1 & \parbox[t]{1.8cm}{6.0, 4.9,
4.2, 3.5} & $\gtrsim 3.8, 3.0$ &
\parbox[t]{5cm}{$L$ entries correspond to the four lattice
spacings, and are the largest of several volumes at each $a$.
${M_{\pi,\rm min}}L\approx3.0$ for the ensemble
at $a\approx0.08\,\fm$. Finite volume effects estimated in {\Ch}PT and
by combined fit to multiple volumes.}
\\[1.0ex]  \\[-1.0ex]
\hline\hline
\end{tabular*}
\caption{Finite volume effects in determinations of $B_K$.
If partially-quenched fits are used,
the quoted $M_{\pi,\rm min}L$ is for lightest valence (RMS) pion.}
}
\end{table}

\begin{table}[!ht]
\addtocounter{table}{-1}
{\footnotesize
\begin{tabular*}{\textwidth}{l @{\extracolsep{\fill}} c c c c l}
\hline\hline \\[-1.0ex]
Collab. & Ref. & $\Nf$ & $L$ [fm] & ${M_{\pi,\rm min}}L$ & Description 
\\[1.0ex] \hline \hline \\[-1.0ex]
RBC/UKQCD 10B & \cite{Aoki:2010pe} & 2+1 & \parbox[t]{1.8cm}{2.7,\,2.8} &  
$\gtrsim 3.1$ % for valence; $\gtrsim 4.0$ for sea
& \parbox[t]{5cm}{$L$ entries correspond to the three lattice spacings.
Finite volume effects estimated using NLO {\Ch}PT.}
\\[7.0ex] \hline \\[-1.0ex]
SWME 10   & \cite{Bae:2010ki} & 2+1 & \parbox[t]{1.8cm}{2.4/3.3, 2.4,2.8} & 
$\gtrsim 3.4$ % for valence; $\gtrsim 3.9$ for sea
& \parbox[t]{5cm}{$L$ entries correspond to the three lattice spacings,
with two volumes for the coarsest spacing. 
Finite-volume error of 0.9\% estimated from difference
obtained these two volumes.}
\\[15.0ex]\hline  \\[-1.0ex]

Aubin 09 & \cite{Aubin:2009jh} & 2+1 & \parbox[t]{1.8cm}{2.4,\,3.4} & 3.5 &
   \parbox[t]{5cm}{$L$ entries correspond to the two lattice spacings.
Keep $m_{\pi}L\;\gtaeq\;3.5$; no
   comparison of results from different volumes; 0.6\% error
   estimated from mixed action {\Ch}PT correction.}
\\[12.0ex] \hline \\[-1.0ex]
RBC/UKQCD 07A, 08 & \parbox[t]{1.0cm}{\cite{Antonio:2007pb,Allton:2008pn}} & 2+1 &
  \parbox[t]{1.8cm}{1.83/2.74} & 4.60 & \parbox[t]{5cm}{Each $L$ entry
  corresponds to a different volume at the same lattice spacing; 1\%
  error from difference in results on two volumes.} 
\\[10.0ex] \hline \\[-1.0ex]
HPQCD/UKQCD 06 & \cite{Gamiz:2006sq} & 2+1 & \parbox[t]{1.8cm}{2.46} &
  \parbox[t]{1.8cm}{4.49} & {Single volume; no error quoted.}
\\  [1.0ex] \hline \\[-1.0ex]

ETM 12D & \cite{Bertone:2012cu} & 2 & \parbox[t]{1.8cm}{2.1,\\
  2.2/2.9,\\ 2.2, 2.6}& \parbox[t]{1.8cm}{5,\\ 3.3/4.3,\\ 3.3, 3.5} &  
  \parbox[t]{5cm}{Each $L$ entry corresponds to a different
  lattice spacing, with two volumes at the second less coarse lattice
  spacing. Results from these two volumes at $M_{\pi} \sim 300$ MeV are
  compatible.} 
\\ [15.0ex] \hline \\[-1.0ex]

ETM 10A & \cite{Constantinou:2010qv} & 2 & \parbox[t]{1.8cm}{2.1,\\
  2.2/2.9,\\ 2.2}& \parbox[t]{1.8cm}{5,\\ 3.3/4.3,\\ 3.3} &  
  \parbox[t]{5cm}{Each $L$ entry corresponds to a different
  lattice spacing, with two volumes at the intermediate lattice
  spacing. Results from these two volumes at $M_{\pi} \sim 300$ MeV are
  compatible.} 
\\ [15.0ex] \hline \\[-1.0ex]
JLQCD 08 & \cite{Aoki:2008ss} & 2 & 1.89 & 2.75 & \parbox[t]{5cm}{Single
  volume; data points with $m_{\rm val} < m_{\rm sea}$ excluded; 5\%
  error quoted as upper bound of PQ{\Ch}PT estimate of the 
  effect.}
\\[10.0ex] \hline \\[-1.0ex]
RBC 04 & \cite{Aoki:2004ht} & 2 & 1.87 & 4.64 & Single volume; no
   error quoted.
\\[1.0ex] \hline \\[-1.0ex]
UKQCD 04 & \cite{Flynn:2004au} & 2 & 1.6 & 6.51 & Single volume; no
   error quoted.
\\[1.0ex] 
\hline\hline
\end{tabular*}
\caption{(cntd.) Finite volume effects in determinations of $B_K$}
}
\end{table}

%%% RENORMALISATION %%%%%%%%%%%%%%%%%

\begin{table}[!ht]
{\footnotesize
\begin{tabular*}{\textwidth}{l @{\extracolsep{\fill}} c c c c l}
\hline\hline \\[-1.0ex]
& & & & running & \\
\rb{Collab.} & \rb{Ref.} & \rb{$\Nf$} & \rb{Ren.} & match. & \rb{Description} 
\\[1.0ex] \hline \hline \\[-1.0ex]
%
%%% SS

ETM 15 & \cite{Carrasco:2015pra} & 2+1+1 & RI & PT1$\ell$ &
  \parbox[t]{5cm}{Uncertainty from RI renormalization estimated at 2\%. Additional error of 0.6\% for
  the conversion to $\msbar$.}
\\ [9.0ex] \hline \\[-1.0ex]

SWME 15A & \cite{Jang:2015sla} & 2+1 & PT1$\ell$ & PT1$\ell$ &
\parbox[t]{5cm}{Uncertainty from neglecting higher orders estimated at
4.4\% by identifying the unknown 2-loop coefficient with result at the
  smallest lattice spacing.}
\\[12.0ex] \hline \\[-1.0ex]

RBC/UKQCD 14B & \cite{Blum:2014tka} & 2+1 & RI & PT1$\ell$ &
\parbox[t]{5cm}{Two different RI-SMOM schemes used to estimate
2\% systematic error in conversion to $\msbar$.}
\\[7.0ex] \hline \\[-1.0ex]

SWME 14 & \cite{Bae:2014sja} & 2+1 & PT1$\ell$ & PT1$\ell$ &
\parbox[t]{5cm}{Uncertainty from neglecting higher orders estimated at
4.4\% by identifying the unknown 2-loop coefficient with result at the
  smallest lattice spacing.}
\\[12.0ex] \hline \\[-1.0ex]

SWME 13A & \cite{Bae:2013tca} & 2+1 & PT1$\ell$ & PT1$\ell$ &
\parbox[t]{5cm}{Uncertainty from neglecting higher orders estimated at
4.4\% (in combination with systematic uncertainty from 
CL and chiral extrapolation fit) by identifying the unknown 2-loop coefficient with result at the
  smallest lattice spacing.}
\\[18.0ex] \hline \\[-1.0ex]

SWME 13 & \cite{Bae:2013lja} & 2+1 & PT1$\ell$ & PT1$\ell$ &
\parbox[t]{5cm}{Uncertainty from neglecting higher orders estimated at
4.4\% by identifying the unknown 2-loop coefficient with result at the
  smallest lattice spacing.}
\\[12.0ex] \hline \\[-1.0ex]
RBC/UKQCD 12A & \cite{Arthur:2012opa} & 2+1 & RI & PT1$\ell$ &
\parbox[t]{5cm}{Two different RI-SMOM schemes used to estimate
2\% systematic error in conversion to $\msbar$.}
\\[7.0ex] \hline \\[-1.0ex]
Laiho 11 & \cite{Laiho:2011np} & 2+1 & RI & PT1$\ell$ &
\parbox[t]{5cm}{Total uncertainty in matching \& running of 3\%.
  Perturbative truncation error in the conversion to $\msbar$, RGI schemes is
  dominant uncertainty.} 
\\[12.0ex] \hline \\[-1.0ex]
SWME 11, 11A & \cite{Kim:2011qg,Bae:2011ff} & 2+1 & PT1$\ell$ & PT1$\ell$ &
\parbox[t]{5cm}{Uncertainty from neglecting higher orders estimated at
4.4\% by identifying the unknown 2-loop coefficient with result at the
  smallest lattice spacing.}
\\[12.0ex] \hline \hline \\[-1.0ex]
\end{tabular*}
\caption{Running and matching in determinations of $B_K$ for $N_f=2+1+1$ and $N_f=2+1$.}
}
\end{table}

\begin{table}[!ht]
\addtocounter{table}{-1}
{\footnotesize
\begin{tabular*}{\textwidth}{l @{\extracolsep{\fill}} c c c c l}
\hline\hline \\[-1.0ex]
& & & & running & \\
\rb{Collab.} & \rb{Ref.} & \rb{$\Nf$} & \rb{Ren.} & match. & \rb{Description} 
\\[1.0ex] \hline \hline \\[-1.0ex]
BMW 11 & \cite{Durr:2011ap} & 2+1 & RI & PT1$\ell$ &
\parbox[t]{5cm}{Uncertainty of 0.05\% in the determination of the
  renormalization factor included. 1\% error estimated due to truncation of perturbative
  matching to $\msbar$ and RGI schemes at NLO.}
\\[15.0ex]  \hline  \\[-1.0ex]
RBC/UKQCD 10B & \cite{Aoki:2010pe} & 2+1 & RI & PT1$\ell$ &
\parbox[t]{5cm}{Variety of different RI-MOM schemes including
non-exceptional momenta. Residual uncertainty of 2\% uncertainty in
running \& matching.}\\[12.0ex] \hline \\[-1.0ex] 
SWME 10  & \cite{Bae:2010ki} & 2+1 & PT1$\ell$ & PT1$\ell$ &
\parbox[t]{5cm}{Uncertainty from neglecting higher orders estimated at
5.5\% by identifying the unknown 2-loop coefficient with result at the
  smallest lattice spacing.}
\\[12.0ex] \hline \\[-1.0ex] 
Aubin 09 & \cite{Aubin:2009jh} & 2+1 & RI & PT1$\ell$ &
  \parbox[t]{5cm}{Total uncertainty in matching \& running of 3.3\%,
  estimated from a number of sources, including chiral extrapolation
  fit ansatz for n.p. determination, strange sea quark mass
  dependence, residual chiral symmetry breaking, perturbative matching
  \& running.}
\\[21.0ex] \hline \\[-1.0ex]
RBC/UKQCD 07A, 08 & {\cite{Antonio:2007pb,Allton:2008pn}} & 
 {2+1} & {RI} & {PT1$\ell$} & {\parbox[t]{5cm}{Uncertainty
 from n.p. determination of ren. factor included in statistical error; 
 2\% systematic error from perturbative matching to $\msbar$
 estimated via size of correction itself.}}
\\[15.0ex] \hline \\[-1.0ex]
HPQCD/UKQCD 06 & {\cite{Gamiz:2006sq}} & {2+1} & {PT1$\ell$}&
    {PT1$\ell$} & {\parbox[t]{5cm}{Uncertainty due to neglecting 2-loop
    order in perturbative matching and running estimated by
    multiplying result by $\alpha^2$.}}
\\[10.0ex] \hline \hline \\[-1.0ex]
\end{tabular*}
\caption{(cntd.) Running and matching in determinations of $B_K$ for $N_f=2+1+1$ and $N_f=2+1$.}
}
\end{table}

\begin{table}[!ht]
{\footnotesize
\begin{tabular*}{\textwidth}{l @{\extracolsep{\fill}} c c c c l}
\hline\hline \\[-1.0ex]
& & & & running & \\
\rb{Collab.} & \rb{Ref.} & \rb{$\Nf$} & \rb{Ren.} & match. & \rb{Description} 
\\[1.0ex] \hline \hline \\[-1.0ex]
ETM 12D & \cite{Bertone:2012cu} & 2 & RI & PT1$\ell$ &
  \parbox[t]{5cm}{Uncertainty from RI renormalization estimated at 2.5\%.}
\\ [4.0ex] \hline \\[-1.0ex]

ETM 10A & \cite{Constantinou:2010qv} & 2 & RI & PT1$\ell$ &
  \parbox[t]{5cm}{Uncertainty from RI renormalization estimated at 2.5\%.}
\\ [4.0ex] \hline \\[-1.0ex]
JLQCD 08 & \cite{Aoki:2008ss} & 2 & RI & PT1$\ell$ &
 \parbox[t]{5cm}{Uncertainty from n.p. determination of ren. factor 
 included in statistical error; 2.3\% systematic error from
 perturbative matching to $\msbar$ estimated via size of correction
 itself.} 
\\[15.0ex] \hline \\[-1.0ex]
RBC 04 & \cite{Aoki:2004ht} & 2 & RI & PT1$\ell$ &
   \parbox[t]{5cm}{Uncertainty from n.p. determination of ren. factor
 included.}
\\[4.0ex] \hline \\[-1.0ex]
UKQCD 04 & \cite{Flynn:2004au} & 2 & PT1$\ell$ & PT1$\ell$ & No error quoted.
\\[1.0ex] 
\hline\hline
\end{tabular*}
\caption{Running and matching in determinations of $B_K$ for $N_f=2$.}
}
\end{table}

\clearpage

\subsubsection{Kaon BSM $B$-parameters}
\label{app-Bi}

%In the following, we summarize the characteristics (lattice actions,
%pion masses, lattice spacings, etc.) of the recent $\Nf=2+1+1$, $\Nf = 2+1$ and
%$\Nf = 2$ runs. We also provide brief descriptions of how systematic
%errors are estimated by the various authors. 

%%%% a -lat. spacing %%%%%%%%%%%%%%%

\begin{table}[!ht]

{\footnotesize
\begin{tabular*}{\textwidth}{l c c c l}
\hline\hline \\[-1.0ex]
Collab. & Ref. & $\Nf$ & $a$ [fm] & Description 
\\[1.0ex] \hline \hline \\[-1.0ex]
%
%%% SS
ETM 15      & \cite{Carrasco:2015pra} & 2+1+1 & 0.09, 0.08, 0.06 & 
 \parbox[t]{6.0cm}{Combined chiral and continuum extrapolation. Systematic errors to $B_{i}$ from about 4\% to 6\% are obtained 
 from the distribution of results over analyses which differ by  
$\cO(a^2)$ effects.}
 \\[12.0ex] \hline \\[-1.0ex] 

SWME 15A & \cite{Jang:2015sla} & 2+1 & 0.12, 0.09, 0.06, 0.045  &
\parbox[t]{6.0cm}{The three finest lattice spacings are used for the combined chiral and continnum extrapolation.  
Residual combined discretization, sea-quark extrapolation
and $\alpha_s$ matching error from about 4.4\% to 9.6\% is reported for $B_{i}$ and is obtained from the difference between linear
  fit in $a^2$, $m_{\rm sea}$ and a fit where $\alpha_s$ dependence is added.}
\\[21.0ex] \hline \\[-1.0ex]

SWME 14C & \cite{Jang:2014aea} & 2+1 & 0.082, 0.059, 0.044  &
\parbox[t]{6.0cm}{Residual combined discretization and sea-quark extrapolationg
error of 1--8\% from difference between linear
  fit in $a^2$, $m_{\rm sea}$  and a constrained nighteen-parameter extrapolation.}
\\[12.0ex] \hline \\[-1.0ex]

SWME 13A & \cite{Bae:2013tca} & 2+1 & 0.09, 0.06, 0.045  &
\parbox[t]{6.0cm}{Residual combined discretization, sea-quark extrapolation
and $\alpha_s$ matching error for $B_i$ varies from 4.5\% to -5.7\% , from difference between linear
  fit in $a^2$, $m_{\rm sea}$ and a fit where $\alpha_s$ dependence is added.}
\\[12.0ex] \hline \\[-1.0ex]

RBC/UKQCD 12E & 
\cite{Boyle:2012qb} & 2+1 & 0.087  &
\parbox[t]{6.0cm}{Computation at only one value of the lattice spacing. 
Estimate for the systematic discretisation error of about 1.5\% based on the 
corresponding estimate from the $B_K$ computation.}
\\[12.0ex] \hline \\[-1.0ex]

ETM 12D      & \cite{Bertone:2012cu} & 2 & 0.1, 0.09, 0.07, 0.05 & 
 \parbox[t]{6.0cm}{Four lattice spacings; Estimates of systematic uncertainties 
 obtained from the half difference of the distance between the finest lattice spacing and the 
 continuum limit.}
 \\[1.0ex]  \\[-1.0ex] 

\hline\hline
\end{tabular*}
\caption{Continuum extrapolations/estimation of lattice artifacts in
  determinations of the BSM $B_i$ parameters.}
}
\end{table}

%%%%%  M_pi MIN  %%%%%%%%%%%%%%%%%%%%%%%%%

\begin{table}[!ht]
{\footnotesize
\begin{tabular*}{\textwidth}{l @{\extracolsep{\fill}} c c c l}
\hline\hline \\[-1.0ex]
{Collab.} & {Ref.} & {$\Nf$} & {$M_{\pi,\rm min}\,[\mev]$} & {Description}  
\\[1.0ex] \hline \hline \\[-1.0ex]
%
%%% SS 
ETM 15 & \cite{Carrasco:2015pra} & 2+1+1 & \parbox[t]{1.8cm}{245,\,239,\,211} & 
  \parbox[t]{6cm}{Each $M_{\pi,\rm min}$ entry corresponds to a
  different lattice spacing. Simultaneous chiral \& continuum
  extrapolations, based on polynomial and {\Ch}PT at NLO, are carried out leads to systematic
  errors of $1.1-2.6$\% depending on the bag parameter.}
  \\[15.0ex]  \hline \\[-1.0ex]

SWME 15A & \cite{Jang:2015sla} & 2+1 & 
\parbox[t]{1.8cm}{222/372,\, 206/174,\, 195/222, \, 206/316}
 & \parbox[t]{6cm}{Valence/sea RMS $M_{\pi,\rm min}$ entries correspond to the
four lattice spacings (the last three are used for the chiral-continuum extrapolation). Chiral extrapolations based on $SU(2)$ staggered
{\Ch}PT at NNLO (with some coefficients fixed by Bayesian priors), and
also including one analytic NNNLO term. Residual error of 0.4-1.2\% depending on the bag parameter from
changing the Bayesian priors and fit method.}
\\[26.0ex] \hline \\[-1.0ex]

SWME 14C & \cite{Jang:2014aea} & 2+1 & 
\parbox[t]{1.8cm}{206/174,\, 195/222,\, 207/316}
 & \parbox[t]{6cm}{Valence/sea RMS $M_{\pi,\rm min}$ entries correspond to the
three lattice spacings. Chiral extrapolations performed via $B_i$-ratios that do not show $SU(2)$
NLO {\Ch}PT contribution and assuming various terms up to NNLO (with some coefficients fixed by Bayesian priors). 
}
\\[18.0ex] \hline \\[-1.0ex]

SWME 13A & \cite{Bae:2013tca} & 2+1 & 
\parbox[t]{1.8cm}{207/243,\, 196/262,\, 207/316}
 & \parbox[t]{6cm}{Valence/sea RMS $M_{\pi,\rm min}$ entries correspond to the
three lattice spacings. Chiral extrapolations performed via $B_i$-ratios that do not show $SU(2)$
NLO {\Ch}PT contribution and assuming various terms up to NNLO (with some coefficients fixed by Bayesian priors). 
Residual error in the valence of about 0.1\% from
doubling the widths of Bayesian priors. In the sea a combined error with the matching procedure of 4.4-5.6\% is reported.}
\\[29.0ex] \hline \\[-1.0ex]

RBC/UKQCD 12E & \cite{Boyle:2012qb} & 2+1 & 
\parbox[t]{1.8cm}{290/290}  &
\parbox[t]{6.0cm}{Chiral extrapolations based on polynomial and {\Ch}PT fits at NLO are carried out. 
  Central values are obtained from polynomial fits. Mild dependence on the quark mass. Systematic 
  uncertainties are estimated to about 4\% for all $B_i$'s.}
\\[15.0ex] \hline \\[-1.0ex]
ETM 12D & \cite{Bertone:2012cu} & 2 & \parbox[t]{1.8cm}{400,
  270,\\ 300, 270} & 
  \parbox[t]{6cm}{Each $M_{\pi,\rm min}$ entry corresponds to a
  different lattice spacing. Simultaneous chiral \& continuum
  extrapolations, based on polynomial and {\Ch}PT at NLO, are carried out. 
 } 
\\ [10.0ex] \hline \hline\\[-1.0ex]
\end{tabular*}
\caption{Chiral extrapolation/minimum pion mass in
  determinations of the BSM $B_i$ parameters.}
}

\end{table}

%%%% Mpi * L %%%%%%%%%%%%%%%%%%%%%%%%%

\begin{table}[!ht]
{\footnotesize
\begin{tabular*}{\textwidth}{l @{\extracolsep{\fill}} c c c c l}
\hline\hline \\[-1.0ex]
Collab. & Ref. & $\Nf$ & $L$ [fm] & ${M_{\pi,\rm min}}L$ & Description 
\\[1.0ex] \hline \hline \\[-1.0ex]
%
%%% SS
ETM 15 & \cite{Carrasco:2015pra} & 2+1+1 & \parbox[t]{1.8cm}{2.1--2.8, 
  2.6, 3.0}& \parbox[t]{1.8cm}{3.5,\,3.2,\,3.2} &  
  \parbox[t]{5cm}{Each $L$ entry corresponds to a different
  lattice spacing, with two volumes at the coarsest lattice
  spacing; results from these two volumes at $M_{\pi} \sim 280$ MeV are
  compatible.} 
\\ [12.0ex] \hline \\[-1.0ex]

SWME 15A & \cite{Jang:2015sla} & 2+1
& \parbox[t]{1.8cm}{2.4--3.4, 2.5--5.8, \,2.9--3.9,\,2.9} & 
$\gtrsim 3.8$ 
& \parbox[t]{5cm}{$L$ entries correspond to the four lattice spacings,
with several volumes in most cases. Finite-volume effects
estimated using NLO $SU(2)$ S{\Ch}PT.}
\\[10.0ex] \hline \\[-1.0ex]

SWME 14C & \cite{Jang:2014aea} & 2+1
& \parbox[t]{1.8cm}{2.8--5.4,\, 2.8--3.8, 2.8} & 
5.6,\,3.7,\,2.9
& \parbox[t]{5cm}{$L$ entries correspond to the three lattice spacings,
with several volumes in most cases. Finite-volume effects
estimated using NLO {\Ch}PT.}
\\[10.0ex] \hline \\[-1.0ex]

SWME 13A & \cite{Bae:2013tca} & 2+1
& \parbox[t]{1.8cm}{2.4--3.4, 2.8--3.3, 2.8} & 
3.5,\,3.3,\,2.9
& \parbox[t]{5cm}{$L$ entries correspond to the three lattice spacings,
with several volumes in most cases. Finite-volume effects
estimated using NLO {\Ch}PT.}
\\[10.0ex] \hline \\[-1.0ex]

RBC/UKQCD 12E & \cite{Boyle:2012qb} 
& 2+1 & \parbox[t]{1.8cm}{2.8} &
$\gtrsim 4.0$ % 
& \parbox[t]{5cm}{ 
The $L$ value corresponds to the unique lattice
spacing. Finite volume effects, estimated using NLO {\Ch}PT are small, as 
it has also been found in the $B_K$ computation,    
and they have thus been  neglected in the final error budget analysis. 
}
\\[18.0ex] \hline \\[-1.0ex]
ETM 12D & \cite{Bertone:2012cu} & 2 & \parbox[t]{1.8cm}{2.1,\\
  2.2/2.9,\\ 2.2, 2.6}& \parbox[t]{1.8cm}{5,\\ 3.3/4.3,\\ 3.3, 3.5} &  
  \parbox[t]{5cm}{Each $L$ entry corresponds to a different
  lattice spacing, with two volumes at the second less coarse lattice
  spacing. Results from these two volumes at $M_{\pi} \sim 300$ MeV are
  compatible.} 
\\ [15.0ex] \hline\hline \\[-1.0ex]
\end{tabular*}
\caption{Finite volume effects in determinations of the BSM $B_i$
  parameters. If partially-quenched fits are used, the quoted
  $M_{\pi,\rm min}L$ is for lightest valence (RMS) pion.}  }
\end{table}

%%% RENORMALISATION %%%%%%%%%%%%%%%%%

\begin{table}[!ht]
{\footnotesize
\begin{tabular*}{\textwidth}{l @{\extracolsep{\fill}} c c c c l}
\hline\hline \\[-1.0ex]
& & & & running & \\
\rb{Collab.} & \rb{Ref.} & \rb{$\Nf$} & \rb{Ren.} & match. & \rb{Description} 
\\[1.0ex] \hline \hline \\[-1.0ex]
%
%%% SS

ETM 15 & \cite{Carrasco:2015pra} & 2+1+1 & RI & PT1$\ell$ &
  \parbox[t]{5cm}{Uncertainty from RI renormalization combined with discretisation effects estimates are reported to be from about 4\% to 6\% . 
  Additional error from 1.8 to 3.9\% (dependind on the bag parameter) for   the conversion to $\msbar$ at the scale of 3 GeV.}
\\ [18.0ex] \hline \\[-1.0ex]

SWME 15A & \cite{Jang:2015sla} & 2+1 & PT1$\ell$ & PT1$\ell$ &
\parbox[t]{5cm}{Uncertainty from neglecting higher orders estimated from about 4.4\% to 9.6\% (depending on the bag parameter)
 by identifying the unknown 2-loop coefficient with result at the
  smallest lattice spacing.}
\\[15.0ex] \hline \\[-1.0ex]

SWME 14C & \cite{Jang:2014aea} & 2+1 & PT1$\ell$ & PT1$\ell$ &
\parbox[t]{5cm}{Uncertainty from neglecting higher orders estimated at
4.4\% by identifying the unknown 2-loop coefficient with result at the
  smallest lattice spacing.}
\\[12.0ex] \hline \\[-1.0ex]

SWME 13A & \cite{Bae:2013tca} & 2+1 & PT1$\ell$ & PT1$\ell$ &
\parbox[t]{5cm}{Uncertainty from neglecting higher orders estimated at
4.4\% (in combination with systematic uncertainty from 
CL and chiral extrapolation fit) by identifying the unknown 2-loop coefficient with result at the
  smallest lattice spacing.}
\\[18.0ex] \hline \\[-1.0ex]

RBC/UKQCD 12E & \cite{Boyle:2012qb} & 2+1 & RI & PT1$\ell$ &
\parbox[t]{5cm}{Computation in RI-MOM scheme. Systematic error from the conversion to $\msbar$
is estimated by taking the half of the difference between the LO and the NLO result.}
\\[12.0ex] \hline \\[-1.0ex]
ETM 12D & \cite{Bertone:2012cu} & 2 & RI & PT1$\ell$ &
  \parbox[t]{5cm}{Uncertainty from RI renormalization estimated at 2.5\%.}
\\ [4.0ex] \hline \hline\\[-1.0ex]
\end{tabular*}
\caption{Running and matching in determinations of the BSM $B_i$ parameters.}
}
\end{table}

%% file: HQ/HQ_app.tex
%=================================================
\subsection{Notes to section \ref{sec:DDecays} on $D$-meson decay constants and form factors}
\label{app:DDecays}
%=================================================

In the following, we summarize the characteristics (lattice actions,
pion masses, lattice spacings, etc.) of the recent $\Nf=2+1+1$, $\Nf = 2+1$ and
$\Nf = 2$ runs. We also provide brief descriptions of how systematic
errors are estimated by the various authors.  We focus on calculations with either preliminary or published quantitative results.

% D leptonics
\input{HQ/AppendixTables/DLapp.tex}

\clearpage

% D semi leptonics
\input{HQ/AppendixTables/DSLapp.tex}

\clearpage

%=================================================
\subsection{Notes to section \ref{sec:BDecays} on $B$-meson decay constants and mixing parameters}
\label{app:BDecays}
%=================================================

In the following, we summarize the characteristics (lattice actions,
pion masses, lattice spacings, etc.) of the recent $\Nf = 2+1+1$, $\Nf=2+1$ and
$\Nf = 2$ runs. We also provide brief descriptions of how systematic
errors are estimated by the various authors.  We focus on calculations with either preliminary or published quantitative results.

% B leptonics
\input{HQ/AppendixTables/BLapp.tex}

\clearpage

% B mixing
\input{HQ/AppendixTables/BBapp.tex}

\clearpage

% B semileptonics

\input{HQ/AppendixTables/BSLapp.tex}

%% file: HQ/AppendixTables/DLapp.tex
%=================================================
\subsubsection{$D_{(s)}$-meson decay constants}
\label{app:fD_Notes}
%=================================================

%%%%%%%%%%%%%%%%%%%%%%%%% CHIRAL EXTRAP %%%%%%%%%%%%%%%%%%%%%%%%%%%%%%%%%%%%%%%%%%%%%%%%%%%%%%%%%%%%%%%%%%%%%%%%%%%%%%%5

\begin{table}[!htb]

%\vspace{-1cm}

{\footnotesize
\begin{tabular*}{\textwidth}{l @{\extracolsep{\fill}} c c c   l}
\hline\hline \\[-1.0ex]
{Collab.} & {Ref.} & {$\Nf$} & {$M_{\pi,\rm min}\,[\mev]$}  & {Description}  
\\[1.0ex] \hline \hline \\[-1.0ex]
\parbox[t]{2cm}{FNAL/MILC~14A} & \cite{Bazavov:2014wgs} & 2+1+1 & \parbox[t]{2cm}{311, 241, 173, 143} &
\parbox[t]{5.8cm}{The lightest pions (not RMS) are around 130 MeV. Analyses are performed either by interpolating to the physical point
or by using HMrAS$\chi$PT formulae to include heavier masses and non-unitary points. The latter procedure gives more accurate, and final, results.}
\\[18.0ex] \hline \\[-2.0ex]
\parbox[t]{2cm}{ETM~13F \\ ETM~14E} & \cite{Dimopoulos:2013qfa,Carrasco:2014poa} & 2+1+1 &  \parbox[t]{2cm}{245, 239, 211}  & \parbox[t]{5.8cm}{$f_{D_s}\sqrt{m_{D_s}}$ in ETM~13F and  $f_{D_s}/m_{D_s}$ in ETM~14E are extrapolated
using both a quadratic and a linear fit in $m_l$ plus $\cO(a^2)$ terms. Then the double ratio $(f_{D_s}/f_D)/(f_K/f_\pi)$ is fitted in continuum HM$\chi$PT,
as no lattice spacing dependence is visible within statistical errors.\\[-1.7ex]}  
\\[18.0ex] \hline \\[-2.0ex]
\parbox[t]{2.3cm}{FNAL/MILC~12B \\FNAL/MILC~13} & \cite{Bazavov:2012dg,Bazavov:2013nfa} & 2+1+1 & \parbox[t]{2cm}{310, 245, 179, 145} &
\parbox[t]{5.8cm}{Chiral and continuum extrapolations are peformed simultaneously. Central values 
are produced  using a fit function quadratic in $a^2$ and linear in the sea-quark mass.
In FNAL/MILC~13 terms of $\cO(a^4)$ are included.}
\\[15.0ex] 
\hline\hline
\end{tabular*}
}
\caption{Chiral extrapolation/minimum pion mass in $N_f=2+1+1$  determinations of the
  $D$ and $D_s$ meson decay constants.  For actions with multiple
 species of pions, masses quoted are the RMS pion masses (where available). 
   The
 different $M_{\pi,\rm min}$ entries correspond to the different lattice
 spacings.} 
\end{table}

\begin{table}[!htb]
{\footnotesize
\begin{tabular*}{\textwidth}{l @{\extracolsep{\fill}} c c c   l}
\hline\hline \\[-1.0ex]
{Collab.} & {Ref.} & {$\Nf$} & {$M_{\pi,\rm min}\,[\mev]$}  & {Description}  
\\[1.0ex] \hline \hline \\[-1.0ex]
$\chi$QCD~14& \cite{Yang:2014sea} & 2+1 & \parbox[t]{2cm}{334, 296} & \parbox[t]{5.8cm}{Chiral and continuum extrapolations
 are peformed simultaneously using linear fits in $m_l$ (and quadratic or including partially quenched chiral logs, in order to assess the systematic error)
plus terms up to $\cO(a^2)$ and $\cO(a^4m_c^4)$.}
\\[15.0ex] \hline \\[-2.0ex]
HPQCD 12A& \cite{Na:2012iu} & 2+1 & \parbox[t]{2cm}{460, 329} & \parbox[t]{5.8cm}{Chiral and continuum extrapolations
 are peformed simultaneously using PQHM$\chi$PT augmented by $a$ dependent terms: $c_0(am_c)^2 + c_1(am_c)^4$.}
\\[9.0ex] \hline \\[-2.0ex]
FNAL/MILC 11 & \cite{Bazavov:2011aa} & 2+1 & \parbox[t]{2cm}{570, 440, 320} &
\parbox[t]{5.8cm}{Chiral and continuum extrapolations are peformed 
simultaneously using HM$\chi$PT for rooted staggered quarks.
Effects of hyperfine and flavour splittings are also included.}
\\[12.0ex] \hline \\[-2.0ex]
PACS-CS 11 & \cite{Namekawa:2011wt} & 2+1 & \parbox[t]{2cm}{152} & \parbox[t]{5.8cm}{Simulations are  
reweighted in the light- and strange-quark masses to the physical point.}
\\[6.0ex] \hline \\[-2.0ex]
HPQCD 10A & \cite{Davies:2010ip} & 2+1 & \parbox[t]{2cm}{542, 460, 329, 258, 334} & \parbox[t]{5.8cm}{Chiral and continuum 
extrapolations are performed simultaneously. Polynomials up to  $\left( \frac{m_{q,sea}-m_{q,phys}}{m_{q,phys}}\right)^2$ for $q=s,l$ and
up to  $(am_{c})^8$ are kept.}
\\[11.0ex] \hline \\[-2.0ex]
HPQCD/UKQCD 07 & \cite{Follana:2007uv} & 2+1 & \parbox[t]{2cm}{542, 460, 329} & \parbox[t]{5.8cm}{Combined chiral and continuum extrapolations using
HM$\chi$PT at NLO augmented by second and third-order polynomial terms in $m_q$ and terms up to $a^4$.}
\\[9.0ex] \hline \\[-2.0ex]
FNAL/MILC 05 & \cite{Aubin:2005ar} & 2+1 &
\parbox[t]{2cm}{$>~440\;, 440\;, 400$} & \parbox[t]{5.8cm}{Chiral extrapolations are first performed at each lattice spacing
uisng NLO HM$\chi$PT for rooted staggered quarks. Lattice artefacts are then extrapolated linearly in $a^2$. }
\\[12.0ex] 
%\hline \\[-2.0ex]
%
\hline\hline
\end{tabular*}
}
\caption{Chiral extrapolation/minimum pion mass in  $N_f=2+1$ determinations of the
  $D$ and $D_s$ meson decay constants.  For actions with multiple
 species of pions, masses quoted are the RMS pion masses (where available). 
   The
 different $M_{\pi,\rm min}$ entries correspond to the different lattice
 spacings.} 
\end{table}
%%%%%%%%%%%%%%%%%%%%%%%%%%
\begin{table}[!htb]
{\footnotesize
\begin{tabular*}{\textwidth}{l @{\extracolsep{\fill}} c c c   l}
\hline\hline \\[-1.0ex]
{Collab.} & {Ref.} & {$\Nf$} & {$M_{\pi,\rm min}\,[\mev]$}  & {Description}  
\\[1.0ex] \hline \hline \\[-1.0ex]
\parbox[t]{2.0cm}{TWQCD~14} & \cite{Chen:2014hva} & 2 & 
\parbox[t]{2cm}{260} & \parbox[t]{5.8cm}{Comparison of
NLO HM$\chi$PT fits for $f_{D_{( \rm s)}}$ and for $f_{D_{( \rm s)}}\sqrt{m_{D_{( \rm s)}}}$
in order to asses systematic error.}
\\[6.0ex] \hline \\[-2.0ex]
\parbox[t]{2cm}{ALPHA~13B} & \cite{Heitger:2013oaa} & 2 &
\parbox[t]{2cm}{190\;, 270} & \parbox[t]{5.8cm}{Linear fits (in $m_\pi^2$ and in $a^2$) and
partially quenched HM$\chi$PT functional forms,
including terms linear in $a^2$,  are used in the 
combined chiral/continuum extrapolation.}
\\[12.0ex] \hline \\[-2.0ex]
\parbox[t]{2cm}{ETM 09\\ ETM 11A\\ ETM~13B} & \cite{Blossier:2009bx,Dimopoulos:2011gx,Carrasco:2013zta}  & 2 & \parbox[t]{2cm}{410, 270, 310, 270}
& \parbox[t]{5.8cm}{$M_{\pi,{\rm min}}$ refers to the charged pions.  NLO $SU(2)$ HM$\chi$PT 
supplemented by terms linear in $a^2$ and in $m_D a^2$ is used in the 
combined chiral/continuum extrapolation. To estimate the systematic due to chiral extrapolation,
once $f_{D_s}\sqrt{m_{D_s}}$ and $f_{D_s}\sqrt{m_{D_s}}/(f_{D}\sqrt{m_{D}})$ 
and once $f_{D_s}\sqrt{m_{D_s}}/f_K$ and 
$f_{D_s}\sqrt{m_{D_s}}/f_K \; \times \;  f_\pi/(f_{D}\sqrt{m_{D}})$
are fitted. In ETM~13 the double ratio $(f_{D_s}/f_D)/(f_K/f_\pi)$ is fitted in HM$\chi$PT. }
\\[25.0ex]  \\[-2.0ex]
\hline\hline
\end{tabular*}
\caption{Chiral extrapolation/minimum pion mass in $N_f=2$ determinations of the
  $D$ and $D_s$ meson decay constants.  For actions with multiple
 species of pions, masses quoted are the RMS pion masses (where available).    The
 different $M_{\pi,\rm min}$ entries correspond to the different lattice
 spacings.} 
}
\end{table}

%%%%%%%%%%%%%%%%%%%%%%%%%%%%% FINITE VOLUME   %%%%%%%%%%%%%%%%%%%%%%%%%%%%%%%%%%%%%%%%%%%%%%%%%%%%%%%%%%%%%%%%%%%%%%%%%%%

\begin{table}[!htb]

%\vspace{-3.4cm}

{\footnotesize
\begin{tabular*}{\textwidth}{l @{\extracolsep{\fill}} c c c c l}
\hline\hline \\[-1.0ex]
Collab. & Ref. & $\Nf$ & $L$ [fm] & ${M_{\pi,\rm min}}L$ & Description 
\\[1.0ex] \hline \hline \\[-1.0ex]
\parbox[t]{2cm}{FNAL/MILC~14A} & \cite{Bazavov:2014wgs} & 2+1+1 & \parbox[t]{1.75cm}{2.38-4.83, 2.90-5.82, 2.95-5.62, 2.94-5.44}   & 
\parbox[t]{1.6cm}{$7.6, 7, 4.9, \\ 3.9 $}  &  \parbox[t]{4.7cm}{3 values of $L$ (2.9, 3.9 and 4.9 fm) at $m_\pi=220$ MeV and $a=0.12$ fm.}
\\[9.0ex] \hline \\[-1.0ex]
\parbox[t]{2cm}{ETM~13F \\ ETM~14E} & \cite{Dimopoulos:2013qfa,Carrasco:2014poa} & 2+1+1 & \parbox[t]{1.75cm}{2.13/2.84, 1.96/2.61, 2.97}   & 
\parbox[t]{1.6cm}{$3.5, 3.2, 3.2 $}  &  \parbox[t]{4.7cm}{The comparison of two different volumes at the two largest lattice spacings indicates that FV effects
are below the statistical errors.} 
\\[9.0ex] \hline \\[-1.0ex]
\parbox[t]{2.6cm}{FNAL/MILC~12B \\ FNAL/MILC~13} & \cite{Bazavov:2012dg,Bazavov:2013nfa} & 2+1+1 & 
	     \parbox[t]{1.75cm}{2.4/4.8, 2.88/5.76, 2.88/5.76, 2.88/5.76} & \parbox[t]{1.6cm}{$7.6, 7$, $4.9, 3.9$}  &
\parbox[t]{4.7cm}{FV errors  estimated in $\chi$PT at NLO and, in FNAL/MILC~12B, by analyzing otherwise identical ensembles with three different spatial sizes at $a=0.12$ fm and $m_l/m_s=0.1$.}  
\\[15.0ex] 
\hline\hline
\end{tabular*}
\caption{Finite volume effects in $N_f=2+1+1$ determinations of the $D$ and $D_s$ meson
 decay constants.  Each $L$-entry corresponds to a different
 lattice spacing, with multiple spatial volumes at some lattice
 spacings. For actions with multiple species of pions, the RMS masses are
used (where available).} 
}
\end{table}

\begin{table}[!htb]
{\footnotesize
\begin{tabular*}{\textwidth}{l @{\extracolsep{\fill}} c c c c l}
\hline\hline \\[-1.0ex]
Collab. & Ref. & $\Nf$ & $L$ [fm] & ${M_{\pi,\rm min}}L$ & Description 
\\[1.0ex] \hline \hline \\[-1.0ex]
$\chi$QCD~14& \cite{Yang:2014sea} & 2+1 &
\parbox[t]{1.75cm}{2.7, 2.7}
& \parbox[t]{1.6cm}{4.6, 4.1}  &
\parbox[t]{4.7cm}{No explicit discussion of FV effects.}  
\\[4.0ex] \hline \\[-1.0ex]
HPQCD 12A& \cite{Na:2012iu} & 2+1 &
\parbox[t]{2.1cm}{2.4/2.8, 2.4/3.4}
& \parbox[t]{1.6cm}{6.7, 4.2}  &
\parbox[t]{4.7cm}{FV errors  estimated by comparing finite and infinite volume $\chi$PT.}  
\\[6.0ex] \hline \\[-1.0ex]
FNAL/MILC 11 & \cite{Bazavov:2011aa} & 2+1 & 
	     \parbox[t]{2.1cm}{2.4, 2.4/2.88, 2.52/3.6} & \parbox[t]{1.6cm}{6.9, 6.4, 5.8 }  &
\parbox[t]{4.7cm}{FV errors  estimated using  finite-volume $\chi$PT.}
\\[4.0ex] \hline \\[-1.0ex]
PACS-CS 11 & \cite{Namekawa:2011wt} & 2+1 &
	     \parbox[t]{2.1cm}{$2.88$} & \parbox[t]{1.6cm}{2.2 (before reweighting)}  &
\parbox[t]{4.7cm}{No discussion of FV effects.}
\\[6.0ex] \hline \\[-1.0ex]
HPQCD 10A & \cite{Davies:2010ip} & 2+1 &
             \parbox[t]{2.1cm}{2.4, 2.4/2.88/3.36, 2.52, 2.88, 2.82} & \parbox[t]{1.6cm}{$6.6, 6.7$, $4.2, 3.8$, 4.8 }  &
\parbox[t]{4.7cm}{FV errors  estimated using finite- vs infinite-volume $\chi$PT.}
\\[6.0ex] \hline \\[-1.0ex]
HPQCD/UKQCD 07 & \cite{Follana:2007uv} & 2+1 & 
	     \parbox[t]{2.1cm}{2.4, 2.4/2.88, 2.52} & \parbox[t]{1.6cm}{6.6, 6.7, 4.2}  &
\parbox[t]{4.7cm}{FV errors  estimated using finite- vs infinite-volume $\chi$PT.}
\\[4.0ex] \hline \\[-1.0ex]
FNAL/MILC 05 & \cite{Aubin:2005ar} & 2+1 & \parbox[t]{1.75cm}{2.8, 2.9, 2.5} & \parbox[t]{1.6cm}{$>6, 6.4,$ $5$ }  &
\parbox[t]{4.7cm}{FV errors estimated to be  $1.5\%$ or less from $\chi$PT.} 
\\[4.0ex] 
\hline\hline
\end{tabular*}
\caption{Finite volume effects in $N_f=2+1$ determinations of the $D$ and $D_s$ meson
 decay constants.  Each $L$-entry corresponds to a different
 lattice spacing, with multiple spatial volumes at some lattice
 spacings. For actions with multiple species of pions, the RMS masses are
used (where available).} 
}
\end{table}

\begin{table}[!htb]
{\footnotesize
\begin{tabular*}{\textwidth}{l @{\extracolsep{\fill}} c c c c l}
\hline\hline \\[-1.0ex]
Collab. & Ref. & $\Nf$ & $L$ [fm] & ${M_{\pi,\rm min}}L$ & Description 
\\[1.0ex] \hline \hline \\[-1.0ex]
\parbox[t]{2.0cm}{TWQCD~14} & \cite{Chen:2014hva} & 2 & 
 \parbox[t]{1.75cm}{1.5} & \parbox[t]{1.6cm}{1.92}  &
\parbox[t]{4.7cm}{No explicit discussion of FV effects.}
\\[4.0ex]  \hline \\[-1.0ex]
\parbox[t]{2cm}{ALPHA~13B} & \cite{Heitger:2013oaa} & 2 &
	     \parbox[t]{1.75cm}{2.1/3.1/4.2, 2.3/3.1 } & \parbox[t]{1.6cm}{4, 4.2}  &
\parbox[t]{4.7cm}{No explicit discussion of FV effects, but $m_\pi L>4$ always.}
\\[4.0ex]  \hline \\[-1.0ex]
\parbox[t]{2cm}{ETM~09\\ ETM~11A\\ ETM~13B} & \cite{Blossier:2009bx,Dimopoulos:2011gx,Carrasco:2013zta} & 2 &
	     \parbox[t]{1.75cm}{2.4, 2.0/2.7, 2.1, 2.6 } & \parbox[t]{1.6cm}{5, 3.7, 3.3, 3.5}  &
\parbox[t]{4.7cm}{FV errors are found to be negligible by comparing results at $m_\pi L = 3.3$ and $m_\pi L=4.3$ for $m_\pi \simeq 310$ MeV.}
\\[8.0ex]  \\[-1.0ex]
\hline\hline
\end{tabular*}
\caption{Finite volume effects in $N_f=2$ determinations of the $D$ and $D_s$ meson
 decay constants.  Each $L$-entry corresponds to a different
 lattice spacing, with multiple spatial volumes at some lattice
 spacings. For actions with multiple species of pions, the RMS masses are
used (where available).} 
}
\end{table}
%%%%%%%%%%%%%%%%%%%%%%%%%%%%%%%% CONT EXTRAP  %%%%%%%%%%%%%%%%%%%%%%%%%%%%%%%%%%%%%%%%%%%%%%%%%%%%%%%%%%%%%%%%%%%%%%%
\begin{table}[!htb]
{\footnotesize
\begin{tabular*}{\textwidth}{l @{\extracolsep{\fill}} c c l l l}
\hline\hline \\[-1.65ex]
Collab. & Ref. & $\Nf$ & $a$ [fm]  & Continuum extrapolation & Scale Setting
\\[.2ex] \hline \hline \\[-1.65ex]
\parbox[t]{2.5cm}{FNAL/MILC~14A} & \cite{Bazavov:2014wgs} & 2+1+1 & 
	     \parbox[t]{.9cm}{0.15, 0.12, 0.09, 0.06} &
\parbox[t]{3.8cm}{Interpolations around the physical light masses used to fix the ratio of quark masses. Subsequent chiral and continuum extrapolations for the charm decay constants
performed simulatenously using different NLO HMrAS$\chi$PT fits.} &
 \parbox[t]{3.4cm}{Relative scale through $F_{4ps}$, the decay constant at valence masses = 0.4 $m_s$ and physical sea-quark masses. Absolute scale set through $f_\pi$; the uncertainty is propagated into the final error.} \\[23.0ex] \hline \\[-1.65ex]
\parbox[t]{2cm}{ETM~13F \\ETM~14E} & \cite{Dimopoulos:2013qfa,Carrasco:2014poa} & 2+1+1 & \parbox[t]{.9cm}{0.09, 0.08, 0.06}  & \parbox[t]{3.8cm}{Chiral and continuum extrapolations  
performed simultaneously by adding an $\cO(a^2)$ term to the chiral fits.} &
\parbox[t]{3.5cm}{Relative scale set through $M_{c's'}$, the mass of a fictitious meson made of valence quarks of mass $r_0m_{s'}\!=0.22$ and
$r_0m_{c'}=2.4$. Absoulte scale through $f_\pi$.} 
\\[18.0ex] \hline \\[-1.65ex]
\parbox[t]{2.2cm}{FNAL/MILC~12B \\ FNAL/MILC~13} & \cite{Bazavov:2012dg,Bazavov:2013nfa} & 2+1+1 & 
	     \parbox[t]{.9cm}{0.15, 0.12, 0.09, 0.06} &
\parbox[t]{3.8cm}{Chiral and continuum extrapolations  performed simultaneously. Central values 
 produced  using a fit function quadratic in $a^2$ and linear in the sea quark mass. In FNAL/MILC~13 terms of $\cO(a^4)$ are included.} &
 \parbox[t]{3.4cm}{Absolute scale set through $f_\pi$; the uncertainty is propagated into the final error.} \\[23.0ex] 
\hline\hline
\end{tabular*}
\caption{Lattice spacings and description of actions used in $N_f=2+1+1$ determinations of the $D$ and $D_s$ meson decay constants.} 
}
\end{table}

\begin{table}[!htb]
{\footnotesize
\begin{tabular*}{\textwidth}{l @{\extracolsep{\fill}} c c l l l}
\hline\hline \\[-1.65ex]
Collab. & Ref. & $\Nf$ & $a$ [fm]  & Continuum extrapolation & Scale Setting
\\[.2ex] \hline \hline \\[-1.65ex]
$\chi$QCD~14 & \cite{Yang:2014sea} & 2+1 & \parbox[t]{.9cm}{0.113, 0.085} &
\parbox[t]{3.8cm}{
Chiral and continuum extrapolations  performed in global fits
including linear terms in $m_l$ and terms up to $\cO(a^2)$ and $\cO(a^4m_c^4)$.} 
& \parbox[t]{3.4cm}{Relative scale set through $r_0$, fixed together with the charm and strange quark masses
using $m_{D_s}$, $m_{D_s^*}$ amd $m_{J/\psi}$ as inputs.}
\\[15.0ex] \hline \\[-1.65ex]
HPQCD 12A& \cite{Na:2012iu} & 2+1 & \parbox[t]{.9cm}{0.12, 0.09} &
\parbox[t]{3.8cm}{
Chiral and continuum extrapolations  peformed 
simultaneously  using PQHM$\chi$PT augmentd by $a$ dependent terms: $c_0(am_c)^2 + c_1(am_c)^4$.
} & \parbox[t]{3.4cm}{Relative scale set through $r_1$; absolute scale from $f_\pi$, $f_{\rm K}$ and the $\Upsilon$ splitting.
Uncertainties from both $r_1$ and $r_1/a$  propagated.}
\\[15.0ex] \hline \\[-1.65ex]
FNAL/MILC 11 & \cite{Bazavov:2011aa} & 2+1 &
             \parbox[t]{.9cm}{0.15, 0.12, 0.09} &
\parbox[t]{3.8cm}{
Chiral and continuum extrapolations  peformed 
simultaneously using one-loop HM$\chi$PT for rooted staggered quarks. Effects of hyperfine and flavour splittings
are also included.} & \parbox[t]{3.4cm}{Relative scale set through $r_1=0.3117(22)$. The error in $r_1$
comes from the spread of different absolute scale determinations using $f_\pi$, $f_{\rm K}$ and the $\Upsilon$ splitting.}
\\[20.0ex] \hline \\[-1.65ex]
PACS-CS 11 & \cite{Namekawa:2011wt} & 2+1 &
             \parbox[t]{.9cm}{$0.09$} & \parbox[t]{3.8cm}{
Cutoff effects from the heavy-quark action estimated by naive power counting to be at the percent level.} 
& \parbox[t]{3.4cm}{Scale set through $m_\Omega$.}
\\[12.0ex] 
\hline \\[-1.65ex]
HPQCD 10A & \cite{Davies:2010ip} & 2+1 &
\parbox[t]{.9cm}{0.15, 0.12, 0.09, 0.06, 0.044} &
\parbox[t]{3.8cm}{
Chiral and continuum 
extrapolations  performed simultaneously. Polynomials 
up to $am_{c}^8$ are kept (even powers only).
} & \parbox[t]{3.4cm}{See the discussion for HPQCD 12A.}
\\[12.0ex] \hline \\[-1.65ex]
HPQCD/UKQCD 07 & \cite{Follana:2007uv} & 2+1 & 
\parbox[t]{.9cm}{0.15, 0.12, 0.09} &
\parbox[t]{3.8cm}{
Combined chiral and continuum extrapolations using
HM$\chi$PT at NLO augmented by second and third-order polynomial terms in $m_q$ and terms up to $a^4$.
} & \parbox[t]{3.4cm}{Scale set through $r_1$ obtained from the $\Upsilon$ spectrum using the 
non-relativistic QCD action for $b$ quarks. Uncertainty propagated among the systematics.}
\\[18.0ex] \hline \\[-1.65ex]
FNAL/MILC 05 & \cite{Aubin:2005ar} & 2+1 &
\parbox[t]{.9cm}{0.175, 0.121, 0.086} &
\parbox[t]{3.8cm}{Most light-quark cutoff effects are removed through
 NLO HM$\chi$PT for rooted staggered quarks. Continuum values are then obtained 
by averaging the $a \approx 0.12$ and $a \approx 0.09$ fm results.}
 &  \parbox[t]{3.4cm}{Scale set through $r_1$ obtained from the $\Upsilon$ spectrum using the 
non-relativistic QCD action for $b$ quarks.}
\\[20.0ex] 
\hline\hline
\end{tabular*}
\caption{Lattice spacings and description of actions used in $N_f=2+1$ determinations of the $D$ and $D_s$ meson decay constants.} 
}
\end{table}

\begin{table}[!htb]
{\footnotesize
\begin{tabular*}{\textwidth}{l @{\extracolsep{\fill}} c c l l l}
\hline\hline \\[-1.65ex]
Collab. & Ref. & $\Nf$ & $a$ [fm]  & Continuum extrapolation & Scale Setting
\\[.2ex] \hline \hline \\[-1.65ex]
\parbox[t]{2.0cm}{TWQCD~14} & \cite{Chen:2014hva} & 2 & 
\parbox[t]{.9cm}{0.061} &
\parbox[t]{3.8cm}{Uncertainties associated to scale setting and discretization effects
estimated by performig the chiral fits once in physical and once in lattice units
($\approx 2 $ MeV on $f_{ D_{\rm s}}$).} & \parbox[t]{3.4cm}{Scale set through the Wilson flow and $r_0$ set to 0.49 fm.}
\\[18.0ex] \hline \\[-2.1ex]
\parbox[t]{2.0cm}{ALPHA~13B} & \cite{Heitger:2013oaa} & 2 &
	     \parbox[t]{.9cm}{0.065, 0.048 } &
\parbox[t]{3.8cm}{Linear fits (in $m_\pi^2$ and in $a^2$) and
partially quenched HM$\chi$PT functional forms,
including terms linear in $a^2$,  are used in the 
combined chiral/continuum extrapolation.} & \parbox[t]{3.4cm}{Scale set through $f_K$.}
\\[18.0ex] \hline \\[-2.1ex]
\parbox[t]{2.0cm}{ETM~09\\ ETM~11A\\ ETM~13B} & \cite{Blossier:2009bx,Dimopoulos:2011gx,Carrasco:2013zta} & 2 &
	     \parbox[t]{.9cm}{0.10, 0.085, 0.065, 0.054 } &
\parbox[t]{3.8cm}{
NLO $SU(2)$ HM$\chi$PT 
supplemented by terms linear in $a^2$ and in $m_D a^2$ is used in the 
combined chiral/continuum extrapolation.} & \parbox[t]{3.4cm}{Scale set through $f_\pi$.}
\\[14.0ex] \\[-2.1ex]
\hline\hline
\end{tabular*}
\caption{Lattice spacings and description of actions used in $N_f=2$ determinations of the $D$ and $D_s$ meson decay constants.} 
}
\end{table}

%%%%%%%%%%%%%%%%%%%%%%%%%%%%%%%%%%%%% OPERATOR RENORMALIZATION  %%%%%%%%%%%%%%%%%%%%%%%%%%%%%%%%%%%%%%%%%

\begin{table}[!htb]
{\footnotesize
\begin{tabular*}{\textwidth}{l @{\extracolsep{\fill}} c c c l}
\hline\hline \\[-1.0ex]
Collab. & Ref. & $\Nf$ &  Ren. & Description
\\[1.0ex] \hline \hline \\[-1.0ex]
\parbox[t]{2cm}{FNAL/MILC~14A} & \cite{Bazavov:2014wgs} & 2+1+1 & 
$-$ &
\parbox[t]{6.8cm}{The axial current is absolutely normalized.} 
\\[1.0ex] \hline \\[-1.0ex]
\parbox[t]{2.5cm}{ETM 13F, 14E} & \cite{Dimopoulos:2013qfa,Carrasco:2014poa} & 2+1+1 & $-$ & 
\parbox[t]{6.8cm}{The axial current is absolutely normalized.} 
\\[1.0ex] \hline \\[-1.0ex]
\parbox[t]{3.1cm}{FNAL/MILC 12B, 13} & \cite{Bazavov:2012dg,Bazavov:2013nfa} & 2+1+1 & 
$-$ &
\parbox[t]{6.8cm}{The axial current is absolutely normalized.} 
\\[1.0ex] \hline \\[-1.0ex]
$\chi$QCD~14 & \cite{Yang:2014sea} & 2+1 & 
RI &
\parbox[t]{6.8cm}{The decay constant is extracted from an exact lattice Ward identity and from the NP renormalized axial
current.}
\\[7.0ex] \hline \\[-1.0ex]
HPQCD 12A& \cite{Na:2012iu} & 2+1 & 
$-$ &
\parbox[t]{6.8cm}{The axial current is absolutely normalized.}
\\[1.0ex] \hline \\[-1.0ex]
FNAL/MILC 11 & \cite{Bazavov:2011aa} & 2+1 &
mNPR &
\parbox[t]{6.8cm}{Two-loop and higher-order perturbative truncation errors estimated to be the full size of the one-loop term.}
\\[7.0ex] \hline \\[-1.0ex]
PACS-CS 11 & \cite{Namekawa:2011wt} & 2+1 &
PT1$\ell$+NP &
\parbox[t]{6.8cm}{Mass dependent part of the renormalization constant of the axial current computed at one-loop; the NP contribution
is added in the chiral limit.}
\\[7.0ex] \hline \\[-1.ex]
HPQCD 10A & \cite{Davies:2010ip} & 2+1 &
$-$ &
\parbox[t]{6.8cm}{The axial current is absolutely normalized.}
\\[1.0ex] \hline \\[-1.0ex]
HPQCD/UKQCD 07 & \cite{Follana:2007uv} & 2+1 & 
$-$&
\parbox[t]{6.8cm}{The axial current is absolutely normalized.}
\\[1.0ex] \hline \\[-1.0ex]
FNAL/MILC 05 & \cite{Aubin:2005ar} & 2+1 &
mNPR &
\parbox[t]{6.8cm}{Errors due to higher order corrections in the perturbative part are estimated to be $1.3\%$.}
\\[4.0ex] \hline \\[-1.0ex]
TWQCD~14 & \cite{Chen:2014hva} & 2 & 
$-$&
\parbox[t]{6.8cm}{The decay constant is extracted from an exact lattice Ward identity.}
\\[4.0ex] \hline \\[-1.0ex]
ALPHA~13B & \cite{Heitger:2013oaa} & 2 &
SF &
\parbox[t]{6.8cm}{NP renormalization and improvement of the axial current ($am$ terms included at 1-loop).}
\\[4.0ex]  \hline \\[-1.0ex]
\parbox[t]{2.7cm}{ETM 09, 11A, 13B} & \cite{Blossier:2009bx,Dimopoulos:2011gx,Carrasco:2013zta} & 2 &
$-$ &
\parbox[t]{6.8cm}{The axial current is absolutely normalized.}
\\[1.0ex] 
\hline\hline
\end{tabular*}
\caption{Operator renormalization in  determinations of the $D$ and $D_s$ meson decay constants.} 
}
\end{table}

%%%%%%%%%%%%%%%%%%%%%%%%%%%%%%%%%%%  HEAVY QUARK TREATMENT %%%%%%%%%%%%%%%%%%%%%%%%%%%%%%%%%%%%%%%%%%%%%%%%%%%

\begin{table}[!htb]
{\footnotesize
\begin{tabular*}{\textwidth}{l @{\extracolsep{\fill}} c c c l}
\hline\hline \\[-1.2ex]
Collab. & Ref. & $\Nf$ &  Action & Description
\\[1.0ex] \hline \hline \\[-1.2ex]
\parbox[t]{2cm}{FNAL/MILC~14A} & \cite{Bazavov:2014wgs} & 2+1+1 & 
HISQ (on HISQ)&
\parbox[t]{6.0cm}{$0.22<am_c<0.84$. Discretization errors estimated to be $\approx $ 1 MeV using the spread of 108 different 
chiral/continuum fits (for example by including or not
some NNLO discretization effects in HMrAS$\chi$PT).}
\\[12.0ex] \hline \\[-1.2ex]
\parbox[t]{2.7cm}{ETM 13F, 14E} & \cite{Dimopoulos:2013qfa,Carrasco:2014poa} & 2+1+1 & tmWil  &
\parbox[t]{6.0cm}{$0.15 \lsim am_c \lsim 0.20$. 
%$D(a_{\rm min})  \geq 2\%$ also when the relative scale is set through $M_{c's'}$.
}
\\[1.0ex] \hline \\[-1.2ex]
\parbox[t]{2.5cm}{FNAL/MILC~12B \\ FNAL/MILC~13} & \cite{Bazavov:2012dg,Bazavov:2013nfa} & 2+1+1 & 
HISQ (on HISQ)&
\parbox[t]{6.0cm}{$0.29<am_c<0.7$. Discretization errors estimated using different fit ans\"atze to be~$\approx 1.5\%$ for $f_{D_{(s)}}$.} 
\\[7.0ex] 
\hline\hline
\end{tabular*}
\caption{Heavy-quark treatment in  $N_f=2+1+1$ determinations of the $D$ and $D_s$ meson decay constants.} 
}
\end{table}

\begin{table}[!htb]
{\footnotesize
\begin{tabular*}{\textwidth}{l @{\extracolsep{\fill}} c c c l}
\hline\hline \\[-1.2ex]
Collab. & Ref. & $\Nf$ &  Action & Description
\\[1.0ex] \hline \hline \\[-1.2ex]
$\chi$QCD~14 & \cite{Yang:2014sea} & 2+1 & 
Overlap on DW &
\parbox[t]{6cm}{$0.29<am_c<0.75$. Heavy-quark discretization errors estimated by including $(am_c)^2$ and $am_c)^4$ terms
in the chiral/continuum extrapolation.}
\\[9.0ex] \hline \\[-1.4ex]
HPQCD 12A& \cite{Na:2012iu} & 2+1 & 
HISQ &
\parbox[t]{6cm}{$0.41<am_c<0.62$. Heavy-quark discretization errors estimated using different fit ans\"atze to be $\approx~1.2\%$.}
\\[6.0ex] \hline \\[-1.4ex]
FNAL/MILC 11 & \cite{Bazavov:2011aa} & 2+1 &
Fermilab  &
\parbox[t]{6cm}{Discretization errors from charm quark estimated through a combination of Heavy Quark and Symanzik Effective Theories to be around 3\% for $f_{D_{(s)}}$ and negligible for the ratio.}
\\[12.0ex] \hline \\[-1.4ex]
PACS-CS 11 & \cite{Namekawa:2011wt} & 2+1 &
Tsukuba &
\parbox[t]{6cm}{$am_c\approx 0.57$. Heavy-quark discretization errors estimated to be at the percent level by power counting.}
\\[6.0ex] \hline \\[-1.2ex]
HPQCD 10A & \cite{Davies:2010ip} & 2+1 &
HISQ &
\parbox[t]{6cm}{$0.193< am_c<0.825$. Heavy-quark discretization errors estimated by changing the fit-inputs to be $\approx~0.4\%$.}
\\[6.0ex] \hline \\[-1.2ex]
HPQCD/UKQCD 07 & \cite{Follana:2007uv} & 2+1 & 
HISQ &
\parbox[t]{6cm}{$0.43<am_c<0.85$.  Heavy-quark discretization errors estimated from the chiral/continuum fits to be $\approx~0.5\%$.
%$\delta(a_{\rm min})$ slightly $>\!1\!$ for $f_{D_s}$.
}
\\[6.0ex] \hline \\[-1.4ex]
FNAL/MILC 05 & \cite{Aubin:2005ar} & 2+1 &
Fermilab  &
\parbox[t]{6cm}{Discretization errors from charm quark estimated via heavy-quark power-counting at 4.2\% for $f_{D_{(s)}}$ and 0.5\% for the ratio.}
\\[6.0ex] 
%\hline \\[-1.4ex]
%
\hline\hline
\end{tabular*}
\caption{Heavy-quark treatment in  $N_f=2+1$ determinations of the $D$ and $D_s$ meson decay constants.} 
}
\end{table}

\begin{table}[!htb]
{\footnotesize
\begin{tabular*}{\textwidth}{l @{\extracolsep{\fill}} c c c l}
\hline\hline \\[-1.2ex]
Collab. & Ref. & $\Nf$ &  Action & Description
\\[1.0ex] \hline \hline \\[-1.2ex]
\parbox[t]{2cm}{TWQCD~14} & \cite{Chen:2014hva} & 2 &
DW &
\parbox[t]{6cm}{$am_c\leq 0.55$. Optimal Domain Wall fermions~\cite{Chiu:2002ir} preserving chiral symmetry.}
\\[4.0ex] \hline  \\[-1.2ex]
\parbox[t]{2cm}{ALPHA~13B} & \cite{Heitger:2013oaa} & 2 &
npSW &
\parbox[t]{6cm}{$am_c\leq 0.28$. Axial current non-perturbatively improved ($\cO(am)$ at 1-loop).}
\\[6.0ex] \hline  \\[-1.2ex]
\parbox[t]{2cm}{ETM 09, 11A, 13B} & \cite{Blossier:2009bx,Dimopoulos:2011gx,Carrasco:2013zta} & 2 &
tmWil &
\parbox[t]{6cm}{$0.16<am_c<0.23$. $D(a_{\rm min})\approx 5\%$ in ETM~09.}
\\[4.0ex]  
\hline\hline
\end{tabular*}
\caption{Heavy-quark treatment in  $N_f=2$ determinations of the $D$ and $D_s$ meson decay constants.} 
}
\end{table}

%% file: HQ/AppendixTables/DSLapp.tex
%=================================================
\subsubsection{$D\to\pi\ell\nu$ and $D\to K\ell\nu$ form factors}
\label{app:DtoPi_Notes}
%=================================================

%%%%%%%%%%%%%%%%%%%%%%%%%%%%%%%% CONT EXTRAP %%%%%%%%%%%%%%%%%%%%%%%%%%%%%%%%%%%%%%%%%%%%%%%%%%%%%%%%%%%%%%%%%%%%%%%
\begin{table}[!ht]

{\footnotesize
\begin{tabular*}{\textwidth}{l c c c l l}
\hline\hline \\[-1.0ex]
Collab. & Ref. & $\Nf$ & $a$ [fm] & Continuum extrapolation & Scale setting 
\\[1.0ex] \hline \hline \\[-1.0ex]
%
%HPQCD 13C & \cite{Koponen:2013tua} & 2+1 & \parbox[t]{1.3cm}{0.09, 0.12}  &
%\parbox[t]{3.8cm}{Modified z-expansion fit combining the continuum and chiral extrapolations and the momentum trans- fer dependence.} & \parbox[t]{3.8cm}{Relative scale $r_1/a$ set from the static-quark potential.  Absolute scale $r_1$ set from several quantities including $f_\pi$, $f_K$, and $\Upsilon$ $2S-1S$ splitting c.f. HPQCD~09B~\cite{Davies:2009tsa}.}
%\\[18.0ex] \hline \\[-2.0ex]
%
HPQCD 10B, 11 & \cite{Na:2010uf,Na:2011mc} & 2+1 & \parbox[t]{1.3cm}{0.09, 0.12}  &
\parbox[t]{3.8cm}{Modified $z$-expansion fit combining the continuum and chiral extrapolations and the momentum transfer dependence. Leading discretization errors from $(am_c)^n$ charm-mass effects (see Table~\ref{tab:DtoPiKHQ}). 
    Subleading $(aE)^n$ discretization corrections 
     estimated to be 1.0\% for both 
    $D\to\pi$ and $D\to K$.} & \parbox[t]{3.8cm}{Relative scale $r_1/a$ set from the static-quark potential.  Absolute scale $r_1$ set from several quantities including $f_\pi$, $f_K$, and $\Upsilon$ $2S-1S$ splitting c.f. HPQCD~09B~\cite{Davies:2009tsa}.  Scale uncertainty estimated to be 0.7\% in $D\to\pi$ and and 0.2\% in $D\to K$.}
\\[32.0ex] \hline \\[-2.0ex]
\parbox[t]{1cm}{FNAL/MILC~04} & \cite{Aubin:2004ej} & 2+1 & 0.12  &
\parbox[t]{3.8cm}{Discretization effects from light-quark sector estimated to be 4\% by power counting. Discretization effects from final-state pion and kaon energies estimated to be 5\%.} & \parbox[t]{3.8cm}{Scale set through $\Upsilon$ $2S-1S$ splitting c.f. HPQCD 03~\cite{Wingate:2003gm}.  Error in $a^{-1}$ estimated to be 1.2\%, but scale error in dimensionless form factor negligible compared to other uncertainties.}
\\[20.0ex] \hline \\[-2.0ex]
ETM 11B & \cite{DiVita:2011py} & 2 & \parbox[t]{1.3cm}{0.068, 0.086, 0.102} &
\parbox[t]{3.8cm}{Discretization errors estimated to be 5\% for $D\to\pi$ and 3\% for $D\to K$ from comparison of results in the continuum limit to those at the finest lattice spacing.}
& \parbox[t]{3.8cm}{Scale set through $f_\pi$ c.f. ETM~07A~\cite{Boucaud:2007uk} and ETM~09C~\cite{Baron:2009wt}.}
\\[18.0ex]
\hline\hline
\end{tabular*}
\caption{Continuum extrapolations/estimation of lattice artifacts in determinations of the $D\to\pi\ell\nu$ and $D\to K \ell\nu$ form factors.}
}
\end{table}
%%%%%%%%%%%%%%%%%%%%%%%%% CHIRAL EXTRAP %%%%%%%%%%%%%%%%%%%%%%%%%%%%%%%%%%%%%%%%%%%%%%%%%%%%%%%%%%%%%%%%%%%%%%%%%%%%%%%
\begin{table}[!ht]
{\footnotesize
\begin{tabular*}{\textwidth}{l @{\extracolsep{\fill}} c  c c  l}
\hline\hline \\[-1.0ex]
{Collab.} & {Ref.} & {$\Nf$} & {$M_{\pi,\rm min}\,[\mev]$}  & {Description}  
\\[1.0ex] \hline \hline \\[-1.0ex]
%
%HPQCD 13C & \cite{Koponen:2013tua} & 2+1 & \parbox[t]{2cm}{390, 390}
%& \parbox[t]{6cm}{Modified z-expansion fit combining the continuum and chiral extrapolations and the momentum transfer dependence.}
%\\[7.0ex] \hline \\[-2.0ex]
%
HPQCD 10B, 11 & \cite{Na:2010uf, Na:2011mc} & 2+1 & \parbox[t]{2cm}{390, 390}
& \parbox[t]{6cm}{Modified $z$-expansion fit combining the continuum and chiral extrapolations and the momentum transfer dependence. Contributions to error budget from light valence and sea-quark mass dependence estimated to be 2.0\% for $D \to\pi$ and 1.0\% for $D \to K$. }
\\[18.0ex] \hline \\[-2.0ex]
FNAL/MILC~04 & \cite{Aubin:2004ej} & 2+1 & \parbox[t]{2cm}{510}
 & \parbox[t]{6cm}{Fit to S$\chi$PT, combined with the Becirevic-Kaidalov ansatz for the momentum
 transfer dependence of form factors. Error estimated to be 3\% for $D \to\pi$ and 2\% for $D \to K$ by comparing fits with and without one extra analytic term.}
\\[15.0ex] \hline \\[-2.0ex]
ETM 11B & \cite{DiVita:2011py} & 2 & \parbox[t]{2cm}{270}
& \parbox[t]{6cm}{$SU(2)$ tmHM$\chi$PT plus Becirevic-Kaidalov ansatz for fits to the momentum transfer dependence of form factors.  Fit uncertainty estimated to be 7\% for $D \to\pi$ and 5\% for $D \to K$
by considering fits with and without NNLO corrections of order $\cO(m_\pi^4)$ and/or higher-order terms through $E^5$, and by excluding data with $E \gtapprox 1$~GeV.}
\\[24.0ex]
\hline\hline
\end{tabular*}
\caption{Chiral extrapolation/minimum pion mass in
  determinations of the $D\to \pi\ell\nu$ and $D\to K \ell\nu$ form factors.  For actions with multiple species of pions, masses quoted are the RMS pion masses.  The different $M_{\pi,\rm min}$ entries correspond to the different lattice spacings.}
}
\end{table}

%%%%%%%%%%%%%%%%%%%%%%%%%%%%% FINITE VOLUME %%%%%%%%%%%%%%%%%%%%%%%%%%%%%%%%%%%%%%%%%%%%%%%%%%%%%%%%%%%%%%%%%%%%%%%%%%%
\begin{table}[!ht]
{\footnotesize
\begin{tabular*}{\textwidth}{l @{\extracolsep{\fill}} c c  c c l}
\hline\hline \\[-1.0ex]
Collab. & Ref. & $\Nf$ & $L$ [fm] & ${M_{\pi,\rm min}}L$ & Description 
\\[1.0ex] \hline \hline \\[-1.0ex]
%
%HPQCD 13C & \cite{Koponen:2013tua} & 2+1 & \parbox[t]{1.8cm}{2.4, 2.4/2.9} & $\gtrsim 3.8$ & 
%\parbox[t]{5cm}{No explicit estimate of FV error, but expected to be small for simulation masses and volumes.}
%\\[7.0ex] \hline \\[-1.0ex]
%
HPQCD 10B, 11 & \cite{Na:2010uf,Na:2011mc} & 2+1 & \parbox[t]{1.8cm}{2.4, 2.4/2.9} & $\gtrsim 3.8$ & 
\parbox[t]{5cm}{Finite volume effects estimated to be 0.04\% for $D\to\pi$ and 0.01\% for $D\to K$ by comparing the ``$m_\pi^2 {\rm log} (m_\pi^2)$" term in infinite and finite volume.}
\\[12.0ex] \hline \\[-1.0ex]
FNAL/MILC 04 & \cite{Aubin:2004ej} & 2+1 & \parbox[t]{1.8cm}{2.4/2.9} & $\gtrsim 3.8$ & 
\parbox[t]{5cm}{No explicit estimate of FV error, but expected to be small for simulation masses and volumes.}
\\[7.0ex] \hline \\[-1.0ex]
ETM 11B & \cite{DiVita:2011py} & 2 & \parbox[t]{1.8cm}{2.2, 2.1/2.8, 2.4} & $\gtrsim 3.7$ &
\parbox[t]{5cm}{Finite volume uncertainty estimated to be at most 2\% by considering
fits with and without the lightest pion mass point at $m_\pi L \approx 3.7$.}
\\[12.0ex]
\hline\hline
\end{tabular*}
\caption{Finite volume effects in determinations of the $D\to \pi\ell\nu$ and $D\to K\ell\nu$ form factors.  Each $L$-entry corresponds to a different lattice
spacing, with multiple spatial volumes at some lattice spacings.  For actions with multiple species of pions, the lightest pion masses are quoted.}
}
\end{table}
%%%%%%%%%%%%%%%%%%%%%%%%%%%%%%%%%%%%% OPERATOR RENORMALIZATION %%%%%%%%%%%%%%%%%%%%%%%%%%%%%%%%%%%%%%%%%
\begin{table}[!ht]
{\footnotesize
\begin{tabular*}{\textwidth}{l @{\extracolsep{\fill}} c c c l}
\hline\hline \\[-1.0ex]
Collab. & Ref. & $\Nf$ & Ren. & Description 
\\[1.0ex] \hline \hline \\[-1.0ex]
%
%HPQCD 13C & \cite{Koponen:2013tua} & 2+1 & --- &
%\parbox[t]{6cm}{Scalar form factor extracted from absolutely normalized scalar current. Vector form factor extracted from both point-split spatial (normalized by requiring $Zf_+(0) = 1$ for flavourless currents and checking mass independence) and local temporal vector currents (normalized by matching to the scalar case using kinematic constraint at zero momentum transfer $f_+(0) = f_0(0)$).}
%\\[20.0ex] \hline \\[-1.0ex]
%
HPQCD 10B, 11 & \cite{Na:2010uf, Na:2011mc} & 2+1 & --- &
\parbox[t]{6cm}{Form factor extracted from absolutely normalized scalar-current matrix element then using kinematic constraint at zero momentum-transfer $f_+(0) = f_0(0)$.}
\\[10.0ex] \hline \\[-1.0ex]
FNAL/MILC~04 & \cite{Aubin:2004ej} & 2+1 & mNPR &
\parbox[t]{6cm}{Size of two-loop correction to current renormalization factor assumed to be negligible.}
\\[7.0ex] \hline \\[-1.0ex]
ETM 11B & \cite{DiVita:2011py} & 2 & --- &
\parbox[t]{6cm}{Form factors extracted from double ratios insensitive to current normalization.}
\\[4.0ex]
\hline\hline
\end{tabular*}
\caption{Operator renormalization in determinations of the $D\to\pi\ell\nu$ and $D\to K \ell\nu$ form factors.}
}
\end{table}
%%%%%%%%%%%%%%%%%%%%%%%%%%%%%%%%%%% HEAVY QUARK TREATMENT %%%%%%%%%%%%%%%%%%%%%%%%%%%%%%%%%%%%%%%%%%%%%%%%%%%
\begin{table}[!ht]
{\footnotesize
\begin{tabular*}{\textwidth}{l @{\extracolsep{\fill}} c c c l}
\hline\hline \\[-1.0ex]
Collab. & Ref. & $\Nf$ & Action & Description 
\\[1.0ex] \hline \hline \\[-1.0ex]
%
%HPQCD 13C & \cite{Koponen:2013tua} & 2+1 & HISQ &
%\parbox[t]{6cm}{Bare charm-quark mass $am_c \sim 0.41$--0.63. No explicit estimate of $(am_c)^n$ errors.}
%\\[4.0ex] \hline \\[-1.0ex]
%
HPQCD 10B, 11 & \cite{Na:2010uf,Na:2011mc} & 2+1 & HISQ &
\parbox[t]{6cm}{Bare charm-quark mass $am_c \sim 0.41$--0.63.  Errors of $(am_c)^n$ estimated within modified $z$-expansion to be 1.4\% for $D\to K$ and 2.0\% for $D \to\pi$.  Consistent with expected size of dominant one-loop cutoff effects on the finest lattice spacing, $\cO(\alpha_S (am_c)^2 (v/c) )\sim 1.6$\%.}
\\[18.0ex] \hline \\[-1.0ex]
FNAL/MILC~04 & \cite{Aubin:2004ej} & 2+1 & Fermilab &
\parbox[t]{6cm}{Discretization errors from charm quark estimated via heavy-quark power-counting to be 7\%.}
\\[7.0ex] \hline \\[-1.0ex]
ETM 11B & \cite{DiVita:2011py} & 2 & tmWil &
\parbox[t]{6cm}{Bare charm-quark mass $am_c \sim 0.17$--0.30.   Expected size of $\cO((am_c)^2)$ cutoff effects on the finest lattice spacing consistent with quoted 5\% continuum-extrapolation uncertainty.}
\\[12.0ex]
\hline\hline
\end{tabular*}
\caption{Heavy quark treatment in determinations of the $D\to\pi\ell\nu$ and $D\to K \ell\nu$ form factors. \label{tab:DtoPiKHQ} 
}}
\end{table}

%% file: HQ/AppendixTables/BLapp.tex
%=================================================
\subsubsection{$B_{(s)}$-meson decay constants}
\label{app:fB_Notes}
%=================================================
%%%%%%%%%%%%%%%%%%%%%%%%%%%% CHIRAL EXTRAP %%%%%%%%%%%%%%%%%%%%%%%
%%%%%%%  Nf = 2+1+1 %%%%%%%
\begin{table}[!htb]
%\vskip-2cm
{\footnotesize
\begin{tabular*}{\textwidth}{l @{\extracolsep{\fill}} c c c   l}
\hline\hline \\[-1.0ex]
Collab. & \hspace{-0.4cm}Ref. & $\Nf$ & $M_{\pi,\rm min}\,[\mev]$  & Description 
\\[1.0ex] \hline \hline \\[-2.0ex]
\parbox[t]{2.8cm}{ETM 13E}
&  \parbox[t]{1.0cm}{\cite{Carrasco:2013naa} } & 2+1+1 & \parbox[t]{1.3cm}{245, 239, 211}
& \parbox[t]{7.0cm}{$M_{\pi,{\rm min}}$ refers to the charged pions.  Linear and NLO (full QCD) 
HM$\chi$PT supplemented by an $a^2$ term  is used for the $SU(3)$ breaking ratios.The chiral fit 
error is estimated from the difference between
the NLO HM$\chi$PT and linear fits with half the difference
used as estimate of the systematic error. The ratio $z_s$ is fit using just linear HM$\chi$PT 
supplemented by an $a^2$ term. }
\\[21.0ex] \hline \\[-2.0ex]
\parbox[t]{2.8cm}{HPQCD 13}
& \parbox[t]{1.0cm}{\cite{Dowdall:2013tga} }  & 2+1+1 & \parbox[t]{1.3cm}{310, 294, 173} & 
\parbox[t]{7.0cm}{Two or three pion masses at each lattice spacing, one each 
with a physical mass GB pion. NLO (full QCD) HM$\chi$PT supplemented by 
generic $a^2$ and $a^4$ terms is used to
interpolate to the physical pion mass.}
\\[12.0ex]
\hline\hline
\end{tabular*}
\caption{Chiral extrapolation/minimum pion mass in determinations of the
  $B$ and $B_s$ meson decay constants for $N_f=2+1+1$ simulations.  
  For actions with multiple species of pions, masses quoted are the RMS pion masses (where available).    
  The different $M_{\pi,\rm min}$ entries correspond to the different lattice
 spacings.} 
}
\end{table}

%%%%%%%  Nf = 2+1 %%%%%%%

\begin{table}[!htb]
%\vskip-0cm
{\footnotesize
\begin{tabular*}{\textwidth}{l @{\extracolsep{\fill}} c c c   l}
\hline\hline \\[-1.0ex]
Collab. & \hspace{-0.4cm}Ref. & $\Nf$ & $M_{\pi,\rm min}\,[\mev]$  & Description
\\[1.0ex] \hline \hline \\[-2.0ex]
\parbox[t]{2.8cm}{RBC/UKQCD~14  \\ RBC/UKQCD 13}
& \parbox[t]{1.0cm}{\cite{Christ:2014uea} \\ \cite{Witzel:2013sla} } & 2+1 & \parbox[t]{1.3cm}{329, 289} &
\parbox[t]{7.0cm}{Two or three light quark masses per lattice spacing.
  In RBC/UKQCD 14, three to four light valence-quark  masses that are heavier than
  the sea-quark masses are also employed to have partially-quenched points.
NLO $SU(2)$ HM$\chi$PT is used.  In RBC/UKQCD 14, the fit with only the
unitary points is the central analysis procedure, and the systematic
errors in the conbined chiral-continuum extrapolation are estimated to
be from $3.1\%$ to $5.9\%$ in the decay constants and the $SU(3)$
breaking ratios.}
\\[29.0ex] \hline \\[-2.0ex]
\parbox[t]{2.8cm}{RBC/UKQCD~14A}
& \parbox[t]{1.0cm}{\cite{Aoki:2014nga} } & 2+1 & \parbox[t]{1.3cm}{327, 289} &
\parbox[t]{7.0cm}{Two or three light quark masses per lattice spacing.
NLO $SU(2)$ HM$\chi$PT is used in the combined chiral-continuum
extrapolation.  The systematic
errors in this extrapolation are estimated to
be $3.54\%$ for $f_{B}$, $1.98\%$ for $f_{B_{s}}$, and $2.66\%$ for $f_{B_{s}}/f_{B}$.}
\\[15.0ex] \hline \\[-2.0ex]
\parbox[t]{2.8cm}{HPQCD 12} &  \parbox[t]{1.0cm}{\cite{Na:2012sp} }  & 2+1 & \parbox[t]{1.3cm}{390, 390} & 
\parbox[t]{7.0cm}{Two or three pion masses at each lattice spacing. NLO (full QCD) 
HM$\chi$PT supplemented by NNLO analytic terms and 
generic $a^2$ and $a^4$ terms is used. The systematic error is estimated by 
varying the fit Ansatz, in particular for the NNLO analytic terms and the $a^{2n}$ terms. }
\\[15.0ex] \hline \\[-2.0ex]
\parbox[t]{2.8cm}{HPQCD 11A} & \parbox[t]{1.0cm}{\cite{McNeile:2011ng}
} & 2+1 & \parbox[t]{1.3cm}{570, 450, 390, 330, 330}  &
\parbox[t]{7.0cm}{One light sea quark mass only at each lattice spacing. The sea-quark mass 
dependence is assumed to be negligible, based on the calculation of $f_{D_s}$ in 
Ref.~\cite{Davies:2010ip}, where the sea quark extrapolation error is estimated as $0.34\%$. } 
\\[12.0ex] \hline \\[-2.0ex]
\parbox[t]{2.8cm}{FNAL/MILC 11}
& \parbox[t]{1.0cm}{\cite{Bazavov:2011aa} } & 2+1 & \parbox[t]{1.3cm}{570, 440, 320} &
\parbox[t]{7.0cm}{Three to five sea-quark masses per lattice spacing, 
and  $9-12$ valence light quark masses per ensemble. NLO partially quenched HMrS$\chi$PT 
including $1/m$ terms and  supplemented by NNLO analytic and $\alpha_s^2a^2$ terms is used. 
The systematic error is estimated by varying the fit Ansatz,  in particular the NNLO analytic 
terms and the chiral scale. }
\\[18.0ex] \hline \\[-2.0ex]
\parbox[t]{2.8cm}{RBC/UKQCD 10C}
& \parbox[t]{1.0cm}{\cite{Albertus:2010nm} } & 2+1 & \parbox[t]{1.3cm}{430} &
\parbox[t]{7.0cm}{Three light quark masses at one lattice spacing. 
NLO $SU(2)$ $\chi$PT is used. The systematic error is estimated
 from the difference between NLO $\chi$PT and linear fits as $\sim 7\%$.}
\\[9.0ex] \hline \\[-2.0ex]
\parbox[t]{2.8cm}{HPQCD 09} &  \parbox[t]{1.0cm}{\cite{Gamiz:2009ku} }
& 2+1 & \parbox[t]{1.3cm}{440, 400} & 
\parbox[t]{7.0cm}{Four or two pion masses per lattice spacing. NLO (full QCD) 
HMrS$\chi$PT supplemented by NNLO analytic terms and 
 $\alpha_s a^2, a^4$ terms is used. The chiral fit error is estimated by varying the fit Ansatz,
in particular, by adding or removing NNLO and discretization terms. }
\\[15.0ex]
\hline\hline
\end{tabular*}
\caption{Chiral extrapolation/minimum pion mass in determinations of the
  $B$ and $B_s$ meson decay constants for $N_f=2+1$ simulations.  
  For actions with multiple species of pions, masses quoted are the RMS pion masses (where available).    
  The different $M_{\pi,\rm min}$ entries correspond to the different lattice
 spacings.} 
}
\end{table}

%%%  Nf = 2 %%%%%%%
\begin{table}[!htb]
%\vskip-2cm
{\footnotesize
\begin{tabular*}{\textwidth}{l @{\extracolsep{\fill}} c c c   l}
\hline\hline \\[-1.0ex]
Collab. & \hspace{-0.4cm}Ref. & $\Nf$ & $M_{\pi,\rm min}\,[\mev]$  & Description 
\\[1.0ex] \hline \hline \\[-2.0ex]
\parbox[t]{2.8cm}{ALPHA 14  \\ ALPHA 13 \\ ALPHA 12A } & 
\parbox[t]{1.0cm}{\cite{Bernardoni:2014fva} \\
    \cite{Bernardoni:2013oda} \\ \cite{Bernardoni:2012ti} } & 
2 & \parbox[t]{1.3cm}{280, 190, 270}
& \parbox[t]{7.0cm}{LO and NLO HM$\chi$PT supplemented by a term linear in $a^2$ are 
used. In ALPHA 13 and ALPHA 12A, the final result is an average between LO and NLO with half the difference
used as estimate of the systematic error.  In ALPHA 14, the NLO fit is
used as the central analysis procedure, and the LO results are used to
estimate the systematic errors ($0.9\%$ MeV for $f_{B_{s}}$, $1.1\%$ for
$f_{B}$, and $1.6\%$ for $f_{B_{s}}/f_{B}$).}
\\[24.0ex]\hline \\[-2.0ex]
\parbox[t]{2.8cm}{ETM 13B, 13C \\ ETM 12B  \\ ETM 11A}  & 
\parbox[t]{1.0cm}{\cite{Carrasco:2013zta,Carrasco:2013iba} \\
  \cite{Carrasco:2012de} \\ \cite{Dimopoulos:2011gx} } & 2 & 
\parbox[t]{1.3cm}{410, 275, 300, 270} &
 \parbox[t]{7.0cm}{$M_{\pi,{\rm min}}$ refers to the charged pions. Linear and NLO (full QCD) 
HM$\chi$PT supplemented by an $a^2$ term is used.The chiral fit error is estimated from the difference between
the NLO HM$\chi$PT and linear fits with half the difference
used as estimate of the systematic error. 
For the static limit calculation in ETM 11A, $\Phi_s^{\rm stat}$ is extrapolated 
assuming a constant in light quark mass.
The ratio $\Phi_s^{\rm stat}/\Phi_{\ell}^{\rm stat}$ is fit using three 
different chiral fit forms (NLO HM$\chi$PT, linear, and quadratic) to estimate the chiral fir error. }
\\[29.0ex] \hline \\[-2.0ex]
\parbox[t]{2.8cm}{ALPHA 11}
& \parbox[t]{1.0cm}{\cite{Blossier:2011dk} }  & 
2 & \parbox[t]{1.3cm}{331, 268, 267}
& \parbox[t]{7.0cm}{Linear and NLO (full QCD) HM$\chi$PT supplemented by a term linear in $a^2$ are 
used. The final result is an average between linear and NLO fits with half the difference
used as estimate of the systematic error. }
\\[12.0ex]\hline \\[-2.0ex]
\parbox[t]{2.8cm}{ETM 09D} & \parbox[t]{1.0cm}{\cite{Blossier:2009hg} } & 2 & \parbox[t]{1.3cm}{410, 275, 300}
& \parbox[t]{7.0cm}{$M_{\pi,{\rm min}}$ refers to the charged pions.  
Linear and NLO (full QCD) 
HM$\chi$PT is used. The final result given by the average of NLO HMChiPT and linear
{\it Ans\"atze} $\pm$ half the difference).}
\\[9.0ex] 
\hline\hline
\end{tabular*}
\caption{Chiral extrapolation/minimum pion mass in determinations of the
  $B$ and $B_s$ meson decay constantsfor $N_f=2$ simulations.  
  For actions with multiple species of pions, masses quoted are the RMS pion masses (where available).    
  The different $M_{\pi,\rm min}$ entries correspond to the different lattice
 spacings.} 
}
\end{table}
%%%%%%%%%%%%%%%%%%%%%%%%%%%% FINITE VOLUME %%%%%%%%%%%%%%%%%%%%%%%

\begin{table}[!htb]
{\footnotesize
\begin{tabular*}{\textwidth}{l @{\extracolsep{\fill}} c c c c l}
\hline\hline \\[-1.0ex]
Collab. & \hspace{-0.4cm}Ref. & $\Nf$ & $L$ [fm] & ${M_{\pi,\rm min}}L$ & Description 
\\[1.0ex] \hline \hline \\[-2.0ex]
\parbox[t]{2.8cm}{ETM 13E} & \parbox[t]{1.0cm}{
  \cite{Carrasco:2013naa}} & 2+1+1 & 
\parbox[t]{1.8cm}{2.84/2.13, 2.61/1.96, 2.97} &
\parbox[t]{1.5cm}{3.5, 3.2, 3.2} &
\parbox[t]{4.9cm}{ FV error estimated how?}
\\[6.0ex] \hline \\[-2.0ex]
\parbox[t]{2.8cm}{HPQCD 13} & \parbox[t]{1.0cm}{\cite{Dowdall:2013tga}} & 2+1+1 & 
\parbox[t]{1.8cm}{2.4/3.5/4.7, 2.9/3.8/5.8, 2.8/5.6}  & 
\parbox[t]{1.5cm}{7.4, 8.6, 4.9} &
\parbox[t]{4.9cm}{ The analysis uses finite-volume $\chi$PT.}
\\[6.0ex] \hline \\[-2.0ex]
\parbox[t]{2.8cm}{RBC/UKQCD~14 \\ RBC/UKQCD 13} & 
\parbox[t]{1.0cm}{ \cite{Christ:2014uea}\\  \cite{Witzel:2013sla}} 
 & 2+1 & 
\parbox[t]{1.8cm}{2.64, 2.75} & \parbox[t]{1.5cm}{4.4, 4.0}& 
\parbox[t]{4.9cm}{In RBC/UKQCD 14, finite-volume effects are
  estimated to be neglegible for $f_{B_{s}}$, $0.4\%$ for $f_{B^{0}}$,
$0.5\%$ for $f_{B^{+}}$ and the $SU(3)$ breaking ratios.}
\\[9.0ex] \hline \\[-2.0ex]
\parbox[t]{2.8cm}{RBC/UKQCD~14A} & \parbox[t]{1.0cm}{\cite{Aoki:2014nga}} & 2+1 & \parbox[t]{1.8cm}{2.74, 2.76}  & 
\parbox[t]{1.8cm}{4.5, 4.0} &
\parbox[t]{4.9cm}{Finite-volume effects are
  estimated to be neglegible for $f_{B_{s}}$, $0.82\%$ for $f_{B}$, and
$1\%$ for $f_{B_{s}}/f_{B}$.}
\\[6.0ex] \hline \\[-2.0ex]
\parbox[t]{2.8cm}{HPQCD 12} & \parbox[t]{1.0cm}{\cite{Na:2012sp}} & 2+1 & \parbox[t]{1.8cm}{2.4/2.9, 2.5/3.6}  & 
\parbox[t]{1.5cm}{5.7, 7.1} &
\parbox[t]{4.9cm}{FV error is taken from Ref.~\cite{Follana:2007uv} for HPQCD's $D$ 
meson analysis, where it was estimated using finite volume $\chi$PT .}
\\[9.0ex] \hline \\[-2.0ex]
\parbox[t]{2.8cm}{HPQCD 11A} & \parbox[t]{1.0cm}{\cite{McNeile:2011ng}} & 2+1 & \parbox[t]{1.8cm}{2.4, 2.4, 2.5, 2.9, 2.9} &
\parbox[t]{1.5cm}{6.9, 5.5, 4.9, 4.8, 4.8}  & 
\parbox[t]{4.9cm}{FV error is assumed to negligible.}
\\[6.0ex] \hline \\[-2.0ex]
\parbox[t]{2.8cm}{FNAL/MILC 11} & \parbox[t]{1.0cm}{\cite{Bazavov:2011aa}} & 2+1 & \parbox[t]{1.8cm}{2.4, 2.4/2.9, 2.5/3.6} &
\parbox[t]{1.5cm}{6.9, 6.4, 5.8}  & 
\parbox[t]{4.9cm}{FV error is estimated using  finite-volume $\chi$PT.}
\\[4.0ex] \hline \\[-2.0ex]
\parbox[t]{2.8cm}{RBC/UKQCD 10C} & \parbox[t]{1.0cm}{\cite{Albertus:2010nm}} & 2+1 & 
\parbox[t]{1.8cm}{1.8} & \parbox[t]{1.5cm}{3.9} & 
\parbox[t]{4.9cm}{FV error estimated using  finite-volume $\chi$PT to be $1\%$
 for $SU(3)$ breaking ratios. }
\\[6.0ex] \hline \\[-2.0ex]
\parbox[t]{2.8cm}{HPQCD 09} & \parbox[t]{1.0cm}{\cite{Gamiz:2009ku}} & 2+1 & \parbox[t]{1.8cm}{2.4/2.9, 2.5} &
\parbox[t]{1.5cm}{6.5, 5.1}  & 
\parbox[t]{4.9cm}{FV error is assumed to negligible.}
\\[1.0ex] \hline \\[-2.0ex]
\parbox[t]{2.8cm}{ALPHA 14 \\ALPHA 13 \\ALPHA 12A \\ALPHA 11} & 
\parbox[t]{1.0cm}{\cite{Bernardoni:2014fva}\\  \cite{Bernardoni:2013oda}\\ \cite{Bernardoni:2012ti}\\ \cite{Blossier:2011dk}} & 2 &
\parbox[t]{1.8cm}{2.4/3.6, 2.1/3.1/4.2, 2.3/3.1} & \parbox[t]{1.5cm}{5.2, 4.1, 4.2}  & 
\parbox[t]{4.9cm}{No explicit estimate of FV errors, but expected to be much smaller than other uncertainties.}
\\[9.0ex] \hline \\[-2.0ex]
\parbox[t]{2.8cm}{ETM 13B, 13C \\ETM 12B \\ETM 11A} & 
\parbox[t]{1.0cm}{ \cite{Carrasco:2013zta,Carrasco:2013iba}\\ \cite{Carrasco:2012de}\\ \cite{Dimopoulos:2011gx}} & 2 &
	     \parbox[t]{1.8cm}{2.4, 2.0/2.7, 2.1, 1.7/2.6} &
	      \parbox[t]{1.5cm}{5.0, 3.7, 3.3, 3.5}  &
\parbox[t]{4.9cm}{FV errors are found to be negligible by comparing results at $m_\pi L = 3.3$ and $m_\pi L=4.3$ for $m_\pi \simeq 310$ MeV.}
\\[9.0ex]
\hline\hline
\end{tabular*}
\caption{Finite volume effects in determinations of the $B$ and $B_s$ meson
 decay constants.  Each $L$-entry corresponds to a different
 lattice spacing, with multiple spatial volumes at some lattice
 spacings.} 
}
\end{table}
%%%%%%%%%%%%%%%%%%%%%%%%%%%% CONT EXTRAP   %%%%%%%%%%%%%%%%%%%%%%%
%%%%%%  Nf = 2+1+1 %%%%%%

\begin{table}[!htb]
%\vskip-2cm
{\footnotesize
\begin{tabular*}{\textwidth}{l @{\extracolsep{\fill}} c c c l l}
\hline\hline \\[-1.0ex]
Collab. & \hspace{-0.4cm}Ref. & $\Nf$ & $a$ [fm]  & Continuum extrapolation & Scale setting
\\[1.0ex] \hline \hline \\[-2.0ex]
\parbox[t]{2.0cm}{ETM 13E} &  \parbox[t]{1.0cm}{\cite{Carrasco:2013naa}} & 2+1+1 & \parbox[t]{0.9cm}{0.89, 0.82, 0.62}
& \parbox[t]{4.3cm}{Combined  continuum and chiral extrapolation, linear in $a^2$.  
%% D ~ 1-3.5\%, delta ~ 1-2
}
& \parbox[t]{4.3cm}{Scale set from $f_\pi$. Scale setting uncertainty included in combined 
statistical and systematic error. }
\\[9.0ex] \hline \\[-2.0ex]
\parbox[t]{2.0cm}{HPQCD 13} & \parbox[t]{1.0cm}{\cite{Dowdall:2013tga}} & 2+1+1 & \parbox[t]{0.9cm}{0.15, 0.12, 0.09}
& \parbox[t]{4.4cm}{Combined  continuum and chiral extrapolation. 
Continuum extrapolation errors estimated to be $0.7\%$. 
%% D ~ 0.5-0.9\%, delta < 1
}
& \parbox[t]{4.3cm}{Scale set from $\Upsilon$(2S-1S)
splitting, see Ref.~\cite{Dowdall:2011wh}.  Scale uncertainty included in statistical error.}
%\\[9.0ex] \hline \\[-2.0ex]
%
\\[9.0ex] 
\hline\hline
\end{tabular*}
\caption{Continuum extrapolations/estimation of lattice artifacts in determinations of 
the $B$ and $B_s$ meson decay constants for $N_f=2+1+1$ simulations.} 
}
\end{table}

%%%%%%%  Nf = 2+1 %%%%%%%

\begin{table}[!htb]
\vspace{-2.4cm}
%\vskip-2cm
{\footnotesize
\begin{tabular*}{\textwidth}{l @{\extracolsep{\fill}} c c c l l}
\hline\hline \\[-1.0ex]
Collab. & Ref. & $\Nf$ & $a$ [fm]  & Continuum extrapolation & Scale setting
\\[1.0ex] \hline \hline \\[-2.0ex]
\parbox[t]{2.8cm}{RBC/UKQCD~14  \\ RBC/UKQCD 13}
& \parbox[t]{0.5cm}{\cite{Christ:2014uea} \\ \cite{Witzel:2013sla}} & 2+1 & \parbox[t]{0.9cm}{0.11, 0.086}
& \parbox[t]{4.3cm}{Combined  continuum and chiral extrapolation with linear in $a^2$ term. 
In RBC/UKQCD 14,  the systematic errors from this procedure are
estimated to be from $3.1\%$ to $5.9\%$ in the decay constants and the
$SU(3)$-breaking ratios.
%$D \sim 2-8\%$, $\delta < 2$
}
& \parbox[t]{4.3cm}{Scale set by the $\Omega$ baryon mass.  In
  RBC/UKQCD 14, scale
  uncertainty estimated to be $1.5\%$ in the decay constants, and
  $0.1\%$ in the $SU(3)$-breaking ratios.}
\\[21.0ex] \hline \\[-2.0ex]
\parbox[t]{2.8cm}{RBC/UKQCD~14A}
& \parbox[t]{0.5cm}{\cite{Aoki:2014nga} } & 2+1 & \parbox[t]{0.9cm}{0.11, 0.086}
& \parbox[t]{4.3cm}{Chiral-continuum extrapolation with
  linear in $a^2$ term is employed, with the systematic errors 
  estimated to be from $1.98\%$ to $3.54\%$ in the decay constants and
  $f_{B_{s}}/f_{B}$.  Discretization errors at $\cO(\alpha_{s} a)$ in
  the static-light system are estimated to be $1\%$
  in the decay constants, and $0.2\%$ in $f_{B_{s}}/f_{B}$. 
%$D \sim 2-8\%$, $\delta < 2$
}
& \parbox[t]{4.3cm}{Scale set by the $\Omega$ baryon mass.}
\\[29.0ex] \hline \\[-2.0ex]
\parbox[t]{2.8cm}{HPQCD 12} & \parbox[t]{0.5cm}{\cite{Na:2012sp} } & 2+1 & \parbox[t]{0.9cm}{0.12, 0.09}
& \parbox[t]{4.3cm}{Combined  continuum and chiral extrapolation. 
Continuum extrapolation errors estimated to be $0.9\%$. 
%$D \sim 1\%$, $\delta < 1$
}
& \parbox[t]{4.3cm}{Relative scale $r_1/a$ from the static-quark potential.  
Absolute scale $r_1$  from $f_\pi$, $f_K$, and $\Upsilon$(2S-1S) 
splitting.  Scale uncertainty estimated to be $1.1\%$.}
\\[15.0ex] \hline \\[-2.0ex]
\parbox[t]{2.8cm}{HPQCD 11A} & \parbox[t]{0.5cm}{\cite{McNeile:2011ng}} & 2+1 & \parbox[t]{0.9cm}{0.15, 0.12, 0.09, 0.06, 0.045}
& \parbox[t]{4.3cm}{ $am_Q \approx 0.2 - 0.85$. Combined  continuum and HQET fit. 
 %Apply cont. criteria over entire range of $am_Q$ used in rating. 
 Continuum extrapolation error estimated by varying the fit ansatz and the included data
 points to be $0.63\%$. Discretization errors appear to decrease with increasing heavy-meson 
 mass. 
 %%%D, delay from fDs analysis okay
 } 
& \parbox[t]{4.3cm}{Relative scale $r_1/a$ from the static-quark potential.  
Absolute scale $r_1$ from $f_\pi$, $f_K$, and 
$\Upsilon$(2S-1S) splitting.  Scale uncertainty estimated to be $0.74\%$.}
\\[23.0ex] \hline \\[-2.0ex]
\parbox[t]{2.8cm}{FNAL/MILC 11} &\parbox[t]{0.5cm}{\cite{Bazavov:2011aa} } & 2+1 & \parbox[t]{0.9cm}{0.15, 0.12, 0.09}
& \parbox[t]{4.3cm}{Combined continuum and chiral extrapolation. Continuum extrapolation 
errors estimated to be 1.3\%.
%$D\sim1\%$, $\delta < 1$
}
& \parbox[t]{4.3cm}{Relative scale $r_1/a$ from the static-quark potential.  
Absolute scale $r_1$ from $f_\pi$, $f_K$, and $\Upsilon$(2S-1S) splitting.  
Scale uncertainty estimated to be 1~MeV.}
\\[15.0ex] \hline \\[-2.0ex]
\parbox[t]{2.8cm}{RBC/UKQCD 10C} &\parbox[t]{0.5cm}{\cite{Albertus:2010nm}}  & 2+1 & \parbox[t]{0.9cm}{0.11}
& \parbox[t]{4.3cm}{One lattice spacing with discretization errors estimated 
by power counting as  3\%.}
& \parbox[t]{4.3cm}{Scale set by the $\Omega$ baryon mass. Combined scale and mass tuning
uncertainties on $f_{B_s}/f_B$ estimated as 1\%}
\\[9.0ex] \hline \\[-2.0ex]
\parbox[t]{2.8cm}{HPQCD 09} & \parbox[t]{0.5cm}{\cite{Gamiz:2009ku} } & 2+1 & \parbox[t]{0.9cm}{0.12, 0.09}
& \parbox[t]{4.3cm}{Combined continuum and chiral extrapolation. Continuum 
extrapolation errors estimated to be 3\%. 
% $D\sim 6\%$, $\delta \sim 1-2$.
}
& \parbox[t]{4.3cm}{Relative scale $r_1/a$ from the static-quark potential.  
Absolute scale $r_1$  from the $\Upsilon$(2S-1S) splitting.  Scale uncertainty
 estimated to be $2.3\%$. }
\\[12.0ex] 
\hline\hline
\end{tabular*}
\caption{Continuum extrapolations/estimation of lattice artifacts in determinations of 
the $B$ and $B_s$ meson decay constants for $N_f=2+1$ simulations.} 
}
\end{table}

%%%% Nf = 2 %%%%%%%%
\begin{table}[!htb]
{\footnotesize
\begin{tabular*}{\textwidth}{l @{\extracolsep{\fill}} l c c l l}
\hline\hline \\[-1.0ex]
Collab. & Ref. & $\Nf$ & $a$ [fm]  & Continuum extrapolation & Scale setting
\\[1.0ex] \hline \hline \\[-2.0ex]
\parbox[t]{2.0cm}{ALPHA 14  \\ ALPHA 13 \\ ALPHA 12A \\ ALPHA 11} & 
\parbox[t]{1.0cm}{\cite{Bernardoni:2014fva} \\
    \cite{Bernardoni:2013oda} \\ \cite{Bernardoni:2012ti} \\ \cite{Blossier:2011dk}}
  & 2 &
\parbox[t]{0.9cm}{0.075, 0.065, 0.048} & 
\parbox[t]{4.3cm}{Combined continuum and chiral extrapolation with linear in  
$a^2$ term. Continuum extrapolation 
errors estimated to be 5~MeV in ALPHA 11.  The
continuum extrapolation with a term linear in $a$ also inverstigated
in ALPHA 14, and within the statistical error no discernable difference was observed.
%Discretization errors appear to be small, all finite lattice spacing lines are within the statistical 
%uncertainty of the cont. extrapolated point. 
%$D ~ 1.6$ and $\delta < 1$. \\
}
& \parbox[t]{4.3cm}{Relative scale set from $r_0$. Absolute scale set from $f_K$. Scale setting uncertainty included in combined statistical and extrapolation error.}
\\[29.0ex] \hline \\[-2.0ex]
\parbox[t]{2.4cm}{ETM 13B, 13C \\ ETM 12B  \\ ETM 11A}  & 
\parbox[t]{1.0cm}{\cite{Carrasco:2013zta,Carrasco:2013iba} \\
  \cite{Carrasco:2012de} \\ \cite{Dimopoulos:2011gx} }& 2 &
	     \parbox[t]{0.9cm}{0.098, 0.085, 0.067, 0.054} &
\parbox[t]{4.3cm}{Combined continuum and chiral extrapolation, with a term linear in $a^2$.
%For $z_s$, $D ~ 1 \%$, $ \delta ~ 1$ at heaviest mass if using the total error. 
ETM 12 and 13 include a heavier masses than ETM 11A. Discretization error included in combined
statistical and systematic error, estimated by dropping the data at the coarsest lattice spacing
as $\sim 0.5 - 1\%$. 
%The ETM 12 paper doesn't include the figure that's needed to evaluate D, delta. \\
}
& \parbox[t]{4.3cm}{Scale set from $f_\pi$. Scale setting uncertainty included in combined 
statistical and systematic error. }
\\[26.0ex] \hline \\[-2.0ex]
\parbox[t]{2.0cm}{ETM 09D} & \parbox[t]{1.0cm}{\cite{Blossier:2009hg} } & 2 &
	     \parbox[t]{0.9cm}{0.098, 0.085, 0.067} &
\parbox[t]{4.3cm}{Combined continuum and chiral extrapolation with a term linear in 
$a^2$. 
}
& \parbox[t]{4.3cm}{Scale set from $f_\pi$. Scale setting uncertainty included in 
combined statistical and systematic error.}
\\[9.0ex] 
\hline\hline
\end{tabular*}
\caption{Continuum extrapolations/estimation of lattice artifacts in determinations of 
the $B$ and $B_s$ meson decay constants for $N_f=2$ simulations.} 
}
\end{table}

%%%%%%%%%%%%%%%%%%%%%%%%%%%% OP RENORM     %%%%%%%%%%%%%%%%%%%%%%%
%
\begin{table}[!htb]
{\footnotesize
\begin{tabular*}{\textwidth}{l @{\extracolsep{\fill}} c c c l}
\hline\hline \\[-1.0ex]
Collab.  &  Ref. & $\Nf$ & Ren. & Description  \\[1.0ex] 
\hline \hline \\[-2.0ex]
ETM 13E  & \cite{Carrasco:2013naa} & 2+1+1 & --, PT1$\ell$ &
\parbox[t]{7.5cm}{The current used for the relativistic decay constants is absolutely normalized. 
 The ratio is constructed from the relativistic decay constant data
 and the heavy-quark pole masses. 
 Ratios of pole-to-$\msbar$ mass  conversion factors are included 
at NLO in continuum perturbation theory.}
\\[15.0ex] \hline \\[-2.0ex]
HPQCD 13 & \cite{Dowdall:2013tga} & 2+1+1 & PT1$\ell$
& \parbox[t]{7.5cm}{The NRQD effective current is matched through $\cO(1/m$) and renormalized using one-loop PT.  Included are all terms though
$\cO(\alpha_s)$,  $\cO(\alpha_s \, a)$, $\cO(\Lambda_{\rm QCD}/M)$, $\cO(\alpha_s/a M)$ ,  $\cO(\alpha_s \, \Lambda_{\rm QCD}/M)$. The dominant error is due unknown $\cO(\alpha_s^2)$ contributions to the current renormalization. The perturbation theory used in this work is the same as in 
HPQCD 09 and 12, but is rearranged to match the mNPR method. Using the fact that the heavy-heavy temporal vector current is normalized, and that the light-light HISQ vector current receives a small one-loop correction, the error is estimated as  $\sim1.4$\%. }
%\\[29.0ex] \hline \\[-2.0ex]
%
\\[32.0ex]  
\hline\hline
\end{tabular*}
\caption{Description of the renormalization/matching procedure adopted
  in the determinations of the $B$ and $B_s$ meson decay constants for
  $N_{f} = 2 + 1+1$ simulations.} 
}
\end{table}

\begin{table}[!htb]
{\footnotesize
\begin{tabular*}{\textwidth}{l @{\extracolsep{\fill}} c c c l}
\hline\hline \\[-1.0ex]
Collab.  &  \hspace{-0.4cm}Ref. & $\Nf$ & Ren. & Description 
\\[1.0ex] \hline \hline \\[-2.0ex]
\parbox[t]{2.8cm}{RBC/UKQCD~14 \\RBC/UKQCD 13}
& \parbox[t]{1.0cm}{ \cite{Christ:2014uea} \\ \cite{Witzel:2013sla}} & 2+1 & mNPR 
& \parbox[t]{7.5cm}{In RBC/UKQCD 14, the error is dominated by the
  perturbative aspect, and is estimated to be $1.7\%$ for the decay
  constants by taking the full size of the
  one-loop correction for the fine lattice. }
\\[9.0ex] \hline \\[-2.0ex]
\parbox[t]{2.8cm}{RBC/UKQCD~14A}
& \parbox[t]{1.0cm}{ \cite{Aoki:2014nga}} & 2+1 & PT1$\ell$ 
& \parbox[t]{7.5cm}{A two-step matching procedure is employed, first from
  QCD to HQET in the continuum at $m_{b}$, then to HQET on the lattice
  at $a^{-1}$ with $\cO(pa)$ and $\cO(m_{q} a)$ errors included.  Both
  matching steps are accurate to one-loop, and the
  running between $m_{b}$ and $a^{-1}$ is performed at two-loop
  accordingly.  The error is estimated using a power-counting argument
to be $6\%$ for the decay constants.}
\\[20.0ex] \hline \\[-2.0ex]
\parbox[t]{2.8cm}{HPQCD 12/09} & \parbox[t]{1.0cm}{\cite{Na:2012sp,Gamiz:2009ku}} & 2+1 & PT1$\ell$
& \parbox[t]{7.5cm}{The NRQD effective current is matched through $\cO(1/m$) and renormalized using one-loop PT.  Included are all terms though
$\cO(\alpha_s)$,  $\cO(\alpha_s \, a)$, $\cO(\Lambda_{\rm QCD}/M)$, $\cO(\alpha_s/a M)$ ,  $\cO(\alpha_s \, \Lambda_{\rm QCD}/M)$. The dominant error is due unknown $\cO(\alpha_s^2)$ contributions to the current renormalization. The authors take the perturbative error as $\sim 2  \rho_0 \, \alpha_s^2$, where $\rho_0$ is the coefficient of the one-loop correction to the leading term, which yields an
error of $\sim4$\%. }
\\[23.0ex] \hline \\[-2.0ex]
\parbox[t]{2.8cm}{HPQCD 11A} & \parbox[t]{1.0cm}{\cite{McNeile:2011ng}} & 2+1 & -- 
& \parbox[t]{7.5cm}{This work uses PCAC together with an absolutely normalized current.}
\\[3.0ex] \hline \\[-2.0ex]
\parbox[t]{2.8cm}{FNAL/MILC 11} & \parbox[t]{1.0cm}{\cite{Bazavov:2011aa}}  & 2+1 & mNPR 
& \parbox[t]{7.5cm}{The authors' estimate of the perturbative errors is comparable in size to the actual 
one-loop corrections. }
\\[3.0ex] \hline \\[-2.0ex]
\parbox[t]{2.8cm}{RBC/UKQCD 10C} & \parbox[t]{1.0cm}{\cite{Albertus:2010nm}} & 2+1 & PT1$\ell$ & 
\parbox[t]{7.5cm}{The static-light current is matched through
 $\cO(\alpha_s a, \alpha_s)$ and renormalized using one-loop
tadpole improved PT. For massless light quarks, the renormalization factors
cancel in the ratio of decay constants.}
\\[12.0ex] \hline \hline 
\end{tabular*}
\caption{Description of the renormalization/matching procedure adopted
  in the determinations of the $B$ and $B_s$ meson decay constants for
  $N_{f} = 2+1$  simulations.} 
}
\end{table}

\begin{table}[!htb]
{\footnotesize
\begin{tabular*}{\textwidth}{l @{\extracolsep{\fill}} c c c l}
\hline\hline \\[-1.0ex]
Collab.  &  \hspace{-0.4cm}Ref. & $\Nf$ & Ren. & Description 
\\[1.0ex] \hline \hline \\[-2.0ex]
\parbox[t]{2.8cm}{ALPHA 14 \\ALPHA 13 \\ALPHA 12A \\ALPHA 11} & 
\parbox[t]{1.0cm}{ \cite{Bernardoni:2014fva}\\ \cite{Bernardoni:2013oda}\\ \cite{Bernardoni:2012ti}\\ \cite{Blossier:2011dk}} & 
2 & NPR & 
\parbox[t]{7.5cm}{The authors use the Schr\"odinger functional for the NP matching. }
\\[9.0ex] \hline \\[-2.0ex]
\parbox[t]{2.8cm}{ETM 13B, 13C \\ETM 12B \\ETM 11A} & 
\parbox[t]{1.0cm}{ \cite{Carrasco:2013zta,Carrasco:2013iba}\\ \cite{Carrasco:2012de}\\ \cite{Dimopoulos:2011gx}} & 
 2 & --, PT1$\ell$ & 
\parbox[t]{7.5cm}{The current used for the relativistic decay constants is absolutely normalized. 
{\bf Interpolation method:} The static limit current renormalization is calculated in one-loop 
mean field improved perturbation theory, there half the correction is used to estimate the
 error.  {\bf Ratio method:} The ratio is constructed from the relativistic decay constant data
 and the heavy-quark pole masses. 
 Ratios of pole-to-$\msbar$ mass  conversion factors are included 
at NLO in continuum perturbation theory.}
\\[26.0ex]  
\hline\hline
\end{tabular*}
\caption{Description of the renormalization/matching procedure adopted
  in the determinations of the $B$ and $B_s$ meson decay constants for
  $N_{f} = 2$ simulations.} 
}
\end{table}
%%%%%%%%%%%%%%%%%%%%%%%%%%%% HQ TREATMENT  %%%%%%%%%%%%%%%%%%%%%%%

\begin{table}[!htb]
%\vskip-1cm
{\footnotesize
\begin{tabular*}{\textwidth}{l @{\extracolsep{\fill}} c c c l}
\hline\hline \\[-2.0ex]
Collab. & \hspace{-0.4cm}Ref. & $\Nf$ &  Action  &  Description
\\[1.0ex] \hline \hline \\[-1.0ex]
\parbox[t]{2.8cm}{ETM 13E}  & \parbox[t]{1.0cm}{\cite{Carrasco:2013naa}} & 2+1+1 & tmWil &
\parbox[t]{7.5cm}{The estimate of the discretization effects is described in the 
continuum table. The relativistic data are 
matched to HQET using NLO continuum PT in an intermediate step, and 
converted back to QCD at the end. 
The error due to HQET matching (estimated by replacing the NLO expressions  with LO)
 is a very small contribution to the systematic error due to the 
heavy quark mass dependence.}
\\[21.0ex] \hline  \\[-2.0ex]
\parbox[t]{2.8cm}{HPQCD 13} & \parbox[t]{1.0cm}{\cite{Dowdall:2013tga}} & 2+1+1 & NRQCD 
& \parbox[t]{7.5cm}{The leading HQ truncation effects are of 
$\cO( \Lambda^{2}_{\rm QCD}/m^{2}_{h})$ and  $\cO(\alpha^{2}_{s} \Lambda_{\rm QCD}/m_h)$, and the errors are at the
subpercentage level.}
\\[6.0ex] \hline \hline
\end{tabular*}
\caption{Heavy quark treatment in $N_f=2+1+1$ determinations of the $B$ and $B_s$ meson decay constants.} 
}
\end{table}

\begin{table}[!htb]
%\vskip-1cm
{\footnotesize
\begin{tabular*}{\textwidth}{l @{\extracolsep{\fill}} c c c l}
\hline\hline \\[-2.0ex]
Collab. & \hspace{-0.4cm}Ref. & $\Nf$ &  Action  &  Description
\\[1.0ex] \hline \hline \\[-1.0ex]
\parbox[t]{2.8cm}{RBC/UKQCD~14 \\RBC/UKQCD 13} & \parbox[t]{1.0cm}{
  \cite{Christ:2014uea} \\ \cite{Witzel:2013sla}} & 2+1 & RHQ & 
\parbox[t]{7.5cm}{In RBC/UKQCD 14, the heavy-quark discretization
  errors are estimated to be $1.7\%$ in the decay constants, and
  $0.3\%$ in the $SU(3)$ breaking ratios.}
\\[6.0ex] \hline \\[-2.0ex]
\parbox[t]{2.8cm}{RBC/UKQCD~14A} & \parbox[t]{1.0cm}{
  \cite{Aoki:2014nga}} & 2+1 & Static & 
\parbox[t]{7.5cm}{Static-limit computation, with $\cO(\Lambda_{{\rm
      QCD}}/m_{h})$ errors estimated to be $10\%$ for the decay
  constants, and $2.2\%$ for $f_{B_{s}}/f_{B}$.}
\\[6.0ex] \hline \\[-2.0ex]
\parbox[t]{2.8cm}{HPQCD 12} & \parbox[t]{1.0cm}{\cite{Na:2012sp}} & 2+1 & NRQCD 
& \parbox[t]{7.5cm}{HQ truncation effects estimated as in HPQCD 09  to be $1.0\%$}
\\[3.0ex] \hline \\[-2.0ex]
\parbox[t]{2.8cm}{HPQCD 11A} & \parbox[t]{1.0cm}{\cite{McNeile:2011ng}} & 2+1 & HISQ
& \parbox[t]{7.5cm}{ The analysis uses a combined continuum and $1/m$ extrapolation. }
\\[3.0ex] \hline \\[-2.0ex]
\parbox[t]{2.8cm}{FNAL/MILC 11} & \parbox[t]{1.0cm}{\cite{Bazavov:2011aa}} & 2+1 & Fermilab
& \parbox[t]{7.5cm}{HQ discretization effects are  
included in the combined chiral and continuum fits, and are estimated by varying the fit Ansatz
and excluding the data at the coarsest lattice spacing to be $\sim 2\%$, consistent with
simple power counting estimates but larger than the residual discretization errors observed
in the data.}
\\[15.0ex] \hline \\[-2.0ex]
\parbox[t]{2.8cm}{RBC/UKQCD 10C} & \parbox[t]{1.0cm}{\cite{Albertus:2010nm}} & 2+1 & Static & 
\parbox[t]{7.5cm}{Truncation effects of $\cO(1/m_h)$ on the $SU(3)$ breaking ratios are
estimated by power counting to be 2\%.}
\\[3.0ex] \hline \\[-2.0ex]
\parbox[t]{2.8cm}{HPQCD 09} & \parbox[t]{1.0cm}{\cite{Gamiz:2009ku}} & 2+1 & NRQCD
& \parbox[t]{7.5cm}{The leading HQ truncation effects are of 
$\cO(\alpha_s \Lambda_{\rm QCD}/m_h)$ due to the
tree-level coefficient of the ${\bm \sigma}\cdot{\rm\bf B}$ term. 
The error is estimated by calculating the $B^*-B$ hyperfine
splitting and comparing with experiment as $1\%$.}
\\[12.0ex] \hline \\[-2.0ex]
\parbox[t]{2.8cm}{ALPHA 14 \\ALPHA 13 \\ALPHA 12A \\ALPHA 11} & 
\parbox[t]{1.0cm}{ \cite{Bernardoni:2014fva}\\ \cite{Bernardoni:2013oda}\\ \cite{Bernardoni:2012ti}\\ \cite{Blossier:2011dk}} & 2 & HQET &
\parbox[t]{7.5cm}{NP improved through $\cO(1/m_h)$. Truncation errors of 
$\cO\left [ (\Lambda_{\rm QCD}/m_h)^2 \right ]$ are not included. \\}
\\[9.0ex] \hline \\[-2.0ex]
\parbox[t]{2.8cm}{ETM 13B, 13C \\ETM 12B \\ETM 11A} & 
\parbox[t]{1.0cm}{ \cite{Carrasco:2013zta,Carrasco:2013iba}\\ \cite{Carrasco:2012de}\\ \cite{Dimopoulos:2011gx}} & 
2 & tmWil &
\parbox[t]{7.5cm}{The estimate of the discretization effects is described in the 
continuum table. In both methods the relativistic data are 
matched to HQET using NLO continuum PT in an intermediate step, and 
converted back to QCD at the end. 
The error due to HQET matching (estimated by replacing the NLO expressions  with LO)
 is a very small contribution to the systematic error due to the 
heavy quark mass dependence. The variation observed from adding 
heavier masses to their data and/or including $1/m_h^3$ terms is 
$0.4-1.3\%$.}
\\[29.0ex]
\hline\hline
\end{tabular*}
\caption{Heavy quark treatment in $N_f=2+1$ and $N_f=2$ determinations of the $B$ and $B_s$ meson decay constants.} 
}
\end{table}

%% file: HQ/AppendixTables/BBapp.tex
%=================================================
\subsubsection{$B_{(s)}$-meson mixing matrix elements}
\label{app:BMix_Notes}
%=================================================
\begin{table}[!ht]
{\footnotesize
\begin{tabular*}{\textwidth}{l @{\extracolsep{\fill}} c c c l l}
\hline\hline \\[-1.0ex]
Collab. & Ref. & $\Nf$ & $a$ [fm] & Continuum extrapolation & Scale setting 
\\[1.0ex] \hline \hline \\[-1.0ex]
RBC/UKQCD 14A& \cite{Aoki:2014nga} & 2+1 & \parbox[t]{0.9cm}{0.11, 0.086} & 
\parbox[t]{4.3cm}{Combined continuum and chiral extrapolation with
		 $SU(2)$ NLO HM$\chi$PT and linear in quark mass
		 both with $\cO(a^2)$ terms.
		 The combined continuum and chiral extrapolation
		 uncertainty is estimated as 2.55, 2.13 and 3.08\% for
		 $f_B\sqrt{B_B}$, $f_{B_s}\sqrt{B_{B_s}}$ and $\xi$
		 respectively.} &
\parbox[t]{4.3cm}{Scale is set using the $\Omega^-$ mass as input 
		 \cite{Aoki:2010dy}. The scale uncertainty is estimated
		 as 1.84, 1.86 and 0.05\% for $f_B\sqrt{B_B}$, 
		 $f_{B_s}\sqrt{B_{B_s}}$ and $\xi$ respectively.}
\\[26.0ex] \hline \\[-2.0ex]
FNAL/MILC 12 & \cite{Bazavov:2012zs} & 2+1 & \parbox[t]{0.9cm}{0.12, 0.09} & 
\parbox[t]{4.3cm}{Combined continuum and chiral extrapolation with
		 NLO rHMS$\chi$PT, NNLO analytic and generic 
		 $\cO(\alpha_s^2 a^2, a^4)$ terms. 
		 Combined statistical, chiral and light-quark discretization error 
		 is estimated, by examining the variation with different fit Ans\"atze
		 to be 3.7\% on $\xi$.} &
\parbox[t]{4.3cm}{Relative scale $r_1/a$ is set via static-quark
		  potential. Absolute scale $r_1=0.3117(22)$ fm is
		  determined \cite{Bazavov:2011aa} through averaging the
		  $f_\pi$ input and the estimate of HPQCD collaboration 
		  \cite{Davies:2009tsa}. The scale uncertainty on $\xi$ is estimated as 0.2\%.} 
\\[23.0ex] \hline \\[-2.0ex]
FNAL/MILC 11A & \cite{Bouchard:2011xj} & 2+1 & \parbox[t]{0.9cm}{0.12, 0.09, 0.06} & 
\parbox[t]{4.3cm}{Combined continuum and chiral extrapolation with
		 NLO rHMS$\chi$PT, NNLO analytic and and generic 
		 $\cO(\alpha_s^2 a^2, a^4)$ terms.} & 
\parbox[t]{4.3cm}{See above. The error in $r_1$ yields a 3\% 
		 uncertainty on $f^2_B B_B$.} 
\\[12.0ex] \hline \\[-2.0ex]
\parbox[t]{2.8cm}{RBC/UKQCD 10C} & \parbox[t]{0.5cm}{\cite{Albertus:2010nm}} & 2+1 & \parbox[t]{0.9cm}{0.11} &
\parbox[t]{4.3cm}{Only one lattice spacing is used. Discretization error
		 is estimated to be 4\% on $\xi$ by power counting.}&
\parbox[t]{4.3cm}{Scale is set using the $\Omega^-$ mass as input 
		 \cite{Allton:2008pn}.The error on $\xi$ due to the combined scale and 
		 light quark mass uncertainties is estimated as 1\%.} 
\\[12.0ex] \hline \\[-2.0ex]
HPQCD 09 & \cite{Gamiz:2009ku} & 2+1 & \parbox[t]{1.1cm}{0.12, 0.09} &
\parbox[t]{4.3cm}{Combined continuum and chiral extrapolation with
		 NLO rHMS$\chi$PT and NNLO analytic terms. Light-quark
		 discretization error
		 is estimated as 3, 2 and 0.3\% for $f_B\sqrt{B_B}$,
		 $f_{B_s}\sqrt{B_{B_s}}$ and $\xi$ respectively.} &
\parbox[t]{4.3cm}{Relative scale $r_1/a$ is set via static-quark
		  potential. Absolute scale $r_1=0.321(5)$ fm is
		  determined through $\Upsilon$ mass \cite{Gray:2005ur}.
		  The error on $f_B\sqrt{B_B}$ due to the scale uncertainty
		  is estimated as 2.3\%.} 
\\[18.0ex] \hline \\[-2.0ex]
HPQCD 06A & \cite{Dalgic:2006gp} & 2+1 & \parbox[t]{1.1cm}{0.12} & 
\parbox[t]{4.3cm}{Only one lattice spacing is used. Light-quark discretization error
		 on $f_{B_s}^2B_{B_s}$ is estimated as 4\% 
		 by power counting.} &
\parbox[t]{4.3cm}{Scale is set using the $\Upsilon$ $2S-1S$ splitting as 
		 input \cite{Gray:2005ur}. The error on $f^2_B B_B$ due to the scale uncertainty
		  is estimated as 5\%.} 
\\[12.0ex]
\hline\hline
\end{tabular*}
\caption{Continuum extrapolations/estimation of lattice artifacts 
 in determinations of the neutral $B$-meson mixing matrix elements for
 $N_f=2+1$ simulations.}
\label{tab:BBcontinum}
}
\end{table}

\begin{table}[!ht]
{\footnotesize
\begin{tabular*}{\textwidth}{l @{\extracolsep{\fill}}c c c l l}
\hline\hline \\[-1.0ex]
Collab. & Ref. & $\Nf$ & $a$ [fm] & Continuum extrapolation & Scale setting 
\\[1.0ex] \hline \hline \\[-1.0ex]
ETM 13B  & \cite{Carrasco:2013zta} & 2 & \parbox[t]{1.1cm}{0.098,
	     0.085, 0.067, 0.054} & 
\parbox[t]{4.3cm}{
	     Combined chiral and continuum extrapolation, with a term linear in $a^2$.
	     Discretization error is estimated by omitting the
	     coarsest lattice as 0.5, 1.7, 1.3 and 1.0 \% for 
	     $B_{B_s}$, $B_B$, $B_{B_s}/B_B$ and $\xi$ respectively.
	     The heavy-quark masses vary in the range $0.13 \lsim  a m_h \lsim 0.85$.} &
\parbox[t]{4.3cm}{See below.} 
\\[26.0ex] 
\hline \\[-2.0ex]
\parbox[t]{2.4cm}{ETM 12A, 12B}  & \cite{Carrasco:2012dd,Carrasco:2012de} & 2 & \parbox[t]{1.1cm}{0.098,
	     0.085, 0.067} & 
\parbox[t]{4.3cm}{Combined chiral and continuum extrapolation, with a term linear in $a^2$.
		Discretization error included in combined statistical, chiral and continuum extrapolation error
		and estimated as $4.5$\%. 
		The heavy-quark masses vary in the range $0.25 \lsim  a m_h \lsim 0.6$.} &
\parbox[t]{4.3cm}{Relative scale $r_0/a$ set from the static-quark
		     potential. Absolute scale set from $f_\pi$. Scale setting uncertainty 
		     included in combined statistical and systematic error.} 
\\[23.0ex] 
\hline\hline
\end{tabular*}
\caption{Continuum extrapolations/estimation of lattice artifacts 
 in determinations of the neutral $B$-meson mixing matrix elements for
 $N_f=2$ simulations.}
\label{tab:BBcontinum2}
}
\end{table}

\begin{table}[!ht]
{\footnotesize
\begin{tabular*}{\textwidth}{l @{\extracolsep{\fill}} c c c  l}
\hline\hline \\[-1.0ex]
Collab. & Ref. & $\Nf$ & {$M_{\pi,\rm min}\,[\mev]$}  & {Description}
\\[1.0ex] \hline \hline \\[-1.0ex]
RBC/UKQCD 14A& \cite{Aoki:2014nga} & 2+1 & \parbox[t]{2cm}{327, 289}
	     % same as coarser 2 lattices of FNAL/MILC 11B
	     &  \parbox[t]{6cm}{Combined continuum and chiral extrapolation with
		 $SU(2)$ NLO HM$\chi$PT and linear in quark mass
		 both with $\cO(a^2)$ terms.
		 The chiral fit error is estimated from 
		 difference between the NLO HM$\chi$PT and linear fits,
		 and further from eliminating the heaviest $ud$ quark
		 mass point.}
\\[18.0ex] \hline \\[-2.0ex]
FNAL/MILC 12 & \cite{Bazavov:2012zs} & 2+1 & \parbox[t]{2cm}{440, 320}
	     % same as coarser 2 lattices of FNAL/MILC 11B
	     &  \parbox[t]{6cm}{Combined continuum and chiral extrapolation with
		 NLO rHMS$\chi$PT and NNLO analytic terms.
		 See the entry in Table~\ref{tab:BBcontinum}.
		 The omission of wrong-spin contributions \cite{Bernard:2013dfa} 
		 in the HMrS$\chi$PT is treated as a systematic error and
		 estimated to be 3.2\% for $\xi$.}
\\[15.0ex] \hline \\[-2.0ex]
FNAL/MILC 11A & \cite{Bouchard:2011xj} & 2+1 & \parbox[t]{2cm}{440, 320, 250} & 
\parbox[t]{6cm}{Combined continuum and chiral extrapolation with
		NLO rHMS$\chi$PT and NNLO analytic terms.}
\\[6.0ex] \hline \\[-2.0ex]
RBC/UKQCD 10C & \cite{Albertus:2010nm} & 2+1 & \parbox[t]{2cm}{430} &
\parbox[t]{6cm}{Linear fit matched with $SU(2)$ NLO HM$\chi$PT at the lightest
		$ud$ mass point is used as the preferred fit. Many different fit Ans\"atze
		are considered. The systematic
		 error is estimated from the difference between the $SU(2)$ 
		 HM$\chi$PT fit described above and a linear fit. }
\\[18.0ex]\hline \\[-2.0ex] 
HPQCD 09 & \cite{Gamiz:2009ku} & 2+1 & \parbox[t]{2cm}{440, 400} &
\parbox[t]{6cm}{Combined continuum and chiral extrapolation with
		NLO rHMS$\chi$PT and NNLO analytic terms.} 
\\[6.0ex] \hline \\[-2.0ex]
HPQCD 06A & \cite{Dalgic:2006gp} & 2+1 & \parbox[t]{2cm}{510} & 
\parbox[t]{6cm}{Two sea $ud$ quark masses $m_{ud}/m_s=0.25$ and $0.5$
		 are used to calculate the matrix element for $B_s$
		 meson at the predetermined value of the strange quark
		 mass. No significant sea quark mass dependence is observed
		 and the value at the lighter sea $ud$ mass is taken as the result.} 
\\[18.0ex] \hline \\[-2.0ex]
\parbox[t]{2.2cm}{ETM 13B\\ ETM 12A,12B}  & \parbox[t]{1cm}{\cite{Carrasco:2013zta}\\ \cite{Carrasco:2012dd,Carrasco:2012de}} & 2 & \parbox[t]{2cm}{410, 275, 300, 270} & 
\parbox[t]{6cm}{$M_{\pi,{\rm min}}$ refers to the charged pions, where 270 MeV
	on the finest lattice only included in ETM 13B. Linear and NLO (full
	QCD) HM$\chi$PT supplemented by an $a^2$ term is used. The chiral fit error is
	estimated from the difference between the NLO HM$\chi$PT and linear fits. }
\\[18.0ex] 
% \hline \\[-2.0ex] 
% %
% ETM 12A, 12B  & \cite{Carrasco:2012dd,Carrasco:2012de} & 2 & \parbox[t]{2cm}{410, 275, 300} & 
% \parbox[t]{6cm}{$M_{\pi,{\rm min}}$ refers to the charged pions. Linear and NLO (full
% 	QCD) HM$\chi$PT supplemented by an $a^2$ term is used. The chiral fit error is
% 	estimated from the difference between the NLO HM$\chi$PT and linear fits. }
% \\[18.0ex] 
%
\hline\hline
\end{tabular*}
\caption{Chiral extrapolation/minimum pion mass
 in determinations of the neutral $B$-meson mixing matrix elements.
 For actions with multiple species of pions, masses quoted are the RMS
 pion masses (where available).  The different $M_{\pi,\rm min}$ entries correspond to the
 different lattice spacings.}
}
\end{table}

\begin{table}[!ht]
{\footnotesize
\begin{tabular*}{\textwidth}{l @{\extracolsep{\fill}} c c c c l}
\hline\hline \\[-1.0ex]
Collab. & Ref. & $\Nf$ & $L$ [fm] & ${M_{\pi,\rm min}}L$ & Description
\\[1.0ex] \hline \hline \\[-1.0ex]
RBC/UKQCD 14A
%\parbox[t]{2.8cm}{RBC/UKQCD 14A} 
& \cite{Aoki:2014nga} & 2+1 & 
\parbox[t]{1.8cm}{2.74, 2.76} & 4.5, 4.0 & 
\parbox[t]{5.4cm}{FV error is estimated from $SU(2)$ $\chi$PT to be 
		  0.76, 0, 1.07\% for $f_B\sqrt{B_B}$,
		   $f_{B_s}\sqrt{B_{B_s}}$ and $\xi$ respectively.} 
\\[6.0ex] \hline \\[-2.0ex]
FNAL/MILC 12 & \cite{Bazavov:2012zs} & 2+1 & 
\parbox[t]{1.8cm}{2.4/2.9, 2.5} & 6.4, 5.1 & 
\parbox[t]{5.4cm}{FV error is estimated to be less than $0.1\%$ for
		     $SU(3)$ breaking ratios from FV HMrS$\chi$PT.} 
\\[6.0ex] \hline \\[-2.0ex]
FNAL/MILC 11A & \cite{Bouchard:2011xj} & 2+1 & 
\parbox[t]{1.8cm}{2.4/2.9, 2.5/2.9/3.6, 3.8} & 6.4, 5.8, 4.9 &
\parbox[t]{5.4cm}{FV error on $f_B\sqrt{B_B}$ is estimated to be less
		  than $1\%$, 
		  which is inferred from the study of the $B$-meson decay
		  constant using FV HM$\chi$PT \cite{Bazavov:2011aa}.} 
\\[9.0ex] \hline \\[-2.0ex]
RBC/UKQCD 10C & \cite{Albertus:2010nm} & 2+1 & 
\parbox[t]{1.8cm}{1.8} & 3.9 & 
\parbox[t]{5.4cm}{FV error estimated through FV HM$\chi$PT as $1\%$ for $SU(3)$ breaking ratios. }
\\[4.0ex]\hline \\[-2.0ex] 
HPQCD 09 & \cite{Gamiz:2009ku} & 2+1 & 
\parbox[t]{1.8cm}{2.4/2.9, 2.5} & 6.4, 5.1 &
\parbox[t]{5.4cm}{No explicit estimate of FV error, but expected to be much smaller than other uncertainties.}
\\[6.0ex] \hline \\[-2.0ex]
HPQCD 06A & \cite{Dalgic:2006gp} & 2+1 & 
\parbox[t]{1.8cm}{2.4} & 6.2 & 
\parbox[t]{5.4cm}{No explicit estimate of FV error, but expected to be much smaller than other uncertainties.}
\\[6.0ex] \hline \\[-2.0ex]
\parbox[t]{2.2cm}{ETM 13B\\ ETM 12A,12B}  &
     \parbox[t]{1cm}{\cite{Carrasco:2013zta}\\ \cite{Carrasco:2012dd,Carrasco:2012de}} & 2 & 
\parbox[t]{1.8cm}{2.4, 2.0/2.7, 2.1, 1.7/2.6} & 
\parbox[t]{1.8cm}{5.0, 3.7, 3.3, 3.5} & 
\parbox[t]{5.4cm}{$L=1.7/2.6\, {\rm fm}$ only included in ETM~13B. FV error is assumed to be negligible based on the study of $D$-meson decay constants in 
Ref.~\cite{Blossier:2009bx}.}
\\[9.0ex] 
\hline\hline
\end{tabular*}
\caption{Finite volume effects 
 in determinations of the neutral $B$-meson mixing matrix elements.
 Each $L$-entry corresponds to a different
 lattice spacing, with multiple spatial volumes at some lattice
 spacings. 
% For staggered actions RMS masses of all tastes are quoted.
% For ETM, {\it charged pion is used for now} 
 For actions with multiple species of pions, masses quoted are the RMS
 pion masses (where available).
}
}
\end{table}

\begin{table}[!ht]
{\footnotesize
\begin{tabular*}{\textwidth}{l @{\extracolsep{\fill}} c c c l}
\hline\hline \\[-1.0ex]
Collab. & Ref. & $\Nf$ & Ren. & Description 
\\[1.0ex] \hline \hline \\[-1.0ex]
RBC/UKQCD 14A & \cite{Aoki:2014nga} & 2+1 & PT$1l$ & 
\parbox[t]{6cm}{Static-light four-quark operators are 
		renormalized with one-loop
		mean field improved PT.
		The errors due to neglected higher order effects 
		are estimated for purely $\cO\alpha_s^2$ to be
		$6$\% on the matrix elements 
		or $1.2$\% on $\xi$ and for $\cO\alpha_s^2 a^2$
		to be $1$\% or $0.2$\% respectively.}
\\[18.0ex] \hline \\[-2.0ex]
FNAL/MILC 12 & \cite{Bazavov:2012zs} & 2+1 & PT$1l$ & 
\parbox[t]{6cm}{One-loop mean-field improved PT
		 is used to renormalize the four-quark operators
		 with heavy quarks rotated to eliminate tree-level $\cO(a)$
		 errors. The error from neglecting higher order corrections 
		 is estimated to be $0.5$\% on $\xi$.} 
\\[15.0ex] \hline \\[-2.0ex]
FNAL/MILC 11A & \cite{Bouchard:2011xj} & 2+1 & PT$1l$ &
\parbox[t]{6cm}{One-loop mean-field improved PT is used to 
                renormalize the four-quark operators
		 with heavy quarks rotated to eliminate tree-level $\cO(a)$
		 errors. The error from neglected higher order corrections 
		 is estimated to be $4$\% on $f_B\sqrt{B_B}$.} 
\\[15.0ex] \hline \\[-2.0ex]
RBC/UKQCD 10C & \cite{Albertus:2010nm} & 2+1 & PT$1l$ & 
\parbox[t]{6cm}{Static-light four-quark operators are 
		 renormalized with one-loop
		 mean field improved PT.
		 The error due to neglected higher order effects 
		 is estimated to be $2.2$\% on $\xi$.}
\\[9.0ex]\hline \\[-2.0ex]
HPQCD 09 & \cite{Gamiz:2009ku} & 2+1 & PT$1l$ &
\parbox[t]{6cm}{Four-quark operators in lattice NRQCD are matched to
		 QCD through order $\alpha_s$, $\Lambda_{\rm QCD}/M$ 
		 and  $\alpha_s/(aM)$ \cite{Gamiz:2008sk} using one-loop PT. 
		 The error due to neglected higher order effects is estimated to 
		 be 4\% on $f_B\sqrt{B_B}$ and 0.7\% on $\xi$.}
\\[15.0ex] \hline \\[-2.0ex]
HPQCD 06A & \cite{Dalgic:2006gp} & 2+1 & PT$1l$ &
\parbox[t]{6cm}{Four-quark operators in lattice NRQCD are matched to
		 full QCD through order $\alpha_s$, $\Lambda_{\rm QCD}/M$ 
		 and  $\alpha_s/(aM)$ \cite{Gamiz:2008sk}. The error  
		 is estimated as $\sim 1\cdot \alpha_s^2$ to be 9\% on
		 $f_{B_s}^2 B_{B_s}$}
\\[9.0ex] \hline \\[-2.0ex]
ETM 13B, 12A, 12B & \cite{Carrasco:2013zta,Carrasco:2012dd,Carrasco:2012de} & 2 & NPR & 
\parbox[t]{6cm}{The bag parameters are nonperturbatively 
	         renormalized in the RI'-MOM scheme. They are calculated
	         as functions of the ($\msbar$) heavy-quark mass (renormalized
	         nonperturbatively in RI/MOM). }
\\[12.0ex] 
%
% ETM 12A, 12B & \cite{Carrasco:2012dd,Carrasco:2012de} & 2 & NPR & 
% \parbox[t]{6cm}{The bag parameters are nonperturbatively 
% 	         renormalized in the RI'-MOM scheme. They are calculated
% 	         as functions of the ($\msbar$) heavy-quark mass (renormalized
% 	         nonperturbatively in RI/MOM). }
% \\[14.0ex] 
%
%  \hline \\[-2.0ex]
% %
% FNAL/MILC 09 & \cite{Evans:2009du} & 2+1 & PT$1l$ & 
% \parbox[t]{6cm}{One-loop tad-pole improved perturbation theory
% 		 is used to renormalized the four quark operators
% 		 with heavy quarks rotated to eliminate the $\cO(a)$
% 		 error. The higher order effects in perturbation
% 		 is estimated to be less than $0.5$\% on $\xi$.}
% \\[16.0ex]
%
\hline\hline
\end{tabular*}
\caption{Operator renormalization 
 in determinations of the neutral $B$-meson mixing matrix elements.
}}
\end{table}

\begin{table}[!ht]
{\footnotesize
\begin{tabular*}{\textwidth}{l @{\extracolsep{\fill}} c c c l}
\hline\hline \\[-1.0ex]
Collab. & Ref. & $\Nf$ & Action & Description
\\[1.0ex] \hline \hline \\[-1.0ex]
RBC/UKQCD 14A& \cite{Aoki:2014nga} & 2+1 & Static & 
\parbox[t]{8cm}{Two different static-quark actions with HYP1 and HYP2
		 smearings are used and the continuum extrapolation is
		 constrained so the two values converges in the limit.
		 The error due to the missing $1/m_b$ corrections is
		 estimated to be $12$\% for individual matrix
		 elements or $2.2$\% on $\xi$ using power-counting.}
\\[15.0ex] \hline \\[-2.0ex]
FNAL/MILC 12 & \cite{Bazavov:2012zs} & 2+1 & Fermilab & 
\parbox[t]{8cm}{The heavy-quark discretization error on $\xi$ is estimated
		to be 0.3 \%. The error on $\xi$ 
		 due to the uncertainty in the $b$-quark mass is
		 are estimated to be 0.4 \%.} 
\\[6.0ex] \hline \\[-2.0ex]
FNAL/MILC 11A & \cite{Bouchard:2011xj} & 2+1 & Fermilab &
\parbox[t]{8cm}{The 
		 heavy-quark discretization error on $f_B\sqrt{B_B}$
		 is estimated as 4\% using power-counting.} 
\\[3.0ex] \hline \\[-2.0ex]
RBC/UKQCD 10C & \cite{Albertus:2010nm} & 2+1 & Static & 
\parbox[t]{8cm}{Two different static-quark actions with Ape and HYP
		 smearings are used. The discretization error
		 on $\xi$ is estimated as  $\sim 4$\% and  the error due to the missing
		 $1/m_b$ corrections  as
		 $\sim 2$\%, both using power-counting.}
\\[9.0ex]\hline \\[-2.0ex]
HPQCD 09 & \cite{Gamiz:2009ku} & 2+1 & NRQCD &
\parbox[t]{8cm}{Heavy-quark truncation errors due to relativistic corrections
		 are estimated to be 2.5, 2.5 and 0.4 \% for $f_B\sqrt{B_B}$,
		 $f_{B_s}\sqrt{B_{B_s}}$ and $\xi$ respectively.}
\\[6.0ex] \hline \\[-2.0ex]
HPQCD 06A & \cite{Dalgic:2006gp} & 2+1 & NRQCD &
\parbox[t]{8cm}{Heavy-quark truncation errors due to relativistic corrections
		 are estimated to be 3\% for $f_{B_s}^2{B_{B_s}}$.}
\\[3.0ex] \hline \\[-2.0ex]
\parbox[t]{2.2cm}{ETM 13B\\ ETM 12A,12B}  &
     \parbox[t]{1cm}{\cite{Carrasco:2013zta}\\ \cite{Carrasco:2012dd,Carrasco:2012de}} & 2 & 
	 tmWil & 
\parbox[t]{8cm}{The ratio method is used to perform an interpolation to
	the physical $b$ quark mass from the simulated
	heavy mass and the known static limit.
	In an intermediate step, the ratios include HQET matching
	factors calculated to tree-level, leading-log, 
	and next-to-leading-log (ETM 13B only)
	in continuum PT.
	The interpolation uses a polynomial up to
	quadratic in the inverse quark-mass. 
	The systematic errors added together with those of the chiral
	fit are estimated as $1.3-1.6$\% for bag parameters for ETM 13B,
	while they are estimated from
	changing the interpolating polynomial as $2$\% and from changing the order of HQET 
	matching factors as $3$\% for ETM 12A and 12B.}
\\[32.0ex]
\hline\hline
\end{tabular*}
\caption{Heavy-quark treatment
 in determinations of the neutral $B$-meson mixing matrix elements.
}}
\end{table}

%% file: HQ/AppendixTables/BSLapp.tex
%=================================================
%\subsubsection{$B\to\pi\ell\nu$ form factor}
\subsubsection{Form factors entering determinations of $|V_{ub}|$ ($B \to \pi l\nu$, $B_s \to K l\nu$, $\Lambda_b \to pl\nu$)}
\label{app:BtoPi_Notes}
%=================================================

%%%%%%%%%%%%%%%%%%%%%%%%%%%%%%%% CONT EXTRAP %%%%%%%%%%%%%%%%%%%%%%%%%%%%%%%%%%%%%%%%%%%%%%%%%%%%%%%%%%%%%%%%%%%%%%%
\begin{table}[!ht]

{\footnotesize
\begin{tabular*}{\textwidth}{l c c c l l}
\hline\hline \\[-1.0ex]
Collab. & Ref. & $\Nf$ & $a$ [fm] & Continuum extrapolation & Scale setting 
\\[1.0ex] \hline \hline \\[-1.0ex]
\SLfnalmilcBpi & \cite{Lattice:2015tia} & 2+1 & \parbox[t]{20mm}{0.045, 0.06, 0.09, 0.12}   &
\parbox[t]{4.2cm}{Fit to HMrS$\chi$PT to remove light-quark discretization errors. Residual heavy-quark discretization errors estimated with power-counting. Total (stat + chiral extrap + HQ discretization + $g_{B^*B\pi}$) error estimated to be 3.1\% for $f_+$ and 3.8\% for $f_0$ at $q^2=20~{\rm GeV}^2$.} & \parbox[t]{30mm}{Relative scale $r_1/a$ set from the static-quark potential.  Absolute scale $r_1$, including related uncertainty estimates, taken from \cite{Bazavov:2011aa}.}
\\[26.0ex] \hline \\[-2.0ex]
\parbox[t]{20mm}{\SLLambdabp} & \cite{Detmold:2015aaa} & 2+1 & \parbox[t]{20mm}{0.0849(12), 0.1119(17)}   &
\parbox[t]{4.2cm}{Joint chiral-continuum extrapolation, combined with fit to $q^2$ dependence of form factors in a ``modified'' $z$-expansion. Systematics estimated by varying fit form and $\cO(a)$ improvement parameter values.} & \parbox[t]{30mm}{Set from $\Upsilon(2S)$--$\Upsilon(1S)$ splitting, cf.~\cite{Meinel:2010pv}.}
\\[21.0ex] \hline \\[-2.0ex]
\SLrbcukqcdBpi & \cite{Flynn:2015mha} & 2+1 & \parbox[t]{20mm}{0.086,0.11}   &
\parbox[t]{4.2cm}{Joint chiral-continuum extrapolation using $SU(2)$ hard-pion HM$\chi$PT.
Systematic uncertainty estimated by varying fit ansatz and form of coefficients, as well as implementing different
cuts on data; ranges from 5.0\% to 10.9\% for $B\to\pi$ form factors, and 2.5\% to 5.1\% for $B_s\to K$.
Light-quark and gluon discretization errors estimated at 1.1\% and 1.3\%, respectively.} & \parbox[t]{30mm}{Scale implicitly set in the light-quark sector using the $\Omega^-$ mass, cf.~\cite{Aoki:2010dy}.}
\\[35.0ex] \hline \\[-2.0ex]
\SLhpqcdBsK & \cite{Bouchard:2014ypa} & 2+1 & \parbox[t]{20mm}{0.09,0.12}   &
\parbox[t]{4.2cm}{Combined chiral-continuum extrapolation using hard-pion rHMS$\chi$PT. (No explicit estimate of discretization effects.)} & \parbox[t]{30mm}{Relative scale $r_1/a$ set from the static-quark potential. Absolute scale $r_1$
set to 0.3133(23)~fm.}
\\[12.0ex] \hline \\[-2.0ex]
HPQCD 06 & \cite{Dalgic:2006dt} & 2+1 & \parbox[t]{20mm}{0.09,0.12}  &
\parbox[t]{4.2cm}{Central values obtained from data at $a=0.12$~fm.  Discretization errors observed to be within the statistical error by comparison with data at $a=0.09$~fm.} &  \parbox[t]{30mm}{Relative scale $r_1/a$ set from the static-quark potential.  Absolute scale $r_1$ set through $\Upsilon$ $2S-1S$ splitting c.f.~HPQCD 05B~\cite{Gray:2005ur}.}
\\[18.0ex]% \hline \\[-2.0ex]
\hline\hline
\end{tabular*}
\caption{Continuum extrapolations/estimation of lattice artifacts in
  determinations of $B\to\pi\ell\nu$, $B_s\to K\ell\nu$, and $\Lambda_b\to p\ell\nu$ form factors.}
}
\end{table}
%%%%%%%%%%%%%%%%%%%%%%%%% CHIRAL EXTRAP %%%%%%%%%%%%%%%%%%%%%%%%%%%%%%%%%%%%%%%%%%%%%%%%%%%%%%%%%%%%%%%%%%%%%%%%%%%%%%%
\begin{table}[!ht]
{\footnotesize
\begin{tabular*}{\textwidth}{l @{\extracolsep{\fill}} c c c   l}
\hline\hline \\[-1.0ex]
{Collab.} & {Ref.} & {$\Nf$} & {$M_{\pi,\rm min}\,[\mev]$}  & {Description}  
\\[1.0ex] \hline \hline \\[-1.0ex]
\SLfnalmilcBpi & \cite{Lattice:2015tia} & 2+1 & \parbox[t]{2cm}{330, 260, 280, 470}
& \parbox[t]{5cm}{Simultaneous chiral-continuum extrapolation and $q^2$ interpolation using NNLO $SU(2)$ hard-pion HMrS$\chi$PT.  Systematic error estimated by adding higher-order analytic terms and varying the $B^*$-$B$-$\pi$ coupling.}
\\[18.0ex] \hline \\[-2.0ex]
\parbox[t]{20mm}{\SLLambdabp} & \cite{Detmold:2015aaa} & 2+1 & \parbox[t]{2cm}{227, 245 (valence pions)}
& \parbox[t]{5cm}{Joint chiral-continuum extrapolation, combined with fit to $q^2$ dependence of form factors in a ``modified'' $z$-expansion. Only analytic NLO terms $\propto (m_\pi^2-m_{\pi,{\rm phys}}^2)$ included in light mass dependence. Systematic uncertainty estimated by repeating fit with added higher-order terms.}
\\[23.0ex] \hline \\[-2.0ex]
\SLrbcukqcdBpi & \cite{Flynn:2015mha} & 2+1 & \parbox[t]{2cm}{289, 329}
& \parbox[t]{5cm}{Joint chiral-continuum extrapolation using $SU(2)$ hard-pion HM$\chi$PT.
Systematic uncertainty estimated by varying fit ansatz and form of coefficients, as well as implementing different
cuts on data; ranges from 5.0\% to 10.9\% for $B\to\pi$ form factors, and 2.5\% to 5.1\% for $B_s\to K$.}
\\[23.0ex] \hline \\[-2.0ex]
\SLhpqcdBsK & \cite{Bouchard:2014ypa} & 2+1 & \parbox[t]{2cm}{295, 260}
& \parbox[t]{5cm}{Combined chiral-continuum extrapolation using hard-pion rHMS$\chi$PT. (No explicit estimate of extrapolation systematics.)}
\\[9.0ex] \hline \\[-2.0ex]
%
%FNAL/MILC 08A & \cite{Bailey:2008wp} & 2+1 & \parbox[t]{2cm}{400, 440}
%& \parbox[t]{5cm}{Simultaneous chiral-continuum extrapolation and $q^2$ interpolation using $SU(3)$ rHMS$\chi$PT.  Systematic error estimated by adding higher-order analytic terms and varying the $B^*$-$B$-$\pi$ coupling.}
%\\[15.0ex] \hline \\[-2.0ex]
%
HPQCD 06 & \cite{Dalgic:2006dt} & 2+1 & \parbox[t]{2cm}{400, 440}
& \parbox[t]{5cm}{First interpolate data at fixed quark mass to fiducial values of $E_\pi$ using the Becirevic-Kaidalov and Ball-Zwicky ans\"atze, then extrapolate data at fixed $E_\pi$ to physical quark masses using $SU(3)$ rHMS$\chi$PT. Systematic error estimated by varying interpolation and extrapolation fit functions.}
\\[24.0ex]% \hline \\[-2.0ex]
%
%\SLalphaBsK & \cite{Bahr:2014iqa} & 2 & \parbox[t]{2cm}{340, 310, 330}
%& \parbox[t]{5cm}{n/a}
%\\[23.0ex]
%%
\hline\hline
\end{tabular*}
\caption{Chiral extrapolation/minimum pion mass in
  determinations of $B\to\pi\ell\nu$, $B_s\to K\ell\nu$, and $\Lambda_b\to p\ell\nu$ form factors.  For actions with multiple species of pions, masses quoted are the RMS pion masses.    The different $M_{\pi,\rm min}$ entries correspond to the different lattice spacings.}
}
\end{table}
%%%%%%%%%%%%%%%%%%%%%%%%%%%%% FINITE VOLUME %%%%%%%%%%%%%%%%%%%%%%%%%%%%%%%%%%%%%%%%%%%%%%%%%%%%%%%%%%%%%%%%%%%%%%%%%%%
\begin{table}[!ht]
{\footnotesize
\begin{tabular*}{\textwidth}{l @{\extracolsep{\fill}} c c c c l}
\hline\hline \\[-1.0ex]
Collab. & Ref. & $\Nf$ & $L$ [fm] & ${M_{\pi,\rm min}}L$ & Description 
\\[1.0ex] \hline \hline \\[-1.0ex]
\SLfnalmilcBpi & \cite{Lattice:2015tia} & 2+1 & \parbox[t]{1.8cm}{2.9, 2.9/3.4/3.8, 2.5/2.9/3.6/5.8, 2.4/2.9} & $\gtrsim 3.8$  &
\parbox[t]{5cm}{FV effects estimated by replacing infinite-volume chiral logs with sums over discrete momenta, found to be negligible.
}
\\[9.0ex] \hline \\[-1.0ex]
\parbox[t]{20mm}{\SLLambdabp} & \cite{Detmold:2015aaa} & 2+1 & \parbox[t]{1.8cm}{2.7, 2.7} & $\gtrsim 3.1$ (valence sector)  &
\parbox[t]{5cm}{FV effect estimated at 3\% from experience on $\chi$PT estimates of FV effects for heavy-baryon axial couplings.
}
\\[9.0ex] \hline \\[-1.0ex]
\SLrbcukqcdBpi & \cite{Flynn:2015mha} & 2+1 & \parbox[t]{1.8cm}{2.8, 2.6} & $4.0, 4.4$  &
\parbox[t]{5cm}{FV effects estimated by correction to chiral logs due to sums over discrete momenta;
quoted 0.3-0.5\% for $f_+$ and 0.4-0.7\% for $f_0$ for $B\to\pi$, and 0.2\% for $f_+$ and 0.1-0.2\% for $f_0$ for $B_s\to K$.
}
\\[15.0ex] \hline \\[-1.0ex]
\SLhpqcdBsK & \cite{Bouchard:2014ypa} & 2+1 & \parbox[t]{1.8cm}{2.5, 2.4/2.9} & $\gtrsim 3.8$  &
\parbox[t]{5cm}{FV effects estimated by shift of pion log, found to be negligible.
}
\\[4.0ex] \hline \\[-1.0ex]
%
%FNAL/MILC 08A & \cite{Bailey:2008wp} & 2+1 & \parbox[t]{1.8cm}{2.4, 2.4/2.9} & $\gtrsim 3.8$  &
%\parbox[t]{5cm}{Estimate FV error to be $0.5\%$ using 1-loop rHMS$\chi$PT.
%}
%\\[4.0ex] \hline \\[-1.0ex]
%
HPQCD 06 & \cite{Dalgic:2006dt} & 2+1 & \parbox[t]{1.8cm}{2.4/2.9} & $\gtrsim 3.8$ & 
\parbox[t]{5cm}{No explicit estimate of FV error, but expected to be much smaller than other uncertainties.}
\\[7.0ex]% \hline \\[-1.0ex]
%
%\SLalphaBsK & \cite{Bahr:2014iqa} & 2 & \parbox[t]{1.8cm}{2.3, 3.1, 2.4} & $4.0, 5.0, 4.0$  &
%\parbox[t]{5cm}{No estimate of FV effects.
%}
%\\[7.0ex]
%
\hline\hline
\end{tabular*}
\caption{Finite volume effects in determinations of $B\to\pi\ell\nu$, $B_s\to K\ell\nu$, and $\Lambda_b\to p\ell\nu$ form factors.  Each $L$-entry corresponds to a different lattice
spacing, with multiple spatial volumes at some lattice spacings.  For actions with multiple species of pions, the lightest masses are quoted.}
}
\end{table}
%%%%%%%%%%%%%%%%%%%%%%%%%%%%%%%%%%%%% OPERATOR RENORMALIZATION %%%%%%%%%%%%%%%%%%%%%%%%%%%%%%%%%%%%%%%%%
\begin{table}[!ht]
{\footnotesize
\begin{tabular*}{\textwidth}{l @{\extracolsep{\fill}} c c c l}
\hline\hline \\[-1.0ex]
Collab. & Ref. & $\Nf$ & Ren. & Description 
\\[1.0ex] \hline \hline \\[-1.0ex]
\SLfnalmilcBpi & \cite{Lattice:2015tia} & 2+1 & mNPR &
\parbox[t]{6cm}{Perturbative truncation error estimated at 1\% with size of 1-loop correction on next-to-finer ensemble.}
\\[7.0ex] \hline \\[-1.0ex]
\parbox[t]{20mm}{\SLLambdabp} & \cite{Detmold:2015aaa} & 2+1 & mNPR &
\parbox[t]{6cm}{Perturbative truncation error estimated at 1\% with size of 1-loop correction on next-to-finer ensemble.}
\\[7.0ex] \hline \\[-1.0ex]
\SLrbcukqcdBpi & \cite{Flynn:2015mha} & 2+1 & mNPR &
\parbox[t]{6cm}{Perturbative truncation error estimated as largest of power counting, effect from value of $\alpha_s$ used, numerical integration. Non-perturbative normalization of flavour-diagonal currents computed by fixing values of ratios of meson two-point functions to three-point functions with an extra current inversion, cf.~\cite{Christ:2014uea}}
\\[21.0ex] \hline \\[-1.0ex]
\SLhpqcdBsK & \cite{Bouchard:2014ypa} & 2+1 & mNPR &
\parbox[t]{6cm}{Currents matched using one-loop HISQ lattice perturbation theory, omitting $\cO(\alpha_s\Lambda_{\rm QCD}/m_b$. Systematic uncertainty resulting from one-loop matching and neglecting $\cO(\Lambda_{\rm QCD}^2/m_b^2$ terms estimated at 4\% from power counting.}
\\[16.0ex] \hline \\[-1.0ex]
%
%FNAL/MILC 08A & \cite{Bailey:2008wp} & 2+1 & mNPR &
%\parbox[t]{6cm}{Perturbative truncation error estimated at 3\% with size of 1-loop correction on finer ensemble.}
%\\[7.0ex] \hline \\[-1.0ex]
%
HPQCD 06 & \cite{Dalgic:2006dt} & 2+1 & PT1$\ell$ &
\parbox[t]{6cm}{Currents included through $\cO(\alpha_S \Lambda_{\rm QCD}/M$, $\alpha_S/(aM)$, $\alpha_S\, a \Lambda_{\rm QCD})$. Perturbative truncation error estimated from power-counting.}
\\[10.0ex]% \hline \\[-1.0ex]
%
%\SLalphaBsK & \cite{Bahr:2014iqa} & 2 & NPR &
%\parbox[t]{6cm}{Implicit non-perturbative current renormalization from non-perturbative HQET setup.}
%\\[10.0ex]
%
\hline\hline
\end{tabular*}
\caption{Operator renormalization in determinations of $B\to\pi\ell\nu$, $B_s\to K\ell\nu$, and $\Lambda_b\to p\ell\nu$ form factors.}
}
\end{table}
%%%%%%%%%%%%%%%%%%%%%%%%%%%%%%%%%%% HEAVY QUARK TREATMENT %%%%%%%%%%%%%%%%%%%%%%%%%%%%%%%%%%%%%%%%%%%%%%%%%%%
\begin{table}[!ht]
{\footnotesize
\begin{tabular*}{\textwidth}{l @{\extracolsep{\fill}} c c c l}
\hline\hline \\[-1.0ex]
Collab. & Ref. & $\Nf$ & Action & Description 
\\[1.0ex] \hline \hline \\[-1.0ex]
\SLfnalmilcBpi & \cite{Lattice:2015tia} & 2+1 & Fermilab &
\parbox[t]{8cm}{Total statistical + chiral extrapolation + heavy-quark discretization + $g_{B^*B\pi}$ error estimated to be 3.1\% for $f_+$ and 3.8\% for $f_0$ at $q^2=20~{\rm GeV}^2$.}
\\[7.0ex] \hline \\[-1.0ex]
\parbox[t]{20mm}{\SLLambdabp} & \cite{Detmold:2015aaa} & 2+1 & Columbia RHQ &
\parbox[t]{8cm}{Discretization errors discussed as part of combined chiral-continuum-$q^2$ fit, stemming from $a^2|\mathbf{p}|^2$ terms.}
\\[4.0ex] \hline \\[-1.0ex]
\SLrbcukqcdBpi & \cite{Flynn:2015mha} & 2+1 & Columbia RHQ &
\parbox[t]{8cm}{Discretization errors estimated by power counting to be 1.8\% for $f_+$ and 1.7\% for $f_0$.}
\\[4.0ex] \hline \\[-1.0ex]
\SLhpqcdBsK & \cite{Bouchard:2014ypa} & 2+1 & NRQCD &
\parbox[t]{8cm}{Currents matched using one-loop HISQ lattice perturbation theory, omitting $\cO(\alpha_s\Lambda_{\rm QCD}/m_b$. Systematic uncertainty resulting from one-loop matching and neglecting $\cO(\Lambda_{\rm QCD}^2/m_b^2$ terms estimated at 4\% from power counting.}
\\[12.0ex] \hline \\[-1.0ex]
%
%FNAL/MILC 08A & \cite{Bailey:2008wp} & 2+1 & Fermilab &
%\parbox[t]{8cm}{Discretization errors in $f+(q^2)$ from heavy-quark action estimated to be 3.4\% by heavy-quark power-counting.}
%\\[4.0ex] \hline \\[-1.0ex]
%
HPQCD 06 & \cite{Dalgic:2006dt} & 2+1 & NRQCD &
\parbox[t]{8cm}{Discretization errors in $f_+(q^2)$ estimated to be $\cO (\alpha_s (a \Lambda_{\rm QCD})^2) \sim  3\%$.  Relativistic errors estimated to be $\cO((\Lambda_{\rm QCD}/M)^2) \sim 1\%$.}
\\[7.0ex]% \hline \\[-1.0ex]
%
%\SLalphaBsK & \cite{Bahr:2014iqa} & 2 & HQET &
%\parbox[t]{8cm}{NP improved through $\cO(1/m_h)$. Truncation errors of $\cO(\Lambda_{\rm QCD}/m_h)^2)$ are not included.}
%\\[7.0ex]
%
\hline\hline
\end{tabular*}
\caption{Heavy quark treatment in determinations of $B\to\pi\ell\nu$, $B_s\to K\ell\nu$, and $\Lambda_b\to p\ell\nu$ form factors.}
}
\end{table}

\clearpage
%=================================================
%\subsubsection{$B\to D \ell\nu$ and $B \to D^* \ell \nu$ form factors and $R(D)$}
\subsubsection{Form factors entering determinations of $|V_{cb}|$ ($B \to D^* l\nu$, $B \to D l\nu$, $B_s \to D_s l\nu$, $\Lambda_b \to \Lambda_c l\nu$) and $R(D)$)}
\label{app:BtoD_Notes}
%=================================================
\vspace{-0.48cm}
%%%%%%%%%%%%%%%%%%%%%%%%%%%%%%%% CONT EXTRAP %%%%%%%%%%%%%%%%%%%%%%%%%%%%%%%%%%%%%%%%%%%%%%%%%%%%%%%%%%%%%%%%%%%%%%%
\begin{table}[!ht]
{\footnotesize
\begin{tabular*}{\textwidth}{l c c c l l}
\hline\hline \\[-1.0ex]
Collab. & Ref. & $\Nf$ & $a$ [fm] & Continuum extrapolation & Scale setting 
\\[1.0ex] \hline \hline \\[-1.0ex]
\parbox[t]{26mm}{\SLhpqcdBD} & \cite{Na:2015kha} & 2+1 & \parbox[t]{20mm}{0.09, 0.12} &
\parbox[t]{4.2cm}{Combined chiral-continuum extrapolation as part of modified $z$-expansion of form factors, which also includes uncertainty related to matching of NRQCD and relativistic currents.}
& \parbox[t]{30mm}{Implicitly set from $r_1$.} 
\\[18ex] \hline \\[-2.0ex]
\parbox[t]{26mm}{\SLfnalmilcBD} & \cite{Lattice:2015rga} & 2+1 & \parbox[t]{20mm}{ 0.045, 0.06, 0.09, 0.12} &
\parbox[t]{4.2cm}{Combined chiral-continuum extrapolation using HMrS$\chi$PT. Form factors fitted to NLO $\chi$PT, with chiral logs taken from staggered version of the Chow-Wise result, modified to include taste-breaking terms. $\cO(a^2)$ terms introduced based on power-counting arguments. Total uncertainty estimated at 0.6\% for $f_+$ and 0.5\% for $f_0$ for the largest recoil.}
& \parbox[t]{30mm}{Relative scale $r_1/a$ set from the static-quark potential.  Absolute scale $r_1$, including related uncertainty estimates, taken from \cite{Bazavov:2011aa}. Uncertainty related to scale setting estimated at 0.2\%.} 
\\[35ex] \hline \\[-2.0ex]
\parbox[t]{26mm}{\SLLambdabc} & \cite{Detmold:2015aaa} & 2+1 & \parbox[t]{20mm}{0.0849(12), 0.1119(17)}   &
\parbox[t]{4.2cm}{Joint chiral-continuum extrapolation, combined with fit to $q^2$ dependence of form factors in a ``modified'' $z$-expansion. Systematics estimated by varying fit form and $\cO(a)$ improvement parameter values.} & \parbox[t]{30mm}{Set from $\Upsilon(2S)$--$\Upsilon(1S)$ splitting, cf.~\cite{Meinel:2010pv}.}
\\[21.0ex] \hline \\[-2.0ex]
\parbox[t]{26mm}{\SLfnalmilcBDstar} & \cite{Bailey:2014tva} & 2+1 & \parbox[t]{20mm}{ 0.045, 0.06, 0.09, 0.12, 0.15} &
\parbox[t]{4.2cm}{Combined chiral-continuum extrapolation using HMrS$\chi$PT. Total uncertainty quoted at 0.5\%.}
& \parbox[t]{30mm}{Relative scale $r_1/a$ set from the static-quark potential.  Absolute scale $r_1$, including related uncertainty estimates, taken from \cite{Bazavov:2011aa}. Uncertainty related to scale setting estimated at 0.1\%.} 
\\[27.0ex] \hline \\[-2.0ex]
\parbox[t]{26mm}{Atoui  13} & \cite{Atoui:2013zza} & 2& \parbox[t]{20mm}{0.054, 0.067, 0.085, 0.098} &
\parbox[t]{4.2cm}{Combined continuum and chiral extrapolation, with linear terms in $a^2$ and $m_{\rm sea}$.
No dependence on $a$ or $m_{\rm sea}$ observed within errors. Stability of results vs fits with no $m_{\rm sea}$
dependence checked.}
& \parbox[t]{30mm}{Scale set through $F_\pi$.} 
\\[18ex]
\hline\hline
\end{tabular*}
\caption{Continuum extrapolations/estimation of lattice artifacts in
  determinations of $B\to D \ell\nu$, $B\to D^* \ell\nu$,  $B_s\to D_s \ell\nu$, and $\Lambda_b\to\Lambda_c\ell\nu$ form factors, and of $R(D)$.}
}
\end{table}

%%%%%%%%%%%%%%%%%%%%%%%%% CHIRAL EXTRAP %%%%%%%%%%%%%%%%%%%%%%%%%%%%%%%%%%%%%%%%%%%%%%%%%%%%%%%%%%%%%%%%%%%%%%%%%%%%%%%
\begin{table}[!ht]
{\footnotesize
\begin{tabular*}{\textwidth}{l @{\extracolsep{\fill}} c c c  l}
\hline\hline \\[-1.0ex]
{Collab.} & {Ref.} & {$\Nf$} & {$M_{\pi,\rm min}\,[\mev]$}  & {Description}  
\\[1.0ex] \hline \hline \\[-1.0ex]
\parbox[t]{26mm}{\SLhpqcdBD} & \cite{Na:2015kha} & 2+1 & \parbox[t]{2cm}{295, 260}
 & \parbox[t]{5cm}{Combined chiral-continuum extrapolation as part of modified $z$-expansion of form factors. Hard-pion $\chi$PT for light mass dependence used to estimate systematic uncertainty to be 1.14\%.}
\\[15.0ex] \hline \\[-2.0ex]
\parbox[t]{26mm}{\SLfnalmilcBD} & \cite{Lattice:2015rga} & 2+1 & \parbox[t]{2cm}{330, 260, 280, 470}
 & \parbox[t]{5cm}{Combined chiral-continuum extrapolation using HMrS$\chi$PT. Form factors fitted to NLO $\chi$PT, with chiral logs taken from staggered version of the Chow-Wise result, modified to include taste-breaking terms. $\cO(a^2)$ terms introduced based on power-counting arguments. Total uncertainty estimated at 0.6\% for $f_+$ and 0.5\% for $f_0$ for the largest recoil.}
\\[29.0ex] \hline \\[-2.0ex]
\parbox[t]{26mm}{\SLLambdabc} & \cite{Detmold:2015aaa} & 2+1 & \parbox[t]{2cm}{227, 245 (valence pions)}
& \parbox[t]{5cm}{Joint chiral-continuum extrapolation, combined with fit to $q^2$ dependence of form factors in a ``modified'' $z$-expansion. Only analytic $NLO$ terms $\propto (m_\pi^2-m_{\pi,{\rm phys}}^2)$ included in light mass dependence. Systematic uncertainty estimated by repeating fit with added higher-order terms.}
\\[23.0ex] \hline \\[-2.0ex]
\parbox[t]{26mm}{\SLfnalmilcBDstar} & \cite{Bailey:2014tva} & 2+1 & \parbox[t]{2cm}{330, 260, 280, 470, 590}
 & \parbox[t]{5cm}{Combined chiral-continuum extrapolation using HMrS$\chi$PT.  Systematic errors estimated by adding higher-order analytic terms and varying the $D^*$-$D$-$\pi$ coupling. Total uncertainty quoted at 0.5\%.}
\\[15.0ex] \hline \\[-2.0ex]
Atoui  13 & \cite{Atoui:2013zza} & 2 & \parbox[t]{2cm}{270, 300, 270, 410}
 & \parbox[t]{5cm}{Combined continuum and chiral extrapolation, with linear terms in $a^2$ and $m_{\rm sea}$.
No dependence on $a$ or $m_{\rm sea}$ observed within errors. Stability of results vs fits with no $m_{\rm sea}$
dependence checked.}
\\[15.0ex]
\hline\hline
\end{tabular*}
\caption{Chiral extrapolation/minimum pion mass in
  determinations of $B\to D \ell\nu$, $B\to D^* \ell\nu$,  $B_s\to D_s \ell\nu$, and $\Lambda_b\to\Lambda_c\ell\nu$ form factors, and of $R(D)$.  For actions with multiple species of pions, masses quoted are the RMS pion masses.  The different $M_{\pi,\rm min}$ entries correspond to the different lattice spacings.}
}
\end{table}
%%%%%%%%%%%%%%%%%%%%%%%%%%%%% FINITE VOLUME %%%%%%%%%%%%%%%%%%%%%%%%%%%%%%%%%%%%%%%%%%%%%%%%%%%%%%%%%%%%%%%%%%%%%%%%%%%
\begin{table}[!ht]
{\footnotesize
\begin{tabular*}{\textwidth}{l @{\extracolsep{\fill}} c c c c l}
\hline\hline \\[-1.0ex]
Collab. & Ref. & $\Nf$ & $L$ [fm] & ${M_{\pi,\rm min}}L$ & Description 
\\[1.0ex] \hline \hline \\[-1.0ex]
\parbox[t]{26mm}{\SLhpqcdBD} & \cite{Na:2015kha} & 2+1 & \parbox[t]{1.8cm}{2.5, 2.4/2.9} & $\gtrsim 3.8$  &
\parbox[t]{4.8cm}{FV effects estimated to be negligible.}
\\[4.0ex] \hline \\[-1.0ex]
\parbox[t]{26mm}{\SLfnalmilcBD} & \cite{Lattice:2015rga} & 2+1 & \parbox[t]{1.8cm}{2.9, 2.9--3.8, 2.5--5.8, 2.4/2.9} & $\gtrsim 3.8$ & 
\parbox[t]{4.8cm}{ FV error estimated to be negligible in \cite{Bailey:2012rr}.}
\\[7.0ex] \hline \\[-1.0ex]
\parbox[t]{26mm}{\SLLambdabc} & \cite{Detmold:2015aaa} & 2+1 & \parbox[t]{1.8cm}{2.7, 2.7} & $\gtrsim 3.1$ (valence sector)  &
\parbox[t]{4.8cm}{FV effect estimated at 1.5\% from experience on $\chi$PT estimates of FV effects for heavy-baryon axial couplings.
}
\\[10.0ex] \hline \\[-1.0ex]
\parbox[t]{26mm}{\SLfnalmilcBDstar} & \cite{Bailey:2014tva} & 2+1 & \parbox[t]{1.8cm}{2.9, 2.9--3.8, 2.4--5.5, 2.4/2.9, 2.4} & $\gtrsim 3.8$ & 
\parbox[t]{4.8cm}{ FV error estimated to be negligible.}
\\[7.0ex] \hline \\[-1.0ex]
Atoui  13 & \cite{Atoui:2013zza} & 2 & \parbox[t]{1.8cm}{1.7/2.6, 2.1, 2.0/2.7, 2.4} & $\gtrsim 3.6$ & 
\parbox[t]{4.8cm}{No volume dependence observed within errors.}
\\[4.0ex]
\hline\hline
\end{tabular*}
\caption{Finite volume effects in determinations of $B\to D \ell\nu$, $B\to D^* \ell\nu$,  $B_s\to D_s \ell\nu$, and $\Lambda_b\to\Lambda_c\ell\nu$ form factors, and of $R(D)$.  Each $L$-entry corresponds to a different lattice spacing, with multiple spatial volumes at some lattice spacings.  For actions with multiple species of pions, the lightest pion masses are quoted.}
}
\end{table}
%%%%%%%%%%%%%%%%%%%%%%%%%%%%%%%%%%%%% OPERATOR RENORMALIZATION %%%%%%%%%%%%%%%%%%%%%%%%%%%%%%%%%%%%%%%%%
\begin{table}[!ht]
{\footnotesize
\begin{tabular*}{\textwidth}{l @{\extracolsep{\fill}} c c c l}
\hline\hline \\[-1.0ex]
Collab. & Ref. & $\Nf$ & Ren. & Description 
\\[1.0ex] \hline \hline \\[-1.0ex]
\parbox[t]{26mm}{\SLhpqcdBD} & \cite{Na:2015kha} & 2+1 & One loop. &
\parbox[t]{6cm}{One-loop matching of currents taken from~\cite{Monahan:2012dq}.}
\\[4.0ex] \hline \\[-1.0ex]
\parbox[t]{26mm}{\SLfnalmilcBD} & \cite{Lattice:2015rga} & 2+1 & mNPR &
\parbox[t]{6cm}{Form factors extracted from ratios of correlators that renormalize with
ratios of current normalizations, computed at one-loop in perturbation theory.
Dependence of renormalization factor on recoil parameter $w$ neglected.
Systematic uncertainty due to perturbative truncation and $w$-dependence estimated
by power counting to 0.7\%.}
\\[24.0ex] \hline \\[-1.0ex]
\parbox[t]{26mm}{\SLLambdabc} & \cite{Detmold:2015aaa} & 2+1 & mNPR &
\parbox[t]{6cm}{Perturbative truncation error estimated at 1\% with size of 1-loop correction on next-to-finer ensemble.}
\\[7.0ex] \hline \\[-1.0ex]
\parbox[t]{26mm}{\SLfnalmilcBDstar} & \cite{Bailey:2014tva} & 2+1 & mNPR &
\parbox[t]{6cm}{Majority of current renormalization factor cancels in double ratio of lattice correlation functions.  Remaining correction calculated with 1-loop tadpole-improved lattice perturbation theory. Systematic uncertainty estimated at 0.4\%.}
\\[15.0ex] \hline \\[-1.0ex]
Atoui  13 & \cite{Atoui:2013zza} & 2 & --- &
\parbox[t]{6cm}{Observables obtained from ratios that do not require renormalization. Checks performed by comparing with results coming from currents that are renormalized separately with non-perturbative $Z_{\rm V}$.}
\\[12.0ex]
\hline\hline
\end{tabular*}
\caption{Operator renormalization in determinations of $B\to D \ell\nu$, $B\to D^* \ell\nu$,  $B_s\to D_s \ell\nu$, and $\Lambda_b\to\Lambda_c\ell\nu$ form factors, and of $R(D)$.}
}
\end{table}
%%%%%%%%%%%%%%%%%%%%%%%%%%%%%%%%%%% HEAVY QUARK TREATMENT %%%%%%%%%%%%%%%%%%%%%%%%%%%%%%%%%%%%%%%%%%%%%%%%%%%
\begin{table}[!ht]
{\footnotesize
\begin{tabular*}{\textwidth}{l @{\extracolsep{\fill}} c c c l}
\hline\hline \\[-1.0ex]
Collab. & Ref. & $\Nf$ & Action & Description 
\\[1.0ex] \hline \hline \\[-1.0ex]
\parbox[t]{26mm}{\SLhpqcdBD} & \cite{Na:2015kha} & 2+1 & \parbox[t]{30mm}{NRQCD for b quark,\\ HISQ for c quark} &
\parbox[t]{6cm}{Discretization errors estimated 
via power counting to be 2.59\%.}
\\[4.0ex] \hline \\[-1.0ex]
\parbox[t]{26mm}{\SLfnalmilcBD} & \cite{Lattice:2015rga} & 2+1 & Fermilab &
\parbox[t]{6cm}{Discretization errors of form factors estimated 
via power counting to be 0.4\%.}
\\[4.0ex] \hline \\[-1.0ex]
\parbox[t]{26mm}{\SLLambdabc} & \cite{Detmold:2015aaa} & 2+1 & Columbia RHQ &
\parbox[t]{6cm}{Discretization errors discussed as part of combined chiral-continuum-$q^2$ fit, stemming from $a^2|\mathbf{p}|^2$ terms.}
\\[7.0ex] \hline \\[-1.0ex]
\parbox[t]{26mm}{\SLfnalmilcBDstar} & \cite{Bailey:2014tva} & 2+1 & Fermilab &
\parbox[t]{6cm}{Discretization errors estimated 
via power counting to be 1\%.}
\\[4.0ex] \hline \\[-1.0ex]
Atoui  13 & \cite{Atoui:2013zza} & 2 & tmWil &
\parbox[t]{6cm}{Results obtained from step-scaling in heavy quark mass via the ratio method. Separate continuum limit extrapolations with mild $a^2$ dependence carried out for each mass point separately. Result at physical value of $m_b$ obtained by interpolation between data region and known exact HQET limit.}
\\[21.0ex]
\hline\hline
\end{tabular*}
\caption{Heavy quark treatment in determinations of $B\to D \ell\nu$, $B\to D^* \ell\nu$,  $B_s\to D_s \ell\nu$, and $\Lambda_b\to\Lambda_c\ell\nu$ form factors, and of $R(D)$.}
}
\end{table}

%% file: Alpha_s/Alpha_s_app.tex
%%%%%%%%%%%%%%%%%%%%%%%%%

\subsection{Notes to section \ref{sec:alpha_s} on the strong coupling $\alpha_{\rm s}$}

\subsubsection{Renormalization scale and perturbative behaviour}

%\eject

%%%%%%%%%%%%%%%%%%%%%%%%%

%%%%%

\begin{table}[!htb]
   \footnotesize
   \begin{tabular*}{\textwidth}[l]{l@{\extracolsep{-0.005\textwidth}}rlllllllll}
\hline\hline \\[-1.0ex]
      Collab. & Ref. & $\Nf$ & $\alpha_{\rm eff}$ & $n_l$ & Description \\
      & & & &  \\
      \hline
      \hline
      & & & &  \\[-0.1cm]

FlowQCD 15    & \cite{Asakawa:2015vta}       & 0        &  
  \parbox[t]{0.09\textwidth}{$0.09$-$0.12$}   & 2 
                  & \parbox[t]{0.555\textwidth}
                    {$\alpha_\msbar(2.63/a)$ computed from
                    the boosted coupling. The physical volume ranges
                    from $2.4 \sim 3.8\,\mbox{fm}$.}  \\
\\[-0.1cm] \hline\\[-1.5ex] 
Sternbeck 12  & \cite{Sternbeck:2012qs}       & 0        &  
  \parbox[t]{0.09\textwidth}{0.11-0.18} & 3
& \parbox[t]{0.555\textwidth}{$\alpha_{\mathrm{T}}(p)$ for $p = 5 - 40$\,GeV.
  Fitted to PT without power corrections.
  ($\beta$ = 6.0, 6.4, 6.7, 6.92.)} \\
\\[-0.1cm] \hline\\[-1.5ex] 
 {Ilgenfritz 10} & \cite{Ilgenfritz:2010gu}   & 0        &
  \parbox[t]{0.09\textwidth}{0.07-0.9} & 3
& \parbox[t]{0.555\textwidth}{$\alpha_{\rm T}(p)$ for $p = 1 - 240$\,GeV.
  ($\beta$ = 5.8, 6.0, 6.2, 6.4, 9.0.) } \\
\\[-0.1cm] \hline\\[-1.5ex] 
Sternbeck 10  & \cite{Sternbeck:2010xu}       & 0        & 
  \parbox[t]{0.09\textwidth}{0.07-0.32} & 3
& \parbox[t]{0.555\textwidth}{$\alpha_{\mathrm{T}}$ for 
  $p = 2.5 - 140$\, GeV, fitted to PT partially on
  very small lattices.} \\
\\[-0.1cm] \hline\\[-1.5ex] 
Brambilla 10  & \cite{Brambilla:2010pp}       & 0        & 
  \parbox[t]{0.09\textwidth}{0.22-0.47} & 3
& \parbox[t]{0.555\textwidth}{$\alpha_\mathrm{qq}(1/r)$ for
  the range $r/r_0 = 0.15 - 0.5$. Fit of $V(r)$ to PT
  with renormalon subtraction and resummation reproduces
  the static potential for $r/r_0 = 0.15 - 0.45$ well.} \\
\\[-0.1cm] \hline\\[-1.5ex] 
{Boucaud~08}  & \cite{Boucaud:2008gn}         & 0        & 
  \parbox[t]{0.09\textwidth}{0.18-0.35} & 3 
& \parbox[t]{0.555\textwidth}{$\alpha_\mathrm{T}(p)$ with 
  $p = 3 - 6$\, GeV. Fitted to PT 
   with $1/p^2$ correction.} \\
\\[-0.1cm] \hline\\[-1.5ex] 
{Boucaud 05}  & \cite{Boucaud:2005gg}         & 0        &
  \parbox[t]{0.09\textwidth}{0.22-0.55} & 3
& \parbox[t]{0.555\textwidth}{$\Lambda_{\widetilde{\rm MOM}_{g,c}}$ using 
  gluon and ghost propagators with $2 \leq \mu \leq 6$\, GeV. 
  Fit to perturbation theory.} \\
\\[-0.1cm] \hline\\[-1.5ex] 
\parbox[t]{0.12\textwidth}{{QCDSF-\\ UKQCD~05}}
              & \cite{Gockeler:2005rv}        & 0        &
  \parbox[t]{0.09\textwidth}{0.10-0.15} & 2
& \parbox[t]{0.555\textwidth}
  {$\alpha_{\msbar}(2.63/a)$ computed from the boosted coupling.} \\        
\\[-0.1cm] \hline\\[-1.5ex] 
CP-PACS 04     & \cite{Takeda:2004xha}         & 0        &      
  \parbox[t]{0.12\textwidth}{0.08-0.28} & 2 
& \parbox[t]{0.5550\textwidth}{$\alpha_{\rm SF}(1/L)$ step-scaling
  functions at $\alpha_{\rm eff} = 0.08, 0.19$, study of continuum limit.
  Agreement of continuum limit with ALPHA~98.} \\
\\[-0.1cm] \hline\\[-1.5ex] 
\parbox[t]{0.14\textwidth}{Boucaud 01A} & \cite{Boucaud:2001st}         &  0       &
  \parbox[t]{0.09\textwidth}{0.18-0.45} & 2
& \parbox[t]{0.555\textwidth}{$\alpha_{\rm MOM}$ with
  $p = 2.5 - 10$\, GeV. Consistency check of $n_l=2$ loop perturbation
  formula with gluon condensate. $\langle A^2 \rangle$ from
  $\alpha_{\rm MOM}$ and gluon propagator are consistent.} \\
\\[-0.1cm] \hline\\[-1.5ex] 
{Soto 01}     & \cite{DeSoto:2001qx}          & 0        &
  \parbox[t]{0.09\textwidth}{ {0.25-0.36, 0.3-0.36, 0.19-0.24 }} & 2
& \parbox[t]{0.555\textwidth}{{$\alpha_{\widetilde{\rm MOM}}$ for
  $p = 3 - 10$\, GeV. Fit with $n_l= 2$ loop formula with gluon condensate.
  (Without condensate does not fit the lattice data.)
  ($\beta$ = 6.0, 6.2, 6.8.)}} \\
\\[-0.1cm] \hline\\[-1.5ex] 
\parbox[t]{0.14\textwidth}{Boucaud 00A} & \cite{Boucaud:2000ey}         &  0       &
  \parbox[t]{0.09\textwidth}{0.35-0.55, 0.25-0.45, 0.22-0.28, 0.18-0.22} & 2
& \parbox[t]{0.555\textwidth}{$\alpha_{\widetilde{\rm MOM}}$ with
  $p = 2 - 10$\, GeV. Fitted to $n_l = 2$ loop perturbation theory
  with power correction.
% 4-loop coefficient is strongly correlated to the power
%  correction coefficient.
 ($\beta$ = 6.0, 6.2, 6.4, 6.8.)} \\
\\[-0.1cm] \hline\\[-1.5ex] 
\parbox[t]{0.14\textwidth}{Boucaud 00B} & \cite{Boucaud:2000nd}         &  0       &
  \parbox[t]{0.09\textwidth}{0.35-0.55, 0.25-0.45, 0.22-0.28, 0.18-0.22} & 2
& \parbox[t]{0.555\textwidth}{$\alpha_{\rm MOM}$ with
  $2 \leq \mu \leq 10$\, GeV. Consistency check of $n_l = 2$ loop
  perturbation formula with gluon condensate.
  $\beta_2^{MOM}=1.5\times \beta_2^{\widetilde{MOM}}$ is needed.
  ($\beta$ = 6.0, 6.2, 6.4, 6.8.)} \\
   &&&&&&&&& \\[-0.1cm]
   \hline
   \hline
\end{tabular*}
\caption{Renormalization scale and perturbative behaviour
         of $\alpha_s$ determinations for $N_f=0$. }
\label{tab_Nf=0_renormalization_a}
\end{table}

\begin{table}[!htb]
\addtocounter{table}{-1}
   \footnotesize
   \begin{tabular*}{\textwidth}[l]{l@{\extracolsep{\fill}}rlllllllll}
\hline\hline \\[-1.0ex]
      Collab. & Ref. & $\Nf$ & $\alpha_{\rm eff}$ & $n_l$ & Description \\
      & & & &  \\
      \hline
      \hline
      & & & &  \\[-0.1cm]
{Becirevic 99A}& \cite{Becirevic:1999uc}      &  0       &
  \parbox[t]{0.09\textwidth}{0.25-0.4} & 2
& \parbox[t]{0.555\textwidth}{$\alpha_{\widetilde{\rm MOM}}$ with
  $p$ = 2.5 - 5.5\, GeV.} \\
\\[-0.1cm] \hline\\[-1.5ex] 
{Becirevic 99B}  & \cite{Becirevic:1999hj}    &  0       &
  \parbox[t]{0.09\textwidth}{0.18-0.25} & 2
& \parbox[t]{0.555\textwidth}{$\alpha_{\widetilde{\rm MOM}}$ from
a single lattice spacing with $p$ = 5.6 - 9.5\, GeV.} \\
\\[-0.1cm] \hline\\[-1.5ex] 
{SESAM 99}    & \cite{Spitz:1999tu}           & 0        &
  \parbox[t]{0.09\textwidth}{0.15} & 1
& \parbox[t]{0.555\textwidth}{$\alpha_\mathrm{V}(3.41/a)$ computed
  from the boosted coupling.} \\
\\[-0.1cm] \hline\\[-1.5ex] 
ALPHA 98      & \cite{Capitani:1998mq}        & 0        & 
  \parbox[t]{0.09\textwidth}{0.07-0.28} & 2
& \parbox[t]{0.5550\textwidth}{$\alpha_{\rm SF}(1/L)$ step scaling,
  agreement with perturbative running ($n_l=2$) for $\alpha_{\rm eff} < 0.15$.} \\
\\[-0.1cm] \hline\\[-1.5ex] 
{Boucaud 98A} & \cite{Boucaud:1998bq}         & 0        &
  \parbox[t]{0.09\textwidth}{0.35-0.5} & 1,2
& \parbox[t]{0.555\textwidth}{$\alpha_\mathrm{MOM}$,  
  with $2.1 \leq \mu \leq 3.9$\, GeV. $n_l=1$ for $\alpha_{\rm MOM}$,
   $n_l=2$ for $\alpha_{\widetilde{\rm MOM}}$.} \\
\\[-0.1cm] \hline\\[-1.5ex] 
{Boucaud 98B} & \cite{Boucaud:1998xi}         & 0        &
  \parbox[t]{0.09\textwidth}{0.27-0.50} & 2
& \parbox[t]{0.555\textwidth}{$\alpha_{\widetilde{\rm MOM}}$
  with $\mu = 2.2 - 4.5$\, GeV.} \\
\\[-0.1cm] \hline\\[-1.5ex] 
Alles 96      & \cite{Alles:1996ka}           &  0       & 
  \parbox[t]{0.09\textwidth}{0.35-0.71} & 1
& \parbox[t]{0.555\textwidth}{$\alpha_{\widetilde{\rm MOM}}(p)$
  with $p=1.8-3.0$\, GeV.} \\
\\[-0.1cm] \hline\\[-1.5ex] 
{Wingate 95}  & \cite{Wingate:1995fd}         & 0        &
  \parbox[t]{0.09\textwidth}{0.15} & 1
& \parbox[t]{0.555\textwidth}{$\alpha_{V}(3.41/a)$ computed
  from the boosted coupling.} \\
\\[-0.1cm] \hline\\[-1.5ex] 
{Davies 94}   & \cite{Davies:1994ei}          & 0        &
  \parbox[t]{0.09\textwidth}{0.15} & 1
& \parbox[t]{0.555\textwidth}{$\alpha_{V}(3.41/a)$ computed from
  the boosted coupling.} \\
\\[-0.1cm] \hline\\[-1.5ex] 
L\"uscher 93 & \cite{Luscher:1993gh}          & 0        & 
  \parbox[t]{0.09\textwidth}{0.09-0.28} & 1
& \parbox[t]{0.5550\textwidth}{$\alpha_{\rm SF}(1/L)$ step scaling,
  agreement with perturbative running  ($n_l=1$) for $\alpha_{\rm eff} < 0.17$.} \\      
\\[-0.1cm] \hline\\[-1.5ex] 
UKQCD 92     & \cite{Booth:1992bm}            & 0        & 
  \parbox[t]{0.09\textwidth}{0.17-0.40} & 1
& \parbox[t]{0.555\textwidth}{$\alpha_\mathrm{qq}(1/r)$ for a single
  lattice spacing. Fit of $\alpha_\mathrm{qq}(1/r)$ to a NLO formula.} \\
\\[-0.1cm] \hline\\[-1.5ex] 
Bali 92     & \cite{Bali:1992ru}              & 0        & 
  \parbox[t]{0.09\textwidth}{0.15-0.35} & 1
& \parbox[t]{0.555\textwidth}{$\alpha_\mathrm{qq}(1/r)$ for the
  lattice spacing used in the analysis. Box size $L\approx1.05$\,fm.
  Fit of $\alpha_\mathrm{qq}(1/r)$ to a NLO formula. $\Lambda_\msbar$ is
  found to depend on the fit-range.} \\
\\[-0.1cm] \hline\\[-1.5ex] 
El-Khadra 92& \cite{ElKhadra:1992vn}          & 0        &
  \parbox[t]{0.09\textwidth}{0.12-0.15}   & 1
%  \parbox[t]{0.09\textwidth}{0.15,0.13,0.12}   & ?
& \parbox[t]{0.555\textwidth}{$\alpha_{\msbar}(\pi/a)$ from 1-loop
  boosted perturbation theory.}\\
   &&&&&&&&& \\[-0.1cm]
   \hline
   \hline
\end{tabular*}
\caption{(contd.) Renormalization scale and perturbative behaviour
         of $\alpha_s$ determinations for $N_f=0$.}
\label{tab_Nf=0_renormalization_b}
\end{table}

%%%%%%%%%%%%%%%%%%%%

\begin{table}[!htb]
   \footnotesize
   \begin{tabular*}{\textwidth}[l]{l@{\extracolsep{\fill}}rlllllllll}
\hline\hline \\[-1.0ex]
      Collab. & Ref. & $\Nf$ & $\alpha_{\rm eff}$ & $n_l$ & Description \\
      &&&&&&& & \\[-0.1cm]
      \hline
      \hline
      &&&&&&&&& \\[-0.1cm]
Karbstein 14 & \cite{Karbstein:2014bsa} & 2 
    & 0.28 - 0.41 & 3
    & \parbox[t]{0.555\textwidth}{$\alpha_V(p)$ for momentum 
                                $1.5 < p < 3.0\,\mbox{GeV}$.
                                Values computed from the 
                                quoted $\Lambda$ parameter with the 2-loop
                                $\beta$ function; larger values ($0.32 - 0.62$)
                                are obtained with 3-loop running. 
%                                $\alpha_\msbar$ is extracted from a 
%                                fit of $V(p)$ to PT with an additional constant.
                                As with ETM 11C
                                central values are taken from 
                                $a = 0.042\,\mbox{fm}$ lattice
                                with $L = 1.3\,\mbox{fm}$ and 
                                $m_\pi = 350\,\mbox{MeV}$.}
\\ \\[-0.1cm] \hline\\[-1.5ex] 
ALPHA 12 & \cite{Fritzsch:2012wq} & 2  & 
\parbox[t]{0.1\textwidth}{see \\ ALPHA 04} & 2
    & \parbox[t]{0.5550\textwidth}{Determination of $\Lambda_\msbar / f_K$ using ALPHA 04}
\\ \\[-0.1cm] \hline\\[-1.5ex] 
 {Sternbeck 12}  & \cite{Sternbeck:2012qs}       & 2        &  
 {0.17-0.23} & 3
& \parbox[t]{0.555\textwidth}{ 
$\alpha_\mathrm{T}$ for $(r_0 p)^2=200-2000$. 
Fit to PT without condensate. Deviation at higher energy is observed.}
\\ \\[-0.1cm] \hline\\[-1.5ex] 
ETM 11C   & \cite{Jansen:2011vv}   & 2 & 
0.26-0.96 & 3 
   & \parbox[t]{0.555\textwidth}{$\alpha_{\rm qq}(1/r)$ as computed by us from $\Lambda_\msbar=315$\,MeV.
\newline Fit of $V(r)$ to PT with renormalon subtraction
         and resummation reproduces the static potential for $r/r_0=0.2-0.6$ well.
         One fit-range, using $r/a=2-4$ at the smallest lattice spacing corresponds to 
         $\alpha_{\rm eff}=0.26-0.40$. 
         In the $\msbar$ scheme one has 
         $\alpha_\msbar(1/r)=0.24 - 0.63$ and for the restricted
         fit $\alpha_\msbar(1/r)=0.24 - 0.36$.
         %{\color{red}A coefficient additional to $\Lambda$ is introduced and adjusted.}
         Central values taken from $a=0.042$\,fm lattice with $L=1.3$\,fm
         and $m_\pi=350$\,MeV.}
\\ \\[-0.1cm] \hline\\[-1.5ex] 
 {ETM 10F}       & \cite{Blossier:2010ky}   & 2              &  
0.24-0.45 & 3
   &  \parbox[t]{0.5550\textwidth}{$\alpha_T$ for momentum up to
2.6 - 5.6\,GeV. Fitted to PT with gluon condensate correction term.}
\\ \\[-0.1cm] \hline\\[-1.5ex]  
 {Sternbeck 10}  & \cite{Sternbeck:2010xu}       & 2       & 
 {0.19-0.38} & 3
   & {$\alpha_\mathrm{T}$ for $1\leq(ap)^2\leq 10$.
     Fitted with $n_l = 3$ loop formula.}
\\ \\[-0.1cm] \hline\\[-1.5ex] 
 {JLQCD 08} & \cite{Shintani:2008ga} & 2                  
&  {0.25-0.30} & 1
& \parbox{0.555\textwidth}{$\alpha_{\overline{\rm MS}}(Q)$ for $0.65<(aQ)^2<1.32$.
   Fit with the perturbative formula with power corrections.}
\\ \\[-0.1cm] \hline\\[-1.5ex] 
\parbox[t]{0.11\textwidth}{QCDSF-\\UKQCD 05}
   & \cite{Gockeler:2005rv}
   & 2
   & \parbox[t]{0.1\textwidth}
        {0.18-0.20}
   & 2
   & \parbox[t]{0.555\textwidth}
        {$\alpha_\msbar(1.4/a)$ computed from the boosted coupling.}
\\ \\[-0.1cm] \hline\\[-1.5ex] 
ALPHA 04 & \cite{DellaMorte:2004bc} & 2 & 
0.078-0.44 & 2 &
\parbox[t]{0.5550\textwidth}{$\alpha_{\rm SF}(1/L)$  step scaling,
agreement with $n_l = 2$ looprunning for $\alpha_s < 0.2$}
\\ \\[-0.1cm] \hline\\[-1.5ex] 
ALPHA 01 & \cite{Bode:2001jv} & 2 &
0.078-0.44 & 2 &
\parbox[t]{0.5550\textwidth}{$\alpha_{\rm SF}(1/L)$  step scaling,
agreement with $n_l=2$ loop running for $\alpha_s < 0.2$}
\\ \\[-0.1cm] \hline\\[-1.5ex] 
{Boucaud 01B}    & \cite{Boucaud:2001qz}    & 2  & 
0.25-0.5   &  3 
   & \parbox[t]{0.5550\textwidth}{$\alpha_{\rm\widetilde{MOM}}$ for momentum up to 7\,GeV.
Fitted with $n_l = 3$ loop formula with and without power correction,
leading to different results for $\Lambda_\msbar^{(2)}$. 
Extrapolation of $\alpha_s(1.3\,\GeV)$ in $N_f$ from $N_f=0, 2$ to
$N_f=3$ is made.}
\\ \\[-0.1cm] \hline\\[-1.5ex] 
 {SESAM 99}
   & \cite{Spitz:1999tu}
   & 2
   & \parbox[t]{0.1\textwidth}
        {0.17}
   & 1
   & \parbox[t]{0.555\textwidth}
        {The boosted coupling $\alpha_{\rm P}(3.41/a)$.}
\\ \\[-0.1cm] \hline\\[-1.5ex] 
 {Wingate 95}
   & \cite{Wingate:1995fd}
   & 2
   & \parbox[t]{0.1\textwidth}
        {0.18}
   & 1
   & \parbox[t]{0.555\textwidth}
        {$\alpha_{\rm V}(3.41/a)$ computed from the boosted coupling.}
\\ \\[-0.1cm] \hline\\[-1.5ex] 
 {Aoki 94}
   & \cite{Aoki:1994pc}
   & 2                   
   & \parbox[t]{0.1\textwidth}
        {0.14}
   & 1
   & \parbox[t]{0.555\textwidth}
        {$\alpha_\msbar(\pi/a)$ computed from the boosted coupling.}
\\ \\[-0.1cm] \hline\\[-1.5ex] 
 {Davies 94} 
   & \cite{Davies:1994ei}  
   & 2
   & \parbox[t]{0.1\textwidth}
        {0.18}
   & 1
   & \parbox[t]{0.555\textwidth}
        {$\alpha_{\rm V}(3.41/a)$ computed from $\ln W_{11}$.}
\\                            
     &&&&&&&&& \\[-0.1cm]
\hline\hline
\end{tabular*}
\caption{Renormalization scale and perturbative behaviour 
of $\alpha_s$ determinations for $N_f=2$.}
\label{tab_Nf=2_renormalization}
\end{table}

%%%%%%%%%%%%%%%%%%%%%%%%%%%%%%%%%%%%%%%%%%%%%%%%%%%%%%%%%%%%%%%%%%%%%%%%%%%%%%%%%%%%%%%%%

\begin{table}[!htb]
   \vspace{3.0cm}
   \footnotesize
   \begin{tabular*}{\textwidth}[l]{l@{\extracolsep{\fill}}rlllllllll}
\hline\hline \\[-1.0ex]
      Collab. & Ref. & $\Nf$ & $\alpha_{\rm eff}$ & $n_l$ & Description \\
      &&&&& \\[-0.1cm]
      \hline
      \hline
      &&&&& \\[-0.1cm]
Bazavov 14  & \cite{Bazavov:2014soa}  & 2+1  & 
              0.19-0.41 & 3 &
%             0.19-0.49 & 
             \parbox[t]{0.54\textwidth}{Update of Bazavov 12
                                        including finer lattices
                                        down to $a = 0.041\,\mbox{fm}$.
%                                        Fit range $r/r_1 = 0.12-0.60$
%                                        ($r/r_0 = 0.08-0.40$).
%%  r0/r1=1.508
                                        Fit range $r/r_1 = 0.12-0.50$
                                        ($r/r_0 = 0.08-0.33$).
                                        Perturbative expansion of the  force $F(r)$ integrated to determine potential.}
\\ \\[-0.1cm] \hline\\[-1.5ex]
Bazavov 12   & \cite{Bazavov:2012ka}  & 2+1  & 
\parbox[t]{0.1\textwidth}{0.23-0.57} & 3 &
\parbox[t]{0.54\textwidth}{$\alpha_{\rm qq}$ computed by us from $\Lambda_\msbar r_0=0.70 $. 
% \newline $\alpha_{\msbar}>0.3$ when  $a/r<0.5$ .
Fit of $V(r)$ to PT with renormalon subtraction
and resummation reproduces the static potential for 
$r/r_0=0.135-0.5$ well. }
\\ \\[-0.1cm] \hline\\[-1.5ex]
 {Sternbeck 12}  & \cite{Sternbeck:2012qs}  & 2+1       &  
 {0.19-0.25}  &  3  & \parbox[t]{0.54\textwidth}{{$\alpha_\mathrm{T}$ for $(p r_0)^2=200-2000$. 
Comparison with 4-loop formula.}}
\\ \\[-0.1cm] \hline\\[-1.5ex]
 {JLQCD 10} & \cite{Shintani:2010ph} & 2+1                  
 & {0.29-0.35} & 2
 &\parbox{0.54\textwidth}{$\alpha_{\overline{\rm MS}}(Q)$ for $0.4<(aQ)^2<1.0$.
     Fit with the perturbative formula with power corrections.}
\\ \\[-0.1cm] \hline\\[-1.5ex] 
 {HPQCD 10}
   & \cite{McNeile:2010ji}
   & 2+1
   &   & 2
   & \parbox[t]{0.54\textwidth}
        {Uses method of section~\ref{s:WL}.
         Update of $r_1$ and $r_1/a$ in HPQCD 08A.}
\\ \\[-0.1cm] \hline\\[-1.5ex]
HPQCD 10    & \cite{McNeile:2010ji}  & 2+1 &  
0.12-0.42   & 2 
   & \parbox[t]{0.54\textwidth}{Uses method of section~\ref{s:curr}.
                $\alpha_\mathrm{eff}$ from $R_4$ and $R_6/R_8$. 
                Fit of $R_n, \;n=4\ldots10$ to PT including $(a m)^{2i}$
                terms with $i\leq 10$; coefficients constrained by priors. } 
% using m_p<2.0 as HPQCD
\\ \\[-0.1cm] \hline\\[-1.5ex]
PACS-CS 09A& \cite{Aoki:2009tf} & 2+1 & 
0.08-0.27 & 2 &
\parbox[t]{0.54\textwidth}{$\alpha_{\rm SF}(1/L)$  step scaling, agreement with 
3-loop running for $\alpha_s \leq 0.27$}
\\  \\[-0.1cm] \hline\\[-1.5ex]
HPQCD 08B    & \cite{Allison:2008xk}  & 2+1 &  
0.38  & 2 &\parbox[t]{0.54\textwidth}{Fit of the ratios to PT
at the charm mass
including $(a m)^{2i}$ terms with $i\leq 2\ldots 4$; coefficients constrained by priors. } 
% using m_p<2.0 as HPQCD
\\ \\[-0.1cm] \hline\\[-1.5ex]
 {HPQCD 08A}
   & \cite{Davies:2008sw}
   & 2+1
   & \parbox[t]{0.1\textwidth}
        {0.15-0.4}
   & 2
   & \parbox[t]{0.54\textwidth}
        {$\alphav(q^*)$ for a variety of short-distance quantities,
         using same method as in HPQCD 05A.}
\\ \\[-0.1cm] \hline\\[-1.5ex]
 {Maltman 08}
   & \cite{Maltman:2008bx}
   & 2+1
   &  & 2
   & \parbox[t]{0.54\textwidth}
        {Re-analysis of HPQCD 05A for a restricted set of
         short-distance quantities with similar results.}
\\ \\[-0.1cm] \hline\\[-1.5ex]
 {HPQCD 05A}
   & \cite{Mason:2005zx}
   & 2+1
   & \parbox[t]{0.1\textwidth}
        {0.2-0.4}
   & 2
   & \parbox[t]{0.54\textwidth}
        {$\alphav(q^*)$ for a variety of short-distance quantities.
        }
\\
      &&&&& \\[-0.1cm]
      \hline
      \hline
\end{tabular*}
\caption{Renormalization scale and perturbative behaviour of $\alpha_s$
         determinations for $N_f=3$.}
\label{tab_Nf=3_renormalization}
\end{table}

%%%%%%%%%%%%%%%%%%%%%%%%%%%%%%%%%%%%%%%%%%%%%%%%%%%%%%%%%%%%%%%%%%%%%%%%%%%%%%%%%%%%%%%%%

\begin{table}[!htb]
   \vspace{3.0cm}
   \footnotesize
   \begin{tabular*}{\textwidth}[l]{l@{\extracolsep{\fill}}rlllllllll}
\hline\hline \\[-1.0ex]
      Collab. & Ref. & $\Nf$ & $\alpha_{\rm eff}$ & $n_l$ & Description \\
      &&&&& \\[-0.1cm]
      \hline
      \hline
      &&&&& \\[-0.1cm]
 HPQCD 14A   &  \cite{Chakraborty:2014aca} & 2+1+1 &  
 0.11-0.33 & 2 
   &  \parbox[t]{0.54\textwidth}{
           Range given for $\alpha_\mathrm{eff}$ from $\tilde R_4$. 
           Fit of ratios $R_n\;n=4\ldots10 $ to perturbation theory
           including $(a m)^{2i}$ terms with $i\leq 10 - 20$ and 
           higher-order perturbative terms;
           coefficients constrained by priors.}
% using m_p<2.0 as HPQCD
 \\ \\[-0.1cm] \hline\\[-1.5ex]
 { {ETM 13D}}       &  \cite{Blossier:2013ioa} & 2+1+1 &  0.26-0.7    & 3 & 
	\parbox[t]{0.54\textwidth}{$\alpha_\mathrm{T}(p)$= for $p=1.6-6.5$ GeV. 
	Update of \cite{Blossier:2012ef}
                                   with improved power law determination.}
 \\ \\[-0.1cm] \hline\\[-1.5ex]
 { ETM 12C}       & \cite{Blossier:2012ef}   & 2+1+1 &
0.24-0.38 & 3 &
	\parbox[t]{0.54\textwidth}{$\alpha_\mathrm{T}(p)$ for $p=1.7-6.8$ GeV. 
	Fit to PT with gluon condensate correction or higher power.}
	 \\ \\[-0.1cm] \hline\\[-1.5ex]
      ALPHA 10A & \cite{Tekin:2010mm} & 4                  & 0.07-0.28 & 2
        & \parbox[t]{0.54\textwidth}{$\alpha_\mathrm{SF}(1/L)$.
        Comparison to PT with 2-, 3-loop $\beta$-function.}
 \\ \\[-0.1cm] \hline\\[-1.5ex]
 {ETM 11D}      & \cite{Blossier:2011tf}   & 2+1+1 & 
	0.24-0.4& 3 &
	\parbox[t]{0.54\textwidth}{$\alpha_\mathrm{T}(p)$ for $p=3.8-7.1$\,GeV
        with H(4)-procedure. Fit to PT with gluon condensate correction.}
 \\ \\[-0.1cm] \hline\\[-1.5ex]
      Perez 10 & \cite{PerezRubio:2010ke} & 4           & 0.06-0.28 & 2
        & \parbox[t]{0.54\textwidth}{$\alpha_\mathrm{SF}(1/L)$.
          Comparison with 1-, 2-, 3-loop $\beta$-function.}\\
      &&&&&&\\[-0.1cm]
      \hline
      \hline
\end{tabular*}
\caption{Renormalization scale and perturbative behaviour of $\alpha_s$
         determinations for $N_f=4$.}
\label{tab_Nf=4_renormalization}
\end{table}

%%%%%%%%%%%%%%%%%%%%%%%%%%%%%%

\clearpage

\subsubsection{Continuum limit}
%\clearpage

%%%%%%%%%%%%%%%%%%%%%%%%%%%%%
\hspace{-1.5cm}

\begin{table}[!htb]
   \footnotesize
   \begin{tabular*}{\textwidth}[l]{l@{\extracolsep{\fill}}rlllllllll}
\hline\hline \\[-1.0ex]
      Collaboration & Ref. & $\Nf$ & $a\,\mu$  & Description\\
      & & & &  \\
      \hline
      \hline
      & & & &  \\[-0.1cm]

FlowQCD 15 & \cite{Asakawa:2015vta}       & 0        &  
      \parbox[t]{0.25\textwidth}{$9$ lattice spacings with 
                                 $a = 0.06$ - $0.02\,\mbox{fm}$.}
      & \parbox[t]{0.39\textwidth}{$w_{0.4}/a$, together
                   with $r_0 = 0.5\,\mbox{fm}$ and conversion factor
                   $r_0/w_{0.4} = 2.587(45)$.} \\      
\\[-0.1cm] \hline\\[-1.5ex]
      Sternbeck 12  & \cite{Sternbeck:2012qs}       & 0        &  
      \parbox[t]{0.25\textwidth}{4 lattice spacings $a\leq 0.1$\,fm}
      & \parbox[t]{0.39\textwidth}{At $\alpha_s$ = 0.18, $ap$ = 2.7, 1.5
                                  for $\beta = 6.0, 6.4$.}\\
\\[-0.1cm] \hline\\[-1.5ex]
      Brambilla 10 & \cite{Brambilla:2010pp} & 0  & 
      \parbox[t]{0.25\textwidth}{ At least 3 lattice spacings 
          with  $0.2 \leq 2a/r \leq 1.1 $ } &
      \parbox[t]{0.39\textwidth}{Extrapolation of potential differences 
          $V(r)-V(0.51r_0)$ linear in $a^2$ performed in \cite{Necco:2001xg}
          with several lattice spacings.} \\
\\[-0.1cm] \hline\\[-1.5ex]
% Necco:2001xg: minimal r/a=1.855, maximal: r_c/a = 9.82
      Ilgenfritz 10 & \cite{Ilgenfritz:2010gu}              & 0       & 
      \parbox[t]{0.25\textwidth}{{$a$ = 0.136, 0.093, 0.068, 0.051\,fm
      ($\beta=$ 5.8, 6.0, 6.2, 6.4), while no value of $a$ is given
      for $\beta = 9.0$}} &
      \parbox[t]{0.39\textwidth}{ {At $\alpha_s = 0.3$, 
      $ap =$ 2.0, 1.4, 1.0, 0.8 ($\beta$ = 5.8, 6.0, 6.2, 6.4). 
      For $\beta = 9.0$ at $ap = 1.4$, $\alpha_s = 0.082$.}
     }\\
\\[-0.1cm] \hline\\[-1.5ex]
      Sternbeck 10  & \cite{Sternbeck:2010xu}       & 0        & 
     \parbox[t]{0.25\textwidth}{{8 lattice spacings $a$ = 0.004 - 0.087\,fm
     ($r_0 = 0.467$\,fm)}}
&  \parbox[t]{0.39\textwidth}{% {$ap\geq 1.7$ throughout} \\
       {$\sqrt{3} < ap < \sqrt{12}$}.} \\
\\[-0.1cm] \hline\\[-1.5ex]
     Boucaud 08    & \cite{Boucaud:2008gn}         
     & 0           & {  $a = 0.1, 0.07, 0.05$\,fm}  
     & \parbox[t]{0.39\textwidth}{At $\alpha_s = 0.3$
                                 the data have $ap = 2.6, 1.9, 1.5$.}\\
\\[-0.1cm] \hline\\[-1.5ex]
     {QCDSF/UKQCD 05}
     & \cite{Gockeler:2005rv}
     & 0                   
     & \parbox[t]{0.25\textwidth}
        {7 lattice spacings with $a = 0.10$ - $0.028\,\mbox{fm}$.}
     & \parbox[t]{0.555\textwidth}
        {$r_0/a$, together with $r_0 = 0.467\,\mbox{fm}$.} \\
\\[-0.1cm] \hline\\[-1.5ex]
    {Boucaud 05}    & \cite{Boucaud:2005gg}         & 0       &
    {$a = 0.1, 0.07, 0.05$\,fm} & 
    {At $\alpha_s \leq 0.3$ $ap = 1.9, 1.4, 1.0$}.\\
\\[-0.1cm] \hline\\[-1.5ex]
    CP-PACS 04 & \cite{Takeda:2004xha} & 0  & 
    \parbox[t]{0.25\textwidth}{ 4 spacings, $a/L = 1/12 - 1/4$.} &
    \parbox[t]{0.39\textwidth}{ Iwasaki and  
    L\"uscher Weisz tree-level improved bulk actions; boundary improvement at 
    tree-level, 1-loop and with two different choices of implementation.} \\
\\[-0.1cm] \hline\\[-1.5ex]
    Soto 01            & \cite{DeSoto:2001qx}       & 0       &  
    {$a = 0.07, 0.05, 0.03$\,fm} 
    & \parbox[t]{0.39\textwidth}{At $\alpha_s \leq 0.3$,
                                the data have $ap = 1.4, 1.0,  0.6$.}\\
\\[-0.1cm] \hline\\[-1.5ex]
    {Boucaud 01A}      & \cite{Boucaud:2001st}      & 0       &   
    {$a = 0.1, 0.07, 0.05, 0.03$\,fm}  & 
    {At $\alpha_s \leq 0.3$ $ap = 1.9, 1.4, 1.0, 0.6$.} \\
\\[-0.1cm] \hline\\[-1.5ex]
    {Boucaud 00A}      &\cite{Boucaud:2000ey}           &  0    & 
    {$a = 0.1, 0.07, 0.05, 0.03$~fm}  &
    {At $\alpha_s \leq 0.3$ $ap = 1.9, 1.4, 1.0,   0.6$.}\\
\\[-0.1cm] \hline\\[-1.5ex]
    {Boucaud 00B}   & \cite{Boucaud:2000nd}      & 0         & 
    {$a = 0.1, 0.07, 0.05, 0.03$\,fm} & 
    {At $\alpha_s \leq 0.3$ $ap = 1.9, 1.4, 1.0, 0.6$.}\\
      &&&&&&\\[-0.1cm]

      \hline
      \hline
\end{tabular*}
\caption{Continuum limit for $\alpha_s$ determinations with $N_f=0$.}
\label{tab_Nf=0_continuumlimit_a}
\end{table}

\begin{table}[!htb]
   \footnotesize
   \begin{tabular*}{\textwidth}[l]{l@{\extracolsep{\fill}
   }rlllllllll}
\hline\hline \\[-1.0ex]
      Collaboration & Ref. & $\Nf$ & $a\,\mu$  & Description\\
      & & & &  \\
      \hline
      \hline
      & & & &  \\[-0.1cm]

 {SESAM 99}
   & \cite{Spitz:1999tu}
   & 0                   
   & \parbox[t]{0.25\textwidth}
        {1 lattice spacing with $a = 0.086\,\mbox{fm}$}
   & \parbox[t]{0.4\textwidth}
        {$\Upsilon$ spectrum splitting.} \\
\\[-0.1cm] \hline\\[-1.5ex]
{Becirevic 99A}  & \cite{Becirevic:1999uc} & 0 &
{$a = 0.07, 0.05$\,fm} & {At $\alpha_s \leq 0.3$ $ap = 1.4, 1.0$.} \\
\\[-0.1cm] \hline\\[-1.5ex]
{Becirevic 99B}  & \cite{Becirevic:1999hj} & 0 &
{$a = 0.1, 0.07, 0.03$\,fm} & 
\parbox[t]{ 0.45\textwidth}{Only $a = 0.03$fm used to extract
$\alpha_s$. At $\alpha_s \leq 0.3$, $ap = 0.6 - 1.5$.} \\
\\[-0.1cm] \hline\\[-1.5ex]
ALPHA 98 & \cite{Capitani:1998mq} & 0  & 
\parbox[t]{0.25\textwidth}{ 4 to 6 spacings, $a/L = 1/12 - 1/5$ 
in step-scaling functions (SSF)} &  
\parbox[t]{0.45\textwidth}{1-loop $\cO(a)$ boundary improvement,
linear extrapolation in $a/L$.
\newline $a/L = 1/8 - 1/5$ for $\alpha_s\leq 0.11$ SSF,
\newline $a/L = 1/12 - 1/5$ for $0.12 \leq \alpha_s\leq 0.20$ SSF.
$L_{\rm max}/r_0$ from \cite{Guagnelli:1998ud},
where several lattice spacings were used.} \\
\\[-0.1cm] \hline\\[-1.5ex]
{Boucaud 98A}    & \cite{Boucaud:1998bq} & 0 &
{$a = 0.1, 0.07, 0.05$\,fm} & {At $\alpha_s \leq 0.3$, $ap = 1.9, 1.4, 1.0$}.\\
\\[-0.1cm] \hline\\[-1.5ex]
Boucaud 98B  & \cite{Boucaud:1998xi} & 0 & 
{$a = 0.1, 0.07, 0.05$\,fm} & {At $\alpha_s \leq 0.3$, $ap = 1.9, 1.4, 1.0$}. \\
\\[-0.1cm] \hline\\[-1.5ex]
Alles 96           & \cite{Alles:1996ka}                &  0        & 
{$a \leq 0.1$\,fm} & {At $\alpha_s = 0.35$, $ap = 1.5$}. \\
\\[-0.1cm] \hline\\[-1.5ex]
{Wingate 95}
   & \cite{Wingate:1995fd}
   & 0                   
   & \parbox[t]{0.25\textwidth}
        {1 lattice spacing with $a = 0.11\,\mbox{fm}$}
   & \parbox[t]{0.4\textwidth}
        {Charmonium $1S$-$1P$ splitting.} \\
\\[-0.1cm] \hline\\[-1.5ex]
 {Davies 94} 
   & \cite{Davies:1994ei}  
   & 0                   
   & \parbox[t]{0.25\textwidth}
        {1 lattice spacing with $a = 0.077\,\mbox{fm}$}
   & \parbox[t]{0.4\textwidth}
        {$\Upsilon$ spectrum splitting.} \\
\\[-0.1cm] \hline\\[-1.5ex]
L\"uscher 93 & \cite{Luscher:1993gh} & 0  & 
\parbox[t]{0.25\textwidth}{4 or 5 lattice spacings, $a/L = 1/12 - 1/5$ 
in step-scaling functions} &  
\parbox[t]{0.45\textwidth}{ 1-loop $\cO(a)$ boundary improvement,
linear extrapolation in $a/L$.
$a/L = 1/8 - 1/5$ for $\alpha_s\leq 0.11$ SSF,
$a/L = 1/10 - 1/5$ for $0.11 \leq \alpha_s\leq 0.22$ SSF,
$a/L = 1/12 - 1/6$ for $0.22 \leq \alpha_s\leq 0.28$ SSF,
$a/L = 1/8.5 - 1/4.5$ for continuum extrapolation of 
$L_{\rm max}/ \sqrt{K}$.} \\
%\newline $a/L = 1/8 - 1/5$ for $\alpha_s\leq 0.11$ SSF,
%\newline $a/L = 1/10 - 1/5$ for $0.11 \leq \alpha_s\leq 0.22$ SSF,
%\newline $a/L = 1/12 - 1/6$ for $0.22 \leq \alpha_s\leq 0.28$ SSF,
%\newline $a/L = 1/8.5 - 1/4.5$ for continuum extrapolation of 
%$L_{\rm max}/ \sqrt{K}$.} \\
\\[-0.1cm] \hline\\[-1.5ex]
UKQCD 92     & \cite{Booth:1992bm}    & 0  &
\parbox[t]{0.25\textwidth}{ One lattice spacing
          with  $0.44 \leq 2a/r \leq 1.6$ } &
\parbox[t]{0.4\textwidth}{No continuum limit.} \\ % have taken the range in fig.1  
\\[-0.1cm] \hline\\[-1.5ex]
Bali 92     & \cite{Bali:1992ru}    & 0  &
\parbox[t]{0.25\textwidth}{One lattice spacing
          with $0.4 \leq 2a/r \leq 1.6 $ } &
\parbox[t]{0.4\textwidth}{No continuum limit.} \\ % have taken everywhere the range in the inlet of fig.5a   
%%--from here Roger
\\[-0.1cm] \hline\\[-1.5ex]
{El-Khadra 92}   
   & \cite{ElKhadra:1992vn} 
   & 0       
   & \parbox[t]{0.25\textwidth}
        {3 lattice spacings with $a = 0.17,\ 0.11,\ 0.08\,\mbox{fm}$}
   & \parbox[t]{0.555\textwidth}
        {Charmonium $1S$-$1P$ splitting.} \\
[-0.1cm]
     &&&&&&&&& \\[-0.1cm]
             \hline
      \hline
\end{tabular*}
\caption{Continuum limit for $\alpha_s$ determinations with $N_f=0$ continued.}
\label{tab_Nf=0_continuumlimit_b}
\end{table}

%%%%%%%%

\begin{table}[!htb]
   \vspace{0.0cm}
   \footnotesize
   \begin{tabular*}{\textwidth}[l]{l@{\extracolsep{\fill}}rlllllllll}
\hline\hline \\[-1.0ex]
      Collab. & Ref. & $\Nf$ &  $a\,\mu$ & Description   \\
      &&&&&&& & \\[-0.1cm]
      \hline
      \hline
      &&&&&&&&& \\[-0.1cm]
Karbstein 14 & \cite{Karbstein:2014bsa} & 2  & 
\parbox[t]{0.1\textwidth}{0.32 -- 1.19 \\ 0.63 -- 1.19 }
    & 
\parbox[t]{0.555\textwidth}{at $p=1.5$\,GeV \\
at $p=3$\,GeV, roughly coincides with $\alpha_s=0.3$.}
\\ \\[-0.1cm] \hline\\[-1.5ex] 
ALPHA 12 & \cite{Fritzsch:2012wq} & 2 &  
\parbox[t]{0.25\textwidth}{$a=0.049, 0.066, 0.076$fm from $f_K$} & 
\parbox[t]{0.46\textwidth}{2-loop $\cO(a)$ boundary improvement, 
linear extrapolation of $L_\mathrm{max} f_\mathrm{K}$ in $a^2$.}
 \\ \\[-0.1cm] \hline\\[-1.5ex]
      Sternbeck 12  & \cite{Sternbeck:2012qs}  & 2        &
     {$a=0.073,0.07,0.06$\,fm}  &  
     {At $\alpha_s = 0.23$, $ap = 2.1, 2.0, 1.7$.}
 \\ \\[-0.1cm] \hline\\[-1.5ex]
ETM 11C   & \cite{Jansen:2011vv}    & 2    & 
\parbox[t]{0.25\textwidth}{$0.30 \leq 2a/r \leq 1.0 $ 
\newline $0.67 \leq 2a/r \leq 1.26 $ when $\alpha_s = 0.3$} &
\parbox[t]{0.46\textwidth}{Four lattice spacings; continuum limit
studied with a particular range in $r$; central result
from the smallest lattice spacing, $a = 0.042$fm.}
 \\ \\[-0.1cm] \hline\\[-1.5ex]
 {ETM 10F}       & \cite{Blossier:2010ky}   & 2              &  
\parbox[t]{0.25\textwidth}{
 {$a=0.05,0.07,0.08$~fm. Different lattice spacings are patched together.}}
&  \parbox[t]{0.45\textwidth}{  At $\alpha_s=0.3$, $ap=1.6, 1.3, 1.1$.} 
 \\ \\[-0.1cm] \hline\\[-1.5ex]
     Sternbeck 10  & \cite{Sternbeck:2010xu}  & 2        &  
     $a=0.068,0.076,0.082$~fm
     &  {At $\alpha_s \leq 0.3$, $ap \geq 1.7$.} 
 \\ \\[-0.1cm] \hline\\[-1.5ex]
 {JLQCD 08} & \cite{Shintani:2008ga} & 2                  
& \parbox[t]{0.25\textwidth}{$a=0.12$ fm  from $r_0=0.49$\,fm} 
& \parbox[t]{0.46\textwidth}{{Single lattice spacing, $0.64 < (aQ)^2 < 1.32$.
At $\alpha_s = 0.3$, $ap = 0.81$.}}
 \\ \\[-0.1cm] \hline\\[-1.5ex]
\parbox[t]{0.11\textwidth}{QCDSF-\\ UKQCD 05}
   & \cite{Gockeler:2005rv}
   & 2
   & \parbox[t]{0.25\textwidth}
        {4 lattice spacings with $a = 0.10$ - $0.066\,\mbox{fm}$}
   & \parbox[t]{0.45\textwidth}
        {$r_0$, together with $r_0 = 0.467\,\mbox{fm}$.}
 \\ \\[-0.1cm] \hline\\[-1.5ex]
ALPHA 04 & \cite{DellaMorte:2004bc} & 2  & 
$a/L=1/8,1/6,1/5,1/4$ & 
\parbox[t]{0.46\textwidth}{ 1-loop (at weak coupling) and
2-loop $\cO(a)$ boundary improvement,
linear extrapolation of % 2-loop improved
SSF in $(a/L)^2$}
 \\ \\[-0.1cm] \hline\\[-1.5ex]
ALPHA 01A & \cite{Bode:2001jv} & 2 & 
$a/L = 1/6, 1/5, 1/4$ & 
\parbox[t]{0.46\textwidth}{1-loop (at weak coupling) and
2-loop $\cO(a)$ boundary improvement,
weighted average of % 2-loop improved
SSF with $a/L=1/5,1/6$.}
 \\ \\[-0.1cm] \hline\\[-1.5ex]
%% -- from here Tetsuya
 {Boucaud 01B}    & \cite{Boucaud:2001qz}    & 2         & 
\parbox[t]{0.25\textwidth}{
 {$a = 0.05, 0.07, 0.09$\,fm. Data at different lattice spacings 
 are patched together}} &
  \parbox[t]{0.46\textwidth}{ 
  {At $\alpha_s = 0.3$, $ap = 1.6, 1.3, 0.9$; plain Wilson action
  with ${\cO}(a)$ errors}.}
 \\ \\[-0.1cm] \hline\\[-1.5ex]
 {SESAM 99}
   & \cite{Spitz:1999tu}
   & 2
   & \parbox[t]{0.25\textwidth}
        {1 lattice spacing with $a = 0.079\,\mbox{fm}$}
   & \parbox[t]{0.45\textwidth}
        {$\Upsilon$ spectrum splitting.}
 \\ \\[-0.1cm] \hline\\[-1.5ex]
 {Wingate 95}
   & \cite{Wingate:1995fd}
   & 2
   & \parbox[t]{0.25\textwidth}
        {1 lattice spacing with $a = 0.11\,\mbox{fm}$}
   & \parbox[t]{0.45\textwidth}
        {Charmonium $1S$-$1P$ splitting.}
 \\ \\[-0.1cm] \hline\\[-1.5ex]
 {Aoki 94}
   & \cite{Aoki:1994pc}
   & 2                   
   & \parbox[t]{0.25\textwidth}
        {1 lattice spacing with $a = 0.10\,\mbox{fm}$}
   & \parbox[t]{0.555\textwidth}
        {Charmonium $1P-1S$ splitting}
 \\ \\[-0.1cm] \hline\\[-1.5ex]
 {Davies 94} 
   & \cite{Davies:1994ei}  
   & 2                   
   & \parbox[t]{0.25\textwidth}
        {1 lattice spacing with $a = 0.08\,\mbox{fm}$}
   & \parbox[t]{0.45\textwidth}
        {$\Upsilon$ spectrum splitting.}
    \\  \\[-0.1cm]
      \hline
      \hline
\end{tabular*}
\caption{Continuum limit for $\alpha_s$ determinations with $N_f=2$.}
\label{tab_Nf=2_continuumlimit}
\end{table}

%%%%%%%%%

\begin{table}[!htb]
   \vspace{0.0cm}
   \footnotesize
   \begin{tabular*}{\textwidth}[l]{l@{\extracolsep{\fill}}rlllllllll}
\hline\hline \\[-1.0ex]
      Collab. & Ref. & $\Nf$ & $a\,\mu$ & Description \\
      &&&&&&& & \\[-0.1cm]
      \hline
      \hline
      &&&&&&& \\[-0.1cm]
Bazavov 14  & \cite{Bazavov:2014soa}  & 2+1  & 
$2a/r = 0.52 - 3.2$ & 
\parbox[t]{0.42\textwidth}{5 lattice spacings; 3 used for determination.
                           At $\alpha_{\rm eff}=0.3$, then $0.86 < a\mu=2a/r < 1.3$.
                           $m_\pi L =2.4,2.6,2.2$ at smallest three lattice spacings
                           of $a=0.060,0.049,0.041$fm respectively
% beta= 7.373, 7.596, 7.825
                           \cite{Bazavov:2014pvz}; adequate coverage of topological
                           sectors is not clear.}
\\ \\[-0.1cm] \hline\\[-1.5ex]
Bazavov 12   & \cite{Bazavov:2012ka} & 2+1  & 
$2a/r=0.6 - 2.0 $  &
\parbox[t]{0.42\textwidth}{7 lattice spacings; 4 lattice spacings with 
\newline $1.14 \leq 2a/r \leq 1.5$ when $\alpha_s(1/r) = 0.3$.
\newline $2a/r = 2$ when $\alpha_s(1/r) = 0.23$ (on the finest lattice).} 
 \\ \\[-0.1cm] \hline\\[-1.5ex]
Sternbeck 12& \cite{Sternbeck:2012qs} & {2+1}       
 & {$a=0.07$\,fm}  &  
   {At $\alpha_s = 0.23$, $ap = 2.1$.}
 \\ \\[-0.1cm] \hline\\[-1.5ex]
HPQCD 10    & \cite{McNeile:2010ji}  & 2+1 & 
\parbox[t]{0.25\textwidth}{$a\mu=2 a\bar m_h = 0.61 - 1.75$}& 
\parbox[t]{0.42\textwidth}{5 lattice spacings;
3 lattice spacings with $1.0 \leq a\mu  \leq 1.5$ 
when $\alpha_{R_4}(\mu)  \leq 0.3$;
3 lattice spacings with $1.0 \leq a\mu  \leq 1.5$ 
when $\alpha_{R_6/R_8}(\mu) \leq 0.33$.
}
 \\ \\[-0.1cm] \hline\\[-1.5ex]
  JLQCD 10 & \cite{Shintani:2010ph} & 2+1                  
 &  $a = 0.11$\,fm from $r_0 = 0.49$\,fm  
 &\parbox[t]{0.42\textwidth}{{Single lattice spacing, $0.4 < (aQ)^2 < 1.0$
 for the momentum fit range. At $\alpha_s = 0.3$, $ap = 0.89$.}}
 \\ & \\[-0.1cm] \hline\\[-1.5ex]
 {HPQCD 10}
   & \cite{McNeile:2010ji}
   & 2+1
   &
   & \parbox[t]{0.555\textwidth}
        {Update of $r_1$ and $r_1/a$ in HPQCD 08A.}
 \\ \\[-0.1cm] \hline\\[-1.5ex]
PACS-CS 09A& \cite{Aoki:2009tf} & 2+1 & 
$a/L=1/8,1/6,1/4$ & 
\parbox[t]{0.42\textwidth}{Tree-level
$\cO(a)$ boundary improvement, which has been seen to behave better
than 1-loop in simulations \cite{Takeda:2004xha}; 
weighted average of $a/L=1/8,1/6$ for step-scaling function
which agrees with a
linear extrapolation in $a/L$ of all data points of the % 2-loop improved
SSF. Linear extrapolation  in $a/L$ of $L_\mathrm{max}m_\rho$ with 
$a/L_\mathrm{max} = 1/8, 1/6, 1/4$.}
 \\ \\[-0.1cm] \hline\\[-1.5ex]
{HPQCD 08B}   & \cite{Allison:2008xk}  & 2+1 &  
\parbox[t]{0.25\textwidth}{ $a\mu=2 a\bar m_h$ = 0.8,\, 1.2,\, 1.7,\, 2.1} & 
\parbox[t]{0.42\textwidth}{4 lattice spacings with heavy quark mass
approximately the charm mass, where $\alpha_{R_4}(\mu) = 0.38$.}
 \\ \\[-0.1cm] \hline\\[-1.5ex]
 {HPQCD 08A}
   & \cite{Davies:2008sw}
   & 2+1
   & \parbox[t]{0.25\textwidth}
        {6 lattice spacings with $a = 0.18$ - $0.045\,\mbox{fm}$}
   & \parbox[t]{0.555\textwidth}
        {$r_1$ using $\Upsilon$ spectrum splitting.}
 \\ \\[-0.1cm] \hline\\[-1.5ex]
 {Maltman 08}
   & \cite{Maltman:2008bx}
   & 2+1
   & \parbox[t]{0.25\textwidth}
        {5 lattice spacings with $a = 0.18$ - $0.06\,\mbox{fm}$}
   & \parbox[t]{0.42\textwidth}
        {Re-analysis of HPQCD 05A with additional lattice
         spacings $a = 0.06$, $0.15\,\mbox{fm}$.}
 \\ \\[-0.1cm] \hline\\[-1.5ex]
 {HPQCD 05A}
   & \cite{Mason:2005zx}
   & 2+1
   & \parbox[t]{0.25\textwidth}
        {3 lattice spacings with $a = 0.18$ - $0.09\,\mbox{fm}$}
   & \parbox[t]{0.555\textwidth}
        {$r_1$ using $\Upsilon$ spectrum splitting.}
\\ 
        &&&&&&&&& \\[-0.1cm]
   \hline
      \hline
\end{tabular*}
\caption{Continuum limit for $\alpha_s$ determinations with $N_f=3$.}
\label{tab_Nf=3_continuumlimit}
\end{table}

%%%%%%%%%%%%%%%

\begin{table}[!htb]
   \vspace{3.0cm}
   \footnotesize
   \begin{tabular*}{\textwidth}[l]{l@{\extracolsep{\fill}}rlllllllll}
\hline\hline \\[-1.0ex]
      Collab. & Ref. & $\Nf$ & $a\,\mu$ & Description \\
      &&&& \\[-0.1cm]
      \hline
      \hline
      &&&& \\[-0.1cm]
HPQCD 14A   &  \cite{Chakraborty:2014aca} & 2+1+1 & 
\parbox[t]{0.25\textwidth}{$a\mu=2 a\bar m_h = 0.78 - 2.09$}& 
\parbox[t]{0.42\textwidth}{4 lattice spacings;
             2 lattice spacings with $ a\mu  \leq 1.5$ 
             and one more lattice spacing with $ a\mu  \lsim 1.6$ 
             when $\alpha_{R_4}(\mu) \leq 0.3$.}
 \\ \\[-0.1cm] \hline\\[-1.5ex]
ETM 13D         &  \cite{Blossier:2013ioa} & 2+1+1 &    
  $a=0.060, 0.068$ fm from $f_\pi$          &For $\alpha_s \leq 0.3$, $ap=1.5,1.7$.
	\parbox[t]{0.555\textwidth}{Update of \cite{Blossier:2012ef}.}
\\ \\[-0.1cm] \hline\\[-1.5ex]      
{ETM 12C}       & \cite{Blossier:2012ef}   & 2+1+1 &  
      $a=0.061, 0.078$ fm from $f_\pi$ & 
      \parbox[t]{0.42\textwidth}{Global fit with $(ap)^2$ discretization effects.
       For $\alpha_s \leq 0.3$, $ap = 1.5, 2.2$.}       
\\ \\[-0.1cm] \hline\\[-1.5ex]      
{ETM 11D}      & \cite{Blossier:2011tf}   & 2+1+1 & 
      $a = 0.061, 0,078$\,fm & {For $\alpha_s \leq 0.3$, $ap = 1.5, 2.0$.}
\\ \\[-0.1cm] \hline\\[-1.5ex]      
      ALPHA 10A & \cite{Tekin:2010mm} & 4                  & 
      $a/L = 1/4, 1/6, 1/8$ & Constant or global linear fit in $(a/L)^2$.
\\ \\[-0.1cm] \hline\\[-1.5ex]      
      Perez 10 & \cite{PerezRubio:2010ke} & 4 & $a/L=1/4, 1/6, 1/8$ &
      \parbox[t]{0.42\textwidth}{Linear extrapolation in $(a/L)^2$.
      1-loop improvement at the boundary.}
\\  &&&&&&&&& \\[-0.1cm]
   \hline
   \hline
\end{tabular*}
\caption{Continuum limit for $\alpha_s$ determinations with $N_f=4$.}
\label{tab_Nf=4_continuumlimit}
\end{table}

%%%%%%%%%

%%%%%%%%%%%%%%%%%%%%%%%%%%%%%